\documentclass{ws-procs961x669}            
\usepackage{ws-toc,ws-multind}

\usepackage{xspace}
\usepackage{lipsum}
\usepackage{array}
\newcommand{\be}{\begin{eqnarray}}
\newcommand{\ee}{\end{eqnarray}}
\usepackage{amsmath,amssymb}
\usepackage{mathtools}

\usepackage{wrapfig}
\usepackage{epsfig}
\usepackage{tcolorbox}
\usepackage{ulem}

\usepackage{mathrsfs}
\usepackage{amsmath}
\usepackage{amsfonts}
\usepackage{array}
\usepackage{verbatim}
\usepackage{epsfig}
\usepackage{graphicx} 
\usepackage{slashed}
\usepackage{tikz}
\usepackage{dsfont}

\newcommand{\bea}{\begin{eqnarray}}
\newcommand{\eea}{\end{eqnarray}}

\newcommand{\bfk}{\mbox{\boldmath $k$}}

\def\kt{k_\perp}
\def\pt{p_\perp}

\newcommand{\bfP}{\mbox{\boldmath $P$}}

\def\pp{p_\perp}

\newcommand{\la}{\lambda}

\newcommand{\da}{\downarrow}
\def\xb{x_{_{\!B}}}

\def\T{_{_T}}

\def\lsim{\mathrel{\rlap{\lower4pt\hbox{\hskip1pt$\sim$}}\raise1pt\hbox{$<$}}}
\def\gsim{\mathrel{\rlap{\lower4pt\hbox{\hskip1pt$\sim$}}\raise1pt\hbox{$>$}}}
\def\nostrocostruttino#1\over#2{\mathrel{\mathop{\kern 0pt \rlap
{\hbox{$#1$}}} \hbox{\kern-.135em $#2$}}}

\usepackage{ws-toc,ws-multind}

\newcommand*{\FigPath}{./Logo/}

\makeindex{author}                     

\begin{document}


\begin{tcolorbox}[colframe=white]
\begin{minipage}{0.2\textwidth}
\includegraphics[width=1.\textwidth]{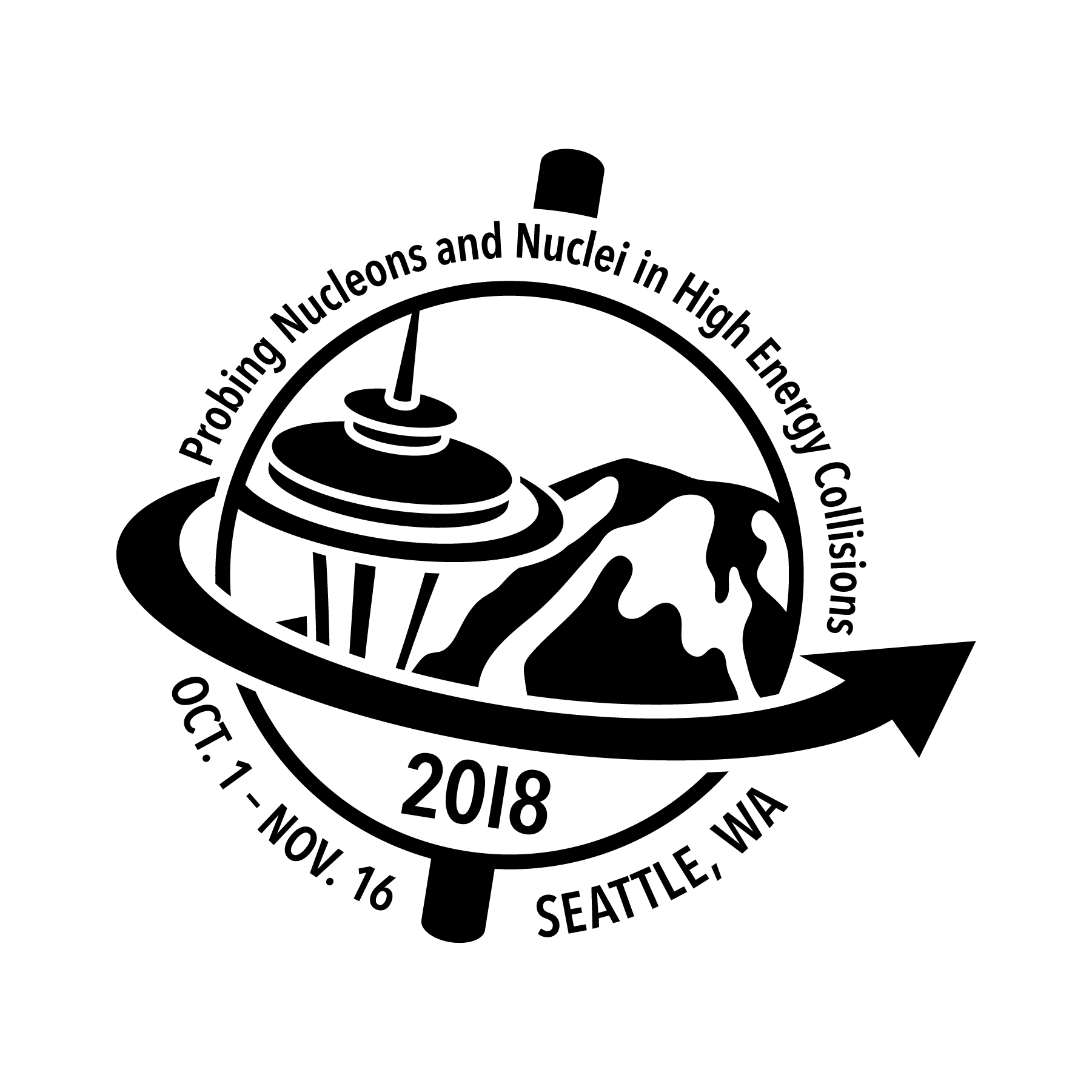}
\end{minipage}
\begin{minipage}{0.7\textwidth}
\Large\bfseries Probing Nucleons and Nuclei\\ in High Energy Collisions
\end{minipage}
\end{tcolorbox}
\vskip -0.5cm 
\hskip -0.2cm \includegraphics[width=1.\textwidth]{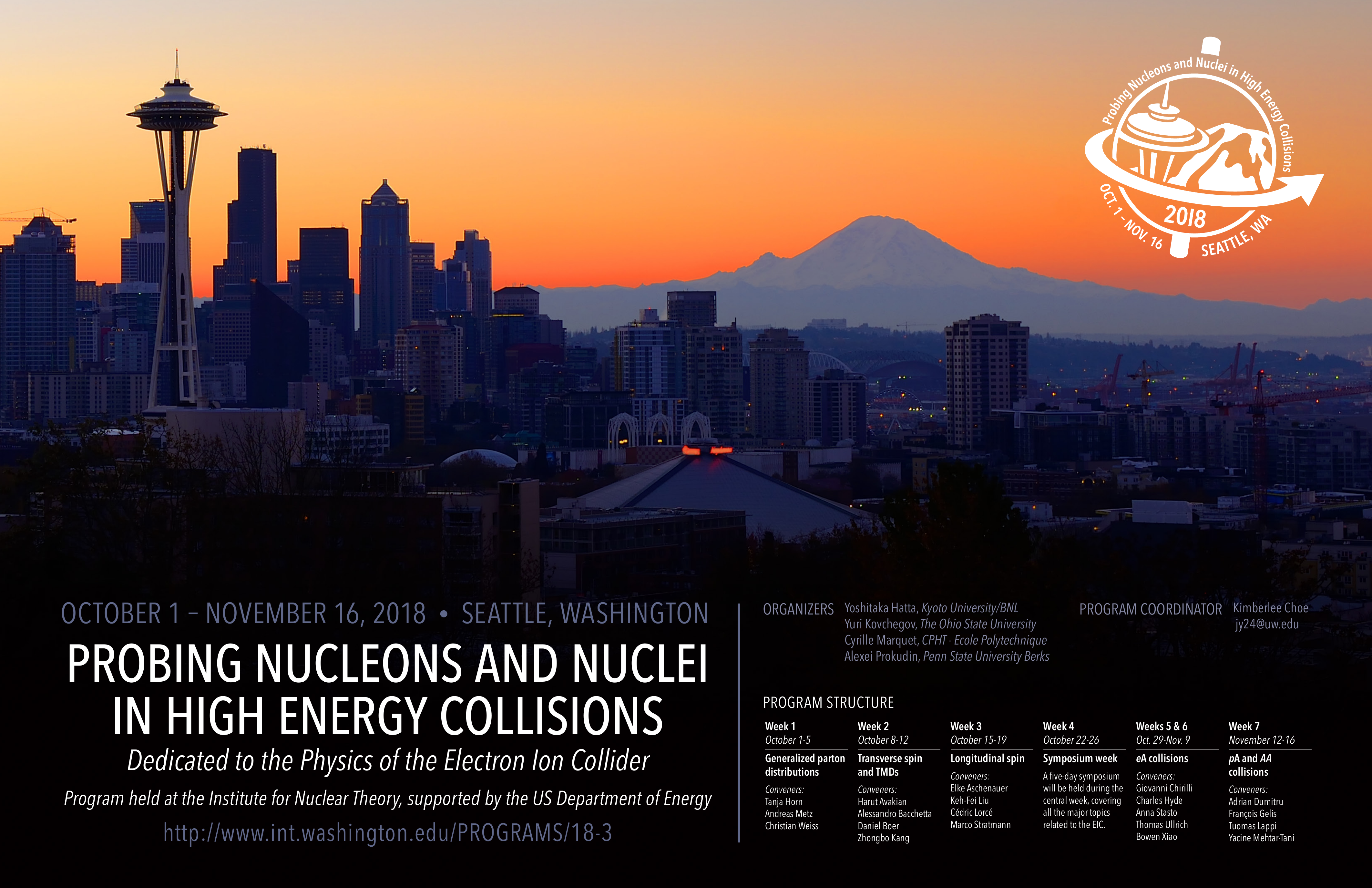}

\author{\bf Editors:}

\author{\bf Yoshitaka Hatta}

\address{Brookhaven National Laboratory, Upton, NY 11973, USA \\
e-mail: yhatta@bnl.gov}

\author{\bf Yuri V. Kovchegov}

\address{Department of Physics, The Ohio State
           University, Columbus, OH 43210, USA \\
e-mail: kovchegov.1@osu.edu}

\author{\bf Cyrille Marquet}

\address{CPHT, CNRS, Ecole Polytechnique, I.P. Paris, 91128 Palaiseau, France \\
e-mail: cyrille.marquet@polytechnique.edu}

\author{\bf Alexei Prokudin}

\address{Division of Science, Penn State University Berks, Reading, Pennsylvania 19610, USA \\
and \\
Theory Center, Jefferson Lab, Newport News, Virginia 23606, USA\\
e-mail: prokudin@jlab.org}


\newpage
 
\wstoc{Preface}{}
\index{author}{Hatta, Y.}
\index{author}{Kovchegov, Y.}
\index{author}{Marquet, C.}
\index{author}{Prokudin, A.}
\title{Preface}

This volume is a collection of proceedings contributions for the 7-week program {\em Probing Nucleons and Nuclei in High Energy Collisions} that was held at the Institute for Nuclear Theory, Seattle, WA, USA, from October 1 until November 16, 2018. The program was dedicated to the 
physics of the Electron Ion Collider (EIC), the world's first polarized electron-nucleon ($ep$) and electron-nucleus ($eA$) collider to be constructed in the USA. The 2015 Nuclear Science Advisory Committee (NSAC) Long Range Plan recommended EIC as the "highest priority for new facility construction". In 2018 the science case for the EIC was strongly endorsed in a report by the US National Academies of Science, Engineering and Medicine. 
The primary goal of the EIC is to establish the precise multi-dimensional imaging of quarks and gluons inside nucleons and nuclei. This includes (i) understanding the spatial and momentum space structure of the nucleon through the studies of TMDs (transverse momentum dependent distributions), GPD (generalized parton distributions) and the Wigner distribution; (ii) determining the partonic origin of the proton spin; (iii) exploring a new quantum chromodynamics (QCD) frontier of ultra-strong gluon fields, with the potential to discover parton saturation --- a new form of gluon matter predicted to be common to all nuclei and nucleons. 

The program brought together both theorists and experimentalists from Jefferson Lab (JLab), Brookhaven National Laboratory (BNL) along with the national and international nuclear physics communities to assess and advance the EIC physics. It summarized the progress in the field since the last INT workshop on EIC in 2010, outlined important new directions for theoretical research in the coming years and proposed new experimental measurements to be performed at the EIC. The topics were organized week-by-week as follows: \\

\noindent
{\bf Spin and Three-Dimensional Structure of the Nucleon:} 

\begin{itemize}
\item[{\bf Week I.}] This week was dedicated to various aspects of GPDs: theory, phenomenology, lattice QCD results, models. Recent progress and efforts towards the EIC were discussed. Particular attention was paid to the D-term and pressure inside the nucleon, transversity GPDs, pion and kaon form factors, PDFs and GPDs, higher twists, GPDs of the light nuclei, and Wigner functions. 
\item[{\bf Week II.}] This week was dedicated to transverse spin structure and the physics of TMDs. New developments in the fields related to the EIC were discussed and summarized. This included improvements on TMD factorization proofs and phenomenological implementations, results of soft collinear effective theories, models, and lattice QCD computations. Sivers and Collins functions received particular attention. We discussed higher-twist effects as well as the evolution of TMDs in $Q^2$ and in energy, including the interplay of small- and large-$x$ physics.
\item[{\bf Week III.}] This week was dedicated to the longitudinal spin structure and the proton PDFs, spin sum rules and the orbital angular momentum (OAM) of the proton. We discussed the experimental and theoretical progress in longitudinal spin physics and possible EIC measurements of large-$x$ PDFs. 
Lattice QCD computations of spin components of the proton were a hot topic, generating intense discussions.   
\end{itemize}

\noindent
{\bf Symposium:}
\begin{itemize}
\item[{\bf Week IV.}] The central symposium took place in week 4. It involved researchers from both the BNL and JLAB, along with the national and international EIC communities, representing all the major topics of the program. \\

\end{itemize}

\noindent
{\bf QCD Interactions inside the Nucleus:} 

\begin{itemize}
\item[{\bf Week V.}]  This week was dedicated to moderate and large-$x$ nuclear structure in $eA$ collisions. The traditional topics of nuclear PDFs, TMDs and GPDs were discussed, along with the new developments in short-range nuclear correlations and in jet physics, as well as in QED radiative corrections.
\item[{\bf Week VI.}] This week was dedicated to small-$x$ physics in $eA$ collisions. Topics included non-linear small-$x$ evolution equations (LO, NLO, NNLO, collinear resummation, etc.), particle production and correlations, diffraction and elastic vector meson production. Discussions were particularly intense when the small-$x$ evolution at NLO took center stage.
 \item[{\bf Week VII.}]  This week was dedicated to studying implications of what we would learn in $eA$ collisions at EIC for our understanding of $pA$ and heavy ion collisions at RHIC and LHC (and {\sl vice versa}). EIC data would help us better constraint initial conditions for heavy ion collisions, both in the event-averaged sense and for the event-by-event fluctuations. Correlation functions measured at EIC may shed light on whether long-range rapidity correlations observed in $pp$, $pA$ and $AA$ collisions at LHC and RHIC are formed in the initial conditions or result from the later-time dynamics. Jet quenching observables in $eA$, $pA$ and $AA$ played a central role. \\

\end{itemize}

\subsection*{Conveners:}

To help us organize and coordinate such an intense and diverse program, we have employed the help of conveners, with a separate group of conveners working on each week. 

\begin{itemize}

\item First week conveners: {Tanja Horn}, Andreas Metz, Christian Weiss
 
\item Second week conveners: Harut Avakian, Alessandro Bacchetta, Daniel Boer, {Zhongbo Kang}

\item Third week conveners: {Elke Aschenauer}, Keh-Fei Liu,  C\'edric Lorc\'e

\item Fifth and sixth  week conveners: {Giovanni Chirilli}, Charles Hyde, {Anna Stasto}, Thomas Ullrich, Bowen Xiao 

\item Seventh week conveners:  Adrian Dumitru, Fran\c{c}ois Gelis, Tuomas Lappi, {Yacine Mehtar-Tani}
\end{itemize}

We (the organizers) are very grateful to all the conveners for their hard work!

\subsection*{Physics Questions Discussed}

The key physics questions addressed by the program were as follows:

\begin{itemize}

\item {\bf How are the sea quarks and gluons, and their spins, distributed in space
and momentum inside the nucleon?}

 GPDs, TMDs and the Wigner distribution allow us to reveal the multi-dimensional nucleon structure in impact parameter and momentum space. The transverse spin polarization of the nucleon can be used as a crucial tool helping us understand non-trivial spin-orbit partonic correlations in the proton.  
Longitudinal spin structure of the nucleon will be definitely explored at the EIC and the measurements at the EIC will allow us to constrain the gluon spin contribution to the spin of the nucleon.

\item {\bf Where does the saturation of gluon densities set in?} 

The large number of partons in a nucleus may result in strong gluon fields leading to the phenomenon of gluon saturation,
also known as the Color Glass Condensate. This universal regime of high-energy QCD
is described by non-linear evolution equations. The program addressed the theoretical and phenomenological progress in our understanding of gluon saturation in $ep$, $eA$, along with proton--nucleus ($pA$) and nucleus--nucleus ($AA$) collisions. 

\item {\bf How does the nuclear environment affect the distribution of quarks and
gluons and their interactions in nuclei?}

Nuclear PDFs, TMDs, and GPDs are interesting and important beyond small-$x$: the large-$x$ structure of nuclei reflects non-trivial non-perturbative QCD dynamics in a cold nuclear matter environment, possibly providing essential information for our understanding of confinement. Cold nuclear matter can serve as a testing ground for the energy loss calculations describing propagation of an energetic quark or gluon in quark-gluon plasma (QGP) created in heavy ion collisions.

\end{itemize}

~\\
Materials of the program and videos of presentations can be found at this URL: {\rm  http://www.int.washington.edu/talks/WorkShops/int\_18\_3/}
~\\
~\\
We thank the Institute for Nuclear Theory at the University of Washington for its kind hospitality and stimulating research environment. We are particularly indebted to Larry McLerran (the INT director), Kimberlee Choe, Cheryl McDaniel, and Linda Vilett for helping us make this program a success. This program was supported in part by the INT's U.S. Department of Energy grant No. DE-FG02- 00ER41132. We are also grateful to the managements of BNL and JLab for the financial support. Many thanks go to Tiffany Bowman from BNL for her tremendous help with the program's poster and logo, which are also displayed in these proceedings.
~\\

We hope these proceedings will be useful to readers as a compilation of EIC-related science at the end of the second decade of the XXI century. We anticipate that significant further progress on the above physics topics will be achieved in the next decades, particularly when EIC is built and starts producing data. 

~\\

\begin{flushright}
Yoshitaka Hatta \\
Yuri Kovchegov \\
Cyrille Marquet \\
Alexei Prokudin \\~\\
(the program organizers)\\~\\
 {\it January 1, 2020} \\

\end{flushright}

\newpage 
\author{\bf List of Conveners}

\author{Elke Aschenauer}
\address{Brookhaven National Laboratory, Upton, NY 11973, USA}

\author{Harut Avakian}
\address{Jefferson Lab, Newport News, Virginia 23606, USA}

\author{Alessandro~Bacchetta}
\address{Dipartimento di Fisica, Universit\`a degli Studi di Pavia, I-27100 Pavia, Italy \\
and Istituto Nazionale di Fisica Nucleare, Sezione di
  Pavia}
  
\author{Daniel Boer}
\address{Van Swinderen Institute for Particle Physics and Gravity, University of
Groningen, Nijenborgh 4, 9747 AG Groningen, The Netherlands} 

\author{Giovanni~A.~Chirilli}
\address{Institut f\"ur Theoretische Physik, Universit\"at Regensburg,\\
	D-93040 Regensburg, Germany} 
	
\author{Adrian~Dumitru}
\address{Department of Natural Sciences, Baruch College, CUNY, \\
17 Lexington Avenue, New York, NY 10010, USA} 
	
\author{Fran\c{c}ois~Gelis}
\address{Institut de physique th\'eorique, Universit\'e Paris Saclay, \\
CNRS, CEA, F-91191 Gif-sur-Yvette, France} 

\author{Tanja~Horn}
\address{Jefferson Lab, Newport News, Virginia 23606, USA}
\address{Catholic University of America, Washington, DC 20064}

\author{Charles Hyde}
\address{Old Dominion University, Norfolk, VA, USA}

\author{Zhong-Bo~Kang}
\address{Department of Physics and Astronomy, University of California, Los Angeles, CA 90095, USA \\
and Mani L. Bhaumik Institute for Theoretical Physics, University of California, Los Angeles, CA 90095, USA}

\author{Tuomas~Lappi}
\address{Department  of  Physics,  University  of  Jyv\"askyl\"a, P.O.
 Box 35, 40014  University  of  Jyv\"askyl\"a,  Finland}
\address{
Helsinki Institute of Physics, P.O. Box 64, 00014 University of Helsinki,
Finland
}

\author{Keh-Fei Liu}
\address{Department of Physics and Astronomy, University of Kentucky, Lexington, KY 40506, USA}

\author{C\'edric~Lorc\'e}
\address{CPHT, CNRS, \'Ecole polytechnique, I.P. Paris, 91128 Palaiseau, France}

\author{Yacine~Mehtar-Tani}
\address{Brookhaven National Laboratory, Upton, NY 11973, USA}

\author{Andreas~Metz}
\address{Department of Physics, Temple University, Philadelphia, PA 19122, USA}

\author{Anna~M.~Stasto}
\address{Department of Physics, The Pennsylvania State University, University Park, PA 16802, USA}

\author{Thomas~Ullrich}
\address{Brookhaven National Laboratory, Upton, NY 11973, USA}

\author{Christian~Weiss}
\address{Jefferson Lab, Newport News, Virginia 23606, USA}

\author{Bo-Wen~Xiao}
\address{Key Laboratory of Quark and Lepton Physics (MOE) and Institute
of Particle Physics,\\ Central China Normal University, Wuhan 430079, China}


\newpage 
\author{\bf List of Authors}

\author{Christine~A.~Aidala$^a$, Elke~Aschenauer$^y$, Fatma~Aslan$^{b,c}$, Alessandro~Bacchetta$^{d}$,
Ian~Balitsky$^{e,f}$, Sanjin~Beni\' c$^g$, Shohini~Bhattacharya$^h$, Mariaelena~Boglione$^i$, Matthias~Burkardt$^b$, Justin~Cammarota$^{j,k}$, Giovanni~A.~Chirilli$^l$, Christopher~Cocuzza$^h$ , Aurore~Courtoy$^m$, Daniel~de~Florian$^n$, Pasquale~Di~Nezza$^o$, Adrian~Dumitru$^p$, Sara~Fucini$^q$, Kenji~Fukushima$^r$, Yulia~Furletova$^{f}$, Leonard~Gamberg$^s$, Oscar~Garcia-Montero$^t$,
Fran\c{c}ois~Gelis$^u$, Vadim~Guzey$^{v,w}$, Yoshitaka~Hatta$^{y}$, Francesco~Hautmann$^z$, Timothy~J. Hobbs$^{f,aa}$, Tanja~Horn$^{f,ba}$, Edmond~Iancu$^{ca}$, Sylvester~Joosten$^{da}$,
Zhong-Bo~Kang$^{ea}$, Raj~Kishore$^{fa}$, Yuri~V.~Kovchegov$^{ga}$, Peter~Kroll$^{ha}$, Kre\v{s}imir~Kumeri\v{c}ki$^{ia,ja}$, Krzysztof~Kutak$^{ka}$, Tuomas~Lappi$^{w,la}$, Huey-Wen~Lin$^{ma}$, Xiaohui~Liu$^{na}$, Simonetta~Liuti$^{oa,o}$, C\'edric~Lorc\'e$^{pa}$, Heikki~M\"antysaari$^{w,la}$, Cyrille~Marquet $^{pa}$, Yiannis~Makris$^{qa}$, Kiminad~A.~Mamo$^{ra}$, Yacine~Mehtar-Tani$^y$, Andreas~Metz$^h$, Zein-Eddine~Meziani$^{da}$ , Gerald~A.~Miller$^{sa}$, Joshua~Miller$^j$,Asmita~Mukherjee$^{fa}$, Pavel~M. Nadolsky$^{aa}$, 
Fredrick~I.~Olness$^{aa}$, Barbara~Pasquini$^{d}$, Bernard~Pire$^{pa}$, Cristian~Pisano$^{ta}$,
Daniel~Pitonyak$^j$, Maxim~V.~Polyakov$^{ua, v}$, Alexei~Prokudin$^{f,s}$, Jian-Wei~Qiu$^{f}$, Marco~Radici$^{d}$, Abha~Rajan$^y$, Sangem~Rajesh$^{ta}$, Matteo~Rinaldi$^q$, Kaushik~Roy$^{y,ra}$, Christophe~Royon$^{va}$, Nobuo~Sato$^{f}$, Marc~Schlegel$^b$, Gunar~Schnell$^{wa}$, Peter~Schweitzer$^{xa}$, Sergio~Scopetta$^q$, Ralf~Seidl$^{ya}$, Kirill~Semenov-Tian-Shansky$^v$ , Andrea~Signori$^{d,f}$, Daria~Sokhan$^{za}$, Anna~M.~Stasto$^{ab}$, Lech~Szymanowski$^{bb}$, Andrey~Tarasov$^{y,ga}$, Dionysios~Triantafyllopoulos$^{cb}$,
Thomas~Ullrich$^{y}$, Raju~Venugopalan$^{y}$, Ivan~Vitev$^{qa}$, Werner~Vogelsang$^{db}$, 
Anselm~Vossen$^{eb, f}$, Bo-Ting~Wang$^{aa}$, Samuel~Wallon$^{fb}$, Kazuhiro~Watanabe$^{f}$,Christian~Weiss$^{f}$, Bo-Wen~Xiao$^{gb}$,  Ho-Ung~Yee$^{hb}$, Yong~Zhao$^{ib}$
}

\address{$^a$ Physics Department, University of Michigan,\\
Ann Arbor, Michigan 48109, USA}

\address{$^b$ Physics Department, New Mexico State University,\\
Las Cruces, NM 88003, U.S.A.
}

\address{$^c$ Department of Physics, University of Connecticut, Storrs, CT 06269, U.S.A.}

\address{$^d$ Dipartimento di Fisica, Universit\`a degli Studi di Pavia, I-27100 Pavia, Italy \\
and Istituto Nazionale di Fisica Nucleare, Sezione di
  Pavia}
  
\address{$^e$ Department of Physics, Old Dominion Univ.,
Newport News, VA23529}

\address{$^f$ Jefferson Lab, Newport News, Virginia 23606, USA}  

\address{$^g$ Yukawa Institute for Theoretical Physics, Kyoto University,\\
Kyoto 606-8502, Japan}

\address{$^h$ Department of Physics, Temple University, Philadelphia, PA 19122, USA}

\address{$^i$ Dipartimento di Fisica, Universit\`a di Torino, INFN-Sezione Torino\\
  Via P. Giuria 1, 10125 Torino, Italy}
  
\address{$^j$ Department of Physics, Lebanon Valley College,\\
Annville, PA 17003, USA}

\address{$^k$ Department of Physics, College of William and Mary,\\
Williamsburg, Virginia 23187, USA}
  
\address{$^l$ Institut f\"ur Theoretische Physik, Universit\"at Regensburg,\\
	D-93040 Regensburg, Germany}  
	
\address{$^m$ Instituto de F\'isica, Universidad Nacional Aut\'onoma de M\'exico\\
Apartado Postal 20-364, 01000 Ciudad de M\'exico, Mexico}
	
 \address{$^n$ International Center for Advanced Studies (ICAS), ECyT-UNSAM,
Campus Miguelete, \\ 25 de Mayo y Francia, (1650) Buenos Aires, Argentina}
 
\address{$^o$ INFN Laboratori Nazionali di Frascati, Frascati (Rome), Italy} 

\address{$^p$ Department of Natural Sciences, Baruch College, CUNY, \\
17 Lexington Avenue, New York, NY 10010, USA}

\address{$^q$ Dipartimento di Fisica e Geologia, University of Perugia and INFN, Perugia Section\\
Perugia, I-06123, Italy}

\address{$^r$ Department of Physics, The University of Tokyo,
7-3-1 Hongo, Bunkyo-ku,\\
Tokyo 113-0033, Japan\\
Institute for Physics of Intelligence (IPI), The University of Tokyo,
7-3-1 Hongo, Bunkyo-ku,\\ 
Tokyo 113-0033, Japan}

\address{$^s$ Division of Science, Penn State University Berks,
Reading, Pennsylvania 19610, USA}

\address{$^t$ Institut f\" {u}r Theoretische Physik, Universit\" {a}t Heidelberg, Philosophenweg 16,\\ 
69120 Heidelberg, Germany}

\address{$^u$  Institut de physique th\'eorique, Universit\'e Paris Saclay, \\
CNRS, CEA, F-91191 Gif-sur-Yvette, France}

\address{$^v$  National Research Center ``Kurchatov Institute'', Petersburg Nuclear Physics Institute (PNPI), 
Gatchina, 188300, Russia}

\address{$^w$  Department  of  Physics,  University  of  Jyv\"askyl\"a, P.O.
 Box 35, 40014  University  of  Jyv\"askyl\"a,  Finland}
 
 \address{$^y$ Brookhaven National Laboratory, Upton, NY 11973, USA}

 \address{$^z$ University of Oxford / University of Antwerp}
 
 \address{$^{aa}$ Department of Physics, Southern Methodist University,\\
 Dallas, TX 75275-0175, U.S.A. }

 \address{$^{ba}$ Catholic University of America, Washington, DC 20064}

\address{$^{ca}$ Institut de Physique Th\`eorique, Universit\'e Paris-Saclay, CNRS, CEA, F-91191 Gif-sur-Yvette, France}

\address{$^{da}$ Argonne National Laboratory,\\
Argonne, IL 60439, USA}

\address{$^{ea}$ Department of Physics and Astronomy, University of California, Los Angeles, CA 90095, USA \\
and Mani L. Bhaumik Institute for Theoretical Physics, University of California, Los Angeles, CA 90095, USA}

\address{$^{fa}$ Department of Physics, Indian Institute of Technology Bombay,
Powai, Mumbai 4000076, India}

\address{$^{ga}$ Department of Physics, The Ohio State
           University, Columbus, OH 43210, USA}
           
\address{$^{ha}$ Fachbereich Physik, Universitaet Wuppertal,\\
D-42097 Wuppertal, Germany}
 
\address{$^{ia}$
    Department of Physics, Faculty of Science, University of Zagreb, 10000 Zagreb, Croatia}

\address{$^{ja}$    
    Institut f\"{u}r Theoretische Physik, Universit\"{a}t Regensburg, D-93040 Regensburg, Germany}

\address{$^{ka}$ Instytut Fizyki Jadrowej Polskiej Akademii Nauk,\\
Krakow, 31-342 Krakow, Poland}

\address{$^{la}$ 
Helsinki Institute of Physics, P.O. Box 64, 00014 University of Helsinki,
Finland
}   

\address{$^{ma}$ Department of Physics and Astronomy, and Computational Mathematics, Science \& Engineering, Michigan State University}

\address{$^{na}$ Center of Advanced Quantum Studies, Department of Physics, Beijing Normal University, \\
Beijing, 100875, China}

\address{$^{oa}$ Physics Department, University of Virginia,\\  382 McCormick Road, VA 22903, USA}    

\address{$^{pa}$ CPHT, CNRS, \'Ecole polytechnique, I.P. Paris, 91128 Palaiseau, France}    

\address{ $^{qa}$ 
Los Alamos National Laboratory, \\
Theoretical Division, MS B283, \\
Los Alamos, NM 87545, USA}

\address{$^{ra}$ Department of Physics and Astronomy, Stony Brook University,\\
Stony Brook, New York 11794-3800, USA}

\address{$^{sa}$ Physics  Department, University of Washington,\\
Seattle, WA 98195-1560, USA}

\address{$^{ta}$ Dipartimento di Fisica, Universit\`a di Cagliari, and INFN, Sezione di Cagliari\\
 Cittadella Universitaria, I-09042 Monserrato (CA), Italy}

 \address{$^{ta}$ 
              Institut f\"ur Theoretische Physik II, 
	      Ruhr-Universit\"at Bochum, D-44780 Bochum, Germany}

\address{$^{va}$ The University of Kansas, Lawrence, USA}

\address{$^{wa}$ Department of Theoretical Physics, University of the Basque Country UPV/EHU,\\
48080 Bilbao, Spain; \\
 IKERBASQUE, Basque Foundation for Science,\\
 48013 Bilbao, Spain}
 
\address{$^{xa}$ Department of Physics, University of Connecticut, 
		Storrs, CT 06269, USA}

\address{$^{ya}$ RIKEN\\
Wako-shi, Saitama-ken 351-0198, Japan}

\address{$^{za}$ School of Physics \& Astronomy, University of Glasgow,\\
Glasgow G12 8QQ, UK}

\address{$^{ab}$ Department of Physics, The Pennsylvania State University, University Park, PA 16802, USA}

\address{$^{bb}$ NCBJ, 02-093 Warsaw, Poland}

\address{$^{cb}$ European Centre for Theoretical Studies in Nuclear Physics and Related Areas (ECT*)
and Fondazione Bruno Kessler, Strada delle Tabarelle 286, I-38123 Villazzano (TN), Italy}

\address{$^{db}$ Institute for Theoretical Physics, T\"ubingen University, \\
Auf der Morgenstelle 14, 72076 T\"ubingen, Germany}	

\address{$^{eb}$ Department of Physics, Duke University
}

\address{$^{fb}$ LPT, CNRS, Univ. Paris-Sud, Universit\'e Paris-Saclay, 91405, Orsay, France {\em \&} \\
Sorbonne Universit\'e, Facult\'e de Physique, 4 place Jussieu, 75252 Paris Cedex 05, France}
      
\address{$^{gb}$ Key Laboratory of Quark and Lepton Physics (MOE) and Institute
of Particle Physics,\\ Central China Normal University, Wuhan 430079, China}

\address{$^{hb}$ Department of Physics, University of Illinois,
Chicago, Illinois 60607, USA}

\address{$^{ib}$ Center for Theoretical Physics, Massachusetts Institute of Technology,\\
Cambridge, Massachusetts 02139, USA}
      


\mastertoc

\newpage
\renewcommand*{\FigPath}{./Logo/} 

\begin{tcolorbox}[colframe=white]
\begin{minipage}{0.2\textwidth}
\includegraphics[width=1.\textwidth]{\FigPath/INT_Workshop_Logo_Final_Black.png}
\end{minipage}
\begin{minipage}{0.7\textwidth}
\wstoc{\bf Week I}{}
\title{Week I}
\end{minipage}
\end{tcolorbox}

\wstoc{Exploring hadron structure with GPDs at EIC: \\
New topics in theory, experiment, interpretation}{Tanja Horn, Andreas Metz, Christian Weiss}
\title{Exploring hadron structure with GPDs at EIC: \\
New topics in theory, experiment, interpretation}
%
%

\author{Tanja Horn$^a$, Andreas Metz$^b$, Christian Weiss$^c$}
\index{author}{Horn, T.}
\index{author}{Metz, A.}
\index{author}{Weiss, C.}

\address{$^a$Catholic University of America, Washington, D.C. 20064, USA \\
$^b$Department of Physics, SERC, Temple University, Philadelphia, PA 19122, USA \\
$^c$Theory Center, Jefferson Lab, Newport News, VA 23606, USA
}
\begin{abstract}
Exploring hadron structure with generalized parton distributions (GPDs) and 
related concepts is an essential part of the physics program of the planned
Electron-Ion Collider (EIC). We discuss new topics in GPD physics
(theory, experiment, interpretation) that have emerged in the last years 
and could be included in the EIC program. This includes: 
(a)~D-term and QCD forces in the nucleon; (b)~Transversity GPDs
in pseudoscalar meson production and high-mass pair production; (c)~Pion and
kaon form factors and GPDs; (d)~Transition distribution amplitudes and
backward meson production; (e)~Neutron GPDs and nuclear modifications;
(f)~Nuclear GPDs and quark/gluon imaging of nuclei;
(g)~Twist-3 GPDs and quark-gluon correlations; (h) Wigner functions and
potential observables. We also comment on future theoretical developments
needed for the GPD program at EIC and emerging connections with other 
areas of nuclear physics. [Conveners' summary, Week 1 of INT Program INT-18-3]
\end{abstract}


\bodymatter


\section{Introduction}
GPDs have emerged as an essential concept in the study of hadron structure in QCD.
They unify the concepts of elastic form factors and parton densities and describe
the spatial distribution of quarks and gluons in hadrons. They also contain 
information on the hadronic matrix elements of certain local QCD operators,
particularly the QCD energy-momentum tensor and the angular momentum of 
quarks and gluons. The experimental determination of the nucleon GPDs synthesizes
results of different types of measurements such as elastic $eN$ scattering (FFs), 
inclusive deep-inelastic $eN$ scattering (PDFs), with the essential ``new'' 
information coming from measurements of hard exclusive processes such as
$e + N \rightarrow e' + M + N$ ($M = \gamma$, meson) or $\gamma + N \rightarrow H + N$ 
($H =$ high-mass dilepton, hadron pair). Measurements of such processes are presently
being performed in dedicated fixed-target experiments at JLab 12 GeV and at CERN COMPASS.

Exploring hadron structure with GPDs is one of the main objectives of
a future EIC. The basic EIC science program as described in the 2011 INT Report \cite{Boer:2011fh}
and the 2012 White Paper \cite{Accardi:2012qut}, and endorsed by the NAS Study, identifies several
measurements related to GPDs, in particular ``quark/gluon imaging'' of the nucleon
with deeply-virtual Compton scattering (DVCS) $e + p \rightarrow e' + \gamma + p$
and exclusive charmonium production $e + p \rightarrow e' + J/\psi + p$.
Beyond these basic measurements, many more possible GPD-related measurements could be 
performed at EIC, realizing the full potential of the GPD concepts and enabling novel
physics studies. Identifying such new measurements, assessing their feasibility,
and integrating them in the EIC science program has become a priority of the EIC community.

The 2018 INT Program ``Probing Nucleons and Nuclei in High Energy Collisions'' 
hosted a 1-week workshop dedicated to the GPD program at EIC. The objectives 
of the meeting were to (a)~identify new GPD-related physics topics that could
be studied at the EIC (concepts, measurements, questions); (b)~outline the theoretical 
developments needed or planned in preparation for the EIC program; (c)~summarize
the connections of GPD physics at EIC with other areas of nuclear physics
(theory, experiment). This article is the Convener's summary of the main results
of the meeting. It reflects what was actually presented and discussed in the week
and is not intended to be a comprehensive summary of all relevant directions in the field. 
For a detailed discussion of the various physics topics we refer to the individual 
contributions in these proceedings and the presentations available online.
\section{New GPD physics topics and EIC measurements}
\subsection{D-Term and QCD forces in the nucleon}
{\it Concepts:} One interesting aspect of GPDs is their connection with the form factors 
of the QCD energy-momentum tensor (EMT), which are of fundamental interest and describe the 
``mechanical properties'' of hadrons in QCD. The second moments ($x$-weighted integrals)
of the unpolarized nucleon GPDs $H$ and $E$ give access to the three nucleon form factors
of the traceless (spin-2) part of the EMT. The first two form factors describe the 
distribution of momentum and angular momentum in the nucleon and have been studied
extensively for many years (Ji sum rule~\cite{Ji:1996nm}, mechanical interpretation, LQCD calculations,
possible model-dependent extraction from exclusive reaction data). The third form factor, 
$D(t)$, related to the D-term in the GPDs~\cite{Polyakov:1999gs}, provides other essential 
information about nucleon structure in QCD and has become an object of great interest 
recently.\cite{Proc:Schweitzer,Polyakov:2018zvc} Its unique properties are: (a)~The ``charge'' $D(0)$ 
can be considered the last unknown global property of nucleon; it is as fundamental as 
the nucleon mass and spin, but contrary to the latter it is not known a priori and
has to be determined experimentally. (b)~The $t$-dependence of $D(t)$ describes the
spatial distribution of forces (pressure) in the nucleon in a 3D interpretation in
the Breit frame.\cite{Polyakov:2002yz}
(c)~The value of $D(t)$ is closely related to dynamical chiral symmetry breaking in QCD, 
the phenomenon which may play an important role for the generation of the nucleon mass.

{\it Measurements:} The D-term can be obtained from the real part of the DVCS amplitude.
It appears as the subtraction constant in the dispersion relation for the DVCS amplitude 
(energy-independent part) and can be determined from the experimental data without modeling or
extraction of the GPDs. This circumstance makes $D(t)$ uniquely accessible and sets it apart 
from the other EMT form factors, which can only be determined through model-dependent GPD
extraction. A first experimental determination of the D-term from JLab data was 
recently reported.\cite{Burkert:2018bqq_1} 
At EIC the D-term could be determined using several types 
of measurements: (a)~DVCS, by extracting the real and imaginary parts of the
DVCS amplitude from the $e + N \rightarrow e' + \gamma + N'$ beam-spin asymmetry
and cross section data, and determining the subtraction constant in the dispersion relation.
In particular, the EIC kinematic coverage would allow one to measure Im(DVCS) at
energies $\nu \sim$ several 10 GeV, fixing the high-energy part of the dispersion
integral and eliminating the associated uncertainty.
(b)~Timelike Compton scattering (TCS), $\gamma + N \rightarrow (\ell^+ \ell^-) + N'$.\cite{Anikin:2017fwu}
The cross-channel process to DVCS is especially sensitive to Re(DVCS). 
In particular, the lepton charge asymmetry is accessible through interchange 
of the leptons in the $\ell^+ \ell^-$ pair.
(c)~Possibly additional timelike processes of (photo)production of a pair of particles 
with large invariant mass

{\it Questions:} (a)~Are there alternative ways to access the D-term at EIC,
e.g.\ through meson production or large-mass pair production, complementing the access
through DVCS/TCS? (b)~Can one give a 2-dimensional light-front interpretation of the
D-term form factor, complementing the 3-dimensional interpretation 
as the pressure distribution in the Breit frame?

\subsection{Transversity or chiral-odd GPDs}
{\it Concepts:} The complete description of the nucleon's partonic structure at the 
twist-2 level involves not only the chiral-even or ``helicity'' GPDs, but also by the
chiral-odd or ``transversity'' GPDs.\cite{Diehl:2001pm} The study of transversity GPDs has become a
subject of great interest and is motivated by the following theoretical properties:
(a)~The transversity GPDs provide new information on nucleon spin structure and spin-orbit
phenomena, such as the deformation of the quark spatial distributions through 
correlations between the position vector (impact parameter) of the quark, the spin of the quark, and (potentially) the spin of the nucleon.
(b)~Comparison between
helicity and transversity GPDs provides insight into dynamical chiral symmetry breaking 
in QCD. (c)~The transversity GPDs are nonsinglets and exhibit only weak scale dependence 
(no mixing with gluons), thus providing clean probes of nonperturbative structure.
(d)~The nucleon tensor charge is calculable in LQCD and of independent interest.
(e)~The transversity GPDs complement the information on the transversity PDFs obtained
from semi-inclusive DIS and polarized proton-proton collisions.

{\it Measurements:}
Transversity GPDs can be probed in hard exclusive processes where the quark-level scattering 
process flips the quark helicity, e.g. through a chiral-odd meson distribution amplitude.
In the last years it was shown that exclusive pseudoscalar meson production ($\pi^0, \eta$) 
at large $x$ ($\gtrsim 0.1$) and moderate $Q^2$ ($\lesssim$ 10 GeV$^2$) proceeds mainly through a 
helicity-flip quark process (involving the twist-3 chiral-odd pion distribution amplitude),
making it possible to use these processes for the study of transversity 
GPDs.\cite{Ahmad:2008hp,Goldstein:2012az,Goloskokov:2011rd,Proc:Kroll} First experimental
determinations of the transversity GPDs have been performed using the JLab 6 GeV $\pi^0$ and
$\eta$ data.\cite{Bedlinskiy:2014tvi} The EIC would allow to perform such measurements
over a much wider kinematic range, extending to $x \sim 10^{-2}$ and $Q^2 \sim$ several 10 GeV$^2$.
The dominance of the chiral-odd mechanism can be (and needs to be) studied with the EIC data themselves, 
using model-independent tests such as the relative size of the longitudinal (L) and transverse (T) amplitudes inferred from the response functions. 
This will definitively answer the question of the reaction mechanism and
permit extraction of the transversity GPDs with high precision.

Another proposed class of processes is the electroproduction of vector meson pairs with large 
invariant mass, $e + N \rightarrow e' + \rho_1 + \rho_2 + N'$\cite{Ivanov:2002jj,Enberg:2006he}
or photon-meson pairs $e + N \rightarrow e' + \gamma + \rho_T + N'$ \cite{Proc:Szymanowski}.
Such measurements would become possible for the first time with the kinematic coverage available 
at EIC. Their experimental feasibility needs to be investigated.

{\it Questions:} (a)~What is the region of dominance of the chiral-odd twist-3 mechanism 
in the T amplitude of pseudoscalar production? What can be inferred theoretically about
the limits of this mechanism? (b)~Can one observe the onset of the asymptotically dominant 
chiral-even twist-2 mechanism in measurements with EIC, e.g.\ by focusing on observables
sensitive to the L amplitudes? 
(c)~How large are subasymptotic corrections to the production mechanism of $\gamma + \rho_T$ pairs? 

\subsection{Pion and kaon form factors, PDFs, GPDs}
{\it Concepts:} While pions and kaons, being Goldstone bosons of (dynamical) chiral symmetry breaking (DCSB) in QCD, occupy a special role in nature, our understanding of their parton structure is very limited in comparison to the nucleon.
However, the EIC can provide very important and unique insights in that field, which go beyond what is presently known from measurements at, for instance, HERA and will also complement results from future experiments at JLab 12.~\cite{Aguilar:2019teb, Proc:Horn}
Specifically, one could aim at measuring PDFs, form factors, and perhaps even GPDs of these mesons.
PDFs of mesons are interesting in their own right.
A comparison between pion and kaon PDFs could also elucidate the role of strangeness in hadron structure.
Moreover, the PDFs of pions and kaons contain very important information about their mass decomposition in QCD.~\cite{}
Form factors depend on the meson distribution amplitudes, which in turn could shed new light on the transition between non-perturbative and perturbative physics and, in particular, on the role played by DCSB.~\cite{Horn:2016rip}
The latter may be critical for understanding the origin of the nucleon mass (see also Sec.~2.1).
Experimental results for PDFs, form factors and GPDs of pions and kaons could also be compared with first-principles calculations in lattice QCD, which represents another key motivation for measuring these quantities.
Further motivations have been reported elsewhere.~\cite{Horn:2016rip, Aguilar:2019teb}

{\it Measurements:} The processes and the corresponding QCD tools for measuring PDFs, FFs and GPDs of hadrons are of course well established.  
It is the lack of a free meson target which represents the major complication in this research field.
But scattering off the nucleon's meson cloud, which in the inclusive case is called Sullivan process, may well allow one to gain information on the parton structure of pions and kaons.
This method was used in the past already.
At EIC the relevant processes would be~\cite{Aguilar:2019teb}:
(a) Inclusive forward hadron production in order to measure meson PDFs.
In the case of the pion one must consider $e + p \to e' + X + n$.
(b) Exclusive meson production for form factor measurements.
A sample process is $e + p \to e' + \pi^+ + n$.
(c) GPDs of mesons could in principle be addressed via more complicated processes such as, for instance, $e + p \to e' + V + \pi^+ + n$  $(V = \rho^0, 
\gamma)$. 

{\it Questions:} (a) A key challenge for all the aforementioned processes is the extraction of the meson-pole contribution which contains the scattering off a free meson one is interested in.
Valuable theoretical studies of this point are available already.~\cite{Qin:2017lcd} 
(b) What is the meson flux factor? In this context, overlap with Drell-Yan measurements such as those proposed for the CERN M2 beam line by the COMPASS++/AMBER Collaboration is expected to provide critical input.
(c) Detailed feasibility studies for EIC (including forward tagging) are needed, but important progress in that regard has already been made.~\cite{Proc:Horn, Aguilar:2019teb}

\subsection{Transition distribution amplitudes}
{\it Concepts:}
Baryon-to-meson transition distribution amplitudes (TDAs) extend the concepts of GPDs and distribution amplitudes.~\cite{Frankfurt:1999fp, Pire:2004ie}
As such they contain new valuable information on the parton structure of hadrons.
TDAs appear in the QCD description of certain exclusive processes in hadron-hadron collisions, and also for backward-angle (small $u$, large $t$) hard exclusive meson production in lepton-nucleon scattering.~\cite{Lansberg:2011aa, Pire:2015kxa}
Related experiments on meson $(\pi, \, \omega)$ electro-production were recently completed at Jefferson Lab~\cite{Park:2017irz, Li:2019xyp}.

{\it Measurements and questions:}
(a) Also at EIC one would look for meson production in the backward direction.~\cite{Szymanowski:2019bnr}
Detailed feasibility studies for this specific kinematics are yet to be performed.
(b) Can one make progress on modeling of TDAs by going beyond the valence quark structure?~\cite{Pasquini:2009ki, Lansberg:2011aa}
(c) Can one further sharpen the physics content of TDAs, perhaps by going to impact parameter space like is often done in the case of GPDs?~\cite{Pire:2019nwa}
(d) Does QCD factorization hold for backward-angle meson production?  In this context it would already be a very important step forward to work out the process at one-loop accuracy.

\subsection{Neutron GPDs and nuclear modifications}
{\it Concepts:} The study of nucleon GPDs requires measurements of hard exclusive
processes on the neutron as well as the proton. Both neutron and proton data are
needed to determine the flavor composition of the GPDs and separate singlet
and nonsinglet GPDs in studies of QCD evolution. The extraction of the 
free neutron GPDs from measurements on light nuclei (D, $^3$He) requires
elimination/correction of nuclear binding effects.~\cite{Proc:Scopetta} 
Detection of the low-energy 
nuclear breakup state (spectator nucleon tagging) controls the nuclear
configuration during the high-energy process and permits a differential
treatment of nuclear binding effects, greatly simplifying the extraction
of free neutron GPDs. The technique also allows one to study nuclear
modifications of the GPDs (EMC effect, shadowing, antishadowing)
in controled nuclear configurations.

{\it Measurements:} Neutron GPDs are measured in quasi-free DVCS on the 
D and $^3$He nuclei in fixed-target experiments at JLab~12.
Spectator tagging with special detectors is planned for inclusive DIS 
on the D (BoNUS) and can be extended to exclusive processes (Marathon),
but is challenging and limited in spectator momentum coverage.
The EIC would enable a comprehensive program of neutron GPD measurements 
and nuclear modification studies, including: 
(a) Measurements of quasi-free DVCS on polarized light nuclei,
$e + A \to e' + \gamma + N' + X(A - 1)$ ($A$ = D, $^3$He);
(b) Measurements of DVCS and meson production on the deuteron
with spectator proton tagging, $e + \textrm{D} \to e' + \gamma + X + p$,
or possibly neutron tagging;
(c) Measurements of DVCS and meson production on $^3$He with
the nuclear breakup measurements, e.g., quasi-two body breakup
$e + ^3\!\textrm{He} \to e' + \gamma + p + \textrm{D}$.
(d) Measurements of the isospin dependence of nuclear modifications
(PDFs/GPDs) using mirror nuclei $^3$He vs $^3$H.

{\it Questions:}
(a) Forward detector requirements for spectator tagging and
nuclear breakup measurements?
(b) Feasibility of neutron spectator tagging (efficiency, resolution)
for bound proton structure measurements?
(c) Magnitude and kinematic dependence of final-state interactions 
for exclusive processes with spectator tagging (theoretical estimates)?
(d) Theoretical calculations of nuclear breakup in high-energy
processes with A$\;\ge\;$3 nuclei (light-front methods, input from
low-energy nuclear structure)?

\subsection{Nuclear GPDs and quark/gluon imaging of nuclei}
{\it Concepts:} Understanding the origin of nuclear binding from QCD is one 
of the prime objectives of nuclear physics. Hard scattering processes on
nuclei allow one to probe the structure of the nucleus directly in terms
of the fundamental QCD degrees of freedom. In particular, coherent exclusive
processes on light nuclei (DVCS, meson productions) can be used to measure 
the nuclear GPDs and construct the transverse distribution of quarks and 
gluons in the the nucleus (``nuclear quark-gluon imaging''). The resulting
distributions can reveal novel spin-orbit effects (deformation of the transverse 
quark/gluon distributions through nuclear spin) and provide new insight into the
dynamics of nuclear shadowing at small $x$ (thickness or impact parameter 
dependence).~\cite{Proc:Guzey} 
Comparison of different transverse distributions (gluons vs. quarks,
spin/flavor dependence) offers direct information on the properties
of the nuclear bound state in QCD.

{\it Measurements:} Exploratory measurements of DVCS on $^4$He are performed 
with JLab 12. The EIC would enable a unique program of quark-gluon
imaging of polarized light nuclei using a variety of processes and targets:
(a) Coherent DVCS on polarized light nuclei, $e + A \to e' + \gamma + A'$, 
using $A$ = D (spin-1), $^3$He (spin-1/2), $^4$He (spin-0) for a comprehensive
study of polarization effects. The $^4$He nucleus as a spin-0 system is
particularly simple and possesses only unpolarized quark and gluon GPDs.
(b) Coherent heavy quarkonium and vector meson production on light nuclei,
$e + A \to e' + V + A'$  $(V = J/\psi, \phi, ...)$, to measure nuclear
gluon GPDs and study the impact parameter dependence of shadowing.

{\it Questions:}
(a) Simulations of coherent nuclear scattering with EIC forward detectors
and quantitative requirements (forward detector resolution, ion beam momentum spread)?
(b) Theoretical calculations of nuclear GPDs using light-front nuclear
structure methods \cite{Proc:Scopetta,Proc:Guzey}.
(c) Hadronic final-state interactions (deviations from color transparency)
in coherent meson production on nuclei.

\subsection{Twist-3 GPDs and quark-gluon correlations}
{\it Concepts:} The study of many-body systems (nuclear, atomic) usually 
proceeds from single-particle densities to multi-particle correlations
(or from 1-body to 2-body operators). Correlations are of special significance
because they provide insight into the microscopic interactions in the system.
In QCD the single-particle densities are the twist-2 PDFs/GPDs, while particle
correlations are encoded in matrix elements of higher-twist operators.
Of particular interest are the twist-3 operators, because (a) their 
renormalization properties are well understood;
(b) their matrix elements can be connected with color forces\cite{Burkardt:2008ps_17_17}
and the QCD vacuum structure (e.g. instantons)\cite{Balla:1997hf};
(c) LQCD calculations of twist-3 matrix elements is possible with non-perturbative 
operator renormalization. The extraction of twist-3 matrix
elements from hard processes is challenging because their contributions to 
observables are usually power-suppressed, and/or because they can only be 
isolated after subtracting kinematic twist-2 contributions (Wandzura-Wilczek approximation).
Experimental twist-3 studies so far have focused on the spin structure function 
$g_2$.\cite{Flay:2016wie,Anthony:2002hy} 
Recently the properties of twist-3 GPDs have come into focus, raising the 
prospect of experimental studies of quark-gluon correlations through 
exclusive processes.

{\it Measurements and questions:} The EIC would greatly expand the possibilities for experimental 
studies of twist-3 quark-gluon correlations. One approach would use inclusive spin 
structure measurements and focus on (a) improving the accuracy of the extraction
of the twist-3 spin structure function $g_2 - g_2^{\textrm{twist-2}}$; in particular,
enabling theoretical analysis of the full $x$-dependence rather than just 
the moment $d_2$; (b) a combined analysis of $g_1$ data (including power corrections)
and $g_2$ data. 
Another approach could focus on studying twist-3 quark-gluon correlations through hard
exclusive processes such as DVCS. The sensitivity of DVCS observables to twist-3 
GPDs and the practical possibility of their extraction need to be investigated.\cite{Aslan:2018zzk}
Also, the interpretation of twist-3 GPDs needs to be developed (zero-mode contributions
in light-front correlations, spatial distributions).

\subsection{Wigner functions in QCD}
{\it Concepts:} Partonic Wigner functions unify the concepts of TMDs and GPDs and encode the most general description of the one-body quark/gluon structure of hadrons in QCD.~\cite{Belitsky:2003nz_29_1, Lorce:2011kd_29, Proc:Pasquini, Talk:Schlegel, Proc:Metz_2}
They may allow for 5D imaging of hadrons, have a direct (and intuitive) connection with quark/gluon orbital angular momentum~\cite{Lorce:2011kd_29, Hatta:2011ku_29, Engelhardt:2017miy_29}, and provide access to spin-orbit correlations that are similar to the ones in atomic systems like the hydrogen atom.
The Fourier transforms of Wigner functions, typically called generalized TMDs~\cite{Meissner:2009ww_29}, appear in the QCD description of certain observables.
Generally, the field of partonic Wigner functions and GTMDs, in which a number of interesting developments have happened over the past few years, may play an important role in the EIC physics program.

{\it Measurements:} At EIC, Wigner functions could in principle be addressed via diffractive exclusive dijet production in both $ep$ and $eA$ scattering.~\cite{Hatta:2016dxp_29, Talk:Yuan}
While this process gives access to Wigner functions for gluons, the only presently known observable for quark Wigner functions requires hadronic collisions.~\cite{Bhattacharya:2017bvs_29}

{\it Questions:} (a) Can one establish a QCD factorization theorem for exclusive processes that are sensitive to transverse parton momenta?  
Adressing this question requires, in particular, a careful analysis of soft gluon radiation.
(b) Can dijet production be measured at EIC with sufficient accuracy?
(c) Can one identify other processes in $ep/eA$ through which Wigner functions could be studied? 
Of special interest in this context are observables that are sensitive to the quark sector.

\section{Theoretical developments for GPD program at EIC}
Several developments in the theory and simulation tools for hard exclusive 
processes are needed in order to support the planned experimental program at EIC.
Here we merely list some of them: 
\\
(a) Develop a DVCS code for amplitudes and cross sections at NLO accuracy,
publicly available, with GPD input in the $x$-representation, 
e.g. within the PARTONS framework\cite{Berthou:2015oaw}.
\\
(b) Perform a quantitative assessment of twist-3 effects in DVCS observables, 
including kinematic and dynamical twist-3 operators (Wandzura-Wilczek, quark-gluon 
correlations), and the possible sensitivity to quark-gluon correlations.
\\
(c) Develop a realistic description of light vector meson production in collinear 
factorization at NLO accuracy, including consistent scale setting and possibly 
$\log(1/x)$ resummation. Develop a method for including finite-size/higher-twist
corrections in meson production, and establish correspondence with LO formulations 
that include finite-size effects in coordinate representation\cite{Frankfurt:1995jw} (finite dipole size)
or momentum representation\cite{Goloskokov:2005sd} (Sudakov form factor).
\\
(d) Explore the applicability of NRQCD methods to exclusive heavy quarkonium 
photo/electroproduction at EIC.\cite{Talk:Qiu} 
Possible points of interest are tests of the 
universality of the NRQCD matrix elements, and better insight into the bound-state
structure and production mechanism through the $Q^2$-dependence.
\section{Connections with other areas}
The study of GPDs and related concepts at EIC has many connections with other fields
of nuclear physics, which could be strengthened and exploited for intellectual gain
or practical uses:

{\it Lattice QCD:} (a) the EMT form factors (second moments of GPDs) are accessible 
as nucleon matrix elements of local operators using standard LQCD methods.
(b) Exploratory calculations of $x$-dependent quasi/pseudo-GPDs are being 
performed.~\cite{Proc:Metz, Bhattacharya:2018zxi, Bhattacharya:2019cme, Alexandrou:2018pbm_17, Lin:2018qky_17, Joo:2019jct}
(c) Precise LQCD calculations of the pion distribution amplitude are available.
(d) Chiral-odd PDFs/GPDs and the tensor charge are comparatively easily calculated 
in LQCD (non-singlets, no mixing with gluons).~\cite{Hasan:2019noy_17, Alexandrou:2019brg}
(e) Twist-3 quark-gluon correlations are computable with non-perturbative renormalization.

{\it Nonrelativistic QCD (NRQCD):} NRQCD matrix elements used in inclusive heavy 
quarkonium production in $pp/pA$ may be used/tested/measured in exclusive production 
in $ep/\gamma p$\cite{Talk:Qiu}

{\it Multiparton interactions in $pp$ collisions (MPI):} (a) The transverse geometry of $pp$ collisions
(impact parameter representation, nucleon GPDs) plays an essential role in the calculation of
MPI rates and the identification of MPI events\cite{Frankfurt:2010ea,Frankfurt:2003td,Talk:Weiss}.
(b) Two-body parton correlations are a natural extension of the concept of GPDs, 
opening the prospect of joint modeling.\cite{Blok:2010ge} 

{\it Ultraperipheral $pA/AA$ collisions (UPCs):} UPCs enable measurements of high-energy $\gamma p/\gamma A$
scattering using the Weizs\"acker-Williams photons created by the field of a heavy nucleus, with
active experimental programs at LHC and RHIC. Hard processes
can be studied through final states with a large mass scale (dijets, heavy quarkonia, etc).
UPCs deliver the highest available energies for electromagnetic processes, $W^2$(UPC at LHC) 
$\sim (10-30) W^2$(HERA) and complement and precede the EIC program. Particular UPC measurements
of interest for EIC are: (a) Measurements of nucleon GPDs with heavy quarkonium photoproduction,
including the search for non-linear effects at $x \lesssim 10^{-5}$; (b) Coherent nuclear processes
as a way to study nuclear shadowing.
%



\newpage 

\renewcommand*{\FigPath}{./WeekI/01_Szymanowski/}

\wstoc{On large mass $\gamma -\gamma$ and $\gamma ~-$ meson photoproduction}{Lech Szymanowski, Bernard Pire, Samuel Wallon}

\title{On large mass $\gamma -\gamma$ and $\gamma ~-$ meson photoproduction }
\author{L. Szymanowski$^*$}
\index{author}{Szymanowski, L.} 

\address{NCBJ, 02-093 Warsaw, Poland\\
$^*$E-mail: Lech.Szymanowski@ncbj.gov.pl }
\author{B. Pire}
\index{author}{Pire, B.}

\address{CPHT, CNRS, \'Ecole polytechnique, I.P. Paris, 91128-Palaiseau, France}
\author{S. Wallon}
\index{author}{Wallon, S.}

\address{LPT, CNRS, Univ. Paris-Sud, Universit\'e Paris-Saclay, 91405, Orsay, France {\em \&} \\
Sorbonne Universit\'e, Facult\'e de Physique, 4 place Jussieu, 75252 Paris Cedex 05, France}
\begin{abstract}
Enlarging the set of hard exclusive reactions to be studied in the framework of QCD collinear
factorization opens new possibilities to access generalized parton distributions (GPDs). We
studied the photoproduction of a large invariant mass photon-photon or photon-meson pair in
the generalized Bjorken regime which may be accessible both at JLab and  at the EIC.
\end{abstract}
\bodymatter
\section{Introduction}\label{aba:sec1_73}
Deeply virtual Compton scattering (DVCS) has proven to be a promising tool to study the three
dimensional arrangement of quarks and gluons in the nucleon \cite{Kumericki:2016ehc_24_8}. The
crossed reaction, the photoproduction of a timelike highly virtual photon which materializes
in a large invariant mass lepton pair (dubbed TCS for timelike Compton scattering) is under
study at JLab. Its amplitude shares many features with the DVCS amplitude
\cite{Berger:2001xd} but with significant and interesting differences \cite{Pire:2011st,
Muller:2012yq, Moutarde:2013qs} due to the analytic behavior in the large scale $Q^2$
measuring the virtuality of the incoming ($q^2 =-Q^2$) or outgoing  ($q^2 =+Q^2$) photon. In
order to enlarge the set of experimental data allowing the extraction of GPDs, we studied the
generalization of TCS to the case of the photoproduction of large invariant mass photon-photon and photon-meson pairs. Although factorization of GPDs from a perturbatively calculable coefficient function has not yet been proven for these processes, they are a natural 
extension of the current picture in the framework of collinear QCD factorization.
\section{$\gamma N \to \gamma \gamma N'$}
\begin{figure}
\includegraphics[width=1.7in]{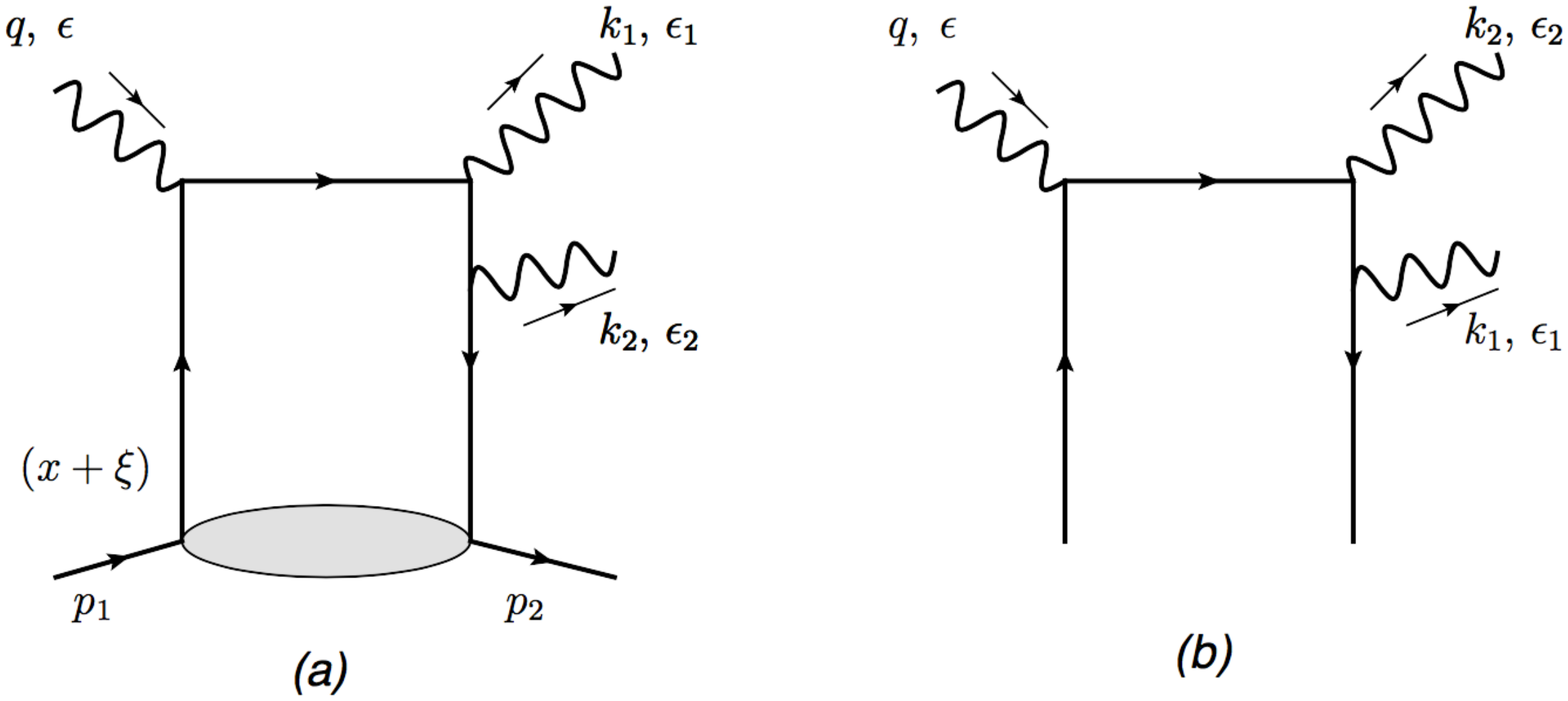}\includegraphics[width=1.7in]{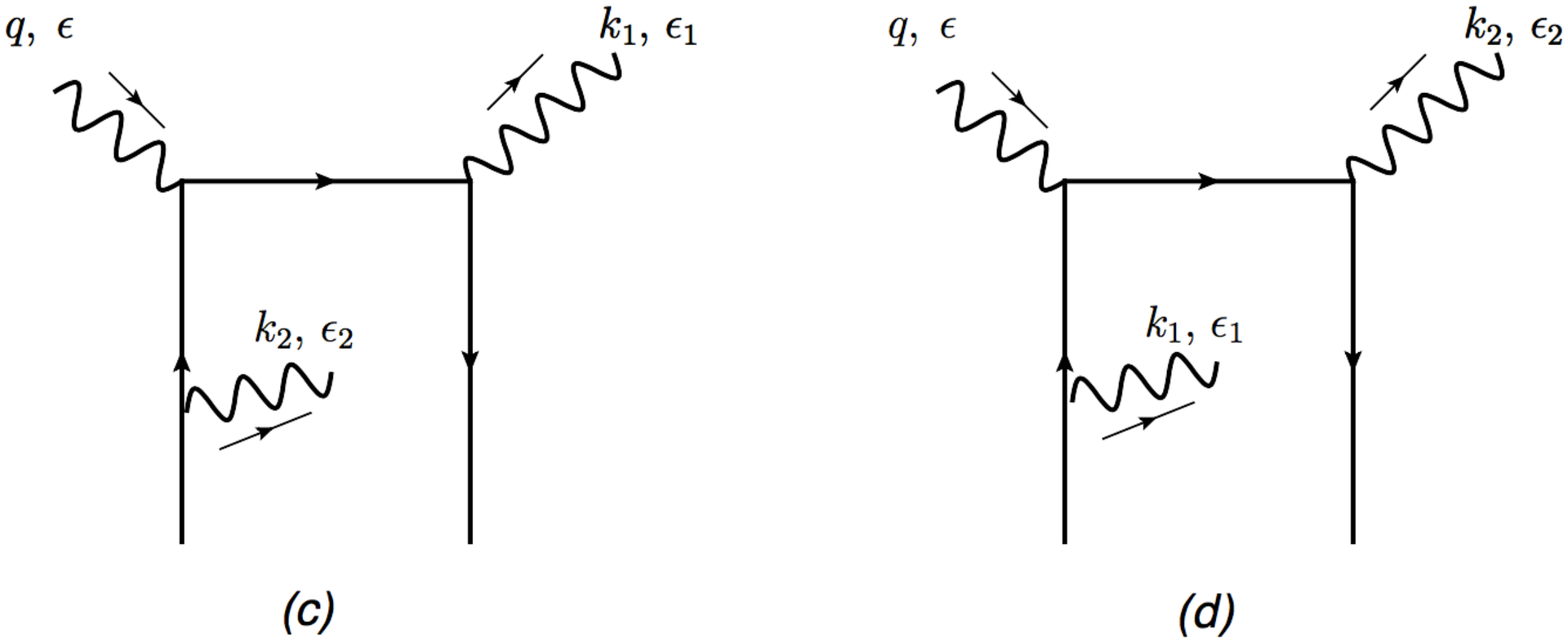}\includegraphics[width=1.7in]{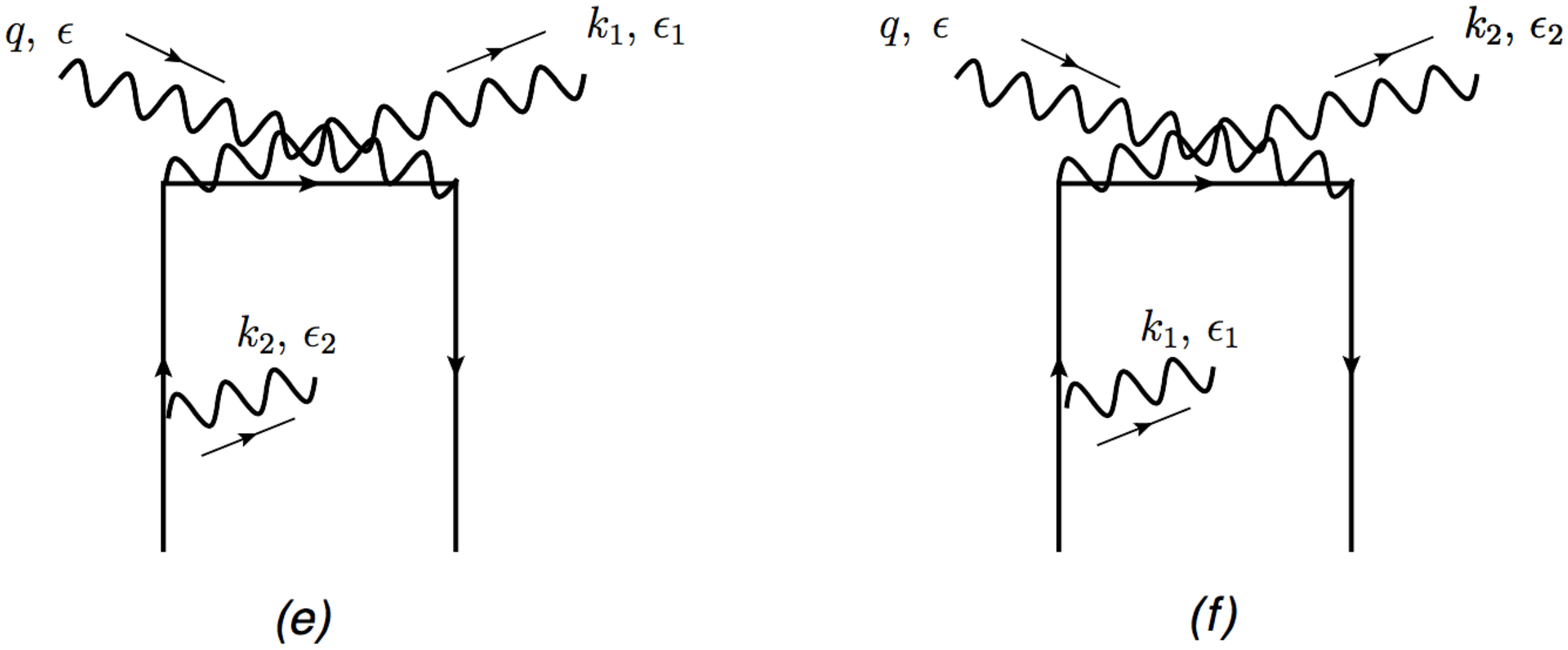}
\caption{Diagrams contributing to the coefficient function for $\gamma\gamma$ production at
the Born level.}
\label{aba:fig1}
\end{figure}
\begin{figure}
\center
\includegraphics[width=2.5in]{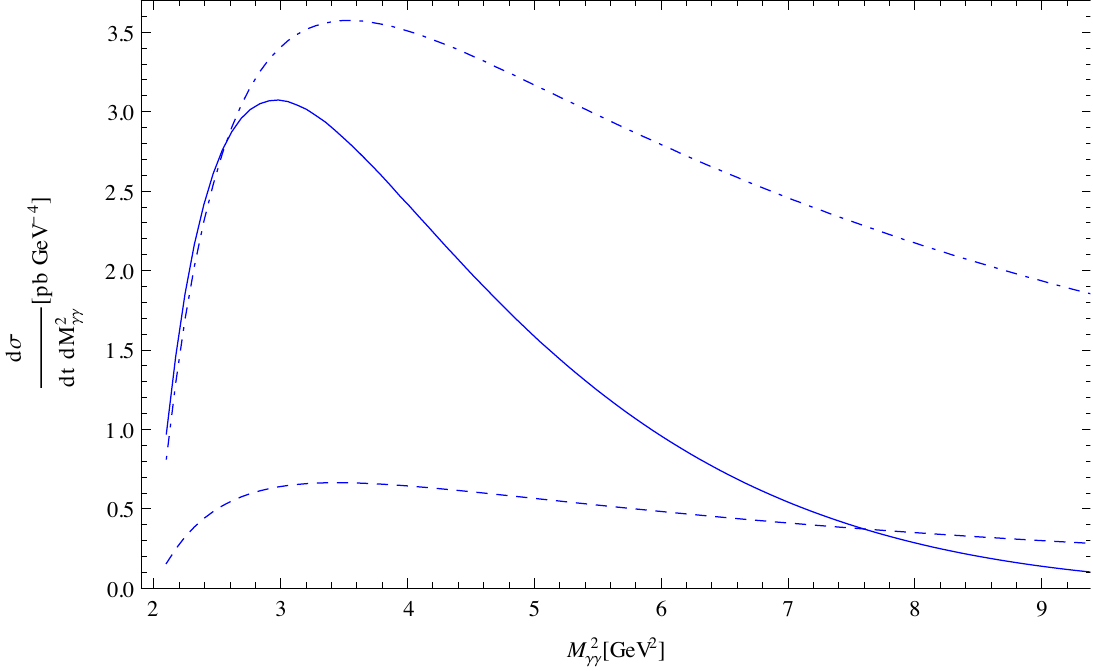}
\caption{$M_{\gamma\gamma}^2$ dependence of the unpolarized differential cross-section for the photoproduction of a diphoton on a proton (left panel) or neutron (right panel) at $t=t_{min}$ and $s_{\gamma N}= 20$ (resp. $100, 10^6$) GeV$^2$ (full, resp. dashed, dash-dotted multiplied by $10^5$).}
\label{gaga}
\end{figure}

The photoproduction of a photon pair \cite{Pedrak:2017cpp}
 shares with DVCS and TCS  the nice feature to be a purely electromagnetic amplitude at
the Born level. Charge parity however selects a complementary set of GPDs, namely the charge
parity - odd GPDs related to the valence part of quark PDFs, with no contribution from the
gluon GPDs. The analytic form of the Born amplitude calculated from the graphs shown on Fig.~\ref{aba:fig1} is very peculiar since the coefficient
function turns out to be proportional to $\delta(x\pm \xi)$ leading through the usual
momentum fraction integration to a scattering  amplitude  proportional to the GPDs taken at
the border values $x=\pm\xi$. For illustration, Fig.~\ref{gaga} displays the  
diphoton invariant squared mass dependence of the unpolarized differential cross section on a proton at $t = t_{min}$ and $s_{\gamma N}= 20$ (resp. $100, 10^6$) GeV$^2$ (full, resp. dashed, dash-dotted multiplied by $10^5$). 

\section{$\gamma N \to \gamma \rho N'$ : the quest for transversity GPDs}
The photoproduction of a $\gamma \rho$ pair \cite{ Boussarie:2016qop}
has the rare feature of being sensitive to chiral-odd transversity quark GPDs at the leading
twist level, because of the leading twist chiral-odd distribution amplitude of the
transversely polarized vector meson. Indeed, except for higher twist amplitudes which suffer
from end-point divergences and heavy meson neutrino production amplitudes
\cite{Pire:2015iza,Pire:2017lfj}  which may be difficult to measure, one needs exclusive
processes with more particules in the final state to access transversity GPDs
\cite{Ivanov:2002jj_1, Enberg:2006he_1, Beiyad:2010cxa, Cosyn:2019eeg}.

\begin{figure}
\includegraphics[width=4.2in]{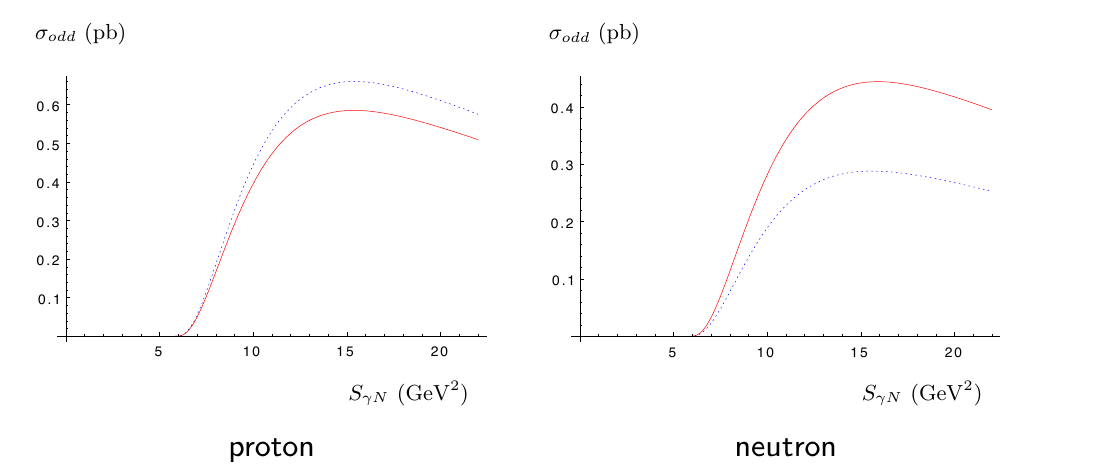} 
\caption{Energy dependence of the integrated cross section for a photon and a   transversely polarized $\rho$ meson production, on a proton (left panel) or neutron (right panel) target. The $\gamma \rho$ pair is required to have an invariant  mass squared larger than $2$ GeV$^2$.
The solid red and dashed blue curves correspond to different parametrization of the transversity GPDs.}
\label{garho}
\end{figure}

We show on Fig.~\ref{garho} the cross section for the production of a transversely polarized $\rho$ in conjunction with a photon, on a proton or a neutron target. The curves show the  sensitivity to the transversity GPD parametrization. Cross sections are sufficiently high for the process to be measurable at JLab \cite{Boussarie:2016qop}.

\section{$\gamma N \to \gamma \pi N'$}

\begin{figure}[h]
\center \includegraphics[width=4in]{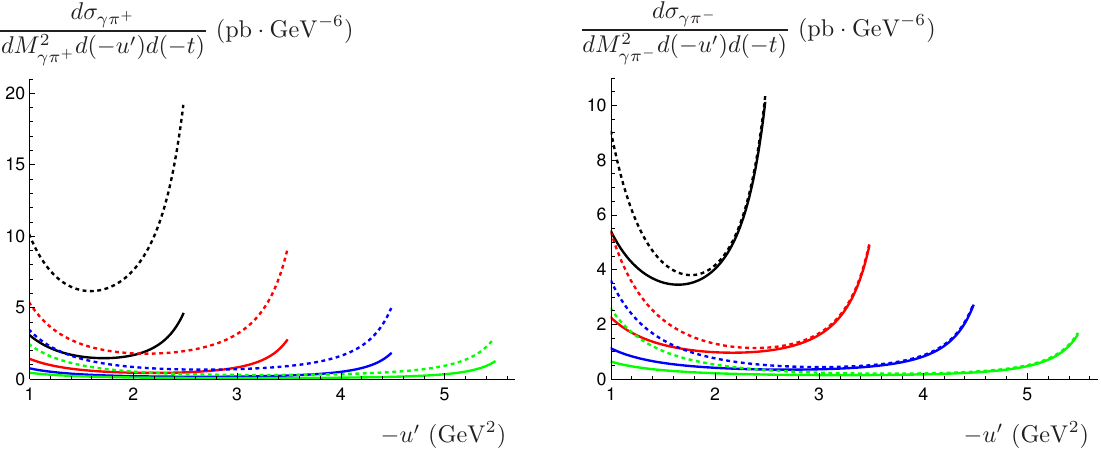}
\caption{ Left panel : the differential cross section for $\gamma \pi^+$  production on a proton target  at $s_{\gamma N} = 20$ GeV$^2$, $t=t_{min}$, and $M^2_{\gamma \pi}= 3$ (resp.$4,5,6$) GeV$^2$ for the black (resp. red, blue, green) curves. The solid and dashed  curves correspond to two different parametrization of the axial GPDs. Right panel : the same curves  for $\gamma \pi^-$  production on a neutron target.
}
\label{gapi}
\end{figure}

Since deep electroproduction of a $\pi$ meson has been shown to resist at moderate $Q^2$ to
leading twist dominance in the factorization framework, it has been tempting to  put the
blame on the peculiar chiral behavior of the higher twist (chiral-odd) pion DA as compared
with the leading twist (chiral even) pion DA. However, the dominance of higher twist contributions may not be a common feature of all exclusive amplitudes involving the pion DA. To check this idea, we propose \cite{Duplancic:2018bum}
 to study  the related process  $\gamma N \to \gamma \pi N'$
where the same pion DAs appear. It turns out that the axial nature of the pion leading twist DA infers a high sensitivity of the amplitudes to the axial GPDs $\tilde H(x,\xi,t)$ as shown on Fig.~\ref{gapi} where the cross sections for the reaction $\gamma p \to \gamma \pi^+ n$ and $\gamma n \to \gamma \pi ^-p$ are displayed for two different sets of axial GPDs. The rates are of the same order as for the $\gamma \rho$ case and we thus expect these reactions to be measurable at JLab.

\section{Conclusions}
The processes discussed here, because of the absence of gluon and sea quark contributions are not enhanced at high photon energy (or small skewness $\xi$) and they are thus more accessible at JLab than at EIC. However, since a high energy electron beam is also an intense source of medium energy quasi real photons ($q^2\approx 0$), with fractions of energy $y = \frac{q.p}{k.p}=0(10^{-3})$, ($k$ and $p$ being the initial electron  and nucleon momenta), one may expect the $\gamma \gamma$ and $\gamma \rho_L$ channels to be accessible at moderate values of $s_{\gamma N}$. Prospects at higher values of $s_{\gamma N}$ (and smaller values of the skewness $\xi$) are brighter for the $\gamma \pi^0$ channel which  benefits from the contributions of small $\xi$ sea-quark and gluon GPDs. 
\section*{Acknowledgments}
We acknowledge the collaboration of R. Boussarie, G.~Duplancic, K.~Passek-Kumericki, A. Pedrak and J. Wagner for the works reported here. L. S. is supported by the grant 2017/26/M/ST2/01074 of the National Science Center in Poland. He thanks the French LABEX P2IO and the French GDR QCD for support. 


\newpage
\renewcommand*{\FigPath}{./WeekI/02_Pire/}
 
\wstoc{Transition Distribution Amplitudes : from JLab to EIC}{Lech Szymanowski, Bernard Pire, Kirill Semenov-Tian-Shansky} 
 
\title{Transition Distribution Amplitudes : from JLab to EIC }
\author{L. Szymanowski$^*$}
\index{author}{Szymanowski, L.}

\address{NCBJ, 02-093 Warsaw, Poland\\
$^*$E-mail: Lech.Szymanowski@ncbj.gov.pl }
\author{B. Pire}
\index{author}{Pire, B.}

\address{CPHT, CNRS, \'Ecole polytechnique, I.P. Paris, 91128-Palaiseau, France}
\author{K.~Semenov-Tian-Shansky}
\index{author}{Semenov-Tian-Shansky, K.}

\address{National Research Centre ``Kurchatov Institute", Petersburg Nuclear Physics Institute, RU-188300 Gatchina, Russia}
\begin{abstract}
Baryon-to-meson Transition Distribution Amplitudes extend both the concepts of generalized parton
distributions  and baryon distribution amplitudes  encoding
 valuable complementary information on the $3$-dimensional hadronic structure. The recent
 analysis of backward meson electroproduction at JLab supports the hope to perform femto-photography of hadrons from a brand new perspective and in particular opens a new domain for EIC experiments.
\end{abstract}
\bodymatter
\section{Introduction}
Baryon-to-meson Transition Distribution Amplitudes~\cite{Pire:2016aqa}  (TDAs) are matrix elements of a three quark operator between a baryon and a meson states.  
Since the corresponding operator carries the quantum numbers of a baryon,
baryon-to-meson TDAs allow the exploration of the  baryonic content of a nucleon.
This is complementary to the information one can obtain from  generalized parton distributions (GPDs), with the operator carrying the quantum numbers of mesons. 
Similarly to the case of GPDs, by Fourier transforming baryon-to-meson TDAs to the
impact parameter space,
one obtains additional insight on the baryon structure in the transverse plane.


The nonlocal three quark (antiquark) operator
on the light cone
($n^2=0$) occurring in the definition of  baryon-to-meson  TDAs reads (gauge links are omitted):
\begin{equation}
\hat{O}^{\alpha \beta \gamma}_{\rho \tau \chi}( \lambda_1 n,\, \lambda_2 n, \, \lambda_3 n)
 =
\varepsilon_{c_{1} c_{2} c_{3}}
\Psi^{c_1 \alpha}_\rho(\lambda_1 n)
\Psi^{c_2 \beta}_\tau(\lambda_2 n)
\Psi^{c_3 \gamma}_\chi (\lambda_3 n),
\label{operators}
\end{equation}
where
$\alpha$, $\beta$, $\gamma$
stand for the quark (antiquark) flavor indices,
$\rho$, $\tau$, $\chi$
denote the Dirac spinor indices and $c_i$ stand for the color indices. Baryon-to-meson  TDAs~
\cite{ Pire:2004ie_1,Pire:2005ax},
share common features both with baryon distribution amplitudes (DAs), introduced as baryon-to-vacuum matrix elements of the same operators,
and with generalized parton distributions (GPDs), since the matrix element in question depends on the longitudinal momentum transfer
$\Delta^+=(p_{\cal M}-p_1) \cdot n$
between a baryon 
$N(p_1)$ 
and a meson 
${\cal M}({p_{\cal M}})$ 
characterized by the skewness variable
$
\xi= -\frac{(p_{\cal M}-p_1) \cdot n}{(p_{\cal M}+p_1) \cdot n}
$
and by the transverse momentum transfer
$\vec \Delta_T$.
 \begin{figure}[h!]
\center
\includegraphics[width=0.3\columnwidth]{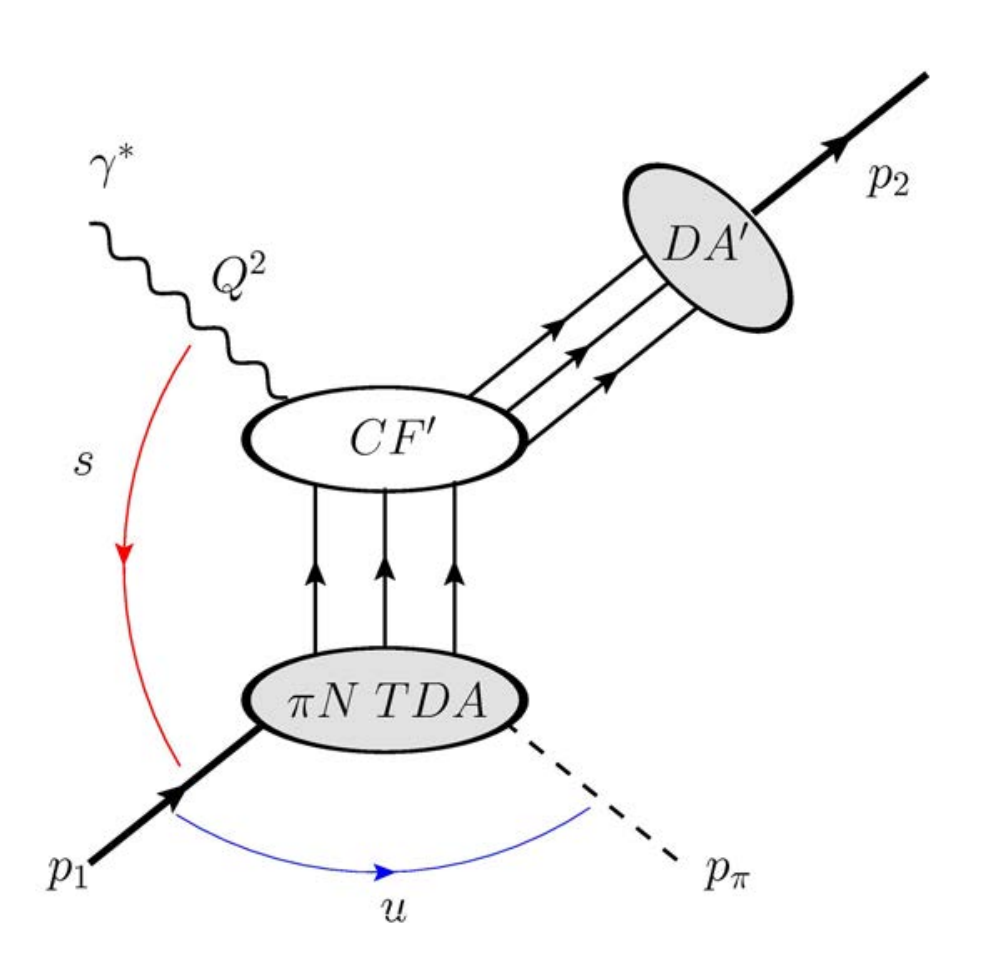}
\caption{ At large $Q^2=-q^2$ for the near-backward angles ({\it i.e.} for small $-u=-(p_1-p_\pi)^2$),
the $\gamma^*(q) N(p_1)\to \pi(p_{\pi}) N'(p_2)$ amplitude factorizes into a convolution of
a coefficient function (CF) with a baryon DA  and a baryon-to-meson TDA.  }
   \label{Fig_1}
\end{figure}
The collinear QCD factorization property, which ensures the validity of the 
GPD-based description of hard exclusive near-forward meson electroproduction reactions, 
may be extended to the complementary near-backward kinematical regime provided that the
$\bar{\Psi} \Psi$ operator is replaced by 
$\hat{O}^{\alpha \beta \gamma}_{\rho \tau \chi}( \lambda_1 n,\, \lambda_2 n, \, \lambda_3 n)$.
This results in the reaction mechanism sketched in Fig.~\ref{Fig_1} with a 
GPD replaced by baryon-to-meson TDA and baryon DA replacing meson DA.    

\section{The physical picture}

\begin{figure}[h!]
\center
\includegraphics[width=0.9\columnwidth]{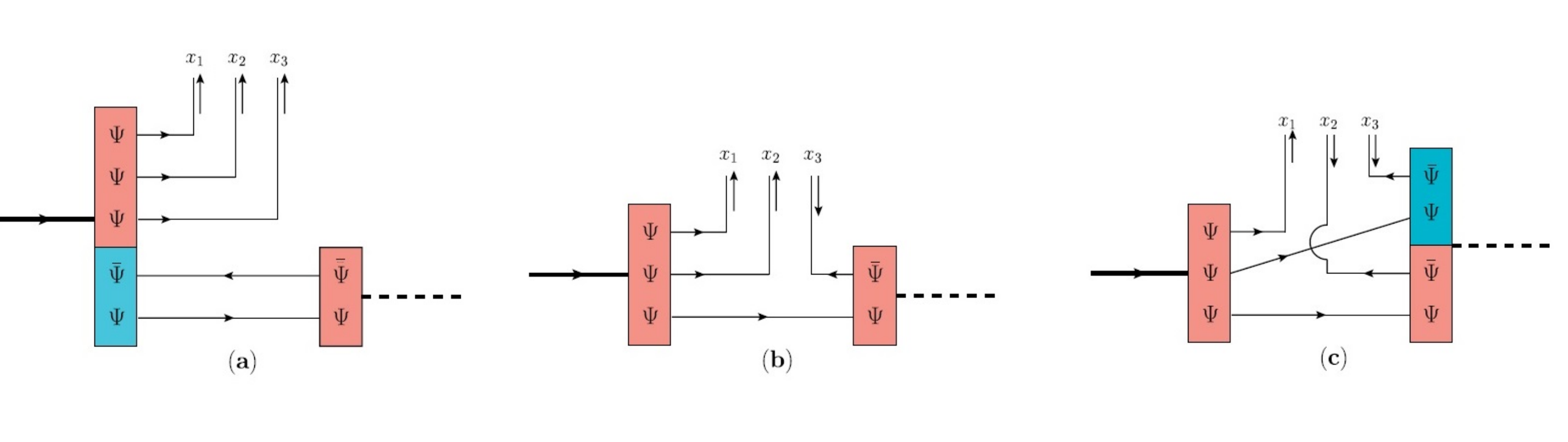}
\caption{Interpretation of baryon-to-meson TDAs at low normalization scale.
    Small vertical arrows show the flow of the momentum.
     {\bf  (a):} Contribution in the ERBL region (all $x_i$ are positive);
    {\bf  (b):} Contribution in the DGLAP I region (one of $x_i$  is negative).
    {\bf  (c):} Contribution in the DGLAP II region (two  $x_i$  are negative).}
   \label{Fig_X}
\end{figure}

The physical picture encoded in baryon-to-meson TDAs is conceptually close to that
contained in baryon GPDs and baryon DAs. Baryon-to-meson TDAs  characterize partonic correlations inside a baryon and give access to the momentum distribution
of the baryonic number inside a baryon. The same operator also defines the nucleon DA,
which can be seen as a limiting case of baryon-to-meson TDAs with the meson state replaced by the vacuum.
In the language of the Fock state decomposition, the leading twist baryon-to-meson TDAs are not restricted to the lowest Fock state
as the leading twist DAs. They rather probe the non-minimal Fock components with an additional
quark-antiquark pair:
\begin{eqnarray}
&&
| {\rm Nucleon} \rangle= |\Psi \Psi \Psi \rangle+ |\Psi \Psi \Psi; \,  \bar{\Psi} \Psi \rangle+....\;; ~~~~
| {\cal M} \rangle= |\bar{\Psi}\Psi \rangle+ |\bar{\Psi}\Psi; \, \bar{\Psi} \Psi \rangle+....\;
\end{eqnarray}
depending on the particular support region in question
(see Fig.~\ref{Fig_X}).

\section{JLab lessons}
\begin{figure}[h!]
\center
\includegraphics[width=0.3\columnwidth]{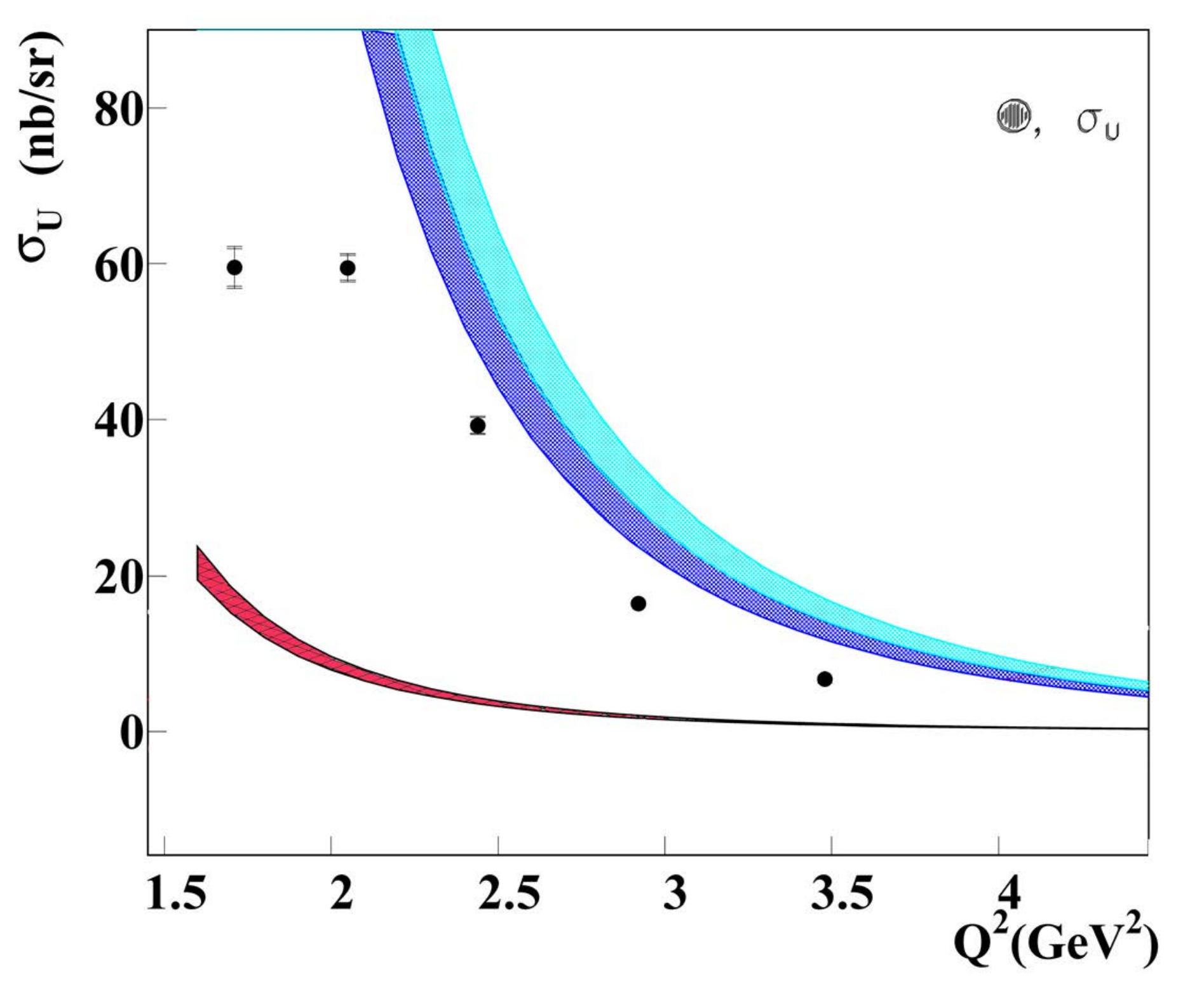} ~~~~~~~~~\includegraphics[width=0.35\columnwidth]{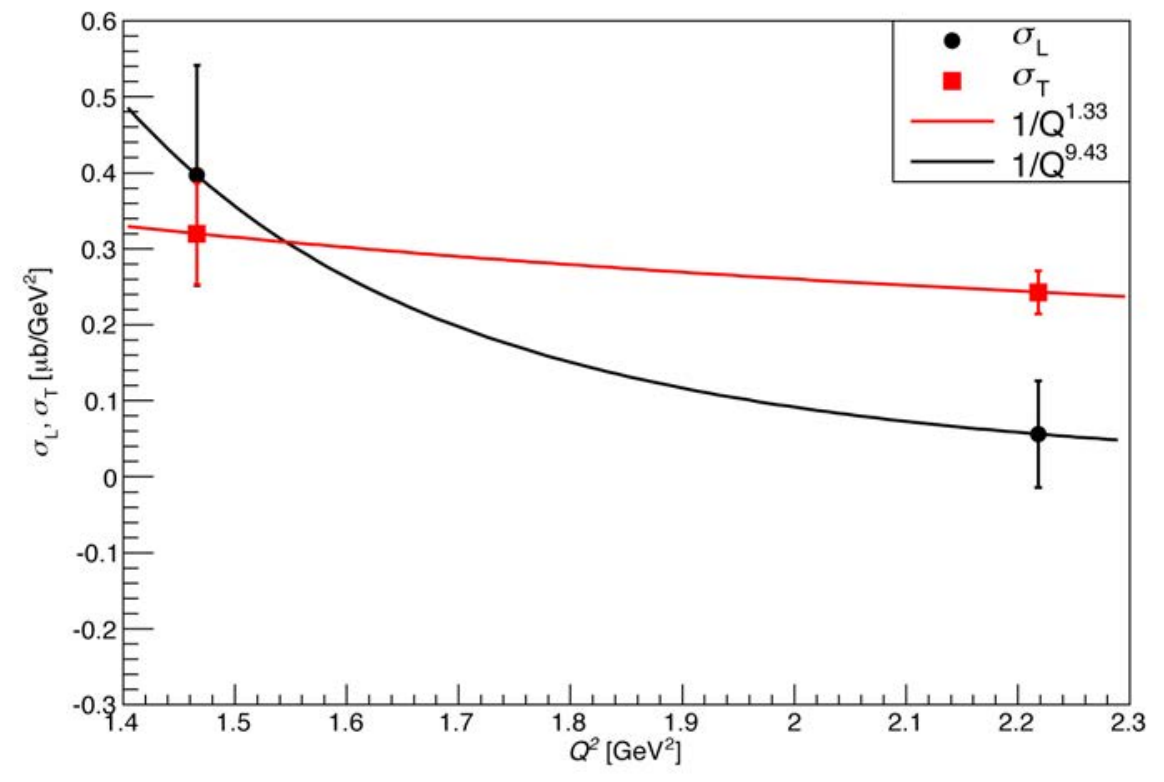}
\caption{Left panel: The $Q^2$ dependence of the $(\gamma^* p \to \pi^+ n)$ backward cross section $\sigma_U = \sigma_T+ \varepsilon \sigma_L$ measured by CLAS \cite{Park:2017irz} (circles) compared to the QCD prediction (colored bands) with three models of TDAs. Right panel: The $Q^2$ dependence of the longitudinal (squares)  and transverse (circles)  cross sections $\sigma_{L,T}(\gamma^* p \to \omega p)$ (in $\mu b$ GeV$^{-2}$) as measured at Hall C in JLab~\cite{Li-Huber}.} 
   \label{Fig_CLAS-Bill}
\end{figure}

The first experimental indications of the validity of the TDA concept have been recently 
presented~\cite{Park:2017irz_1, Li-Huber}. 
The left panel of Fig.~\ref{Fig_CLAS-Bill} shows the results of the $Q^2$-dependence of $\sigma_T + \epsilon~\sigma_L$ ($\varepsilon$ is the virtual photon linear polarization parameter) obtained by the CLAS collaboration at JLab for the  $e p \to e^\prime n \pi^+$ reaction in relatively large $Q^2$ ($>1.7$ GeV$^2$) and
small-$|u|$ domain ($\langle -u \rangle =0.5$ GeV$^2$) above the resonance region ($W^2=(q+p_1)^2> 4\,{\rm GeV}^2$), {\it i.e.} close to the ``near backward'' kinematical
regime  in which the TDA formalism is potentially applicable.
As anticipated, the cross sections has a strong $Q^2$-dependence.
The data are compared to the theoretical predictions of $\sigma_T$
from the cross channel nucleon pole exchange $\pi N$ TDA model suggested in Ref.~\cite{Pire:2011xv}.
The curves show the results of three  theoretical calculations
using different input phenomenological solutions for nucleon DAs
with their uncertainties represented by the bands. The crucial point is that the TDA formalism involves a dominance of the transverse amplitude.
Therefore, in order to be able to claim the validity of the TDA approach it
is necessary to separate $\sigma_T$ from $\sigma_L$  and check
that $\sigma_T \gg \sigma_L$.
This goal has been fulfilled by the Hall C experiment~\cite{Li-Huber} at JLab, which measured the reaction $e p \to e^\prime p^\prime \omega$ in a similar kinematic range. The data shown on  the right panel of  Fig.~\ref{Fig_CLAS-Bill} show indeed that $\sigma_T$ dominates over  $\sigma_L$ for sufficiently large values of $Q^2$, as anticipated by the collinear QCD factorization approach~\cite{Pire:2015kxa_1}.

\section{Perspectives for EIC}
TDAs are opening a new window on the study of the $3$-dimensional structure of nucleons
and recent experimental analysis of  backward meson electroproduction  hints that this concept may be applicable at moderate values of $Q^2$.
Clearly a more detailed experimental analysis  is required for the confirmation of  relevance of TDAs concept for analysis of hard processes. Testing the validity of the collinear factorized description of hard backward meson electroproduction reactions at the energies of Jlab@12 GeV  will help to elaborate a unified and consistent approach for hard exclusive reactions. Moreover, backward DVCS is also a very interesting channel to be explored; it is  a source of new information on the D-term form factor analytically continued to large $-t$. Let us stress also that TDAs are natural concepts to be used for the description of nuclear break-up reaction - such as deuteron electrodissociation - which may be interesting to visualize the partonic content of light nuclei. Let us also note that  the \={P}ANDA experiment at GSI-FAIR~\cite{Lutz:2009ff} will provide opportunities to access the cross conjugated counterparts of the reaction depicted on Fig. 1 and test the universality of baryon-to-meson 
TDAs~\cite{Lansberg:2012ha, Pire:2013jva}. 

Although detailed predictions have not yet been worked out for higher energies, one can anticipate that studies at the electron-ion collider (EIC) will allow this new domain physics to be further explored. Higher $Q^2$ should be accessible in a domain of moderate $\gamma^* N$ energies, i.e. rather small values of the usual $y$ variable and not too small values of $\xi$. The peculiar EIC kinematics, as compared to fixed target experiments, allows in principle a thorough analysis of the backward region pertinent to TDA studies. More phenomenological prospective studies are clearly needed.

\section*{Acknowledgments}
 L. S. is supported by the grant 2017/26/M/ST2/01074 of the National Science Center in Poland. He thanks the LABEX P2IO and the  GDR QCD for support. 

\newpage
%

\renewcommand*{\FigPath}{./WeekI/03_Scopetta/}

\wstoc{Coherent deeply virtual Compton scattering off He nuclei}{Sara Fucini, Matteo Rinaldi and Sergio Scopetta} 

\title{Coherent deeply virtual Compton scattering off He nuclei}

\author{Sara Fucini, Matteo Rinaldi and Sergio Scopetta$^*$}
\index{author}{Fucini, S.}
\index{author}{Rinaldi, M.}
\index{author}{Scopetta, S.}

\address{Dipartimento di Fisica e Geologia, University of Perugia and INFN, Perugia Section\\
Perugia, I-06123, Italy\\
$^*$E-mail: 
sergio.scopetta@unipg.it
}

\begin{abstract}
The status of realistic calculations of nuclear generalized parton distributions,
entering the theoretical description of coherent 
deeply virtual Compton scattering off nuclei, is reviewed for trinucleons and for $^4$He, also in view
of forthcoming measurements at the Jefferson Laboratory and at the future Electron Ion Collider.
\end{abstract}


\bodymatter

\section{Introduction}

Nuclear modifications of the nucleon parton structures,
discovered by the European Muon Collaboration 
\cite{Aubert:1983xm} several decades ago, cannot be explained by means of inclusive measurements only.
One of the possible ways out is to perform nuclear
imaging, now possible for the first time
through deeply virtual Compton scattering (DVCS) and deeply-virtual meson production, using the tool of generalized parton distributions
(GPDs) (see Refs \cite{Dupre:2015jha,Cloet:2019mql} for recent reports).
The comparison of the transverse spatial quark and gluon distributions in nuclei
or bound nucleons (to be obtained in coherent or incoherent DVCS, respectively) to
the corresponding quantities in free nucleons,
will allow ultimately a pictorial representation of the EMC effect.
The relevance of non-nucleonic degrees of freedom, 
as addressed in Ref. \cite{Berger:2001zb},
or the 
change of size for bound nucleons, will be observed. 
The most discussed sector of the EMC effect is
the valence region at intermediate $Q^2$, which
will be investigated by 
Jefferson Lab (JLab) at 12~GeV.
For the lightest nuclei, $^2$H, $^3$He, $^4$He,
sophisticated calculations of conventional effects,
although challenging, are possible.
This would allow one to distinguish them
from exotic ones, likely responsible for
the observed EMC behavior.
Without realistic benchmark calculations, 
making use of wave functions obtained as
exact solution of the Schr\"odinger equation
using realistic nucleon nucleon potentials
and three-body forces whenever appropriate,
the interpretation of experimental data is difficult.
Among few-body nuclei, in this talk we will concentrate
in three- and four-body systems.

\begin{figure}[t]
\hspace{3.cm}
\includegraphics[scale=0.35,angle=0]{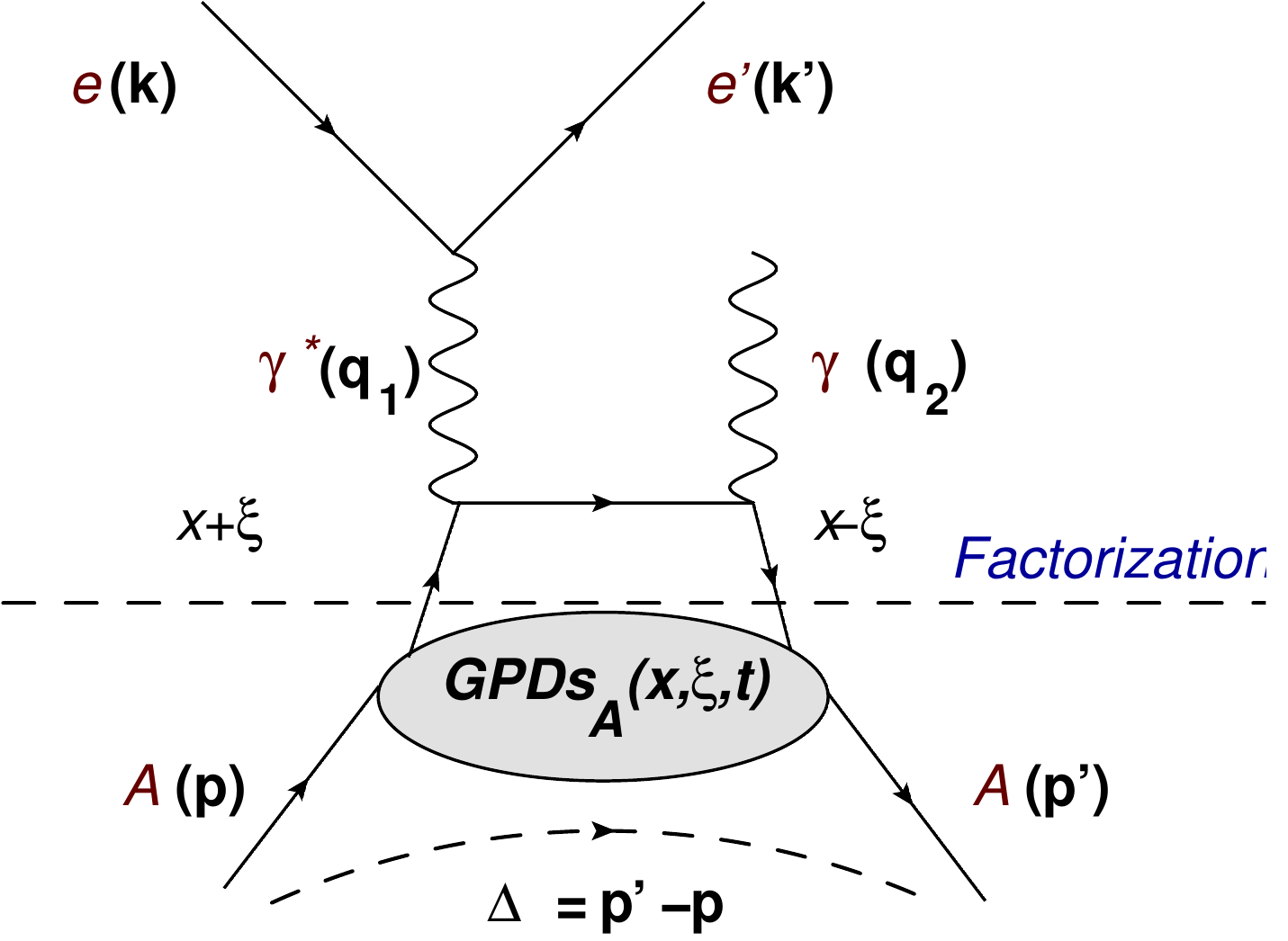}
\caption{
Coherent DVCS process off a nucleus $A$ in the handbag approximation.
}
\label{dvcsinco}
\end{figure}
\section{Coherent DVCS off $^3$He}

For the coherent channel of DVCS, the one where
the nucleus does not break up (see Fig. 1 for a
representation of the process with a generic $A$ nucleus, in the handbag approximation, i.e., with the interaction occurring on a leading quark),
due to very small cross sections,
the measurements addressed in the Introduction are very difficult.
In between the $^2$H nucleus, very interesting for its rich spin structure and for the the possible extraction of neutron information, and
$^4$He, ideal to study nuclear effects, being deeply-bound, scalar and isoscalar, with a simple
description of its spin and flavor structure,
$^3$He provides an opportunity
to study the $A$ dependence of nuclear effects, and
it could give easy access to neutron polarization properties,
due to its specific spin structure.
In addition, being isospin-$1/2$, it guarantees
that flavor dependence of nuclear effects can be studied, 
in particular if parallel measurements 
on $^3$H targets,
likely possible at the Electron Ion Collider (EIC), were performed \cite{Scopetta:2009sn}.
\begin{figure}[t]
\vskip -.9cm
\centering\includegraphics[scale=0.30]{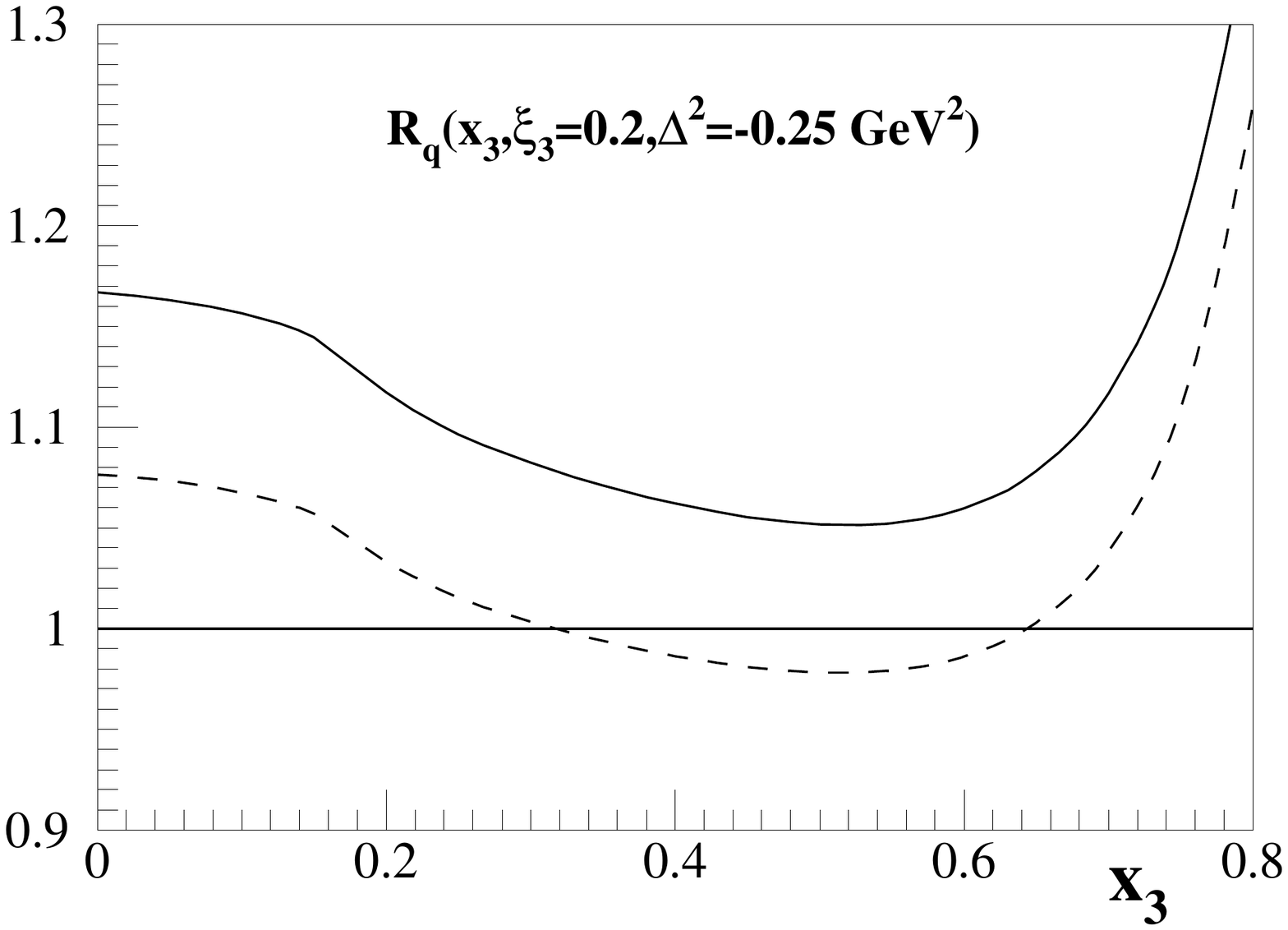}
\vskip -3.9cm
\caption{ Estimate of nuclear effects on $^3$He GPD
for the $d (u)$ flavor, full (dashed) line. The effect is given by the difference of the curves with 
respect to one.}
\label{alu}
\end{figure}
A complete realistic study of leading twist 
DVCS requires the evaluation of nuclear GPDs.
From the theoretical point of view, conventional effects for
nuclear systems are seen in a plane wave impulse approximation (IA)
analysis, i.e., with the struck quark belonging to one nucleon in the target, and disregarding possible final state interaction effects between this nucleon and the remnants.  This requires the evaluation of realistic non-diagonal spectral
functions \cite{Scopetta:2004kj}.
For $^3$He, a complete analysis using the AV18
nucleon-nucleon (NN) potential is available
\cite{Scopetta:2009sn,Rinaldi:2012pj,
Rinaldi:2012ft,Rinaldi:2014bba,Scopetta:2004kj}.
Nuclear GPDs are found to be sensitive to details of the used NN interaction.
In particular, nuclear effects are found to grow with the momentum transfer to the target, $\Delta^2$, and with the longitudinal momentum asymmetry of the process
(parametrized by the so-called skewness variable, $\xi$). In $^3$He, nuclear effects are found to be bigger for the $d$ flavor than for the $u$  one (see Fig. 2), a prediction of a realistic impulse approximation (IA), where also violations of nuclear charge symmetry are considered in the AV18 NN interaction, which could
be tested experimentally. Besides, the dependence on the excitation energy of the nuclear recoiling system in the IA description, parametrized by the so-called removal energy, is found to be bigger in nuclear GPDs than in inclusive observables.
Anyway, it is also found that close to the forward limit the information on neutron polarization can be safely extracted from 
$^3$He data and workable extraction formulae have been proposed in this sense\cite{Rinaldi:2012pj,
Rinaldi:2012ft,Rinaldi:2014bba}.
Measurements for $^3$He and $^3$H
are not planned, but could be considered
as extensions of the impressive ALERT detector
project at JLab 12
\cite{Armstrong:2017zqr,Armstrong:2017zcm}, at least in the unpolarized 
sector. Polarized measurements, which could
access neutron angular momentum information \cite{Rinaldi:2012pj,
Rinaldi:2012ft,Rinaldi:2014bba}, seem
very unlikely at JLab, due to the difficulty of arranging
a polarized target and a recoil detector in the same experimental setup, but are in principle accessible
at the EIC, where
the extension to lower
$x$ regions will be also possible \cite{Accardi:2012qut_2}. 

\section{Coherent DVCS off $^4$He}

Despite the difficulty of measuring coherent
DVCS off nuclei, due to small cross-sections, the first data for coherent DVCS off $^4$He
collected at JLab in the 6~GeV setup
have been published \cite{Hattawy:2017woc}.
A new impressive program is on the way at JLab,
carried on by the CLAS collaboration with the ALERT detector project 
\cite{Armstrong:2017zqr,Armstrong:2017zcm}.
\begin{figure}[t]
\centering\includegraphics[scale=0.24]{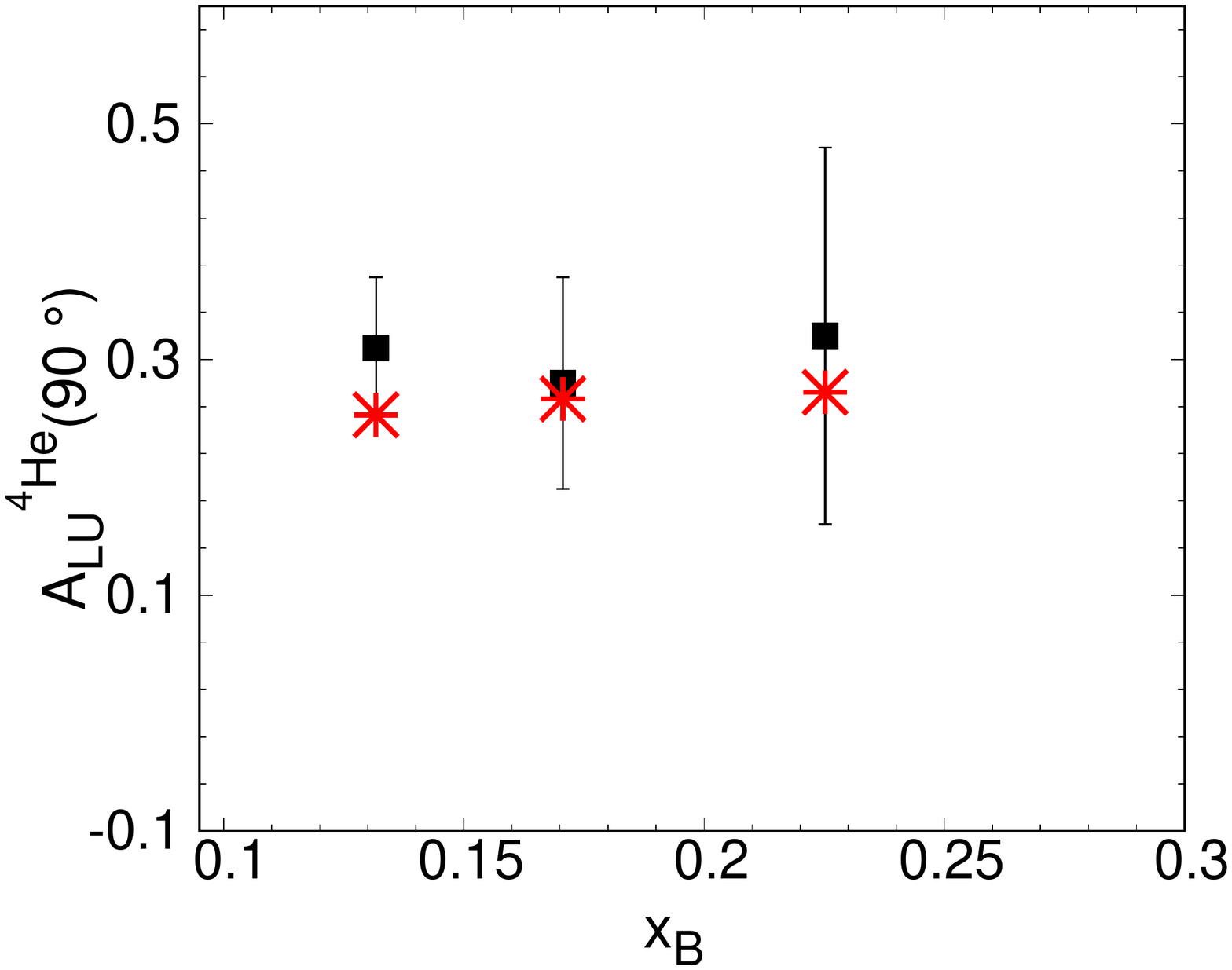}
\vskip -.5cm
\caption{ $^4$He azimuthal
(with respect to the angle $\phi$ between lepton and nuclear planes)
beam-spin asymmetry $A_{LU}(\phi)$: results of Ref. \cite{Fucini:2018gso} (red stars) compared with data
(black squares)
\cite{Hattawy:2017woc}.
\label{alu1}}
\end{figure}
A study for DVCS off $^4$He with nuclear ingredients of the same quality of those summarized above 
for $^3$He is still missing and should be done, to update 
existing calculations,
performed long time ago\cite{Guzey:2003jh,Liuti:2005gi}. 
The evaluation of a realistic spectral functions of $^4$He,
using state-of-the-art NN potentials,
will require in particular the wave function of a nuclear three-body scattering state, 
which is a really challenging few-body problem.
An encouraging calculation has been recently
performed for coherent DVCS off $^4$He \cite{Fucini:2018gso}, with the aim to
describe the CLAS data
\cite{Hattawy:2017woc},
as a relevant intermediate
step towards a rigorous realistic evaluation.
A model of the nuclear non-diagonal spectral function, 
based on the momentum distribution
corresponding to the AV18 NN interaction
\cite{PhysRevC.67.034003}, has been used in the actual IA calculation. 
In particular the spectral function 
is exact in its ground state part (when the remnant is a bound three-body system)
and modelled in the complicated excited sector.
As a test of the procedure, typical results
are found for the nuclear form factor
and for nuclear parton distributions, in proper limits.
Nuclear GPD and the actual observable, 
the so-called,
Compton form factor
(CFF) are evaluated using
a well known GPD model to take into account the nucleonic information~\cite{Goloskokov:2011rd_8}. 
As can be seen in Fig.~\ref{alu1}, a very good agreement is found with the data, for the so-called beam-spin asymmetry, theoretically obtained in terms of the CFF, in turn evaluated from the GPD.
One can conclude that 
a careful analysis of the reaction mechanism in terms of
basic conventional ingredients is successful and that
the present experimental accuracy does not require
the use of exotic arguments, such as dynamical off-shellness.
More refined nuclear calculations will be certainly necessary for the expected improved accuracy of the next generation of experiments 
at JLab, with the 12 GeV electron beam and high luminosity, and, above all, at the EIC.
Very recent results for the incoherent channel are reported in \cite{Fucini:2019xlc},
where an encouraging comparison with
data from JLab \cite{Hattawy:2018liu} is presented.

This work was supported in part by the STRONG-2020 project of the European Union’s Horizon 2020
research and innovation programme under grant agreement No 824093, and by
the project ``Deeply Virtual Compton Scattering off $^4$He", in the programme FRB of the University
of Perugia.
  



\newpage
%


\renewcommand*{\FigPath}{./WeekI/04_Kumericki/Figs}

\renewcommand\Re{\operatorname{\mathfrak{Re}}}
\renewcommand\Im{\operatorname{\mathfrak{Im}}}

\wstoc{Extraction of DVCS form factors with uncertainties}{Kre\v{s}imir Kumeri\v{c}ki} 
\title{Extraction of DVCS form factors with uncertainties}

\author{Kre\v{s}imir Kumeri\v{c}ki}
\index{author}{Kumeri\v{c}k, K.}

\address{
    Department of Physics, Faculty of Science, University of Zagreb, 10000 Zagreb, Croatia\\
    Institut f\"{u}r Theoretische Physik, Universit\"{a}t Regensburg, D-93040 Regensburg, Germany\\
     E-mail: kkumer@phy.hr}

\begin{abstract}
    We discuss recent attempts to extract deeply virtual Compton scattering 
    form factors with emphasis on their uncertainties, which turn out to
    be most reliably provided by method of neural networks.
\end{abstract}


\section{Introduction}

Partonic structure of the nucleon, as encoded by generalized parton distributions
(GPDs), is essentially non-perturbative. As such, main avenue
to its determination is extraction from experimental data,
mostly from measurements of deeply virtual Compton scattering (DVCS),
which is a subprocess of electroproduction of real photon off nucleon.
Still, more than a decade after the first such fitting attempts\cite{Kumericki:2007sa_153},
we have only partial phenomenological knowledge of GPDs. (Recent review is
available in Ref.~\citenum{Kumericki:2016ehc_54}.)
Furthermore, although assessment of uncertainties is indispensable part
of any quantitative scientific result, authors of global GPD fits usually hesitated to 
discuss error bands of extracted functions.
It was understood that standard simple propagation of experimental uncertainties is
not enough. GPD functions depend in
a rather unknown manner on three kinematic variables (average
and transferred parton longitudinal momentum fractions, $x$ and $\xi$, 
and nucleon momentum transfer squared $t$), which makes the
problem very complex from the data-analysis standpoint, and
the very choice of fitting parametrization introduces unknown and
possibly dominant uncertainty.

\section{DVCS subtraction constant}

\begin{figure}[h]
\begin{center}
    \includegraphics[width=0.9\textwidth]{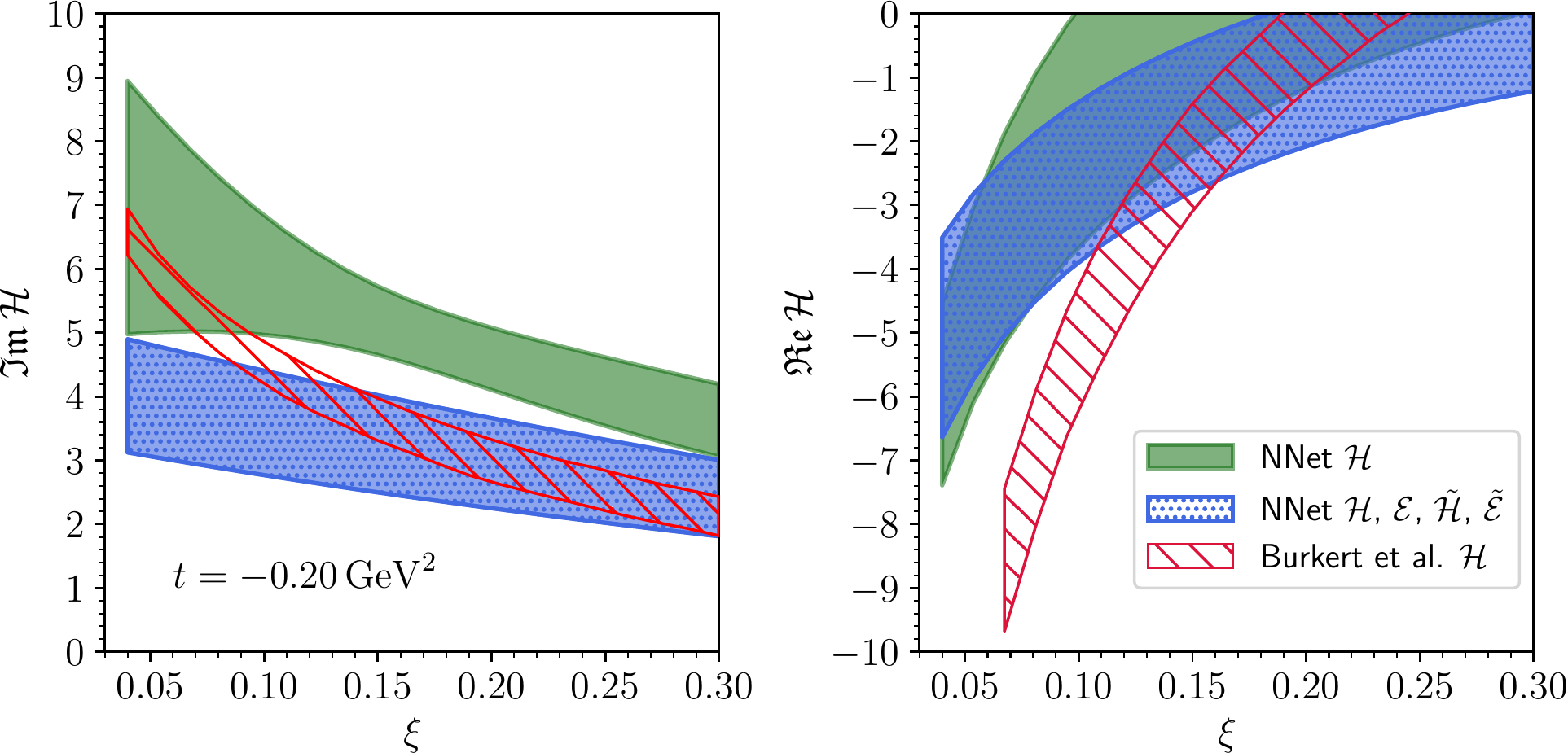}
\end{center}
\caption{Imaginary and real part of CFF $\mathcal{H}$ resulting from
fitting different parametrizations to the essentially the same data.}
\label{fig:nat}
\end{figure}

Important role of the choice of the parametrization may be illustrated by recent
attempts to determine the subtraction constant $\Delta(t)$ of DVCS dispersion
relation, 
\begin{equation}
\Re\mathcal{H}(\xi,t) = \Delta(t) + \frac{1}{\pi} {\rm P.V.}
\int_{0}^{1} {\rm d}x \left(\frac{1}{\xi-x} - \frac{1}{\xi+x}\right)\Im\mathcal{H} (x,t) \,,
\label{eq:disp}
\end{equation}
that is of great phenomenological interest since it is
closely related to the pressure in the nucleon\cite{Polyakov:2002yz_21_2,Teryaev:2005uj_150}.
Compton form factor (CFF) $\mathcal{H}(\xi,t)$ in Eq. (\ref{eq:disp})
is a convolution of
GPD $H(x,\xi,t)$ with the known hard scattering amplitude and is, being dependent
on two variables only, more easy extraction target.
Still, $\Delta(t)$ resulting from fits to CLAS DVCS 
data\cite{Jo:2015ema}, came out with very different uncertainty estimation,
depending on whether relatively rigid ansatz\cite{Kumericki:2009uq} for $\mathcal{H}$ was
used\cite{Burkert:2018bqq_2} or it was parametrized by completely
flexible neural networks\cite{Kumericki:2019ddg_156}, see Fig. \ref{fig:nat}.

\section{Neural network fits}
In the lack of general procedure for assessment of systematic uncertainties
coming from the choice of fitting ansatz, one convenient approach
is using the parametrization by neural networks,
which is known \emph{not} to introduce any such systematic error.
After the early proof of concept\cite{Kumericki:2011rz}, first global
neural network determination of CFFs was reported in Ref. \citenum{Moutarde:2019tqa},
demonstrating the power of this approach.

Similarly, in the framework of neural net approach, we attempted 
to address the question
of which of the four leading order CFFs, $\mathcal{H}$, $\mathcal{E}$,
$\tilde{\mathcal{H}}$, and $\tilde{\mathcal{E}}$ (or, more accurately,
eight sub-CFFs which are the real and imaginary parts of these four),
can be reliably extracted from the given data.
To this end, we used the stepwise regression method proposed in
Ref. \citenum{Kumericki:2013br}, where the number of sub-CFFs is gradually
increased and all combinations are tried, until there is no statistically
significant improvement in the description of the data.
Representative subset of global DVCS data was used, with
various beam and target, spin and charge asymmetries measured
by HERMES\cite{Airapetian:2012mq,Airapetian:2008aa,Airapetian:2010ab},
and helicity independent and dependent cross-sections measured
by Hall A and CLAS JLab collaborations\cite{Defurne:2015kxq,Jo:2015ema},
where JLab data was Fourier-transformed, so that we fitted to the
total of 128 harmonics.

\begin{figure}[h]
\begin{center}
\includegraphics[width=0.95\textwidth]{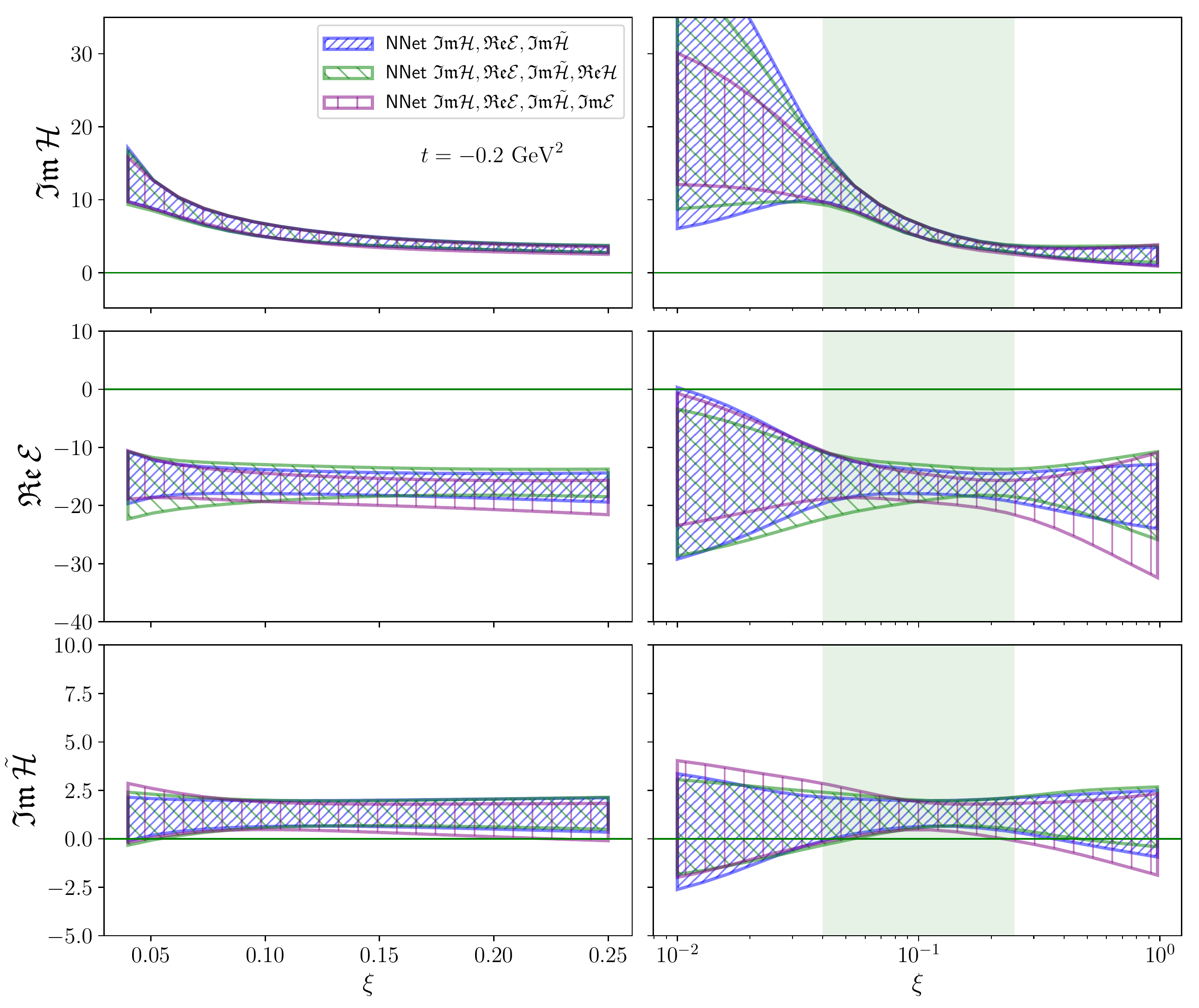}%
\end{center}
\caption{Neural network extraction of dominant CFFs from DVCS
data. Results for various sets of CFFs are consistent in the data
region (left) and also when extrapolated outside of the data region
(right). Dispersion relation constraints were \emph{not} used.}
\label{fig:left}
\end{figure}

\begin{figure}[h]
\begin{center}
\includegraphics[width=0.92\textwidth]{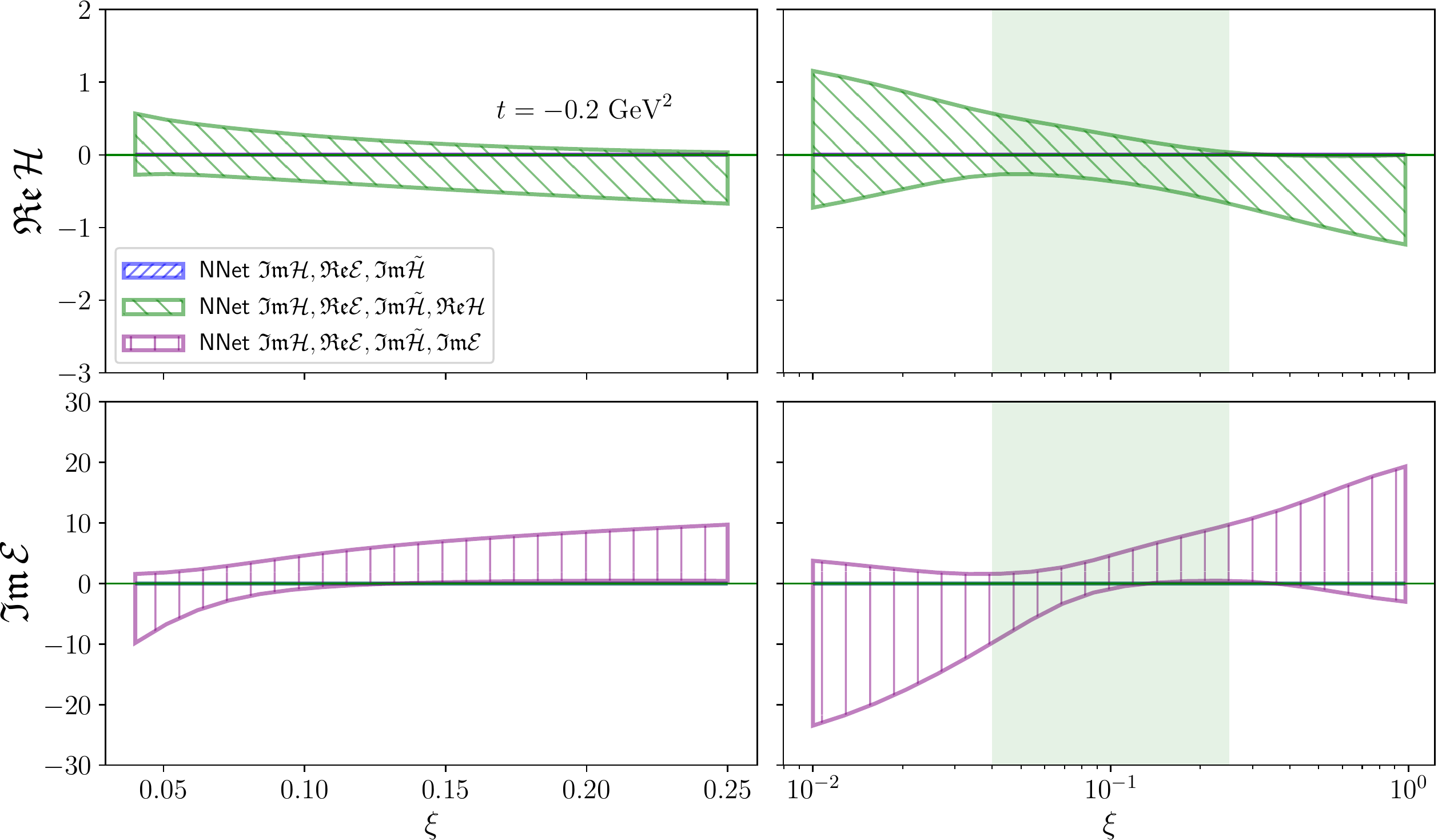}%
\end{center}
\caption{Extracted $\Re\mathcal{H}$ or $\Im\mathcal{E}$ are
mostly consistent with zero, but their addition to the model
improves description of the data from $\chi^{2}/{\rm npts}$
= 103.1/128 to 97.4/128 or 96.4/128, respectively.}
\label{fig:right}
\end{figure}

Results, displayed on Figs. \ref{fig:left} and \ref{fig:right}, show
that from the present data only $\Im\mathcal{H}$, $\Im\tilde{\mathcal{H}}$,
and $\Re\mathcal{E}$ can be reliably extracted, with maybe some
ambiguous hints of $\Re\mathcal{H}$ or $\Im\mathcal{E}$.
This is similar to the conclusions of Ref. \citenum{Kumericki:2013br}, which
used method of local fits (which is also resistant to the problem
of choice of the ansatz function).

\section{Conclusion}
How to reliably determine uncertainties of GPD or CFF functions
extracted by fitting of ansatz function is an important open
question for this area of research. At the moment, the best
confidence is provided by the method of neural networks.

\section*{Acknowledgments}

This work was supported by 
QuantiXLie Centre of Excellence through grant KK.01.1.1.01.0004,
and European Union's Horizon 2020
research and innovation programme under grant agreement No 824093.





\newpage
\renewcommand*{\FigPath}{./WeekI/05_Schweitzer/Figs}
\wstoc{Form factors of the energy-momentum tensor}{Maxim V.~Polyakov, Peter Schweitzer} 

\title{Form factors of the energy-momentum tensor}

\author{Maxim V.~Polyakov}
\address{Petersburg Nuclear Physics Institute, 
	      Gatchina, 188300, St.~Petersburg, Russia\\
              Institut f\"ur Theoretische Physik II, 
	      Ruhr-Universit\"at Bochum, D-44780 Bochum, Germany}
\index{author}{Polyakov, M.}	      

\author{Peter Schweitzer}
\address{Department of Physics, University of Connecticut, 
		Storrs, CT 06269, USA}
\index{author}{Schweitzer, P.}

\begin{abstract}
Matrix elements of the energy momentum tensor (EMT) bear fundamental
information like mass, spin and $D$-term of a particle which is
the ``last unknown global property.'' Recent progress on EMT form factors 
of hadrons, their interpretations and applications as well as the 
experimental status is given.
\end{abstract}


\bodymatter

\section{Introduction}

Matrix elements of the EMT\cite{Kobzarev:1962wt}
yield fundamental particle properties like mass and spin as well
as the $D$-term \cite{Polyakov:1999gs_7} which is 
related to the stress-tensor components of the EMT and gives access to 
mechanical properties of the system \cite{Polyakov:2002yz_7,Polyakov:2018zvc_7}.
EMT form factors can be accessed through studies of generalized 
parton distributions (GPDs) in hard exclusive reactions
\cite{Mueller:1998fv_7,Goeke:2001tz}. While 
a model-independent extraction of GPDs is a challenging long-term
task, accessing information on the $D$-term may be possible sooner
thanks to its relation to the subtraction constant in fixed-$t$ 
dispersion relations in deeply virtual Compton scattering (DVCS) 
\cite{Polyakov:2002yz_7,Teryaev:2005uj_7}.
The physics of EMT form factors has important applications.
The purpose of this article is to provide an overview of 
the latest developments and experimental status. 

\section{EMT form factors of hadrons}
\label{Sec-2:EMT-in-QCD-general-definitions}

The nucleon form factors of the symmetric EMT 
$\hat T_{\mu\nu}=\hat T_{\mu\nu}^Q+\hat T_{\mu\nu}^G$
are defined as  
\begin{align}
    \langle p^\prime,s^\prime| \hat T_{\mu\nu}(0) |p,s\rangle
    = \bar u{ }^\prime\biggl[
      A(t)\,\frac{\gamma_{\{\mu} P_{\nu\}}}{2}
    + B(t)\,\frac{i\,P_{\{\mu}\sigma_{\nu\}\rho}\Delta^\rho}{4m}
    + D(t)\,\frac{\Delta_\mu\Delta_\nu-g_{\mu\nu}\Delta^2}{4m}
    \biggr]u\,,
    \label{Eq:EMT-FFs-spin-12}
\end{align}
with $P= \frac12(p^\prime + p)$, $\Delta = p^\prime-p$, $t=\Delta^2$,
$a_{\{\mu} b_{\nu\}}=a_\mu b_\nu + a_\nu b_\mu$ and a covariant normalization
of the states is used with the nucleon spinors $\bar u(p,s)\, u(p,s) =2\,m$. 
Spin-0 hadrons like the pion have only the 2 total EMT form factors
$A(t)$ and $D(t)$. Hadrons with spin $J\ge 1$ have more form factors
\cite{Cosyn:2019aio_7,Polyakov:2018rew}.

The quark and gluon QCD operators $\hat T_{\mu\nu}^a$ ($a=Q,\,G$)
are each gauge invariant. Their form factors $A^a(t,\mu)$, etc
depend on renormalization 
scale $\mu$ and additional form factors appear, e.g.\ as the structure
${m}\,{\bar c}^a(t,\mu)\,g_{\mu\nu}$ in (\ref{Eq:EMT-FFs-spin-12}),
with $\sum_a\bar{c}^a(t,\mu)=0$.

\section{Relation to GPDs and 2D interpretation}

GPDs provide a practical way to access EMT form factors through the DVCS 
process $eN\to e^\prime N^\prime \gamma$ or hard exclusive meson production.
For the nucleon the second Mellin moments of unpolarized GPDs yield 
\begin{align}
	\int{\rm d}x\;x\, H^a(x,\xi,t) = A^a(t) + \xi^2 D^a(t), 
        \;\;\; 
        \int{\rm d}x\;x\, E^a(x,\xi,t) = B^a(t) - \xi^2 D^a(t).
    	\label{Eq:GPD-Mellin}
\end{align}
The Fourier transform 
$H^a(x,b_\perp)=\int d^2\Delta_\perp/(2\pi)^2 \,
e^{-i\vec{\Delta}_\perp \vec{b}_\perp}\,H^a(x,\xi,-\vec{\Delta}^2_\perp)|_{\xi=0}$
is the probability to find a parton carrying the momentum 
fraction $x$ and located at the distance $b_\perp$ from the 
hadron's (transverse) center-of-mass on the lightcone
\cite{Burkardt:2000za_7}.
The 2D interpretation of EMT form factors was also discussed
\cite{Lorce:2018egm_7}.

\section{The static EMT and 3D interpretation}

In the Breit frame characterized by $P=(E,0,0,0)$ and $\Delta=(0,\vec{\Delta})$ 
with $t=-\vec{\Delta}^2$ and $E=\sqrt{m^2+\vec{\Delta}{ }^2/4}$ one can
define the static EMT \cite{Polyakov:2002yz_7}
\begin{equation}\label{EQ:staticEMT}
	T^a_{\mu\nu}(\vec{r},\vec{s})=\int \frac{d^3 \vec{\Delta}}{(2\pi)^3 2E}\ 
	e^{{-i} \vec{r\Delta}} \langle p'| \hat{T}^a_{\mu\nu}(0)|p\rangle,
\end{equation}
where $\vec{s}$ is the polarization vector of the states 
$|p\rangle$, $|p^\prime\rangle $ in the respective rest frames.
The 00-component of (\ref{EQ:staticEMT}) is the energy density
which only can be defined for the total system, and yields
$\int d^3r \,T_{00}(r) = m$. Decomposition of the nucleon mass in 
terms of quark and gluon contributions was discussed in
\cite{Ji:1994av_7}.
The $0k$-components yield 
the spatial distribution of the nucleon spin density
$J_a^i(\vec{r},\vec{s})=\epsilon^{ijk}r^jT_a^{0k}(\vec{r},\vec{s})$.
This 3D density has a monopole term \cite{Polyakov:2002yz_7}, and a 
quadrupole term \cite{Lorce:2017wkb_7}  which are related to each other
\cite{Schweitzer:2019kkd}.
The $ij$-components of (\ref{EQ:staticEMT}) define the stress tensor 
which can be decomposed in contributions from shear forces $s(r)$ 
and pressure $p(r)$ as follows \cite{Polyakov:2002yz_7} 
\begin{align}\label{Eq:stress-tensor-p-s}
	T^{ij}(\vec{r}) = \biggl(
	\frac{r^ir^j}{r^2}-\frac13\,\delta^{ij}\biggr) s(r)
	+ \delta^{ij}\,p(r)\,.
\end{align}
For the nucleon the 3D interpretation is subject to small relativistic 
corrections \cite{Hudson:2017xug} and becomes exact in the large-$N_c$ limit.
The shear forces can be defined separately for quarks and gluons in terms of 
$D^{Q,G}(t)$. For the ``partial'' pressures from quarks and gluons one also 
needs $\bar{c}^{Q,G}(t,\mu)$.

EMT conservation relates $s(r)$ and $p(r)$ as
$\frac23\,s^\prime(r)+\frac2r\,s(r)+p^\prime(r)=0$. Notice that $s(r)=0$ 
would imply $p(r)=\;$constant and isotropic matter, cf.\ 
(\ref{Eq:stress-tensor-p-s}). Thus, the shear forces are responsible 
for structure formation \cite{Lorce:2018egm_7}. Another consequence of 
EMT conservation is the von Laue condition \cite{von-Laue}, 
\begin{equation}\label{Eq:von-Laue}
\int_0^\infty {\rm d}r\,r^2p(r)=0,
\end{equation} 
implying that $p(r)$ must have at least one node. In all model 
studies so far $p(r)$ was found positive in the inner region 
(repulsion towards outside) and negative in the outer region
(attraction towards inside). 
The $D$-term can be expressed in two equivalent ways as 
$D=-\,\frac{4}{15}\,m\int d^3r\;r^2\,s(r)= m \int d^3 r\;r^2\, p(r)$.
The stress tensor in (\ref{Eq:stress-tensor-p-s}) has two eigenvalues
related to normal ($dF_r$) and tangential ($dF_\phi,\, dF_\theta$) forces 
\begin{equation}\label{Eq:force-spherical-components}
	\frac{dF_r}{dS_r}=\frac 23 s(r)+p(r), \quad 
	\frac{dF_\theta}{dS_\theta}=\frac{dF_\phi}{dS_\phi}=-\frac 13 s(r)+p(r)
\end{equation}
with eigenvectors $\vec{e}_r$ and $\vec{e}_{\theta,\phi}$. The degeneracy 
is lifted for spin $J\ge 1$.
In a stable system the normal force $dF_r{/dS_r}=\frac 23 s(r) +p(r)>0$.
Otherwise the system would collapse. This mechanical stability requirement 
can be written as $\int_0^R dr \,r^2 p(r)>0$ (for any $R$), thus
complementing  the von Laue condition (\ref{Eq:von-Laue}). 
It also determines the $D$-term of a stable system to be negative 
\cite{Perevalova:2016dln},
$D < 0$.
The positivity of $\frac 23 s(r) +p(r)$ allows us 
to define the mechanical radius \cite{Polyakov:2018guq}
\begin{equation}
\label{eq:mechanicalradius}
	\langle r^2\rangle_{\rm mech} 
	=  \frac{\int d^3r\  r^2\ \left[\frac 23 s(r) +p(r)\right]}
		{\int d^3r\ \left[\frac 23 s(r) +p(r)\right]}
	= \frac{6 D}{\int_{-\infty}^0 dt\ D(t)} \, .
\end{equation}
Interestingly the mechanical radius is given by an ``anti-derivative'' 
of $D(t)$ at $t=0$ unlike e.g.\ 
the proton mean charge square radius 
$\langle r^2\rangle_{\rm charge}=6G_E^\prime(0)/G_E(0)$ given 
in terms of the electric form factor $G_E(t)$. 

One can also consider forces in lower-dimensional subsystems 
\cite{Polyakov:2018zvc_7}.
The 2D pressure $p^{(2D)}(r)=-\frac13 s(r)+p(r)$ satisfies 
$\int_0^\infty dr\,r\,p^{(2D)}(r)=0$ and corresponds to the 
tangential forces in (\ref{Eq:force-spherical-components}). 
Similarly the 1D pressure $p^{(1D)}(r)=-\frac43 s(r)+p(r)$ satisfies 
$\int_0^\infty dr\,p^{(1D)}(r)=0$. Generically, for a spherically symmetric
mechanical system in $nD$ one can express its pressure and shear forces 
in terms of pressure in $kD$ spherical subsystem as follows:
\begin{align}\label{eq:nDvskD}
  &p^{(nD)}(r)=\frac kn\ p^{(kD)}(r)+\frac{k (n-k)}{n}\;\frac{1}{r^k}
  \int_0^r dr^\prime\ r^{\prime\,k-1} p^{(kD)}(r^\prime), \\
  &s^{(nD)}(r)=-\frac {k}{n-1}\ p^{(kD)}(r)+\frac{k^2}{n-1}\;\frac{1}{r^k}
  \int_0^r dr^\prime\ r^{\prime\,k-1} p^{(kD)}(r^\prime).
\end{align}
Such relations can be useful, e.g.\ in holographic approaches to QCD.
The concepts can be generalized to higher spins \cite{Polyakov:2018rew}.
The energy density and pressure in the center of a hadron are given by
\cite{Polyakov:2018zvc_7}
\begin{equation}\label{eq:T00(0)+p(0)}
	T_{00}(0)=\frac{m}{4\pi^2}\  
    	\int_{-\infty}^0 dt\ \sqrt{-t}\ \biggl[
        A(t) - \frac{t}{4m^2}D(t)\biggr]\,, \quad
	p(0)=\frac{1}{24\pi^2 m} 
   	\int_{-\infty}^0 dt\ \sqrt{-t}\ \; t D(t).
\end{equation}

\section{\boldmath The $D$-term in theory and experiment}

In contrast to the constraints $A(0)=1$ and $B(0)=0$ resulting from 
properties of the states under Lorentz transformations 
\cite{Lowdon:2017idv}, the form factor 
$D(t)$ is not constrained (not even at $t=0$) by general principles. 
It is not related to external properties like Lorentz transformations 
but reflects internal dynamics inside the hadron. The value $D=D(0)$
is therefore not known for (nearly) any particle. 

In free field theories one finds $D=-1$ for free Klein Gordon fields
\cite{Kobzarev:1962wt,Hudson:2017xug}, and $D=0$ for free Dirac fields
\cite{Hudson:2017oul}. 
For Goldstone bosons of spontaneous chiral symmetry breaking it is 
predicted in the chiral limit that $D_{\rm Goldstone}=-1$ 
from soft pion theorems for EMT form factors \cite{Novikov:1980fa}
or pion GPDs\cite{Polyakov:1999gs_7}. Corrections due to finite masses 
are expected to be small for pions and larger for kaons and 
$\eta$ \cite{Donoghue:1991qv,Hudson:2017xug}.

For large nuclei in the liquid drop model \cite{Polyakov:2002yz_7} 
$p(r)=p_0 \theta(r-R) - \frac{p_0 R}{3} \delta(r-R)$ and
$s(r)=\gamma\delta(r-R)$ with nucleus radius $R=R_0A^{1/3}$ and
surface tension $\gamma$ related by the Kelvin 
relation $p_0=2 \gamma/R$ \cite{Kelvin}. It is 
predicted $\langle r^2\rangle_{\rm mech}=\frac 35 R^2$ and
$D =-\frac45\,m\,\gamma\,\frac{4\pi}{3}\,R^4 \propto A^{7/3}$
which is supported in Walecka model \cite{Guzey:2005ba}.

\begin{figure}[t!]
\centering
\includegraphics[height=4cm]{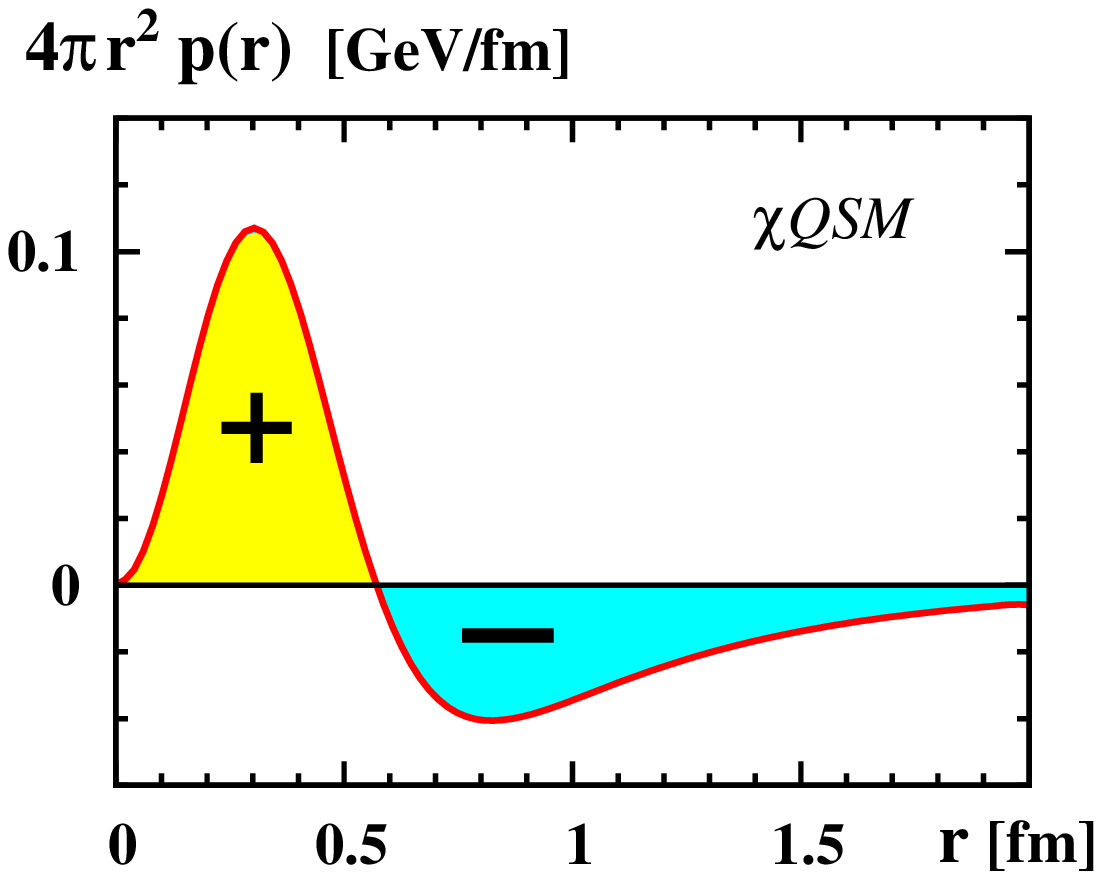}
\caption{\label{FIG-02:pressure-CQSM}
	Pressure in chiral quark soliton ($\chi$QSM) \cite{Petrov:1998kf} 
        and realization of the von Laue condition (\ref{Eq:von-Laue}).}
\end{figure}

The $D$-term of the nucleon was studied in the chiral quark soliton 
model \cite{Petrov:1998kf} which predicts $D\approx-3.5$ 
and $\langle r^2\rangle_{\rm mech}\approx0.75\,\langle r^2\rangle_{\rm charge}$.
Studies were also reported in Skyrme models including nuclear 
medium corrections \cite{Cebulla:2007ei},
bag model \cite{Ji:1997gm},
a Nambu--Jona-Lasinio diquark approach \cite{Freese:2019eww}, 
using dispersion relations \cite{Pasquini:2014vua},
chiral perturbation theory \cite{Belitsky:2002jp},
lattice QCD \cite{Hagler:2003jd},
and QCD lightcone sum rules \cite{Anikin:2019kwi}.
$D$-terms of mesons\cite{Freese:2019bhb}, $Q$-balls \cite{Mai:2012yc},
photons \cite{Friot:2006mm} and 
$\Delta$-resonance\cite{Perevalova:2016dln} were also studied.

A first extraction of the quark contribution to the pion $D$-term from 
the BELLE data \cite{Masuda:2015yoh} on $\gamma^*\gamma\to 2 \pi^0$ 
gave \cite{Kumano:2017lhr} $D^Q(0)\approx -0.75$ with
unestimated uncertainties. For the $D$-term of the nucleon 
phenomenological fits indicate that $D^Q<0$ with large uncertainties
\cite{Kumericki:2007sa_7}.
The $D$-term can be accessed in DVCS with help of fixed-$t$ dispersion 
relations \cite{Polyakov:2002yz_7,Teryaev:2005uj_7}
which relate the real and imaginary parts of the complex DVCS Compton form 
factors with a subtraction constant $\Delta(t,\mu)$ related to $D^Q(t,\mu)
=\frac{2}{5}\;\Delta(t,\mu)/(e_u^2+e_d^2)=\frac{18}{25}\,\Delta(t,\mu)$
under certain assumptions (large-$N_c$ limit, $\mu\to\infty$). 
An analysis of the JLab data \cite{Girod:2007aa} performed 
under such assumptions and additional constraints gave a first insight 
on  $\Delta(t,\mu)$ of the nucleon \cite{Burkert:2018bqq}. 
Relaxing these assumptions and constraints at the current 
stage yields much larger uncertainties \cite{Kumericki:2019ddg_7}
though the method in principle works. 

\section{Applications and Conclusions}

The EMT form factors have important applications including hard
exclusive reactions, the description of hadrons in strong gravitational 
fields, hadronic decays of heavy quarkonia \cite{Novikov:1980fa}, 
and the description of exotic hadrons with hidden charm as hadroquarkonia
\cite{Eides:2015dtr,Perevalova:2016dln}.

Unlike the EMT form factors $A(t)$ and $B(t)$ related
to the generators of the Poincar\'e group and ultimately to the
mass and spin of a particle, the form factor $D(t)$ is related 
to the internal forces and opens a new window for studies of the
hadron structure and visualization of internal hadronic forces.

\section{Acknowledgements}
This work was supported by NSF (grant no.\ 1812423)
and by CRC110 (DFG).


\newpage
%

\renewcommand*{\FigPath}{./WeekI/06_Kroll/}

 \newcommand{\nn}{\nonumber}
\renewcommand{\be}{\begin{equation}}
\renewcommand{\ee}{\end{equation}}
\newcommand{\ba}{\begin{eqnarray}}
\newcommand{\ea}{\end{eqnarray}}
\newcommand{\req}[1]{(\ref{#1})}
\def\={\,=\,}
\newcommand{\ci}[1]{\cite{#1}}
\renewcommand{\lsim}{\raisebox{-4pt}{$\,\stackrel{\textstyle
                                                         <}{\sim}\,$}}
\renewcommand{\gsim}{\raisebox{-4pt}{$\,\stackrel{\textstyle
                                                         <}{\sim}\,$}}

\def\mev{~{\rm MeV}}
\def\gev{~{\rm GeV}}
\def\kev{~{\rm keV}}
\def\ale{\alpha_{\rm elm}}
\def\als{\alpha_{\rm s}}
\def\eps{\epsilon}
\def\veps{\varepsilon}
\def\LQCD{\Lambda_{\rm QCD}}
\def\xbj{x_{\rm Bj}}
\def\muF{\mu^2_F}
\def\muR{\mu^2_R}
\def\muO{\mu^2_0}
\def\xb{\bar{x}}
\def\taub{\bar{\tau}}
\newcommand{\tw}{\textwidth}
\def\vk{{\bf k}_{\perp}}
\def\vl{{\bf l}_{\perp}}
\def\vd{{\bf \Delta}_\perp}
\def\vbs{{\bf b}}
\def\vb0{{\bf b}_0}
\def\vdabs{\Delta_\perp}
\def\xbj{x_{\rm Bj}}
\newcommand{\sla}{\hspace*{-0.20cm}/}
\newcommand{\Sla}{\hspace*{-0.22cm}/}
\renewcommand{\da}{{distribution amplitude}}
\newcommand{\wf}{wavefunction}
\newcommand{\tr}[1]{{\bf #1}_\perp}
\newcommand{\ov}[1]{\overline#1}
\def\sh{\hat{s}}
\def\uh{\hat{u}}
\def\th{\hat{t}}
\def\={\,=\,}

\wstoc{GPDs from meson electroproduction and applications}{Peter Kroll}

\title{GPDs from meson electroproduction and applications}

\author{Peter Kroll}

\address{Fachbereich Physik, Universitaet Wuppertal,\\
D-42097 Wuppertal, Germany\\
E-mail: pkroll@uni-wuppertal.de}
\index{author}{Kroll, P.}

\begin{abstract}
  This article briefly reviews an attempt to extract the generalized parton distributions
  (GPDs) from meson electroproduction data and discusses their applications such as their
  use in calculations of other hard exclusive processes, their behavior in the transverse
  position space or the evaluation of the parton angular momenta.
\end{abstract}

\bodymatter
    {~}\\
    {~}\\
The GPDs which encode the soft physics, are not calculable from first principles as yet
with the exception of some of their lowest moments evaluated from lattice QCD.
Thus, the GPDs have to be modeled or parametrized in order to fit experimental data on
meson-electroproduction data. Very profitable is the extraction of the GPDs  from experiment
in analogy to the determination of the familiar parton densities (PDFs).
In a series of papers \ci{GK1,GK2,GK3,GK5,GK6} we have attempted such a GPD analysis using
meson electroproduction data ($\rho^0$, $\phi$, $\pi^{+,0}$ and $\eta$). The GPDs are constructed
from the double distribution representation \ci{rad} where the double distribution is assumed
to be a product of a zero-skewness GPD and a weight function that generates the skewness ($\xi$)
dependence. The zero-skewness GPDs are parametrized as
\be
K_i(x,\xi=0,t) \= k_i(x) \exp{[tf_i(x)]}
\label{eq:ansatz}
\ee
where the $\xi=t=0$ limit, $k_i(x)$, is a PDF  in some cases (for the GPDs $H, \widetilde{H}, H_T$) or
parametrized like the PDFs. For small $-t$ it suffices to employ a simple profile
function  that bears similarity to the $t$-dependence of a Regge-pole contribution
\be
f_{Ri}(x)\=\alpha'_i\ln{x} + B_i
\label{eq:regge-profile}
\ee
where $\alpha'_i$ and $B_i$ are parameters.
For large $-t$ (larger than about $1\,\gev^2$) a more complicate profile function is required, e.g.\ \ci{DFJK4,DK13}
(similar profiles are proposed in \ci{brodsky,moutarde}):
\be
f_{DKi}(x)\=(\alpha'_i\ln{x} + B_i)(1-x)^3 + A_ix(1-x)^2\,.
\label{eq:DK-profile}
\ee

The GPDs are to be convoluted with suitable subprocess amplitudes which in principle can be
computed perturbatively. It turns out however that the leading-twist amplitudes  
do not suffer. From experiment  \ci{H1,dufurne} we learned that there are strong contributions from
transversely polarized virtual photons. Moreover,
also a leading-twist calculation of the longitudinal amplitude fails except at very large
values of $Q^2$ ( larger than about $50\,\gev^2$). In fact, the transverse size of the meson
cannot be ignored as has been pointed out in \ci{frankfurt} long time ago. We have have modeled
its effect by quark transverse momenta in the subprocess whereas the emission and reabsorption of
partons by the proton is treated collinearly to the proton momenta. Sudakov suppressions are also
taken into account. Since the Sudakov factor, $S$,
involves a resummation of all orders of perturbation theory in next-to-leading-log approximation
\ci{botts} which can only effectively be performed in the impact parameter ($\vbs_\perp$) space, canonically
conjugated to the quark transverse momentum space, we are forced to work in a $b$-space
\be
{\cal H}_{0\lambda,0\lambda} \= \int d\tau d^2\vbs_\perp \hat{\Psi}(\tau,b^2_\perp)
    \hat{\cal F}_{0\lambda,0\lambda}(x,\xi,\tau,Q^2,\vbs_\perp)
    \als(\mu_R) \exp{[-S(\tau,\vbs_\perp,Q^2)]}
\ee
where $\hat{\cal F}$ and $\Psi$ denote the Fourier transforms of the hard scattering kernel and the
meson's light-cone wave function, respectively. For the latter one a Gaussian in $\vbs_\perp$ is used with
a parameter that desribes the transverse size of the meson. This approach
also allows to calculate the amplitudes for transitions from a transverse photon, $\gamma^*_T$, to a likewise
polarized vector meson, $V_T$, as well as those for $\gamma_T^*\to V_L(P)$ transitions. In the latter case a
twist-3 meson wave function is required.  

The current state of the GPD extraction is the following:\\
The GPDs $H$ and $E$ for valence quarks are determined in an analysis of the nuclear form factors
\ci{DFJK4,DK13} and are subsequently used in the analyses of the electroproduction data. From the
longitudinal cross section data on $\rho^0$ and $\phi$ production for $Q^2\geq 3\,\gev^2$ and $W\geq 4\,\gev$
($\xi\lsim 0.1$, $-t\lsim 0.5\,\gev^2$) $H$ for gluons and sea quarks is fixed \ci{GK2}. The contribution
from $E$ is negligible in the longitudinal cross section at small $-t$, the other GPDs do not contribute. The
transverse size parameter, $a_V$, in the meson wave function are adjusted to the data. With $E$ for valence
quarks and $H$ for quarks and gluons at disposal also the transverse cross sections for $\rho^0$ and
$\phi$ production as well as many spin density matrix elements can be computed \ci{GK3}.

From the analysis of cross section and polarization data on $\pi$ and $\eta$ production \ci{HERMES,CLAS}
we learned about the GPDs ${\widetilde H}, H_T$ and $\bar{E}_T$ for valence quarks \ci{GK5,GK6}. There is no clear
signal in the data for contributions from $\tilde E$. For $\pi^+$ production there is also an important
contribution from the pion pole which is treated as a one-pion exchange contribution. The available data
on meson elecroproduction do not provide information on the GPDs ${\widetilde E}, \tilde E, H_T$ and $\bar{E}_T$
for gluons and sea quarks; $E_T$ and $\tilde E_T$ are yet completely unknown.

\begin{figure}[ht]
\centering  
\includegraphics[width=0.28\tw]{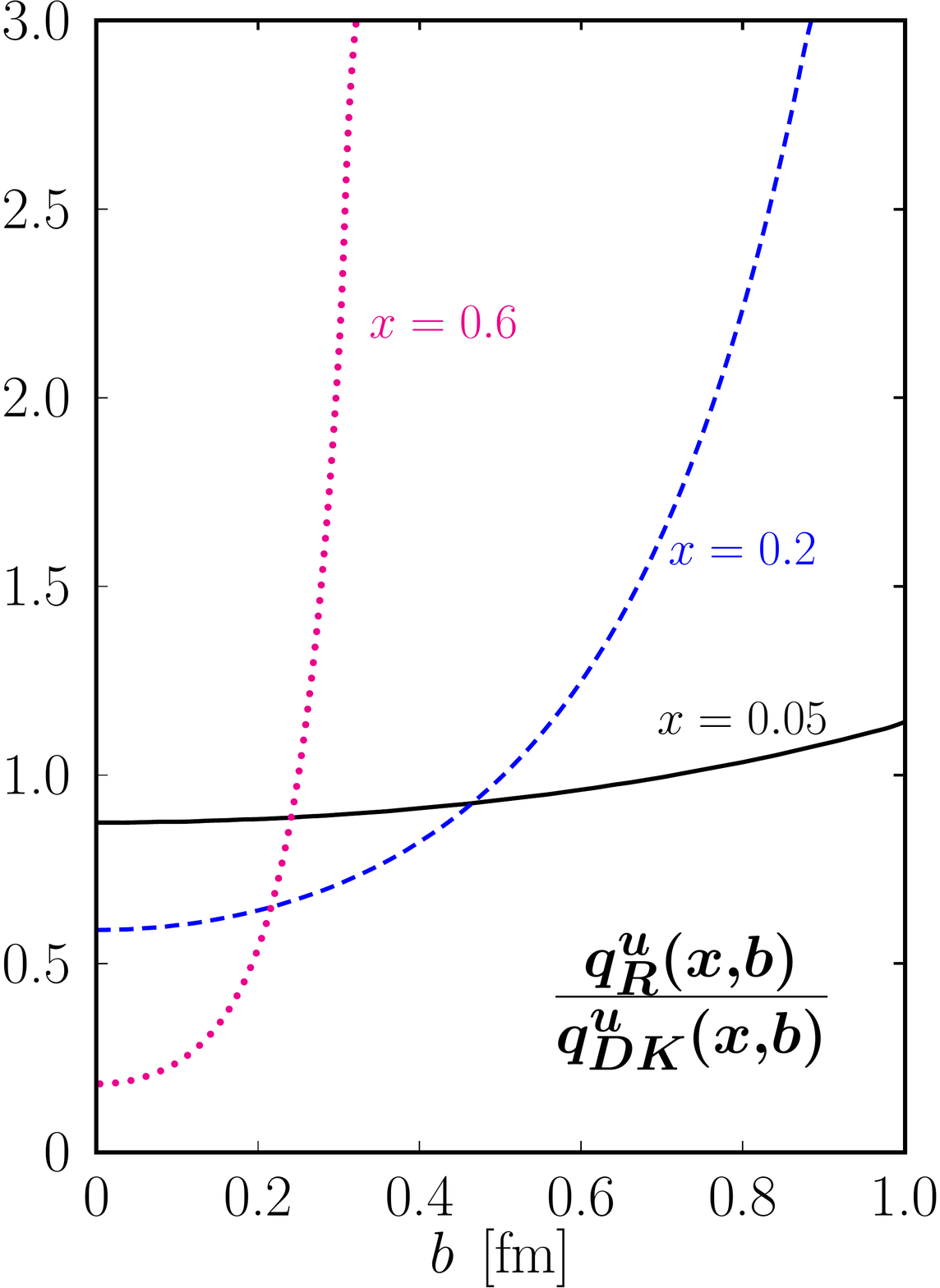} \hspace*{0.2\tw}  
\includegraphics[width=0.28\tw]{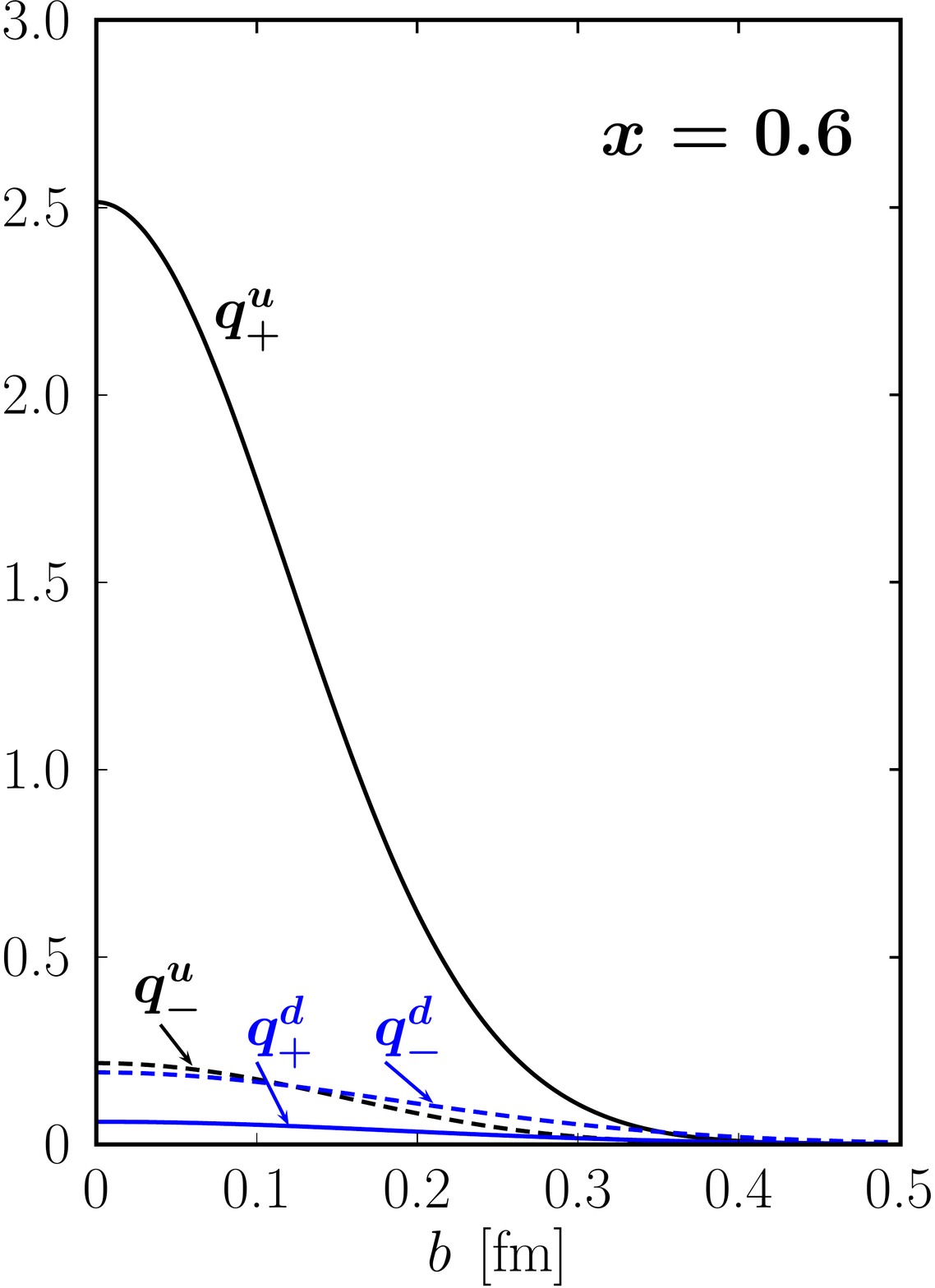}
\caption{\label{fig:1} Left: The ratio of $q_v^u(x,b^2)$ for the Regge-like profile
function \req{eq:regge-profile} and for \req{eq:DK-profile} at $x=0.05 (0.2, 0.6)$ solid 
(dashed, dotted) line. The scale is $2\,\gev$.}
\caption{\label{fig:2} Right: The impact parameter distributions for valence
    quarks with definite helicities at $x=0.6$ (in fm${}^{-2}$). $q_{\pm}(x,b^2)$ is evaluated 
    from the profile function \req{eq:DK-profile}. The scale is $2\,\gev$.} 
\end{figure}

The {\it Fourier transform (FT)} of the zero-skewness GPD \req{eq:ansatz} reads ($t=-\Delta^2_\perp$)
\be
k^i(x,\vbs)=\int \frac{d^2\vd}{(2\pi)^2} e^{-i\vbs\cdot \vd}K_i(x,\xi=0,t)
           =\frac1{4\pi}\,\frac{k_i(x)}{f_i(x)} \exp{[-b^2/(4f_i(x))]}\,.
\ee
It has been shown \ci{burkhardt,haegler} that these FTs, or better combinations of these FTs, have a density
interpretation. For instance, $q^a(x,\vbs)$, the FT of $H^a(x,\xi=0,t)$, describes the density of unpolarized
quarks of flavor $a$ in an unpolarized proton while $q^a_{\pm}=1/2[q^a(x,\vbs)\pm \Delta q^a(x,\vbs)]$ is the
density of flavor $a$ quarks with helicity (anti-) parallel to the proton helicity ($\Delta q$ is the FT of
$\widetilde H$). It can readily been shown \ci{DK13,kroll17} that the small $-t$ electroproduction data
($-t\lsim 1\,\gev^2$) are only sensitive to the GPDs in the small $x$-region ($x\lsim 0.1$). For their large
$x$-behavior information from sources other than electroproduction is required.
As an example the ratio of the FTs, $q^u$, of $H^u$ obtained  with either the profile function
\req{eq:regge-profile} or \req{eq:DK-profile}, is shown in Fig.\ \ref{fig:1}. Both the versions of $H^u$ describe
equally well the electroproduction data but they drastically differ at large $x$ and large $-t$. Only with
the profile function \req{eq:DK-profile} is  $H^u$ well in agreement with the large $-t$ proton form factor data.
In Fig.\ \ref{fig:2} the functions $q_{\pm}$ for valence quarks are displayed
at $x=0.6$. In the case at hand the required large $-t$ information comes from the nucleon's axial form factor
and from the  helicity correlation $K_{LL}$ in wide-angle Compton scattering \ci{kroll17}.

\begin{figure}[ht]
\centering  
\includegraphics[width=0.34\tw]{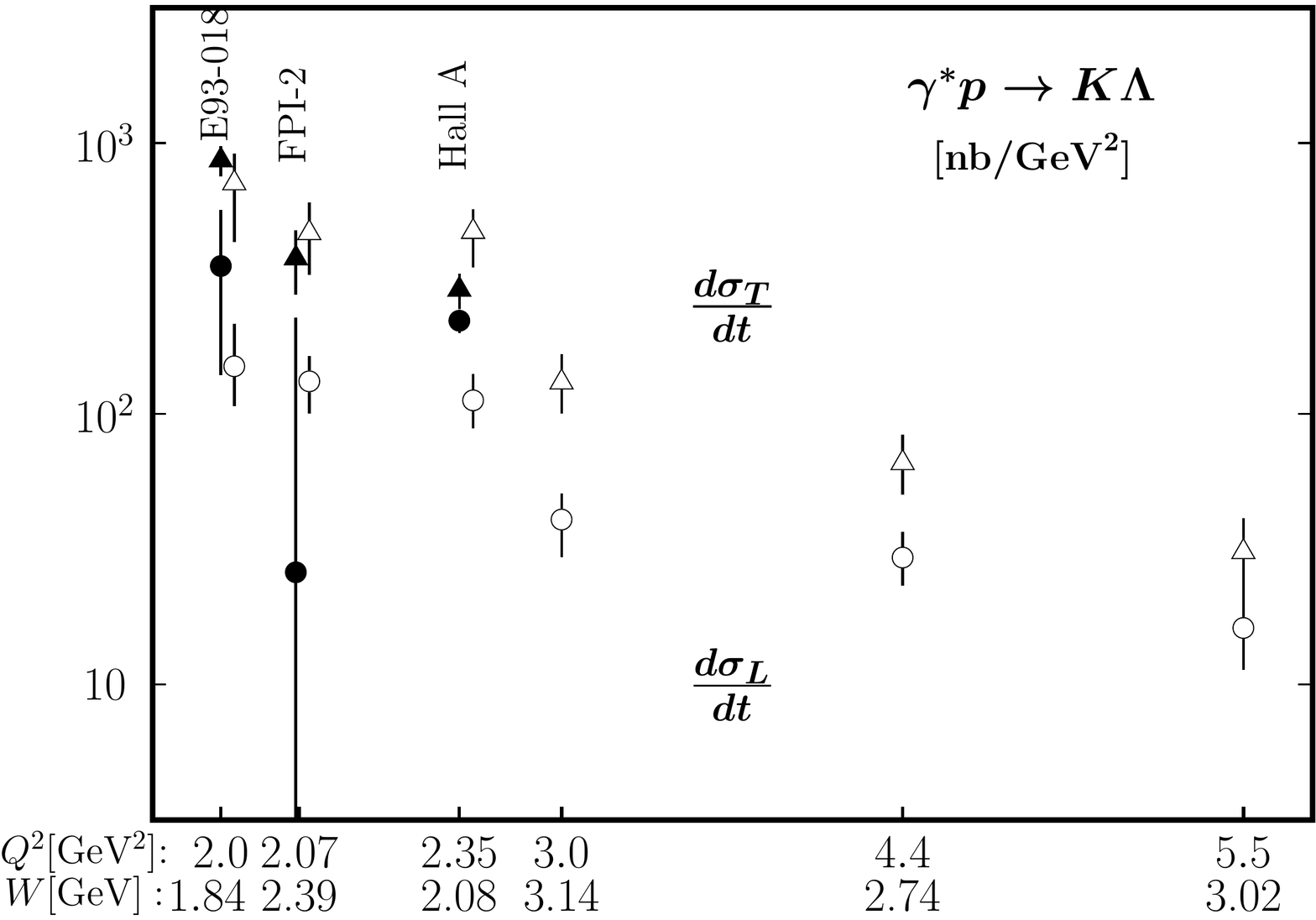} \hspace*{0.15\tw}  
\includegraphics[width=0.33\tw]{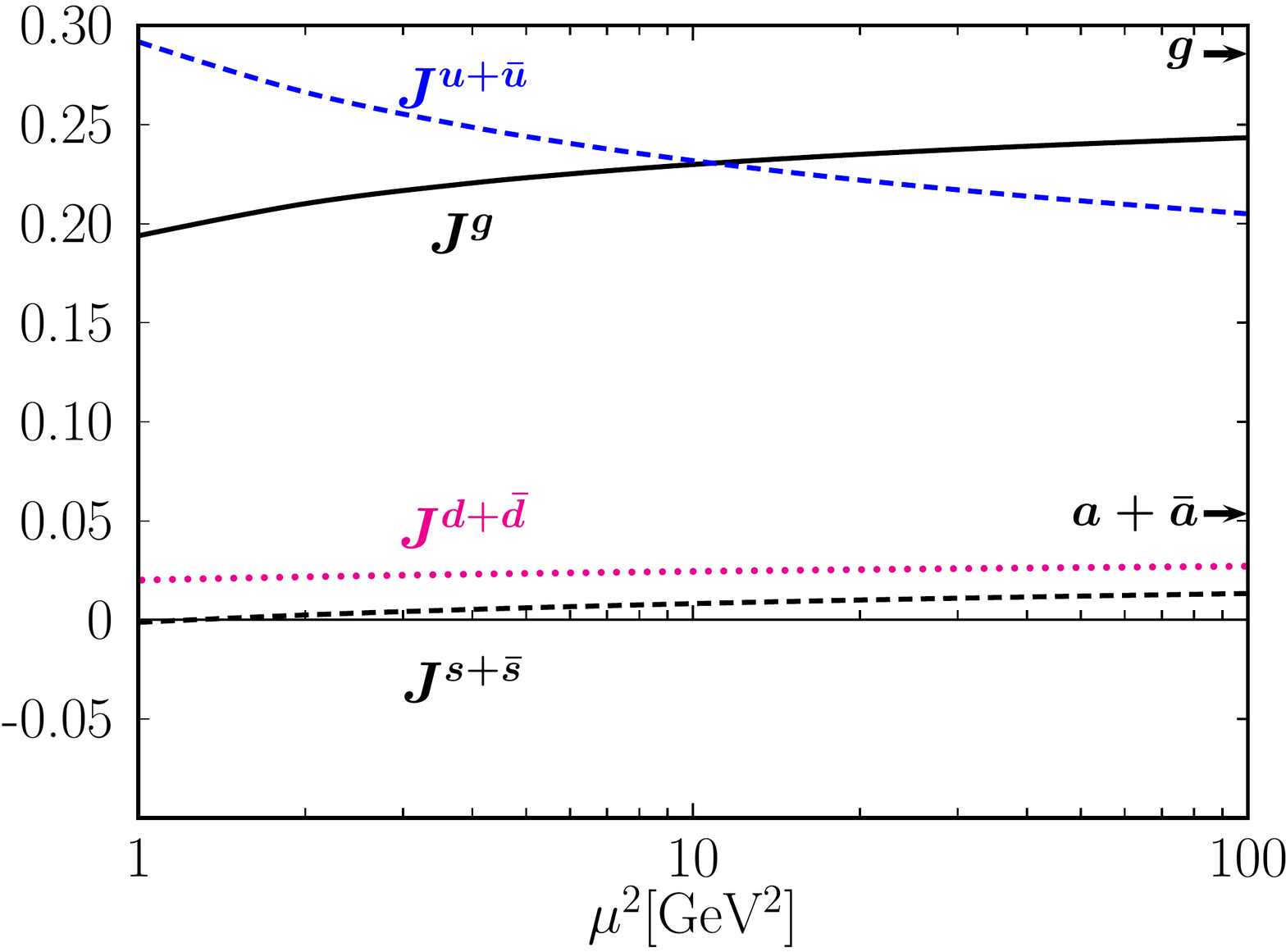}
\caption{\label{fig:3} Left: The forward longitudinal (circles) and transverse (triangles) cross section for
  $\gamma^*p\to K^+\Lambda$. Full (open) symbols represent the  data (predictions). References to data can be
  found in \ci{kroll18}.}
\caption{\label{fig:4} Right: The evolution of the parton angular momenta. The second moments of $H$ ($q^a_{20}$, $g_{20}$)
are evaluated from the ABM11 (NLO) PDFs. } 
\end{figure}

\vspace*{-0.05\tw}
The {\it universality} property of the GPDs, i.e.\ their process independence, now allows to calculate
several other hard exclusive processes. Among them are lepton-pair production in exclusive processes
(time-like DVCS \ci{pire}, $\pi^-p\to l^+l^-n$ \ci{GK9}) and neutrino meson production \ci{pire2,schmidt}
for which no data are available so far. Also $\omega$ and kaon production have been predicted \ci{DK7,kroll18}
and reasonable agreement with the available data has been found. An example is shown in Fig.\ \ref{fig:3}.
Another process which can now be calculated free of parameters, is DVCS. A leading-twist calculation \ci{sabatie}
leads to very good agreement with the DVCS cross section and polarization data except for Jlab kinematics.
In this region higher-twist corrections, at least those of kinematical origin are required \ci{braun}.
Interestingly, some modulations of the DVCS asymmetry measured with a transversely polarized target
provide some, admittedly rough, information on the GPD $E$ for sea quarks (assuming a flavor symmetric sea).
This information in combination with positivity bounds and the sum rule for the second moments, $e^a_{20}$, of $E$
allows for an evaluation of the {\it parton angular momenta}
\be
J^a\=\frac12\big[q^q_{20}+e^a_{20}\big]\,, \qquad J^g\=\frac12\big[g_{20}+e^g_{20}\big]\,.
\ee
The results for $J$ in dependence on the scale are displayed in Fig.\ \ref{fig:4}.

{\it Summary:} I briefly reported on the present status of the extraction of the GPDs from meson electroproduction
data. Needless to say that this first attempt requires improvements. More and more accurate data
which will come from Jlab12 and perhaps from the EIC, will help. With the GPDs at disposal one can
predict other hard space- and time-like exclusive processes and 
study many propertie of the GPDs.


\newpage
%


\wstoc{Lattice calculation of the generalized parton distributions with the LaMET approach}{Yong Zhao}

\title{Lattice calculation of the generalized parton distributions with the LaMET approach}

\author{Yong Zhao}
\index{author}{Zhao, Y.}

\address{Center for Theoretical Physics, Massachusetts Institute of Technology,\\
Cambridge, Massachusetts 02139, USA\\
www.mit.edu}
\address{Physics Department, Brookhaven National Laboratory, Bldg. 510A, \\
Upton, NY 11973, USA\\
$^*$E-mail: yzhao@bnl.gov\\
www.bnl.gov}

\begin{abstract}
The generalized parton distributions (GPDs) can be calculated from the corresponding quasi-GPDs through the large-momentum effective theory approach. Using operator product expansion, we strictly derive the factorization formula which matches the quasi-GPDs to GPDs. We also calculate the one-loop matching coefficients for the quasi-GPDs in the $\overline{\rm MS}$ and a regularization-independent momentum subtraction scheme. These results will lay the foundation for a systematic extraction of GPDs from lattice QCD.
\end{abstract}


\bodymatter

\section{Introduction}\label{sec:intro}


Over the past few years, lattice QCD has made remarkable progress towards the calculation of the $x$-dependence of parton distribution functions (PDFs). Among them, the large-momentum effective theory (LaMET)~\cite{Ji:2013dva_63,Ji:2014gla_14} has gained the most interest and been implemented in the lattice calculation of isovector PDFs of different spin structures (for reviews see Refs.~\citenum{Monahan:2018euv,Cichy:2018mum,Zhao:2018fyu}). The factorization formula for the PDF has been proven~\cite{Ma:2014jla,Ma:2017pxb,Izubuchi:2018srq}, and the perturbative matching coefficients in the factorization formulas have been derived for quasi-PDFs~\cite{Xiong:2013bka,Izubuchi:2018srq}. A nonperturbative renormalization of the quasi-PDF in the regularization-independent momentum subtraction (RI/MOM) scheme has been performed on the lattice~\cite{Alexandrou:2017huk,Chen:2017mzz_14}, and their matching to the $\overline{\rm MS}$ PDF has been well understood at one-loop order~\cite{Constantinou:2017sej,Stewart:2017tvs,Liu:2018uuj_14}. 

As for the GPDs, the factorizaiton of quasi-GPDs into GPDs have been studied at one-loop order for the isovector quark case with different spin structures~\cite{Ji:2015qla,Xiong:2015nua}, and the one-loop matching coefficients were obtained in a transverse momentum cutoff scheme. Due to the linear power divergence in the quasi-GPDs, this scheme is not suitable in the lattice implementation. Therefore, the renormalization and matching of the nonsinglet quark quasi-GPDs are re-examined in the RI/MOM scheme~\cite{Liu:2019urm}, with the matching coefficient calculated at one-loop order. In this writing, we present the derivation and results in Ref.~\citenum{Liu:2019urm}.

\section{Definitions and conventions}\label{sec:GPD}
The parent function for the nonsinglet quark GPDs, which we call parent-GPD for simplicity, is defined from the Fourier transform of the off-forward matrix element of a light-cone correlator, 
\begin{equation}\label{eq:GPD}
F(\bar\Gamma,x,\xi,t,\mu)\!
=\!\! \int \frac{d\zeta^-}{4\pi}e^{-i x\zeta^- P^+}\!\langle P''S''| \bar\psi\big(\!\frac{\zeta^-}{2}\!\big)\bar\Gamma\lambda^a W_+\big(\!\frac{\zeta^-}{2}\!,\!-\!\frac{\zeta^-}{2}\!\big)\psi\big(\!-\!\frac{\zeta^-}{2}\!\big)|P'S'\rangle\,,
\end{equation}
where $x\in [-1,1]$, the light-cone coordinates $\zeta^\pm=(\zeta^t\pm\zeta^z)/\sqrt{2}$, and the hadron state $|P'S'\rangle$ ($|P''S''\rangle$) is denoted by its momentum and spin. The parent-GPD is defined in the $\overline{\rm MS}$ scheme and $\mu$ is the renormalization scale.
The kinematic variables are defined as
\begin{align}\label{eq:kinematic_variable}
\Delta\equiv P''-P',\;\; t\equiv \Delta^2,\;\; \xi\equiv -\frac{P''^+-P'^+}{P''^++P'^+}=-\frac{\Delta^+}{2P^+}\,,
\end{align}
where without loss of generality we choose a particular Lorentz frame so that the average momentum
$P^\mu\equiv (P''^\mu+P'^\mu)/2=(P^t,0,0,P^z)$,
and only consider the case with $0<\xi<1$.
$\bar\Gamma=\gamma^+$, $\gamma^+\gamma_5$, and $i\sigma^{+\perp}=\gamma^\perp\gamma^+$ correspond to the unpolarized, helicity, and transversity parent-GPDs, respectively.
$\lambda$ is a Gell-Mann matrix in flavor space, e.g., $\lambda^3$ corresponds to flavor isovector ($u-d$) distribution. $W_+\big(\frac{\zeta^-}{2},\!-\frac{\zeta^-}{2}\big)$ is a lightlike Wilson line connecting $-\zeta^-/2$ to $\zeta^-/2$.

The GPDs are defined as form factors of the parent-GPD, which we do not specify here. As we shall see, the matching coefficient is universal for all the GPDs defined from the same bilinear operator.

To calculate the quark GPDs within LaMET, we consider a quark quasi-parent-GPD defined from an equal-time correlator:
\begin{equation}\label{eq:quasiGPD}
\widetilde F(\Gamma,x,\widetilde{\xi},t,P^z,\widetilde\mu)\!=\! {2P^z\over N}\int \frac{dz}{4\pi}e^{i xz P^z}\langle P''S''|\bar\psi\big(\frac{z}{2}\big)\Gamma\lambda^a W_z\big(\!\frac{z}{2},\!-\!\frac{z}{2}\!\big)\psi\big(\!-\!\frac{z}{2}\!\big)|P'S'\rangle\,,
\end{equation}
where $\widetilde{\mu}$ is the renormalization scale in a particular scheme, {and $N$ is a normalization factor that depends on the choice of $\Gamma$. In order to minimize operator mixing on lattice~\cite{Constantinou:2017sej,Green:2017xeu_14,Chen:2017mie}, we choose $\Gamma=\gamma^t$, $\gamma^z\gamma_5$, and $i\sigma^{z\perp}$ for the unpolarized, helicity, and transversity quasi-GPDs, which all correspond to the same normalization factor $N=2P^t$. $W_z(z_2,z_1)$ is a spacelike Wilson line that connects $z_1$ to $z_2$.

The kinematic variables are similar to those in Eq.~(\ref{eq:kinematic_variable}) except that the ``quasi" skewness parameter
\begin{align}
\widetilde{\xi}=-\frac{P''^z-P'^z}{P''^z+P'^z}=-\frac{\Delta^z}{2P^z} =  \xi +\mathcal{O}\left({M^2\over P_z^2}\right)\,,
\end{align}
which is equal to $\xi$ up to power corrections. From now on we will replace $\widetilde \xi$ with $\xi$ by assuming that the power corrections are small.

\section{Operator product expansion and the factorization formula}\label{sec:OPE}

In this section, we derive the explicit form of the factorization formula for the quasi-GPDs using the OPE of the nonlocal quark bilinear $\widetilde{O}(\Gamma,z)=\bar\psi\Gamma W_z\psi$. The same method has been used for the ``lattice cross section''~\cite{Ma:2017pxb} and quasi-PDF~\cite{Izubuchi:2018srq}, which are both forward matrix elements of a nonlocal gauge-invariant operator. In the off-forward case, the matrix element of $\widetilde{O}(\Gamma,z)$ at leading twist approximation is
\begin{align}\label{eq:ope-twist2}
\langle P'|\widetilde{O}(\gamma^z,z,\mu)|P\rangle  = & 2P^z\sum_{n=0}^\infty C_n ({\mu}^2 z^2){\cal F}_n(-zP^z)\sum_{m=0}^n{\cal B}_{nm}(\mu)\ \xi^n\nonumber\\
&\times\! \int_{-1}^1 dy C_m^{3/2}\big({y\over \xi}\big)\! F(\gamma^+\!,y,\xi,t,\mu) + \mathcal{O}\big({M^2\over P_z^2},{t\over P_z^2},z^2\Lambda_{\rm QCD}^2\big)\,,
\end{align}
where ${\cal F}_n(-zP^z)$ are partial wave polynomials whose explicit forms are known in the conformal OPE~\cite{Braun:2007wv}, and $C_m^{3/2}(x)$ are Gegenbauer polynomials. ${\cal B}_{nm}(\mu)$ diagonalize the renormalization group equations for the conformal operators~\cite{Braun:2003rp,Mueller:1993hg}. The higher-twist terms contribute to $\mathcal O(z^2\Lambda_{\rm QCD}^2)$. Then we can Fourier transform Eq.~\ref{eq:ope-twist2} from $z$ to $xP^z$ to obtain the quasi-GPD and its factorization formula,
\begin{align}
\widetilde F(\gamma^z\!,x,\xi,t,P^z\!,\mu) = & \int_{-1}^1 {dy\over |\xi|}\bar{C}_{\gamma^z}\big({x\over \xi},\! {y\over \xi},\! {\mu \over \xi P^z }\big)F(\gamma^+\!,y,\xi,t,\mu) + \mathcal{O}(1/P_z^2)\,,\nonumber\\
=& \int_{-1}^1 {dy\over |y|}C_{\gamma^z}\big({x\over y},\! {\xi\over y},\! {\mu \over yP^z }\big)F(\gamma^+\!,y,\xi,t,\mu) + \mathcal{O}(1/P_z^2)\,,\label{eq:fact2}
\end{align}
where the defintions of the two matching coefficients $\bar{C}$ and $C$ are given in Ref.~\citenum{Liu:2019urm}. The second form in Eq.~\ref{eq:fact2} was postulated in Refs.~\citenum{Ji:2015qla,Xiong:2015nua}. Based on Eq.~\ref{eq:fact2}, we can infer that the matching coefficients for all the quasi-GPDs must be the same. Similar formulas can also be derived for the helicity and transversity cases.

The factorization formulas in Eq.~\ref{eq:fact2} are similar to the evolution equations for the GPDs~\cite{Mueller:1998fv_61,Ji:1996nm_62}. Notably, at zero skewness $\xi=0$, we have
\begin{equation} \label{eq:zeroskew}
\widetilde F(\gamma^z,x,0,t,P^z,\mu) 
= \int_{-1}^1 {dy\over |y|}C_{\gamma^z}\left({x\over y}, 0, {\mu \over yP^z }\right)F(\gamma^+,y,0,t,\mu) + \mathcal{O}(1/P_z^2)\,,
\end{equation}
where the matching kernel $C_{\gamma^z}(x/y,0,\mu/(yP^z))$ is exactly the same matching coefficient for the $\overline{\rm MS}$ quasi-PDF~\cite{Izubuchi:2018srq}, even when $t\neq0$. Moreover, in the forward limit $\xi\to0$ and $t\to0$, Eq.~\ref{eq:zeroskew} is exactly the factorization formula for the $\overline{\rm MS}$ quasi-PDF~\cite{Izubuchi:2018srq}.

On the other hand, in the limit $\xi\to1$ and $t\to0$, we obtain the factorization formula for the quasi distribution amplitude (DA),
\begin{equation}
\widetilde F(\gamma^z,x,1,t=0,P^z,\mu) =  \int_{-1}^1 dy\ \bar{C}_{\gamma^z}\big(x, y, {\mu \over P^z }\big)F(\gamma^+,y,1,t=0,\mu) + \mathcal{O}(1/P_z^2)\,,
\end{equation}
whose explicit form has been postulated in Refs.~\citenum{Zhang:2017bzy,Xu:2018mpf,Liu:2018tox}.

\section{Renormalization and matching in the RI/MOM scheme}\label{sec:renormalization}

Since the UV divergence of the quasi-GPD only depends on the operator $\widetilde O(\Gamma,z)$, not on the external states, we can choose the same renormalization factors as those for the quasi-PDF to renormalize the quasi-GPDs. These renormalization factors have been nonperturbatively calculated on the lattice~\cite{Alexandrou:2017huk,Chen:2017mzz_14}, and their contributions to the matching as counter-terms have also been calculated at one-loop order in continumm perturbation theory~\cite{Constantinou:2017sej,Stewart:2017tvs,Liu:2018uuj_14}.

When the hadron momentum $P^z$ is much greater than $M$ and $\Lambda_{\rm QCD}$, the RI/MOM quasi-GPD can be matched onto the $\overline{\rm MS}$ GPD through the factorization formula~\cite{Stewart:2017tvs,Izubuchi:2018srq}
	\begin{equation}\label{eq:matching2}
	\widetilde F(\Gamma,x,\xi,t,P^z\!,p_R)\!=\!\int_{-1}^1 \frac{dy}{|y|} C_\Gamma\big(\frac{x}{y},\!\frac{\xi}{y},\!r,\!\frac{yP^z}{\mu},\!\frac{yP^z}{p_R^z}\big)\! F(\bar\Gamma,y,\xi,t,\mu)+ {\cal O}(1/P_z^2)\,,
	\end{equation}
where $r=(p_R)^2/(p^z_R)^2$, and
\begin{equation}
C_\Gamma\big(x,\xi,r,\!\frac{p^z}{\mu},\!\frac{p^z}{p^z_R}\big)\!=\!\delta(1-x)+C^{(1)}_B\big(\Gamma,x,\xi,\!\frac{p^z}{\mu},\!{\mu\over \mu'}\big)-C^{(1)}_{CT}\big(\Gamma,x,r,\!\frac{p^z}{p^z_R},\!\frac{\mu_R}{\mu'}\big)\,,
\end{equation}
with 
\begin{equation}
C^{(1)}_{B}\big(\Gamma,x,\xi,\frac{p^z}{\mu}\!,\!{\mu\over \mu'}\big)\!=\!f_1\big(\Gamma,x,\xi,\!\frac{p^z}{\mu}\big)_+\! +\! \delta_{\Gamma,i\sigma^{z\perp}}\delta(1-x)\frac{\alpha_s C_F}{4\pi}\!\left[\!-\frac{1}{\varepsilon}\!+\!\ln\left(\frac{\mu^2}{\mu'^2}\right)\right]\,,
\end{equation}
where the subscript $B$ denotes ``bare" for the quasi-GPD, and $\varepsilon$ regulates the ultraviolet (UV) divergence. The results for $f_1$ can be found in Ref.~\citenum{Liu:2019urm}. The matching coefficients reduce to those for the quasi-PDFs~\cite{Izubuchi:2018srq,Liu:2018hxv_14} when $\xi=0$ even if $t\neq 0$, as well as those for the DAs~\cite{Liu:2018tox} by the replacement $\xi\to1/(2y-1)$, $x/\xi\to 2x-1$, and $p^z\to p^z/2$~\cite{Ji:2015qla}.

As we argued in Sec. \ref{sec:renormalization}, we can use the renormalization factor for the quasi-PDF to renormalize the quasi-GPD, which leads to the same one-loop RI/MOM counterterm~\cite{Stewart:2017tvs,Liu:2018uuj_14}.
Finally, with the replacements $x\to x/y$, $\xi\to \xi/y$, and $p^z\to yP^z$~\cite{Stewart:2017tvs,Izubuchi:2018srq}, we obtain the RI/MOM matching coefficient,
\begin{align}
C^{(1)}_\Gamma\big({x\over y},{\xi\over y},r,\!\frac{yP^z}{\mu},\!\frac{yP^z}{p^z_R}\big)=&\left[f_1\big(\Gamma,\!{x\over y},\!{\xi\over y},\!\frac{yP^z}{\mu}\big)-\left|\frac{y P^z}{p^z_R}\right|f_2\big(\Gamma,\frac{y P^z}{p^z_R}\big({x\over y}-1\big)+1,r\big)\right]_+\nonumber\\
&+\delta_{\Gamma,i\sigma^{z\perp}}\delta\left(1-{x\over y}\right)\frac{\alpha_s C_F}{4\pi}\ln\left(\frac{\mu^2}{\mu_R^2}\right)+\mathcal{O}(\alpha_s^2)\,,
\end{align}
where $f_2(\Gamma,x,r)$ for different spin structures can be found in Refs.~\citenum{Liu:2018uuj_14,Liu:2018hxv_14}.

\section{Conclusion}\label{sec:conclusion}

Within the framework of LaMET, we have derived the one-loop matching coefficients that match the isovector quark quasi-GPDs in the RI/MOM scheme to GPDs in the $\overline{\rm MS}$ scheme for different spin structures. We also presented
a rigorous derivation of the factorization formula for isovector quark quasi-GPDs based on OPE. As a result, for quasi-GPDs with zero skewness the matching coefficient is the same as that for the quasi-PDF. Our results will be used to extract the isovector quark GPDs from lattice calculations of the corresponding quasi-GPDs.

\section*{Acknowledgments}
This work was supported by the U.S.\ Department of Energy, Office of Science,
Office of Nuclear Physics, from DE-SC0011090, DE-SC0012704 and within the framework of the TMD Topical Collaboration.




\newpage
%

\renewcommand*{\FigPath}{./WeekI/08_Guzey/}

\wstoc{Nuclear shadowing in exclusive processes}{Vadim Guzey}

\title{Nuclear shadowing in exclusive processes}

\author{Vadim Guzey$^*$}
\index{author}{Guzey, V.}

\address{National Research Center ``Kurchatov Institute'', Petersburg Nuclear Physics Institute (PNPI), 
Gatchina, 188300, Russia\\
Department  of  Physics,  University  of  Jyv\"askyl\"a, P.O.
 Box 35, 40014  University  of  Jyv\"askyl\"a,  Finland\\
 Helsinki Institute of Physics, P.O.  Box  64,  00014 
 University  of  Helsinki,  Finland\\
$^1$E-mail: guzey\_va@pnpi.nrcki.ru}

\address{Department  of  Physics,  University  of  Jyv\"askyl\"a, P.O.
 Box 35, 40014  University  of  Jyv\"askyl\"a,  Finland}

\begin{abstract}

We show that exclusive photoproduction of $J/\psi$ in ultraperipheral collisions of heavy ions at the LHC constrains
the nuclear gluon density at small $x$ and gives evidence of large nuclear gluon shadowing.

\end{abstract}


\bodymatter

\section{Introduction}
\label{sec:intro}

Nuclear parton distribution functions (nPDFs) are fundamental quantities of QCD, which describe quark and gluon 
distributions in nuclei as a function of the light-cone momentum fraction $x$ at a resolution scale $\mu$. 
These  distributions quantify so-called cold nuclear matter effects (deviations of nPDFs from the sum of free proton and 
neutron PDFs)
 and are also needed for quantitative estimates of 
the onset of the gluon saturation, which are essential for phenomenology of 
hard processes with nuclei at the Relativistic Heavy Ion Collider (RHIC), the Large Hadron Collider (LHC), and 
the future Electron-Ion Collider (EIC), the Future Circular Collider (FCC), and the Large Hadron-Electron Collider (LHeC).

Within the framework of the QCD collinear factorization, nPDFs are determined using available data on lepton deep inelastic
scattering (DIS) on fixed nuclear targets and selected hard processes in deuteron-nucleus scattering at RHIC (inclusive pion production) and proton-nucleus scattering at the LHC (gauge boson and dijet production).
Unfortunately, it does not constrain well nPDFs, which as a result are known with large uncertainties.
\cite{deFlorian:2003qf,Hirai:2007sx,Eskola:2009uj,deFlorian:2011fp,Kovarik:2015cma_47,Khanpour:2016pph,Eskola:2016oht_46} 
This is especially acute for small $x$, where nPDFs are expected to be suppressed due to nuclear shadowing.
The improvement in the determination of nPDFs using the LHC Run 2 data is rather modest.\cite{Eskola:2019dui,Eskola:2019zqy} In the future, nPDFs -- especially the poorly known gluon density -- will be constrained very well at an EIC taking advantage of its wide kinematic coverage and measurements of the longitudinal 
$F_L^{A}(x,Q^2)$ and the charm $F_2^{c}(x,Q^2)$ nuclear structure functions.\cite{Aschenauer:2017oxs}

Another possibility to constrain nPDFs at small $x$ is provided by ultraperipheral collisions (UPCs) of heavy ions at the LHC, 
which give an opportunity to study real photon-nucleus scattering at unprecedentedly high energies.\cite{Baltz:2007kq}.
In particular, coherent photoproduction of $J/\psi$ on nuclei directly probes the nuclear gluon density.\cite{Ryskin:1992ui_44,Guzey:2013xba,Guzey:2013qza}
 In the remainder of this contribution, we explain how this process can be used to obtain new constraints on the nuclear gluon density at small $x$.

\section{Gluon nuclear shadowing from coherent $J/\psi$ photoproduction on nuclei}
\label{sec:gluon}

In UPCs heavy ions interact at large impact parameters $b \gg 2 R_A$ ($R_A$ is the effective radius of colliding nuclei)
so that the strong interaction is suppressed and the interaction proceeds via the exchange of quasi-real photons (the Weizs\"acker-Williams method of equivalent photons). The cross section of coherent $J/\psi$ photoproduction in UPCs of ions $A$ reads
\begin{equation}
\frac{d\sigma_{AA \to AA J/\psi}(y)}{dy}=N_{\gamma/A}(y) \sigma_{\gamma A \to J/\psi A}(y)+N_{\gamma/A}(-y) \sigma_{\gamma A \to J/\psi A}(-y) \,,
\label{eq:upc}
\end{equation}
where $N_{\gamma/A}(y)$ is the photon flux known from QED; $\sigma_{\gamma A \to J/\psi A}$ is the photoproduction cross section;
$y=\ln[W_{\gamma p}^2/(2 \gamma_L m_N M_{J/\psi})]$ is the $J/\psi$ rapidity; $W_{\gamma p}$ is the invariant photon-nucleon energy.
 Since each ion can serve as a source of photons and as a target,
there are two terms in Eq. (\ref{eq:upc}) corresponding to contributions of high-energy and low-energy photons, respectively.

In the leading logarithmic approximation (LLA) of perturbative QCD and the non-relativistic (static) approximation for the charmonium wave function, the cross section of exclusive $J/\psi$ photoproduction is proportional to the gluon density squared.\cite{Ryskin:1992ui_44}
Applying this to the nuclear target, one obtains\cite{Guzey:2013xba,Guzey:2013qza}
\begin{equation}
\sigma_{\gamma A \to J/\psi A}(W_{\gamma p})=\kappa_{A/N}^2 \frac{d\sigma_{\gamma p \to J/\psi p}(W_{\gamma p},t=0)}{dt} 
\left[\frac{g_A(x,\mu^2)}{A g_N(x,\mu^2)}\right]^2 \Phi_A(t) \,,
\label{eq:upc2}
\end{equation}
where $g_A(x,\mu^2)$ and $g_N(x,\mu^2)$ are the nucleus and nucleon gluon densities, respectively;
$x=M_{J/\psi}^2/W_{\gamma p}^2$;
$d\sigma_{\gamma p \to J/\psi p}/dt$ is the cross section on the proton, which has been measured in the relevant kinematics at HERA and by the LHCb collaboration; $\Phi_A(t)=\int dt |F_A(t)|^2$ is calculated using the nuclear form factor
$F_A(t)$; the parameter $\kappa_{A/N}=0.9-0.95$ corrects for the effect of skewness, see the discussion in Sec.~\ref{sec:discussion}.
It is useful to define the nuclear suppression factor of $S_{Pb}$,
\begin{equation}
S_{Pb}(W_{\gamma p})=\left[\frac{\sigma_{\gamma A \to J/\psi A}(W_{\gamma p})}{\sigma_{\gamma A \to J/\psi A}^{\rm IA}(W_{\gamma p})} \right]^{1/2}=\kappa_{A/N} \frac{g_A(x,\mu^2)}{A g_N(x,\mu^2)} \,,
\label{eq:upc3}
\end{equation}
where $\sigma_{\gamma A \to J/\psi A}^{\rm IA}$ is the cross section calculated in the impulse approximation, which implies no nuclear
modifications of nPDFs. The left-hand side of Eq. (\ref{eq:upc3}) can be evaluated model-independently using the
LHC UPC data and parametrizations of $d\sigma_{\gamma p \to J/\psi p}(W_{\gamma p},t=0)/dt$  and $F_A(t)$; 
the gluon shadowing ratio $R_g=g_A(x,\mu^2)/[A g_N(x,\mu^2)]$ in
the right-hand side of Eq. (\ref{eq:upc3}) is predicted by global QCD fits of nPDFs and the leading twist model of nuclear shadowing\cite{Frankfurt:2011cs}. 
Figure~\ref{fig:S_Pb} shows the comparison of values of $S_{Pb}$ extracted from the ALICE and CMS data with the theoretical predictions, see Ref.~\citenum{Guzey:2013qza} for details. As one can see from the figure, the good agreement with ALICE and CMS data on coherent $J/\psi$ photoproduction in Pb-Pb UPCs at 2.76 TeV gives direct evidence of large gluon nuclear shadowing, 
$R_g(x=0.001,\mu^2=3 \ {\rm GeV}^2) \approx  0.6$.

\begin{figure}[t]%
\begin{center}
\includegraphics[width=4.in]{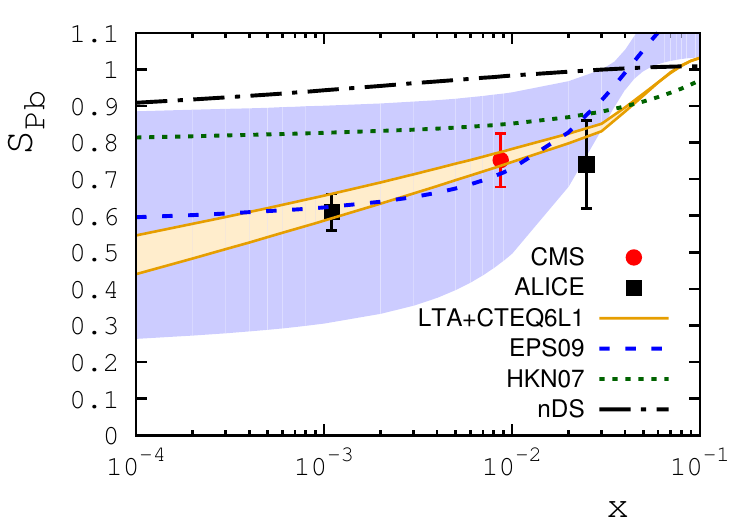}
 \caption{The nuclear suppression factor of $S_{Pb}$ for coherent $J/\psi$ photoproduction in Pb-Pb UPCs at 2.76 TeV: the values 
 extracted from the ALICE and CMS data vs. predictions of global QCD fits of nPDFs and the leading twist nuclear shadowing
 model, see Ref.~\citenum{Guzey:2013qza} for details. }
\label{fig:S_Pb}
\end{center}
\end{figure}

\section{Discussion and conclusions}
\label{sec:discussion}

There has been a wealth of theoretical investigations of exclusive $J/\psi$ photoproduction in pQCD with the aim to go beyond the 
LLA used in Ref.~\citenum{Ryskin:1992ui_44}.
 To stay within the collinear
factorization framework, we note the recent next-to-leading order (NLO) calculation in the framework of generalized parton distributions (GPDs).\cite{Ivanov:2004vd,Jones:2016ldq}
Not only NLO corrections were found to be very large at small $x$, but also GPDs in general differ from PDFs. Hence, 
it remains a challenge to rigorously employ the UPC data on exclusive $J/\psi$ photoproduction  on nuclei in global fits on nPDFs.
In our analysis, we used a model-dependent connection between the gluon GPD and the usual gluon density, 
which resulted in the  factor of $\kappa_{A/N}$ in Eqs. (\ref{eq:upc2}) and (\ref{eq:upc3}).
At the same time, photoproduction of dijets in UPCs probes nPDFs directly, but at significantly higher values of the resolution scale, where nuclear modifications are smaller.\cite{Guzey:2018dlm,Guzey:2019kik}

In conclusion, exclusive photoproduction of charmonia and inclusive photoproduction of dijets in heavy-ion UPCs give an excellent 
opportunity to constrain nPDFs at small $x$ in the nuclear shadowing region.


\newpage
%

\renewcommand*{\FigPath}{./WeekI/09_Miller/}
 
\newcommand{\trho}{{\tilde \rho}}\def\cG{{\cal G}}
\def\g{\gamma}\def\r{\rho}\def\L{\Lambda}\def\a{\alpha}\def\bfB{{\bf B}}\def\d{\delta}\def\bfb{{\bf b}}\def\l{\lambda}\def\th{\theta}\def\tz{{\tilde z}}\def\z{\zeta}\def\b{\beta}
\renewcommand{\bea}{\begin{eqnarray}}
\renewcommand{\eea}{\end{eqnarray}}
 \newcommand{\bfR}{{\bf R}}\renewcommand{\tr}{\mathrm{tr\, }}\newcommand{\bcl}[3]{b^\dagger_{#1,#2,#3}}\newcommand{\bal}[3]{b_{#1,#2,#3}}\newcommand{\bra}[1]{\Bigl<#1\Bigl|}
\newcommand{\ket}[1]{\Bigr|#1\Bigr>}\def\bfk{{\bf k}}

\def\D{\Delta}

\wstoc{Color Correlations in the  Proton}{Gerald~A.~Miller}

\title{Color Correlations in the  Proton}

\author{Gerald~A.~Miller$^*$ }
\index{author}{Miller, G.}

\address{Physics  Department, University of Washington,\\
Seattle, WA 98195-1560, USA\\
$^*$E-mail:miller@uw.edu \\
www.washington.edu}

\begin{abstract}
A general QCD light front formalism to compute many-body color charge correlation functions due to quarks  in the proton was constructed~\cite{Dumitru:2018vpr}. These enable new studies of color charge distributions in the nucleon. The analogies between such correlation functions   to electric and magnetization charge densities in the proton and  also to nucleon-nucleon correlations in the nucleus are discussed.  Including  the color charge correlations leads naturally to the removal of infrared divergences that occurs in two- and three-gluon exchange interactions in $q\bar q$-proton scattering.  Extensions to include  gluonic color charge correlations are  discussed.  \end{abstract}


\bodymatter

\section{Introduction}\label{aba:sec1_9}
This talk is concerned with a basic question:  Where is the color charge located in a nucleon, or in a
nucleus?
A related question: Is the color charge distribution the same as the charge distribution? We know that this cannot be the case because the integral of the color charge density must vanish because of color neutrality.  
Instead one must be concerned with matrix elements of powers of the color charge density operators such as $\rho^2$ and $\rho^3$.  Thus moments  must be constructed and understood.  This talk is a new way of looking at nucleon and nuclear structure. The formalism~\cite{Dumitru:2018vpr} can be thought of as an extension  of the 
  Mclerran-Venugopalan (MV) model for relativistic heavy ion physics~\cite{McLerran:1993ni_9,McLerran:1993ka_9,McLerran:1994vd_9} to nucleon structure. 

\section{Electromagnetic Charge Densities and Correlations}
I set the stage by discussing electromagnetic charge densities within the light-front formalism~\cite{Miller:2007uy,Carlson:2007xd,Miller:2009qu,Miller:2010nz_9}. 
The electromagnetic current density is given by $J^\mu=\sum_q e_q \bar q \gamma^\mu q$ where $q$ represents the quark flavor. One forms the current density from the $J^+$ operator acting in an eigenstate of transverse position $\bfR$:
\bea
\rho_\infty(x^-,\bfb)=\langle p^+,\bfR,\l|\sum_qe_qq_+^\dagger(x^-,\bfb)q_+(x^-,\bfb)|p^+,\bfR,\l\rangle,
\eea
where the longitudinal (transverse) spatial coordinates are $x^-\propto(t-z)$ (\bfb), and $q_+\propto\gamma^0\gamma^+q$ is the independent quark-field operator.   The value of $p^+$ must be very large because of the inherent sum over all of the proton's transverse momenta.  The transverse density can then be constructed  from the Dirac form factor $F_1$ as:
\bea \r(b)=\int dx^-\r_\infty(x^-,\bfb)=\int {QdQ\over 2\pi}F_1(Q^2)J_0(Qb).\eea
  Results are shown in the cited references and in the talk   posted on the INT website  http://www.int.washington.edu/talks/WorkShops/{\rm int}$_{-}18_{-}3$.
  
  Another question can be asked: given a $u$-quark at a position $(x^-,\bfb)$, what is the probability $P(\D x^-,\D\bfb)$ that a $d$-quark is a distance $(\D x^-,\D\bfb)$ away? This given by
  \bea P(\Delta x^-, \Delta \bfb)=\langle p^+{\bf R},\lambda|\int dx^-d^2\bfb \rho_u (x^-,\bfb) \rho_d (x^-+\Delta x^-,\bfb+\Delta \bfb)|p^+{\bf R}\lambda\rangle \eea
where
$\rho_q(x^-,\bfb)\equiv e_q q^\dagger_+(x^-,\bfb) q_+(x^-,r_\perp)$.  This represents a correlation function, the matrix-element of a two-quark operator that  enters in the evaluation of the two-photon-exchange matrix element. The matrix element of Eq.(3) is analogous to the short-ranged nucleon-nucleon correlations that are now under investigation. See {\it e.g.}  \cite{Hen:2016kwk,Cruz-Torres:2017sjy}.  Interactions between high energy particles and the proton are governed by two-gluon exchanges, so that probing quark-quark correlations might be easier than with two-photon exchanges.
\section{Color Charge  Density Operator }
The color charge density operator is given by 
\bea\rho^{ a} (x)=
\bar\psi_{i,f}(x)\, \gamma^+ \,\psi_{j,f}(x) (t^a)_{ij}+\,\rm{ gluon\,terms},\eea
where $a$ is the color index and $t^a$, $a=1-8$
are the generators of the fundamental representation of color-SU(3)
normalized as $\tr t^a t^b=\delta^{ab}/2$. 
The interesting matrix elements  in the proton are
$\langle \,{\rm }|\rho^a(x)|{\rm } \rangle, \langle \,{\rm }|\rho^a(x)\rho^b(y)|{\rm } \rangle, \langle \,{\rm }|\rho^a(x)\rho^b(y)\rho^c(y)|{\rm } \rangle$. The use of moments of the color charge density originated in the MV model, which shows how 
observables may be computed in terms of moments of $\rho^a$.

A few details are presented. When using light-front dynamics, quark-fields  at $x^+=0$  are  expanded in terms of    creation  and  destruction operators. The present evaluation ignores the presence of anti-quarks in the proton. Then using $r=(x^-,\bfb)$  
\bea
\rho^a(r) = 2 P^+\sum_{\lambda,\lambda'} \int {d x_q d^2q \over  {16\pi^3\sqrt{x_q}}}
\, \bcl{q}{i}{\lambda} e^{i q\cdot r}
\int {d x_p d^2p \over {16\pi^3\sqrt{x_p}}}\, \bal{p}{j}{\lambda'}
 e^{-i p\cdot r} \, (t^a)_{ij}\, \delta_{\lambda \lambda'}.\eea 
It is interesting to evaluate this operator in  the infinite momentum frame (IMF), $P^+\to\infty$. Then, the expression 
$ \lim_{P^+\to\infty} P^+ e^{i(x_q-x_p)P^+ r^-}$ appears. Taking the limit carefully~\cite{Dumitru:2018vpr}, we found that $\r^a$ contains a factor $2\pi \delta(x_p-x_q)\delta(r^-) $. Thus in the IMF the color-charge-density operator takes on the characteristics of a very thin disk.
In particular, the 
color charge per unit area is given by matrix elements of 
\bea
\rho^a(x^-,\bfb) = \d(x^-)\int \frac{d^2k}{(2\pi)^2}
\, e^{i\bfk\cdot\bfb}
 \int_0^\infty {dq^+\over q^+}\int {d^2q\over 16\pi^3}\sum_\lambda
\bcl{x_q,\vec q-\vec k}{i}{\lambda} \, \bal{x_q,\vec q}{j}{\lambda} \, (t^a)_{ij} .
\eea
The two-dimensional Fourier transform $\rho^a(x^-,\vec k_\perp) $ is used also.
 The notation \bea
\langle {\bf O}\,\rangle_{K_\perp} = \frac{\bra{P^+,\vec K_\perp} \, {\bf O}\, \ket{P^+, \vec P_\perp=0}}{\langle K|P\rangle}
\eea is used in the following.

\section{Color Charge Correlations}
Evaluations were made using light front wave functions for a 3-quark Fock state \cite{Dumitru:2018vpr}. The first result is that 
$ \langle \rho^a(x^-,\bfb)\rangle_{K_\perp}=0$ as expected for a color singlet. But the obvious result raises a question. Consider the matrix element of $\r^a$ for a 
Fock space component of proton: $|3q,{\rm G}\rangle$. Would the matrix element still vanish? The sum of quark and gluon densities must vanish, so any non-zero contribution to the color charge density from the quarks would be cancelled by the contribution from the gluons. That it is highly likely that  $\langle3q, {\rm G}| \rho^a(x^-,\bfb)|3q, {\rm G}\rangle$ would not vanish can be seen immediately  by considering color SU(2). In this case, the situation is analogous to that of the nucleon's pion cloud, which gives the neutron a non-vanishing charge density even though the total charge is zero \cite{Miller:2002ig}.
 
 The next step was to evaluate the two-quark color charge density:
$
\langle\, \tilde \rho^a(\vec K_\perp -\vec k_\perp) \, \trho^b(\vec k_\perp) \,\rangle_{K_\perp} =  \frac{1}{2}\,\delta^{ab}( \cG_1(\vec K_\perp)- \cG_2(\vec k,\vec K_\perp))\equiv \frac{1}{2}\,\delta^{ab}\cG(\vec k_\perp,K_\perp)$. The first term occurs when both density operators act on the same quark, and the second occurs when the action is on two different quarks. The limit of forward scattering is $\vec K_\perp=0$ and then   
$\cG(\vec k,0)=1-\cG_2(\vec k,0) $. Note that  taking $k_\perp=0$ yields $\cG(0,0) =0$, a consequence of color neutrality,   necessary to suppress   infrared divergences that would come from gluon propagators. 

A simple three-quark wave function \cite{Frank:1995pv} was used to evaluate $\cG(\vec k,0)$. The result is shown in Fig.~1.
 \begin{figure}
\includegraphics[width=3.in]{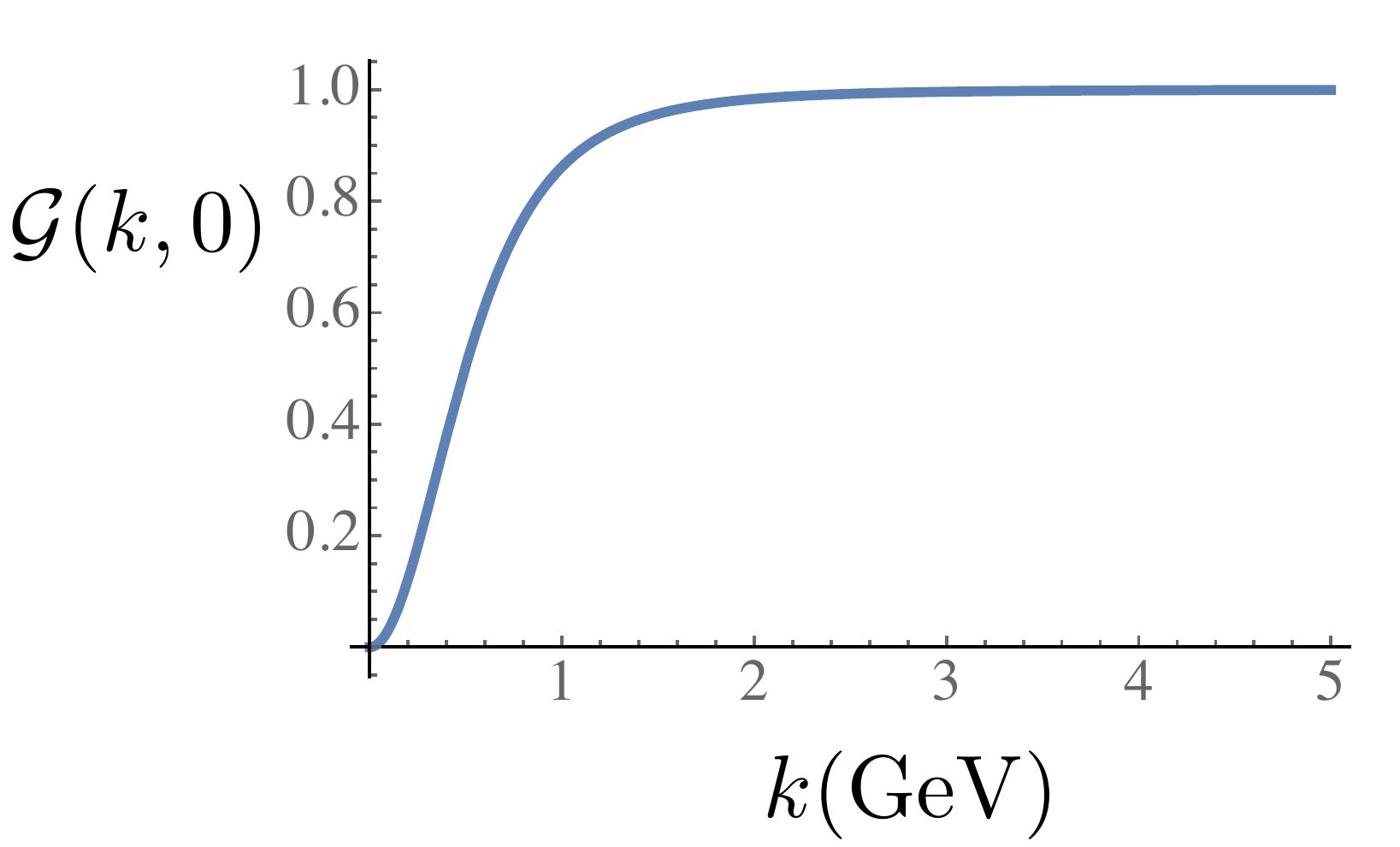}
\caption{$\cG(\vec k,0)$. Cancellation at $k=0$ is needed to prevent infrared divergences from appearing.}
\label{aba:fig1}
\end{figure}

 Space limitations prevent me from saying much more, but it's all included in~\cite{Dumitru:2018vpr}. See also \cite{Dumitru:2019qec} for an interesting application of the formalism.

\section{Summary}
  Ref.~\cite{Dumitru:2018vpr} provides a new way of looking at proton structure that involves 
using  moments of the  color charge density operator. 
Quadratic and cubic correlation functions  in the proton have been constructed for 3-quark light-front  wave functions.  The formalism is general so that evaluations can be made for 
  more complicated wave functions. 
The quadratic correlator ($\r^a\r^b$)   corresponds to Pomeron exchange, and the cubic correlator ($\r^a\r^b\r^c)$ corresponds  to Odderon exchange. 
The present formalism complements the standard GPD formalism.

\section*{Acknowledgments}
I thank  A.~Dumitru  and R.~Venugopalan for many interesting discussions needed to  develope Ref.~\cite{Dumitru:2018vpr}.
I   thank the  MIT-LNS,  the Southgate Fellowship of Adelaide University,, the Bathsheba de Rothchild Fellowship of Hebrew University, the Shaoul Fellowship of Tel Aviv University, the Physics Division of Argonne National Laboratory and the U.S. DOE, Office of Science, Office of Nuclear Physics under Award No. DE-FG02-97ER-41014 for support that enabled this work. 


\newpage
%

\renewcommand*{\FigPath}{./WeekI/10_Metz/}

\wstoc{Model Calculations of Euclidean Correlators}{Shohini~Bhattacharya, Christopher~Cocuzza, Andreas~Metz}
\title{Model Calculations of Euclidean Correlators}

\author{Shohini~Bhattacharya, Christopher~Cocuzza, Andreas~Metz$^*$}
\index{author}{Bhattacharya, S.}
\index{author}{Cocuzza, C.}
\index{author}{Metz, A.}

\address{Department of Physics, Temple University, Philadelphia, PA 19122, USA \\
$^*$E-mail: metza@temple.edu}

\begin{abstract}
Studying light-cone parton distribution functions (PDFs) through Euclidean correlators in lattice QCD is currently a very active field of research. 
In particular, the parton quasi-distributions (quasi-PDFs) have attracted a lot of attention. 
Quasi-PDFs converge to their respective standard distributions in the limit of infinite hadron momentum. 
We explore the quasi-distribution approach for twist-2 generalized parton distributions (GPDs) in a frequently used
diquark spectator model. 
Our main focus is to test how well the quasi-distributions agree with their light-cone counterparts for finite hadron momenta. 
We also discuss model-independent results for moments of quasi-distributions.
\end{abstract}


\bodymatter

\section{Introduction}
Quasi-PDFs put forward by Ji~\cite{Ji:2013dva_114} are at the forefront of numerical calculations of the partonic structure of strongly interacting systems.
They are defined through spatial (Euclidean) correlation functions and thus are directly calculable in lattice QCD. 
We have investigated this new approach by calculating (eight) twist-2 GPDs in a scalar diquark model (SDM)~\cite{Bhattacharya:2018zxi_116,Bhattacharya:2019cme_118}. 
In this short write-up of the talk, we mainly concentrate on the unpolarized quasi-GPD $H_{\rm Q}$ which corresponds to the standard GPD $H$. 
All the numerical features discussed subsequently are robust and not specific to this distribution function.
Related model calculations of Euclidean correlators are available as well~\cite{Gamberg:2014zwa, Bacchetta:2016zjm, Nam:2017gzm, Broniowski:2017wbr, Hobbs:2017xtq, Broniowski:2017gfp, Xu:2018eii}.

\section{Quasi-GPDs: Definition and Analytical Results}
Quasi-GPDs are defined through an equal-time spatial correlation function~\cite{Ji:2013dva_114, Bhattacharya:2018zxi_116},
\begin{equation}
\frac{1}{2} \int \frac{dz^{3}}{2\pi}  e^{ik \cdot z} \langle p',\lambda '| \bar{\psi} \Big( -\tfrac{z}{2} \Big) \, \Gamma \, {\cal W}_{\rm Q} \Big( -\frac{z}{2}, \frac{z}{2} \Big) \psi \Big( \frac{z}{2} \Big)|p, \lambda \rangle \bigg |_{z^{0}=0, \vec{z}_{\perp}=\vec{0}_{\perp}} \, ,
\label{e:corr_quasi_GPD}
\end{equation}
where $\cal W_{\rm Q}$ denotes a Wilson line. 
The unpolarized quasi-GPDs $H_{\rm Q(0/3)}$ and $E_{\rm Q(0/3)}$ are defined through the choice $\Gamma=\gamma^{0/3}$.
More details on the definition of quasi-GPDs can be found elsewhere~\cite{Bhattacharya:2018zxi_116, Bhattacharya:2019cme_118}.
Quasi-GPDs are functions of four kinematical variables: the average parton momentum fraction $x = \tfrac{k^{3}}{P^{3}}$, the (standard) skewness $\xi$, $t = (p - p')^2$ (or the transverse momentum transfer $|\vec{\Delta}_{\perp}|$), and the average $z$-component of the hadron momentum $P^{3}$. 
Note that $x$ differs from $\tfrac{k^{+}}{P^{+}}$ that appears for light-cone GPDs. 
The support for the quasi-GPDs is given by $- \infty < x < \infty$. 
We also use the quantity $\delta = \sqrt{1+ \frac{M^{2}-t/4}{(P^{3})^{2}}}$ (with $M$ denoting the hadron mass), which shows up in the relation $P^{0} = \delta P^{3}$.

We now consider quasi-GPDs in the scalar diquark model (SDM), whose main parameters are the nucleon-quark-diquark coupling $g$, the quark mass $m_q$, and the diquark mass $m_s$.
Here we quote the analytical result for the quasi-GPD $H_{\rm Q}$~\cite{Bhattacharya:2018zxi_116, Bhattacharya:2019cme_118},
\begin{equation}
H_{{\rm Q}(0)}(x, \xi, t; P^3) = \frac{i \, g^2 P^3}{(2\pi)^4} \int dk^0 \, d^2\vec{k}_\perp \, \frac{N_{H(0)}}{D_{\rm GPD}} \,,
\label{e:HQ}
\end{equation}
where the numerator reads
\begin{eqnarray}
N_{H(0)} & = & \delta (k^0)^2 - \frac{2}{P^3} \bigg[ \, x (P^3)^2 - m_q M - x \, \frac{t}{4} - \frac{1}{2} \, \delta \xi t \, \frac{\vec{k}_\perp \cdot \vec{\Delta}_\perp}{\vec{\Delta}_\perp^2} \bigg] k^0 
\nonumber \\[0.1cm]
&& + \, \delta \bigg[ \, x^2 (P^3)^2 + \vec{k}^{2}_\perp + m_q^2 + (1 - 2x) \, \frac{t}{4} - \delta \xi t \, \frac{\vec{k}_\perp \cdot \vec{\Delta}_\perp}{\vec{\Delta}_\perp^2} \bigg] \,.
\end{eqnarray}
When performing $\int dk^{0}$ via contour integration, one can verify that the position of the $k^{0}$-poles never switch half planes. 
This implies that the quasi-GPDs are continuous functions of $x$, which differs from light-cone GPDs which can be discontinuous at the cross-over points $x=\pm \, \xi$~\cite{Aslan:2018zzk_120}.
Also, we have verified the consistency check that all quasi-GPDs reduce to the corresponding light-cone GPDs for $P^3 \to \infty$.

\begin{figure}[t]
\begin{center}
\includegraphics[width = 5.5cm]{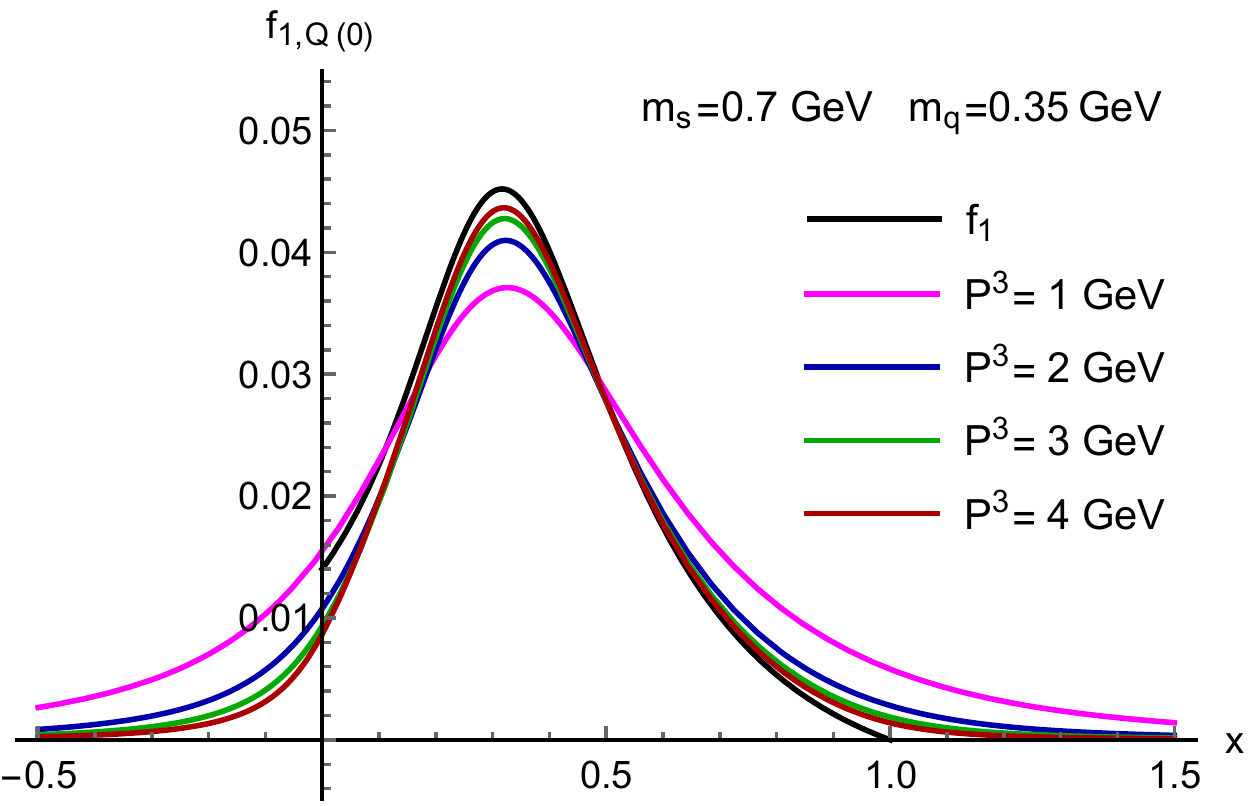}
\hspace{0.8cm}
\includegraphics[width = 5.5cm]{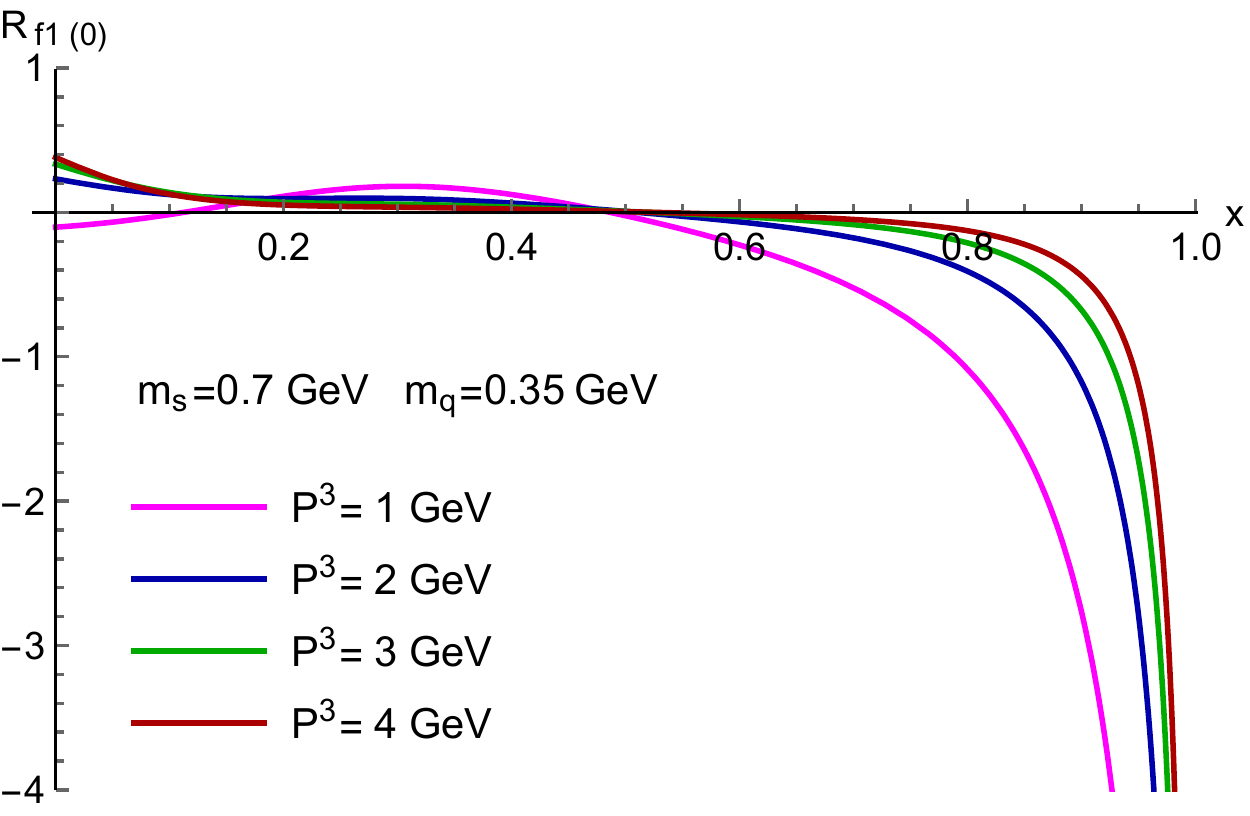}
\end{center}
\caption{Left panel: Quasi-PDF $f_{1,\rm Q(0)}$ as a function of $x$ for different values of $P^3$.
The black curve represents the light-cone PDF $f_1$. 
Right panel: Relative difference between $f_{1,\rm Q(0)}$ and $f_1$ as a function of $x$ for different values of $P^3$.}
\label{f:f_1}
\end{figure}

\section{Numerical Results in Scalar Diquark Model}
Our numerical results are largely insensitive to variations of the model parameters~\cite{Bhattacharya:2018zxi_116, Bhattacharya:2019cme_118}. 
We first discuss the unpolarized quasi-PDF $f_{1,\rm Q(0)}$, which is the forward limit of $H_{\rm Q(0)}$ --- see Fig.$\,$\ref{f:f_1}. 
For larger values of $P^{3}$, there is a good agreement between quasi and light-cone PDF over a wide range of $x$. 
However, considerable discrepancies appear as $x \rightarrow 0$ and $x \rightarrow 1$. 
The discrepancies around $x = 0$ can be expected~\cite{Bhattacharya:2018zxi_116}.
The relative difference $R_{f1(0)}(x; P^3) = \frac{f_1(x) - f_{1,{\rm Q}(0)}(x; P^3)}{f_1(x)}$ in Fig.$\,$\ref{f:f_1} better illustrates, in particular, the discrepancies at large $x$, which are partly due to the mismatch between the parton momentum fractions appearing in the light-cone PDFs and the quasi-PDFs~\cite{Bhattacharya:2019cme_118}.
\begin{figure}[t]
\begin{center}
\includegraphics[width = 5.5cm]{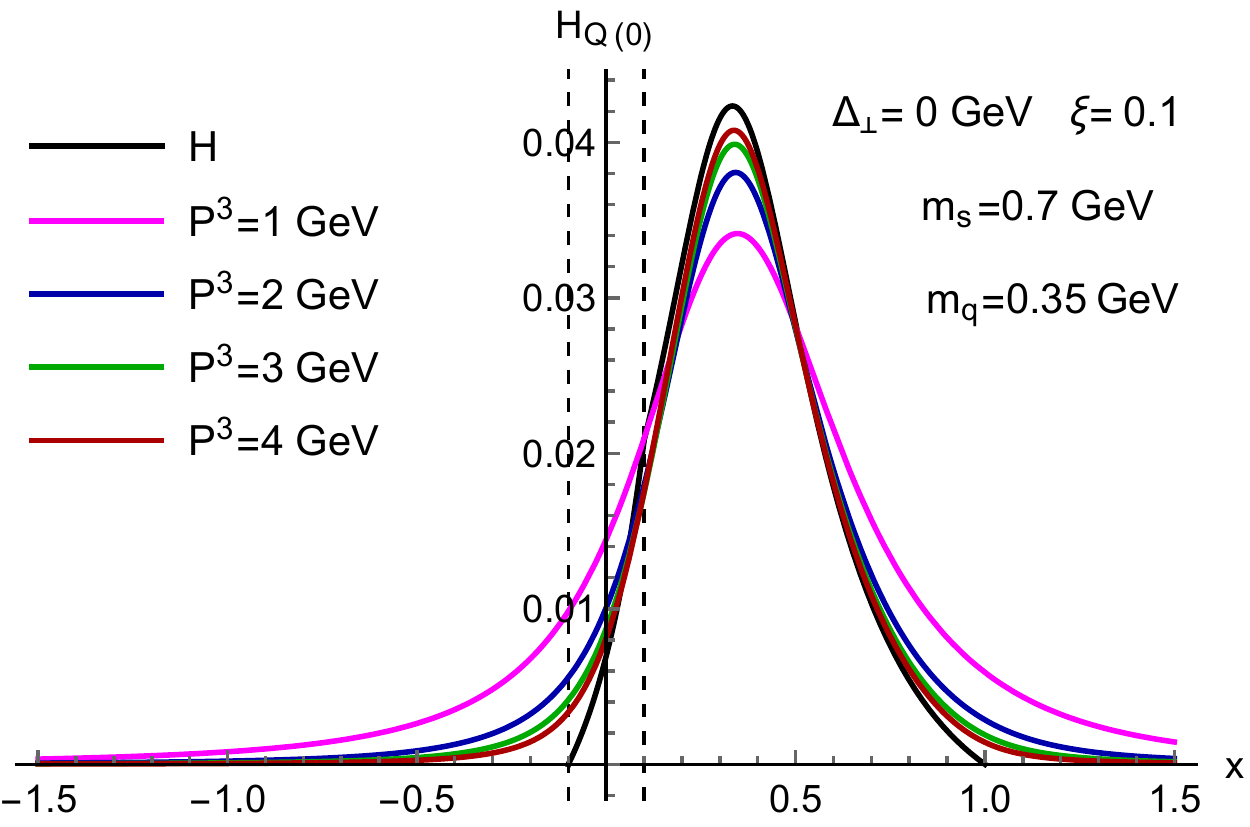}
\hspace{0.8cm}
\includegraphics[width = 5.5cm]{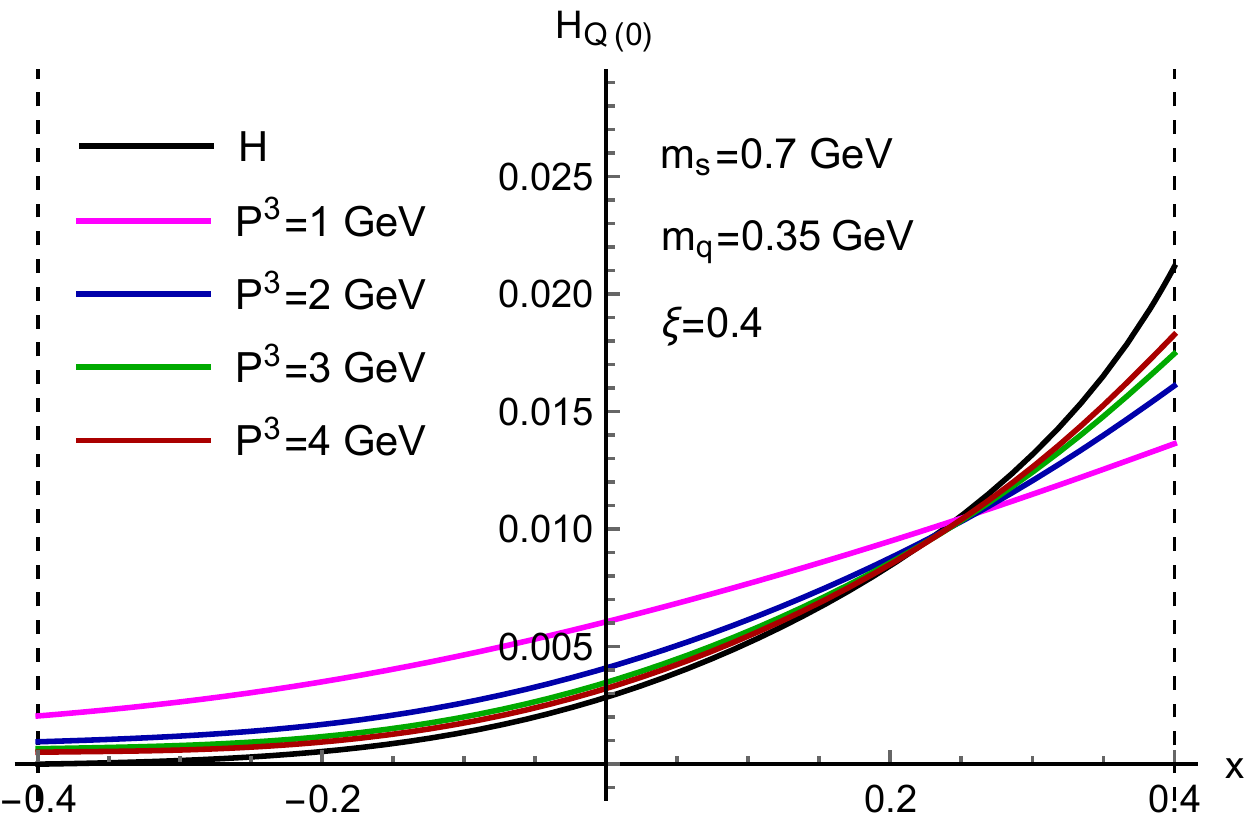}
\end{center}
\caption{Left panel: Quasi-GPD $H_{\rm Q(0)}$ as a function of $x$ for $\xi = 0.1$ and different values of $P^3$.  The limits of the ERBL region are indicated by vertical dashed lines.
Right panel: $H_{\rm Q(0)}$ as a function of $x$ in the ERBL region for different values of $P^3$ and for $\xi = 0.01$.
Black curves represents the light-cone GPD $H$. }
\label{f:f_2}
\end{figure}

\begin{figure}[!]
\begin{center}
\includegraphics[width = 5.5cm]{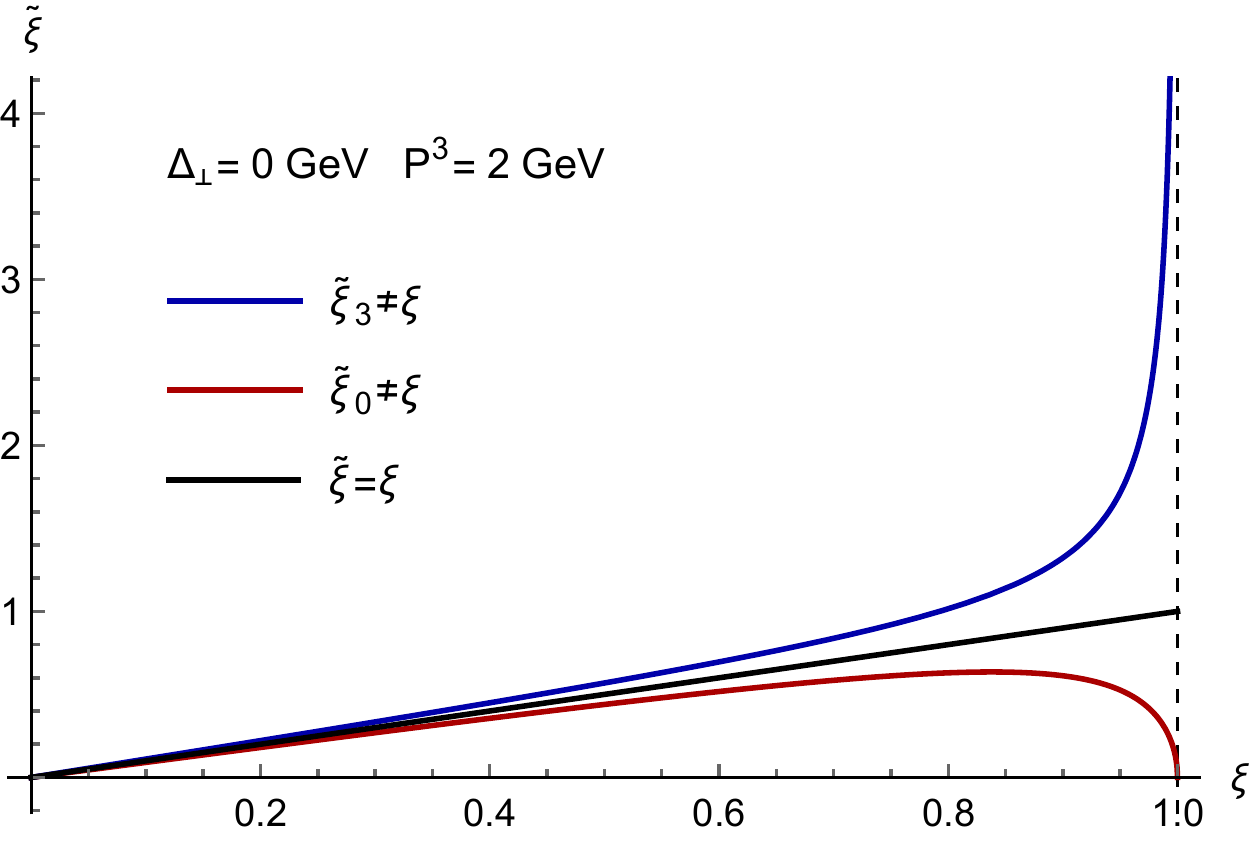}
\hspace{0.8cm}
\includegraphics[width = 5.5cm]{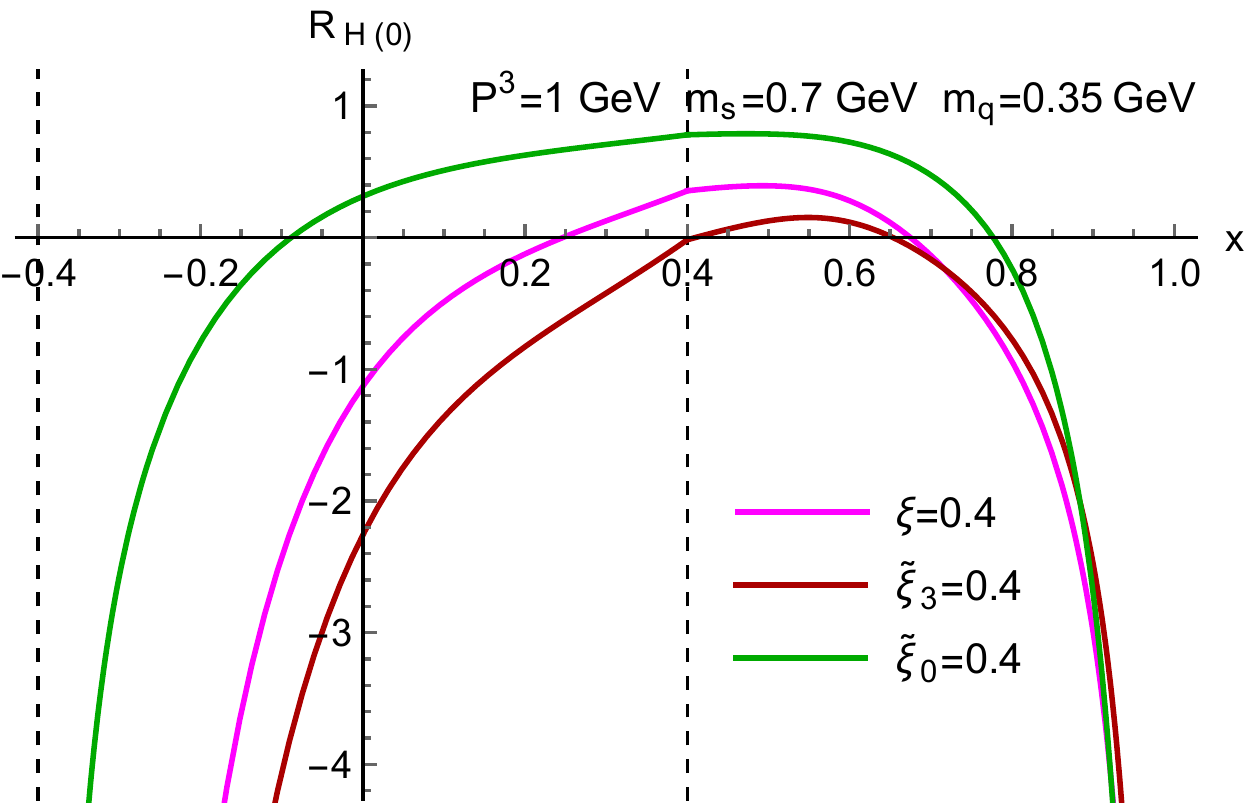}
\end{center}
\caption{Left panel: Comparison of the skewness variables $\xi$ , $\tilde{\xi}_3$ and $\tilde{\xi}_0$ for $P^3 = 2\,\textrm{GeV}$ and $|\vec{\Delta}_\perp| = 0\,\textrm{GeV}$.
Right panel: Relative difference between $H_{\rm Q(0)}$ and $H$ for three different definitions of the skewness variable.}
\label{f:f_3}
\end{figure}

The quasi-GPD $H_{\rm Q(0)}$ for $\xi=0.1$ is shown in Fig.$\,$\ref{f:f_2}.
Generally, we have explored the range $0.01 \leq \xi \leq 0.4$. 
The convergence problem at large $x$ persists for all the quasi-GPDs.
There is a tendency of the discrepancies at large $x$ to increase when $\xi$ gets larger, where the significance of this feature depends on the GPD under consideration~\cite{Bhattacharya:2018zxi_116, Bhattacharya:2019cme_118}.
In Fig.$\,$\ref{f:f_2} we also show $H_{\rm Q(0)}$ for just the ERBL region for $\xi=0.4$.
For large $\xi$, we observe a good agreement between quasi-GPDs and the light-cone GPDs for a large fraction of the ERBL region. 
This outcome suggests that lattice calculations could provide very valuable information in the ERBL region.
On the other hand, for small $\xi$ one finds significant deviations between the quasi-GPDs and the light-cone GPDs in the ERBL region~\cite{Bhattacharya:2018zxi_116, Bhattacharya:2019cme_118}. 
So far we have used the same skewness variable $\xi$ for the light-cone GPDs and the quasi-GPDs. 
But for the quasi-GPDs one could also consider the ``quasi-skewness" variables $\tilde{\xi}_{3} = - \frac{\Delta^{3}}{2P^{3}}$ and $\tilde{\xi}_{0} = - \frac{\Delta^{0}}{2P^{0}}$.
Though they differ from $\xi$ by a higher-twist effect, the numerical difference can be large as shown in Fig.$\,$\ref{f:f_3}.
As also shown in Fig.$\,$\ref{f:f_3}, ignoring the higher-twist effect and using $\tilde{\xi}_{3}$ leads to a better convergence of $H_{\rm Q(0)}$
for most of the DGLAP region. 
The same holds for most (spin-dependent) quasi-GPDs though not all~\cite{Bhattacharya:2018zxi_116, Bhattacharya:2019cme_118}.

\section{Moments of Quasi-Distributions}
It is instructive to consider Mellin moments of quasi-GPDs. 
Including a flavor index `$q$' one finds the model-independent relation~\cite{Bhattacharya:2019cme_118}
\begin{equation}
\int\limits_{-1}^{1} dx \, H^{q}(x,\xi,t) \! = \! \int\limits_{-\infty}^{\infty} dx \, \frac{1}{\delta} \, H^{q}_{\rm Q(0)}(x,\xi,t;P^3) 
\! = \! \int\limits_{-\infty}^{\infty} dx \, H^{q}_{\rm Q(3)}(x,\xi,t;P^3) \! = \! F^{q}_{1}(t) \,,
\label{e:FF_1}
\end{equation}
where $F_{1}$ is the Dirac form factor. 
It is remarkable that the $P^3$-dependence of $H_{\rm Q}$ drops out in the lowest moment.
However, one must divide $H_{\rm Q(0)}$ by the factor $\delta$ in order to arrive at this result. 
Similar considerations apply for the other quasi-GPDs, where the need of properly including the factor $\delta$ has been overlooked in most of the previous papers on quasi-distributions~\cite{Bhattacharya:2019cme_118}.
For the second moment of the quasi-GPDs we find, in close analogy to the celebrated expression $\int\limits_{-1}^{1} dx \, x \, \big(H^{q}(x,\xi ,t)+E^{q}(x,\xi, t)\big) = A^{q}(t)+B^{q}(t)$ where $A^{q}(0)+B^{q}(0) = J^q$ is the total angular momentum for the quark flavor `$q$', 
\begin{equation}
\int\limits_{-\infty}^{\infty} dx \, x \, \big(H^{q}_{\rm Q(3)}(x,\xi ,t; P^3)+E^{q}_{\rm Q(3)}(x,\xi, t; P^3)\big)  =  A^{q}(t)+B^{q}(t) \,.
\label{e:smoment_1} 
\end{equation}
The second moment of $H_{\rm Q(3)} + E_{\rm Q(3)}$ is therefore directly related to the angular momentum of quarks, while the corresponding relation for $H_{\rm Q(0)} + E_{\rm Q(0)}$ contains in addition a higher-twist contributions~\cite{Bhattacharya:2019cme_118}.
Generally, the model-independent expressions for the moments of the quasi-distributions may be useful for studying the systematic uncertainties of results from lattice QCD, especially due to the fact that the $P^{3}$-dependence of the moments is either computable or nonexistent.

\section*{Acknowledgments}
This work has been supported by the National Science Foundation under grant number  PHY-1812359, and by the U.S.~Department of Energy, Office  of  Science, Office of Nuclear Physics, within the framework of the TMD Topical Collaboration.



\newpage



\renewcommand*{\FigPath}{./WeekI/11_Horn/}

\wstoc{Experimental Investigations of Hadron Structure}{Tanja Horn}
\title{Experimental Investigations of Hadron Structure}

\author{Tanja Horn$^*$}
\index{author}{Horn, T.}

\address{Catholic University of America, Washington, DC 20064}
\address{Jefferson Lab, Newport News, VA, 23606, $^*$E-mail: hornt@cua.edu}
 

\begin{abstract}
Pions and kaons occupy a special role in nature and thus have a central role in our current description of nucleon and nuclear structure. The pion is the lightest quark system, with a single valence quark and a single valence antiquark. It is also the particle responsible for the long range character of the strong interaction that binds the atomic nucleus together. A general belief is that the rules governing the strong interaction are left-right, {\em{i.e.}} chirally, symmetric.  If this were true, the pion would have no mass. The chiral symmetry of massless Quantum Chromodynamics (QCD) is broken dynamically by quark-gluon interactions and explicitly by inclusion of light quark masses, giving the pion and kaon mass. The pion and kaon are thus seen as the key to confirm the mechanism that dynamically generates nearly all of the mass of hadrons and central to the effort to understand hadron structure.
\end{abstract}

\section{Introduction}

This talk is based on References~\cite{Aguilar19,Horn16} and  focused on measurements using the Sullivan process and includes the current status, upcoming measurements at the 12 GeV Jefferson Lab, and future prospects at the Electron-Ion Collider. Specific measurements discussed are those of the pion and kaon form factors, as well as pion and kaon parton distribution functions and pion generalized distribution functions. 

\section{Role of the proton's pion cloud}

The electron deep-inelastic-scattering off the meson cloud of a nucleon target is called the Sullivan process.  The Sullivan process can provide reliable access to a meson target as t becomes space-like  if  the  pole  associated  with  the  ground-state  meson  ($t$-pole)  remains  the  dominant feature of the process and the structure of the related correlation evolves slowly and smoothly with virtuality. The experiments will provide data covering a range in $-t$, particularly low $-t$, to check if these conditions are satisfied empirically, and compare with phenomenological and theoretical expectations. Theoretically, a recent calculation~\cite{Qin18} explored the circumstances under which these conditions should be satisfied and found to $-t$=0.6 GeV2,all changes in pion structure are modest so that a well-constrained experimental analysis should be reliable. Similar analyses for the kaon indicate that Sullivan processes can provide a valid kaon target for $-t \leq 0.9\,$GeV$^2$.

\section{Pion and Kaon Form Factors}

The elastic electromagnetic form factors of the charged pion and kaon, $F_\pi(Q^2)$ and $F_K(Q^2)$, are a rich source of insights into basic features of hadron structure, such as the roles played by confinement and DCSB in fixing the hadron's size, determining its mass, and defining the transition from the strong- to perturbative-QCD domains.

Studies during the last decade, based on JLab 6-GeV measurements, have generated confidence in the reliability of pion electroproduction as a tool for pion form factor extractions. Measurements in Hall C~\cite{Horn06} confirmed that with a photon virtuality of 2.45 GeV$^2$, one is still far from the resolution region where the pion behaves  like  a  simple  quark/anti-quark  pair, i.e. far  from  the  "asymptotic"  limit.   However, this perception is based on the assumption that the asymptotic form of the pion's valence quark parton distribution amplitude (PDA) is valid at $Q^2$=2.45 GeV$^2$.  Modern calculations show that  this  discrepancy could  be  explained  by  using  a  pion  valence  quark  PDA  evaluated  at  a  scale appropriate to the experiment.  Confirming these calculations empirically would be a great step towards our understanding of QCD.

Forthcoming measurements at the 12-GeV JLab~\cite{E12-19-006} taking advantage of the new SHMS system and particle identification detectors~\cite{Aerogel} will deliver pion form factor data to $Q^2$=6.0 GeV$^2$ with high precision,  and  to  8.5  GeV$^2$ with  somewhat  larger  experimental  and  theoretical  uncertainties. Extractions of the kaon form factor may be possible up to $Q^2$=5.5 GeV$^2$ from completed 12-GeV experiment E12-09-011~\cite{E12-09-011}. The Electron-Ion Collider provides the facilities to extend measurements of the pion form factor even further probing deep into the region where $F_\pi(Q^2)$ exhibits strong sensitivity to both emergent mass generation via DCSB and the evolution of this effect with scale. Shown in Fig.~\ref{fig:SF-JLAB-EIC} projected EIC pion form factor data as extracted from a combination of electron-proton and electron-deuteron scattering, each with an integrated luminosity of $20\,{\rm fb}^{-1}$ -- black stars with error bars. Also shown are projected JLab 12-GeV data from a Rosenbluth-separation technique -- orange diamonds and green triangle. The long-dashed green curve is a monopole form factor whose scale is determined by the pion radius.

\section{Validation of the exclusive reaction mechanism}

To validate the meson factorisation theorems and potentially extract flavour separated GPDs from experiment, one has measure the separated longitudinal (L) and transverse (T) cross sections and their $t$ and $Q^2$ dependencies. Only L/T separated cross sections can unambigously show the dominance of longitudinal or transverse photons and allow one to determine possible correlations in $t$ and $Q^2$. The onset of factorisation for light mesons may be expected earlier than for heavier ones. Recent calculations predict the onset for pions and kaons in the 5-10 GeV$^2$ regime, a region accessible with 12 GeV JLab experiments~\cite{E12-19-006,E12-09-011}. Experiment E12-19-006~\cite{E12-19-006} will  provide  essential  constraints  on  Generalized  Parton  Distributions  (GPDs) central to the 12 GeV program. If $\sigma_T$ is confirmed to be large, it could subsequently allow for a detailed investigation of transversity GPDs.  If, on the other hand, $\sigma_L$ is measured to be larger than expected, this would allow for probing the usual GPDs. Furthermore, the Neutral Particle Spectrometer~\cite{NPS} will allow for precision (coincidence) cross section measurements of neutral particles ($\gamma$ and $\pi^0$)~\cite{E12-13-010}. 

\section{Pion and Kaon PDFs}

The mass of the pion is roughly 140\,MeV, of the kaon 493\,MeV, and of the proton 939\,MeV. In the chiral limit, the mass of the proton is entirely given by the trace anomaly in QCD. The mass of the pion has, in this same limit, either no contribution at all from the trace anomaly or, more likely, a cancellation of terms occurs in order to ensure the pion is massless. Beyond the chiral limit, a decomposition of the proton mass budget has been suggested, expressing contributions from quark and gluon energy and quark masses. With the various quark (flavor) and gluon distributions in the proton reasonably well known, the largest uncertainty here lies with the trace anomaly contribution. For the pion, further guidance on the magnitude of quark and gluon energy can, as for the proton, be determined from measurements of the pion and kaon structure functions, with resulting constraints on quark and gluon PDFs, over a large range of $x_\pi$ and momentum-transfer squared, $Q^2$. This is accessible for the EIC, roughly covering down to $x_\pi = 10^{-3}$ at $Q^2 = 1\,$GeV$^2$ and up to $x_\pi=1$ at $Q^2 = 1000\,$GeV$^2$. 

The projected brightness for a high-luminosity EIC is nearly three orders-of-magnitude above that of HERA, $10^{34}\,$e-nucleons/cm$^2$/s versus $10^{31}\,$e-nucleons/cm$^2$/s, with an acceptance essentially covering the full forward region. With a suitable detector configuration, access to high $x_\pi$ ($\to 1$) will be possible, allowing overlap with fixed-target experiments \cite{Keppel:2015, Keppel:2015B, McKenney:2015xis}.  Overlap with Drell-Yan measurements such as those proposed for the CERN M2 beam line by the COMPASS++/AMBER collaboration will settle the unknown pion flux factor associated with the Sullivan process measurements. A sample EIC extraction of valence quark, sea quark and gluon PDFs in the pion, at a scale $Q^2 =10\,$GeV$^2$ is shown in Fig.~\ref{fig:SF-JLAB-EIC}. The extraction is done with the following assumptions on PDFs: the $u$ PDF equals the $\bar d$ PDF in the pion and the $\bar u$ PDF is the same as the other sea quark PDFs ($d$, $s$ and $\bar s$). The extraction at $x_\pi  < 10^{-2}$, at this $Q^2$ scale, is constrained by the existing HERA data.

Kaon structure was not studied at HERA; but the ratio of kaon structure function (under the condition of a $\Lambda$ detection) to the proton structure function at small $-t$ is similar to that for the pion Sullivan process ($\sim 10^{-3}$).   Hence, one would anticipate both pion and kaon structure function measurements as functions of $(–t, x, Q^2)$ at a high-luminosity ($10^{34}$ or more) EIC to be of similar statistical precision as the well-known, textbook HERA proton $F_2$ structure function measurements.  One should therefore be able to constrain the gluon distributions in the pion and kaon.  
\begin{figure}[t]
\centering
\begin{minipage}{0.4\linewidth}
\hspace*{-1cm}
\includegraphics[width=1.0\linewidth]{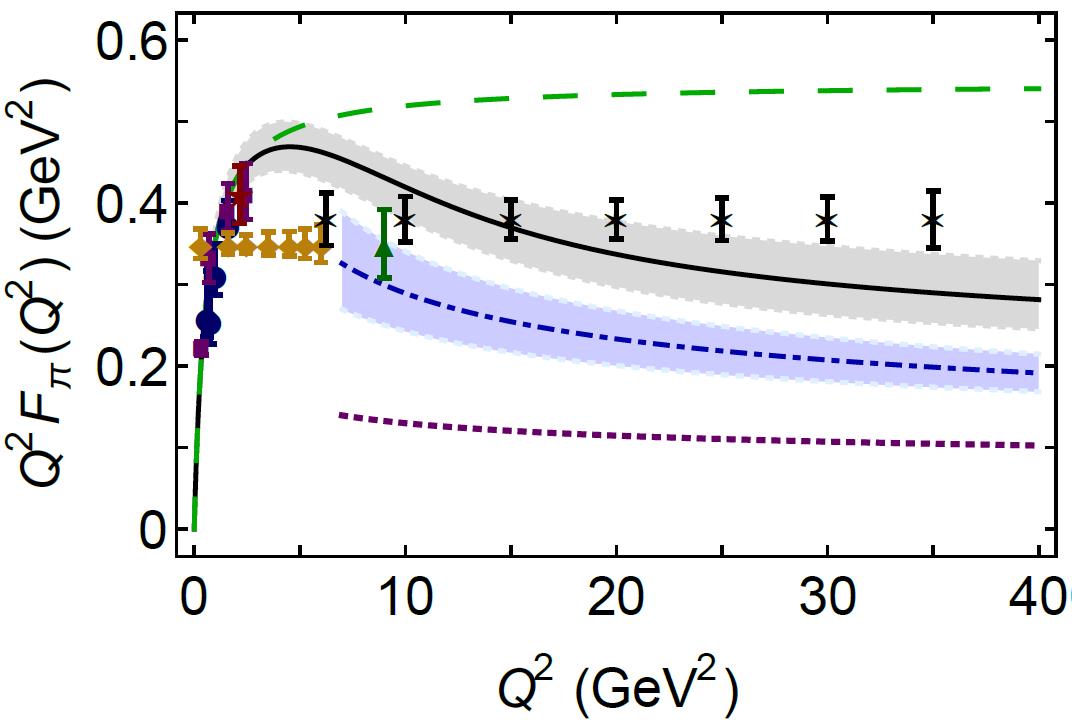}
\end{minipage}
\begin{minipage}{0.4\linewidth}
\includegraphics[width=1.2\linewidth]{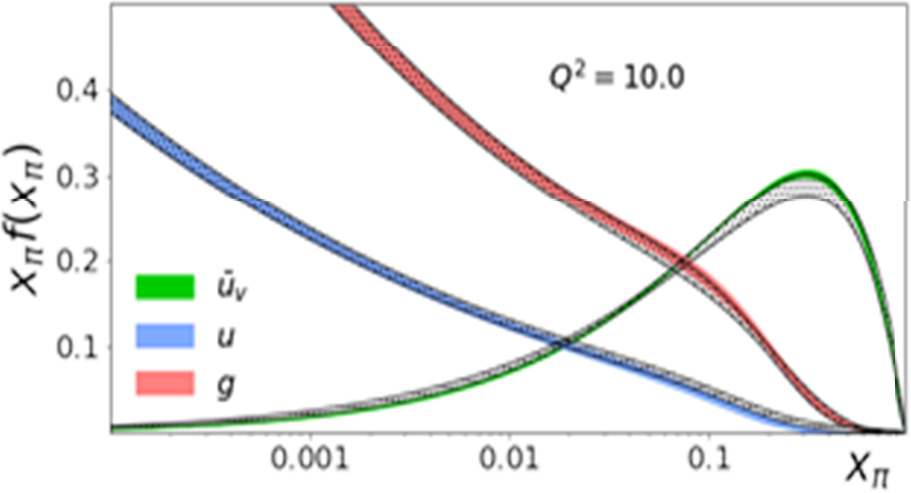}
\end{minipage}
\caption{Projections for pion form factor (left) and structure function (right) measurements at JLab and EIC.}
\label{fig:SF-JLAB-EIC}
\end{figure}

\section{Summary}
  
A striking feature of the strong interaction is its emergent $1$-GeV mass-scale, as exhibited in the masses of protons, neutrons and numerous other everyday hadronic bound states. In sharp contrast, the energy associated with the gluons and quarks confined inside the strong interaction's Nambu-Goldstone bosons, such as the pion and kaon, is not so readily apparent. Pion and kaon structure can be measured at JLab 12 GeV and EIC through the Sullivan process, which necessarily means mesons are accessed off-shell. Nevertheless, recent experimental and phenomenological work strongly indicates that, under certain achievable kinematic conditions, the Sullivan process provides reliable access to a true meson target. JLab 12 GeV will dramatically improve the pion and kaon electroproduction data set making possible the extraction of meson form factors and the interpretation of 12 GeV JLab GPD data. The planned Electron-Ion Collider provides unique opportunities to map pion and kaon structure over a wide kinematic range. Measurement of pion and kaon structure functions and their generalized parton distributions will render insights into quark and gluon energy contributions to hadron masses. Measurement of the charged-pion form factor up to $Q^2 \approx 35\,{\rm GeV}^2$, which can be quantitatively related to emergent-mass acquisition from dynamical chiral symmetry breaking.

\section*{Acknowledgments}
This work was supported in part by National Science Foundation grant PHY1714133.


\newpage
%


\renewcommand*{\FigPath}{./WeekI/12_Pasquini/}

\newcommand{\uvec}[1]{\boldsymbol{#1}}

\newcommand{\ud}{\mathrm{d}}
\newcommand{\uL}{\mathcal{L}}
\newcommand{\uM}{\mathcal{M}}
\newcommand{\uPcal}{\mathcal{P}}
\newcommand{\uZ}{\mathcal{Z}}
\newcommand{\uQcal}{\mathcal{Q}}
\newcommand{\uN}{\mathcal{N}}
\newcommand{\uTr}{\mathrm{Tr}}
\newcommand{\uSp}{\mathrm{Sp}}
\newcommand{\uslash}{/\!\!\!}
\newcommand{\dslash}{\partial\!\!\!/}
\newcommand{\uk}{\boldsymbol{k}}
\newcommand{\uD}{\mathcal{D}}
\newcommand{\uV}{\boldsymbol{V}}
\newcommand{\uVcal}{\mathcal{V}}
\newcommand{\sgn}{\text{sgn}}
\newcommand{\barpsi}{\overline{\psi}}

\newlength\savedwidth
\renewcommand{\arraystretch}{1.3}
\newcommand\whline{\noalign{\global\savedwidth\arrayrulewidth
\global\arrayrulewidth 1pt}%
\hline
\noalign{\global\arrayrulewidth\savedwidth}}

\wstoc{Wigner functions and nucleon structure}{Barbara Pasquini,  C\'edric Lorc\'e}

\title{Wigner functions and nucleon structure}

\author{Barbara Pasquini$^*$}
\index{author}{Pasquini, B.}

\address{Dipartimento di Fisica, Universit\`a degli Studi di Pavia, I-27100 Pavia, Italy and
Istituto Nazionale di Fisica Nucleare, Sezione di
  Pavia,  I-27100 Pavia, Italy\\
$^*$E-mail: barbara.pasquini@unipv.it}
    
    \author{C\'edric  Lorc\'e}
    \address{
          Centre de Physique Th\'eorique, \'Ecole polytechnique, CNRS, Universit\'e Paris-Saclay, F-91128 Palaiseau, France}
\index{author}{Lorc\'e, C.}

\begin{abstract}
We discuss the leading-twist quark Wigner distributions in the nucleon, 
both in the T-even and T-odd sectors, considering all the possible configurations of the quark and nucleon
polarizations. We identify the basic multipole structures associated with each distribution in the transverse
phase space, providing a transparent interpretation of the spin-spin and spin-orbit correlations of quarks and
nucleons encoded in these functions.
\end{abstract}


\bodymatter

\section{Wigner functions: definition and properties}\label{aba:sec1}
The concept of Wigner distributions has been borrowed from quantum mechanics to study the partonic  structure of the nucleon in the phase-space.
The six dimensional
version of these phase-space distributions has
been discussed for the first time  in Refs.~\cite{Ji:2003ak_15,Belitsky:2003nz_20}, using a definition  in the Breit
frame  whose physical interpretation is plagued by relativistic corrections. To avoid this problem, one can
use  the
light-front formalism by integrating over the longitudinal
spatial dimension and then
 introduce 
 five-dimensional
phase-space distributions~\cite{Lorce:2011kd_20}, which depend on two position and three momentum coordinates.
In this case the Wigner distributions appear as the two-dimensional Fourier transforms of generalized transverse momentum dependent parton distributions (GTMDs)~\cite{Meissner:2008ay_15,Meissner:2009ww_20,Lorce:2013pza_15}, which reduce to generalized parton distributions (GPDs) and transverse momentum dependent parton distributions (TMDs) in particular limits. However, the GTMDs contain richer physics
than TMDs and GPDs combined, as they carry information about the correlations between
the quark momentum $(x,\uvec k_T)$ and transverse space position $\uvec b_T$, which cannot be accessed
by separately studying TMDs or GPDs. Currently, the
best hope to access the GTMDs is for the gluon contribution in the low-$x$ ~\cite{Hagiwara:2017fye_15,
Hatta:2016dxp_20,Hatta:2016aoc_15,Ji:2016jgn_15,Boussarie:2018zwg_15} and intermediate-$x$~\cite{Bhattacharya:2018lgm_15} regimes,
while a single process has been identified sofar for the quark sector~\cite{Bhattacharya:2017bvs_20}.

The quark GTMD correlator is defined as~\cite{Meissner:2009ww_20,Lorce:2013pza_15}
\begin{equation}\label{GTMDcorr-def}
W^{ab}_{\Lambda'\Lambda}\equiv\int\ud k^-\int\frac{\ud^4z}{(2\pi)^4}\,e^{ik\cdot z}\,\langle P+\tfrac{\Delta}{2},\Lambda'|\overline\psi_b(-\tfrac{z}{2})\,\mathcal W\,\psi_a(\tfrac{z}{2})|P-\tfrac{\Delta}{2},\Lambda\rangle|_{z^+=0},
\end{equation}
where $\mathcal W$ is an appropriate Wilson line ensuring color gauge invariance, $k$ is the quark average four-momentum conjugate to the quark field separation $z$, and $|p,\Lambda\rangle$ is the spin-$1/2$ target state with four-momentum $p$ and light-front helicity $\Lambda$. A proper definition of the GTMDs should include also a soft-factor contribution~\cite{Echevarria:2016mrc_15}. However, it is not relevant for the following multipole analysis.

We choose to work in the symmetric frame defined by $P^\mu=\tfrac{p'^\mu+p^\mu}{2}=[P^+, P^-,\uvec 0_T]$. At leading twist and for a spin $1/2$ target, one can interpret
\begin{equation}\label{GTMDcorr}
W_{\uvec S\uvec S^q}=\tfrac{1}{8}\sum_{\Lambda',\Lambda}(\mathds 1+\uvec S\cdot\uvec \sigma)_{\Lambda\Lambda'}\,\uTr[W_{\Lambda'\Lambda}\Gamma_{\uvec S^q}],
\end{equation}
with $\Gamma_{\uvec S^q}=\gamma^++S^q_L\,\gamma^+\gamma_5+S^{qj}_T\,i\sigma^{j+}_T\gamma_5$,
as the GTMD correlator describing the distribution of quarks with polarization $\uvec S^q$ inside a target with polarization $\uvec S$.

The corresponding phase-space distribution is obtained by Fourier transform~\cite{Lorce:2011kd_20}
\begin{equation}\label{phase-space}
\rho_{\uvec S\uvec S^q}(x,\uvec k_T,\uvec b_T;\hat P,\eta)=\int\frac{\ud^2\Delta_T}{(2\pi)^2}\,e^{-i\uvec\Delta_T\cdot\uvec b_T}\,W_{\uvec S\uvec S^q}(P,k,\Delta)\big|_{\Delta^+=0},
\end{equation}
and can be interpreted as giving the quasi-probability of finding a quark with polarization $\uvec S^q$, transverse position $\uvec b_T$ and light-front momentum $(xP^+,\uvec k_T)$ inside a spin-$1/2$ target with polarization $\uvec S$~\cite{Lorce:2011kd_20}. The parameter $\eta$ indicates whether $\mathcal W$ goes to $+\infty^-$ or $-\infty^-$. Because of the hermiticity property of the GTMD correlator~\eqref{GTMDcorr-def}, these phase-space distributions are always real-valued, consistently with their quasi-probabilistic interpretation. 
There are 16 independent polarization configurations~\cite{Lorce:2011kd_20,Lorce:2013pza_15} corresponding to 16 independent linear combinations of GTMDs~\cite{Meissner:2009ww_20,Lorce:2013pza_15}. 
By construction,
the real and imaginary parts of these GTMDs have opposite behavior
under naive time-reversal ($\mathsf T$) transformation. Similarly, each Wigner distribution can
be separated into naive $\mathsf T$-even and $\mathsf T$-odd contributions, 
$
\rho_{\uvec S\uvec S^q}=\rho_{\uvec S\uvec S^q}^{e}+\rho_{\uvec S\uvec S^q}^{o},
$
with $\rho_{\uvec S\uvec S^q}^{e,o}(x,\uvec k_T,\uvec b_T;\hat P,\eta)=\pm\rho_{\uvec S\uvec S^q}^{e,o}(x,\uvec k_T,\uvec b_T;\hat P,-\eta)=\pm \rho_{\uvec S\uvec S^q}^{e,o}(x,-\uvec k_T,\uvec b_T;-\hat P,\eta)$.
We can interpret the $\mathsf T$-even contributions as describing the intrinsic distribution of quarks inside the target, whereas
the naive $\mathsf T$-odd contributions describe how initial- and final-state interactions modify these distributions.

The relativistic phase-space distribution is linear in $\uvec S$ and $\uvec S^q$ 
\begin{eqnarray}
\rho_{\uvec S\uvec S^q}
&=&\rho_{UU}+S_L\,\rho_{LU}+S^q_L\,\rho_{UL}+S_LS^q_L\,\rho_{LL}\nonumber
\\
&&+S^i_T\,(\rho_{T^iU}+S^q_L\,\rho_{T^iL})+S^{qi}_T\,(\rho_{UT^i}+S_L\,\rho_{LT^i})+S^i_TS^{qj}_T\,\rho_{T^iT^j},
\end{eqnarray}
and can further be decomposed into two-dimensional multipoles in both $\uvec k_T$ and $\uvec b_T$ spaces~\cite{Lorce:2015sqe_15}.  While there is no limit in the multipole order, parity and
time-reversal give constraints on the allowed multipoles. It is therefore more
appropriate to decompose the Wigner distributions  $\rho_X$  as follows
\begin{align}
\rho_{X}(x,\uvec k_T,\uvec b_T;\hat P,\eta)&=\sum_{m_k,\, m_b}\rho^{(m_k,m_b)}_{X}(x,\uvec k_T,\uvec b_T;\hat P,\eta), \qquad (X=UU, LU,\dots)\ ,\nonumber\\
\rho^{(m_k,m_b)}_{X}(x,\uvec k_T,\uvec b_T;\hat P,\eta)&=B^{(m_k,m_b)}_{X}(\hat k_T,\hat b_T;\hat P,\eta)\,C^{(m_k,m_b)}_{X}[x,\uvec k^2_T,(\uvec k_T\cdot\uvec b_T)^2,\uvec b^2_T],
\nonumber
\label{Decomposition}
\end{align}
where $B^{(m_k,m_b)}_{X}$ are the basic (or simplest) multipoles allowed by parity ($\mathsf P$) and time-reversal symmetries, multiplied by  the coefficient functions $C^{(m_k,m_b)}_{X}$, which depend on $\mathsf P$- and $\mathsf T$-invariant variables only. The couple of integers $(m_k,m_b)$ gives the basic multipole order in both $\uvec k_T$ and $\uvec b_T$ spaces.
All the contributions $\rho_X$ can be understood as encoding all the possible correlations
between target and quark angular momenta, see Table~\ref{angcorr}.

\begin{table}[t!]
\begin{center}
\tbl{Correlations between target polarization ($S_L,\uvec S_T$), quark polarization ($S^q_L,\uvec S^q_T$) and quark OAM ($\ell^q_L,\uvec{\ell}^q_T$) encoded in the various phase-space distributions $\rho_X$. 
}
{\begin{tabular}{@{\quad\!}c@{\quad}|@{\quad}c@{\quad}c@{\quad}c@{\quad}c@{\quad\!}}\whline
$\rho_X$&$U$&$L$&$T_x$&$T_y$\\
\hline
$U$&$\langle 1\rangle$&$\langle S^q_L\ell^q_L\rangle$&$\langle S^q_x\ell^q_x\rangle$&$\langle S^q_y\ell^q_y\rangle$\\
$L$&$\langle S_L\ell^q_L\rangle$&$\langle S_LS^q_L\rangle$&$\langle S_L\ell^q_LS^q_x\ell^q_x\rangle$&$\langle S_L\ell^q_LS^q_y\ell^q_y\rangle$\\
$T_x$&$\langle S_x\ell^q_x\rangle$&$\langle S_x\ell^q_xS^q_L\ell^q_L\rangle$&$\langle S_xS^q_x\rangle$& $\langle S_x\ell^q_xS^q_y\ell^q_y\rangle$\\
$T_y$&$\langle S_y\ell^q_y\rangle$&$\langle S_y\ell^q_yS^q_L\ell^q_L\rangle$& $\langle S_y\ell^q_yS^q_x\ell^q_x\rangle$&$\langle S_yS^q_y\rangle$\\
\whline
\end{tabular}}
\end{center}
\label{angcorr}
\end{table}
\section{Results and relation to the orbital angular momentum}

In order to obtain a two-dimensional representation  of the Wigner functions,  we integrate these phase-space distributions over $x$ and 
discretize the polar coordinates of $\uvec b_T$.
We also set $\eta=+1$ and choose $\hat P\equiv\uvec P/|\uvec P|=\uvec e_z=(0,0,1)$ so that $\hat b_T=(\cos\phi_b,\sin\phi_b,0)$ and $\hat k_T=(\cos\phi_k,\sin\phi_k,0)$. The resulting transverse phase-space distributions are then represented as sets of distributions in $\uvec k_T$-space 
\begin{equation}
\rho_X(\uvec k_T|\,\uvec b_T)=\int\ud x\,\rho_X(x,\uvec k_T,\uvec b_T;\hat P=\uvec e_z,\eta=+1)\big|_{\uvec b_T\text{ fixed}},
\end{equation}
with the origin of axes lying on circles of radius $|\uvec b_T|$ at polar angle $\phi_b$ in impact-parameter space.
This representation of transverse phase space has the advantage of making the multipole structure
in both $\uvec{k}_T$ and $\uvec{b}_T$ spaces particularly clear in a model independent way. 
For example, the basic multipole $B^{(m_k,m_b)}_X$
 will be
represented by a $m_k$-pole in transverse-momentum space at any transverse position $\uvec{b}_T$, with the
orientation determined by $m_b$ and $\phi_b = \mathrm{arg}\,\hat{b}_T$. 
The nucleon structure information from various model calculations 
is encoded in the weight of the coefficient functions $C^{(m_k,m_b)}_{X}$. 
In the following, we will show results from the light-front constituent quark model~\cite{Lorce:2011dv} only for a couple of multipole structures only. The complete discussion can be found in Ref.~\cite{Lorce:2015sqe_15}.
In particular, we choose   to represent only eight points in impact-parameter space lying on a circle with radius $|\uvec b_T|=0.4$ fm and $\phi_b=k\pi/4$ with $k\in \mathbb{Z}$. Furthermore, the $\uvec k_T$-distributions are normalized to the absolute maximal value over the whole circle in impact-parameter space.

The simplest multipole is  with $m_k=m_b=0$. It appears in $\rho^e_X$ with~$X=UU,LL,TT$ associated to the respective spin structures $1$, $S_LS^q_L$ and $(\uvec S_T\cdot\uvec S^q_T)$. These spin-spin correlations survive integration over $\uvec k_T$ and $\uvec b_T$; they are respectively related to $(H,\tilde H,H_T)$ in the GPD sector and to $(f_1,g_1,h_1)$ in the TMD sector. Contrary to these GPDs and TMDs, $\rho^e_X$ is not circularly symmetric, see Fig.~\ref{fig1}. The reason is that $\rho^e_X$ also contains information about the correlation between $\uvec k_T$ and $\uvec b_T$ (encoded in the coefficient functions $C^e_X$ through the $(\hat b_T\cdot\hat k_T)^2$ dependence) which is lost under integration over $\uvec k_T$ or $\uvec b_T$~\cite{Lorce:2011kd_20}. 
\begin{figure}[t!]
\centerline{\includegraphics[width=6cm]{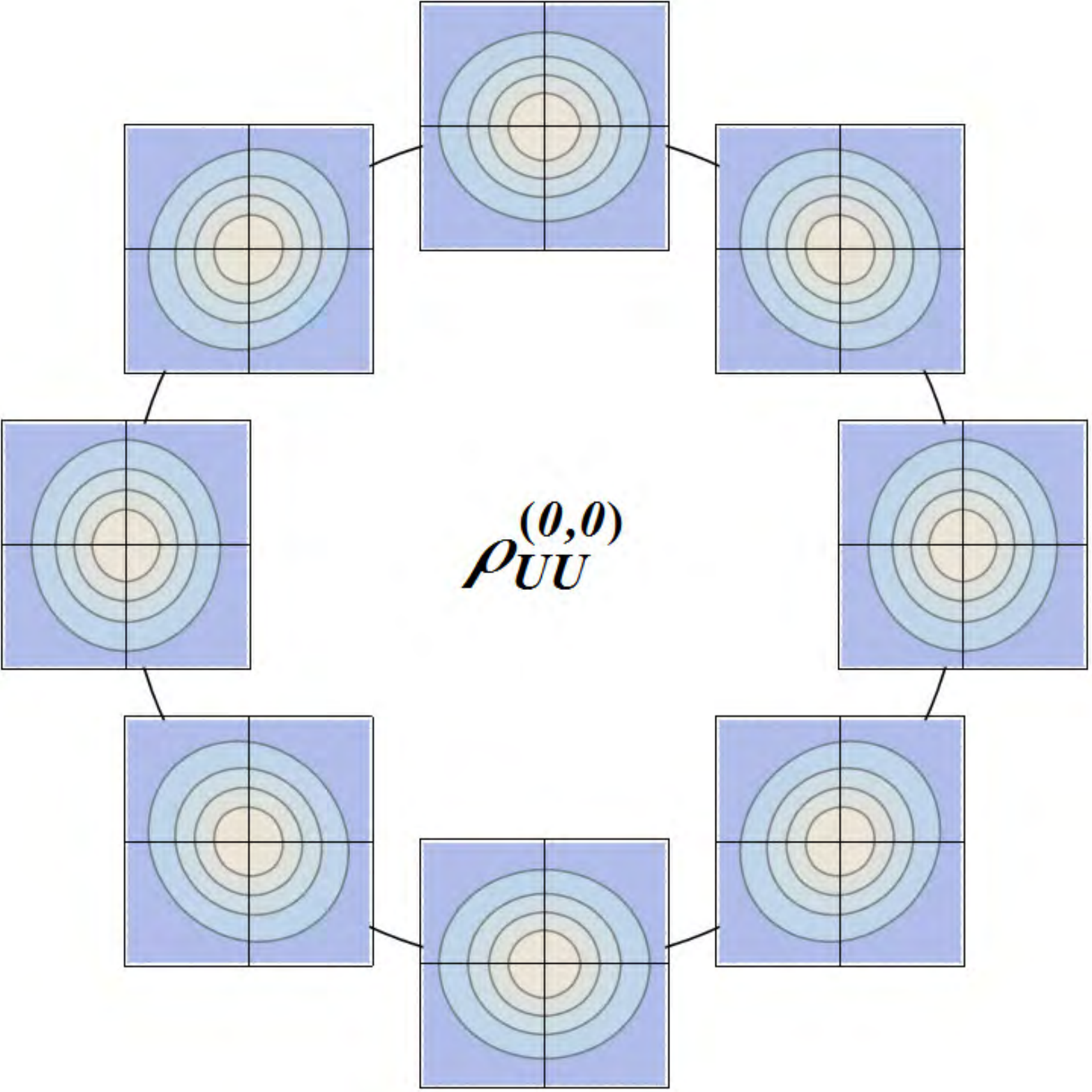}\hspace{0.5cm}\includegraphics[width=6cm]{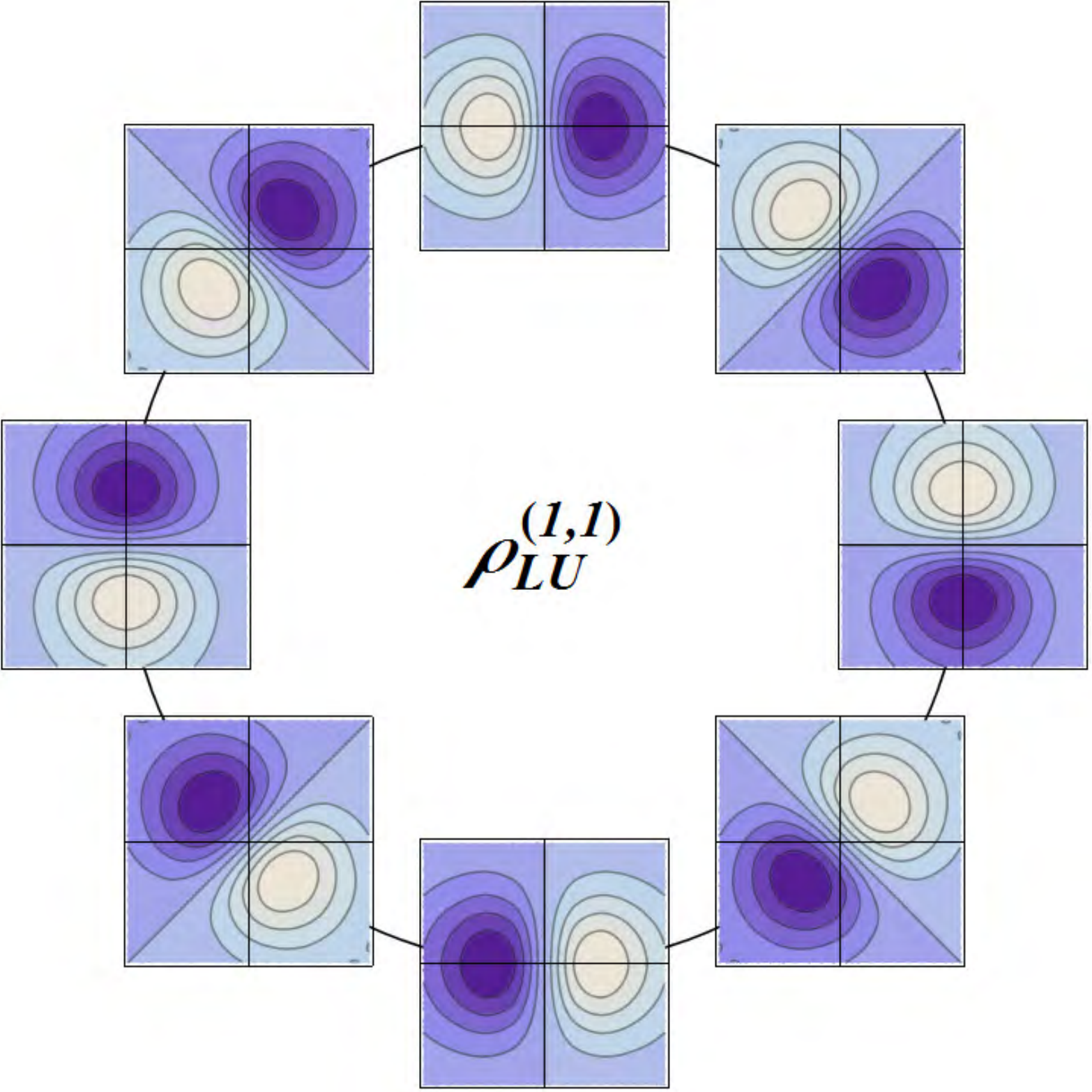}}
\caption{The $(0,0)$ (left) and $(1,1)$ (right) multipoles appearing in  $\rho^e_{UU}$ and $\rho^e_{LU}$, respectively. See text for more details. \label{fig1}}
\end{figure}

The next multipoles we discuss  here are the ones with $m_k=m_b=1$. Since $m_k=m_b$, these multipoles are invariant under rotation about the longitudinal direction. They appear in $\rho^e_X$ with $X=UL,LU$ and $\rho^o_X$ with $X=UU,LL,TT,TT'$. In $\rho^e_X$ and $\rho^o_{TT'}$, the $\uvec k_T$-dipole is oriented along the polar direction $S^q_L(\hat b_T\times\hat k_T)_L$, $S_L(\hat b_T\times\hat k_T)_L$ and $(\uvec S_T\times\uvec S^q_T)_L(\hat b_T\times\hat k_T)_L$.
In $\rho^o_{X}$ with $X =UU,LL,TT$, the $\uvec k_T$-dipole is oriented along the radial direction
$(\hat b_T \cdot\hat k_T )$, $S_LS^q_L
(\hat b_T \cdot \hat k_T )$ and $(\uvec S_T \cdot \uvec S^q_T )(\hat b_T \cdot \hat k_T )$.
None of these structures survive integration over $\uvec k_T$ or $\uvec b_T$, and therefore represent completely new information
which is not accessible via GPDs or TMDs at leading twist.
 Clearly, $\rho^e_X$ is related to the orbital motion of quarks correlated with the longitudinal polarization~\cite{Lorce:2011kd_20,Hatta:2011ku_20,Lorce:2011ni,Kanazawa:2014nha,Lorce:2014mxa_15}. 
In particular,
 the $\uvec k_T$-dipole in $\rho^{(1,1)}_{LU}$ signals the presence of a net longitudinal component of quark orbital angular momentum (OAM) correlated with the target longitudinal polarization $S_L$ (see Fig.~\ref{fig1}).
By reversing the target longitudinal polarization $S_L$, one reverses also the orbital flow. The  coefficient function $C^{(1,1)}_{LU}$ then gives  the amount of longitudinal quark OAM in a longitudinally polarized target $\langle S_L\ell^q_L\rangle$~\cite{Lorce:2011kd_20}.
As a matter of fact, the quark/gluon OAM  can be obtained from the  $\rho_{LU}$ distribution 
as~\cite{Lorce:2011kd_20,Lorce:2011ni,Kanazawa:2014nha}
\begin{eqnarray}
L_z^{q,g}=\int{\rm d}x\, {\rm d}^2 \uvec k_T\, {\rm d}^2 \uvec b_T\left(\uvec b_T\times \uvec k_T\right)_z \rho_{LU}(x,\uvec k_T,\uvec b_T;\hat P=\uvec e_z,\eta).
\label{OAM-wigner}
\end{eqnarray}
This relation 
coincides with treating the Wigner functions as if they were classical distributions,
with the quark/gluon OAM calculated from the integral over phase space of the quark/gluon distribution in a longitudinally
polarized nucleon multiplied by the naive OAM operator $(\uvec b_T\times \uvec k_T)_z$.
Depending on the shape of
the Wilson line, one obtains different definitions for the  OAM in Eq.~\eqref{OAM-wigner}~\cite{Burkardt:2015qoa,Liu:2015xha}. The gauge-invariant canonical
OAM is obtained from Eqs.~\eqref{GTMDcorr-def}-\eqref{phase-space} using a staple-like gauge link connecting the points $-z/2$ and $+z/2$ via
the intermediary points $-z/2 + \eta\infty^-$ and $z/2 + \eta\infty^-$ by straight lines. In the light-front
gauge $A^+=0$, this definition reduces to the Jaffe-Manohar OAM~\cite{Jaffe:1989jz_15}, irrespective of whether the staple is future or past-pointing.
If we connect the points $-z/2$ and $+z/2$ by a direct straight Wilson line, we
obtain the kinetic OAM associated with the OAM operator corresponding to the Ji decomposition~\cite{Ji:2012sj_15}
 and to the Ji-Xiong-Yuan~\cite{Ji:2012sj_15} definition of the gauge-invariant gluon OAM.
The $x$-dependent integrand in Eq.~\eqref{OAM-wigner} can also be interpreted as density of total longitudinal~OAM~\cite{Lorce:2011kd_20,Lorce:2011ni}.

In terms of GTMDs, 
Eq.~\eqref{OAM-wigner} can also be written as (using the notation of~Ref.~\cite{Meissner:2009ww_20})
 \begin{eqnarray}
L_z=-\int{\rm d}x\, {\rm d}^2 \uvec k_T\,\frac{\uvec k_T^2}{M^2} F_{1,4}(x,0,\uvec k^2_T,0,0).
\end{eqnarray}
This relation has been exploited  to provide the first results in lattice QCD for the canonical OAM in comparison with the kinetic OAM~\cite{Engelhardt:2017miy_20}.




\newpage

\renewcommand*{\FigPath}{./WeekI/13_Aslan/}

\wstoc{ Lorentz invariance relations for  twist-3 quark distributions}{Fatma P. Aslan, Matthias Burkardt}

\title{ Lorentz invariance relations for  twist-3 quark distributions}

\author{Fatma Aslan}
 \address{Department of Physics, University of Connecticut, Storrs, CT 06269, U.S.A.}
\index{author}{Aslan, F.}

\author{Matthias Burkardt}
\index{author}{Burkardt, M.}
 \address{Department of Physics, New Mexico State University,
Las Cruces, New Mexico,  88003-0001, USA}
\date{\today}

\begin{abstract}
We calculate twist-3 parton ditribution functions (PDFs) using cut and uncut diagrams. Uncut diagrams lead to a Dirac delta function term. No such term appears when cut diagrams are used. We show that a $\delta(x)$  is necessary to satisfy the Lorentz invariance relations of  twist-3 PDFs, except for the Burkhardt-Cottingham sum rule in QCD.
\end{abstract}


\bodymatter

\vspace*{-.1cm}
\section{Introduction} \label{s:intro}


In the scalar diquark model (SDM) and quark target model (QTM), twist-3 generalized parton distributions (GPDs) exhibit discontinuities at the points where the DGLAP and ERBL regions meet ($x =\pm\xi$)\cite {Aslan:2018zzk_567}.  In the forward limit, these discontinuities can grow into Dirac delta functions ($\delta(x)$) \cite{Aslan:2018zzk_567,Aslan:2018tff}. 
While none of the twist-2 PDFs exhibit these types of singularities in both models, all twist-3 PDFs, with the exception of $g_2(x)$ in the QTM, contain such singularities.
As we will show in section \ref{section:sumrules}, this $\delta(x)$  is necessary to satisfy the Lorentz invariance relations and the sum rules for twist-3 PDFs, except the Burkhardt-Cottingham sum rule in QCD \cite{Burkhardt:1970ti}.

This paper is organized as follows; in section \ref{section:LIR}, we investigate the two methods to calculate the PDFs, 'cut' and 'uncut' diagrams, and show that there is difference between the two approaches and one violates Lorentz invariance relations (LIR). Violations of sum rules involving higher twist PDFs is investigated in section \ref{section:sumrules}.


\vspace*{-.2cm}
\section{Lorentz invarince relations}\label{section:LIR}

Lorentz invariance, applied to the integral $I^\mu \equiv \int d^4k \hspace{.15cm}  \dfrac{k^{\mu}}{(k^2-m^2)^2}\delta[(P-k)^2- \lambda^2]$
implies $I^\mu \propto P^\mu$ as $P^\mu$ is the only 4-vector in this problem. Thus for $I^{\mu}$ the appropriate Lorentz invariance relation (LIR) reads,
\begin{equation}\label{LIR0}
\dfrac{I^+}{P^+}-\dfrac{I^-}{P^-}=0.
\end{equation}
\begin{figure}[!h]
	\begin{minipage}[t]{4cm}
		\includegraphics[scale=0.4]{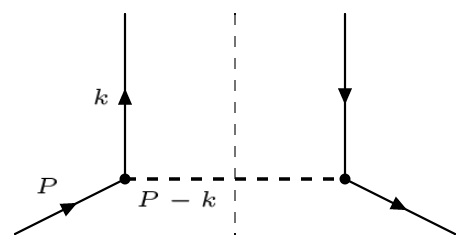}
		\caption{Cut diagram}
		\label{fig:cut}
	\end{minipage}
	\hspace{3cm}
	\begin{minipage}[t]{4cm}
		\includegraphics[scale=0.4]{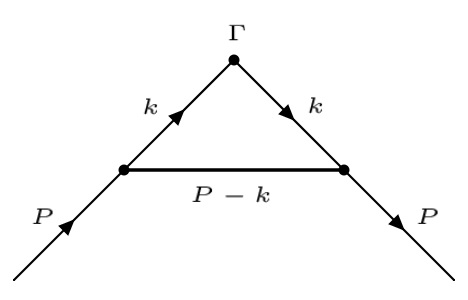}
		\caption{Uncut diagram}
		\label{fig:uncut}
	\end{minipage}
\end{figure}
 In the following, we will analyze this LIR in the SDM (in which the three valence quarks of the nucleon are considered to be in a bound state of a single quark and a scalar diquark)  using both
cut diagrams  and using uncut diagrams. We pretend that we are analysing a PDF where the factor
$k^\mu$ arises from the Dirac numerators.

In the
forward limit, the model can be represented using a cut diagram as in  FIG.\ref{fig:cut} or  an uncut diagram as in FIG. \ref{fig:uncut}.  
Using cut diagrams, the spectator propagator is replaced by $\delta\left( (p-k)^2-\lambda^2\right)$ thus enforcing the mass-shell
condition. Using uncut diagrams for the spectator line, the usual Feynman propagator is used and the energy integrals are performed using complex contour integration - picking up the pole of the spectator propagator.
Naively, the two methods should thus yield the same results. However, a subtle difference may arise at $x=0$ corresponding to infinite light-cone energy for the active quark.  In the literature, PDFs are calculated using either diagram. As we will show that, even though, both methods are equivalent and yield identical PDFs for nonzero $x$, a difference manifests itself at higher orders. This difference originates from the $\delta(x)$ term  which is present only in the higher order PDFs and  revealed when an uncut diagram is used. Such a term is not present in the calculations made by using cut diagrams! As we shall show, the $\delta(x)$ term is essential to satisfy the LIR involving twist-3 distributions and therefore the right approach is  to calculate higher order PDFs is to use uncut diagrams.

\vspace*{-.3cm}
\subsection{Cut Diagrams} \label{section:cut}

When cut diagrams are used,  $I^{\mu}\equiv I^{\mu}_{cut}$  is obtained as,
\begin{equation}\label{Imu}
I^{\mu}_{cut}\equiv P^+
\int_0^1 dx \hspace{.15cm} \int d^4k \hspace{.15cm}  \delta(k^+-xP^+) \dfrac{k^{\mu}}{(k^2-m^2)^2}\delta[(P-k)^2-\lambda^2].
\end{equation}
Here $P$  is the nucleon, $k$ is the quark momentum, $M, m, \lambda$ are the nucleon, quark and scalar diquark mass respectively.
%
The $k^+$ integral in Eq. (\ref{Imu}) is evaluated using $ \delta(k^+-xP^+) $,  while the $k^-$ intergral is evaluated by using the identity
\begin{equation}
\delta[(P-k)^2-\lambda^2]=\dfrac{1}{2P^+(1-x)}\delta\bigg(k^--\dfrac{M^2}{2P^+}+\dfrac{k_{\perp}^2+\lambda^2}{2P^+(1-x)}\bigg).
\end{equation}
Consequently, one finds for  $\mu=+$  and $\mu=-$ respectively,
\begin{equation}
\dfrac{I^+_{cut}}{P^+}\equiv
\dfrac{1}{2}\int{}d^2k_{\perp}\int_0^1{}dx \dfrac{x(1-x)}{(k_{\perp}^2+\omega)^2},
\end{equation}\vspace*{-.2cm}
\begin{equation}
\dfrac{I^-_{cut}}{P^-}\equiv
\dfrac{1}{P^-}\int{}\!\!\!d^2k_{\perp}\int_0^1{}\!\!\!dx\dfrac{M^2(1-x)-k_{\perp}^2-\lambda^2}{4P^+(k_{\perp}^2+\omega)^2},
\end{equation} where, 
$
\omega=-x(1-x)M^2+(1-x)m^2+x\lambda^2.
$
Hence, the LIR(\ref{LIR0}) is violated
\begin{eqnarray}\label{Difference}
\!\!\!\!\!\!\dfrac{I^+_{cut}}{P^+}-\dfrac{I^-_{cut}}{P^-}=\dfrac{1}{2M^2}\!\!\!\int{}\!\!d^2k_{\perp}\!\!\!\int_0^1{}\!\!\!\!dx \dfrac{-(1-x)^2M^2+k_{\perp}^2+\lambda^2}{(k_{\perp}^2+\omega)^2}=\dfrac{1}{2M^2}\!\!\int{}\!\!d^2k_{\perp}\dfrac{1}{k_{\perp}^2+m^2}.
\end{eqnarray}

\subsection{Uncut Diagrams}

However when  uncut diagrams are used, where $I^{\mu}\equiv I^{\mu}_{uncut}$ is defined as,
\begin{equation}\label{Imuncut}
I^{\mu}_{uncut}\equiv P^+
\int_0^1 dx \int d^4k \hspace{.15cm} \delta(k^+-xP^+) \dfrac{k^{\mu}}{(k^2-m^2+i\epsilon)^2}\dfrac{i}{[(P-k)^2-\lambda^2+i\epsilon]}.
\end{equation}


The $k^+$ integral  in (\ref{Imuncut}) is again taken using $ \delta(k^+-xP^+)$. However, in this case, the $k^-$ integral is taken using residue method.
For $\mu=+$, we obtain,
\begin{equation}\label{I+uncut}
\dfrac{I^+_{uncut}}{P^+}=\pi\int{}d^2k_{\perp}\int_0^1{}dx\dfrac{x(1-x)}{(k_{\perp}^2+\omega)^2}.
\end{equation}
For $\mu=-$, before taking the $k^-$ integral, we use the algebraic identity to rewrite the term in the numerator,
\begin{eqnarray}\label{kminus}
k^-= \dfrac{M^2}{2P^+}-\dfrac{(k_{\perp}^2+\lambda^2)}{2P^+(1-x)}-\dfrac{[(P-k)^2-\lambda^2]}{2P^+(1-x)}.
\end{eqnarray}
The last term in Eq.(\ref{kminus}) cancels the spectator propagator in the denominator leading to two different types of $k^-$ integrals in the expression for $I^-_{uncut}$,
\begin{eqnarray}\label{Iminus2}
I^-_{uncut}\!
=\!\dfrac{i}{2P^+}\!\! \int_0^1{}\!\!\!\!\dfrac{dx}{1-x}\!\!\int{}\!\!d^2k_{\perp} \!\Big\{\int \!dk^-\!  \dfrac{M^2(1-x)-k_\perp^2-\lambda^2}{(k^2-m^2+i\epsilon)^2[(P\!-\!k)^2-\lambda^2+i\epsilon]}
 \nonumber \\
-\int\! \!\dfrac{dk^-}{(k^2-m^2+i\epsilon)^2}\Big\}
\end{eqnarray}
  The $k^-$ integral in Eq.(\ref{Iminus2}), leads to a delta function   \cite{Yan:1973qg},
\begin{eqnarray}
\int{}\dfrac{dk^-}{(k^2-m^2+i\epsilon)^2}=\dfrac{i\pi}{k_{\perp}^2+m^2}\delta(k^+).
\end{eqnarray}
Using this result, and taking the $k^-$ integrals in Eq.(\ref{Iminus2}) we obtain,
\begin{align}\label{I-uncut}
\dfrac{I^-_{uncut}}{P^+}=\dfrac{\pi}{2P^{+2}}\int{}d^2k_{\perp}\Big[\int_0^1{}dx\dfrac{M^2(1-x)-k_{\perp}^2-\lambda^2}{(k_{\perp}^2+\omega)^2}+\dfrac{1}{k_{\perp}^2+m^2}\Big],
\end{align}
which equals Eq. (\ref{I+uncut}), i.e. the
LIR (\ref{LIR0})  is satisfied when uncut diagrams are used.

The reason one method results in a violation of the  LIR while other does not
is the appearance of $\delta(x)$ term which is revealed only when an uncut diagram is used. Cut diagrams do not include the point $x=0$, and miss the $\delta(x)$ term at this point.


\section{Violation Of Sum Rules}\label{section:sumrules}

The point $x=0$ is not experimentally accessible in DIS  since  it corresponds to $ P \cdot q \rightarrow \infty$ and thus $\delta (x)$ cannot be seen. Any relation involving a twist 3 PDF containing a $\delta(x)$  would appear to be violated. Nevertheless, there  is no doubt in the validity of these sum rules because they are direct consequences of Lorentz invariance.  Therefore, the violation of the sum rules from the experimental data would provide an indirect evidence on the existence of the Dirac delta functions.

 The most famous Lorentz invariance relation between a twist 2 PDF ($g_1(x)$) and a twist 3 PDF ($g_T(x)$) is the
 Burkhardt-Cottingham sum rule \cite{Burkhardt:1970ti},
\begin{eqnarray}\label{BCsumrule}
\int_{-1}^1{}dx g_1 (x)=\int_{-1}^1{}dx g_T (x).
\end{eqnarray}
A similar relation is the $h$-sum rule, where the l.h.s. is equal to the tensor charge
\begin{eqnarray}\label{hsumrule}
\int_{-1}^1{}dx h_1 (x)=\int_{-1}^1{}dx h_L (x).
\end{eqnarray}

Another sum rule including a twist-3 PDF is the $\sigma$-term sum rule,  which provides a relation between quark mass $m$ and nucleon mass $M$,

\begin{eqnarray}\label{esumrule}
\int_{-1}^1{}dx e (x)=\dfrac{1}{2M}\langle P| \overline{\psi}(0)\psi(0)|P\rangle=\dfrac{d}{dm}M.
\end{eqnarray}
If any of the twist 3 PDFs above contain a $\delta (x)$ term, experimental measurements would not be able to confirm the sum rule in ) and claim their violation.


\section{Summary and Discussion}\label{section:sum}



Twist-3 PDFs contain a $\delta(x)$ in both QTM and SDM  only with the exception of $g_2 (x)$ in the QTM. These $\delta(x)$ terms are not related to the twist-2 (WW) parts of the twist-3 PDFs but contributes both the qgq correlation and mass terms. Since $x=0$ is not experimentally accesible, violations of the sum rules containing twist-3 PDFs and GPDs from the experimental data would provide an indirect evidence on the existence of these $\delta(x)$ contributions.

Using cut or uncut diagrams to calculate PDFs from models leads to the same result for $x\neq 0$. However, there is a difference between the two approaches for higher orders. Cut diagrams exclude $x=0$ point and hence miss the $\delta(x)$ terms. Therefore higher twist distributions calculated with cut digrams do not satisfy LIR. In order to restore it, one needs to include $x=0$ point by using an uncut diagram.

{\bf Acknowledgments}
This work was partially supported by the DOE under  Grant  No.DE-FG03-95ER40965.   F.A. was partially also supported by the DOE Contract No. DE-AC05-06OR23177, under which
Jefferson Science Associates, LLC operates Jefferson Lab.





\newpage
%


\wstoc{Deeply Virtual Compton Scattering: status of experiments at Jefferson Lab and COMPASS}{Daria Sokhan}
\title{Deeply Virtual Compton Scattering: status of experiments at Jefferson Lab and COMPASS}

\author{Daria Sokhan}
\index{author}{Sokhan, D.}

\address{School of Physics \& Astronomy, University of Glasgow,\\
Glasgow G12 8QQ, UK\\
E-mail: daria.sokhan@glasgow.ac.uk}

\begin{abstract}
Deeply Virtual Compton Scattering (DVCS) is considered as the ``golden channel" for accessing Generalised Parton Distributions, which contain information on the 3D imaging of the nucleon, the composition of its spin and pressure distributions within it. We present a summary of the recent, ongoing and future DVCS measurements in the valence quark region accessible at Jefferson Lab and in the sea-quark region reached by COMPASS, and their implication for the study of nucleon structure.
\end{abstract}


\bodymatter

\section{Deeply Virtual Compton Scattering}\label{aba:sec1}
DVCS is a process in which a lepton scatters from a quark within a nucleon, minimally perturbing it, and a high energy photon is produced as a result. At leading-order and leading-twist (twist-2, where twist is given by powers of $\sqrt{Q^{-2}}$ in the scattering amplitude), DVCS provides access to four chiral-even GPDs, each a function of $Q^2$, $x$ (the nucleon's longitudinal momentum fraction carried by the struck quark), $\xi$ (half the change in $x$ as a result of the scattering) and $t$ (squared four-momentum change of the nucleon): $E^q$, $\tilde{E^q}$, $H^q$ and $\tilde{H^q}$. These are accessible indirectly through the complex Compton Form Factors (CFFs), the real parts of which are given by an integral of the corresponding GPD across $x$ and parametrise the DVCS cross-section, beam-charge and double-spin asymmetries, while the imaginary parts are the values of the GPD at the points where $x = \xi$ and parametrise single-spin asymmetries\cite{BMJ}.  Different CFFs dominate different observables -- a wide range of measurements across a large kinematic range is therefore required to fully constrain GPDs. 

The $eN \rightarrow e'N'\gamma$ interaction, however well reconstructed, necessarily includes the Bethe-Heitler (BH) process, where the photon is radiated by the incoming or scattered electron. BH interferes with DVCS at the amplitude level but its contribution can be calculated from the well-known Form Factors and QED. Since BH usually dominates over that of DVCS, the BH-DVCS interference term in the cross-section is often used to extract the DVCS amplitude.

\section{Experimental facilities: Jefferson Lab and COMPASS}\label{aba:sec2}
Jefferson Lab (JLab)  in Virginia, USA, has recently upgraded its electron beam accelerator from 6 GeV to 12 GeV maximum energy, with 11 GeV deliverable to the three experimental halls focussed on the study of nucleon structure: A, B and C.  The detectors in the three halls enable complementary measurements of electron scattering to be performed. Hall B housed CLAS (CEBAF Large Acceptance Spectrometer) and now houses the upgraded CLAS12, which provides almost full angular coverage and detection of charged and neutral particles at luminosities of $10^{35}$ cm$^{-2}$s$^{-1}$ (a factor of 10 higher than CLAS). DVCS measurements from Hall B include cross-sections and asymmetries across a very wide range of phase space.  Halls A and C, in contrast, contain movable high-resolution spectrometer arms, which are combined with calorimeter arrays for the detection of photos in DVCS. The recoiling nucleon is usually reconstructed through missing mass.  As such, the phase space covered is limited, however the instrumentation can accept $10^3$ higher currents than hall B, resulting in high-precision measurements of cross-sections at well-defined kinematic points.

While JLab provides access to the valence region, the COMPASS (Compact Muon and Proton Apparatus for Structure and Spectroscopy) experiment, which uses a tertiary 160 GeV polarised $\mu^{+/-}$ beam from the SPS beamline at Cern, enables measurement in the sea-quark region. The DVCS programme at COMPASS-II, which is in operation since 2012, uses an unpolarised liquid $H_2$ target and includes one month of data-taking in 2012 and six months in 2016-17. Spectrometers, time-of-flight detectors and calorimeters allow a fully exclusive reconstruction of the DVCS process.
    
\section{JLab in 6 GeV era}
The first high-precision DVCS cross-sections on the proton, which are dominated by the real part of the $\mathbf{H}$ CFF, were measured in Hall A at fixed Bjorken-x, $x_B = 0.36$, and three $Q^2$ points (1.5, 1.9 and 2.3 GeV$^2$).  The observed rough scaling of the CFFs across the limited range of $Q^2$ supported the assumption that factorisation holds at this moderate $Q^2$.  The strong deviation of the unpolarised BH-DVCS cross-section from BH calculation indicated that the pure DVCS term in the cross-section could be separable from the interference term. This was enabled in a dedicated experiment with beam energies $E_e = 4.5$ and $5.6$ GeV using a generalised Rosenbluth technique which made use of the fact that the pure DVCS term in the cross-section scales as $E_e^2$, while the BH-DVCS interference term scales as $E_e^3$.  The simultaneous fit could be significantly improved by the inclusion of either higher-order or higher-twist effects. While the sensitivity of the data was insufficient to resolve between these two scenarios, it suggested that we may be seeing effects of gluon vertices in the interaction\cite{Max}.

While flavour separation of GPDs may be achieved through related processes such as hard-exclusive meson production, in DVCS it is only possible via measurements on both proton and neutron targets. Rosenbluth separation of cross-section terms for DVCS on the neutron (using a liquid $D_2$ target) is currently underway in Hall A. It follows a first measurement of beam-spin asymmetry in neutron-DVCS\cite{HallAn}, which is dominated by the GPD $\mathbf{E}$, the least known and least constrained of the CFFs, but which presents particular interest in the study of the spin composition of the nucleon. The total angular momentum, $J_q$, for each quark-flavour can be expressed in terms of integrals over GPDs $\mathbf{H}$ and $\mathbf{E}$ via Ji's relation\cite{BMJ}. In combination with the known contribution of intrinsic quark spin to that of the nucleon, a determination of $J_q$ would reveal the completely unknown contribution of quark orbital angular momentum.  

An extensive experimental programme using CLAS and both unpolarised liquid $H_2$\cite{Jo} and longitudinally polarised $NH_3$ targets\cite{Silvia} yielded cross-sections, beam-spin (dominated by the imaginary part of $\mathbf{H}$), target-spin and double-spin asymmetries (both dominated by $\Im({\tilde{\mathbf{H}}})$ and $\Im({\mathbf{H}})$), binned finely across a wide range of $x_B$ (0.1-0.5), $Q^2$ (1-4 GeV$^2$) and $t$ (0.1-0.5 GeV$^2$). The polarised target experiment in particular enabled a simultaneous fit to all three asymmetries, at the same kinematic points. Since the imaginary part of a CFF gives the GPD at a particular value of $x$, it provides the most direct access to tomographic information on the nucleon: a steep slope in its exponential dependence on $t$ indicates a smaller range in transverse spatial distribution. Extraction of CFFs was carried out both in local and global fits. One of the main local fits applied to the observables minimised model-dependence by setting CFFs as free parameters with ranges within $\pm5$ times the VGG model predictions for them\cite{VGG}. Global fits used models constrained by other measurements to simultaneously fit the entire available world data-set, for example using the PARTONS framework\cite{PARTONS}. While a mapping of spatial distributions vs momentum fraction $x$ is still heavily model-dependent, there is an indication that valence quarks are more centered in the nucleon, while sea quarks are spread across a wider transverse area\cite{Dupre}. Constraints from the measurements at JLab at 11 GeV and in the quark-gluon sea region using the EIC are crucial to complete 3D tomographic imaging of the nucleon.  

A relation of GPD $\mathbf{H}$ to mechanical properties of the nucleon, encoded in the Gravitational Form Factors of its energy-momentum tensor, provides access to previously completely unprobed aspects of nucleon structure, such as pressure and shear forces\cite{Polyakov}. While the constraints of current data are insufficient for a model-independent extraction of pressure distributions\cite{VolKum}, the additional measurements from JLab at 11 GeV and EIC will enable this radically novel perspective on nucleon structure.

\section{JLab in 11 GeV era}\label{aba:sec4}    
The programme for halls A and C in the new, extended regions of phase space made accessible by the JLab energy increase makes use of the upgraded spectrometer arms for the detection of scattered electrons and includes experiments with an unpolarised liquid $H_2$ target. With measurements at $E_e$ of 6.6, 8.8 and 11 GeV, scans in $Q^2$ ($\sim2-10$) and $x_B$ ($\sim0.2-0.6$), the aim is to separate the azimuthal, energy and helicity dependencies of the pure DVCS and BH-DVCS interference terms across a wide kinematic range\cite{HallA11}. The precision of the data will greatly constrain the extraction of CFFs and shed light on possible higher-twist or higher-order effects. Hall A started taking data in spring 2017. The first CLAS12 experiment ran in 2018 with an unpolarised $LH_2$ target. DVCS asymmetries and cross-sections are currently under analysis. The programme likewise includes measurements at 6.6 and 8.8 GeV, at 11 GeV on an unpolarised $LD_2$ target to allow access to neutron-DVCS (the beam-spin asymmetry in which is particularly sensitive to different contributions of $J_q$ in this kinematic regime) and at 11 GeV on longitudinally polarised $NH_3$ and $ND_3$ targets, for access to target-spin and double-spin asymmetries in proton- and neutron-DVCS. A transversely polarised $HD$ target is under development, which will give access to $\mathbf{E}$ on the proton -- crucial to enable flavour-separation in combination with beam-asymmetry measurements in neutron-DVCS. The partonic structure of nuclei will be studied in coherent DVCS on $^4He$ and $d$ using a recoil detector ALERT currently under construction, which will be integrated with CLAS12\cite{HallB11}.  

\section{COMPASS}\label{aba:sec5} 
The DVCS programme at COMPASS has been focussed on extraction of cross-sections in the sea-quark regime. While BH provides almost all the signal at very low $x_B$ ($<0.01$), DVCS begins to dominate at $x_B> 0.03$. 2012 data was used to extract a slope parameter for the exponential dependence of the DVCS cross-section on $t$, which, within the framework of the model, can be converted into a transverse extension of the partons\cite{COMPASS}. Analysis of the 2016-17 data, which will increase the statistics by a factor of 10, is underway. COMPASS-II will come to an end in 2021 and proposals are under development for measurements with a transversely-polarised target in the future COMPASS-III/AMBER programme. The study of the deep quark-gluon sea awaits the construction of the Electron-Ion Collider, which will provide invaluable constraints on quark and gluon GPDs in the low-$x$ region.


\newpage
%


\wstoc{
Heavy Quarkonium Production at the EIC Energies
}{Jian-Wei Qiu}

\title{
Heavy Quarkonium Production at the EIC Energies
}

\author{Jian-Wei Qiu$^*$}
\index{author}{Qiu, J.}

\address{Theory Center, Jefferson Lab,\\
12000 Jefferson Avenue, Newport News, VA 23606, U.S.A.\\
$^*$E-mail: jqiu@jlab.org}

\begin{abstract}
Heavy quarkonium production offers not only unique perspectives into the formation of QCD bound states, but also an excellent probe sensitive to both the short-distance as well as long-distance dynamics of QCD, due to its multiple and well-separated momentum scales.  In this talk, I discuss how the production of heavy quarkonia could fulfill this dual role at the proposed Electron-Ion Collider. 
\end{abstract}


\bodymatter

%
\section{Introduction}\label{sec1:intro}

The production of heavy quarkonia is still one of the most fascinating subjects in strong interaction physics after the discovery of J/$\psi$ over 45-year ago \cite{Brambilla:2010cs}. With the heavy quark mass $m_Q \gg \Lambda_{\rm QCD}$, the inclusive production of a pair of heavy quarks is an essentially perturbatively calculable process, which provides a controllable short-distance probe, while the subsequent evolution of the pair into a quarkonium is nonperturbative.  Different treatments of the nonperturbative transformation from a heavy quark pair to a bound quarkonium have led to various theoretical models for quarkonium production, most notably, the non-relativistic QCD (NRQCD) model \cite{Brambilla:2004wf,Brambilla:2010cs}.  Although the NRQCD treatment of heavy quarkonium production~\cite{Bodwin:1994jh} is by far the most theoretically sound and phenomenologically successful \cite{Braaten:1996pv,Petrelli:1997ge,Kramer:2001hh,Brambilla:2010cs}, there have been notable surprises and anomalies in describing new data \cite{Lansberg:2019adr}.  

With its high luminosity, polarized lepton and hadron beams, as well as beams of heavy ions, the proposed Electron-Ion Collider (EIC) could be a new and ideal facility to produce and study heavy quarkonia.  By varying the momentum of a produced heavy quarkonium, due to the time dilation, we could control the formation time of this heavy quarkonium from a pair of produced heavy quarks, allowing us to study how this pair to interact with the medium around it during its formation process.  With the perturbative nature of producing the heavy quark pairs, momentum dependence of heavy quarkonium production in lepton-nuclei collisions at the EIC energies could offer a controllable observable to use nuclei as the femtometer sized detectors to probe the underline mechanism of color neutralization and formation of QCD bound states.  With the dominance of gluon initiated subprocesses, exclusive production of heavy quarkonia in the lepton-hadron diffractive scattering at the EIC energies, both near and away from the threshold of producing the quarkonia, could provide a unique way to explore the hadron's gluonic structure.  With the ability to control the hard scale in lepton-hadron scattering, relative to the heavy quark mass, heavy quarkonium production at the EIC energies could provide wealthy opportunities to study QCD and strong interaction dynamics.

%
\section{Inclusive production and factorization}\label{sec2:production}

In order to produce a heavy quarkonium, 
the energy exchange in the collisions has to be larger
than the invariant mass of the produced quark pair ($\ge 2m_Q$),
and therefore, the pairs are effectively produced at a distance scale
$\Delta r \le 1/2m_Q \le 0.1$~fm for charmonia or 0.025~fm for 
bottomonia.  Since the binding energy of a heavy quarkonium of 
mass $M\sim 2m_Q$ is much less than heavy quark mass, 
$(M^2-4m_Q^2)/4m_Q^2\ll 1$, the 
transition from the pair to a meson is sensitive to 
soft physics.  The quantum interference between the production
of the heavy quark pairs and the transition process 
is powerly suppressed by the heavy quark mass, and 
the production rate for a heavy quarkonium state, $H$, 
up to corrections in powers of $1/m_{Q}$, 
could be factorized as, 
\begin{equation}
\sigma_{A+B\rightarrow H+X} 
\approx 
\sum_{n} \int d\Gamma_{Q\bar{Q}}\
\sigma_{A+B\rightarrow Q\bar{Q}[n]+X}(\Gamma_{Q\bar{Q}}, m_Q) \
F_{Q\bar{Q}[n]\rightarrow H}(\Gamma_{Q\bar{Q}}) 
\label{qq-fac}
\end{equation}
with a sum over possible $Q\bar{Q}[n]$ states and an integration 
over available $Q\bar{Q}$ phase space $d\Gamma_{Q\bar{Q}}$ 
for producing a bound quarkonium.
In Eq.~(\ref{qq-fac}), $\sigma_{A+B\rightarrow Q\bar{Q}[n]+X}$  
represents the production of a pair of {\it on-shell} heavy quarks and 
should be calculable in perturbative QCD \cite{Collins:1985gm}; and
the $F_{Q\bar{Q}[n]\rightarrow H}$ represent a set of nonperturbative transition probabilities 
for the pair of heavy quark ($\psi$) and antiquark ($\chi$) in the state $Q\bar{Q}[n]$
to transform into a quarkonium state $H$, and are proportional to the Fourier transform 
of following non-local matrix elements
\begin{equation}
\sum_N
\langle 0|\chi^\dagger(y_1)\, {\mathcal K}_n\, \psi(y_2)|H+N\rangle
\langle H+N| \psi^\dagger(\tilde{y}_2)\, {\mathcal K}'_n\,
             \chi(\tilde{y}_1)|0\rangle\, ,
\label{F-nonlocal}
\end{equation}
where $y_i (\tilde{y}_i)$, $i=1,2$ are coordinates, and
${\mathcal K}_{n}$ and ${\mathcal K}'_{n}$ are 
local combinations of color and spin matrices for the $Q\bar{Q}$  
state $[n]$.  A proper insertion of Wilson lines to make
the operators in Eq.~(\ref{F-nonlocal}) gauge invariant is implicit \cite{Nayak:2005rt}.  
The debate on the production mechanism has been effectively focusing on 
what is the best ``approximation" for the $F_{Q\bar{Q}[n]\rightarrow H}$ to 
represent the transition from the pair to the meson.  
Predictive power of the factorization in Eq.~(\ref{qq-fac}) relies on the
universality or process independence of the non-local function,
$F_{Q\bar{Q}[n]\rightarrow H}(\Gamma_{Q\bar{Q}})$ or its various approximations.

The factorization formula in Eq.~(\ref{F-nonlocal}) is an approximation and
can be violated by any leading power soft gluon interactions or color exchange
between the $F_{Q\bar{Q}[n]\rightarrow H}$ and initial-state hadron $A$ (and/or $B$)
or any measured final-state hadron(s).
Such factorization breaking interaction are suppressed
if the ``invariant mass" of the two-body system made of the observed heavy quarkonium 
and any identified hadron(s) is so much larger than the mass of 
heavy quarkonium.  In the lepton-hadron scattering at the EIC energies, this
necessary condition effectively limits to the events in which the quarkonium
has a large relative transverse momentum away from the beam hadron, or 
it has a large rapidity and/or transverse momentum difference from any observed 
final-state hadron.

%
\section{Transverse momentum dependence of quarkonium production}\label{sec1:matching}

Transverse momentum distribution of heavy quarkonium produced in a lepton-hadron collisions 
depends on the choice of frame.  In the lepton-hadron head-on frame, if the quarkonium 
transverse momentum $p_T \gg m_Q$, the produced heavy quarkonium should be 
sufficiently far away from the beam jet and the factorization in Eq.~(\ref{qq-fac})
should be valid~\cite{Kang:2014tta}
\begin{eqnarray}
E_p\frac{d\sigma^{\rm PQCD}_{l+h\to H(p)+X}}{d^3p}
&\approx &
\sum_{f} E_c\frac{d\hat{\sigma}_{l+h\to f(p_c)+X}}{d^3p_c}
\otimes 
D_{f\to H}(z;m_Q)
\nonumber \\
&\ &  \hspace{-10mm} +\ 
\sum_{[Q\bar{Q}(\kappa)]}
E_c\frac{d\hat{\sigma}_{l+h\to [Q\bar{Q}(\kappa)](p_c)+X}}{d^3p_c}
\otimes
{\cal D}_{[Q\bar{Q}(\kappa)]\to H}(z,u,v;m_Q)\, ,
\label{eq:pqcd_fac}
\end{eqnarray}
where the 1$^{\rm st}$ term on the right-hand side is the leading power in $1/p_T$ 
and the 2$^{\rm nd}$ gives the dominant contribution at the 1st sub-leading power, and
$p_c=p/z$ in the 1$^{\rm st}$, $p_c=p_Q+p_{\overline{Q}}$ in the 2$^{\rm nd}$, and 
$p_Q=up/z$ and $p_{\overline{Q}}=(1-u)p/z$ for $Q\bar{Q}$ in the amplitude and 
$u\to v$ for the $Q\bar{Q}$ in the complex conjugate amplitude.  
In Eq.~(\ref{eq:pqcd_fac}), the universal fragmentation functions $D_{f\to H}$
and ${\cal D}_{[Q\bar{Q}(\kappa)]\to H}$ resum the large logarithmic contributions 
in powers of $\ln(p_T^2/m_Q^2)$ by solving the evolution equations~\cite{Kang:2014tta}, and 
their input distributions at the scale $\sim m_Q$ could be factorized into the universal 
non-perturbative $F_{Q\bar{Q}[n]\rightarrow H}$ as in Eq.~(\ref{qq-fac}) or its 
approximated sum of long-distance local matrix elements (LDMEs) in 
NRQCD~\cite{Ma:2013yla}.

When $p_T^2\to m_Q^2 \gg \Lambda^2_{\rm QCD}$, the factorization formalism in
Eq.~(\ref{qq-fac}) might still be true, and we introduce the following matching formalism~\cite{Lee:2020},
\begin{equation}
E_p\frac{d\sigma_{l+h\to H(p)+X}}{d^3p}
\approx  
E_p\frac{d\sigma^{\rm PQCD}_{l+h\to H(p)+X}}{d^3p} 
- E_p\frac{d\sigma^{\rm PQCD-Asym}_{l+h\to H(p)+X}}{d^3p}
+ E_p\frac{d\sigma^{\rm NRQCD}_{l+h\to H(p)+X}}{d^3p} 
\label{eq:matching}
\end{equation}
where $d\sigma^{\rm PQCD-Asym}$ is given by $d\sigma^{\rm PQCD}$ in
Eq.~(\ref{eq:pqcd_fac}) with the fragmentation functions evaluated in 
NRQCD factorization at the fixed order.  When $p_T \gg m_Q$, the none-logarithmic 
mass dependent terms in $d\sigma^{\rm NRQCD}$ vanish, and 
$d\sigma_{l+h\to H(p)+X}\to d\sigma^{\rm PQCD}$ since 
$d\sigma^{\rm PQCD-Asym}$ cancels $d\sigma^{\rm NRQCD}$.  On the other hand,
when $p_T \to m_Q$, the high power resummed $\ln(p_T^2/m_Q^2)$-type logarithmic contribution 
in $d\sigma^{\rm PQCD}$ is less important and $d\sigma^{\rm PQCD-Asym}$ cancels 
$d\sigma^{\rm PQCD}$ and $d\sigma_{l+h\to H(p)+X}\to d\sigma^{\rm NRQCD}$.

%
\section{Exclusive production of heavy quarkonia}\label{sec4:threshold}

As emphasized in the EIC White Paper~\cite{Accardi:2012qut_890}, the exclusive production of J/$\psi$ sufficiently away from the threshold in $l+h\to l' + {\rm J/}\psi +h'$ is an excellent probe of gluon generalized parton distributions (GPDs), from which we could derive the spatial density distribution of gluons inside a bound hadron. Although we do not expect the same factorization to work near the threshold, we find that the exclusive $\Upsilon$ production in $l+h\to l' + \Upsilon +h'$ could have a new type of factorization even at the threshold to isolate valuable information sensitive to the QCD trace anomaly that is very important for us to understand the origin of hadron mass.  With the scattered lepton $l'$ measured, we can uniquely fix the kinematic for $\gamma^*(q)+h(p)\to \Upsilon +h(p')$, from which we found that $|t/s|\propto m_h/m_\Upsilon\ll 1$, where $t=(p-p')^2$ and $s=(p+q)^2$, the strong ordering of scales necessary for the factorization of $\langle p|F^2|p'\rangle$ from the hard part to produce the $\Upsilon$.  More detailed study is underway.

%
\section{Summary}

The validity of factorization requires the hard scale to be much larger than the typical hadronic scale $\sim 1/$fm, and consequently, single scale hard probes are not very sensitive to the rich physics at the hadronic scale.  With its well-separated multiple momentum scales, production of heavy quarkonia could offer excellent opportunities for studying hadron structure and QCD dynamics at the proposed Electron-Ion Collider. 

{\it Acknowledgments:}\ This work is supported by Jefferson Science Associates, LLC under U.S. DOE Contract number: DE-AC05-06OR23177.


%


\newpage
\renewcommand*{\FigPath}{./Logo/} 

\begin{tcolorbox}[colframe=white]
\begin{minipage}{0.2\textwidth}
\includegraphics[width=1.\textwidth]{\FigPath/INT_Workshop_Logo_Final_Black.png}
\end{minipage}
\begin{minipage}{0.7\textwidth}
\wstoc{\bf Week II}{}
\title{Week II}
\end{minipage}
\end{tcolorbox} 
 
%

 
\wstoc{Week 2: Transverse momentum and transverse spin}{Alessandro Bacchetta, Zhong-Bo Kang}
\title{Week 2: Transverse momentum and transverse spin}

\author{Alessandro Bacchetta$^*$}
\index{author}{Bacchetta, A.}

\address{Dipartimento di Fisica, Universit\`a degli Studi di Pavia, I-27100 Pavia, Italy \\
and Istituto Nazionale di Fisica Nucleare, Sezione di
  Pavia,  I-27100 Pavia, Italy\\
$^*$E-mail: alessandro.bacchetta@unipv.it}

\author{Zhong-Bo Kang$^\dagger$}\
\index{author}{Kang, Z.}

\address{Department of Physics and Astronomy, University of California, Los Angeles, CA 90095, USA \\
and Mani L. Bhaumik Institute for Theoretical Physics, University of California, Los Angeles, CA 90095, USA \\
$^\dagger$E-mail: zkang@physics.ucla.edu}

\begin{abstract}
The study of transverse momentum dependent parton distributions (TMDs) features prominently among the goals of the EIC program. During Week 2, several topics related to TMDs have been discussed, offering complementary information with respect to the original EIC white paper. In this introduction to Week 2, we mention the main ideas and include some developments that occurred even after the conclusion of the program.
\end{abstract}


\bodymatter

\section{Introduction}

Since the writing of the 2011 INT Report and the EIC White Paper in 2012, an impressive amount of results has been obtained in the experimental, theoretical, and phenomenological study of TMDs. 

Experimental results have been collected in SIDIS by the HERMES and COMPASS Collaborations and by experiments at Jefferson Lab, in $e^+e^-$ by BaBar, BELLE and BESIII, and in proton-proton collisions
by RHIC experiments. New polarized Drell-Yan (DY) measurements have taken place at COMPASS~\cite{Aghasyan:2017jop_568} and RHIC~\cite{Aschenauer:2016our}, while the E1039/SpinQuest collaboration at Fermilab will soon start the data taking~\cite{Chen:2019hhx}. Jefferson Lab has just started the data-taking phase after the 12 GeV upgrade~\cite{Dudek:2012vr_568}. The Large Hadron Collider has become an important source of information on TMDs via Drell-Yan, $Z,~W^\pm$ production data, as well as via measurements of jets and their substructure.

In parallel, theoretical studies on TMDs have also seen remarkable progress in the past few years. First of all, in the study of the quark TMDs, which can be probed through the traditional standard processes (SIDIS, Drell-Yan and $e^+e^-$), perturbative precision has been improved enormously, from the usual next-to-leading order (NLO) to next-to-next-to-leading (NNLO) since 2016~\cite{Echevarria:2016scs,Luo:2019hmp}, up to next-to-next-to-next-to-leading (NNNLO) order in 2019~\cite{Luo:2019szz}. Second of all, gluon TMDs have now also been studied extensively~\cite{Echevarria:2015uaa,Boer:2015vso,Luo:2019bmw}. For example, unpolarized gluon TMDs and gluon Sivers functions have been studied through the heavy quark pair production at the EIC~\cite{Boer:2016fqd_568,Zheng:2018awe}. Thirdly, new opportunities to probe TMDs beyond those standard processes have been proposed, which typically involve jets. For example, One of the main directions is to use internal structure of jets, such as transverse momentum distribution of hadrons inside a fully reconstructed jet to extract TMD fragmentation functions. Lastly, processes that could lead to potential TMD factorization breaking have been studied recently in the community, both in theory~\cite{Buffing:2018ggv_568} and in experiment~\cite{Aidala:2018bjf_568}.

Concurrently, as we will review in details below, the phenomenological study of TMDs have gain even more active developments, due to the community's collective efforts. Besides several leading groups are making outstanding progress worldwide, US Department of Energy (DOE) has funded a topical group -- ``Topical Collaboration for the Coordinated Theoretical Approach to Transverse Momentum Dependent Hadron Structure in QCD'', in short TMD collaboration, which organizes efforts in the community to make coherent progress in the study of TMDs.

\section{Unpolarized TMDs}

In 2017, data from SIDIS and Drell-Yan processes were combined for the first time to extract TMDs 
at next-to-leading logarithmic (NLL) accuracy.\cite{Bacchetta:2017gcc}
Ref.~\cite{Bertone:2019nxa_568} took into consideration only Drell-Yan data, reaching NNLL accuracy. An extraction of pion TMDs has been attempted for the first time~\cite{Vladimirov:2019bfa}. Very recently, an extraction based on SIDIS and Drell-Yan data at NNLL~\cite{Scimemi:2019cmh_568} and an extraction based on Drell-Yan data at N$^3$LL were presented~\cite{Bacchetta:2019sam_568}. These results were unthinkable only a few years ago. The increase in perturbative accuracy will be extremely useful to make predictions for the EIC. Vice versa, precise measurements at the EIC will provide useful information also for high-precision analyses at hadron colliders.
Apart from the differences, all analyses seem to agree in stating that unpolarized TMDs cannot be parametrized by simple functional forms (i.e., simple Gaussians, without $x$ dependence). 

In spite of these successes, some problems have been identified. For instance, 
calculations of SIDIS and Drell-Yan at high transverse momentum, $q_T \sim Q$, where the collinear QCD approach should be applicable, drastically underestimate the available data~\cite{Gonzalez-Hernandez:2018ipj,Bacchetta:2019tcu}. Another problem is that SIDIS multiplicities at low transverse momentum seem to be significantly underestimated by the TMD description when going at higher perturbative accuracy~\cite{Gonzalez:2019}. However, this is in contradiction with the results of Ref.~\cite{Scimemi:2019cmh_568}.
In general, it is necessary to better clarify the regions of applicability of the TMD factorization formalism, as discussed, e.g., in Ref.~\cite{Gonzalez:2019}. The situation at the EIC should be much more favorable than in fixed-target experiments, as the hard scale $Q$ at the EIC would be large enough to be well in the perturbative region. 

Another open question is the dependence of unpolarized TMDs on partonic flavor. In Ref.~\cite{Signori:2013mda_568} it was shown that there is room for a significant flavor dependence, i.e., different flavors may have different distributions in transverse momentum space. Since then, however, all analyses assumed no flavor dependence. This question is relevant not only for a better understanding of nonperturbative QCD effects, but also because it may have an impact on the determination of the $W$ boson mass~\cite{Bacchetta:2018lna_568}. 
In order to address this question, it will be important to collect data from different beams at the EIC. 

\section{Sivers function and polarized TMDs}

The Sivers function was identified as one of the ``golden measurements'' of the EIC. It requires nonzero partonic orbital angular momentum and it is predicted to change sign in Drell-Yan compared to SIDIS, due to the different effect of the initial-state interactions compared to the final-state ones~\cite{Collins:2002kn_568}. 
It has been extracted from SIDIS data
by several groups, with consistent results~\cite{Anselmino:2010bs,Collins:2005ie,Vogelsang:2005cs,Bacchetta:2011gx,Echevarria:2014xaa_568}. Only one analysis has taken into account the role of TMD evolution~\cite{Echevarria:2014xaa_568}, also because the available data are only from fixed-target SIDIS measurements~\cite{Airapetian:2009ae,Adolph:2014zba,Aghasyan:2017jop_568,Adamczyk:2015gyk}. The EIC will change the situation entirely and will make it necessary to apply the complete TMD formalism at the highest possible level of accuracy.

First measurements of the Sivers effect in Drell-Yan processes have been reported by the STAR~\cite{Adamczyk:2015gyk} and COMPASS~\cite{Aghasyan:2017jop_568} experiments. They are compatible with the predicted sign change, but within large uncertainties~\cite{Anselmino:2016uie_567}. Surprisingly, however, they seem to be in better agreement with predictions that do {\em not} take into account TMD evolution. This is an open issue that will be clarified with EIC measurements, where a large lever arm in $Q$ would allow a detailed test/constraint of TMD evolution.

Apart from the Sivers function, there are other six polarized TMDs and all of them can be studied at EIC. The more information we can gather about partonic densities, the better we can check QCD predictions and constrain models. Together with other observables such as GPDs, a comprehensive study of TMDs will eventually lead to a much deeper understanding of the nucleon and of QCD. Among the functions that may become particularly relevant, we mention the so called Boer-Mulders function, $h_{1}^{\perp}$, which is involved in unpolarized scattering processes, and the helicity TMD, which generalizes the standard helicity collinear PDF and represents the best way to test the TMD formalism with spin.  

\section{Transversity}

Transversity was identified as a ``silver measurement'' of the EIC. So far, transversity has been extracted using data from single-hadron-inclusive DIS and two-hadron-inclusive DIS. In the first case, the TMD formalism has been used, where transversity is combined with the so-called Collins fragmentation function~\cite{Anselmino:2015sxa,Kang:2015msa,Lin:2017stx}. In the second case, the collinear formalism has been used, and transversity is combined with a dihadron fragmentation function~\cite{Radici:2018iag}. 

The integral of transversity over the light-cone momentum fraction $x$ is related to the tensor charge of the nucleon. This quantity is of particular relevance for a couple of reasons. First of all, it can be computed by lattice QCD in a reliable way, offering the chance to test lattice results (see \cite{Alexandrou:2019brg_567} for the most recent determination). Secondly, the knowledge of the tensor charge is useful to search for physics beyond the Standard Model~\cite{Courtoy:2015haa}. Phenomenological extractions of the tensor charge are currently affected by large error bars~\cite{Radici:2018abr}. However, at present 
the up and down contributions are {\em not} in good agreement with lattice predictions \cite{Radici:2018abr}. 
A description of the data can be achieved \cite{Lin:2017stx} with a value of the isovector combination $u-d$ compatible with lattice computations, but without reproducing the individual values of $u$ and $d$. 
Recent studies extend the uncertainties on the tensor charge extraction by including the possibility of violating the Soffer positivity bound on transversity\cite{Benel:2019mcq,DAlesio:2020vtw}, but discrepancies with lattice still persist.

The EIC will be crucial to extend the measurements of transversity to the low-$x$ region and thus decrease extrapolation errors in the determination of the tensor charge.

\section{TMD fragmentation functions}

The knowledge of TMD fragmentation functions is essential for TMD studies at the EIC, since they appear inside convolutions in any semi-inclusive observable. They offer the unique opportunity to understand spin and TMD dynamics in the hadronization process, which could be different from those in TMD parton distribution functions. For example, while Sivers function changes sign from SIDIS to DY process, the Collins fragmentation function is expected to be universal in SIDIS and $e^+e^-$ process. In addition, TMD fragmentation functions could provide novel insights into the hadronization of colored quarks and gluons.

In order to have an independent determination of TMD fragmentation functions, data from $e^+e^-$ colliders are necessary, with a level of precision similar to the EIC.
A pioneering measurement has been presented in Ref.~\cite{Seidl:2019jei_568}, where the transverse momentum of a final-state hadron is measured with respect to the thrust axis. These data have been used for a parton-model based extraction of TMD fragmentation functions~\cite{Soleymaninia:2019jqo}, similar to an earlier work~\cite{Boglione:2017jlh}. To further understand such data, one might consider a factorization formalism~\cite{Jain:2011iu}, which is also differential in the thrust variable. Another interesting observation of this measurement is that the Gaussian widths for the hadron transverse momentum distributions are increasing with the light-cone momentum fraction $z$ until around 0.6
before decreasing at even higher $z$, suggesting once again that TMDs cannot be parametrized by naive simple Gaussians. The decreasing in large-$z$ region is quite interesting, pointing into the requirement of additional dynamical effect, one of which could be threshold resummation in $z$. 

\section{Higher-twist functions}

TMD factorization formalism applies in the kinematic region when $q_T\ll Q$, while collinear factorization is valid in the region when $q_T\sim Q$. These two frameworks are closely connected, and same is true for the TMDs and the associated collinear multi-parton correlation functions. The well-known example is the relation between the quark Sivers function and the so-called Qiu-Sterman function, which is a higher-twist (twist-3) quark-gluon-quark correlation functions. Through an Operator Product Expansion (OPE), the quark Sivers function can be written as a convolution of a coefficient function and the Qiu-Sterman function. Such type of relations apply in general to all the spin-dependent TMDs. 

Because of the close connection, one can in principle perform global analysis of the spin asymmetry data, to extract the TMDs and the associated collinear higher-twist functions simultaneously. Earlier attempts along this direction have been performed in Refs.~\cite{Kang:2011hk,Kang:2012xf_568} based on TMD parton distribution functions, as well as in Refs.~\cite{Kanazawa:2014dca_568,Gamberg:2017gle} which explore the role of TMD fragmentation functions. A work~\cite{Pitonyak:2019} was discussed and presented in this week, where a global fit of the Sivers effect in SIDIS, Collins effect in SIDIS, Collins effect in $e^+e^-$, and $A_N$ in proton-proton collisions, was performed, and thus the corresponding TMD parton distribution functions, TMD fragmentation functions, and the collinear higher-twist quark-gluon-quark correlations are extracted. This is a remarkable progress, showing that an ultimate understanding of all spin asymmetries within a unified framework is possible. 

Apart from collinear higher-twist functions, also higher-twist TMDs can be introduced. TMD factorization up to subleading twist has not been proven, but a formula has been conjectured to generalize parton-model results~\cite{Bacchetta:2019qkv}, suggesting a possible path toward a full factorization proof.

\section{TMDs and jets}
In high energy collisions, quarks and gluons lead to the formation of highly energetic collimated sprays of hadrons observed in the detectors which are known as jets~\cite{Sterman:1977wj}. This is particularly true at the highest energy collider~--~the LHC, which produces a lot of jets for us to study. With the advances in experimental techniques, and corresponding advances in theoretical understanding over time, jets have become precision tools for studying the parton structure of matter. Jets are guaranteed to contribute at the EIC to a variety of key electron-proton and electron-nucleus physics topics. There are outstanding progress in studying jets at the EIC, at LO~\cite{Kang:2011jw}, NLO~\cite{Hinderer:2015hra}, NNLO~\cite{Abelof:2016pby}, and N$^3$LO~\cite{Currie:2018fgr}. 

Jets are very useful tools for probing TMDs. Two ways of using jets have been investigated in the community. In the first case, one studies the transverse momentum imbalance between a particle and a jet in either proton-proton or lepton-proton collisions. For example, Ref.~\cite{Liu:2018trl_568} studies lepton-jet azimuthal angular correlation in deep inelastic scattering as a unique tool for unpolarized quark TMD and the quark Sivers function. Another study focuses on photon-jet imbalance in proton-proton collision~\cite{Buffing:2018ggv_568}, and provides opportunities to probe potential TMD factorization breaking effects~\cite{Collins:2007nk_568,Rogers:2010dm_568}.

In the second case, one relies on the internal structure of the jets, or called jet substructure. Of particular interesting measurement is the so-called jet fragmentation function measurement, which has been studied at the LHC~\cite{Adamczyk:2017wld_568,Chatrchyan:2012gw,Aad:2011td,Aaij:2017fak}. For example, one might study the transverse momentum distribution of hadrons with respect to the jet axis inside a fully reconstructed jet. It has been shown~\cite{Kang:2017glf} that such distribution can be related to TMD fragmentation functions, and thus serves as a novel way of constraining TMD fragmentation functions, including the spin-dependent ones such as Collins fragmentation functions~\cite{Kang:2017btw}. One might even apply modern jet tools such as soft-drop grooming to constrain non-perturbative contribution in the TMD fragmentation functions~\cite{Makris:2017arq_568}. 

\section{Gluon TMDs}
As EIC is designed to probe the gluons, the study of gluon TMDs is an outstanding topic and has seen tremendous progress. For example, heavy quark pair production at the EIC has been proposed to explore unpolarized gluon TMDs and gluon Sivers functions~\cite{Boer:2010zf_568,Boer:2016fqd_568}. An experimental simulation at the EIC kinematics for such a process has been carried out~\cite{Zheng:2018awe}, in which other channels such as dijet production at the EIC has also been studied. Other SIDIS processes that have been investigated are $J/\Psi$ production~\cite{Godbole:2014tha,Mukherjee:2016qxa,Bacchetta:2018ivt_568} and $J/\Psi+$jet production~\cite{DAlesio:2019qpk_568}.   
From the experimental side, COMPASS provided some interesting information on gluon Sivers function through high-$p_T$ hadron pairs in SIDIS~\cite{Adolph:2017pgv}. 
The EIC will be crucial to eventually make precise measurements on (un)polarized gluon TMDs. 

It is worthwhile to mention that despite great progress, most studies on gluon TMDs at EIC are parton-model based. In the future, an important direction would be promote these studies with the full glory of QCD and TMD factorization framework, with which one can then investigate in details the energy evolution and universality properties of gluon TMDs (see recent discussion of TMD factorization in quarkonium production\cite{Echevarria:2019ynx,Fleming:2019pzj}).

\section{Acknowledgments}
The program of Week II was prepared in collaboration with D. Boer and H. Avakian. This work is supported by the European Research Council (ERC) under the European Union's Horizon 2020 research and innovation program (Grant No.~647981, 3DSPIN), and by the National Science Foundation under Grant No.~PHY-1720486.



\newpage 
%

\renewcommand*{\FigPath}{./WeekII/01_Asmita/}

\wstoc{Probing Gluon Sivers Function in Inelastic Photoproduction of
$J/\psi$ at the EIC}{Raj Kishore, Asmita Mukherjee, Sangem Rajesh}
\title{Probing Gluon Sivers Function in Inelastic Photoproduction of
$J/\psi$ at the EIC}

\author{Raj Kishore and Asmita Mukherjee$^*$}
\index{author}{Kishore, R.}
\index{author}{Mukherjee, A.}

\address{Department of Physics, Indian Institute of Technology Bombay,
Powai, Mumbai 4000076, India\\
$^*$E-mail: asmita@phy.iitb.ac.in}

\author{Sangem Rajesh}
\index{author}{Rajesh, S.}

\address{Dipartimento di Fisica, Universit\`a di Cagliari, Cittadella Universitaria, 
I-09042 Monserrato (CA), Italy
and INFN, Sezione di Cagliari, Cittadella Universitaria, I-09042 Monserrato (CA), Italy}

\begin{abstract}
We present a recent calculation of the single spin asymmetry in inclusive
photoproducton of $J/\psi$ at the future EIC, that can be used to probe the
gluon Sivers function.
\end{abstract}


\bodymatter

\section{Introduction}\label{intro}
$J/\psi$ production in $ep$ and $pp$ collisions is known to be an effective
tool to probe the gluon TMDs, as the contribution comes at leading order
(LO) through $\gamma g$ and $gg$ initiated processes. One of the most
interesting gluon TMDs is the gluon Sivers function (GSF) which probes the
coupling of the intrinsic transverse momenta of the gluons with the
transverse spin of the nucleon. The Sivers function \cite{Sivers:1989cc_223,
Sivers:1990fh} is a time reversal odd
(T-odd) object and initial and final state interactions play an important
role in the Sivers asymmetry. The gluon Sivers function for
any process can be written as a linear combination of two gluon Sivers
functions, one containing a C-even operator (f-type)  and the other
C-odd (d-type) \cite{Buffing:2013kca}. Compared to the quark Sivers function, much less is known
about the gluon Sivers function, apart from a positivity bound
\cite{Mulders:2000sh_181}. In this talk,
we present a recent calculation \cite{Rajesh:2018qks}
of single spin asymmetry in inelastic photoproduction of $J/\psi$ at the
future EIC.

\section{Calculation of the asymmetry}     

The process considered is
\be
e(l)+p^\uparrow(P) \rightarrow J/\psi (P_h) +X 
\ee
where the quantities within brackets are the momenta. We use the kinematics where the interaction takes place 
through the exchange of an (almost) real photon $\gamma(q)+ g(k) \rightarrow J/\psi(P_h) +g(p_g)$.  
We consider only the direct photon contribution and contribution from the resolved photon is eliminated by imposing a cut on the variable $z= {P \cdot P_h \over P\cdot q}$, which is the energy fraction transferred from the photon to the $J/\psi$ in the rest frame of the proton. 
The inelasticity variable  $z$ for inclusive photoproduction can be measured in experiments by Jacquet-Blondel method.  
The leading order (LO) process $ \gamma+g \rightarrow J/\psi$ contributes at $z=1$ (see
\cite{Mukherjee:2016qxa_47})
for the calculation of Sivers asymmetry in electroproduction) and we
use a cutoff $z<0.9$ to remove this contribution. Also contribution to
$J/\psi$ production from gluon and heavy quark fragmentation was removed  
by imposing a cut on $P_T$, which is the transverse momentum of the
$J/\psi$. We assume TMD factorization for the process considered and
generalized parton model (GPM) with the inclusion of the intrinsic transverse
momenta. We use NRQCD \cite{Bodwin:1994jh_359} to calculate the production of $J/\psi$. The $c{\bar
c}$ pair can be produced in color singlet (CS) or color octet (CO) state. In
$eP$ collision, non-zero asymmetry can be observed only if the $c {\bar c}$
pair is produced in the CO state \cite{Yuan:2008vn}. The SSA is defined as            

\begin{eqnarray}\label{asy}
A_N=\frac{d \sigma^{\uparrow}-d 
\sigma^{\downarrow}}{d \sigma^{\uparrow}+d \sigma^{\downarrow}},
\end{eqnarray}

where $d\sigma^{\uparrow}$ and $d\sigma^{\downarrow}$ are the  differential cross-sections 
measured when one of the particle is transversely polarized up ($\uparrow$) and down
($\downarrow$), respectively,  with respect to the scattering plane.

We consider the inclusive process   $e(l)+p^{\uparrow}(P)\rightarrow 
J/\psi(P_h)+X$. The virtual photon radiated by the initial electron is
almost real, $q^2=-Q^2 \approx 0$. The numerator and the denominator of the
asymmetry are given by 
\be\label{d3}
 \begin{aligned}
 d\sigma^{\uparrow}-d\sigma^{\downarrow}={}&\frac{d\sigma^{ep^{\uparrow}\rightarrow J/\psi X}}{dzd^2{\bm 
P}_T}-\frac{d\sigma^{ep^{\downarrow}\rightarrow J/\psi X}}{dzd^2{\bm P}_T}\\
={}&\frac{1}{2z(2\pi)^2}\int dx_\gamma dx_g  
d^2{\bm k}_{\perp g}
f_{\gamma/e}(x_\gamma)\Delta^N f_{g/p^{\uparrow}}(x_g,{\bm k}_{\perp 
g})\\
&\times\delta(\hat{s}+\hat{t}+\hat{u}-M^2)\frac{1}{2\hat{s}}|\mathcal{M}_{\gamma+g\rightarrow J/\psi +g}|^2,
\end{aligned}
\ee
and 
\be\label{d4}
\begin{aligned}
d\sigma^{\uparrow}+d\sigma^{\downarrow}={}&\frac{d\sigma^{ep^{\uparrow}\rightarrow  J/\psi X}}{dzd^2{\bm 
P}_T}+\frac{d\sigma^{ep^{\downarrow}\rightarrow  J/\psi X}}{dzd^2{\bm P}_T}=2\frac{d\sigma}{dzd^2{\bm P}_T}\\
={}&\frac{2}{2z(2\pi)^2}\int dx_{\gamma} dx_g 
d^2{\bm k}_{\perp g}
f_{\gamma/e}(x_\gamma) f_{g/p}(x_g,{\bm k}_{\perp 
g})\\
&\times\delta(\hat{s}+\hat{t}+\hat{u}-M^2)\frac{1}{2\hat{s}}|\mathcal{M}_{\gamma+g\rightarrow J/\psi +g}|^2.
\end{aligned}
\ee 

$x_\gamma$ and $x_g$ are the light-cone momentum fractions of the photon and gluon respectively. 
The Weizs$\ddot{a}$ker-Williams distribution function, $f_{\gamma/e}(x_\gamma)$, describes the density of photons inside the
electron. The $J/\psi $ production rate is calculated in NRQCD based color
octet model. 
We follow the approach given in \cite{Boer:2012bt} and the details of the calculation can be
found in \cite{Rajesh:2018qks}.  Here, we report on some of our numerical results in the
kinematics of the future planned EIC.   
\section{Results}
For the numerical estimates of the SSA we assume Gaussian parametrization of
unpolarized TMDs and best fit parameters from \cite{alesio} for the gluon Sivers
function. These are denoted by SIDIS1 and SIDIS2, respectively. Also,
following \cite{Boer:2003tx}  we parametrize the GSF in terms of $u$ and $d$ quark Sivers
functions \cite{Anselmino:2016uie}. Two different choices in this line are labeled as BV-a and
BV-b. For the dominating channel of photon-gluon fusion, contribution to the
numerator of the SSA comes mainly from GSF. As the heavy quark pair is
produced unpolarized, there is no contribution from the Collins function. 
Figs 1 and 2 show plots of the SSA for $\sqrt{s}= 100$ ~GeV and $45$ ~ GeV
respectively which will be possible at the future EIC. The asymmetry depends
strongly on the parametrization for the GSF, it is positive for SIDIS1 and
SIDIS2 and negative for BV-a and BV-b parametrizations. The magnitude of the
asymmetry is largest for BV-b. We have incorporated contributions from the 
 ${^{3}}{S}{_1}^{(8)}$, ${^{1}}{S}{_0}^{(8)}$ and 
${^{3}}{P}{_{J(0,1,2)}}^{(8)}$ in the asymmetry.
Numerical estimates of the unpolarized cross
section shows that the data from HERA can be explained if both CS and CO
contributions are incorporated.

\begin{figure}[]
\begin{minipage}[c]{0.99\textwidth}
\small{(a)}\includegraphics[width=5.5cm,height=5.5cm,clip]{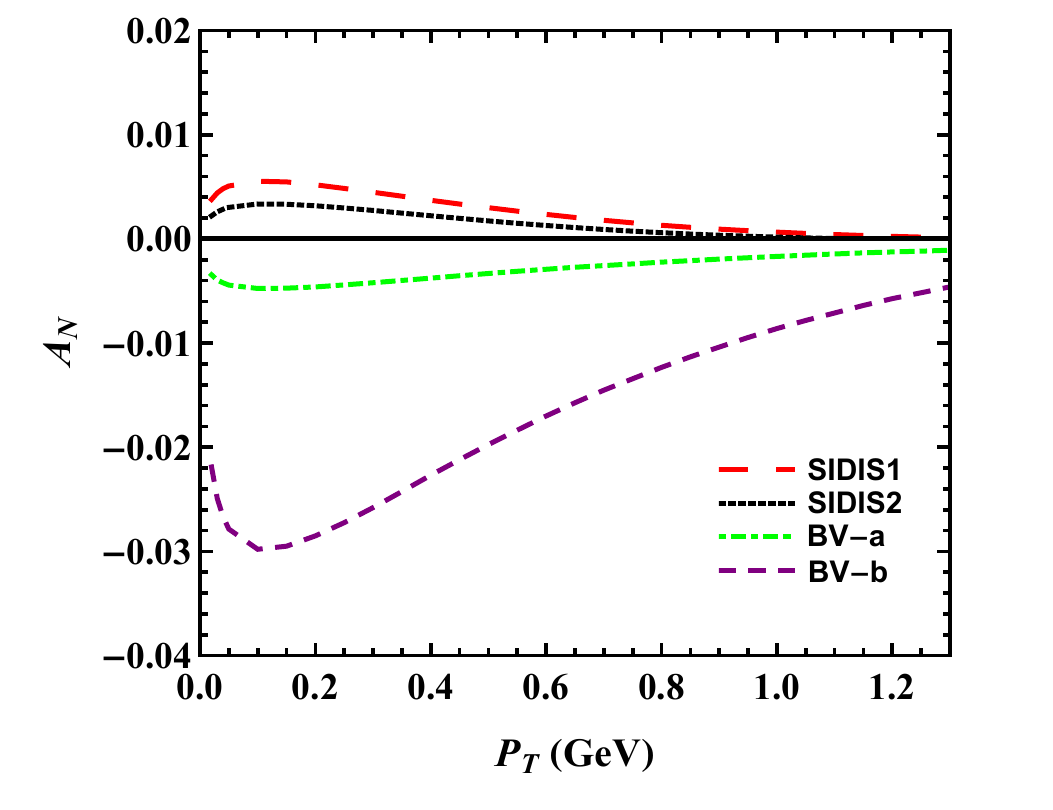}
\hspace{0.1cm}
\small{(b)}\includegraphics[width=5.5cm,height=5.5cm,clip]{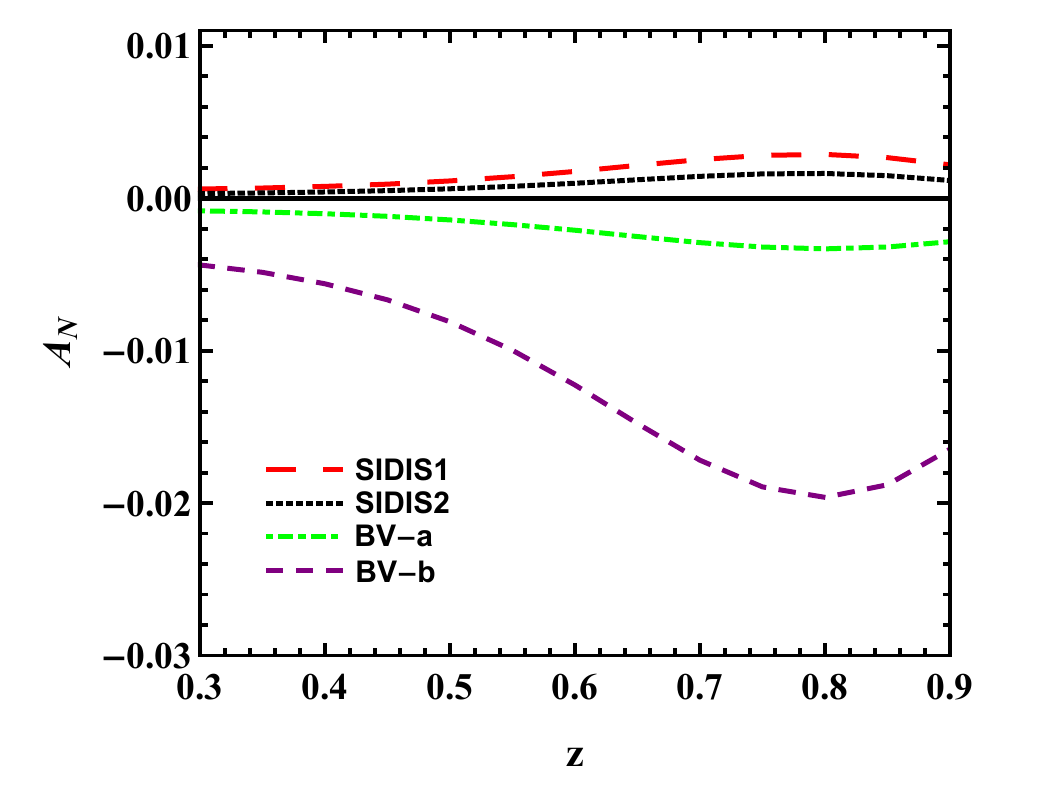}
\end{minipage}
\caption{\label{fig3}Single spin asymmetry  in $e+p^{\uparrow}\rightarrow J/\psi +X$
process as function of 
(a) $P_T$ (left panel) and  (b) $z$ (right panel) at $\sqrt{s}=100$ GeV (EIC)
\cite{Rajesh:2018qks}. The integration ranges are 
$0<P_{T}\leq1$ GeV and $0.3<z<0.9$. For convention of lines see the text.}
\end{figure}
\begin{figure}[]
\begin{minipage}[c]{0.99\textwidth}
\small{(a)}\includegraphics[width=5.5cm,height=5.5cm,clip]{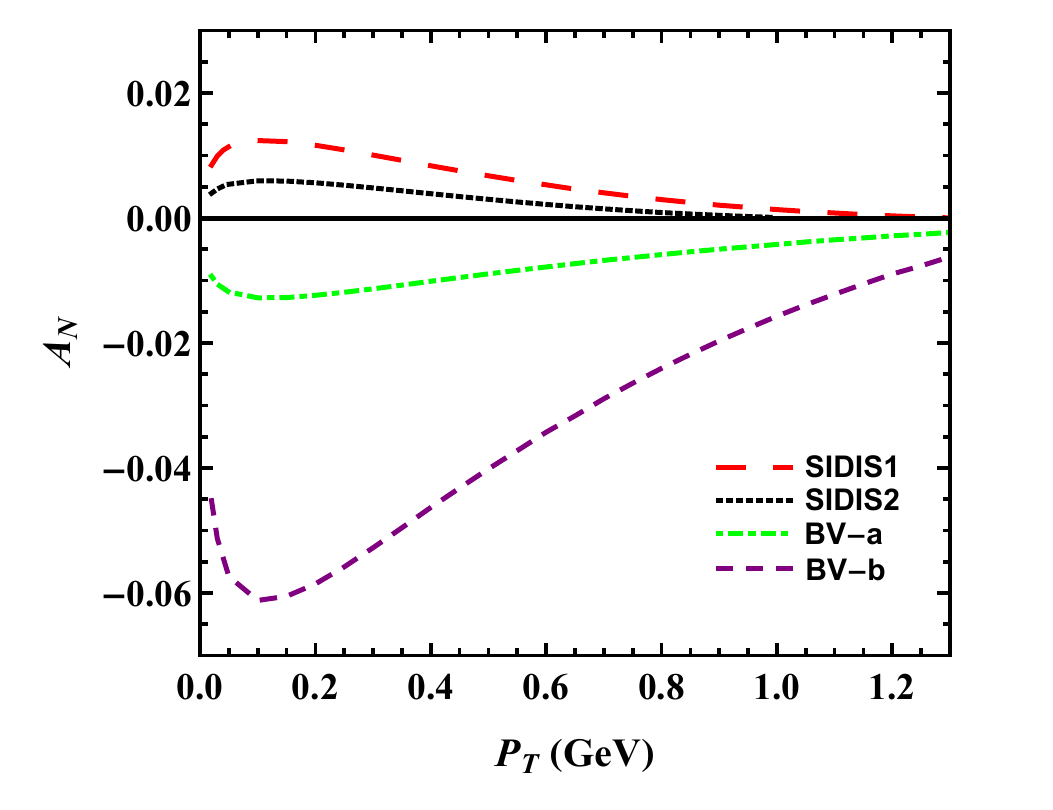}
\hspace{0.1cm}
\small{(b)}\includegraphics[width=5.5cm,height=5.5cm,clip]{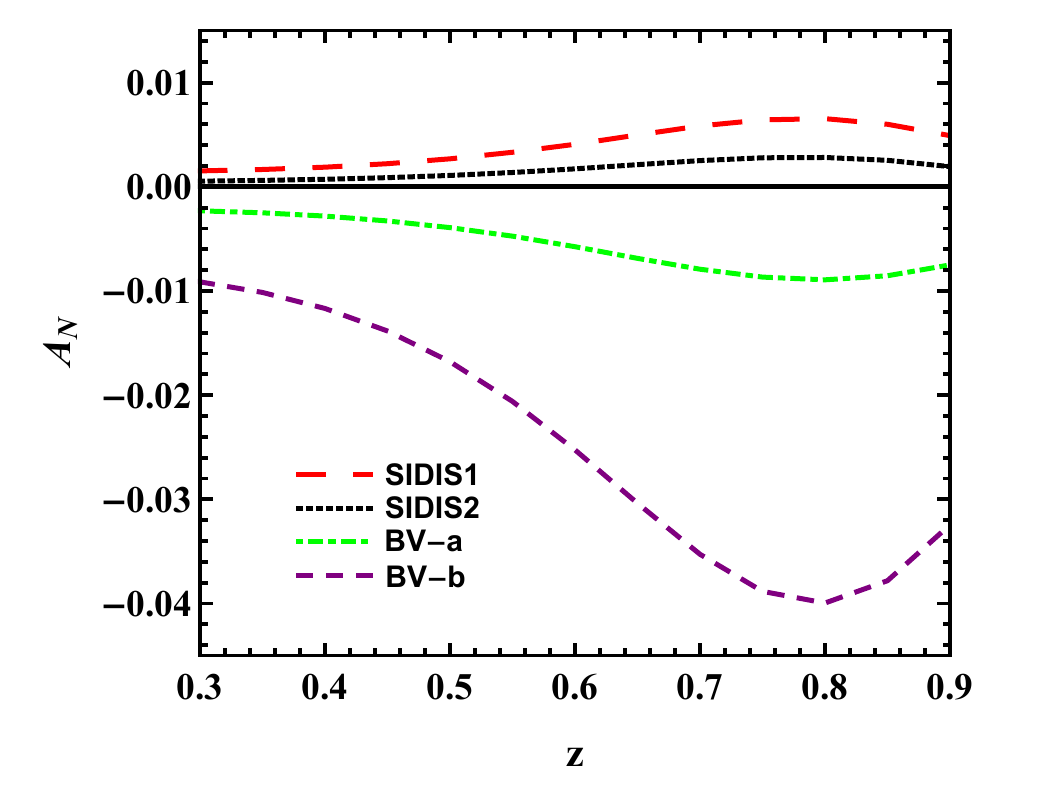}
\end{minipage}
\caption{\label{fig3}Single spin asymmetry  in $e+p^{\uparrow}\rightarrow J/\psi +X$
process as function of 
(a) $P_T$ (left panel) and  (b) $z$ (right panel) at $\sqrt{s}=45$ GeV (EIC). The 
integration ranges are  $0<P_{T}\leq1$ GeV and $0.3<z<0.9$
\cite{Rajesh:2018qks}. For convention of lines see the text.}
\end{figure}

\section{Conclusion}

We have presented a recent calculation of the SSA in inclusive
photoproduction of $J/\psi$ production in the kinematics of the future EIC. 
A sizable asymmetry in NRQCD based color octet model is reported, for the
range $0 < P_T < 1 $ ~ GeV and $0.3 < z < 0.9$. This asymmetry can give
direct access to the GSF.           

\section{Acknowledgement}
AM would like to thank the organizers of INT program "Probing Nucleons and
Nuclei in High Energy Collisions", October 1- November 16, 2018 for the
invitation.



\newpage
%
 
\renewcommand*{\FigPath}{./WeekII/02_Pitonyak}

\wstoc{Global Analysis of Transverse-Spin Observables}{Daniel~Pitonyak, Justin~Cammarota, Leonard~Gamberg, Zhongbo~Kang, Joshua~Miller, Alexei~Prokudin, Nobuo~Sato} 
\title{Global Analysis of Transverse-Spin Observables}

\author{Daniel~Pitonyak$^a$, Justin~Cammarota$^{a,b}$, Leonard~Gamberg$^c$, Zhongbo~Kang$^d$, Joshua~Miller$^a$, Alexei~Prokudin$^{c,e}$, Nobuo~Sato$^{e,f}$}
\index{author}{Pitonyak, D.}
\index{author}{Cammarota, J.}
\index{author}{Gamberg, L.}
\index{author}{Kang, Z.}
\index{author}{Miller, J.}
\index{author}{Prokudin, A.}
\index{author}{Sato, N.}

\address{$^a$Department of Physics, Lebanon Valley College,\\
Annville, PA 17003, USA}

\address{$^b$Department of Physics, College of William and Mary,\\
Williamsburg, Virginia 23187, USA}

\address{$^c$Division of Science, Penn State University Berks,\\
Reading, Pennsylvania 19610, USA}

\address{$^d$Department of Physics and Astronomy, University of California,\\
Los Angeles, California 90095, USA\\
Mani L. Bhaumik Institute for Theoretical Physics\\
Los Angeles, California 90095, USA}

\address{$^e$Theory Center, Jefferson Lab,\\
Newport News, Virginia 23606, USA}

\address{$^f$Department of Physics, Old Dominion University,\\
Norfolk, Virginia 23529, USA}

\begin{abstract}
We report preliminary results on a global fit of the Sivers effect in semi-inclusive deep-inelastic scattering (SIDIS), Collins effect in SIDIS, Collins effect in semi-inclusive annihilation (SIA), and $A_N$ in proton-proton collisions.  All of these transverse-spin observables are driven by the same quark-gluon-quark correlations, which allows for such a global analysis to be performed.  
\end{abstract}


\bodymatter

\section{Introduction}\label{s:Intro}
Transverse single-spin asymmetries (TSSAs) arise in reactions where a single hadron carries a transverse polarization.  Some of the most widely studied of these observables are the Sivers effect in semi-inclusive deep-inelastic scattering (SIDIS), Collins effect in SIDIS, and Collins effect in semi-inclusive annihilation (SIA) -- see Ref.~\citenum{Perdekamp:2015vwa} for a review.  These processes are sensitive to transverse momentum-dependent functions (TMDs) and offer insight into the 3-dimensional structure of hadrons.  Another TSSA that has been investigated for many years, commonly denoted as $A_N$, is single-inclusive proton-proton collisions (e.g., $p^\uparrow p\to h\,X$) --  see Ref.~\citenum{Pitonyak:2016hqh} for a review.  Unlike the Sivers and Collins effects, $A_N$ is sensitive to collinear twist-3 non-perturbative functions that contain information about quark-gluon-quark correlations in hadrons.  As we will elaborate more on in Sec.~\ref{s:Obs}, at large transverse momentum, the Sivers and Collins TMDs can be written in terms of some of these quark-gluon-quark correlators.  Moreover, because of this connection, one can include {\it all} TSSA data in a global fit, which we will discuss in Sec.~\ref{s:Pheno}.  In Sec.~\ref{s:Concl} we will summarize our work and comment on the future outlook for this research.

\section{The Observables and Non-Perturbative Functions}\label{s:Obs}
The observable $A_N$ in $p^\uparrow p\to \pi\,X$ for many years was thought to be driven dominantly by an initial-state effect given by the Qiu-Sterman (QS) function $F_{FT}(x,x)$. 
We now believe, due to our work in Refs.~\citenum{Kanazawa:2014dca,Gamberg:2017gle_55}, that $A_N$ in $p^\uparrow p\to \pi\,X$ is actually dominated by two collinear fragmentation functions (FFs), $H_1^{\perp(1)}(z)$ and $\tilde{H}(z)$, coupled to the transversity function $h_1(x)$. The functions $H_1^{\perp(1)}(z)$ and $\tilde{H}(z)$ are particular integrals over $z_1$ of a quark-gluon-quark FF, $\hat{H}^{\Im}_{FU}(z,z_1)$.  The function $H_1^{\perp(1)}(z)$ is also the first moment of the Collins FF $H_1^\perp(z,p_\perp)$.

In the Collins-Soper-Sterman (CSS) formalism, as updated in Ref.~\citenum{Collins:2011zzd_201}, a generic $b$-space TMD $\tilde{F}(x,b)$ is written in terms of a collinear function (calculated at small-$b$ through the operator product expansion (OPE)) and an exponential involving the so-called perturbative ($P$) and non-perturbative ($NP$) Sudakov factors:
$\tilde{F}(x,b;Q) \sim F_{OPE}(x;\mu_b)\exp[-S_P(b;\mu_b,Q)-S^{\tilde{F}}_{NP}(b,Q)]$,
where $\mu_b$ is a $b$-dependent scale.  For example, for the transversity TMD $h_1(x,k_T)$, $F_{OPE}(x)$ at leading order (LO) is the standard collinear transversity function $h_1(x)$.  For the Sivers function, $F_{OPE}(x)$ at LO is the QS function $F_{FT}(x,x)$~\cite{Aybat:2011ge_55,Echevarria:2014xaa}, while for the Collins function $F_{OPE}(z)=H_1^{\perp(1)}(z)$~\cite{Kang:2010xv}.  Note that $h_1(x)$, $F_{FT}(x,x)$, and $H_1^{\perp(1)}(z)$ are the same functions discussed above for $A_N$.  That is, the Sivers effect in SIDIS, Collins effect in SIDIS, and Collins effect in SIA contain the same non-perturbative objects as $A_N$ in $p^\uparrow p\to \pi\,X$.  Therefore, we can perform a global fit of all these observables to extract information on quark-gluon correlations in hadrons.

\begin{figure}[h]
\centering \includegraphics[scale=0.235]{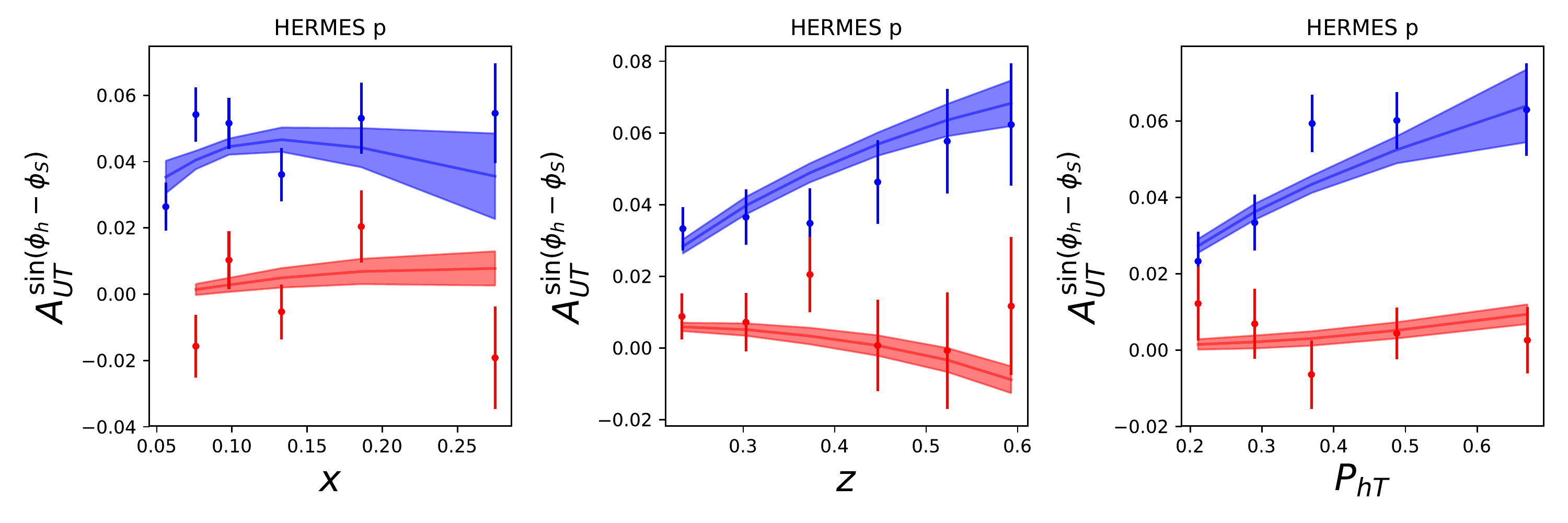} \includegraphics[scale=0.235]{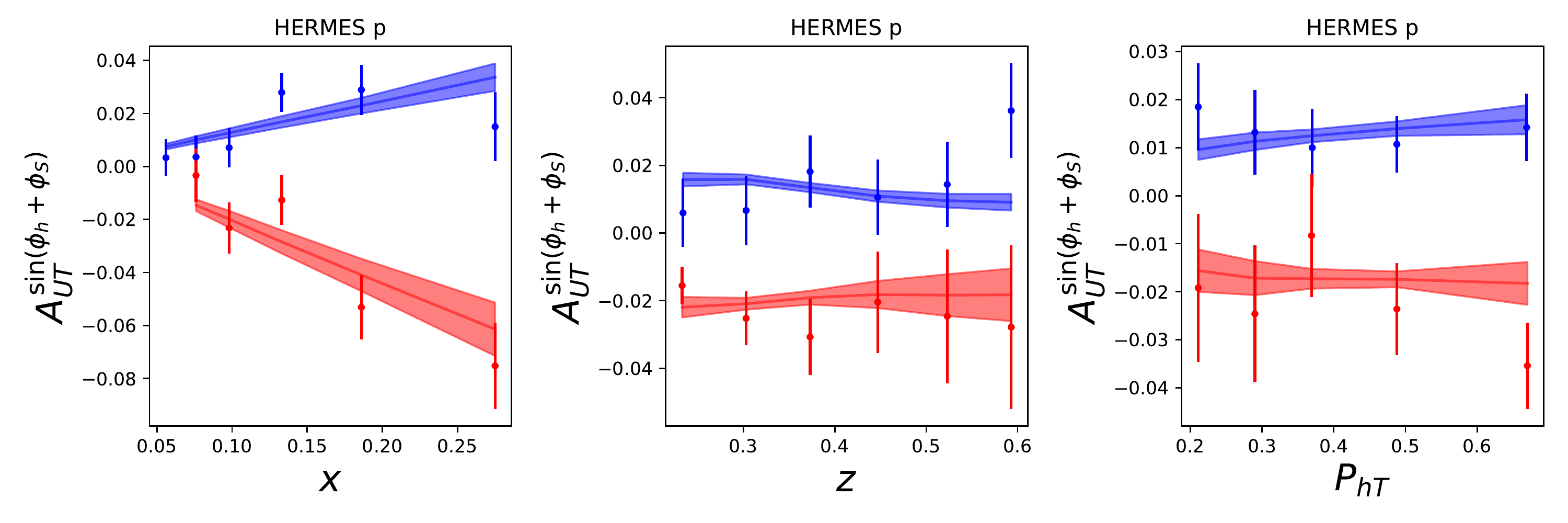} \\
\hskip 0.4cm \includegraphics[scale=0.225]{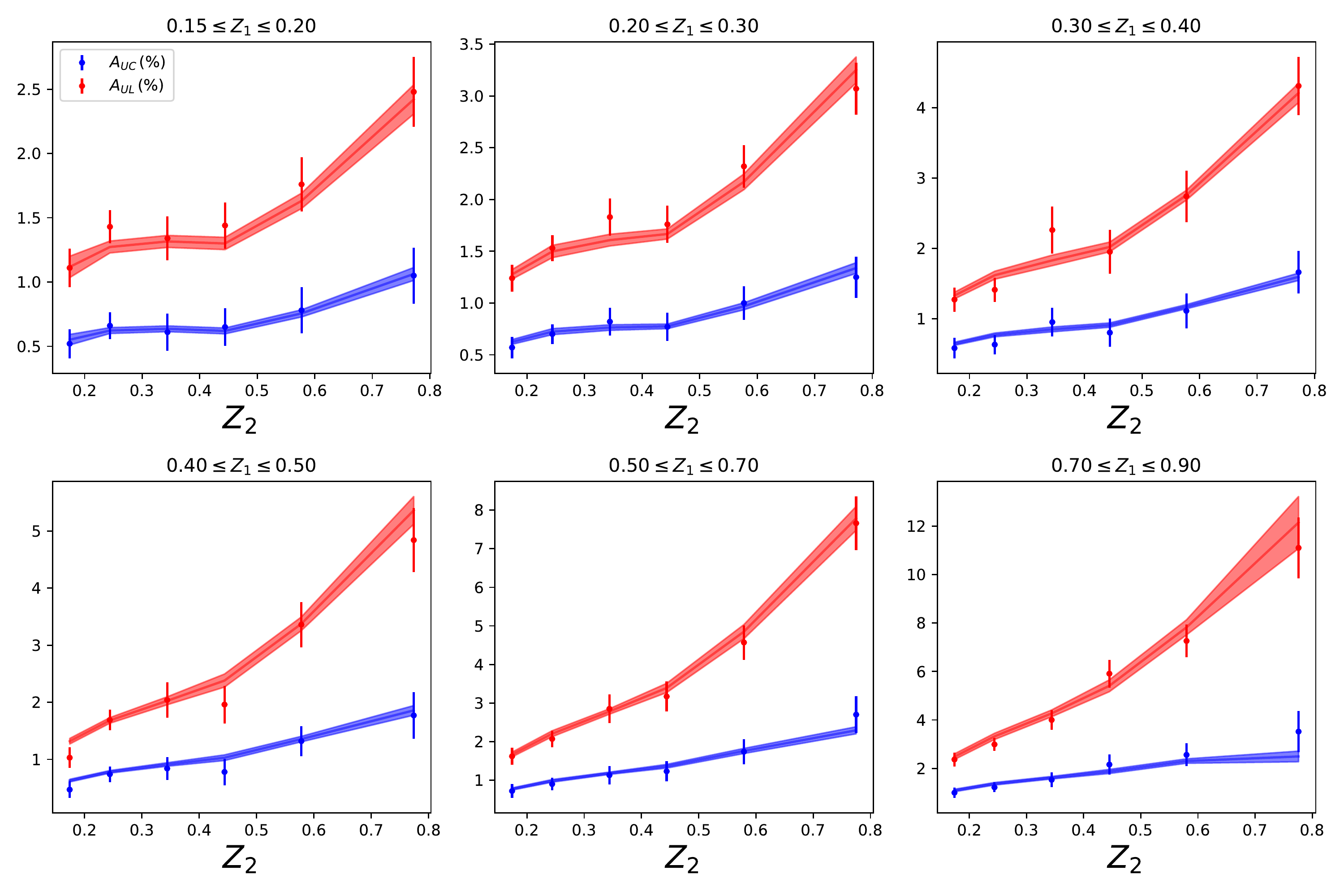} 
\caption{Comparison of the theoretical curves from our global fit to HERMES measurements of the Sivers effect (first panel) and Collins effect (second panel) in SIDIS.  The red is for $\pi^-$ and the blue is for $\pi^+$.  Also shown is the comparison to BaBar measurements of the Collins effect in SIA (bottom two panels).  The red is for $A_{UL}$ and the blue for $A_{UC}$.} \label{f:SIDIS_SIA}
\end{figure}


\begin{figure}[h]
\centering 
\includegraphics[scale=0.275]{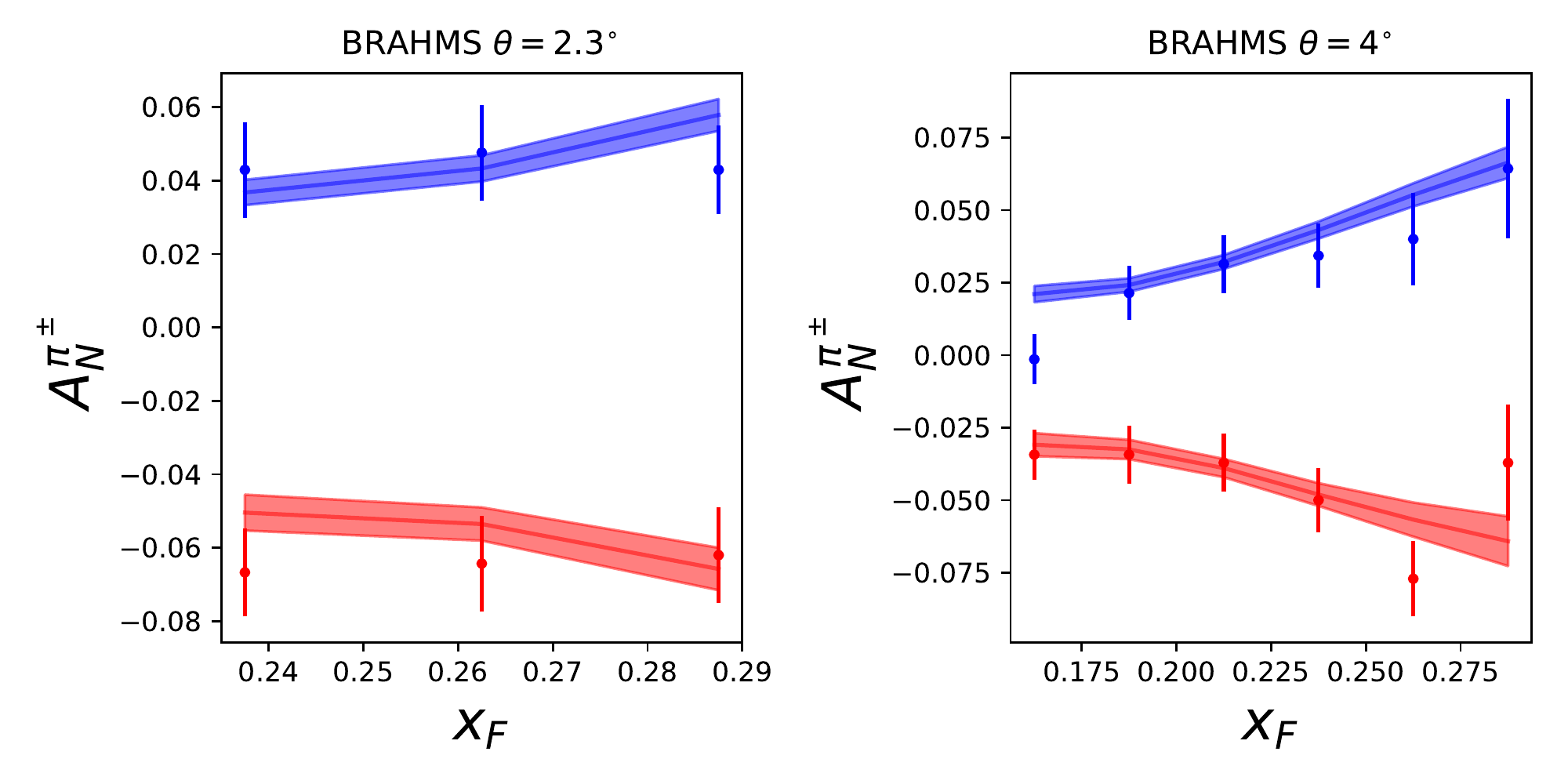} 
\includegraphics[scale=0.275]{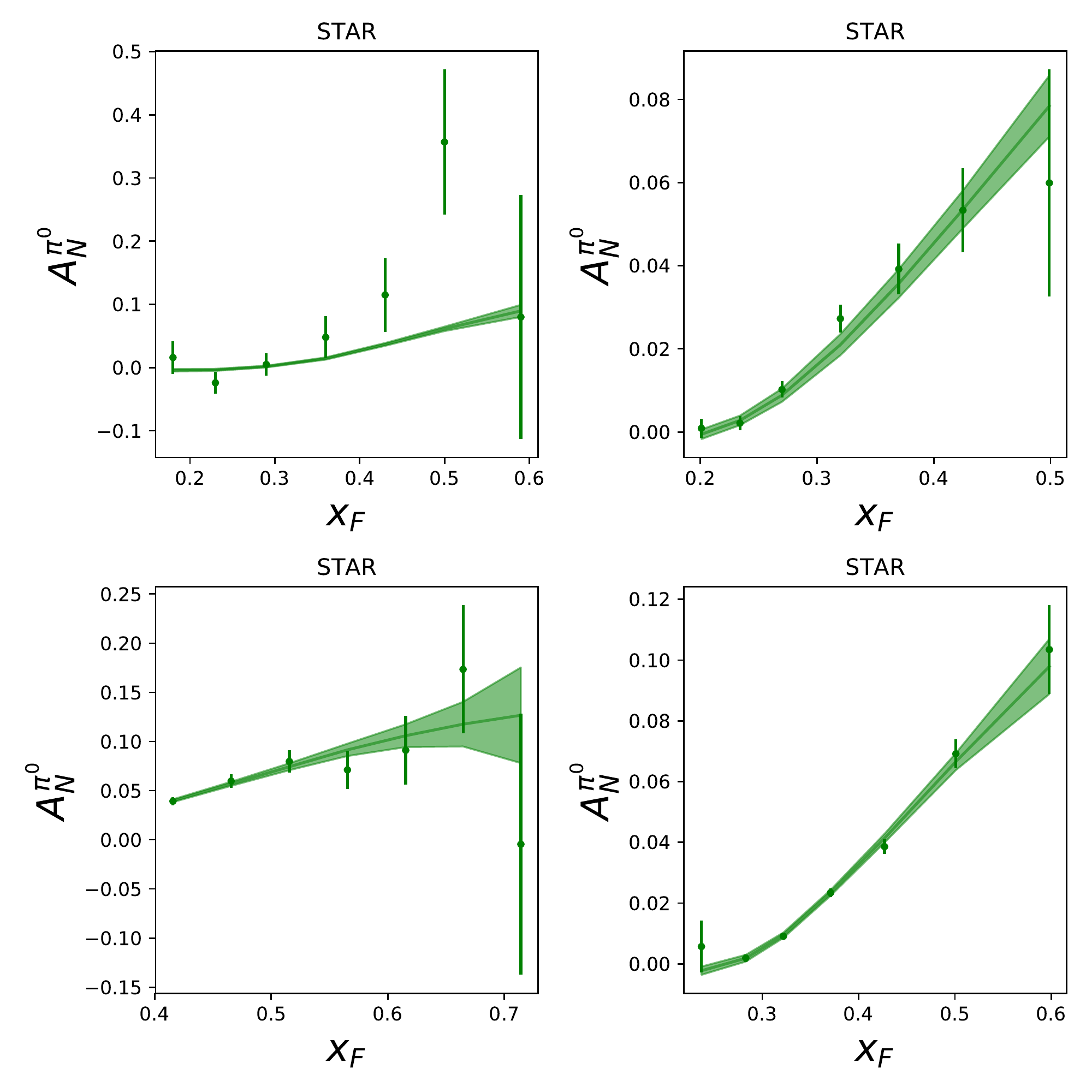} 
\caption{Comparison of the theoretical curves from our global fit to the BRAHMS (left) and STAR (right) measurements of $A_N$.  The red is for $\pi^-$, blue for $\pi^+$, and green for $\pi^0$.} \label{f:AN}
\end{figure}

\section{Phenomenology}\label{s:Pheno}
The global fit involves Sivers and Collins measurements from HERMES~\cite{Airapetian:2009ae_237_189} and COMPASS~\cite{Alekseev:2008aa_195,Adolph:2014zba_213},  Collins measurements from Belle~\cite{Seidl:2008xc_207} and BaBar~\cite{TheBABAR:2013yha_201}, and $A_N$ measurements from RHIC~\cite{Lee:2007zzh,Arsene:2008mi,Abelev:2008qb,Adamczyk:2012qj,Adamczyk:2012xd}.  We parameterize the TMDs generically using a Gaussian ansatz for the transverse momentum dependence along with their respective connection to a collinear function $F_{OPE}(x)$, i.e., $F(x,k_T)\sim F_{OPE}(x)\exp(-k_T^2/\langle k_T^2 \rangle)$.  Again, the functions $F_{OPE}(x)$ also show up in $A_N$.  Our fit therefore allows us to simultaneously extract the functions $F_{OPE}(x)$ and the widths $\langle k_T^2 \rangle$.  Some sample plots from this preliminary fit are shown in Figs.~\ref{f:SIDIS_SIA},\ref{f:AN}.  One sees that we are able to describe all the transverse-spin data relatively well.

\section{Conclusions}\label{s:Concl}
We have demonstrated that a simultaneous fit of all transverse-spin data (both TMD and collinear) is possible.  Future work will look to incorporate full CSS evolution for the TMDs and the proper twist-3 evolution for the relevant collinear functions.  This study will better constrain the non-perturbative functions that underlie transverse-spin observables, which is crucial when more precise data becomes available from JLab-12 and a future Electron-Ion Collider (EIC).  Since both these facilities involve lepton-proton collisions, an updated analysis of $A_N$ in $\ell\,p^\uparrow \to h\,X$ will also be imperative and require the collinear twist-3 functions extracted in this work.

\newpage 
\renewcommand*{\FigPath}{./WeekII/03_Courtoy}

\wstoc{Pion nucleus Drell-Yan process and parton transverse momentum in the pion}{Aurore~Courtoy} 
\title{Pion nucleus Drell-Yan process and parton transverse momentum in the pion
	}

\author{Aurore~Courtoy$^*$}
\index{author}{Courtoy, A.}

\address{Instituto de F\'isica, Universidad Nacional Aut\'onoma de M\'exico\\
Apartado Postal 20-364, 01000 Ciudad de M\'exico, Mexico\\
$^*$E-mail: aurore@fisica.unam.mx}

\begin{abstract}
During the INT-18-3 workshop, we presented an analysis of unpolarized Drell-Yan
pair production in pion-nucleus scattering with a particular focus into the pion Transverse Momentum Distributions in view of the future Electron Ion Collider. 
The  transverse distributions of
the pion calculated in a Nambu--Jona-Lasinio framework, with Pauli-Villars regularization, were used. 
The pion Transverse Momentum Distributions evolved  up to next-to-leading logarithmic accuracy is then tested against the transverse momentum spectrum
of dilepton pairs up to a transverse momentum of 2 GeV. 
  We found a  fair agreement with available
pion-nucleus data. 
This contribution joins common efforts from the TMD  and the pion structure communities for the Electron Ion Collider.
\end{abstract}


\bodymatter

\section{Introduction}
The structure of the pion is intrisically related to the dynamics of QCD. Non-perturbative approaches that incorporate chiral symmetry have been successfully leading towards a more general understanding of the pion structure.  
While there are less experiments involving pion observables than, {\it e.g.}, proton's, an Electron Ion Collider (EIC) will be capable of  providing more experimental insights, as described in Ref.~\cite{Aguilar:2019te_352b}. In these proceedings, we will focus on a calculation of the pion Transverse Momentum Distributions (TMDs) in the Nambu--Jona-Lasinio (NJL) model, whose results fully embody chiral symmetry.

The pion-induced Drell-Yan process has played a central role in the early determination of the pion distributions, especially through the data available in Ref.~\cite{Conway:1989fs}.   
This process is described by $h_1(p_1) + \; h_2(p_2)\to \gamma^*(q)+X$
in which a virtual photon is produced with large 
invariant mass-squared $Q^2$ in the 
collisions of two hadrons at a center-of-mass energy $s = (p_1 + p_2)^2$, with $p_{1,2}$ the four momentum of hadrons $h_{1,2}$, 
 $h_1$  being a pion and $h_2$ a proton. 
 Beyond collinear approaches, the  unintegrated cross-sections characterize the spectrum of transverse momentum of the virtual photon, $q_T$. In the kinematical regime in which $q_T$ is of order $\Lambda_{\mbox{\tiny QCD}}$, that is  small {\it w.r.t.} $Q$, such an effect is accounted into transverse momentum of the partons through TMDs. The departure from collinearity is here a highly non-perturbative effect, originated in the internal dynamics of the parent hadron.
While there exist model predictions as well as phenomenological analyses, the implementation of the transverse momentum factorization theorems has added in complexity in globally fitting and phenomenologically determining TMDs.

In Ref.~\cite{Ceccopieri:2018nop}, we presented one of the first analyses of the $\pi N$ DY process in terms of the modern TMD formulation. Our method focuses on investigating the DY cross section from the perspective of the dynamics of the pion as expressed in the Nambu--Jona-Lasinio (NJL) model~\cite{ns}.

\section{Pion dynamics in Drell-Yan cross sections}
When $q_T^2$ becomes small compared to $Q^2$, large logarithmic 
corrections of the form of $\alpha_s^n \log^m(Q^2/q_T^2)$ with $0 \leq m \leq 2n-1$
appear in fixed order 
results, being $n$ the order of the perturbative calculation. 
These large logarithmic corrections can be resummed to all
orders by using the Collins-Soper-Sterman (CSS) formalism~\cite{CSS}.
In this limit, the 
cross-section, differential in ${\bf q}_T$, can be written as~\cite{Ceccopieri:2014qha} 
\begin{eqnarray}
\label{CSS}
\frac{d\sigma}{dq_T^2 d\tau dy}&=& \sum_{a,b} \sigma_{q\bar{q}}^{(LO)}  
\int_0^\infty db \frac{b}{2} J_0(b \, q_T)\,S_q(Q,b)\, S_{NP}^{\pi p}(b)   \nonumber\\
&& \Big[ (f_{a/\pi}\otimes C_{qa})\left(x_1,\frac{b_0^2}{b^2}\right) 
(f_{b/p}\otimes C_{\bar{q}b})\left(x_2,\frac{b_0^2}{b^2}\right)+ q \leftrightarrow  \bar q \Big]\,,
\end{eqnarray}
where $b_0=2 e^{-\gamma_e}$, the symbol $\otimes$ stands for convolution 
and $\sigma_{q\bar{q}}^{(LO)}$ is the leading-order 
total partonic cross section for producing a lepton pair.
We use the nuclear PDFs 
of Ref.~\cite{nCTEQ} for the proton collinear PDFs. 
The cross section in Eq.~(\ref{CSS}) is also differential in $\tau=Q^2/s$ and $y$, the rapidity of the DY pair. 
The large logarithmic corrections are exponentiated 
in $b$-space in the Sudakov perturbative form factor, $S_q(Q,b)$.
The non-perturbative factor, $S_{NP}^{\pi p}$, contains all the information about the non-perturbative ${\bf k}_T$ behavior. Models for hadron's structure might describe the true theory but at one specific value of the RGE scale, $Q_0$. In that sense, they incorporate ---here, in the chiral limit---  a factorized ${\bf k}_T$  and Bjorken $x$ behavior in a fashion that can be resumed as 
\begin{eqnarray}
f^{q/\pi}(x_{\pi}, b; Q_0^2)&=& q(x_{\pi}; Q_0^2) \,S_{NP}^{\pi}(b)\,=\,q(x_{\pi}; Q_0^2)  \exp\{\ln S_{NP}^{\pi}(b)\}~;
\label{chi_sudakov}
\end{eqnarray}
while the full  TMD parton densities should involve a further $Q^2$ dependence, that is also called   {\it non-perturbative} evolution. Such an evolution has been parameterized in the past, for the proton-proton DY, as
\begin{eqnarray}
\label{fnp}
S_{NP}^{pp}(b)
&=&\exp\{-[a_1 + a_2 \ln (M/(3.2 \,\mbox{GeV})) + a_3 \ln(100 x_1 x_2)] b^2\}\,,
\end{eqnarray}
where $a_1$ plays an equivalent r\^ole to $\ln S_{NP}^{\pi}(b)$ in Eq.~(\ref{chi_sudakov}) and the other parameters, determined  {\it e.g.} in Ref.~\cite{KN05},  reflect a $Q^2$ evolution, {\it i.e.} through the invariant mass $M$ dependence, as well as an unfactorized $(x, {\bf k}_T)$ term.

Our ansatz, incorporating the NJL results, is 
\begin{equation}
\label{prescript_us}
S_{NP}^{\pi p}(b)=S_{NP}^{\pi}(b) \, \sqrt{S_{NP}^{pp}(b)} \,,
\end{equation}
where $S_{NP}^{\pi}(b)$ is given by the model and is non-gaussian, {\it i.e.}
\begin{eqnarray}
S_{NP}^{\pi}(b)
&=& \frac{3}{2 \pi^2} {\left( m \over f_\pi  \right )^2}~
\sum_{i=0,2} c_i K_0 ( m_i \, b)\,,
\label{bsp}  
\end{eqnarray}
and the square root of $S_{NP}^{pp}(b)$ is  given 
in Eq.~(\ref{fnp}).
 
 At very large values of $b$, the perturbative form factor needs to include a {\it taming} through the so-called $b^{\star}$-prescription.
By  splitting the perturbative
form factor, we can use distinct $b_{max}$ on 
the proton and pion side with $b_\star(b,b_{max})=b/\sqrt{1+\Big(\frac{b}{b_{max}}\Big)^2} $ and the respective $b_{max}^p=1.5$GeV$^{-1}$ and  $b_{max}^{\pi}=b_0/Q_0=2.44$GeV$^{-1}$, with $Q_0$ ---the hadronic scale of the model--- evaluated at NLO through a minimization procedure for the collinear PDF on the integrated DY cross-sections. That hadronic scale so determined represents the only free parameter of our approach.

\begin{figure}[h]
\begin{center}
\includegraphics[width=5in]{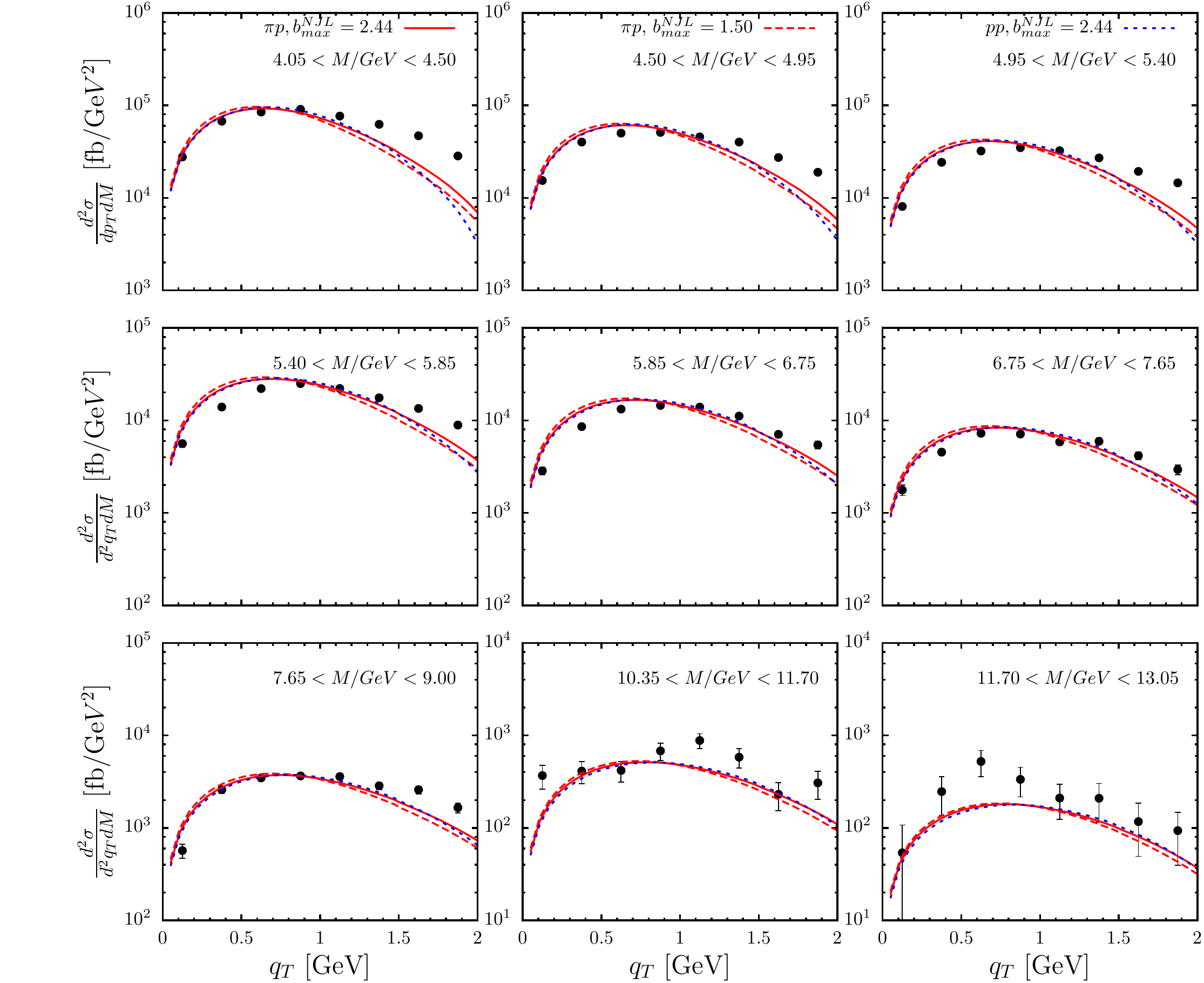}
\end{center}
\caption{Our model results compared to cross sections
in various invariant mass bins of the DY lepton pair  integrated in $0<x_F<1$. $b_{max}$ is given in unit GeV$^{-1}$.}
\label{Fig:DYqTNJL}
\end{figure}

 The results  for the lepton pair $q_T$-spectra of Ref.~\cite{Conway:1989fs}, measured in $\pi W$ collisions, are shown in Fig.~\ref{Fig:DYqTNJL}. The plots range up to $q_T \sim 2$ GeV, range for which the chosen proton description holds~\cite{Ceccopieri:2018nop}. Both red, full and dashed, curves correspond to results using Eq.~(\ref{prescript_us}) with, respectively, the proposed regulator value $b_{max}^{\pi}=2.44$ GeV$^{-1}$ and $b_{max}^{\pi}=1.5$ GeV$^{-1}$ demonstrating the stability of our results at small values of $q_T$. The short-dashed blue curve corresponds to a different ansatz: Eq.~(\ref{fnp}) is used for both hadrons, still with different $b_{max}$ values. At low-$q_T$, the difference between the two ans\"atze is quantitatively small. There seems to be a reduced sensitivity of the data 
to non-perturbative structure. However, this is a first analysis for which no {\it non-perturbative evolution} of the type  Eq.~(\ref{fnp}) has been included, since it is not inherent to the NJL approach used here.

\section{Conclusions}

We have synthetized the analysis of the DY pair production
in pion-nucleus scattering~\cite{Ceccopieri:2018nop}, in which we tested the outcome of a NJL approach for the pion TMD plugged in the CSS framework  at next-to-leading logarithmic accuracy against the differential transverse momentum
spectra of DY pairs produced in $pA$ collisions. 
Further analyses require an extension of the current approach to include a $Q^2$ dependence. This will be particularly relevant for both the pion structure and the TMD formalism that will be addressed at the EIC.

\section*{Acknowledgments}

The author thanks her colleagues and co-authors of the original publication. This work has been funded by UNAM through the PIIF project Perspectivas en F\'isica de Part\'iculas y Astropart\'iculas as well as  Grant No. DGAPA-PAPIIT IA102418.

\bibliographystyle{ws-procs961x669}

\newpage 
%

\renewcommand*{\FigPath}{./WeekII/04_Kutak/}

\renewcommand{\kt}{{\mathbf k}}
\newcommand{\qt}{{\mathbf q}} 
\renewcommand{\pt}{{\mathbf p}}
\newcommand{\ptt}{\mathbf{\tilde{p}}}
\newcommand{\qtt}{\mathbf{\tilde{q}}}
\newcommand{\Pbb}{\mathbb{P}}

\wstoc{Towards generalization of low x evolution equations}{Krzysztof Kutak}
\title{Towards generalization of low x evolution equations}

\author{Krzysztof Kutak$^*$}
\index{author}{Kutak, K.}

\address{Instytut Fizyki Jadrowej Polskiej Akademii Nauk,\\
Krakow, 31-342 Krakow, Poland\\
$^*$E-mail: krzysztof.kutak@ifj.edu.pl\\
www.ifj.edu.pl}
\begin{abstract}
I present an calculation of transverse momentum dependent real
emission contributions to splitting functions within
$k_T$-factorization~\cite{Gituliar:2015agu,Hentschinski:2016wya,Hentschinski:2017ayz}, giving special attention to the gluon-to-gluon splitting. The calculation is performed in a formalism that generalizes the framework
of~\cite{Curci:1980uw,Catani:1994sq}.
\end{abstract}


\bodymatter

\section{Introduction}\label{aba:sec1_362}
Parton distributions functions (PDFs) are crucial elements of collider
phenomenology. In presence of a hard scale $M$ with
$M \gg \Lambda_{\text{QCD}}$,
factorization theorems allow to express cross-sections as convolutions of parton
densities (PDFs) and hard matrix elements, where the latter are calculated
within perturbative QCD~\cite{Collins:1984xc}. Here we are focused on parton  densities which depend on transversal momentum.
Such PDFs arise naturally in regions of phase space characterized by a hierarchy
of scales. A particularly interesting example is provided by the so called low $x$
region, where $x$ is the ratio of the hard scale $M^2$ of the process and the
center-of-mass energy squared $s$. The low $x$ region corresponds therefore
to the hierarchy $s \gg M^2 \gg \Lambda_{\text{QCD}}^2$. We follow here a proposal initially outlined
in~\cite{Catani:1994sq}. There, the splitting functions which are basic blocks allowing to construct PDFs have been constructed following the definition of
DGLAP splittings by Curci-Furmanski-Petronzio (CFP)~\cite{Curci:1980uw}.
The authors of~\cite{Catani:1994sq} were able to generalize the CFP framework and define a Transversal Momentum Dependent (TMD) gluon-to-quark splitting function $\tilde{P}_{qg}$, both
exact in transverse momentum and longitudinal momentum fraction.%
    \footnote{Hereafter, we will use the symbol $\tilde{P}$ to indicate a
    transverse momentum dependent splitting function.}
Following observation of~\cite{Hautmann:2012sh} in \cite{Gituliar:2015agu} the $\tilde{P}_{gq}$, and
$\tilde{P}_{qg}$ and $\tilde{P}_{qq}$~were obtained. 
The computation of the gluon-to-gluon splitting $\tilde{P}_{gg}$ required
a further modification of the formalism used in~\cite{Catani:1994sq,Gituliar:2015agu}
as has been developed in~\cite{Hentschinski:2017ayz}.
In this contribution I summarize the most relevant results obtained in this work.

\section{Calculation of gluon-to-gluon splitting}
The calculation of the $\tilde{P}_{gg}$ splitting function follows the methods
used in ~\cite{Gituliar:2015agu,Hentschinski:2017ayz}.
In the high energy kinematics the momentum of the incoming parton is off-shell
and given by: $k^\mu = y p^\mu + k_\perp^\mu$, 
where $p$ is a light-like momentum defining the
direction of a hadron beam and $n$ defines the axial gauge with $n^2=p^2=0$
(that is necessary when using CFP inspired formalism).
Additionally, the outgoing momenta is parametrized as:
$q^\mu = x p^\mu + q_\perp^\mu + \frac{q^2+\qt^2}{2x p\cdot n} n^\mu$ and we also use:
$\qtt = \qt - z \kt$ with $z=x/y$.

The TMD splitting function, $\tilde{P}_{gg}$, is defined as 
\begin{equation}
\begin{split}
\label{eq:TMDkernelDefINI}
\hat K_{gg} \left(z, \frac{\kt^2}{\mu^2}, \epsilon \right) &=
z \int \frac{d^{2 + 2 \epsilon} {\qt}}{2(2\pi)^{4+2\epsilon}}
     \underbrace{\int d q^2 \, \mathbb{P}_{g,\,\text{in}} \otimes
                 \hat{K}_{gg}^{(0)}(q, k) \otimes \mathbb{P}_{g,\,\text{out}}}
                    _{\tilde{P}_{gg}^{(0)} \left(z, \kt, \qtt, \epsilon \right)}
     \,\Theta(\mu_F^2+q^2),
\end{split}
\end{equation}
with $\mathbb{P}_{g}$ being appropriate projection operators and $\hat{K}_{gg}$
the matrix element contributing to the kernel.

In order to calculate $\tilde{P}_{gg}$ splitting one needs to extend formalism
of~\cite{Catani:1994sq,Gituliar:2015agu} to the gluon case. This was achieved
in~\cite{Hentschinski:2017ayz} by generalizing definition of projector operators
and defining appropriate generalized 3-gluon vertex that is gauge invariant in
the presence of the off-shell momentum $k$. Definition of the generalized vertex
follows from application of the spinor helicity methods to the high-energy
factorization~\cite{vanHameren:2012uj,vanHameren:2012if,vanHameren:2013csa,vanHameren:2014iua,vanHameren:2015bba,vanHameren:2016bfc}
and can be obtained by summing the diagrams of Fig.~\ref{fig:pggVertex},
giving:
\begin{equation}
\label{eq:ggg_vertex} 
\Gamma^{\mu_1\mu_2\mu_3}_{g^*g^*g}(q,k,p') = \mathcal{V}^{\lambda
  \kappa \mu_3}(-q,k,-p') \, {d^{\mu_1}}_{\lambda} (q)\,
{d^{\mu_2}}_{\kappa}(k) +\, d^{\mu_1\mu_2}(k)\, \frac{q^2 n^{\mu_3}}{ n\cdot p'} -
d^{\mu_1\mu_2}(q)\, \frac{k^2 p^{\mu_3}}{ p\cdot p'} \, ,
\end{equation}
with $\mathcal{V}^{\lambda \kappa \mu_3}(-q,k,-p')$ being the ordinary 3-gluon QCD vertex,
and $d^{\mu_1\mu_2}(q)=-g^{\mu_1\mu_2}+\frac{q^{\mu_1}n^{\mu_2}+q^{\mu_2}n^{\mu_1}}{q\cdot n}$
is the numerator of the gluon propagator in the light-cone gauge.
More details on the exact procedure can be found in~\cite{Hentschinski:2017ayz}.
\begin{figure}[ht]
\begin{center}
\includegraphics[scale=0.8]{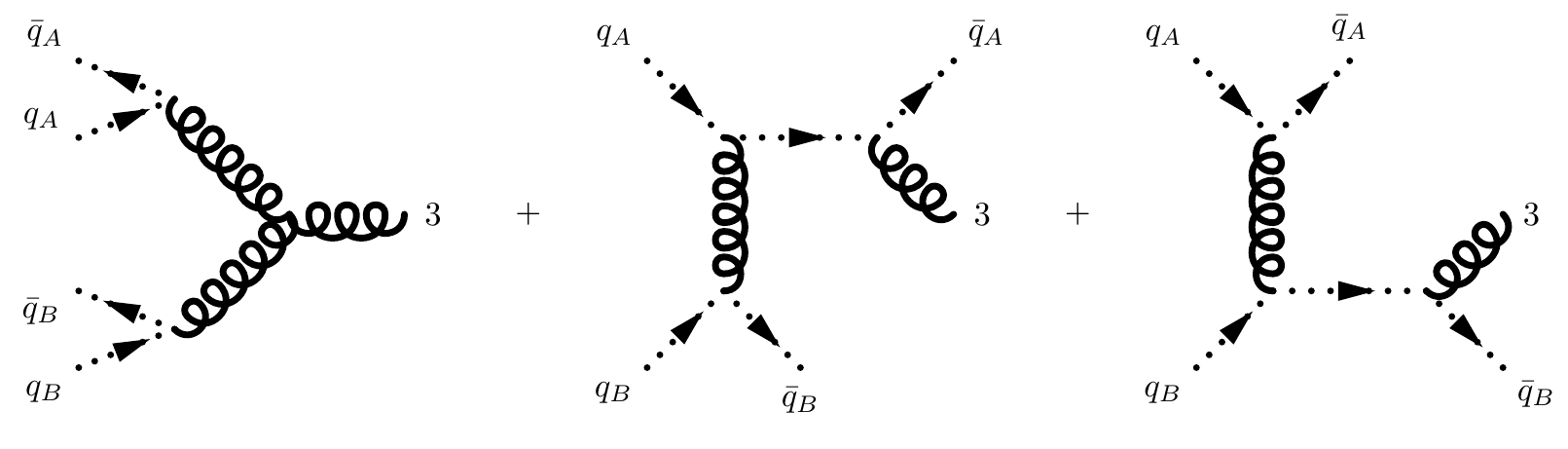}
\vspace{-0.5cm}
\caption{Feynman diagrams contributing to the amplitude used to define the generalized 3-gluon vertex.}
\label{fig:pggVertex}
\end{center}
\end{figure}

The definition of the projectors is more involved. We need to ensure that in the collinear
limit the new projectors reduce to the ones introduced by the CFP~\cite{Curci:1980uw}
and that the appropriate high-energy limit for the gluon splitting is also obtained.
A natural approach is to modify only the incoming projector $\mathbb{P}_{g,\,\text{in}}$, 
as the kinematics of the incoming momentum is more general, and keep the collinear outgoing
projector $\mathbb{P}_{g,\,\text{out}}$ unchanged (as the kinematics of the outgoing momentum is the same).
This is what was done by Catani and Hautmann~\cite{Catani:1994sq,Catani:1990eg} and later by
us~\cite{Gituliar:2015agu} when defining the projector for the calculation of the quark splitting
functions. In that case we used a transverse projector: $k_\perp^\mu k_\perp^\nu / k_\perp^2$
which, however, can not be used anymore since it does not guarentee gauge invariance of the 3-gluon vertex
$\Gamma^{\mu_1\mu_2\mu_3}_{g^*g^*g}$. Instead we use, longitudinal projector given by:
$\mathbb{P}_{g,\,\text{in}}^{\mu\nu} = y^2p^{\mu}p^{\nu}/k_{\perp}^2$. However, in order to satisfy
the requirements of $\mathbb{P}_{g}^s=\mathbb{P}_{g,\,\text{in}}\mathbb{P}_{g,\,\text{out}}$ being
a projector operator ($\mathbb{P}_{g}^s\otimes\mathbb{P}_{g}^s=\mathbb{P}_{g}^s$) we also need to modify
the outgoing projector. In the end we end up with the following gluon projectors:
\begin{equation}
\Pbb_{g,\,\text{in}}^{\mu\nu} =
-y^2\, \frac{p^{\mu} p^{\nu}}{k_\perp^2} \, , 
\quad\quad
\Pbb_{g,\,\text{out}}^{\mu\nu}  = 
-g^{\mu\nu} + \frac{k^\mu n^\nu + k^\nu n^\mu }{k\cdot n} - k^2\, \frac{n_\mu n_\nu}{(k\cdot n)^2} \, . 
\label{eq:HE_proj_new}
\end{equation}
One can check now that $\mathbb{P}_{g}^s\otimes\mathbb{P}_{g}^s=\mathbb{P}_{g}^s$ holds.
Also the new outgoing projector is consistent with the collinear case and one can show that 
in the collinear limit the difference between $y^2\, p^\mu p^\nu/k_\perp^2$ and
$k^\mu_\perp k^\nu_\perp/k_\perp^2$ vanishes when contracted into the relevant vertices.
More details are provided in~\cite{Hentschinski:2017ayz}.

\section{Results}
Using the elements introduced in the previous section the transverse momentum
dependent gluon-to-gluon splitting function has been computed. The result  is:
\begin{align}
  \label{eq:ggsplitting}
  \tilde{P}_{gg}^{(0)} (z, \qtt, \kt)
 &=
 2 C_A\, \bigg\{
 \frac{\qtt^4}{\left(\qtt-(1-z)\kt\right)^2[{\qtt^2+z(1-z)\kt^2}]} 
\bigg[\frac{z}{1-z} + \frac{1-z}{z}  +
\notag \\
& \hspace{-1.5cm}  + 
(3-4z) \frac{\qtt \cdot \kt}{ \qtt^2} + z(3-2z) \frac{\kt^2}{\qtt^2}
\bigg] + \frac{(1 + \epsilon)\qtt^2 z(1-z) [2 \qtt \cdot \kt + (2z -1)
  \kt^2]^2}{2 \kt^2 [\qtt^2+z(1-z)\kt^2]^2} \bigg\} \, ,
\end{align}
Now we can explicitly check the corresponding kinematic limits. 
In the collinear case this is straightforward, since the transverse integral in
Eq.~\eqref{eq:TMDkernelDefINI} is specially adapted for this limit. 
In particular, one easily obtains the real part of the DGLAP gluon-to-gluon splitting 
function:
\begin{align}
\label{eq:7}
\lim_{\kt^2 \to 0} \bar{P}_{ij}^{(0)}\left(z, \frac{\kt^2}{\qtt^2}\right) &= 2\, C_A\, \left[ \frac{z}{1-z} + \frac{1-z}{z} + z\,\left(1-z\right) \right] \, .
\end{align}
In order to study the high-energy and soft limit it is convenient to change to the following variables:
$\ptt = \frac{\kt - \qt}{1-z} = \kt - \frac{\qtt}{1-z}$, then in the high-energy limit ($z\to0$)
we obtain:
\begin{align}
  \label{eq:8}
  \lim_{z\to 0} \hat K_{gg} \left(z, \frac{\kt^2}{\mu^2}, \epsilon, \alpha_s \right)  
&=
\frac{\alpha_s C_A}{\pi (e^{\gamma_E}\mu^2)^\epsilon}\int \frac{d^{2 + 2 \epsilon} \ptt}{\pi^{1 + \epsilon}} \Theta\left(\mu_F^2 - (\kt - \ptt)^2\right) \frac{1}{\ptt^2} \notag
\end{align}
where the term under the integral is easily identified as the real
part of the LO BFKL kernel.
Additionally, one can check that in the angular ordered region of phase space,
where $\ptt^2 \to 0$, we reproduce the
real/unresummed part of the CCFM kernel~\cite{Ciafaloni:1987ur,Catani:1989yc,CCFMd}:
$\frac{1}{z} + \frac{1}{1-z} + \mathcal{O}\left(\frac{\ptt^2}{\kt^2}\right)$.

\section*{Acknowledgments}
The project is partially supported by
Narodowe Centrum Nauki with grant DEC-2017/27/B/ST2/01985.

\newpage 
%

\renewcommand*{\FigPath}{./WeekII/05_Xiaohui/Figs}
 
\def\bea#1\eea{\begin{align}#1\end{align}}

\wstoc{Jet TMDs}{Xiaohui Liu}
 \title{
Jet TMDs}

\author{Xiaohui Liu$^*$ 
}
\index{author}{Liu, X.}

\address{Center of Advanced Quantum Studies, Department of Physics, Beijing Normal University, \\
Beijing, 100875, China,\\
$^*$E-mail: xiliu@bnu.edu.cn}



\begin{abstract}
This article reviews the opportunities of applying jets in probing internal (un)-polarized transverse structures of hadrons and nulcei/nucleons. 
\end{abstract}


\bodymatter

\section{Introduction}\label{aba:intro}
Jets and their substructures play important roles in modern collider physics. On the experimental side, due to the large production rates,
jets can probe physics at scales crossing several orders of magnitude with well-controlled statistic errors, for instance from ${\cal O}(10)\>{\rm GeV}$ to ${\cal O}(1)\>{\rm TeV}$ at the LHC. Advanced jet substructure techniques further extend the region to sub-GeV regime, makes them ideal tools for studying the internal hadron structures. Recently proposed jet substructure operations, including jet grooming, trimming, substantially reduce the sensitivity to the soft contamination from multiple-parton interactions. On the theory side, thanks to the rapid developments in perturbation calculations, many collider jet processes, such as single jet inclusive production at the EIC~\footnote{see this proceeding.},  are now known to the next-to-next-to-leading order accuracy in $\alpha_s$ leading to a typical (sub)-percent level theoretical uncertainty. All order resummation out of the first principle at next-to-leading logarithmic accuracy (NLL) and beyond are also feasible for jet processes involving large hierarchies. These developments for collider jet physics  make us confident in extracting interesting internal (un)-polarized transverse structures of hadrons and nulcei/nucleons by using jets in hard probes, just like what we do with the semi-inclusive DIS, Drell-Yan and $e^+e^-$ annihilation processes. In the remaining of this article, we will review the possible avenues toward the TMD physics using jet or jet substructure observables.

\section{Jet TMDs}\label{aba:Jtmd}
\subsection{inside-jet TMDs}
As the first example, we consider the transverse momentum distribution of a hadron $h$ within fully reconstructed jets in $pp$ collisions, $pp \to ({\rm jet}h)X$, where the transverse momentum is determined with respect to the jet axis and the jets are constructed using the anti-$k_T$ algorithm with the radius parameter $R$. Typically $R$ is chosen to be $0.4 - 0.8$. With the small $R$ limit, the cross section for this process is found to have the factorized form~\cite{Kang:2017glf_224}
\bea
\frac{\mathrm{d}\sigma^{pp\to({\rm jet} h)X} }
{\mathrm{d}p_T \mathrm{d}\eta \mathrm{d}z_h \mathrm{d}^2j_\perp}
= \sum_{a,b,c}f_a(x_a)\otimes f_b(x_b) \otimes H^c_{ab}(x_i;\eta,p_T/z,\mu)\otimes {\cal G}_c^h(z,z_h,j_\perp) \,,
\eea
where 
 $H$ the hard functions for producing an energetic parton $c$ in the hard-scattering by the parton $a$ and $b$. ${\cal G}$ is the semi-inclusive TMD fragmenting jet functions which describes the formation of a jet out of the parton $c$ with the observed hadron $h$ inside . The jet function ${\cal G}$ encodes all the information on the hadron transverse momentum $j_\perp$. When $j_\perp \ll p_T$, the semi-inclusive jet function can be further factorized 
 \bea
 {\cal G} = {\cal H}_{c\to i}(R) \int \mathrm{d}^2k_\perp \mathrm{d}^2\lambda_\perp \delta^{(2)}\left( z_h \lambda_\perp + k_\perp - j_\perp
 \right) D_{h/i}(z_h,k_\perp) S_i(\lambda_\perp)\,.
  \eea
 Here ${\cal H}$ is related to the out-of-jet radiations. The soft function $S_i$ describes the soft radiations with momentum of order $j_\perp$ and $D_i$'s are the standard TMD fragmentation functions (TMDFFs). Here we have suppressed all the scale dependence including both the factorization scale $\mu$ and the rapidity scale $\nu$. When $j_\perp \gg \Lambda_{\rm QCD}$, the TMDFFs can be further matched onto the collinear PDFs. All components in the factorization theorem including the TMD matching coefficients have been calculated to NLO and thus the NLL resummation in both $\ln R$ and $\ln j_\perp$ are realized. 

 \begin{figure}[h]%
\begin{center}
 \parbox{2.1in}{\includegraphics[width=2.2in]{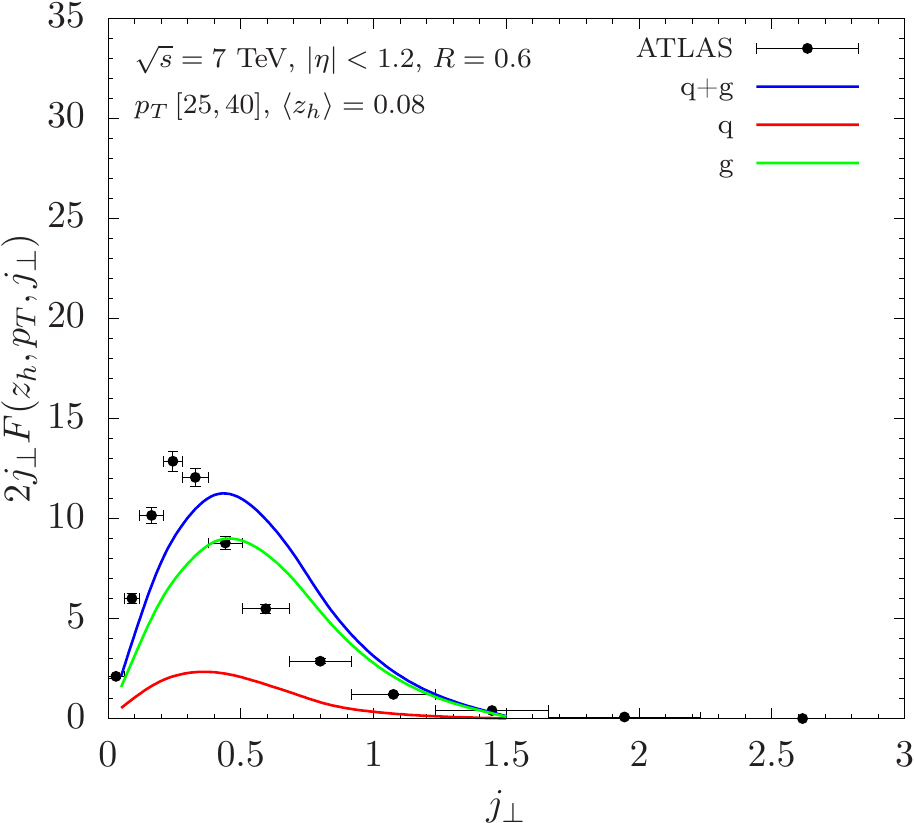}}
 \hspace*{26pt}
 \parbox{2.1in}{\includegraphics[width=2.2in]{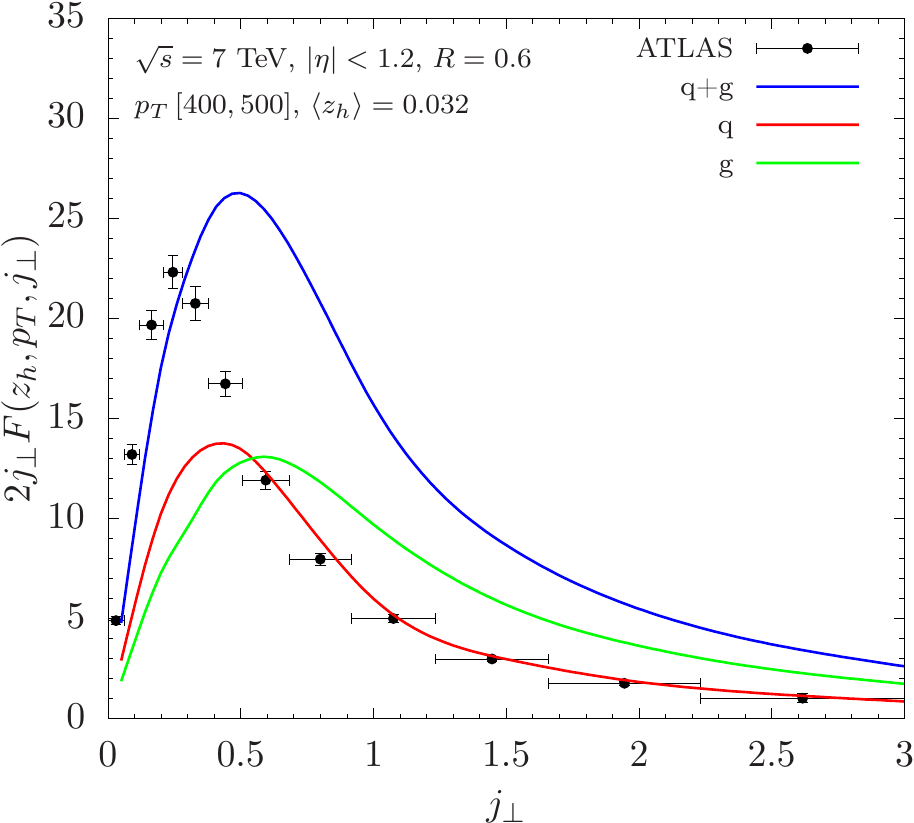}}
 \caption{Breakdown of the hadron $j_\perp$-distributions inside jets (blue) into quark initiated (red) and gluon initiated (green) TMDFF channels for different jet $p_T$ ranges.
 The plots are taken from Ref.~\cite{Kang:2017glf_224}.}
\label{fig: jperp}
\end{center}
\end{figure}

In fig.~\ref{fig: jperp}, we show the breakdown of the hardron $j_\perp$ distribution into the quark and the gluon TMDFF components. The breakdown is valid in the TMD Sudakov regime. We see that for low jet transverse momentum, the gluon TMD dominates over the quark contributions by a factor of ${\cal O}(10)$ in $pp$ collisions. Therefore, the low jet transverse momentum region will serve the “golden channel” to extract the gluon TMDFF. As a qualitative comparison, we also include the ATLAS data~\cite{Aad:2011sc}. However we note that  the experimental data are presented for the $z_h$-integrated hadron distribution, i.e. with $z_h$ integrated from $0$ to $1$. We suggest  a future measurement of the hadron $j_\perp$ with binning $z_h$ to allow for a direct comparison between the theory and the experiments to benefit the extraction of the TMDFFs.
We further notice that inside-jet hadron distributions have been applied to a recent investigation of the Collins azimuthal asymmetry in transversely polarized $pp$ collisions~\cite{Kang:2017btw_218}.

\subsection{jet imbalance}
Now we turn to the jet imbalance in $pp$~\cite{Buffing:2018ggv} and $ep(A)$ colllisions~\cite{Liu:2018trl}. 
\subsubsection{probing the factorization breaking effects}
For $pp$ collision, we consider $p(p_1, s_\perp) + p(p_2)\rightarrow \text{jet} + \gamma$, where the transverse momenta of the jet and the photon are given by $p_{J\perp}$ and $p_{\gamma\perp}$, respectively. Here the transverse momentum is measured with
respect to the beam axis in the center-of-mass frame of
the colliding protons. We focus on the small imbalance case in which 
$| \vec{q}_\perp| \equiv \vec{p}_{\gamma\perp}+ \vec{p}_{J\perp} \ll |\vec{p}_{\gamma\perp} |\,, | \vec{p}_{J\perp}|$.  The factorization theorem for this process in the small $q_\perp$ region is known to break-down. However its size and thus its impact on the phenomenological studies are un-known. 

To probe the size of the factorization breaking effects, a novel framework without the Glauber modes based on the generalized TMD factorization along with adding additional soft contributions are derived for both un-polarized and polarized case. For instance, for the polarized case, the proposed framework reads~\cite{Buffing:2018ggv}
\bea
\frac{d\Delta\sigma}{d {\cal PS}} =&  \epsilon^{\alpha\beta}s_\perp^\alpha \sum_{a,b,c} \int d\phi_J \int \prod_i^4 d^2\vec{k}_{i\perp}  \delta^{(2)}(\vec{q}_\perp - \sum_i^4 \vec{k}_{i\perp})
\nonumber \\
& 
\hspace{-7mm} \times \frac{k_{1\perp}^{\beta}}{M}    f_{1T,a}^{\perp\,\rm SIDIS}(x_a, k_{1\perp}^2) f_b^{\rm unsub}(x_b,k_{2\perp}^2) 
\nonumber \\
&
\hspace{-7mm} \times 
S_{n{\bar n}n_J}(\vec{k}_{3\perp})
S^{cs}_{n_J}(\vec{k}_{4\perp},R) 
  H^{\text{Sivers}}_{a b\rightarrow c \gamma}(p_\perp)  J_{c}(p_\perp R) \, ,
 \label{Sivers_factorization} 
\eea
which is sensitive to the Silvers function. The process dependent color factors are included in the hard coefficient $ H^{\text{Sivers}}$. 
We note that the framework is self-consistent and can systematically improve its predictive accuracy. Currently, the resummation is achieved at the NLL accuracy level. 
By controlling the jet $p_{J\perp}$, we change the sensitivity to the non-perturbative TMD effects. 
Therefore any difference between the future data and the outcome of this framework will shed light on the size of the factorization breaking.  If the sizes were small, the process can be excellent candidate for extracting the gluon TMDs.

\subsubsection{jet TMDs at the future EIC}
Similar process can be investigated at the EIC, but now we study the imbalance between the jet and the lepton ($ \ell'(k_\ell)$) in the center of mass frame in
$\ell (k)+A(P)\to \ell'(k_\ell)+\text{Jet} (P_J)+X $, where $A$ can be (un)-polarized proton or nuclei/nucleon. One can derive the all-order factorization  theorem for this process for both polarized and un-polarized cases. Also the $P_T$-broadening generated by cold nuclear matter effects can be incorporated systematically~\cite{Liu:2018trl}. 
 
 \begin{figure}[h]%
\begin{center}
 \parbox{2.3 in}{\includegraphics[width=2.2in]{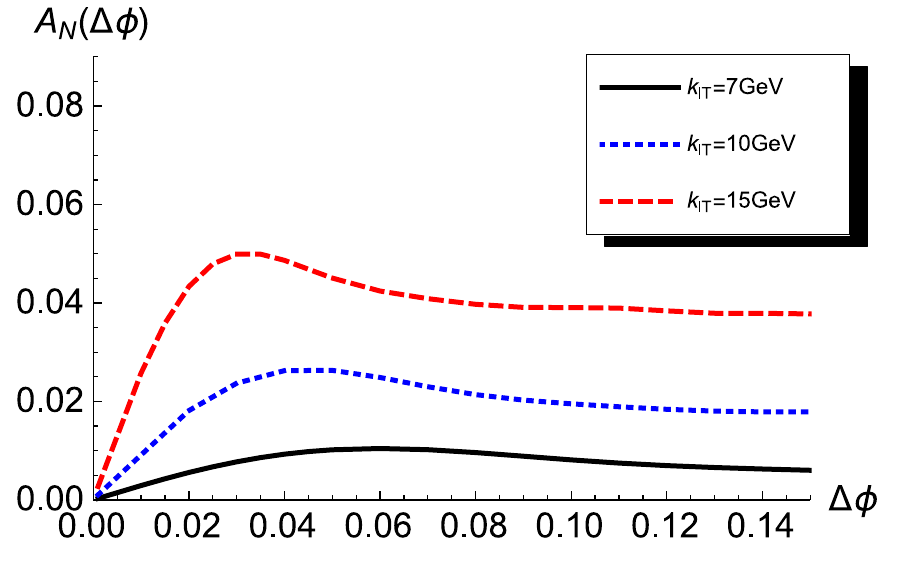}}
 \hspace*{15pt}
 \parbox{2.1in}{\includegraphics[width=2.2in]{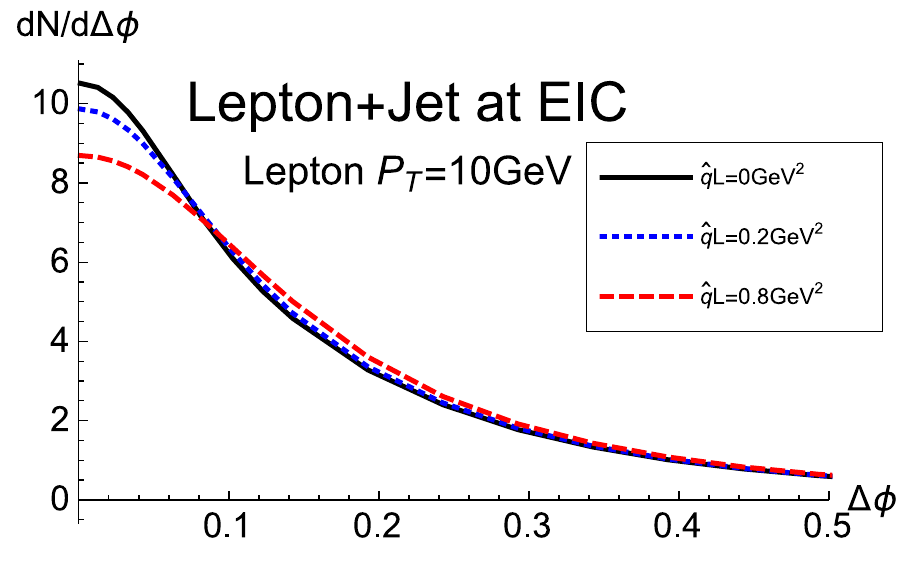}}
 \caption{Left panel: the single spin asymmetry in transversely polarized $ep$ collision. Right panel: $P_T$-broadening in $eA$ collision. 
 The plots are taken from Ref.~\cite{Liu:2018trl}.}
\label{fig:eic}
\end{center}
\end{figure}
The left panel of fig.~\ref{fig:eic} shows the single spin asymmetry as a function of the azimuthal angle $\Delta \phi$ distribution. By changing the lepton transverse momentum, we probe different range in $x$ of the Sivers function. For $k_{\ell,T} = 15\>{\rm GeV}$, we found the single spin asymmetry can be as large as $5\%$ which should be observable in the future EIC which typically has sufficient statistics in the lepton-nucleon frame. 

The right panel in fig.~\ref{fig:eic} represents the sensitivity of the azimuthal angular correlation as a function of $\Delta \phi$ to different values of $\hat qL$ in the range of a theoretical estimate for cold
nuclear matter~\cite{Baier:1996sk}. This correlation is expected to be able to 
investigate at the future EIC.

\section{Conclusions}
 Several examples have been presented in this article to highlight the opportunities that the jet brings to the TMD physics for both un-polarized and polarized cases. With increasing experimental accuracy and available high precision theoretical tools, we believe jets will show their advantages in extracting non-perturbative structures of the hadron or nucleon/nuclei by hard probes in the near future.

\section{Acknowledgments}
X. L. is supported by the National Natural
Science Foundation of China under Grant No. 11775023
and the Fundamental Research Funds for the Central
Universities.




\newpage

\renewcommand*{\FigPath}{./WeekII/06_Yiannis/}
\newcommand{\zcut}{z_{\text{cut}}}
\newcommand{\bmat}[1]{\boldsymbol{#1}}

\wstoc{TMDs through jets and quarkonia}{Yiannis Makris}
\title{
TMDs through jets and quarkonia}

\author{Yiannis Makris}
\index{author}{Makris, Y.}

\address{ 
Los Alamos National Laboratory, \\
Theoretical Division, MS B283, \\
Los Alamos, NM 87545, USA \\
E-mail: yiannism58@gmail.com}

\begin{abstract}
In this talk, focusing on groomed jet and quarkonium production and decay  processes, we discuss the factorization and resummation of large logarithms at the level of cross section involving TMDs. For groomed jets we analyze the measurement of hadronic transverse momentum with respect to the groomed-jet-axis and show that this observable can be used for probing the non-perturbative part of the TMD (rapidity) evolution kernel. For the case of quarkonium we present a new framework for factorization of quarkonium production and decay TMD related  processes. In this new approach the factorization theorems are written in terms of the new TMD quarkonium shape functions. We work in the framework of effective field theories and particularly soft collinear effective theory and non-relativistic quantum chromodynamics.   
\end{abstract}


\bodymatter
\section{Hadronization within groomed jets }\label{sec:groomed_jets}
In the recent years there were many studies\cite{Bain:2016rrv, Makris:2017arq, Makris:2018npl} proposing the use of in-jet hadronization as a probe to TMD physics. Here we discuss the measurement of hadronic transverse momentum inside a groomed jet, particularly the energy fraction, $z_{h} = E_{h}/E_{J}$ and the transverse momentum  w.r.t. the groomed-jet-axis, $p_{Th}$, of an identified hadron $h$ within a groomed jet. The jet is clustered using anti-$k_T $ algorithm and then groomed using modified mass drop tagging (mMDT) algorithm\cite{Dasgupta:2013ihk, Dasgupta:2013via}.  The mMDT algorithm (or its generalization known as soft-drop) removes contaminating soft radiation from the jet by constructing an angular ordered tree of the jet through the Cambridge/Aachen (C/A) clustering algorithm, and removing the branches at the widest angles which fail an energy requirement, $E > \zcut E_{\text{branch}} $. As soon as a branch is found that passes, this branch is declared the groomed jet, and all constituents of the branch are the groomed constituents. What is remarkable about the procedure, is that it gives a jet with essentially zero angular area, since at large angles, all collinear energetic radiation is to be found at the center of the jet, and \emph{no cone is actually imposed to enclose this core}. One simply finds the branch whose daughters are sufficiently energetic. Formally the daughters could have any opening angle, though their most likely configuration is collinear.

We use soft-collinear effective theory (SCET) to factorize the relevant cross section. The factorization  for this observable\cite{Larkoski:2014wba} is given in terms of the quark/gluon jet fraction, $F_i$, which are specific to the experimental configuration, and the jet function, $J_i$,
 \begin{equation}
 \label{eq:g-fact}
 \frac{d\sigma}{d\vec{p}_{J}d\mathcal{M}} = \sum_i F_{i}( \mathcal{C}, p^{\mu}_{J}, R, \zcut) J_i( E_J,R, \zcut , \mathcal{M})\;,
 \end{equation}
 where $\mathcal{C}$ denotes collectively the experimental constraints and $\mathcal{M}$ the measurements imposed on the grooomed constituents.  For $e^+ e^- \to \text{di-jet}$ processes  the quark/gluon fraction, $F_i$, can be calculated perturbatively  and for DIS can be written in terms of a convolution of the  partonic matching coefficient and the collinear PDFs. For hadronic collisions a perturbative calculation of  the fractions $F_i$ within SCET suffers from factorization breaking effects,  but since they are independent of the measurement $\mathcal{M}$ they could be extracted from one process and used in an other (in a similar manner to the PDFs).

For the case of an identified hadron in jet and for $\mathcal{M} = \{ \bmat{q}_T, z_h\}$, where the small transverse momentum, $\bmat{q}_{T} = \bmat{p}_{Th}/z_{h}$, is defined w.r.t. the groomed jet axis, then eq.~(\ref{eq:g-fact}) with $R\sim 1$ becomes:  
 \begin{equation}
 \frac{d\sigma}{d\vec{p}_{J}d \bmat{q}_T dz_h} = \sum_i F_{i}( \mathcal{C}, R, p^{\mu}_{J}, \zcut) \mathcal{G}_{i/h}( E_J, \zcut , \bmat{q}_T, z_h)\;,
 \end{equation}
 up to power-corrections where $\mathcal{G}_{i/h}$ is the groomed TMD fragmenting jet function (gTMDFJF) which is further refactorized in the context of SCET$_+$ into a collinear-soft function, $S_{i}^{\perp}$, and the purely collinear term\footnote{In the language of SCET the term ``collinear" corresponds to the contribution from modes that satisfy collinear scaling. Not to be confused with the integrated distributions, usually present in the collinear factorization }, $D_{i/h}^{\perp}$. The purely collinear term is simply the so called ``un-subtracted TMD fragmentation functions".  The factorized expression can then be written conveniently in the impact parameter space, 
 \begin{equation}
 \mathcal{G}_{i/h}( E_J, \zcut , \bmat{b}, z_h) = S_{i}^{\perp} (E_J \zcut, \bmat{b} ; \nu) D_{i/h}^{\perp}(E_J, z_h, \bmat{b};\nu)\;,
 \end{equation}
 where we have explicitly included the rapidity scale, $\nu$, dependence which is introduced by the necessary regulator for rapidity  divergences. We use the formalism of the rapidity renormalization group (RRG) to handle the rapidity scale dependence and to resum rapidity logarithms.  Identifying as $Q$ and $Q \zcut$ as the collinear and collinear-soft canonical rapidity scales we suggest the logarithmic derivative of the cross-section (w.r.t. $\ln \zcut$) as a probe to the rapidity anomalous dimension.  We demonstrate the sensitivity  of this observable to the non-perturbative aspects of the rapidity anomalous dimension using four different parameterizations (for the details of the parameterizations see the original paper~\cite{Makris:2017arq}).  As an example we study the process $e^+ +  e^- \to (u\to\pi^+ ) + X $. The results are shown in figure~\ref{fig:plt-1}. 
 
\begin{figure}[h!]
\begin{center}
\includegraphics[width= 0.65 \textwidth]{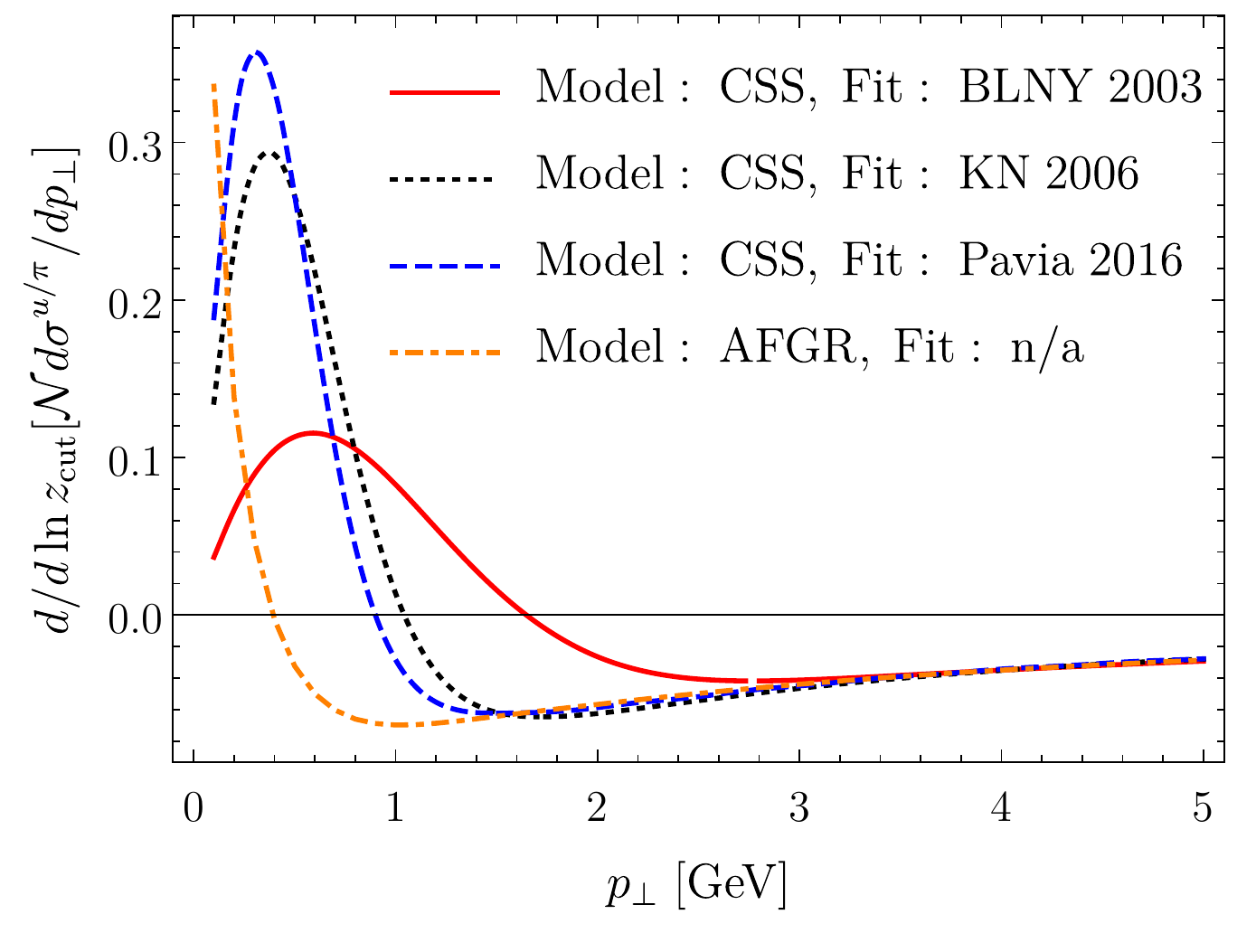}
\end{center}
\caption{The logarithmic derivative of TMDFJF for three different models. All models agree in the perturbative regime but show significant differences in the non-perturbative region.}
\label{fig:plt-1}
\end{figure}

From figure~\ref{fig:plt-1} we see that the logarithmic derivative of the groomed TMD distribution shows strong discrepancy between various modes at small values of the transverse momentum. At moderate values, $p_T \gtrsim 5 \; \text{(GeV)}\gg  \Lambda_{\text{QCD}} $,  the observable becomes insensitive to the non-perturbative models, as expected since all models should merge to the perturbative result in this range of the spectrum.

\section{A new EFT approach to quarkonium production at small $p_T$ }\label{sec:quarkonium}

 Among others, some of the processes proposed to study gluon TMDs in collider experiments are inclusive, exclusive, and associated quarkonium production processes. The goal of this talk is to emphasize many difficulties that arise when studying observables  sensitive to soft radiation through quarkonium production  or decay processes. In collider physics, quarkonia are studied within the framework of NRQCD factorization~\cite{Caswell:1985ui, Bodwin:1994jh_356} where the cross section is written as a sum of products of short distance matching coefficients and the corresponding long distance matrix elements (LDMEs). 

The NRQCD factorization approach was used extensively in quarkonium phenomenology, but despite the many successes,  NRQCD factorization  is only effective when the quarkonium is produced with relatively large transverse momenta. Intuitively, in this kinematic regime, emissions of soft and ultra-soft gluons from the heavy quark pair cannot alter the large transverse momentum of the quarkonium state. Ignoring these soft emissions, the quarkonium transverse momentum is then determined from the hard process alone and the infrared (IR) divergences that are present in perturbative calculations of the partonic hard process are absorbed into the non-perturbative LDMEs and collinear matrix elements (PDF/FF). However, when quarkonia are produced with small transverse momenta, this soft gluon factorization assumption must be relaxed.  The quarkonium photo/lepto-production was studied\cite{Beneke:1997qw, Fleming:2003gt, Fleming:2006cd} in this region sensitive to soft radiation where NRQCD factorization break down and it was found that promoting the LDMEs into quarkonium shape functions is necessary for correctly accounting the soft radiation from the heavy quark pair.

Here we propose a new EFT in which the treatment of soft and collinear modes is similar to SCET. We refer this new EFT as SCET$_Q$ and within this effective theory we demonstrate how we can obtain factorization theorems for quarkonium production and decay.  

A crucial role in our analysis plays the diagrammatical derivation of the relevant partonic operators. For example in a generic hard process, the matching onto the S-wave operators is given by   
\begin{equation}
\mathcal{O}_{\Gamma}(\text{S-wave}) = \psi^{\dag}S_v^{\dag} \Gamma S_v \chi\;,
\end{equation}
where $S_v$ is a time-like (soft-gluon) Wilson line. This suggests the presence of time-like Wilson line in the soft sector. We confirm that for color-octet quarkonia this Wilson line is  necessary for the IR-finiteness  of the fixed order cross section. We showed that such diagrammatical analysis strongly constrained the form of operators that contribute to the hard sector of the SCET$_Q$ Lagrangian. Also this allowed us to relate the matching coefficients of  specific operators, which naively may seem unrelated, at all orders in perturbation theory.  We further demonstrate that the resulting operators are invariant under collinear, soft, and ultra-soft gauge transformations.



\newpage 
%

\renewcommand*{\FigPath}{./WeekII/07_Boglione/Figs}

\renewcommand{\kt}{k_\perp}
\renewcommand{\pt}{p_\perp}

\wstoc{TMD Phenomenology:\\
Recent developments in the 
global analyses of polarized TMDs}{Mariaelena Boglione}
\title{TMD Phenomenology:\\
Recent developments in the 
global analyses of polarized TMDs
}

\author{Mariaelena Boglione}
\index{author}{Boglione, M.}

\address{Dipartimento di Fisica, Universit\`a di Torino, INFN-Sezione Torino\\
  Via P. Giuria 1, 10125 Torino, Italy\\
E-mail: elena.boglione@to.infn.it}

\begin{abstract}
Establishing a precise multi-dimensional imaging of quarks and gluons inside nucleons
is one of the primary goals of the Electron Ion Collider project. 
I will present a personal choice of recent developments in the phenomenological mapping of 
the momentum space structure of the nucleon, through the studies of transverse momentum 
dependent parton distributions, with special focus on the polarized cases, 
where correlations between spin and partonic momenta are more relevant.
\end{abstract}


\bodymatter

\section{Introduction}\label{sec:intro}

In the last decades several experiments have been performed to unravel the 3D-structure of nucleons in terms 
of their partonic degrees of freedom. Transverse momentum dependent parton distributions (TMDs) allow us to 
build a map of the hadron structure in momentum space, which will ultimately lead to a full knowledge 
of its composition and internal dynamics in terms of elementary constituents.

TMDs encode the confined phase of hadrons. Consequently, they cannot be computed in pQCD, but have to be modeled 
and extracted phenomenologically from experimental data. However, when properly defined within the realm of TMD 
factorization, TMDs are universal and, therefore, process independent. 

TMDs can be studied in different hadronic processes, mainly in Drell-Yan scattering, which allows TMD parton 
distribution functions (PDFs) to be extracted, $e^+e^-$ annihilations, where the parton hadronization mechanism can 
be investigated through TMD fragmentation functions (FFs), and finally in Semi-Inclusive Deep Inelastic 
Scattering (SIDIS), where both TMD PDFs and FFs play a crucial role.
Recently a great experimental effort has been brought forward by the HERMES, COMPASS and JLab Collaborations in collecting 
a large amount of unpolarized and polarized SIDIS data, covering rather different kinematic regimes, and an equally 
strong effort has been performed by several communities of theorists and phenomenologists, who have worked together in the  
pioneering work of extracting the TMDs. However, the real break through in this field is expected to occur when a new, 
purposely designed experimental facility will come to life: the Electron Ion Collider~\cite{Accardi:2012qut_8}(EIC). 
Its experimental set up and 
kinematic coverage have been engineered in such a way to complement and extend the $Q^2$ and $x$ coverage, towards larger 
$Q^2$ and smaller $x$ values, allowing to reach an unprecedented precision in mapping the 3D hadron structure.

  There are eight independent TMD PDFs~\cite{Boglione:2015zyc}, among which the distribution of unpolarized partons in an unpolarized proton 
and the distribution of unpolarized partons in a transversely polarized proton, the so-called Sivers function, and 
the distribution of transversely polarized partons in a transversely polarized proton, usually referred to as ``transversity''.
Although all eight functions play a fundamental role in the study of spin and transverse momentum correlations, 
I will focus only on the TMDs mentioned above, which have recently been the object of extensive phenomenological 
analyses, together with the Collins function, a TMD FF related to the hadronization of transversely polarized partons into 
spinless final hadrons.

\section{The Sivers function}

Very recently, a novel extraction of the Sivers function from SIDIS asymmetry measurements was presented~\cite{Boglione:2018dqd},  
where all available SIDIS data from HERMES~\cite{Airapetian:2009ae_232_184}, 
JLab~\cite{Qian:2011py}, COMPASS-deuteron~\cite{Alekseev:2008aa_190} and COMPASS-proton experiment~\cite{Adolph:2016dvl} were included.
The increased statistics and precision of these new sets of data, together with a finer binning in $Q^2$ as well as in $x$, 
has allowed a critical re-analysis of the uncertainties affecting the extraction procedure, and a thorough study of the subtle 
interplay among experimental errors, theoretical uncertainties and model-dependent constraints. Moreover, we have used a simpler 
and more transparent parametric form of the Sivers function, in an attempt to extract the Sivers function on the sole 
information provided by experimental data. 
We have also assessed the flavour content of the Sivers function 
and its separation into valence and sea contributions.
We found that the existing data can only resolve unambiguously the total $u$-flavour (valence + sea) contribution, 
while leaving all others largely undetermined.
Any attempt to separate valence from sea contributions, 
resulted in a decreased quality of the fit, due to a lack of information in the 
experimental data presently available.

For this analysis we have performed two best fits: the first was a basic fit, which we referred to as the ``reference fit'', 
based on the most simple parametric form which could reproduce the main features of the Sivers function; the second fit included 
two extra free parameters, to make the parametrization more flexible in the small-$x$ region, in such a way that possible sea 
contributions to the $u$ and $d$ flavours could be accounted for. 
Although we could not separate sea from valence contributions within the Sivers first moments, this approach allowed us to obtain a 
much more realistic estimate of the uncertainties affecting the extracted functions at small values of $x$. 
Drawing well-founded conclusions on the low-$x$ and large-$x$ kinematic regimes will only become possible when 
new experimental information will become available 
from dedicated experiments like the EIC~\cite{Accardi:2012qut_8}, 
or JLab12~\cite{Dudek:2012vr}. 
Our results are shown in Fig.~\ref{fig:ref-vs-alpha}.
%
 \begin{figure}[ht]%
 \begin{center}
   \vspace*{-9pt}
   \parbox{2.1in}{\includegraphics[width=2in]{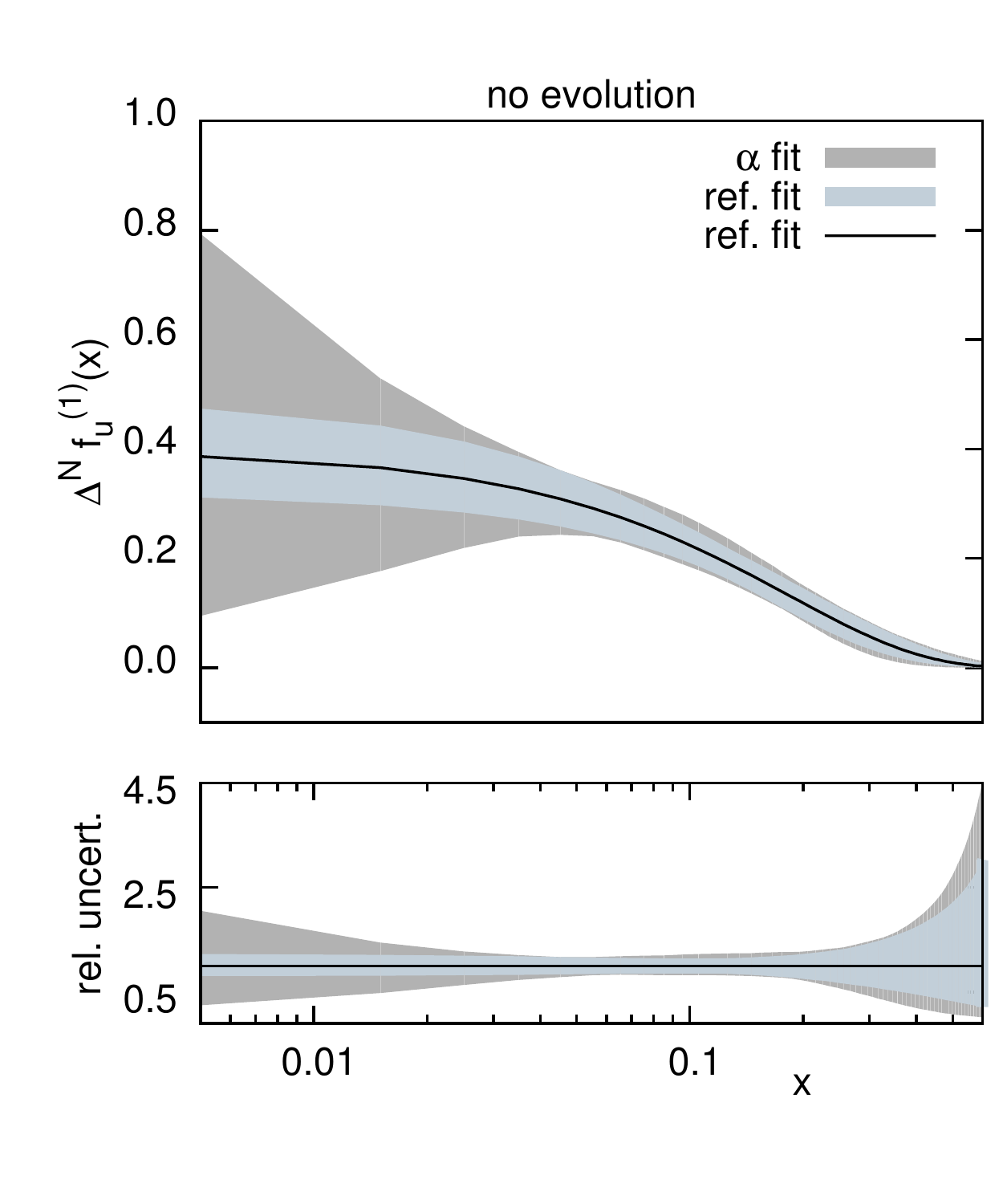}}
   \hspace*{8pt}
   \parbox{2.1in}{\includegraphics[width=2in]{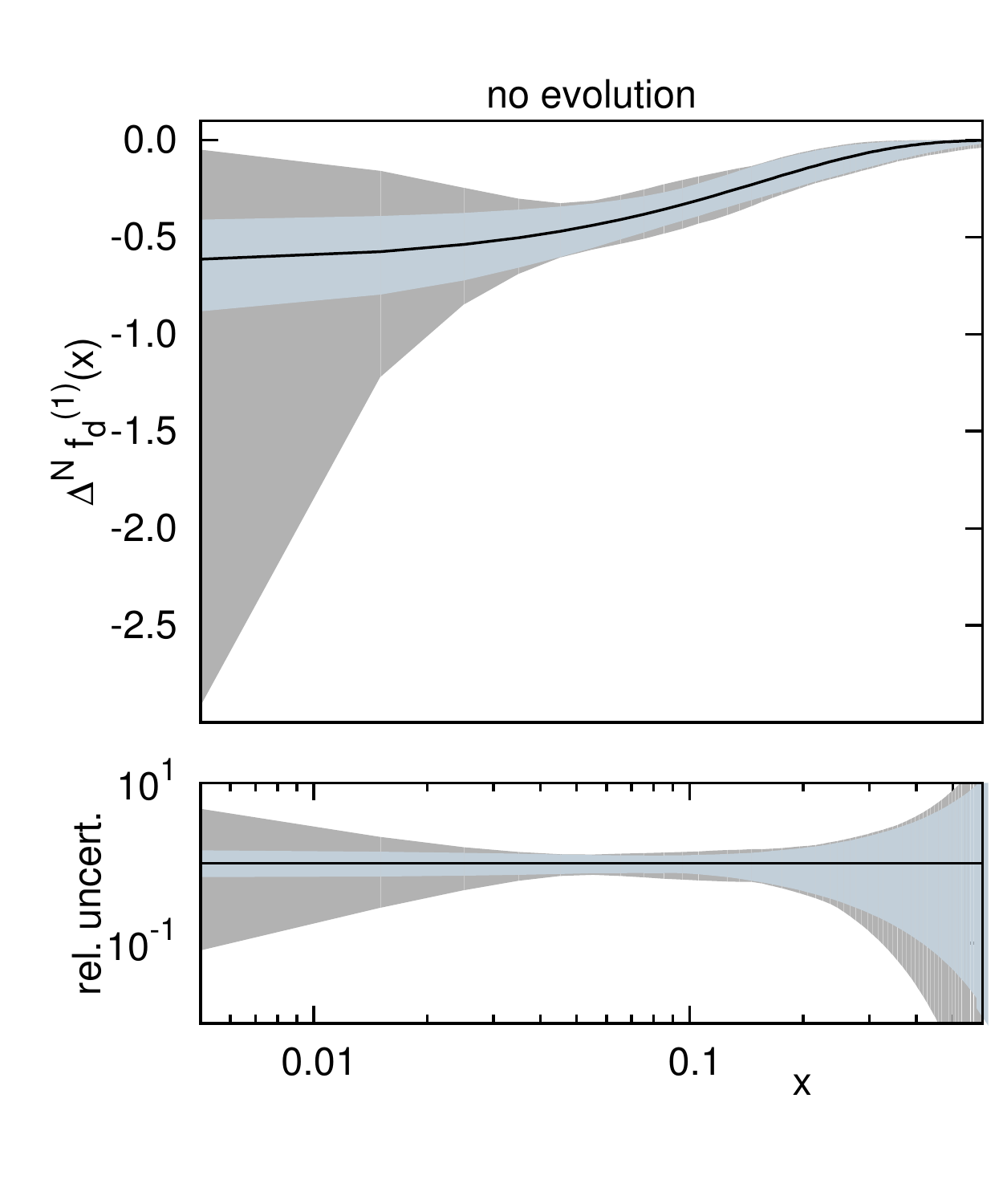}}
   \\
 \parbox{2.1in}{\includegraphics[width=2in]{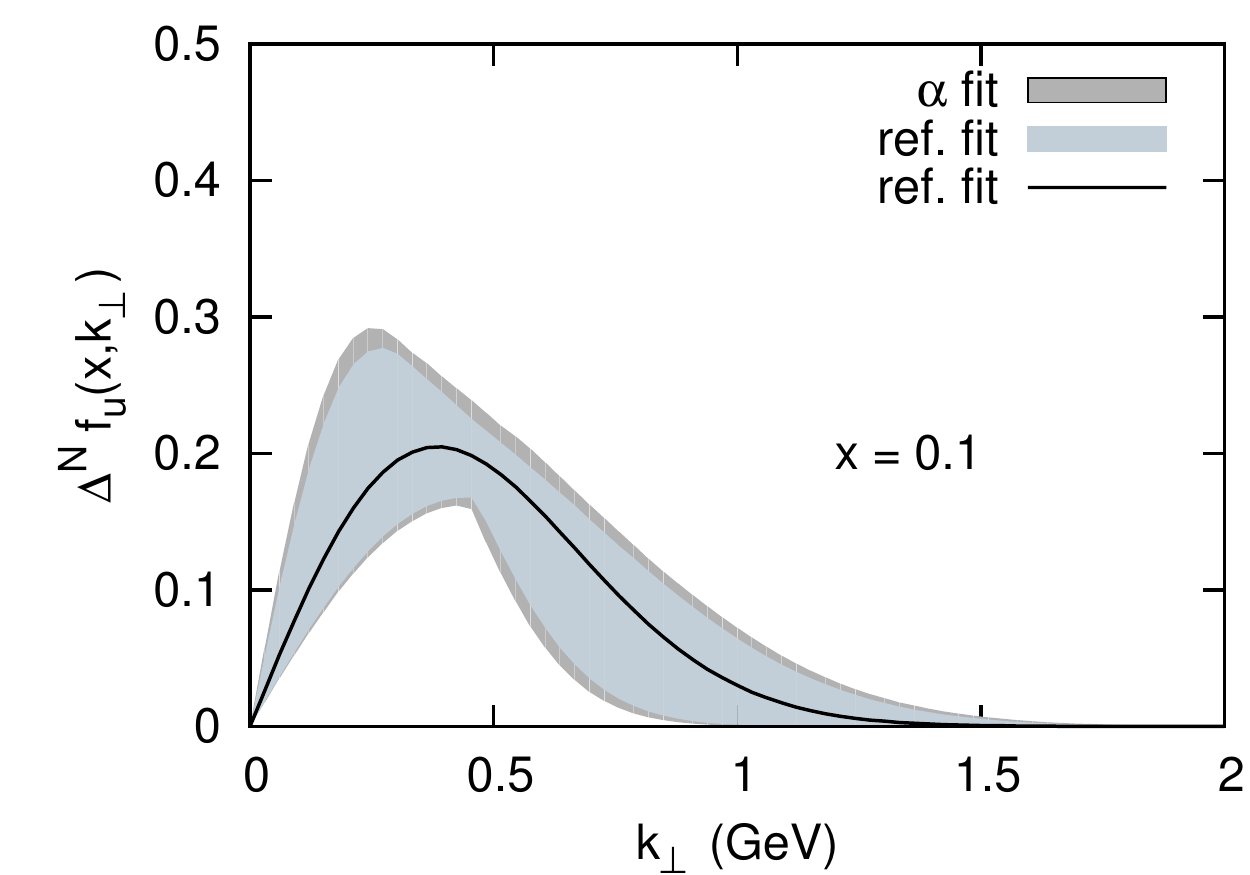}}
   \hspace*{8pt}
   \parbox{2.1in}{\includegraphics[width=2in]{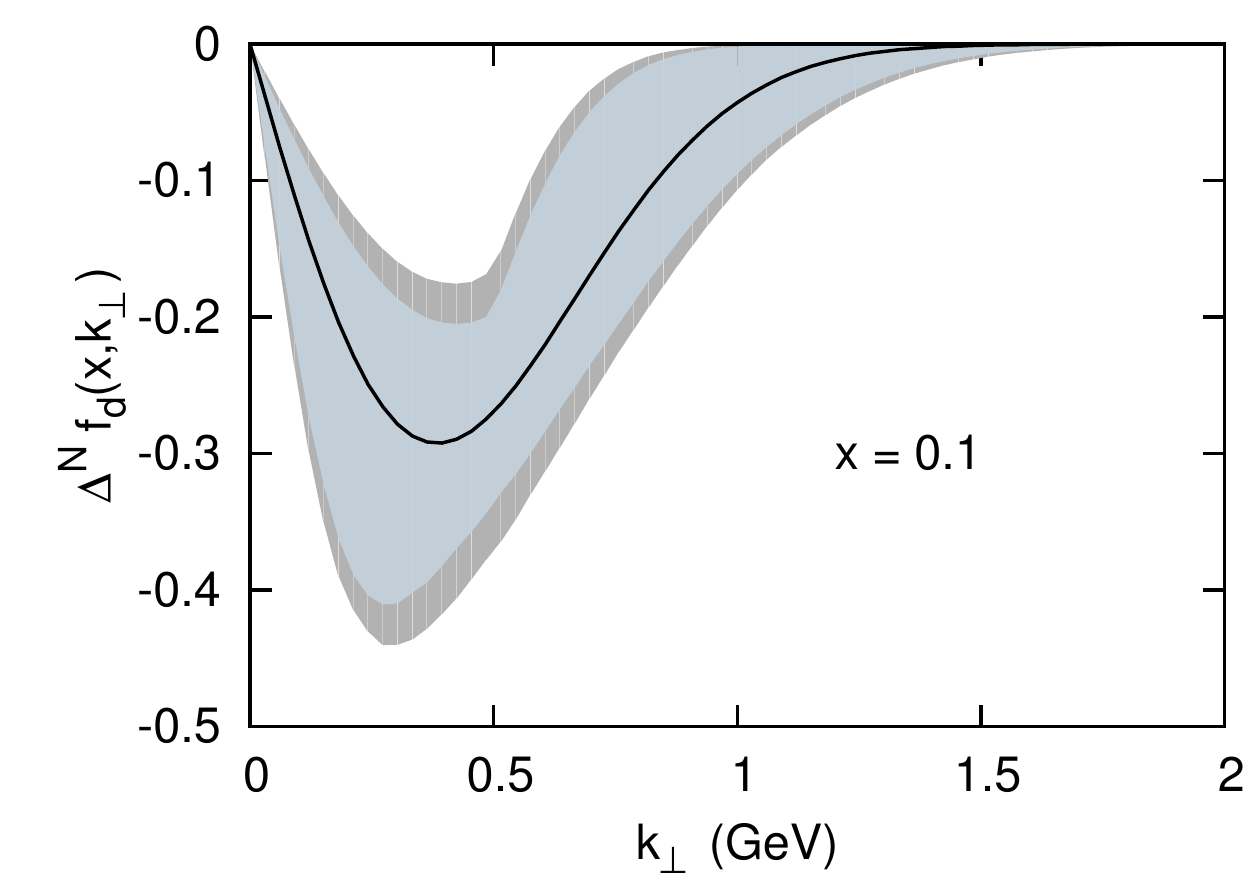}}  
    \caption{The extracted Sivers distributions for $u = u_v + \bar u$ and  
 $d = d_v + \bar d$. Upper panels: the first moments of the 
 Sivers function
 are shown versus $x$. 
 Middle panel: relative uncertainties, 
 given by the ratio between the upper/lower border of the uncertainty bands and the 
 best-fit curve for the reference fit. 
 Lower panel: the Sivers functions 
 is shown versus $\kt$, at $x=0.1$. 
 Here we have no $Q^2$ dependence. 
 The shaded bands correspond to our estimate of $2\sigma$ C.L.
 In all panels, the light blue bands correspond to the uncertainties of the reference fit, 
 while the large gray bands correspond to the 
 uncertainties for the fit which includes also the $\alpha_u$ and $\alpha_d$ 
 parameters\cite{Boglione:2018dqd}.
 }%
\label{fig:ref-vs-alpha} 
\end{center}
 \end{figure}

Let us stress that also the unpolarized TMDs play a crucial role in the extraction of the Sivers function, 
as they appear as denominators of the Sivers single spin asymmetries used in the fit.
In fact, different assumptions about these functions can alter results significantly. Differently from previous 
analyses, here the $\kt$-widths of the unpolarized functions were chosen to ensure the best possible description  
of the available unpolarized SIDIS data. This released the tension encountered when trying to simultaneously 
fit COMPASS and HERMES data on the Sivers asymmetry
which could, for instance, lead to inadvertently over-fit the data by adding more parameters in order 
to reduce an artificially large $\chi^2$. This type of complications make it evident how critical it is to 
obtain a better knowledge of the unpolarized functions, not only for this but also for any other
SIDIS asymmetries.

A considerable part of our work was devoted to the study of scale dependence effects. 
We considered three different scenarios:  no-evolution, collinear twist-3 evolution and TMD-evolution.
Collinear twist-3 evolution, which proceeds only through $x$ while not affecting the $k_\perp$ dependence of the Sivers 
function, was found to be quite fast. 
In fact, when spanning the range of $\langle Q^2 \rangle$ values covered by the experimental data, the extracted Sivers 
function shows variations that are larger than the error band for the reference fit. 
%
Signals of TMD evolution, which instead affects mostly the $k_\perp$ dependence of the Sivers function 
turned out to be  more elusive.
Our attempts to estimate them resulted in a rather poor determination of the $g_2$ parameter, 
which regulates the logarithmic variation of the $k_\perp$ width with $Q^2$.
Our best fit delivered a very small value of $g_2$, with a large uncertainty.
This does not mean that TMD evolution is slow. In fact, within the large uncertainty bands corresponding to this extraction, 
there is room for quite a large variety of different $Q^2$ behaviors.
Unfortunately, the available experimental information is presently too limited to determine $g_2$ 
with a satisfactory precision.

\section{Transversity and the Collins function}

Issues related to TMD evolution are also particularly relevant when dealing with the phenomenological 
extraction of other TMDs, like transversity and the Collins functions. As these two functions appear 
convoluted together in the SIDIS cross sections, it is very difficult to extract 
them separately. 
In 2007 we first proposed the first simultaneous extraction of transversity and the Collins 
function~\cite{Anselmino:2007fs}, exploiting both SIDIS and $e^+e^- \to h_1 h_2 X$ experimental data. 
Ever since many other authors have followed the similar paths, 
proposing more and more refined analyses of the available experimental
data~\cite{Anselmino:2013vqa_50,Kang:2014zza,Kang:2015msa_50,Bacchetta:2015ora,Anselmino:2015sxa_50,Anselmino:2015fty}. 

All these analyses found high quality best fit results, but one may wonder how TMD evolution might affect 
these results, as the typical $Q^2$ scales of the SIDIS and  $e^+e^- \to h_1 h_2 X$ experiments 
are very different ($Q^2 \sim 3$ GeV$^2$ for SIDIS and $Q^2 \sim 100$ GeV$^2$ for $e^+e^-$). 
To answer this question it is interesting to compare two analyses which follow different philosophies: 
one~\cite{Anselmino:2015sxa_50} in which no TMD evolution is taken into account, 
and the other~\cite{Kang:2015msa_50} 
where the $Q^2$ dependence of the $\kt$ and $\pt$ distributions are accounted for. Both analyses exploited the same 
data sets, from the BaBar~\cite{TheBABAR:2013yha_196} and BELLE~\cite{Seidl:2008xc_202}, 
HERMES~\cite{Airapetian:2010ds_108} and COMPASS~\cite{Adolph:2014zba_208} experiments.
 
Plots showing the transversity and Collins functions as obtained with~\cite{Kang:2015msa_50} and 
without~\cite{Anselmino:2013vqa_50} TMD evolution are presented in Fig.~\ref{fig:transversity}. 
It is easy to notice the similarity between 
the two curves, well within the uncertainty bands in the kinematic regions covered by experimental data. 
Notice, however, that as the asymmetries measured by BaBar and BELLE are double ratios, 
this similarity might simply be due to the cancellations of strong evolution 
effects between numerators and denominators.

 \begin{figure}[ht]%
 \begin{center}
   \vspace*{-6pt}
   \parbox{2.1in}{\includegraphics[width=2in]{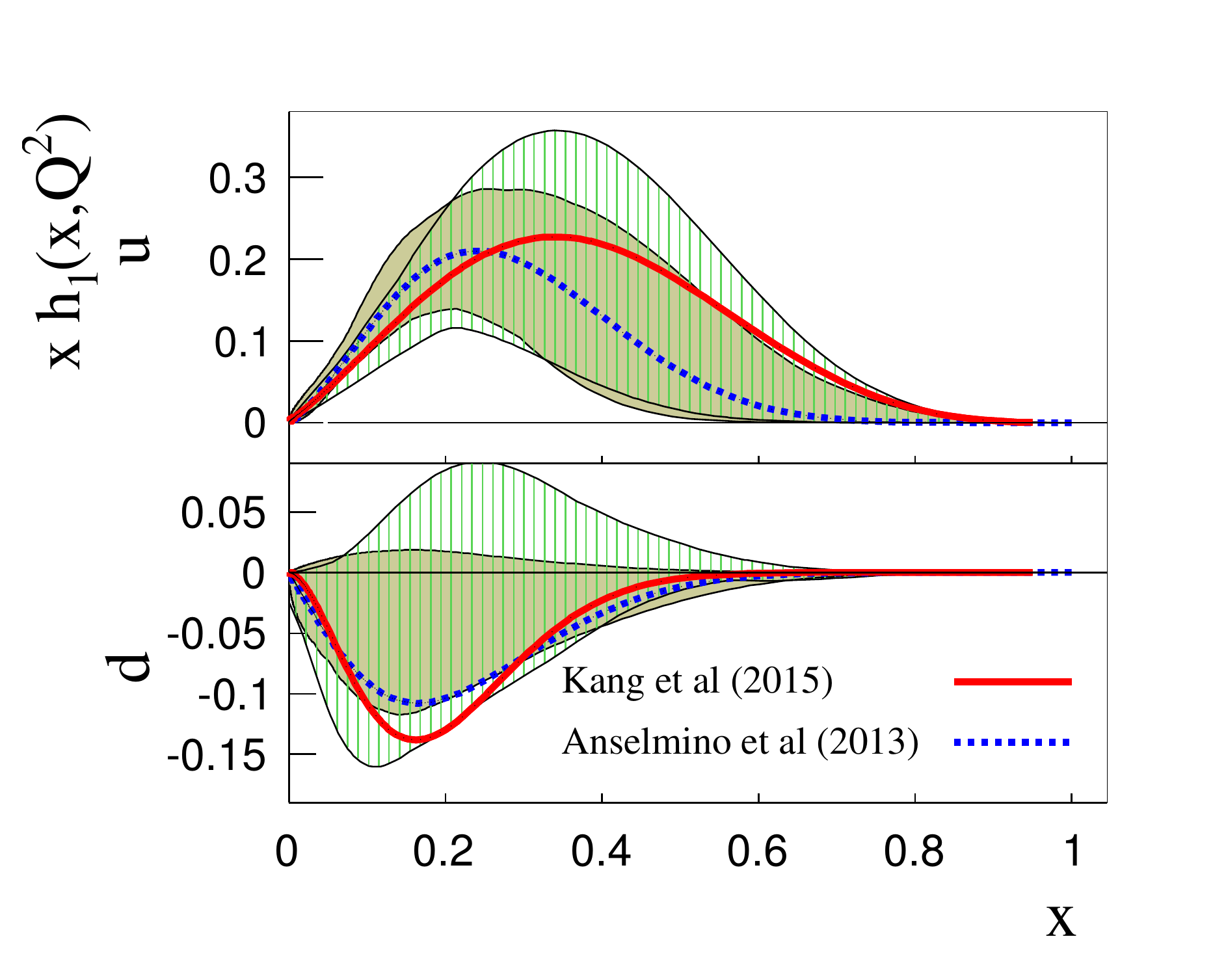}}
   \hspace*{6pt}
   \parbox{2.1in}{\includegraphics[width=2in]{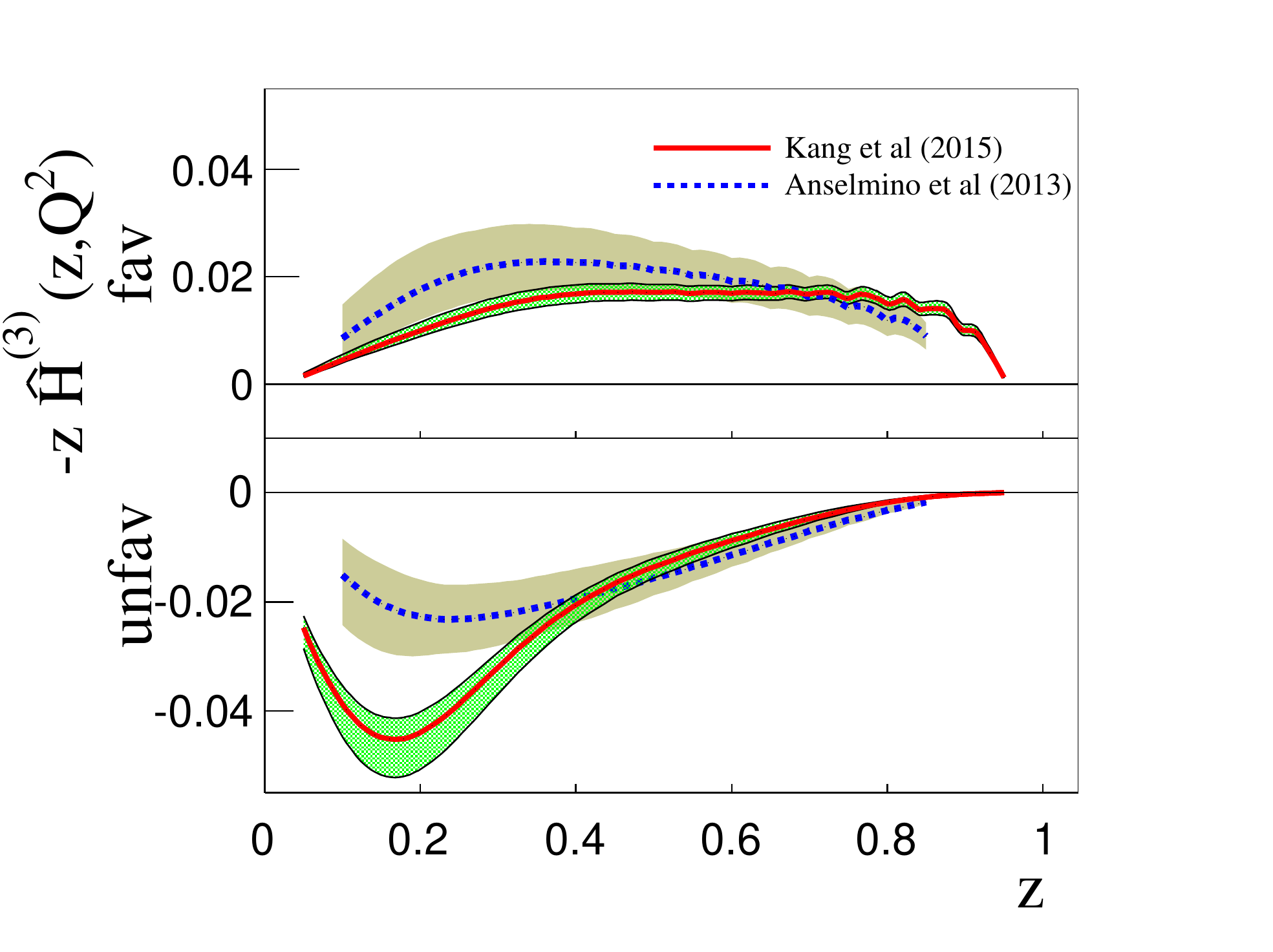}}
 \caption{Transversity distribution for up and down quarks and Collins fragmentation functions 
 for favoured and disfavoured flavour contributions. A comparison of the extractions  
 with~\cite{Kang:2015msa_50} and without~\cite{Anselmino:2013vqa_50} TMD evolution is presented. 
 The band corresponds to the uncertainty of the extraction. 
 }
 \label{fig:transversity}
 \end{center}
  \end{figure}

Interestingly, the BESIII Collaboration~\cite{Ablikim:2015pta} has recently measured 
the same Collins asymmetries observed by BaBar and Belle, but at the
lower energy  $\sqrt{s} = Q = 3.65$
GeV.  Comparisons with the asymmetries obtained at these low $Q^2$ values, much more similar 
to those measured in SIDIS, might help in assessing the importance of TMD evolution effects. 
It is therefore important to check how a model in which the $Q^2$ dependence of the TMD Gaussian 
width is not included~\cite{Anselmino:2015sxa_50} can describe these new sets of measurements, 
and compare these results with the description obtained by using a TMD evolution 
scheme~\cite{Kang:2015msa_50}. 
As in the previous case, there is a striking similarity between the predictions obtained 
with~\cite{Kang:2015msa_50} and without~\cite{Anselmino:2015sxa_50} TMD evolution, which turn out 
to give almost identical asymmetries for remarkably different values of $Q^2$.
In fact, although the typical broadening in $\pt$ of the Collins TMD as $Q^2$ increases is clearly visible, 
the TMD evolution in the low $Q^2$ range is rather slow.

At this stage, it is quite difficult to draw any clear-cut conclusion: 
despite the sizeable difference in $Q^2$ among the different sets of
$e^+e^-$ data
differences among the measured 
asymmetries are mild and 
could be explained solely by the different kinematical configurations and cuts.
Predictions obtained with and without TMD evolution are both in qualitatively 
good agreement with the present BESIII measurements, indicating that present data are little sensitive to 
the $Q^2$ dependence in the transverse momentum distribution.

There are indeed other mechanisms where the Collins TMD can be explored, for example studying the 
transverse momentum distribution of single hadron production in jets off proton-proton 
scattering~\cite{Kang:2017btw_214,Kang:2017glf_220,DAlesio:2017bvu}.
These analyses provide evidence that the Collins function extracted in these independent processes is well consistent
with that extracted by fitting $e^+e^-$ and SIDIS data simultaneously.
Moreover, they confirm that the experimental data presently available do
not show signals of strong evolution effects, and cannot resolve calculations
done with or without TMD evolution.



\newpage 
%


 
\wstoc{Measurements of transverse-momentum distributions in semi-inclusive deep-inelastic scattering}{Gunar Schnell} 
\title{Measurements of transverse-momentum distributions in semi-inclusive deep-inelastic scattering}

\author{Gunar Schnell\footnote{Gratefully acknowledging support by the INT.}}
\index{author}{Schnell, G.}

\address{Department of Theoretical Physics, University of the Basque Country UPV/EHU,\\
48080 Bilbao, Spain; \\
 IKERBASQUE, Basque Foundation for Science,\\
 48013 Bilbao, Spain\\
$^*$E-mail: gunar.schnell@desy.de\\
www.ehu.es}

\begin{abstract}
In this contribution, a brief overview of existing measurements of transverse-momentum distributions in semi-inclusive deep-inelastic scattering from HERMES and COMPASS is given. An even briefer outlook on medium-term possibilities will be provided and how all those lay a path to a future Electron-Ion Collider.
\end{abstract}


\bodymatter

\section{Introduction}\label{schnell:intro}

The past 20 years have seen a remarkable rise of experimental evidence for a rich and non-trivial multi-dimensional structure of hadrons. I will not even attempt to give an equal or even in-depth account of all of the measurements and related observables, this is left rather to reviews.\cite{Avakian:2019drf}
The selection of results will thus be a personal one, though demonstrating a huge success story. 

Two deep-inelastic scattering (DIS) experiments have played particular roles: on one side the HERMES experiment, which used the 27.5~GeV polarized electron/positron beam of HERA in combination with a variety of polarized and unpolarized pure gas targets. HERMES took data from 1995 until the shutdown of HERA in 2007. While personally biased, it is probably fair to claim that HERMES has played a pioneering role in the field of transverse-momentum distributions (TMDs), and in fact is still contributing to the field.\cite{Airapetian:2019mov, Airapetian:2018rlq}
Slightly shifted in time, but still overlapping with HERMES, is the COMPASS experiment, utilizing (mainly) the 160~GeV polarized muon beam of the North Area M2 beam line at CERN. For the study of TMDs, mainly the data taken with polarized NH$_{3}$ or $^{6}$LiD targets has been used. However, a very large data set with an unpolarized liquid-H target was taken in 2016/17. While motivated for the measurement of deeply virtual Compton scattering, it will allow detailed studies of TMDs as well. In contrast to HERMES, the COMPASS has also a fruitful hadron-beam program. In the context of TMDs, the measurements of target-spin asymmetries in the Drell-Yan process stick out.~\cite{Aghasyan:2017jop}

But let us go back twenty years, to more or less the beginning of experimental evidence from semi-inclusive DIS. 
At the 1999 DIS conference, results from HERMES on the azimuthal distribution of hadrons about the virtual-photon direction were presented~\cite{Avakian:1999rr}, which demonstrated the non-trivial role of transverse momentum and correlations with the nucleon (or parton) spin. In this particular case, the azimuthal asymmetries were measured using a longitudinally polarized hydrogen target. There are various aspects worthwhile to recall: (i) It took quite an effort to convince an experimental collaboration that was geared towards measuring the $g_{1}$ structure function and quark helicity distributions to look at such exotic ideas of azimuthal asymmetries arising from spin-orbit correlations. It was well worth the effort and should serve as a prime example of being open to seemingly ``weird'' ideas. (ii) At that time, an idea by Sivers to attribute single-spin asymmetries observed in polarized hadron-hadron collision to correlations between the transverse momentum of quarks and the (transverse) polarization of the nucleon\cite{Sivers:1989cc_218} struggled still with plenty of skepticism and opposition\cite{Collins:1992kk_42}. Most attempts to explain the data were those focused on the so-called Collins effect, an intriguing naive-T-odd spin-momentum correlation in the hadronization process. However, and this brings me to (iii) a courageous---and I would also say seminal---interpretation brought back to the stage the idea of Sivers of a naive-T-odd distribution function.\cite{Brodsky:2002cx}  And while the original observable measured by HERMES was a somewhat messy one, being a subleading-twist asymmetry potentially receiving a multitude of different contributions\cite{Bacchetta:2006tn_42}, it spurred ample theoretical activity with numerous important outcomes. It wasn't until a few years later when finally both the Sivers and Collins effects were both unambiguously shown to exist.\cite{Airapetian:2004tw}

\section{Experimental status of TMDs}

At leading twist, there are eight TMDs required to describe the nucleon structure. The Sivers function is one of them and, as explained above, one that was shown to be non-zero already early on in this young endeavor of measuring TMDs. They all lead to different azimuthal and polarization dependences of the semi-inclusive DIS cross section. By now, all of them have been investigated albeit to different levels of precision and flavor dependence, the latter through variation of targets and final-state hadrons. While not the first one, the {\bf TMD of unpolarized quarks in unpolarized nucleons} profits from the largest data sets on hand. Results on transverse-momentum dependent multiplicities are available for charged pions and kaons from H and D targets\cite{Airapetian:2012ki} as well as for unidentified charged hadrons from a $^{6}$LiD target.\cite{Aghasyan:2017ctw} The second TMD that does not require polarization of either beam or target is the {\bf Boer-Mulders} distribution. It gives rise to a \(\cos2\phi\) modulation\footnote{For more detailed definitions of angles and asymmetries, the reader might consult the {\it Trento Conventions}.\cite{Bacchetta:2004jz}} of the cross section that has been extracted again both by HERMES and COMPASS for the same target and final-state hadron choices as for the multiplicities.\cite{Airapetian:2012yg,Adolph:2014pwc} Clearly non-vanishing modulations are present, however, interpretation of those is hampered by possible contributions from the Cahn effect. This exhausts the list of leading-twist TMDs that can be probed with unpolarized targets. Nevertheless, still using an unpolarized target but at subleading-twist, a \(\cos\phi\) modulation (again related for instance to the Cahn effect) can be measured as well as the \(\sin\phi\) modulation of the beam-helicity asymmetry \(A_{LU}\). The previous has been extracted in parallel to the determination of the \(\cos2\phi\) modulation and found to be non-zero, the latter as well and for a similar set of target and final-hadron choices (though also results for final-state protons and anti-protons have become available recently\cite{Airapetian:2019mov}).
The longitudinal-target-spin asymmetry \(A_{LU}\) provides a way to measure the {\bf worm-gear \(h_{1L}^{\perp}\)} through the  \(\sin2\phi\) modulation, while the  \(\sin\phi\) modulation discussed in the introduction is a subleading-twist effect. While the latter was found to be non-zero, the existing COMPASS and HERMES measurements of the \(\sin2\phi\) modulation are consistent with zero.
Longitudinal beam and target polarization allows for the measurement of the {\bf helicity} TMD. It is the only one besides the unpolarized TMD that does not lead directly to a azimuthal modulation and has been measured both at COMPASS and HERMES with at max weak transverse-momentum dependence. Measurements of the subleading \(\cos\phi\) modulation of the longitudinal double-spin asymmetry are consistent with zero.
Turning to transverse target polarization allows one to assess the four remaining leading-twist TMDs as well as various subleading ones. The most prominent ones are the Sivers and Collins asymmetries arising from the {\bf Sivers} and {\bf transversity} TMDs (the latter in conjunction with the Collins fragmentation function). They have been measured on a hydrogen target at HERMES for pions, charged kaons,\cite{Airapetian:2009ae_224_176,Airapetian:2010ds_100} and lately also for protons and anti-protons (still only available preliminary) and at COMPASS on NH$_{3}$ and $^{6}$LiD targets for unidentified hadrons as well as charged pions and kaons.\cite{Alekseev:2008aa_182,Adolph:2014zba_200}  Both effects do not require polarized beams and are found to be non-zero, though mainly for the  COMPASS  NH$_{3}$  and the HERMES data. Nevertheless,  $^{6}$LiD is important for the flavor separation, which is the prime motivation for additional data taking with such target  in 2021. The {\bf Pretzelosity} TMD leads to another distinctive modulation of the transverse-target asymmetry. Preliminary results are consistent with zero both at COMPASS and HERMES for the same target and final-state combinations used for the Sivers and Collins asymmetries. The same applies to the subleading-twist \(\sin(2\phi - \phi_{S})\) contribution to the transverse-target asymmetry \(A_{UT}\). On the other hand, the subleading-twist  \(\sin\phi_{S}\) modulation, which under certain assumptions is linked to the Collins effect, has been found to be sizable for proton targets at both COMPASS and HERMES for unidentified charged hadrons and charged pions, respectively.
Last but not least, the double-spin asymmetry  \(A_{LT}\) provides a channel to constrain the second {\bf worm-gear} TMD, \(g_{1T}^{\perp}\), and preliminary results point to a non-vanishing effect (in contrast to the vanishing asymmetries of the worm-gear \(h_{1L}^{\perp}\)).

\section{Path to the EIC}

With possibly the exception of the case of the unpolarized TMD, most of the measurements are still very much limited in precision. Moreover, while the beam energies are certainly quite different at COMPASS and HERMES, the ranges in the virtuality of the photon, \(Q^{2}\), is not sufficient to perform precision TMD evolution studies. 
The Jefferson Lab 12 GeV program will result in unprecedented statistical precision for a number of the modulations discussed. However, the beam energy and thus the range in \(Q^{2}\) will be even lower. Some of the observables, e.g., the beam-helicity asymmetry could be analyzed using HERA data at a much larger \(Q^{2}\), though the situation at both H1 and ZEUS is challenging more than a decade after the shut-down of those experiments. Furthermore, those experiments were not really designed for semi-inclusive DIS with limited hadron particle identification. Hadron-hadron collision and electron-positron annihilation will provide additional and complementary data, however, not always as theoretically clean as DIS, limited in precision, or sensitive to fragmentation functions only.
 
Currently, solely the EIC has the potential to unravel in detail the multi-dimensional structure of the nucleon \cite{Accardi:2012qut_5}. The very wide kinematic coverage and a suitable detector will allow detailed semi-inclusive DIS measurements. Polarization and a variety of light ions in the initial state will be mandatory for such program -- all more or less part of the presently foreseen program. And who knows, may one or the other unexpected result trigger a similar wealth of activity as the one presented twenty years ago at DIS'99.



\newpage 
%

\renewcommand*{\FigPath}{./WeekII/09_Vossen/Figs}

\wstoc{Di-hadrons and polarized $\Lambda$'s as novel probes of the nucleon structure}{Anselm Vossen}
\title{Di-hadrons and polarized $\Lambda$'s as novel probes of the nucleon structure}

\author{Anselm Vossen$^*$}
\index{author}{Vossen, A.}

\address{Department of Physics, Duke University and Jefferson Lab,\\
$^*$E-mail: anselm.vossen@duke.edu\\
}

\begin{abstract}
Considering final states with additional degrees of freedom, such as relative orbital angular momentum or polarization, allows a more targeted access to properties of the nucleon structure then would be possible using only single, spinless, hadrons like pions or kaons. This contribution discusses recent results and future opportunities using di-hadron correlations and polarized $\Lambda$ production.
\end{abstract}


\bodymatter
\section{Introduction}
A focus of the EIC physics program is the extraction of the three-dimensional structure of the proton as encoded in TMDs and GPDs~\cite{Accardi:2012qut_7}. At leading twist there are eight TMDs. Beyond, at Twist-3 there are three additional PDFs, if one only considers collinear functions~\cite{}. While the interpretation of TMD PDFs is fairly straightforward in the parton model, they encode spin-momentum correlations of partons in the proton, twist3 pdfs have no partonic interpretation. However, strong theoretic interest, spurred in part by the observation of large signals, led to significant insight into twist3 PDFs over the last decade~\cite{}. They are related to handbag diagrams where additional gluons are exchanged, and thus sensitive to correlations of the quark and gluon fields inside the nucleon. This leads to connections between twist3 pdfs and the force experienced by a struck quark inside the proton while it traverses the gluon background field~\cite{Burkardt:2008ps_49_49}. Via Wandzura-Wilczek relations~\cite{Wandzura:1977qf_52}, some can also be related to integrated TMDs~\cite{}. 
There is a wealth of information on these functions that can be extracted from the SIDIS cross-section.  However, the cross-section also receives contribution from kinematic and higher twist effects, like the Cahn effect. Therefore it is not surprising that it is not always possible to separate the different contributions to the SIDIS cross-section of single, spinless hadron production.  A prominent example for this issue is the extraction of the Boer-Mulders function~\cite{Boer:1997nt_49} from SIDIS which is measured in scatterering off an unpolarized target. By utilizing additional degrees of freedom in the final state, in our case angular momentum, more intricate aspects of the nucleon structure can be disentangled. This contribution will focus on fragmentation functions (FFs) describing the production of final states with angular momentum, either as spin polarization, as in the case of $\Lambda$ production, or relative angular momentum, as in the case of di-hadron production. After a short introduction to these FFs, we will focus on recent and planned measurements. 
Although polarized $\Lambda$ and di-hadron FFs (DiFFs) are important to learn about the proton structure, it is also important to point out, that measuring them is interesting by themselves. They are the crossed channel analogues of PDFs, and follow a similar structure. Therefore, analogous insights into QCD can be gained and spin-orbit correlation in hadronization can be explored. Unlike PDFs, FFs cannot be calculated from the lattice, therefore measurements are indispensable.


\section{Spin-polarized FFs}
TMD PDFs are often organized in a table according to the polarization of the parent nucleon and the probed quark. TMD FFs can be organized analogously as shown in Table~\ref{t:quark_TMDs}~\cite{Metz:2016swz}. Here $q$ is the polarization of the outgoing quark and $H$ is the polarization of the outgoing hadron. For practical purposes, this hadron will usually be a $\Lambda$ due to its self-analyzing decay.

\begin{table}
\tbl{Interpretation of TMD FFs for quarks. The columns indicate the quark polarization --- unpolarized (U), longitudinally polarized (L), transversely polarized (T). The rows indicate the hadron polarization.}
{\begin{tabular}{|c|c|c|c|}
\hline 
$\textrm{H} \, \big\backslash \, \textrm{q}$ & $\textrm{U}$ & $\textrm{L}$ & $\phantom{aa} \textrm{T}_{\phantom{M_{M}}}^{\phantom{M^{M}}}$ \\
\hline 
$\textrm{U}$ & $D_1^{h/q}$ & & $ H_1^{\perp \, h/q}$ \\
\hline
$\textrm{L}$ &  & $G_1^{h/q}$ & $\phantom{a} H_{1L}^{\perp \, h/q}$ \\
\hline
$\textrm{T}$ & $\phantom{a} D_{1T}^{\perp \, h/q} \phantom{a}$ & $\phantom{a} G_{1T}^{h/q} \phantom{a}$ & $\phantom{a} H_1^{h/q} \quad H_{1T}^{\perp \, h/q} \phantom{a}$ \\
\hline
\end{tabular}}
\label{t:quark_TMDs}
\end{table}
These function can for example be measured in correlation measurements in $e^+e^-$\cite{Pitonyak:2013dsu} and have probabilistic interpretations analogues to the corresponding PDFs. The spin dependent FFs $G_1^h$ and $H_1^h$ play a role in the longitudinal and transverse spin transfer to the $\Lambda$, whereas the polarizing FF $D_{1T}^\perp$ is thought to play a role in the observation of transversely polarized $\Lambda$'s in $pp$ collisions~(see Refs.~\citenum{Panagiotou:1989sv,Metz:2016swz} and references therein), which helped motivate a  still ongoing research in transverse single spin asymmetries. It is the analogue of the Sivers PDF and, similarly, poses interesting questions about universality~\cite{Boer:2010ya}.

\section{Correlations of pairs of unpolarized hadrons}
When producing two unpolarized hadrons, the resulting system can have relative angular momentum, such that one can assign a polarization similar to the spin polarized final hadron states as well. The final state is often parametrized by the relative momentum vector $\vec{R}=\vec{P}_{h1}-\vec{P}_{h2}$, the momentum sum vector $\vec{P_h}=\vec{P}_{h1}+\vec{P}_{h2}$, the azimuthal angles vs the scattering plane of those vectors in the Breit system~\cite{Bacchetta:2002ux,Bacchetta:2003vn} $\Phi_R$, $\Phi_h$ as well as the decay angle $\theta$ in the two-hadron center-of-mass system.
Note, that an infinite number of angular momentum numbers are possible, leading in principle to a an infinite tower of functions~\cite{Gliske:2014wba}. For the lowest angular momentum states, the functions can be ordered according to table~\ref{tab:diHadFFs}. See Refs.~\citenum{Bacchetta:2002ux,Bacchetta:2003vn} for details on each function. Here we want to highlight just some interesting properties of these functions. Due to the additional degree of freedom, it is possibible to access the transversity $h_1$ in a collinear framework using $H_1^\sphericalangle$ as a spin analyzer. The additional degree of freedom allows a more targeted access to PDFs in the SIDIS cross-section. Two examples are the access to $e$ using $H_1^\sphericalangle$ and access to the Boer-Mulders PDF through $\bar{H}_1^\perp$.
Analogous to fragmentation in jets, in TMD DiFFs the intrinsic transverse momentum of the initial state can decouple from the intrinsic transverse momentum generated in hadronization.
The "worm-gear" DiFF $G_1^\perp$ is also of interest, since it is linked to spin-orbit correlations in hadronization and NJL-model predictions exist that would allow to connect measurements to microscopic dynamics, at least in said model~\cite{Matevosyan:2017alv}.
\begin{table}
\tbl{Interpretation of Di-hadron FFs for quarks. 
The colums indicate the quark polarization --- unpolarized (U), longitudinally polarized (L), transversely polarized (T). 
The rows indicate the polarization of the hadron pair.}
{
\begin{tabular}{|c|c|c|c|}
\hline 
$\textrm{H} \, \big\backslash \, \textrm{q}$ & $\textrm{U}$ & $\textrm{L}$ & $\phantom{aa} \textrm{T}_{\phantom{M_{M}}}^{\phantom{M^{M}}}$ \\
\hline 
$\textrm{U}$ & $D_1^{h_1h_2/q}$ & & $ H_1^{\perp \, h_1h_2/q}$ \\
\hline
$\textrm{L}$ &  &  &  \\
\hline
$\textrm{T}$ &  & $\phantom{a} G_{1}^{\perp\,h_1h_2/q} \phantom{a}$ & $\phantom{a} H_1^{\sphericalangle h_1 h_2/q} \quad \bar{H}_{1}^{\sphericalangle \, h_1h_2/q} \phantom{a}$ \\
\hline
\end{tabular}}

\label{tab:diHadFFs}
\end{table}
\section{Recent Results}
In this section, some recent results will be highlighted, starting with results on $\Lambda$ polarization. To shed more light on this mechanism behind the observation of transversely polarized $\Lambda$'s in unpolarized $pp$ collisions and address questions of universality, measurements in SIDIS and $e^+e^-$ would be important. However, until recently, limits in luminosity and available energy in SIDIS, and the dilutions due to evolution at high $\sqrt{s}$ in the $e^+e^-$ case precluded any precision meausurement. 
This changed with the measurement of the transverse polarization of $\Lambda$'s at Belle~\cite{Guan:2018ckx}. However, the comparison with theory predictions~\cite{Boer:2010ya}, in particular using associated production still shows significant discrepancies, even in the sign, which need to be understood. Measuring $\Lambda$ polarization at the EIC are an exciting prospect. In the meantime, planned measurements at CLAS12~\cite{Vossen:2018nkg} could give a first glimpse at this quantity, albeit limited by the kinematic limits with the given beam energy.

\begin{figure}
    \centering
    \includegraphics[width=0.49\textwidth]{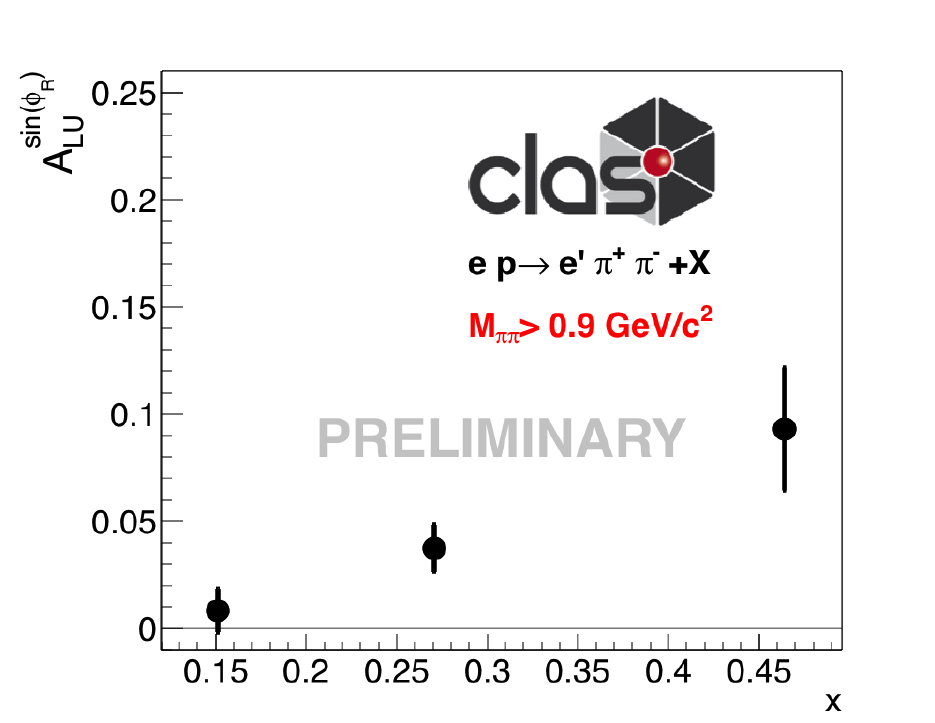}\includegraphics[width=0.49\textwidth]{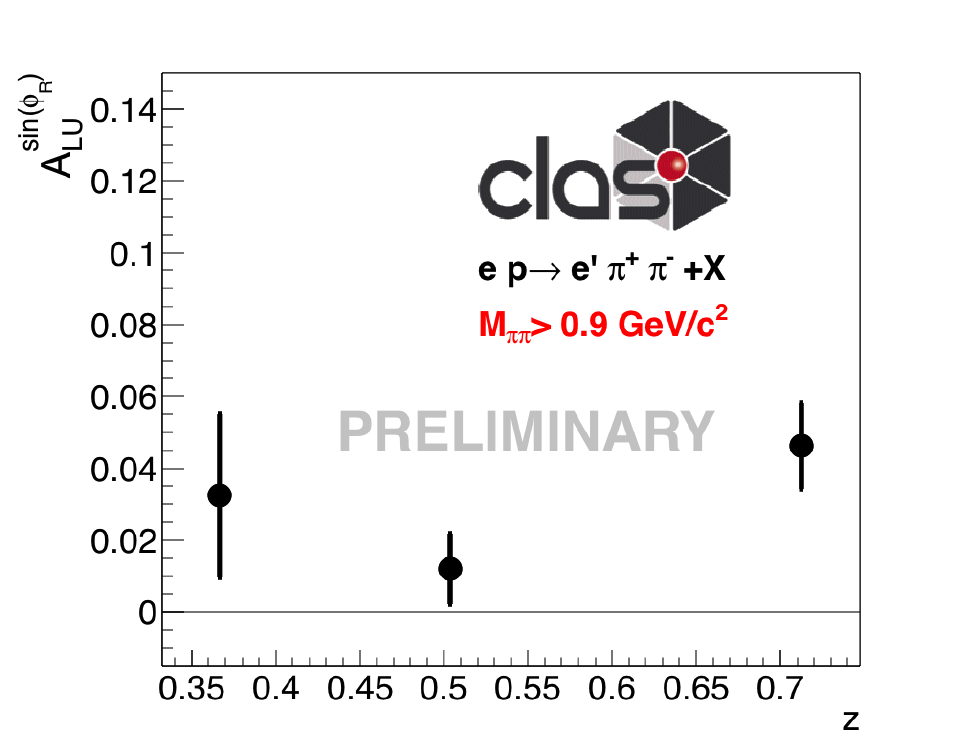}
    \caption{Preliminary results on $A_{LU}^{\sin\Phi_R}$ for $\pi^+\pi^-$ pairs from the CLAS12 experiment. Asymmetries are shown vs $x$ and $z$. A constraint on the invariant mass of the hadron pair of $M_{\pi\pi}>0.9$~GeV/c$^2$ is applied.}
    \label{fig:ex}
\end{figure}

Results on di-hadron correlations into light mesons are available from SIDIS~\cite{Airapetian:2008sk, Adolph:2012nw,Adolph:2014fjw}, $e^+e^-$~\cite{Vossen:2011fk_49} and $pp$~\cite{Adamczyk:2015hri,Adamczyk:2017wld}.
A focus has been the extraction of transversity in a collinear framework. Particularly in $pp$ collisions di-hadron correlations are advantageous compared to single hadron measurements and the STAR results were the first signal of transversity observed in $pp$ and have been included in a first global fit of transversity~\cite{Radici:2018iag_49_49}. 
Another example of the advantages of di-hadron correlations are measurements sensitive to the twist3 PDF $e$~\cite{Bacchetta:2003vn,Efremov:2002qh}.
In the single hadron cross-section, $e$ contributes to a $\sin\Phi_h$ asymmetry in beam-spin asymmetries off an unpolarized target. But there are three more terms contributing that are not necessary small as illustrated by the interpretation of the structure function $F_{LU}^{\sin\Phi_h}$ at twist3~\cite{Mulders:1995dh_49}:
\begin{equation}
   F_{LU}^{\sin\Phi_h}=\frac{2M}{Q}\mathcal{I}\left[-\frac{\vec{k}_T\vec{P}_{h\perp}}{M_h}\left(    xeH_1^\perp + \frac{M_h}{Mz}f_1 \tilde{G}^\perp \right) 
   +\frac{{p}_T\vec{P}_{h\perp}}{M}\left(xg^\perp D_1 + \frac{M_h}{Mz}h_1^\perp \tilde{E} \right) \right].
    \label{eq:singleHadEq}
\end{equation}
Here $\vec{k}_T$ and $\vec{P}_{h\perp}$ are the (intrinsic) transverse momenta of the quark and hadron, which are convoluted (indicated by the function $\mathcal{I}$). The symbols $x$ and $z$ are the usual SIDIS variables, $e$ and $g^\perp$ are the twist3 PDFs, whereas $f_1$ and $h_1^\perp$ are the leading twist unpolarized and Boer-Mulders PDFs. Four fragmentation functions appear, the Collins FF $H_1^\perp$, the unpolarized $D_1$ and two, a-priori unknown twist3 FFs, $\tilde{G}^\perp$ and $\tilde{E}$. 
Comparing with the corresponding asymmetry in two-hadron production~\cite{Bacchetta:2003vn} in eq.~\ref{eq:twoHads} shows the number of terms reduced by half expression becomes colinear. Additionally, the second term containing the twist3 DiFF $\tilde{G}^\sphericalangle$ can potentially be insulated considering the different kinematic dependencies. There are also arguments why this term might be small~\cite{Courtoy:2014ixa}. 
\begin{equation}
F_{LU}^{\sin\Phi_R}=-x\frac{|\vec{R}|\sin\theta}{Q}\left[\frac{M}{m_{hh}}x e^q (x) H_1^{\sphericalangle q}+\frac{1}{z} f_1^q(x) \tilde{G}^{\sphericalangle q} \right]
    \label{eq:twoHads}
\end{equation}

A measurement of this asymmetry has recently been performed on a small initial dataset from CLAS12 and is shown in Fig.~\ref{fig:ex}. Even given the small dataset (about 3\% of the planned final data), a significant asymmetry is observed.

\section{Summary and Outlook}
The expansion of our knowledge of FFs from the production of single, spinless hadrons to polarized hadrons and di-hadron correlations will be important to study spin-orbit correlations in hadronization as well as access the structure of the nucleon. In particular for $\Lambda$ production, the EIC will be the first time, precision measurements will be done in SIDIS.
For di-hadron production, a host of interesting measurements already exist, but further measurements are needed to exploit the full potential of these channels. The EIC can expand this program to TMD di-FFs to access certain aspects of the nucleon structure, such as the Boer-Mulders function, more cleanly and explore FFs that are not allowed for single unpolarized hadron fragmentation, such as the "worm-gear" function $G_1^\perp$.
Given the additional degrees of freedom, DiFFs are also more complex objects than single-hadron FFs. The full kinematic dependence has not been explored yet, in particular the dependence on the decay angle $\theta$ and remains an important task at the EIC. As with the extraction of FFs in general, measurements in $e^+e^-$ and SIDIS data are needed. Therefore the data of Belle II~\cite{Kou:2018nap}, which succeeded Belle with the goal of collecting 50 times the integrated luminosity over approximately one decade, will be an important input to maximize the scientific impact of the EIC.
Further experimental results are also needed from $pp$, in particular to constrain the unpolarized DiFFs of gluons, which are currently a major contributor to the uncertainties of the global fit to transversity~\cite{Radici:2018iag_49_49}.

\vskip 24pt 
\noindent
{\bf Acknowledgements}
This material is based upon work supported by the U.S. Department of Energy, Office of Science, Office of Nuclear Physics under Award Numbers DE-SC0019230 and DE-AC05-06OR23177.



\newpage 
%

\renewcommand*{\FigPath}{./WeekII/10_Seidl/}

\wstoc{TMDs and Fragmentation functions in $e^+e^-$ - and relation to the EIC}{Ralf Seidl}
\title{TMDs and Fragmentation functions in $e^+e^-$ - and relation to the EIC}

\author{Ralf Seidl$^*$}
\index{author}{Seidl, R.}

\address{RIKEN\\
Wako-shi, Saitama-ken 351-0198, Japan\\
$^*$E-mail: rseidl@riken.jp\\
www.riken.jp}

\begin{abstract}
  Fragmentation functions can be cleanly obtained from $e^+e^-$ annihilation. In the recent years various measurements related to unpolarized, polarized and transverse-momentum dependent fragmentation functions have become available from the Belle, BaBar and BESIII experiments. These fragmentation functions are absolutely essential in extracting the spin and flavor structure of the nucleon and will play an important role in fulfilling the scientific goals of the electron-ion collider.  
\end{abstract}


\bodymatter

\section{Introduction}\label{aba:sec1_354}
The formation of final state hadrons from high-energetic, asymptotically free partons is described by fragmentation functions, FFs. Just as parton distribution functions, FFs are nonperturbative objects which cannot be calculated from first principles and even lattice QCD calculations cannot provide FFs. However, FFs are universal such that the same FFs can appear in different processes such as $e^+e^-$ annihilation into hadrons, semi-inclusive deeply inelastic lepton-nucleon scattering or p-p collisions. Therefore the FFs extracted from one process can be used in turn to gain additional knowledge about the nucleon structure. In particular the spin and flavor information can be obtained via FFs and is even essential in accessing chiral-odd distribution functions such as transversity.
\section{Unpolarized fragmentation}
\subsection{Single hadrons}
Single hadron cross sections for pions, kaons \cite{martin,Lees:2013rqd} and protons \cite{Seidl:2015lla} have been available for a while from both Belle and Babar collaborations. They played a very significant role in the recent global FF fits \cite{deFlorian:2017lwf,Bertone:2018ecm,Sato:2016wqj} as most of the other data from $e^+e^-$ annihilation came from much higher c.m.~ energies and therefore the B factory data provided and important lever arm. Together with other new data from different processes they allowed a significant improvement in the determination of unpolarized FFs particularly for light quark flavors and also gluons. 
Additionally also various new measurements for production cross sections of heavier particles, particularly various hyperons and charmed baryons have been extracted at Belle \cite{Niiyama:2017wpp}.
\subsection{Di-hadrons}
One limitation of single hadron measurements in $e^+e^-$ annihilation is that they are only sensitive to the charge square weighted sum over all accessible flavors of FFs. One way to overcome this limitation is to measure two hadrons in the final state in opposite hemispheres where one can expect one hadron to be produced from one quark and the other from the corresponding initial anti-quark. By then studying same and opposite charge sign combinations, one can then select different combinations of favored and disfavored FFs.
These results were extracted from the Belle experiment for pion and kaon combinations \cite{Seidl:2015lla} where one indeed can find different behavior for the two charge combinations. While for example both pion pair combinations are almost comparable at small fractional energies, the differ substantially at large fractional energies where the disfavored FFs should be suppressed. 
When one studies two hadrons within the same hemisphere, they are likely to originate from the same parton. This has been experimentally confirmed. In this regime di-hadron FFs are of interest. Particularly the $z$ and invariant mass dependence of unpolarized di-hadron FFs is important as an unpolarized baseline for the previous measurements related to the interference FF as both are needed for transversity extractions. Belle has provided precise di-hadron cross sections for pion and kaon combinations \cite{Seidl:2017qhp}. The various prominent resonances such as $K_S$, $\rho$,$K^*$, $\Phi$, as well as $D^0$ and others are reflected in the cross sections.
\section{Transverse spin and momentum dependent fragmentation}
\subsection{Chiral-odd fragmentation functions}
The Collins related asymmetries for various charge combinations for pion pairs have been available now for close to a decade from Belle\cite{bellecollins,Seidl:2008xc_200}. Since then, BaBar has confirmed these results and extended them toward extracting the explicit transverse momentum dependence \cite{babarcollins}. In addition, they have also published Collins asymmetries for pion-kaon and kaon pairs \cite{babark}. The BESIII collaboration has also provided Collins asymmetries\cite{bes3}.
In terms of chiral-odd di-hadron FFs, Belle has also provided the corresponding asymmetries related to the interference FF for unlike sign pion pairs as a function of fractional energy and invariant mass \cite{Vossen:2011fk_48}. Both types of FF results have since been used in global fits to extract transversity distributions from SIDIS, pp and $e^+e^-$ data \cite{Kang:2017btw_212,Anselmino:2015sxa_48,Radici:2018iag_48_48}.
\subsection{Unpolarized TMD FFs}
The latest new result is the measurement of the explicit transverse momentum dependence of single charged hadron cross sections by the Belle collaboration \cite{Seidl:2019jei}. Only one pion, kaon or proton was selected inclusively and its transverse momentum was calculated relative to the axis of the event shape variable thrust. As thrust approximates the quark-antiquark axis for a clean two-jet event it serves as a good reference axis. The thrust value however, also affects how collimated an event was and thus the transverse momentum may be correlated with the thrust value. Therefore the cross sections were extracted not only differential in fractional energy and transverse momentum, but also in bins of thrust. The raw hadron yields were corrected for particle misidentification, backgrounds from two-photon, $\Upsilon (4S)$ decays and $\tau$ pair production and momentum and axis smearing. Also the acceptance and reconstruction efficiencies were corrected for as well as the effects of initial state photon radiation. The cross sections either including or excluding weak decays as obtained from MC were both provided for analyzers. In addition to the cross sections, the small transverse momentum region was fitted by a Gaussian distribution. As can be seen in Fig.~\ref{fig}, the widths of pions and kaons behave similarly within systematics while the widths for protons are substantially smaller, probably due to the differences between mesons and baryons. For mesons, the widths increase until about 0.6 before they decrease again to high fractional energies.   
\begin{figure}[h]%
\begin{center}
\includegraphics[width=7.5cm]{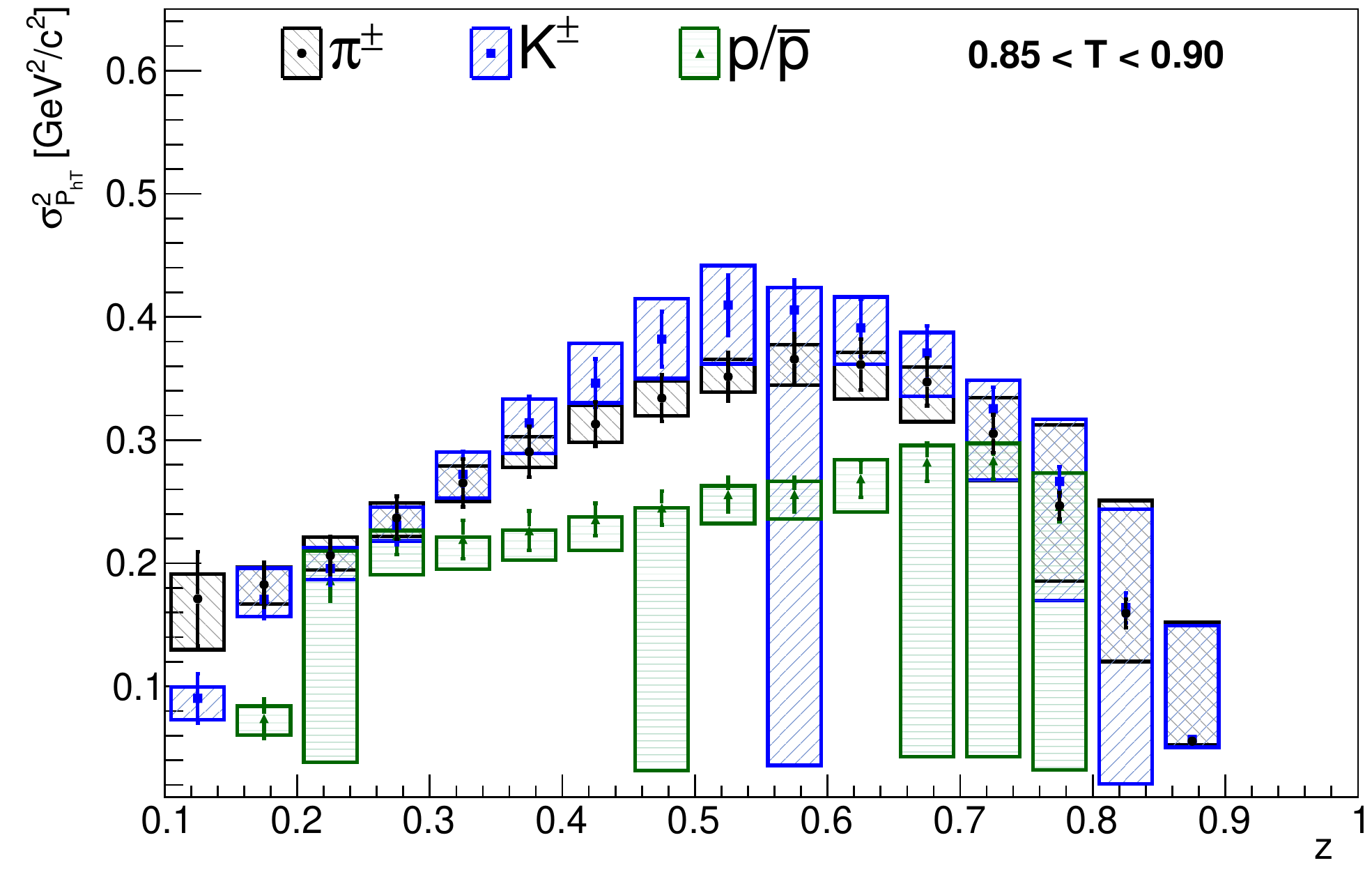}
  \caption{Transverse momentum widths for pions, kaons and protons as a function of fractional energy $z$ for an intermediate thrust bin. }%
  \label{fig}
\end{center}
\end{figure}
\section{Outlook toward the EIC}
Various further FF related results can be obtained from $e^+e^-$ annihilation at Belle and its successor BelleII as well as potentially other colliders envisioned for the future. There is the possibility to obtain FF information for various resonances as well as considerations to actively make use of initial state photon radiation events by detecting the photons and using them to vary the otherwise fixed c.m.~ energy. This would allow to study explicitly the TMD evolution of transverse momentum dependent FFs and help with one of the most pressing theoretical uncertainties in the understanding of TMD variables to date.
Generally, all the already extracted measurements as well as those to come are absolutely essential to successful scientific output of an EIC. Even the simplest flavor dependent extraction of parton helicities will rely on FFs. For nearly all TMD observables at an EIC transverse momentum and/or transverse spin dependent FFs are needed. So, without the rich FF related data provided by $e^+e^-$ colliders, particularly the B-factories, the whole semi-inclusive physics program at an EIC would not be possible.



\newpage 
\wstoc{Power corrections to TMD factorization}{Andrey Tarasov}
\title{Power corrections to TMD factorization}
\author{Andrey Tarasov}
\index{author}{Tarasov, A.}

\address{Physics Department, Brookhaven National Laboratory,\\
Upton, NY 11973, U.S.A.\\
Department of Physics, The Ohio State University,\\
Columbus, OH 43210, U.S.A.\\
E-mail: tarasov.3@osu.edu}

\begin{abstract}
In this paper we discuss calculation of power corrections to the transverse momentum dependent (TMD) factorization which describes production of a particle with small transverse momentum in a hadron-hadron collision. As an example we consider power corrections to the TMD factorization formula for Z-boson production. We demonstrate that in the leading order in $N_c$ power corrections can be expressed in terms of leading-twist TMDs which is a consequence of QCD equations of motion.
\end{abstract}


\bodymatter

\section{Introduction}
In the TMD factorization approach \cite{Collins:2011zzd_189, Collins:1981uw_213, Collins:1984kg_207, Ji:2004wu_201, GarciaEchevarria:2011rb_195,Bauer:2000yr,Bauer:2001yt,Bauer:2002nz,Becher:2010tm,Chiu:2012ir} a differential cross section of particle production in a hadron-hadron collision at small momentum transfer is defined by convolution of two TMD distributions with a cross section of the final particle production by two partons of colliding hadrons:  
\begin{align}
{d\sigma\over  d\eta d^2q_\perp}~=~\sum_f\!\int\! d^2b_\perp e^{i(q,b)_\perp}
{\cal D}_{f/A}(x_A,b_\perp,\eta){\cal D}_{f/B}(x_B,b_\perp,\eta)\sigma(ff\rightarrow H),
\label{TMDf}
\end{align}
where $\eta$ is the rapidity, $q$ is the momentum of the produced particle in the hadron frame, ${\cal D}_{f/A}(x,b_\perp,\eta)$ is the TMD density of  a parton $f$  in hadron $A$, and $\sigma(ff\rightarrow H)$ is the cross section of production of particle $H$ in the scattering of two partons. The factorization formula (\ref{TMDf}) is valid in the limit of an infinitesimally small transverse momentum of the produced particle $q_\perp$. For phenomenological applications it is important to understand how small the transverse momentum should be to make Eq. (\ref{TMDf}) applicable. With increasing transverse momentum $q_\perp$ power corrections to the TMD factorization (\ref{TMDf}) become important. In our recent publication \cite{Balitsky:2017gis_165} we calculated power corrections $\sim {q_\perp^2\over Q^2}$ for Z-boson production which are determined by quark-gluon operators.\footnote{In the leading order Z-boson production was studied in \cite{Landry:1999an,Qiu:2000hf_45,Landry:2002ix,Mantry:2010bi,Su:2014wpa,DAlesio:2014mrz,Catani:2015vma,Bacchetta:2017gcc_45,Scimemi:2017etj}.} Here we review those results. In particular we show that in the leading order in $N_c$ matrix elements of the relevant quark-gluon operators can be expressed in terms of leading-power quark TMDs by QCD equations of motion, see Ref. \cite{Mulders:1995dh_43}.

\section{Z-boson production in TMD factorization}
Let us consider Z-boson production in the Drell-Yan reaction:
\begin{align}
h_A(p_A) + h_B(p_A) \to Z(q) + X \to l_1(k_1) + l_2(k_2) + X,
\end{align}
where $h_{A,B}$ denote the colliding hadrons, and $l_{1, 2}$ the outgoing lepton pair with total momentum $q = k_1 + k_2$.

The relevant term of the Lagrangian for the fermion fields $\psi_i$ describing coupling between fermions  and Z-boson is ($s_W\equiv\sin\theta_W$, $c_W\equiv\cos\theta_W$)
\begin{equation}
{\cal L}_Z~=~ \int\! d^4x~J_\mu Z^\mu(x),~~~~~~~J_\mu~=~-{e\over 2s_Wc_W}\sum_i {\bar \psi}_i\gamma_\mu(g_i^V-g_i^A\gamma_5)\psi_i,
\end{equation}
where sum goes over different types of fermions, and coupling constants $g_i^V=(t_3^L)_i-2q_is^2_W$ and $g_i^A=(t_3^L)_i$ are defined by week isospin $(t_3^L)_i$ of the fermion $i$, see Ref. \cite{Patrignani:2016xqp}. Here we take into account only $u,d,s,c$ quarks and $e, \mu$ leptons. We consider all fermions to be massless.

The differential cross section of Z-boson production with subsequent decay into $e^+e^-$ (or $\mu^+\mu^-$) pair is
\begin{align}
{d\sigma\over dQ^2 dy dq_\perp^2}
~=~{e^2Q^2\over 192s s_W^2c_W^2}
{1-4s_W^2+8s_W^4\over (m_Z^2-Q^2)^2+\Gamma_Z^2m_Z^2}[-W(p_A,p_B,q)],
\label{Zfact}
\end{align}
where we defined the ``hadronic tensor''  $W(p_A,p_B,q)$ as 
\begin{align}
W(p_A,p_B,q)~&\stackrel{\rm def}{=}{1\over (2\pi)^4}\!\int\! d^4x~e^{-iqx}
\langle p_A,p_B|J_\mu(x)J^\mu(0) |p_A,p_B\rangle.
\label{W}
\end{align}

For an arbitrary parton's momentum we use 
Sudakov decomposition $p=\alpha p_1+\beta p_2+p_\perp$, where $p_1$ and $p_2$ are light-like vectors close to $p_A$ and $p_B$, and the notations $x_\bullet\equiv x_\mu p_1^\mu$ and $x_\ast\equiv x_\mu p_2^\mu$ 
for the dimensionless light-cone coordinates ($x_\ast=\sqrt{s\over 2}x_+$ and $x_\bullet=\sqrt{s\over 2}x_-$). We use metric $g^{\mu\nu}~=~(1,-1,-1,-1)$, so 
that $p\cdot q~=~(\alpha_p\beta_q+\alpha_q\beta_p){s\over 2}-(p,q)_\perp$ where $(p,q)_\perp\equiv -p_iq^i$.

To define factorization we introduce rapidity factorization cutoffs $\sigma_a$ and $\sigma_b$ and separate quark and gluon fields into three sectors: 
``projectile'' fields $A_\mu, \psi_A$ 
with $|\beta|<\sigma_a$, 
``target'' fields $B_\mu, \psi_B$ with $|\alpha|<\sigma_b$ and ``central rapidity'' fields $C_\mu,\psi_C$ with $|\alpha|>\sigma_b$ and $|\beta|>\sigma_a$.

With such separation of fields Z-boson with large $Q^2$ is produced in interaction of $C$-fields. Our goal is to integrate over central fields $C$ and present an amplitude of Z-boson production in the factorized form (\ref{TMDf}), i.e. as a product of a projectile matrix element of $A$ fields (TMDs of the projectile) and a target matrix element of $B$ fields (TMDs of the target). To do that we ``freeze'' projectile and target fields and calculate a sum of diagrams of $C$ fields in their external background.

\section{Power counting for background fields}
As we discussed in the previous section, to present the cross-section of $Z$-boson production (\ref{Zfact}) in a factorized form we want to sum diagrams of $C$ fields in the background of fields $A$ and $B$. Since we do not explicitly integrate over fields $A$ and $B$ which define non-perturbative structure of TMDs, we may assume that they satisfy Yang-Mills equations
\begin{align}
iD\!\!\!\!/_A \psi_A~=~0,~~~D_A^\nu A_{\mu\nu}^a~=~g\sum_f\bar\psi^f_A\gamma_\mu t^a\psi^f_A,
\label{YMs}
\end{align}
where $A_{\mu\nu}\equiv\partial_\mu A_\nu-\partial_\nu A_\mu-ig[A_\mu, A_\nu]$, 
$D_A^\mu\equiv (\partial^\mu-ig[A^\mu,)$ and similarly for $B$ fields. 

It is convenient to choose a gauge where $A_\ast=0$ for projectile fields and $B_\bullet=0$ for target fields.
The rotation from
a general gauge to this gauge is performed by a matrix $\Omega(x_\ast,x_\bullet,x_\perp)$ satisfying boundary conditions
\begin{equation}
\Omega(x_\ast,x_\bullet,x_\perp)~\stackrel{x_\ast\rightarrow -\infty}{\rightarrow}~[x_\bullet,-\infty_\bullet]_x^{A_\ast},~~~~~
\Omega(x_\ast,x_\bullet,x_\perp)~\stackrel{x_\bullet\rightarrow -\infty}{\rightarrow}~[x_\ast,-\infty_\ast]_x^{B_\bullet},
\label{Omega}
\end{equation}
where $A_\ast(x_\bullet,x_\perp)$ and  $B_\bullet(x_\ast,x_\perp)$ are projectile and target fields in an arbitrary gauge and $[x_\bullet, y_\bullet]^{A_\ast}_z$ denotes a gauge link constructed from $A$ fields ordered along a light-like line $[x_\bullet, y_\bullet]^{A_\ast}_z = P\exp\big\{\frac{2ig}{s}\int^{x_\bullet}_{y_\bullet} dz_\bullet A_\ast (z_\bullet, z_\perp)\big\}$ and similarly for $[x_\ast, y_\ast]^{B_\bullet}_z$.

We estimate a relative strength of Lorentz components of projectile and target fields in this gauge as 
\begin{align}
&p\!\!\!/_1\psi_A(x_\bullet,x_\perp)~\sim~m_\perp^{5/2}, ~~\gamma_i\psi_A(x_\bullet,x_\perp)~\sim~m_\perp^{3/2},~~p\!\!\!/_2\psi_A(x_\bullet,x_\perp)~\sim~s\sqrt{m_\perp},
\nonumber\\
&A_\bullet(x_\bullet,x_\perp)~\sim~m_\perp^2,~~A_i(x_\bullet,x_\perp)~\sim m_\perp
\label{fildz}
\end{align}
and similarly for $B$ fields. Here $m_\perp$ is a scale of order of $m_N$ or $q_\perp$. For the 
purpose of power counting we will not distinguish between $m_N$ and $q_\perp$ so we
introduce $m_\perp$ which may be of order of $m_N$ or $q_\perp$ depending on matrix element. Using estimations for relative strength of fields' components (\ref{fildz}) we can expand diagrams of $C$ fields calculated in the background of target and projectile in powers of ${q_\perp^2\over Q^2}$ and as a result extract the structure of the leading power correction to the TMD factorization formula (\ref{TMDf}).

\section{Classical equations of motion for $C$-fields}
Resummation of Feynman diagrams for $C$-fields in the background of $A$ and $B$
in the tree approximation reduces to finding fields $C_\mu$ and $\psi_C$ as solutions of Yang-Mills equations for the action
 $S_C~=~S_{\rm QCD}(C+A+B, \psi_C + \psi_A + \psi_B)-S_{\rm QCD}(A, \psi_A)-S_{\rm QCD}(B, \psi_B)$. The solution of equations of motion which we need corresponds to the sum of set of diagrams
in a background field $A + B$ with {\it retarded} Green functions. The solution in terms of retarded Green functions gives fields $C_\mu$ and $\psi_C$ which vanish at $t\rightarrow -\infty$. Thus, we need to solve usual classical YM equations
\begin{equation}
\mathbb{D}^\nu \mathbb{F}^a_{\mu\nu}~=~\sum_fg\bar{\Psi}^f t^a\gamma_\mu \Psi^f,~~~\mathbb{P}\!\!\!/ \Psi^f~=~0,
\label{kleqs}
\end{equation}
where
\begin{align}
&\mathbb{A}_\mu = C_\mu + A_\mu + B_\mu,\ \ \ \Psi^f = \psi^f_C + \psi^f_A + \psi^f_B,
\nonumber\\
&\mathbb{P}_\mu~\equiv~i\partial_\mu+gC_\mu+gA_\mu+gB_\mu, ~~~~
\mathbb{F}_{\mu\nu}~=~\partial_\mu \mathbb{A}_\nu-\mu\leftrightarrow\nu-ig[\mathbb{A}_\mu, \mathbb{A}_\nu],
\end{align}
with boundary conditions
\begin{align}
&\mathbb{A}_\mu(x)\stackrel{x_\ast\rightarrow -\infty}{=}A_\mu(x_\bullet,x_\perp),~~
\Psi(x)\stackrel{x_\ast\rightarrow -\infty}{=}\psi_A(x_\bullet,x_\perp),
\nonumber\\
&\mathbb{A}_\mu(x)\stackrel{x_\bullet\rightarrow -\infty}{=}B_\mu(x_\ast,x_\perp),~~
\Psi(x)\stackrel{x_\bullet\rightarrow -\infty}{=}\psi_B(x_\ast,x_\perp)
\label{inicondi}
\end{align}
following from $C_\mu,\psi_C\stackrel{t\rightarrow -\infty}{\rightarrow} 0$.
These boundary conditions reflect the fact that at $t\rightarrow -\infty$ we have only incoming hadrons with $A$ and $B$ fields.

As discussed above, for our case of particle production with ${q_\perp\over Q}\ll 1$ it is possible to find 
an approximate solution of  (\ref{kleqs}) as a series in this small parameter using relative strength of background fields (\ref{fildz}). 
We will solve Eqs. (\ref{kleqs}) iteratively,  order by order in perturbation theory,  starting from the 
zero-order approximation in the form of the sum of projectile and target fields
\begin{align}
\mathbb{A}_\mu^{[0]}(x)~=~A_\mu(x_\bullet,x_\perp)+B_\mu(x_\ast,x_\perp), ~~\Psi^{[0]}(x)~=~\psi_A(x_\bullet,x_\perp)+\psi_B(x_\ast,x_\perp)
\label{trials}
\end{align}
and improving it by calculation of Feynman diagrams with retarded propagators in the background fields (\ref{trials}).

The first step is the calculation of the linear term for the trial configuration (\ref{trials}). 
The quark part of the linear term has a form
\begin{align}
&L_\psi~\equiv~\mathcal{P}\!\!\!/ \Psi^{[0]}~=~L_\psi^{(0)}+L_\psi^{(1)},~~L_\psi^{(0)}~=~g\gamma^iA_i\psi_B + g\gamma^iB_i\psi_A,
\label{kvlinterm}
\end{align}
where $\mathcal{P}_\bullet~=~i\partial_\bullet+gA_\bullet$, $\mathcal{P}_\ast~=~i\partial_\ast+gB_\ast$, $\mathcal{P}_i~=~i\partial_i+gA_i+gB_i$ are operators in external zero-order fields (\ref{trials}).
Here we denote the order of expansion in the parameter ${m_\perp^2\over s}$ by $(...)^{(n)}$,
and the order of perturbative expansion is labeled by $(...)^{[n]}$ as usual. 
The power-counting estimates for linear term in Eq. (\ref{kvlinterm}) comes from Eq. (\ref{fildz})
in the form $L_\psi^{(0)}~\sim~m_\perp^{5/2}$, $L_\psi^{(1)}~\sim~m^{9/2}_\perp/s$.

With the linear terms (\ref{kvlinterm}), the first terms in our perturbative solution of Eq. (\ref{kleqs}) is
\begin{align}
&\Psi^{[1]}(x)~=~-\!\int\! d^4z~(x|{1\over \mathcal{P}\!\!\!\!/}|z)L_\psi(z)
\label{kvarkfildz}
\end{align}
for quark fields and a similar expression for gluon fields, see Refs.~\cite{Balitsky:2017flc_171,Balitsky:2017gis_165}.

Now we expand the classical  quark fields (\ref{kvarkfildz}) in powers of  ${p_\perp^2\over p_\parallel^2}\sim{m_\perp^2\over s}$, we obtain
\begin{align}
\Psi(x)~=~\Psi^{[0]}(x)+\Psi^{[1]}(x)+\dots~=~\Psi_A^{(0)}+\Psi_B^{(0)}
+\Psi_A^{(1)}+\Psi_B^{(1)}+\dots,
\label{reordering}
\end{align}
where
\begin{align}
\Psi_A^{(0)}~=~\psi_A+\Xi_{2A},~~~~
\Xi_{2A}~=~-{gp\!\!\!/_2\over s}\gamma^iB_i{1\over \alpha+i\epsilon}\psi_A,
\label{fildz0}
\end{align}
and similarly for $B$ fields. The same procedure can be performed for gluon fields as well.

\section{Leading power corrections at  $s\gg Q^2\gg q_\perp^2$\label{sec:lhtc}}
Our method is relevant to calculation of power corrections
at any $s,Q^2\gg q_\perp^2,m^2_N$. However, the expressions can be greatly simplified in the physically interesting case $s\gg Q^2\gg q_\perp^2$ which we consider here.

After integration over central fields in the tree approximation for hadronic tensor (\ref{W}) we obtain
\begin{equation}
\hspace{-1mm}
W(\alpha_q,\beta_q,x_\perp)~\equiv~
{2\over s}\!\int\! \frac{dx_\bullet dx_\ast}{(2\pi)^4} ~e^{-i\alpha_qx_\bullet-i\beta_q x_\ast}
\langle p_A, p_B|{\cal J}_\mu(x_\bullet,x_\ast,x_\perp){\cal J}^\mu(0)|p_A, p_B\rangle,
\nonumber
\end{equation}
where ${\cal J}^\mu~=~{\cal J}^\mu_{AA}+{\cal J}^\mu_{BB}+{\cal J}^\mu_{AB}+{\cal J}^\mu_{BA}$ with
\begin{align}
{\cal J}^\mu_{AB}~=~{e\over 4s_Wc_W}\big[-\bar\Psi_{Au}{\breve \gamma}^\mu\Psi_{Bu}-\bar\Psi_{Ac}{\breve \gamma}^\mu\Psi_{Bc}
+\bar\Psi_{Ad}{\breve \gamma}^\mu\Psi_{Bd}+\bar\Psi_{As}{\breve \gamma}^\mu\Psi_{Bs}\big]
\label{kaljs}
\end{align}
and similarly for ${\cal J}^\mu_{AA}$, ${\cal J}^\mu_{BB}$ and ${\cal J}^\mu_{BA}$. 
Hereafter we use a notation ${\breve \gamma}_\mu\equiv\gamma_\mu(a-\gamma_5)$ where $a$ is one of $a_{u,c}=(1-{8\over 3}s_W^2)$ or $a_{d,s}=(1-{4\over 3}s_W^2)$ depending on quark's flavor.

The quark fields are given by a series in the parameter
${m_\perp^2\over s}$, see Eqs. (\ref{reordering}) and (\ref{fildz0}), where $\Psi$ can be any of $u,d,s$ or $c$ quarks. Accordingly, the currents (\ref{kaljs}) can be expressed as a series in this parameter, e.g.
\begin{align}
{\cal J}^{(0)\mu}_{AB}~=~{e\over 4s_Wc_W}\big[-\bar\Psi^{(0)}_{Au}{\breve \gamma}^\mu\Psi^{(0)}_{Bu}-\bar\Psi^{(0)}_{Ac}{\breve \gamma}^\mu\Psi^{(0)}_{Bc}
+\bar\Psi^{(0)}_{Ad}{\breve \gamma}^\mu\Psi^{(0)}_{Bd}+\bar\Psi^{(0)}_{As}{\breve \gamma}^\mu\Psi^{(0)}_{Bs}\big].
\label{kaljeis}
\end{align}
The leading-power term as well as the leading in $N_c$ power correction come from convolutions ${\cal J}_{AB}^{(0)\mu}(x){\cal J}^{(0)}_{BA\mu}(0)$ and ${\cal J}_{BA}^{(0)\mu}(x){\cal J}^{(0)}_{AB\mu}(0)$. The leading-power term which yields the TMD factorization result (\ref{TMDf}) has a structure $({\bar \psi}_A{\breve \gamma}^\mu\chi_A)({\bar \chi}_B{\breve \gamma}_\mu\psi_B)$, while the first power correction arises from a term $\big(\bar\psi_A(x){\breve \gamma}_\mu\Xi_{1B}(x)\big)\big(\bar\psi_B(0){\breve \gamma}^\mu\Xi_{2A}(0)\big)$.

Using explicit expressions for $\Xi_{A,B}$ fields, see Eq. (\ref{fildz0}), we can calculate leading in $N_c$ power corrections to the TMD factorization:
\begin{align}
&W(p_A,p_B,q)=-{e^2\over 8s_W^2c_W^2N_c}\!\int\! d^2k_\perp\bigg[\Big\{(1+a_u^2)
\Big[1-2{(k,q-k)_\perp\over Q^2}\Big]
\label{resultN}\\
&\times f_{1u}(\alpha_z,k_\perp)\bar{f}_{1u}(\beta_z,q_\perp-k_\perp) + 2(a_u^2-1){k_\perp^2(q-k)_\perp^2\over m^2_NQ^2}
h^\perp_{1u}(\alpha_z,k_\perp){\bar h}^\perp_{1u}(\beta_z,q_\perp-k_\perp) 
\nonumber\\
&+(\alpha_z\leftrightarrow\beta_z)\Big\}
+\Big\{u\leftrightarrow c\Big\}+\Big\{u\leftrightarrow d\Big\}+\Big\{u\leftrightarrow s\Big\}\bigg]\Big(1
+O\big({m_\perp^2\over s}\big)+O\big({1\over N_c}\big)\Big).
\nonumber
\end{align}
where unity in the square brackets in the first line in Eq. (\ref{resultN}) corresponds to the leading term of the TMD factorization while all other terms are leading power corrections suppressed by a ration $\sim {q_\perp^2\over Q^2}$. Note that in the leading order in $N_c$ power corrections are expressed in terms of leading power functions $f_1$ and $h_1^\perp$ which is a consequence of equations of motion for background fields (\ref{YMs}), details of calculation can be found in Refs. \cite{Balitsky:2017flc_171,Balitsky:2017gis_165}. From Eq. (\ref{resultN}) one can estimate power corrections to be of few percent of the leading power result at $q_\perp\sim {1\over 4}Q$ which surprisingly agrees with phenomenological estimations made in Ref. \cite{Scimemi:2017etj} obtained by comparing leading-order fits to experimental data.

\section*{Acknowledgments}
The author is grateful to  S. Dawson,  A. Prokudin, T. Rogers, 
 R. Venugopalan, and A. Vladimirov for valuable discussions. This material is based upon work 
 supported by the U.S. Department of Energy, Office of Science, Office of Nuclear Physics under contract DE-SC0012704 and by U.S. DOE grants DE-FG02-97ER41028 and DE-SC0004286. The work is also supported by the Center for Frontiers in Nuclear Science (CFNS) at the Stony Brook University and Brookhaven National Laboratory.



\newpage 
%


\renewcommand*{\FigPath}{./WeekII/12_Aidala/Figs}

\newcommand{\pout}{\mbox{$p_{\rm{out}}$}\xspace}
\newcommand{\pttrig}{\mbox{$p_T^{\rm{trig}}$}\xspace}
\newcommand{\ptassoc}{\mbox{$p_{T}^{\rm assoc}$}\xspace}
\newcommand{\dphi}{\mbox{$\Delta\phi$}\xspace}
\newcommand{\xe}{\mbox{$x_E$}\xspace}
\newcommand{\Ncoll}{\mbox{$N_{\rm coll}$}\xspace}

\wstoc{Searching for TMD-factorization breaking in $p+p$ and $p$+A collisions:\\
Color interactions in QCD}{Christine A. Aidala}
\title{Searching for TMD-factorization breaking in $p+p$ and $p$+A collisions:\\
Color interactions in QCD }

\author{Christine A. Aidala$^*$}
\index{author}{Aidala, C.}

\address{Physics Department, University of Michigan,\\
Ann Arbor, Michigan 48109, USA\\
$^*$E-mail: caidala@umich.edu}

\begin{abstract}
Dihadron and isolated direct photon-hadron angular correlations have been measured in $p+p$ and $p$+A collisions to investigate possible effects from transverse-momentum-dependent factorization breaking due to color exchange between partons involved in the hard scattering and the proton remnants. The correlations are sensitive to nonperturbative initial-state and final-state transverse momentum $k_T$ and $j_T$ in the azimuthal nearly back-to-back region $\dphi\sim\pi$. In this region, transverse-momentum-dependent evolution can be studied when several different hard scales are measured. To have sensitivity to small transverse momentum scales, nonperturbative momentum widths of \pout, the out-of-plane transverse momentum component perpendicular to the trigger particle, are measured. To quantify the magnitude of any transverse-momentum-dependent factorization breaking effects, calculations will need to be performed for comparison.
\end{abstract}


\bodymatter

\section{Introduction}
One of the frontiers in QCD research is the study of color flow in hadronic interactions. The predicted modified universality of $PT$-odd and $T$-odd correlations in the proton when probed via semi-inclusive deep-inelastic scattering (SIDIS) versus Drell-Yan\cite{Collins:2002kn} is due to different color flow  in these two processes, mediated by gluon exchange between a parton involved in the hard scattering and a proton remnant.  Because of these different color interactions, $PT$-odd and $T$-odd correlations in the proton are predicted to have an opposite sign in the two processes.  The extension of these ideas to hadroproduction of hadrons led to the prediction of TMD-factorization breaking\cite{Bomhof:2006dp,Collins:2007jp,Collins:2007nk,Rogers:2010dm}, with the predicted effect also known as color entanglement.  With gluon exchanges between partons involved in the hard scattering and hadron remnants in both the initial and final state, color flow paths are introduced that cannot be described as flow in the two exchanged gluons separately.  This is a consequence of QCD specifically as a non-Abelian gauge theory, i.e.~due to the fact that gluons themselves carry color.  While breaking of factorization is in fact the rule in QCD interactions, and processes that factorize are the exception, the 2010 prediction by Rogers and Mulders\cite{Rogers:2010dm} is important in that it describes a well-controlled way of breaking factorization, and it in fact goes beyond this, implying novel QCD states of quantum correlated partons across colliding hadrons.  Independent parton distribution functions in the two protons can no longer be used.

In order to search experimentally for the TMD-factorization breaking and color entanglement predicted by Rogers and Mulders, several components are necessary.  
\begin{itemize}
	\item An observable sensitive to a nonperturbative transverse momentum scale as well as a hard interaction scale, such that the TMD-factorization framework would nominally apply.  
	\item Two initial hadrons, such that gluon exchange can occur between a parton involved in the hard scattering and the remnant of the other hadron.
	\item At least one measured final-state hadron, such that gluon exchange can occur between a scattered parton and either remnant.
\end{itemize}

\section{Results}
In proton-proton collisions at PHENIX, such an observable has been developed and measured\cite{Adare:2016bug,Aidala:2018bjf,Aidala:2018eqn}: the out-of-plane momentum component \pout in nearly back-to-back photon-hadron and hadron-hadron production; see Fig.~\ref{fig:kTandpout}a.  The transverse momentum of the "trigger" direct photon or neutral pion, \pttrig, serves as a proxy for the hard scale of the interaction; the out-of-plane momentum component distributions have been measured at several different hard interaction scales.  Figure~\ref{fig:kTandpout}b shows the \pout distributions for both photon-hadron and dihadron correlations, with the shape of the distributions well described by a Gaussian around \pout$=0$, the nonperturbative region.  The distributions transition outward to power-law tails generated by hard (perturbative) gluon radiation.  Note that the curves shown are fits, not phenomenological calculations.  In principle, the measurements can be compared to theoretical calculations in the TMD framework assuming that factorization holds in order to search for TMD-factorization breaking via deviations from the magnitudes, widths, and/or dependence on the hard scale.  The change in the nonperturbative width as a function of the hard scale is of particular interest because the Collins-Soper evolution equation\cite{CS:1981, CS:1982,css_evolution} describing the evolution of TMD distributions with hard scale comes directly from the proof of TMD-factorization\cite{Collins:2012ss}.  While an initial measurement of the Gaussian \pout widths as a function of \pttrig\cite{Adare:2016bug} suggested that they decreased for increasing hard scale, which would be contrary to Collins-Soper-Sterman evolution, the decrease was subsequently understood to be due to the fragmentation kinematics of the away-side particle.  When the fragmentation kinematics are taken into account, the nonperturbative widths instead increase with \pttrig\cite{Aidala:2018bjf}, which is qualitatively similar to what is predicted and observed for Drell-Yan and SIDIS, processes for which TMD-factorization holds.  The Gaussian width of \pout as a function of \pttrig when controlling for the fragmentation kinematics is shown in Fig.~\ref{fig:widthsvspttrigandNcoll}a.  Similar measurements were performed in proton-aluminum and proton-gold collisions\cite{Aidala:2018eqn}, motivated by the idea that the color field of a nuclear remnant might be stronger and lead to larger TMD-factorization breaking effects.  While there is no conclusive evidence for TMD-factorization breaking in these measurements, a broadening of the nonperturbative transverse momentum widths in nuclear collisions was observed, as shown in Fig.~\ref{fig:widthsvspttrigandNcoll}b.

\def\figsubcap#1{\par\noindent\centering\footnotesize(#1)}

\begin{figure}[h]%
	\begin{center}
		\parbox{2.2in}{\includegraphics[width=2.2in]{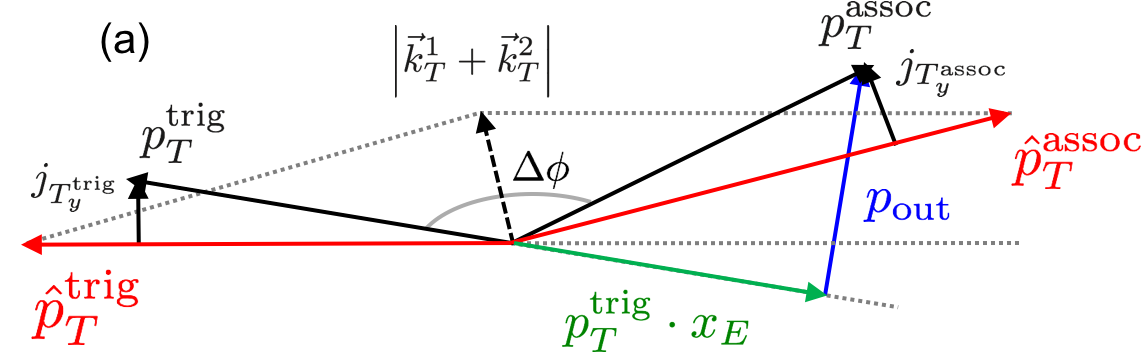}\figsubcap{a}}
		\hspace*{4pt}
		\parbox{2.2in}{\includegraphics[width=2.2in]{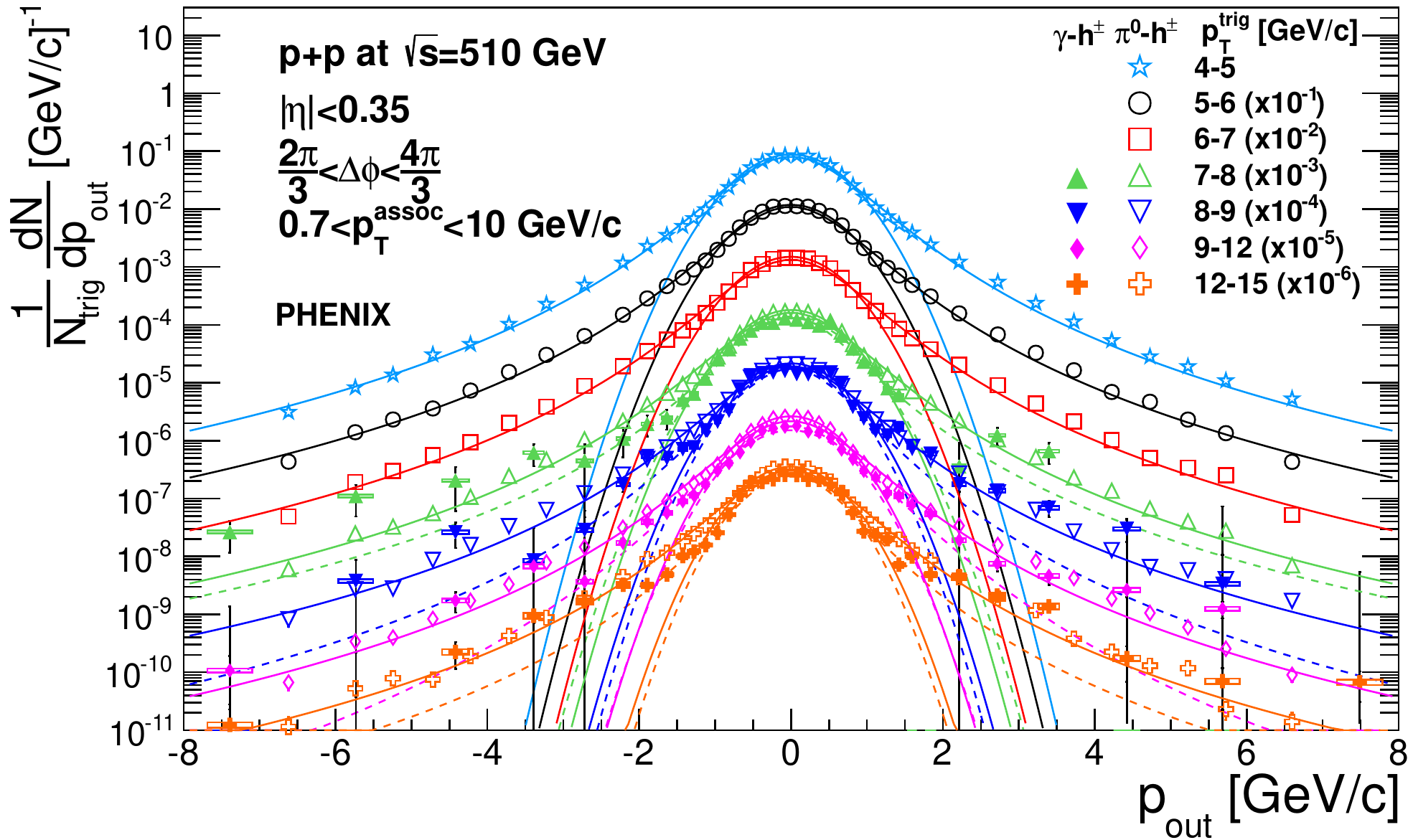}\figsubcap{b}}
		\caption{(a) Hard scattering kinematics of nearly back-to-back dihadrons. Two hard-scattered partons, shown in red, are acoplanar due to the initial-state $\vec{k}_T^1$ and $\vec{k}_T^2$ of the colliding partons. The partons result in a trigger and associated hadrons with \pttrig and \ptassoc. The quantity $x_E$ approximates the momentum fraction $z$ of the final-state away-side hadron\cite{Aidala:2018bjf}.  (b) Per-trigger yields of charged hadrons as a function of \pout. The 
			dihadron and direct photon-hadron distributions are fit with Gaussian functions at 
			small \pout and Kaplan functions over the whole range, showing the 
			transition from nonperturbative behavior generated by initial-state $k_T$ 
			to perturbative behavior generated by hard gluon radiation\cite{Adare:2016bug}.}%
		\label{fig:kTandpout}
	\end{center}
\end{figure}

\def\figsubcap#1{\par\noindent\centering\footnotesize(#1)}
\begin{figure}[h]%
\begin{center}
  \parbox{2.2in}{\includegraphics[width=2.2in]{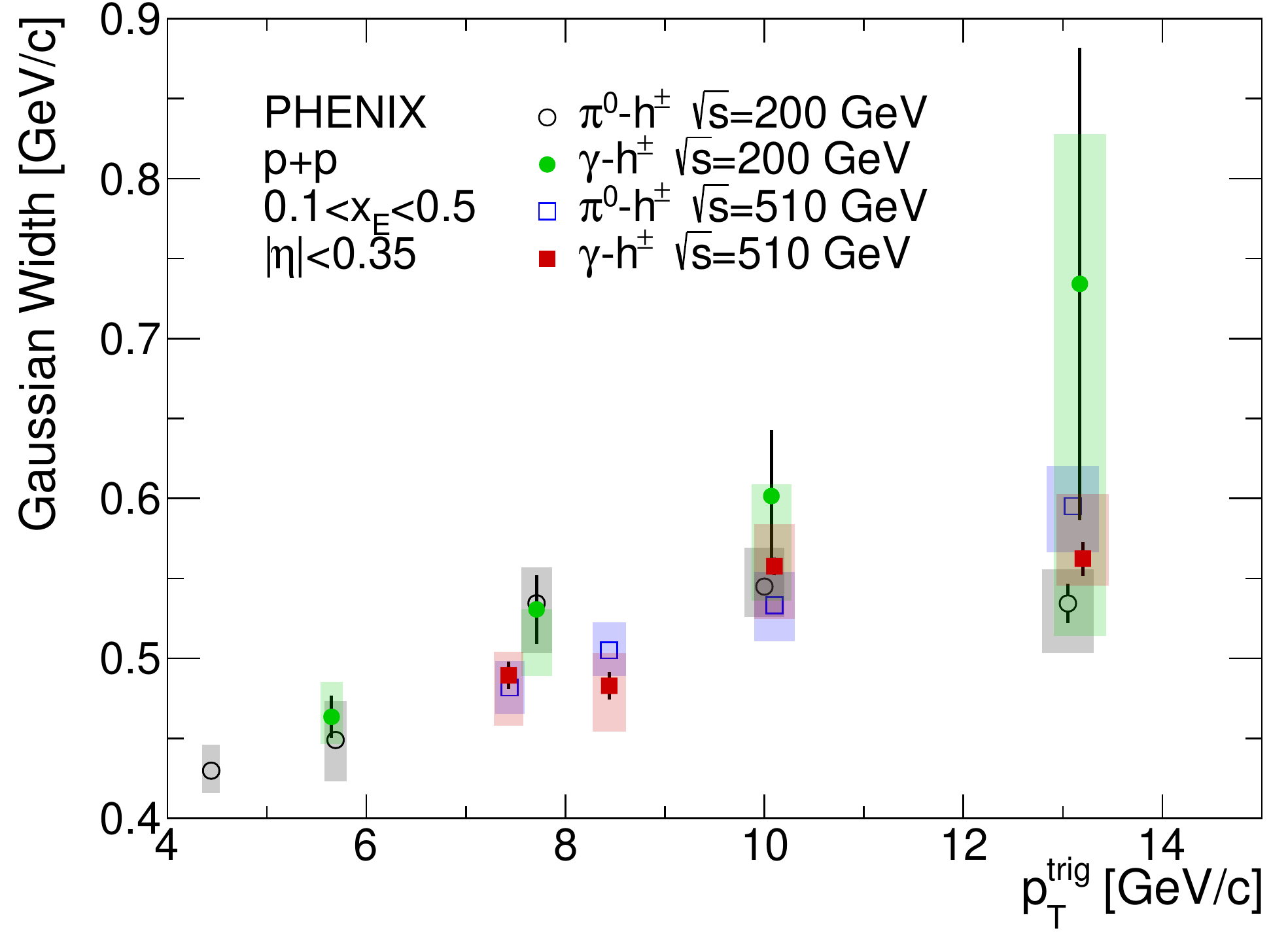}\figsubcap{a}}
  \hspace*{4pt}
  \parbox{2.2in}{\includegraphics[width=2.2in]{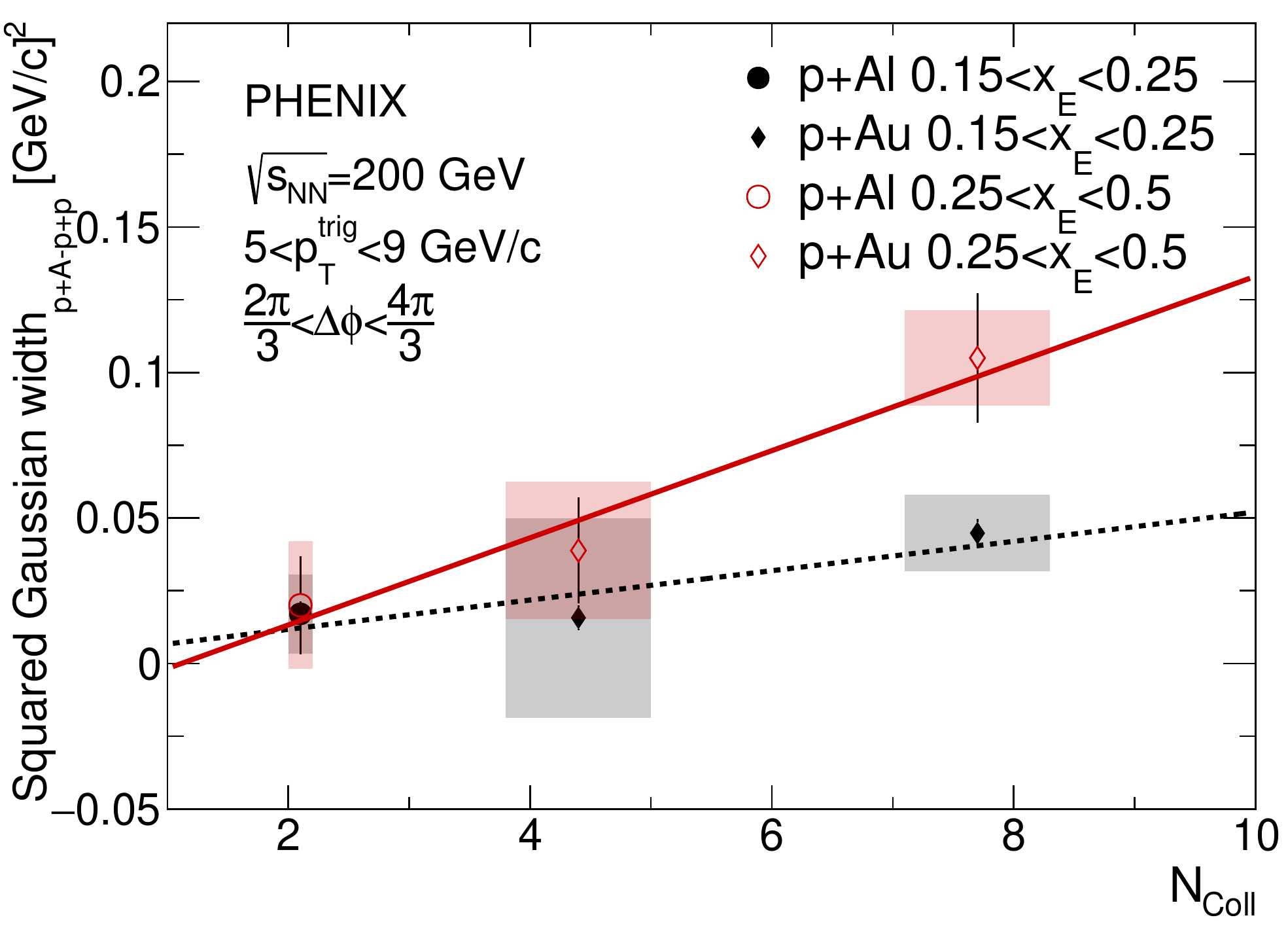}\figsubcap{b}}
  \caption{(a) The Gaussian widths extracted from the \pout distributions in both $\sqrt{s}=$~200 GeV and $\sqrt{s}=$~510 GeV $p+p$ collisions shown as a function of \pttrig\cite{Aidala:2018bjf}. (b) The Gaussian width differences between $p$+A and $p+p$ shown in two \xe bins as a function of the mean number of binary nucleon-nucleon collisions \Ncoll\cite{Aidala:2018eqn}. Linear fits are shown for each \xe bin.}
  \label{fig:widthsvspttrigandNcoll}
\end{center}
\end{figure}

\section{Outlook}
Follow-up measurements are underway in $Z$-hadron, dihadron, and $Z$-jet correlations at the LHCb experiment; $Z$-jet correlations constrain the momentum fractions $x_1$, $x_2$ of the colliding partons and remove sensitivity to any details of hadronization.  Analogous measurements of Drell-Yan and $Z$ dileptons are also planned at LHCb for direct comparison to the $Z$-jet results in overlapping kinematics.  

The community is in the early years of exploring color flow in hadronic scattering processes, and there will remain a great deal to learn from measurements at the future Electron-Ion Collider (EIC).  Color coherence in the form of increased soft hadron production in the region between a high-transverse-momentum hadron or jet and the beam could for example be studied, following earlier measurements in $e^+e^-$ and hadronic collisions (see e.g.~Ref.~\citenum{Chatrchyan:2013fha} and references therein).  In general, color flow is tightly connected to color neutralization and hadronization.  With excellent coverage across the current and target fragmentation regions, at the EIC it may also be possible to examine color connections between these regions by performing more exclusive measurements.   More broadly speaking, studying different hadronization mechanisms, such as vacuum hadronization through string-breaking-type processes, hadronization in a nuclear medium, and recombination of partons moving nearby in phase space in the target region, may provide further insights into color flow and different color-neutralization mechanisms.

\section*{Acknowledgments}
The author acknowledges support from the Office of Nuclear Physics in the Office of Science of the Department of Energy under Grant No.~DE-SC0013393.




\newpage 
%
\newcommand{\sT}{{\scriptscriptstyle T}}
\renewcommand{\d}{\mathrm{d}}
\newcommand{\xB}{x_{\scriptscriptstyle B}}
\renewcommand*{\FigPath}{./WeekII/13_Pisano/}

\wstoc{Gluon TMDs and Opportunities at an EIC}{Cristian Pisano}

\title{Gluon TMDs and Opportunities at an EIC}

\author{Cristian Pisano}
\index{author}{Pisano, C.}

\address{Dipartimento di Fisica, Universit\`a di Cagliari, and INFN, Sezione di Cagliari\\
 Cittadella Universitaria, I-09042 Monserrato (CA), Italy\\
$^*$E-mail: cristian.pisano@ca.infn.it}

\begin{abstract}
 
We show how transverse momentum dependent gluon distributions  could be probed at a future Electron-Ion Collider through the analysis of transverse momentum spectra and azimuthal asymmetries in the inclusive electroproduction of $J/\psi$ and $\Upsilon$ mesons. The maximum values of these asymmetries, obtained in a model-independent way by imposing the positivity bounds of the polarized gluon distributions, suggest the feasibility of the proposed measurements.

\end{abstract}


\bodymatter

\section{Introduction}

Transverse momentum dependent distributions (TMDs) of unpolarized and polarized gluons represent a novel way of exploring the structure of the proton. They encode information on the transverse motion of partons and the correlation between spin and partonic transverse momenta~\cite{Mulders:2000sh_187}, providing a more complete description than the usual parton distribution functions, integrated over transverse momenta. Gluon TMDs are also of great interest because of their intrinsic process dependence, due to their gauge link structure: an unambiguous verification of this property will represent an important confirmation of our actual knowledge of the TMD formalism and nonperturbative QCD in general. 

At present, almost nothing is experimentally known about gluon TMDs. However, many proposals have been put forward to access them, mainly by looking at transverse momentum distributions and azimuthal asymmetries for bound or open heavy-quark pair production, both in lepton-proton and in proton-proton collisions. In particular,  the process $e\, p \to e^\prime \,Q \,\overline{Q}\, X$, with $Q$ being either a charm or a bottom quark, has been proposed as a tool to probe gluon TMDs at the future Electron-Ion Collider (EIC), which will be based in the US. In a series of papers~\cite{Boer:2010zf,Pisano:2013cya,Boer:2016fqd} it has been shown that five different gluon TMDs contribute to the unpolarized and transversely polarized cross sections, through specific azimuthal modulations. The measurements of properly defined azimuthal moments would allow to single out and  extract all the distributions.  Moreover, attention has been paid to the small-$x$ behavior of these observables and to their  process dependence properties, by relating them to analogous observables defined in proton-proton collisions.  

Along the same lines, in the following we describe a complementary analysis for the case in which two heavy quarks produced in a semi-inclusive deep inelastic scattering  process (SIDIS) form a bound state, specifically a $J/\psi$ or a $\Upsilon$ meson~\cite{Bacchetta:2018ivt,Mukherjee:2016qxa_53}.

\section{$J/\psi$ and $\Upsilon$ production in SIDIS}
We consider the process $e\, p^{\uparrow} \to e \,J/\psi \,(\Upsilon)\, X$, where the initial proton is polarized with polarization vector $\bm S$ and $q_\sT \equiv \vert \bm q_\sT\vert$, the quarkonium momentum transverse w.r.t.\ the lepton plane, is small compared to its mass $M_{\cal Q}$ and to the virtuality  $Q$  of the photon exchanged in the reaction.
In a reference frame in which the proton and the photon move along the $\hat z$-axis and the azimuthal angles are measured w.r.t.\ to the lepton plane, the cross section has the following structure
\begin{equation}
\frac{\d\sigma}
{\\d y\,\d\xB\,\d^2\bm{q}_{\sT}} \equiv \d\sigma (\phi_S, \phi_\sT) =    \d\sigma^U(\phi_\sT)  +  \d\sigma^T (\phi_S, \phi_\sT)  \,,
\label{eq:cs}
\end{equation}
where $y$ and $\xB$ are the inelasticity and Bjorken variables, respectively, while $\phi_S$ and $\phi_\sT$ denote the azimuthal angles 
of the transverse vectors $\bm S_\sT$ and $\bm q_\sT$. The unpolarized cross section reads
\begin{align}
\d\sigma^U
  & =  {{\cal N}}\, \bigg [ A^U  f_1^g (x, \bm q_\sT^2 )+  \frac{\bm q_\sT^2}{M_p^2}\, B^U\, h_1^{\perp\, g} (x, \bm q_\sT^2 ) \cos 2 \phi_\sT  \bigg ] \,,
\label{eq:csU}
\end{align}
with ${\cal N} =   4 \pi^2 {{\alpha^2 \alpha_se_Q^2}}/[{ y\,  Q^2\, M_{\cal Q}(M_{\cal Q}^2+Q^2)}]$, $e_Q$ being the fractional electric charge of the quark $Q$, and $M_p$ the proton mass. Moreover,  $f_1^g $ is the unpolarized gluon TMD distribution and  $h_1^{\perp\,g} $ is the distribution of linearly polarized gluons inside an unpolarized proton. The amplitudes $A^U$ and $B^U$ can be calculated within the framework of nonrelativistic QCD (NRQCD). As depicted in Fig.~\ref{fig:fd-lo}, at leading order (LO) in the strong coupling constant $\alpha_s$, the partonic subprocess that contributes to $J/\psi$ production is $\gamma^*g \to Q \overline Q  [ ^{2S+1}L_J^{(8)} ] $. The spectroscopic notation indicates that the $Q \overline Q$ pair forms a bound state with  spin $S$, orbital angular momentum $L$, total angular momentum $J$ and color octet configuration (8). We find that $A^U$ and $B^U$ depend on the relevant long-distance matrix elements (LDMEs) $\langle 0 \vert {\cal O}_8^{J/\psi} (^1 S_0)\vert 0 \rangle$ and $\langle 0 \vert {\cal O}_8^{J/\psi} (^3 P_J)\vert 0 \rangle$, with $J=0,1,2$. Similarly, we obtain
\begin{figure}[t]
\begin{centering}
\includegraphics[trim={2cm 23cm 0cm 3cm},clip,scale=1]{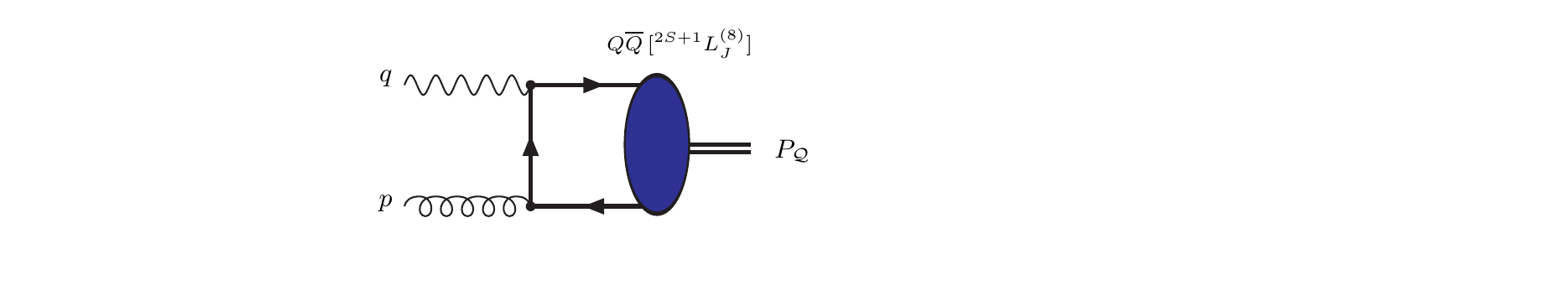}
\par\end{centering}
\caption{LO Feynman diagram for the partonic process $\gamma^* (q) \,+ \, g(p)\to {\cal Q} (P_{\cal Q})$, with ${\cal Q} = J/\psi$ or $\Upsilon$. The crossed diagram, in which the directions of the arrows are reversed, has to be considered as well. The $^1S_0^{(8)}$ and $^3P^{(8)}_{J}$ configurations, with $J=0,1,2$, are the only nonzero contributions. }
\label{fig:fd-lo}
\end{figure}
\begin{align}
\d\sigma^T & =   {\cal N}\,\vert \bm S_\sT\vert \,\frac{\vert\bm q_\sT\vert }{M_p} \bigg \{A^T f_{1T}^{\perp\,g} (x, \bm q_\sT^2 ) \sin(\phi_S -\phi_\sT)  + B ^T \,\left [ h_{1}^{g} (x, \bm q_\sT^2 )\, \sin(\phi_S+\phi_\sT)  \right .
\nonumber \\ 
& \qquad \qquad \qquad \qquad -\left .  \,\frac{\bm q_\sT^2}{2 M_p^2}\, h_{1\sT}^{\perp\,g} (x, \bm q_\sT^2 )\, \sin (\phi_S - 3 \phi_\sT)  \right ] \bigg \} \, ,
\label{eq:csT}
\end{align}
where $f_{1T}^{\perp\,g}$ is the gluon Sivers function, while $h_{1}^{g}$ and $h_{1\sT}^{\perp\,g}$ are chiral-even distributions of linearly polarized gluons inside a transversely polarized proton. They are all $T$-odd, {\it i.e.}\ they would be zero in absence of initial or final state interactions. Therefore, they contribute to the cross section only because the underlying production mechanism is the color-octet one. 

In order to single out the different azimuthal modulations, each one corresponding to a different gluon TMD, we introduce the following azimuthal moments
\begin{align}
A^{W(\phi_S,\phi_\sT)} & \equiv 2\,\frac {\int   \d \phi_S \, \d \phi_\sT \, W(\phi_S,\phi_\sT)\,\d\sigma (\phi_S,\,\phi_\sT)}{\int  \d \phi_S\,  \d \phi_\sT \,\d\sigma (\phi_S, \phi_\sT)}\,.
\label{eq:mom}
\end{align}
Furthermore, by taking $W= \cos2\phi_\sT$, we define $A^{\cos 2\phi_\sT} \equiv 2 \langle \cos 2\phi_\sT \rangle $.  The maximum values of such asymmetries, obtained from the positivity bounds of the TMDs, are presented in Fig.~\ref{fig:cos2phisat} in a kinematic region accessible at the EIC. They turn out to be sizable, but depend very strongly on the specific set of the adopted LDMEs. We point out that a measurement of the ratios
\begin{align}
\frac{A^{\cos 2\phi_\sT }}{A^{\sin(\phi_S+\phi_\sT)}}  \!= \! \frac{\bm q_\sT^2}{M_p^2}\, \frac{h_{1}^{\perp\,g}}{h_{1}^{g}}\,, \,
\frac{A^{\sin(\phi_S-3\phi_\sT)}}{A^{\cos 2\phi_\sT }} \!=\!- \frac{\vert \bm q_\sT\vert}{2 M_p}\, \frac{h_{1 \sT}^{\perp\,g}}{h_{1}^{\perp\,g}}\,, \,
\frac{A^{\sin(\phi_S-3\phi_\sT)}}{A^{\sin(\phi_S+\phi_\sT)}}\!  =\!- \frac{\bm q_\sT^2}{2 M_p^2}\, \frac{h_{1T}^{\perp\,g}}{h_{1}^{g}}
\label{eq:ratioA3}
\end{align}
would directly probe the relative magnitude of the various TMDs, without any dependence on the LDMEs.

\begin{figure}[t]
\begin{centering}
\hspace*{-0.5cm}
\includegraphics[clip,scale=.65]{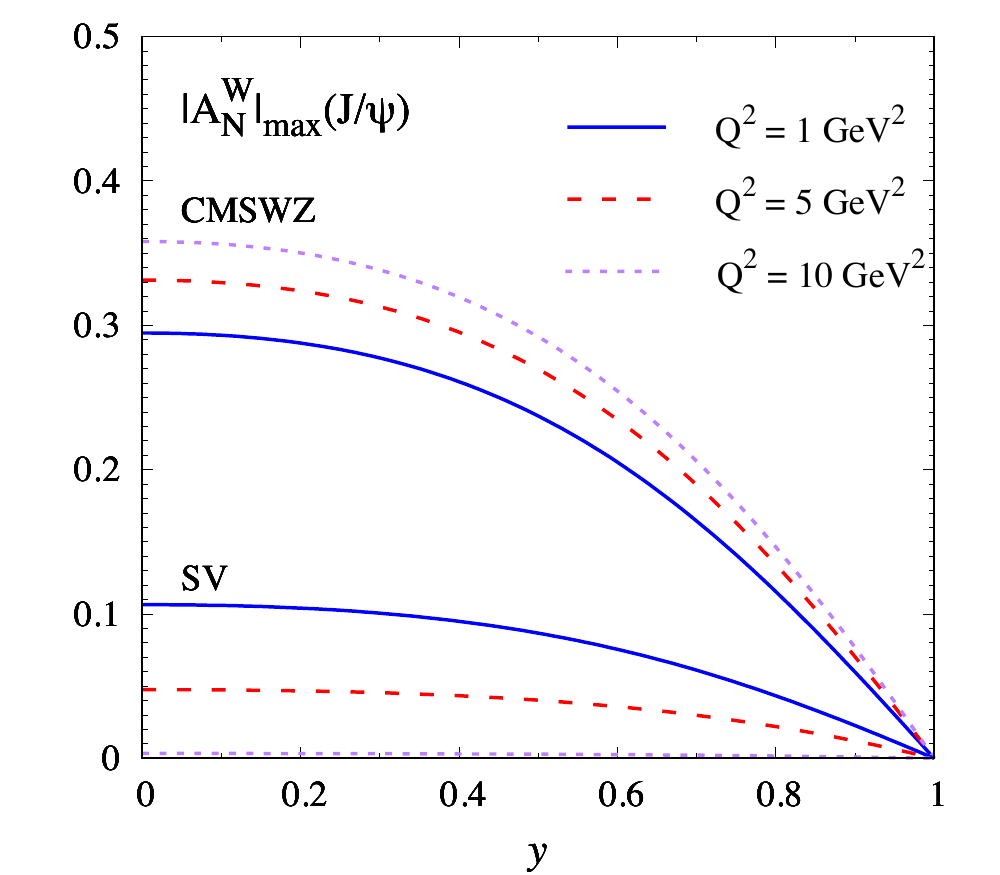}\includegraphics[clip,scale=.65]{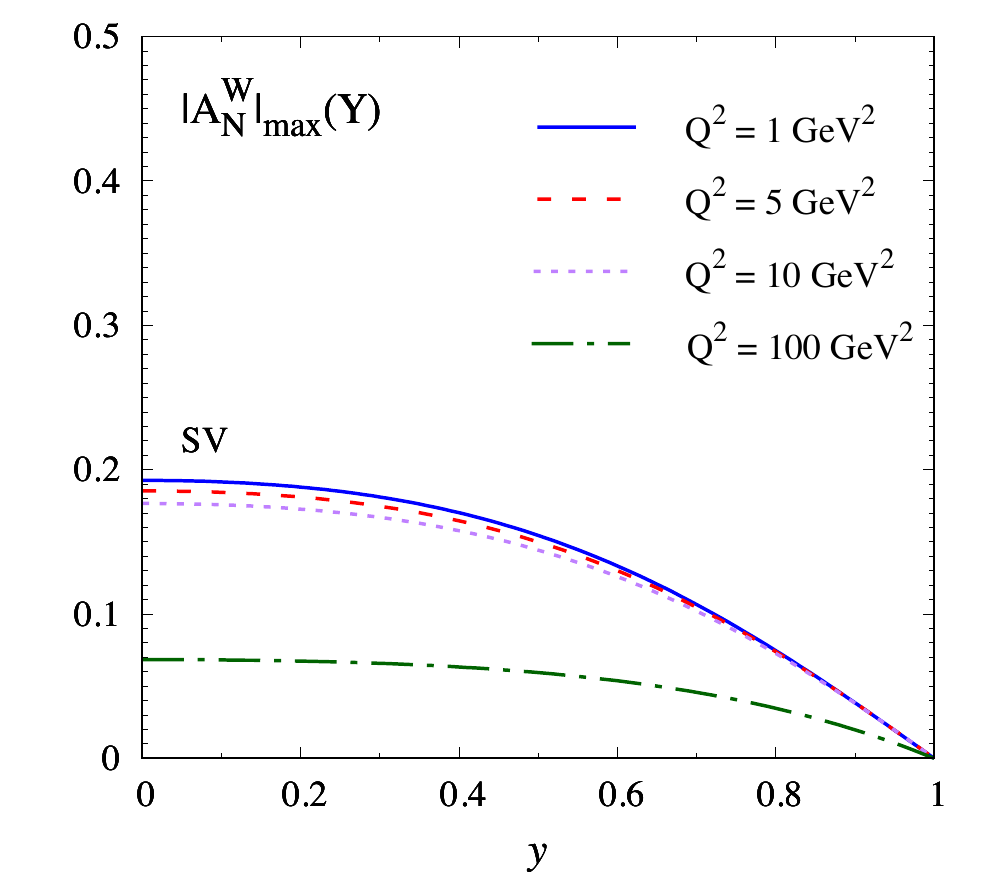}
\par\end{centering}
\caption{ Maximal $\langle \cos 2\phi_\sT\rangle$ and  $A_N^W$  asymmetries, with $W =  \sin (\phi_S +\phi_\sT),\, \sin (\phi_S - 3 \phi_\sT)$, for $J/\psi$ (left panel) and  $\Upsilon$ (right panel) production in SIDIS. The labels SV and CMSWZ refer to the adopted LDME sets~\cite{Chao:2012iv,Sharma:2012dy}.}
\label{fig:cos2phisat} 
\end{figure}
%
\section{Conclusions}

We have presented the LO expressions of the azimuthal asymmetries for $J/\psi$ and $\Upsilon$ production in SIDIS processes, obtained in the framework of NRQCD and assuming TMD factorization. Our results are valid when the transverse momentum of the quarkonium state is much smaller than its invariant mass, and can be used to gather information on gluon TMDs. To this aim, we have proposed the measurement of ratios of asymmetries in which the $\langle0\vert{\cal O}_{8}^{J/\psi}(^{1}S_{0})\vert0\rangle$ and $\langle0\vert{\cal O}_{8}^{J/\psi}(^{3}P_{0})\vert0\rangle$ long-distance matrix elements cancel out. Moreover, these asymmetries can shed light on the mechanism underlying quarkonium production in a totally novel way, for example if we compare them with the ones for $e\, p \to e^\prime \,Q \,\overline{Q}\, X$, at the same hard scale in order not to include TMD evolution~\cite{Bacchetta:2018ivt}. The method consists in the definition of other ratios in which the TMDs cancel out. Hence one would directly probe the two color-octet matrix elements,  which are still poorly known. 

 To conclude, we point out that the study of $J/\psi+$jet production at the EIC has been proposed very recently as an additional tool to access gluon TMDs~\cite{DAlesio:2019qpk}. In this case the soft scale is given by the total transverse momentum of the $J/\psi$+jet pair, required to be much smaller than its invariant mass. The advantage would be that, by varying the invariant mass of the pair, one could access a wide range of scales, having the opportunity to map out TMD evolution.

\bibliographystyle{ws-procs961x669}


\newpage 
%

\newcommand\3[1]{\boldsymbol{#1}}

\renewcommand{\T}[1]{\boldsymbol{#1}_{\text{T}}}
\newcommand{\Tsc}[1]{#1_{\text{T}}}

\def\asy{\mathop{\text{asy}}}

\newcommand\BIBjour[1]{\emph{#1}}
\newcommand\BIBvol[1]{\textbf{#1}}
\newcommand{\bsc}{b_*(b_c(b_T))}
\newcommand{\bsca}{b_{*c_{T}}}
\newcommand{\bct}{b_{c_{T}}}

\newcommand{\mubar}{\bar\mu}

\newcommand{\crd}{\color{red}}
\definecolor{dpmagenta}{rgb}{0.8, 0.0, 0.8}
\newcommand{\cdm}{\textcolor{dpmagenta}}
\newcommand{\lgc}{\textcolor{blue}}
\newcommand{\lgcc}{\textcolor{green}}
\newcommand{\lgco}{\textcolor{orange}}
\newcommand{\no}{\nonumber \\}
\newcommand{\bmax}{b_{\rm max}}
\newcommand{\bmin}{b_{\rm min}}
\newcommand{\barmutwo}{\bar{\mu}^2}
\newcommand{\gp}{\Gamma_{\uparrow/\, A}}
\newcommand{\gm}{\Gamma_{\downarrow/\, A}}
\newcommand{\tf}{\tilde{f}}
\newcommand{\tF}{\tilde{F}}
\newcommand{\tD}{\tilde{D}}
\newcommand{\td}{\tilde{d}}
\renewcommand{\ba}{\begin{array}}
\renewcommand{\ea}{\end{array}}
\newcommand{\beq}{\begin{equation}}
\newcommand{\eeq}{\end{equation}}
\newcommand{\bss}{\odot}
\newcommand\bstar{\3{b}_*}
\newcommand\bstarsc{b_*}
\newcommand\qstar{\Tsc{q}^*}
\newcommand\bone{b_c}
\newcommand\mubstar{\mu_{\bstarsc}}
\newcommand\muQ{\mu_Q}
\newcommand{\Qin}{Q_{\rm init}}
\newcommand{\appor}[1]{{\rm T}_{\rm #1}}
\newcommand\bstarstar{\3{b}_{**}}
\newcommand\mub{\mu_b}
\newcommand{\bT}{\3{{b}_T}}
\newcommand{\cs}[2]{ \Gamma_{#2}(\Tsc{q},Q_{#1}) }
\newcommand{\csb}[2]{ \tilde{\Gamma}^{#2}(\Tsc{b},Q_{#1}) }
\newcommand{\as}[2]{{\rm AY}^{#2}(\Tsc{q},Q_{#1})}
\newcommand{\asnew}[2]{{\rm AY}_{\rm New}^{#2}(\Tsc{q},Q_{#1};\eta,C_5)}
\newcommand{\fixo}[2]{{\rm FO}^{#2}(\Tsc{q},Q_{#1})}
\newcommand{\TT}[2]{W^{#2}(\Tsc{q},Q_{#1})}
\newcommand{\TTnew}[2]{W_{\rm New}^{#2}(\Tsc{q},Q_{#1};\eta,C_5)}
\newcommand{\TTa}[2]{W_{\rm a}^{#2}(\Tsc{q},Q_{#1};\eta,C_5)}
\newcommand{\TTb}[2]{\tilde{W}^{#2}(\Tsc{b},Q_{#1})}
\newcommand{\YY}[2]{Y^{#2}(\Tsc{q},Q_{#1})}
\newcommand{\YYnew}[2]{Y_{\rm New}^{#2}(\Tsc{q},Q_{#1};\eta,C_5)}
\newcommand{\mucirc}{\bar{\mu}}
\newcommand{\mucircm}{\mu_c}
\newcommand{\TTbst}[2]{\tilde{W}^{#2}(\bstar(\T{b}{}),Q_{#1},S)}
\newcommand{\jbar}{\bar{\jmath}}
\newcommand{\TTbtemp}[2]{\tilde{W}^{#2}(\T{b}{},Q_{#1},S)}
\newcommand{\TTbsttemp}[2]{\tilde{W}^{#2}(\bstar(\T{b}{}),Q_{#1},S)}
\newcommand{\TTbstmod}[2]{\tilde{W}^{#2}(\bstar^{\rm mod}(\T{b}{}),Q_{#1},S)}
\newcommand{\TTbc}[2]{\tilde{W}^{#2}({\bm b}_{\rm c}(\T{b}{}),Q_{#1},S)}
\newcommand{\TTbcns}[2]{\tilde{W}^{#2}({b}_{\rm c}(\Tsc{b}{}),Q_{#1})}
\newcommand{\TTbcnsB}[2]{\tilde{W}^{#2}({b}_{\rm c}(\Tsc{B}{}),Q_{#1})}
\newcommand{\TTbcnsmin}[2]{\tilde{W}^{#2}(\bmin',Q_{#1})}
\newcommand{\TTbstc}[2]{\tilde{W}^{#2}(\bstar({\bm b}_{\rm c}(\T{b}{})),Q_{#1},S_A,S_B)}
\newcommand{\arrowcom}[1]{\textcolor{red}{\textbf{$\Longrightarrow$ #1}} \\}
\newcommand{\arrowcomtwo}[1]{\textcolor{blue}{ \textbf{$\Longrightarrow$ #1}} \\}
\newcommand{\arrowcomthree}[1]{\textcolor{magenta}{ \textbf{$\Longrightarrow$ #1}} \\}
\def\??#1{\mbox{}\\\arrowcom{#1}}
\long\def\JC#1{{\color{brown} #1}}
\long\def\TR#1{{\color{blue} #1}}
\def\cc#1{\mbox{}\\\arrowcomthree{#1}}

\def\NS#1{{\color{blue}#1}}
\def\NSX#1{{\color{blue} \sout{#1}}}

\newcommand\pORm[2]{\genfrac{}{}{0pt}{2}{+#1}{-#2}}

\wstoc{Matching collinear and transverse momentum dependent observables in the CSS formalism}{Leonard~Gamberg, Andreas~Metz, Daniel~Pitonyak, Alexei~Prokudin}
\title{Matching collinear and transverse momentum dependent observables in the CSS formalism}
\author{ Leonard~Gamberg$^a$, Andreas~Metz$^b$, Daniel~Pitonyak$^c$, Alexei~Prokudin$^{a,d}$}
\address{$^a$Division of Science, Penn State University Berks,
Reading, Pennsylvania 19610, USA}
\address{$^b$Department of Physics, SERC, Temple University, Philadelphia, PA 19122, USA}
\address{$^c$Department of Physics, Lebanon Valley College,
Annville, PA 17003, USA}
\address{$^d$Theory Center, Jefferson Lab,
Newport News, Virginia 23606, USA}

\index{author}{Gamberg, L.}
\index{author}{Metz, A.}
\index{author}{Pitonyak, D.}
\index{author}{Prokudin, A.}

\begin{abstract}
 We report on the improved Collins-Soper-Sterman (iCSS) formalism
  presented in~\cite{Collins:2016hqq,Gamberg:2017jha}, in particular for the case of polarized observables, such as the Sivers effect in semi-inclusive deep-inelastic scattering. 
   We also outline how this study can be   extended beyond leading order to address the matching of   collinear PDFs to transverse momentum integrated TMDs.

\end{abstract}


\bodymatter

\section{Introduction \label{sec:intro}}
One of the primary goals of transverse-momentum-dependent (TMD) factorization theorems, which rely largely on the work of Collins, Soper, and Sterman (CSS)~\cite{Collins:1984kg_210,Collins:2011zzd_192},
is to describe the cross section as a function of the
transverse momentum  $q_T$
point-by-point, from small $q_T\sim m$ (where $m$ is a typical hadronic mass scale), to large $q_T \sim Q$ ($Q$ is a large momentum which sets the hard scale).
 To achieve this, CSS organized the cross section in an additive form,
\begin{equation} \label{e:SIDIS_WY}
 \Gamma(\T{q},Q) \hspace{-0.05cm}\equiv \hspace{-0.05cm}\frac{d\sigma} {dxdyd\phi_sdzd\phi_h(z^2dq_T^2)}\hspace{-0.05cm} =\hspace{-0.05cm} W(\T{q},Q,S)+Y(\T{q},Q,S)+{\cal O}((m/Q)^{c}) \, ,
\end{equation}
where $\Tsc{q}= |\T{q}|$ and $-Q^2$ are the transverse momentum and virtuality, respectively, of the virtual photon,  $S$ is a 4-vector for the spin of the target and  $\phi_S$ the azimuthal angle of its transverse component, $\phi_h$ is the azimuthal angle of the produced hadron, and $x,\,y,\,z$ are the other standard SIDIS kinematic variables~\cite{Bacchetta:2006tn_46}, and a constant $c>0$; we suppress the $x$ and $z$ dependence in the arguments of $\Gamma$, $W$, and $Y$  for brevity.
 The $W$-term factorizes into TMD parton distribution functions (PDFs) and  fragmentation functions (FFs), and is valid for $q_T\ll Q$.  The $Y$-term  serves  as a correction for larger $q_T$ values and is expressed in terms of collinear PDFs and FFs~\cite{Collins:1984kg_210}. 
 This formalism is  designed to be valid to leading power in $m/Q$
 uniformly in $q_T$, where $m$ is a typical hadron mass scale. The advantages of the $W+Y$ 
decomposition are clearest when this ratio is 
sufficiently small that TMD factorization is valid to good accuracy,
while $m/\Tsc{q}$ is also sufficiently small that collinear
factorization is simultaneously valid. However, at lower phenomenologically interesting values of
$Q$, neither of these ratios is necessarily very small. This issue led the authors of Ref.~\cite{Collins:2016hqq} to develop an improved version of the original CSS $W+Y$ construction (referred  to as iCSS) for the case of the unpolarized SIDIS cross section.
Given the  focus on the 3D structure of hadrons through spin-dependent observables, we extended the iCSS formalism~\cite{Gamberg:2017jha}  to the
the Sivers~\cite{Sivers:1989cc_222} and
Collins~\cite{Collins:1992kk_46} 
transverse single-spin asymmetries (TSSAs),  where,  in particular, we revisited the  relation between the (TMD) Sivers function and the (collinear) 3-parton correlator  Qiu-Sterman function~\cite{Efremov:1981sh,Qiu:1991pp}, in the context of the iCSS formalism.  Here we discuss  the importance of these results with regard to the interpretation of TMDs  and how to extend these results beyond leading order.

\section{TMD Factorization \& Evolution}\label{s:physical}
In the CSS factorization formalism, the TMD evolution of the $W$-term in (\ref{e:SIDIS_WY}) is carried out in   $b_T$-space 
where the Fourier transform (FT) $\tilde{W}(\T{b},Q,S)$ of  $W(\T{q},Q,S)$ is $W(\T{q},Q,S) = \int\! d^2\T{b}/(2\pi)^2 \,e^{i\T{q}\cdot\T{b}}\,\tilde{W}(\T{b},Q,S)$, which can be expanded in the following structures~\cite{Boer:2011xd,Aybat:2011ge_46},
\begin{equation} \label{e:Wb_old}
 \tilde{W}(\T{b},Q,S) = \tilde{W}_{\rm UU}(b_T,Q) - iM_P\,\epsilon^{ij}b_T^iS_T^j\,\tilde{W}_{\rm UT}^{\rm siv}(b_T,Q) + \dots\;,
\end{equation}
with $M_P$ the mass of the target and the epsilon tensor defined with $\epsilon^{12} = 1$. The factorized unpolarized and Sivers structure functions read~\cite{Boer:2011xd,Aybat:2011ge_46,Collins:2011zzd_192},
\begin{align}
  \tilde{W}_{\rm UU}(b_T,Q) &= \sum_j H_j(\mu_Q,Q)\,\tilde{f}_1^{j}(x,b_T;Q^2,\mu_Q)\,\tilde{D}_1^{h/j}(z,b_T;Q^2,\mu_Q), \label{e:WbUU_old}\\
  \tilde{W}_{\rm UT}^{\rm siv}(b_T,Q) &= \sum_j H_j(\mu_Q,Q)\,\tilde{f}_{1T}^{\perp(1)j}(x,b_T;Q^2,\mu_Q)\,\tilde{D}_1^{h/j}(z,b_T;Q^2,\mu_Q)\,,\label{e:WbUT_old}
\end{align}
respectively, where 
 $M_P^2\tilde{f}_{1T}^{\perp(1)j}(x,b_T;Q^2,\mu_Q)\equiv  -\partial\tilde{f}_{1T}^{\perp j}(x,b_T;Q^2,\mu_Q)/ \partial \ln{b_T}$~\cite{Boer:2011xd,Aybat:2011ge_46}.  Following the CSS procedure in~\cite{Collins:2011zzd_192,Collins:2014jpa_162}, the FT TMDs (FFs \& PDFs) are written as solutions of the Collins-Soper and renormalization group equations, where the  $W$-term can be expressed as, 
\begin{equation}
 \tilde{W}(b_T,Q)= \tilde{W}^{\rm OPE}(b_*(b_T),Q)\tilde{W}^{\rm NP}(b_T,Q), 
\end{equation}
and $b_*(b_T) \equiv b_T/ \sqrt{1+b_T^2/b_{max}^2}$, $\mu_{b_*} \equiv C_1/b_*(b_T)\,,$  where $b_{max}$ separates small and large $b_T$, and $C_1$ is a constant chosen to allow for perturbative calculations of $b_*(b_T)$-dependent quantities
without large logarithms~\cite{Collins:2014jpa_162}.  Note that $b_*(b_T)$ freezes at $b_{max}$ when $b_T$ is large so that $b_*(b_T)$ resides in the perturbative region.  Further, since the unpolarized and Sivers $b_T$-space structure functions
are restricted to small $b_T$, they can expand them in an OPE in terms of twist-2 collinear functions~\cite{Boer:2011xd,Aybat:2011ge_46,Collins:2011zzd_192}.

\section{Improved CSS (iCSS)}\label{s:icss}
Owing to the fact that  the integral over all $\T{q}$ of 
$W(\T{q},Q,S)$ vanishes (operationally equivalent to the limit of $\Tsc{b}\to 0$ of $\tilde{W}(\T{b},Q,S)$ in the CSS framework), the integral of 
$\Gamma(\T{q},Q)$ over all $\T{q}$ in Eq.~(\ref{e:SIDIS_WY}) results in a mismatch of orders in $\alpha_S(Q)$ of the leading contributions
 on the l.h.s.~and r.h.s.~of the equation~\cite{Collins:2016hqq}. A similar analysis applies 
 for the Sivers contribution to $W(\T{q},Q,S)$~\cite{Gamberg:2017jha}.  From these results, one observes~\cite{Gamberg:2017jha} that at leading order the integrals over $\T{k}$ 
 of the unpolarized functions 
vanish upon integration over transverse momentum,
\begin{align}
\int\!d^2\T{k}\,f_1^{j}(x,k_T;Q^2,\mu_Q)&=\tilde{f}_1^{j}(x,b_T\to 0;Q^2,\mu_Q)=0 \label{e:f1_col_old}\, ,
\end{align}
and likewise the first moment of the Sivers function,
\begin{align}
\int\!d^2\T{k}\,\frac{k_T^2} {2M_P^2}\,f_{1T}^{\perp j}(x,k_T;Q^2,\mu_Q)
=\tilde{f}_{1T}^{\perp(1)j}(x,b_T\to 0;Q^2,\mu_Q)=0 \label{e:Siv_col_old}\,.
\end{align}
A  consequence of (\ref{e:f1_col_old})--(\ref{e:Siv_col_old}) is that the physical interpretation of integrated TMDs is lost.  For example, it has been stated that the  l.h.s.~of Eq.~(\ref{e:Siv_col_old}) is average transverse momentum of unpolarized quarks in a transversely polarized spin-$\frac{1}{2}$ target~\cite{Boer:1997nt_46,Burkardt:2003yg}.  Clearly, such a statement is not true in the original CSS framework.

In order to explore the limits of this interpretation, as well as  deal with the large logarithms that arise as $b_T\to 0$  which results in  the vanishing of $\tilde{W}(b_T,Q)$  
 the authors of \cite{Collins:2016hqq,Gamberg:2017jha} introduce a cut off $b_T$ at $O(1/Q)$ by replacing $b_T$ with 
$b_c(b_T) = \sqrt{b_T^2+b_{min}^{2}}$, where $b_{min} \equiv b_0/(C_5 Q)$, 
  and where
 $b_0\equiv 2\exp(-\gamma_E)$, and $C_5$ is chosen to fix the   proportionality between $b_c(0)$ and $1/Q$. In terms of $\tilde{W}(\T{b},Q,S)$ this modification is to be understood as
\begin{align}
    \label{e:Wb}
  \tilde{W}(\T{b},b_c(b_T),Q,S) &\equiv \tilde{W}_{\rm UU}(b_c(b_T),Q) - iM_P\,\epsilon^{ij}b_T^iS_T^j\,\tilde{W}_{\rm UT}^{\rm siv}(b_c(b_T),Q) ,
\end{align} 
where now $\bstarsc$ becomes
$b_*(b_c(b_T)) = \sqrt{\left(b_T^2+b_{min}^{2}\right)/ \left(1+b_T^2/b_{max}^2+b_{min}^{2}/b_{max}^2\right)}$
and the renormalization scale is now,  $\bar{\mu} \equiv C_1/b_*(b_c(b_T))$. A new $W$-term that respects factorization is established~\cite{Collins:2016hqq} and 
 the $q_T$-differential cross section (\ref{e:SIDIS_WY}) is modified accordingly (see Ref.~\cite{Collins:2016hqq} for details).

These improvements formally resolve the problems of implementing  the collinear 
limit, that is, integrating $\Gamma(\T{q},Q)$ and
the TMD functions, over transverse momentum, while respecting TMD factorization. For example, upon $q_T$ integration 
we  recover the expected leading order relations between TMD and collinear quantities~\cite{Collins:2016hqq,Gamberg:2017jha}. 
In terms of the momentum-space functions we find~\cite{Gamberg:2017jha}, 
\begin{align}
  \int d^2\T{k}\, f_1^{j}(x,k_T;Q^2,\mu_Q;C_5)&=\tilde{f}_1^{j}(x,b_{min};Q^2,\mu_Q)\nonumber \\
  &=f_1^{j}(x;\mu_c) + {\cal O}(\alpha_s(Q)) + {\cal O}((m/Q)^p) \label{e:unpol_col} 
  \end{align} 
  \begin{align}
  \int d^2\T{k}\,\frac{k_T^2} {2M_P^2}\,f_{1T}^{\perp j}(x,k_T;Q^2,\mu_Q;C_5)
    &=\tilde{f}_{1T}^{\perp(1)j}(x,b_{min};Q^2,\mu_Q)
    \nonumber \\
  &=\frac{-T^j_F(x,x;\mu_c)} {2M_P}+ {\cal O}(\alpha_s(Q)) + {\cal O}((m/Q)^{p'}) \label{e:Siv_col}. 
\end{align}
Note, due to the $b_T \to b_c(b_T)$ modification, the above integrals on the l.h.s. are UV finite,  and for $Q\gg m$ renormalized collinear functions on the  r.h.s.
In particular, the operator defining the TMD $f_{1T}^{\perp j}(x,k_T;Q^2,\mu_Q;C_5)$ on the l.h.s.~of (\ref{e:Siv_col}) includes a UV renormalization factor and a soft factor, along with non-light-like Wilson lines (see, e.g., Refs.~\cite{Collins:2011zzd_192,Aybat:2011ge_46}).  
Further, an important consequence of Eqs.~(\ref{e:Siv_col}) is that the ``na\"{i}ve'' operator definition interpretation of TMDs is restored at LO.  For example, one can determine the average transverse momentum of unpolarized quarks in a transversely polarized spin-$\frac{1}{2}$ target according to~\cite{Boer:1997nt_46,Burkardt:2003yg}
$\langle k_T^i (x;\mu) \rangle_{UT} =  
\frac{1} {2}\epsilon^{ij}S_T^j\,T_F(x,x;\mu)\,$.
 Therefore, it is the Qiu-Sterman function which fundamentally is related to average transverse momentum, and, due to Eq.~(\ref{e:Siv_col}), the first $k_T$-moment of the Sivers function (within the iCSS formalism) retains this interpretation at LO.  
Note that the interpretation of the Qiu-Sterman function 
 is compatible with the understanding of the average transverse force acting on quarks in a transversely polarized spin-$\frac{1}{2}$ target~\cite{Burkardt:2008ps_46_46}.  Moreover, relation (\ref{e:Siv_col}) made it possible to connect TSSAs in different processes (e.g., the Sivers effect in SIDIS and $A_N$ in proton-proton collisions) and has been used routinely in phenomenology (see, e.g., Refs.~\cite{Metz:2012ui,Kang:2012xf,Gamberg:2013kla}).  The incorporation of evolution in the TMD correlator through the original CSS formalism breaks the ``na\"{i}ve'' relations between TMDs and collinear functions (see Eqs.~(\ref{e:f1_col_old})--(\ref{e:Siv_col_old})).  Nevertheless, the modifications implemented in the iCSS framework allow one to preserve these identities at LO.

\section{iCSS and NLO Corrections}\label{sec:concl}
While Eqs.~(\ref{e:unpol_col})-(\ref{e:Siv_col}) imply a scale dependent, renormalized connection between TMDs and collinear PDFs beyond leading order through the explicit dependence on the scale $\mu_c$,  in order to establish a rigorous connection of TMDs and collinear PDFs beyond leading order, it is  necessary to carry out the transverse momentum integration of the TMD factorized cross sections beyond leading order in $\alpha_s$ and $q_T\sim Q$.  
Assuming that such a connection can be established, one can unambiguously  address the matching of TMDs to collinear PDFs which in principle relies on the matching of TMD and collinear factorization~\cite{Collins:2016hqq}.  

It is tempting, however, to explore the matching of TMDs to collinear PDFs beyond leading order  by considering the NLO corrections to the  FT TMDs, and in turn taking the collinear limit as has been carried out in the SCET, and background field factorization frameworks~\cite{Scimemi:2019gge}, 
but here in the iCSS framework.  Using the universal definition of the CSS TMDs~\cite{Collins:2017oxh} 
 and with the improvements outlined here~\cite{Collins:2016hqq,Gamberg:2017jha} we can rigorously explore this limit at NLO.  In this regard, consider the FT TMD~\cite{Gamberg:2017jha},
\begin{align}
\tilde{f}_1^{j}(x,b_c(b_T);Q^2,\mu_Q) = 
&\, \tilde{C}^{\rm pdf}_{j/j'}(x/\hat{x},\bsca;\bar{\mu}^2,\bar{\mu},\alpha_s(\bar{\mu}))\otimes f_1^{j'}(\hat{x};\bar{\mu}) \nonumber\\ &\times e^{\tilde{K}(\bsca;\bar{\mu})\ln\left(\frac{\sqrt{\xi}} {\bar{\mu}}\right)+\int_{\bar{\mu}}^{\mu_Q} \frac{d\mu^{\prime}}{\mu^{\prime}}
\left[\gamma(\alpha_s(\mu^\prime);1)-\ln\left(\frac{\sqrt{\xi}} {\mu^{\prime}}\right)\gamma_K(\alpha_s(\mu^\prime))\right]}
\nonumber\\ &\times
e^{-g_{\rm pdf}(x,\bct;Q_0,b_{max})-g_K(b_c(b_T);b_{max})\ln\left(\frac{Q}{Q_0}\right)} \,,
\label{CFT}
\end{align}
where $\bsca\equiv\bsc$, $\bct\equiv b_c(b_T)$, and $\tilde{C}^{\rm pdf}_{j/j'}(x/\hat{x},\bsca;\bar{\mu}^2,\bar{\mu},\alpha_s(\bar{\mu}))\otimes f_1^{j'}(\hat{x};\bar{\mu})$ is shorthand notation for the small $b_T$ OPE convolution integral over $\hat{x}$. 

Since the l.h.s.~of Eq.~\ref{CFT} is independent of the renormalization scale $\mubar$, we can derive at NLO in the strong coupling the DGLAP equation by taking the derivative  with respect to the intrinsic scale $\mubar$ in the limit $\bsca\rightarrow b_{min}$ ($\mubar\rightarrow \mucircm $); we obtain
\begin{equation}
\frac{df_1^{j}(x,\bar{\mu})}{d\ln\barmutwo}\Bigg|_{\mubar\rightarrow \mucircm}\hspace{-0.3cm}=
\frac{\alpha_s(\mucircm)}{ 2\pi} \left[\sum_{j'}P_{jj'}(x/\hat{x})\otimes f_1^{j'}(x,\mucircm)
+ P_{jg}(x/\hat{x})\otimes f_1^{g}(x,\mucircm)\right] ,
\end{equation}
in a scheme where $C_5=1$ and $b_0=C_1$, where $P_{jj'}$ and $P_{jg}$ are the quark-quark and quark-gluon splitting functions to order $\alpha_s$.  It is worth mentioning that this procedure implies a specific normalization for matching the collinear PDFs and the TMDs which is fixed by TMD factorization.   A similar analysis 
for the first moment of 
the Sivers function in the iCSS framework is under consideration.  Such a matching of collinear PDFs and TMDs from the iCSS framework implies a more general matching
at the level of the cross section, 
 as discussed in Ref.~\cite{Collins:2016hqq}.

\section{Conclusions}
We have reviewed  the 
iCSS factorization framework   presented in~\cite{Collins:2016hqq,Gamberg:2017jha}, in particular for the case of transversely polarized observables, such as the Sivers effect. We  demonstrate how one recovers the expected leading-order relation between the TMD Sivers function and the collinear twist-3 Qiu-Sterman function.  This allows one to obtain the collinear twist-3 cross section from a (weighted) $q_T$-integral of the differential cross section.  
We have also outlined how this study can be  extended beyond leading order to  address the matching of  collinear PDFs to transverse momentum integrated TMDs.  A fully consistent analysis 
will require matching the $\Tsc{q}$ integrated cross section in SIDIS, Drell Yan, and $e^+\, e^-$ to their collinear factorized forms beyond leading order.  Clarifying the matching of TMD and collinear observables is a high priority for  theory and
phenomenology~\cite{Ji:2006ub,Bacchetta:2008xw,Accardi:2012qut_6} as it relates to global fits of 
TMD and collinear transverse-spin observables~\cite{Metz:2012ui,Kang:2012xf,Gamberg:2013kla,Gamberg:2017gle_46}. A future Electron Ion Collider will help to provide precise data to test these assertions.
  
  \vspace{0.3cm}
  \noindent
 {\bf {Acknowledgments}:}
 This work is supported by the U.S. Department of Energy, 
 Office of Nuclear Physics under Award No. DE-FG02-07ER41460 (L.G.), the DOE Contract No.~DE-AC05-06OR23177 (A.P.), under which Jefferson Science Associates, LLC operates Jefferson Lab; and by the National Science Foundation under Contract No. PHY-1516088 (A.M.), No. PHY-1623454 (A.P.), and within the framework of the TMD Topical Collaboration.



\newpage 
%
\renewcommand*{\FigPath}{./WeekII/15_Radici}
\wstoc{Transversity distribution and its extraction}{Marco Radici} 

\def\figsubcap#1{\par\noindent\centering\footnotesize(#1)}
 

\title{Transversity distribution and its extraction}

\author{M. Radici}
\index{author}{Radici, M.}

\address{INFN - Sezione di Pavia \\
Pavia, I-27100, Italy\\
E-mail: marco.radici@pv.infn.it \\
www.pv.infn.it}

\begin{abstract}
We describe the extraction of the transversity distribution involving a global analysis of pion-pair production in deep-inelastic scattering and in proton-proton collisions with one transversely polarized proton. The extraction relies on di-hadron fragmentation functions, which are taken from the analysis of the corresponding pion-pair production in electron-positron annihilation. We discuss the impact of {\tt COMPASS} pseudodata from a future measurement of deep-inelastic scattering on transversely polarized deuterons. We discuss the potential incompatibility with lattice about the tensor charge, which is an important ingredient in the exploration of possible signals of new physics.
\end{abstract}


\bodymatter

\vspace{-0.02cm}

\section{Introduction}
\label{s:intro}

We access the chiral-odd transversity $h_1$ at leading twist by considering the semi-inclusive production of two hadrons with small invariant mass. The chiral-odd partner is represented by the di-hadron fragmentation function $H_1^\sphericalangle$, which encodes the correlation between the transverse polarization of the fragmenting quark and the transverse relative momentum of the two detected hadrons. The main advantage of this strategy is the possibility of working in the standard framework of collinear factorization; appropriate factorization theorems ensure that the simple product $h_1 H_1^\sphericalangle$ is the universal elementary mechanism in all considered hard processes. 

By analyzing data for the semi-inclusive production of $(\pi^+ \pi^-)$ pairs in deep-inelastic scattering on transversely polarized proton and deuteron targets, as well as in proton-proton collisions with a transversely polarized proton, the valence components of $h_1$ were extracted for the first time from a global fit similarly to what is done for the unpolarized PDF (see Ref.~\citenum{Radici:2018iag_51_51} and references therein). Information on the unknown $H_1^\sphericalangle$ were independently obtained from data on the azimuthal correlation of $(\pi^+ \pi^-)$ pairs produced in back-to-back jets in $e^+ e^-$ annihilation. 

Here, we discuss the impact of {\tt COMPASS} pseudodata for a future measurement on deuterons~\cite{Bradamante:2018ick}. We also explore the potential tensions between the calculation of the first Mellin moment of the extracted $h_1^q$ for flavor $q$, the so-called tensor charge $\delta q$, and the average result of the corresponding calculation on lattice.

\section{Formalism}
\label{s:theory}

We use a functional form for the valence component $q_v$ of the transversity $h_1^{q_v} (x,Q_0^2)$ at the starting scale $Q_0^2 = 1$ GeV$^2$,  that must satisfy the Soffer bound at any $x$ and any scale $Q^2$ and whose Mellin transform must be analytically calculable. The latter is a well known workaround to speed up the computation of the matrix element for proton-proton collisions during the fit. Hence, our functional form is the product of two terms. The first one is a Mellin-transformable parametric function that fits the Soffer bound~\cite{Radici:2018iag_51_51}. The second term is a normalized linear combination of Cebyshev polynomials up to order 3, that depends on 10 fitting parameters. Some of these parameters are constrained by controlling the low$-x$ divergence such that the first Mellin moment $\delta q$ is finite~\cite{Radici:2018iag_51_51}. 

The statistical uncertainty of the global fit is studied at the 90\% confidence level using the bootstrap method. The gluon channel of di-hadron fragmentation is not constrained by the leading-order analysis of $e^+ e^-$ data. We parametrize this uncertainty by allowing for the three options  $D_1^g Q_0^2) = 0$, or $D_1^u (Q_0^2) / 4$, or $D_1^u (Q_0^2)$. For each option, we employ 200 replicas. Hence, we have in total 600  replicas. For the global fit of 32 independent data points with 10 fitting parameters, we obtain $\chi^2$/d.o.f. $= 1.76 \pm 0.11$~\cite{Radici:2018iag_51_51}. 

\vspace{-0.2cm}

\begin{figure}
\begin{center}
\includegraphics[width=0.4\textwidth]{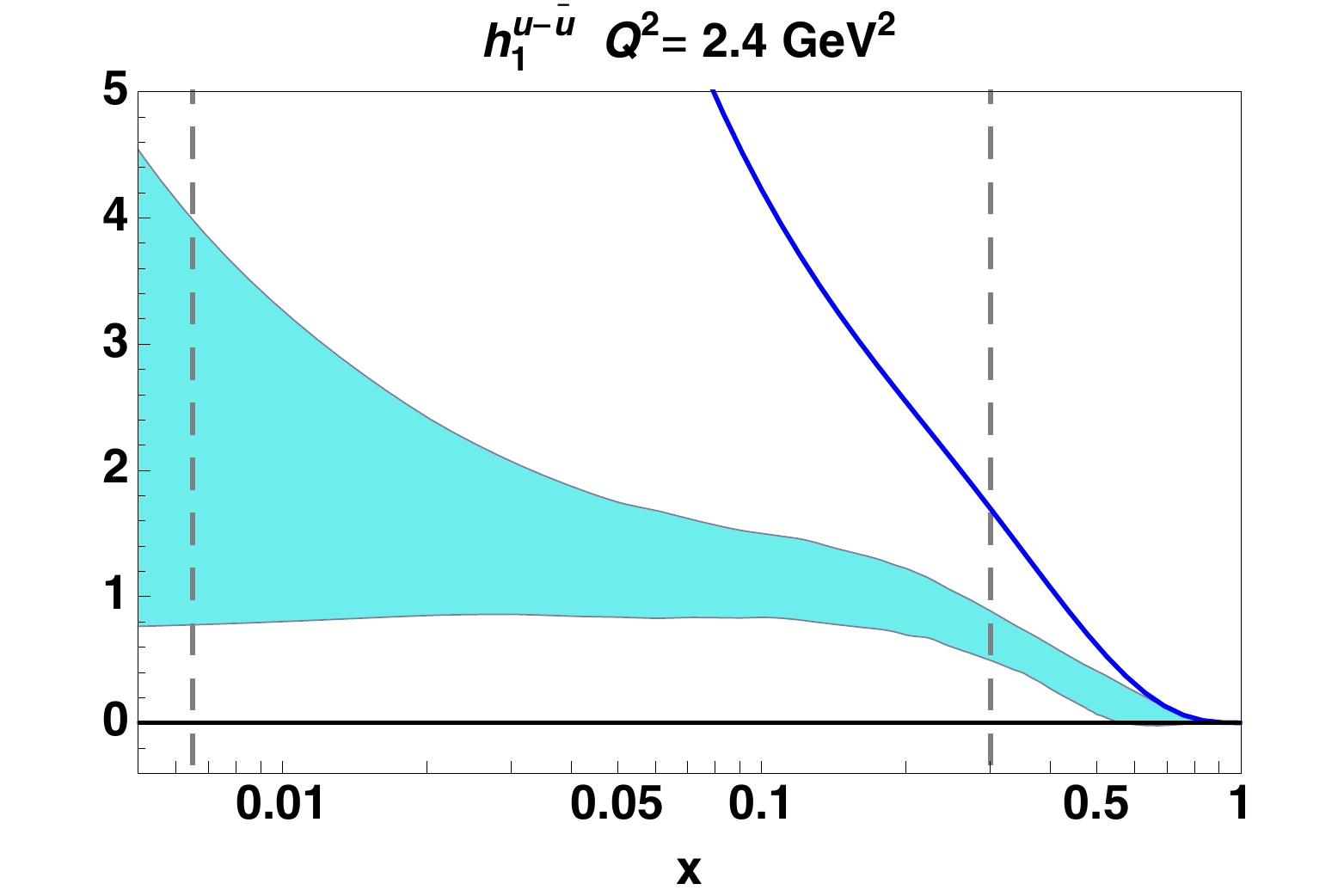} \hspace{0.1cm} 
\includegraphics[width=0.4\textwidth]{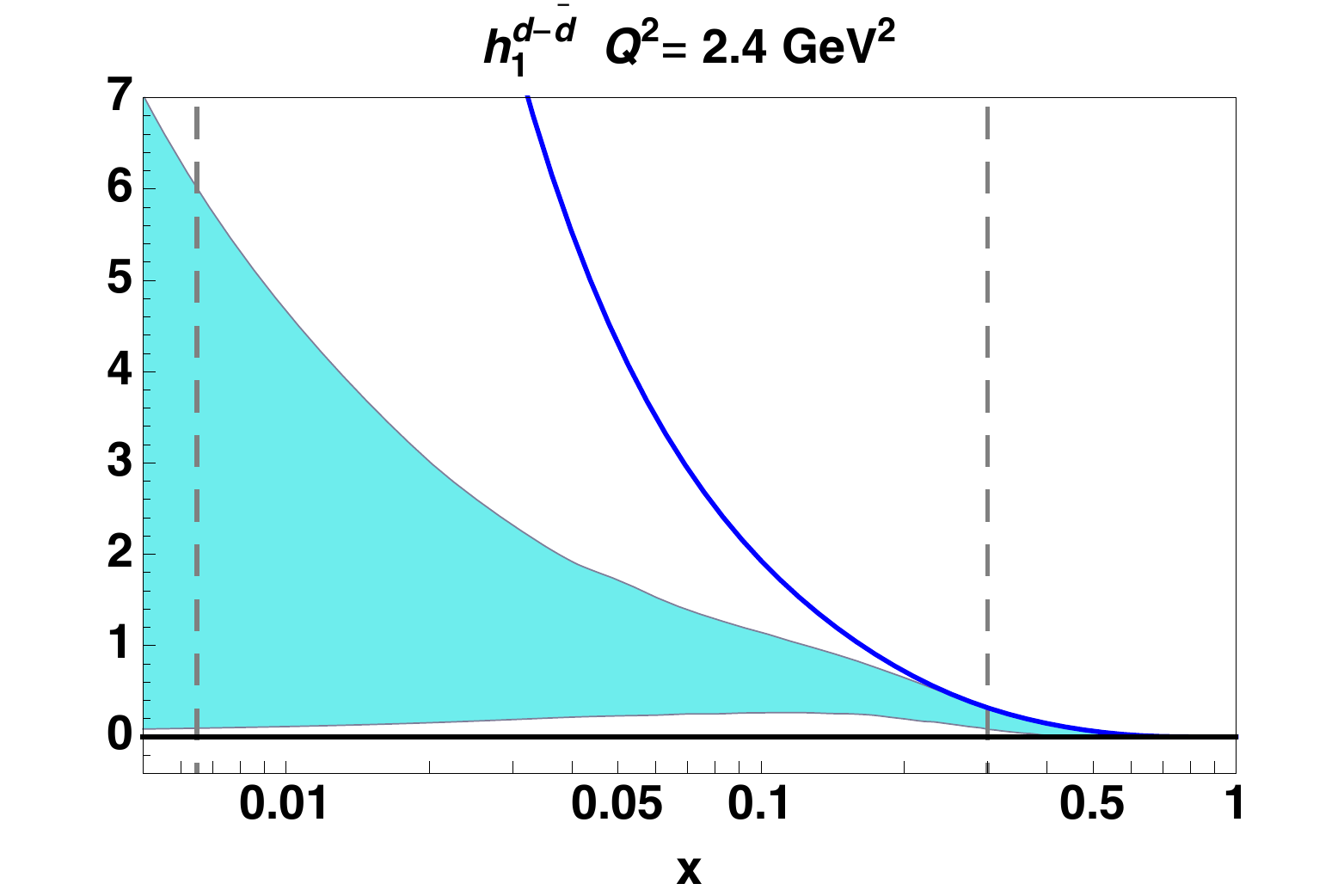} \\
\includegraphics[width=0.4\textwidth]{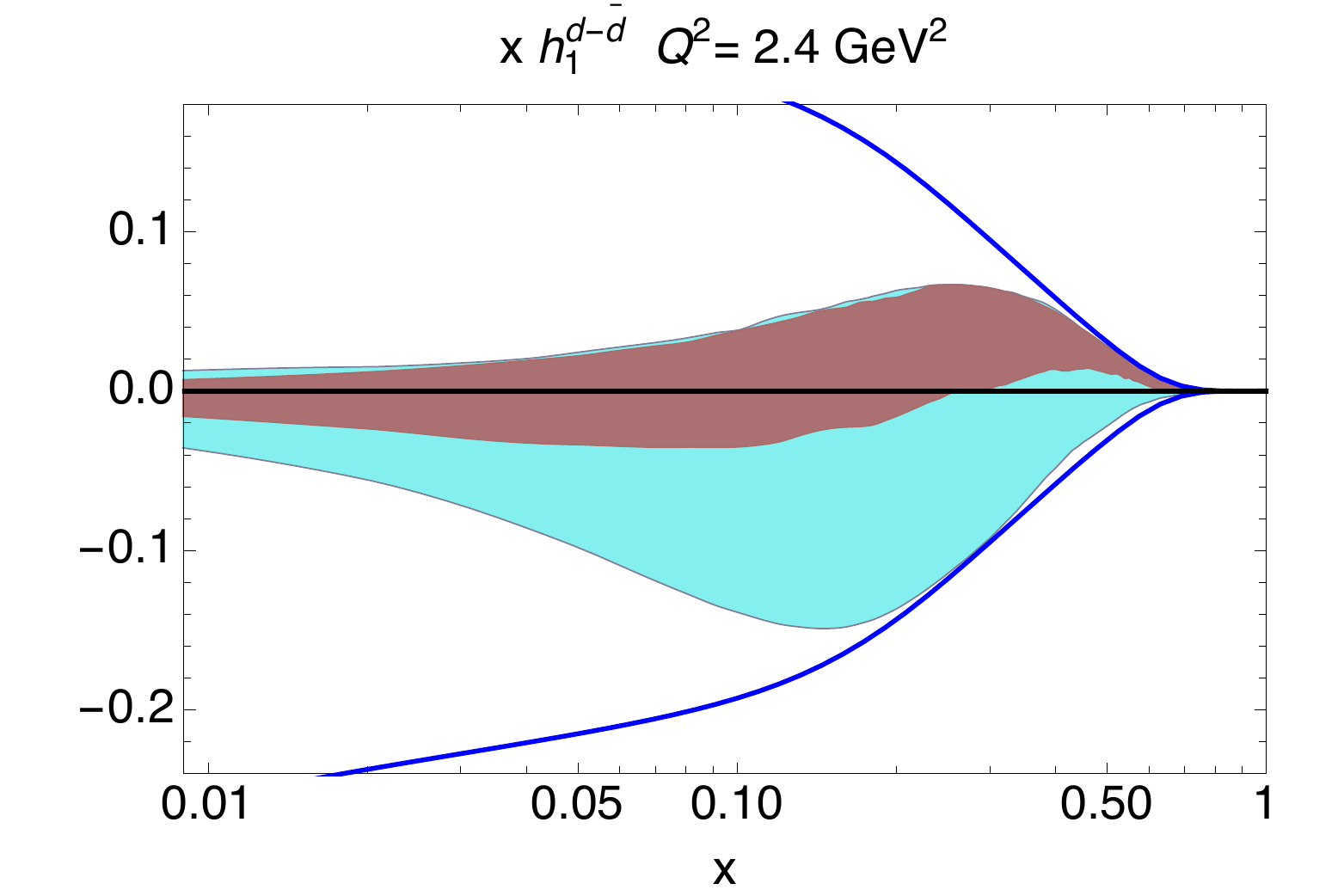} \hspace{0.1cm} 
\includegraphics[width=0.4\textwidth]{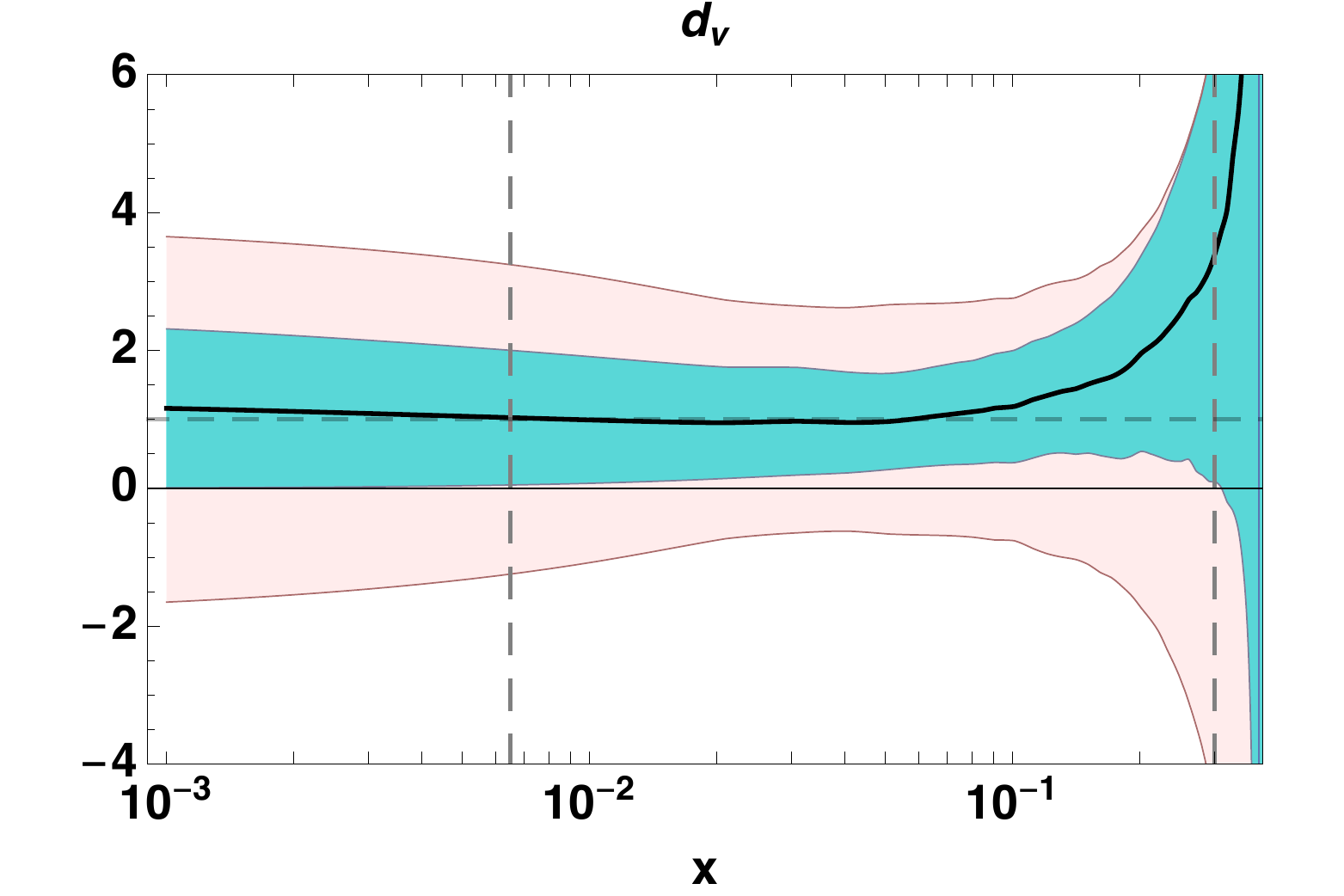}
\end{center}
\vspace{-0.4cm}
\caption{Upper panels: the transversity $h_1$ as a function of $x$ at $Q^2 = 2.4$ GeV$^2$ for valence up $u_v$ (left) and down $d_v$ (right). Solid (blue) lines represent the Soffer bounds. Lower left panels: the $x\, h_1^{d_v}$, lighter (cyan) band in the same conditions as before, darker band when including only the $D_1^g (Q_0^2) = 0$ option (see text). Right panel: the $h_1^{d_v}$ normalized to the central value of the uncertainty band in upper right panel. Lighter (pink) band for our global fit, darker (cyan) band when including also pseudodata from {\tt COMPASS} on deuteron target~\cite{Bradamante:2018ick}.}
\label{f:h1}
\end{figure}
\vspace{-0.5cm}

\section{Results}
\label{s:figures}

In Fig.~\ref{f:h1}, in the upper panels the transversity $h_1$ is displayed as a function of $x$ at $Q^2 = 2.4$ GeV$^2$ for valence up $u_v$ (left) and down $d_v$ (right) quarks. The solid (blue) lines represent the Soffer bounds. The vertical dashed lines show the experimental data coverage $0.0065 \leq x \leq 0.35$. The uncertainty band corresponds to the 90\% confidence level of the total 600 replicas (see Sec.~\ref{s:theory}). For $x\lesssim 0.0065$, the replicas are unconstrained and tend to spread over a larger range. The uncertainty is mostly driven by the behavior of $D_1^g$. In fact, in the lower left panel the $x\, h_1^{d_v}$ is shown in the same conditions and with the same notations. The darker band is the result when using only the $D_1^g (Q_0^2) = 0$ option, while the lighter (cyan) band corresponds to including all options. Such a strong sensitivity is not evident for the up quark. This can be understood by considering that for proton-proton collisions the gluon effects are active already at leading order and up and down contributions have equal weight, whilst for deep-inelastic scattering the down quark is suppressed and gluon effects appear only at subleading order. For the very same reason, the deuteron is more sensitive to the down contribution than the proton, and better data on this target would help in reducing the uncertainty. This is confirmed by the plot in the lower right panel, where the light (pink) band indicates our result for $h_1^{d_v}$ and the darker (cyan) band shows the effect of including also {\tt COMPASS} pseudodata from a future measurement on (transversely polarized) deuterons~\cite{Bradamante:2018ick}. The bins are the same as in the previous measurement but the statistic is much higher. All the replicas are drawn normalized to the average value of the band displayed in the corresponding upper right panel. In the range covered by experimental data, the gain in precision can be almost a factor 2. Finally, data on $(\pi^+ \pi^-)$ multiplicities in proton-proton collisions would be also very useful in reducing the uncertainty on $h_1^{d_v}$.

\vspace{-0.2cm}

\begin{figure}
\begin{center}
\includegraphics[width=0.5\textwidth]{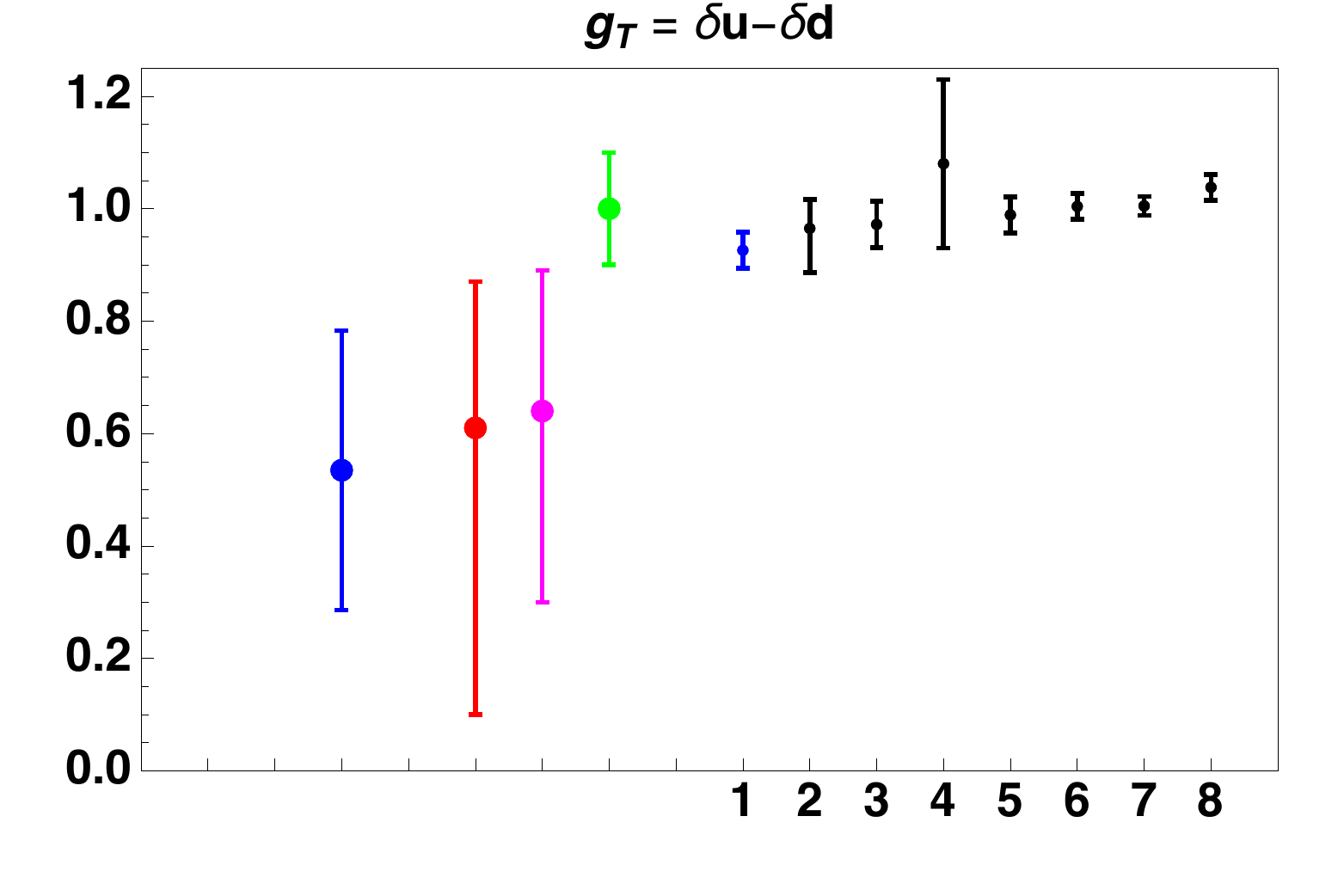} \hspace{0.1cm} 
\includegraphics[width=0.35\textwidth]{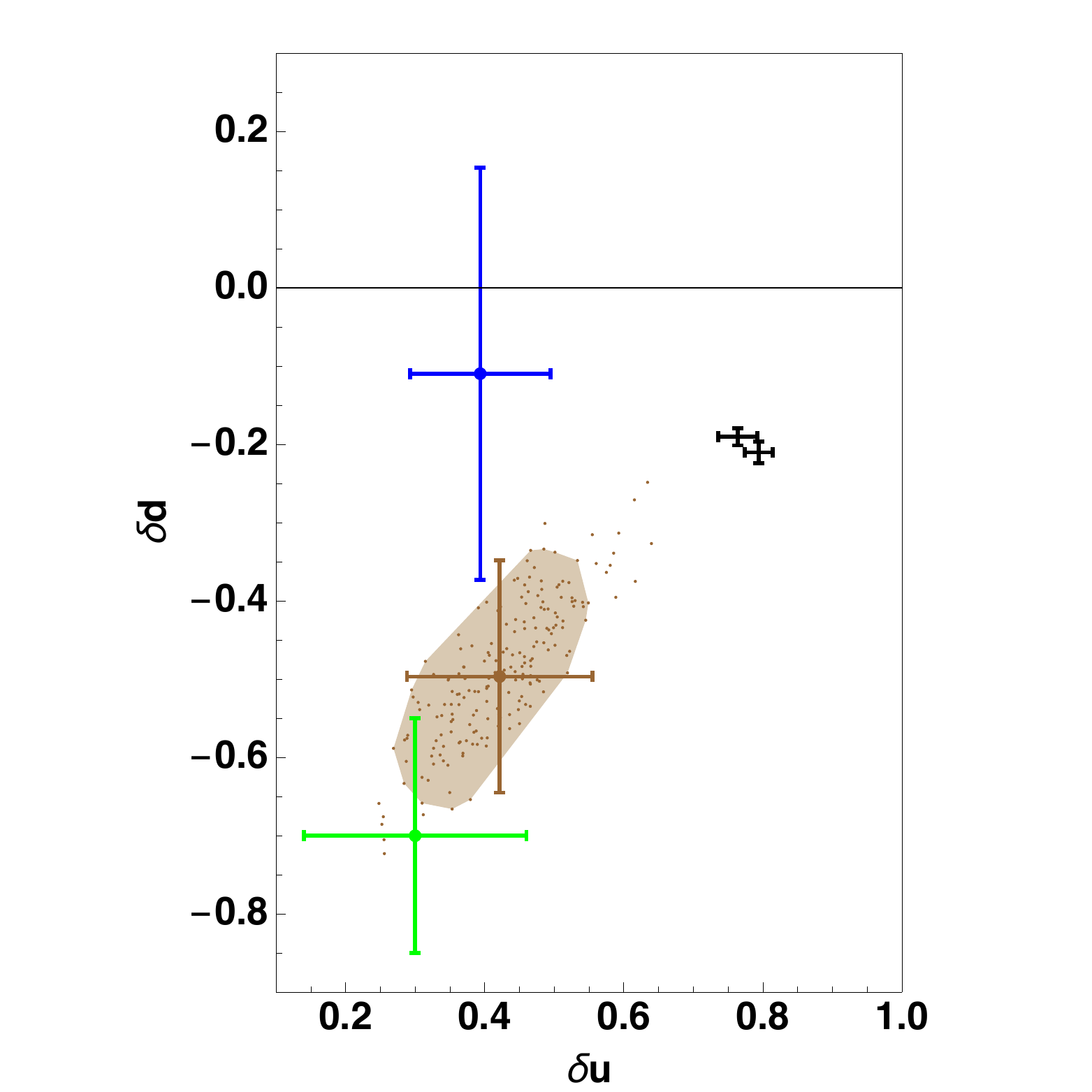}
\end{center}
\vspace{-0.4cm}
\caption{Left panel: the isovector tensor charges $g_T$ at $Q^2 = 4$ GeV$^2$. From left to right: result from our global fit (blue point), from Ref.~\citenum{Kang:2015msa_51} (red), from Ref.~\citenum{Anselmino:2013vqa_51} (magenta), from Ref.~\citenum{Lin:2017stx_137} (green). Lattice points with labels 1,\ldots ,8 from Refs.~\citenum{Alexandrou:2019brg_99_93,Harris:2019bih,Hasan:2019noy_51,Yamanaka:2018uud,Gupta:2018lvp,Alexandrou:2017qyt,Bali:2014nma,Green:2012ej}, respectively. Right panel: tensor charges $\delta d$ vs. $\delta u$. Bottom (green) cross from Ref.~\citenum{Lin:2017stx_137}, top (blue) cross for our global fit, middle (brown) cross with shaded area when including constraint to reproduce average lattice $g_T$. Small (black) crosses on the right from lattice: left from Ref.~\citenum{Gupta:2018lvp}, right from Ref.~\citenum{Alexandrou:2017qyt}.}
\label{f:gT}
\end{figure}
\vspace{-0.6cm}

\subsection{Tensor charge}
\label{s:gT}

In the left panel of Fig.~\ref{f:gT}, we compare the various results for $g_T = \delta u - \delta d$. From left to right, the first four points are based on  phenomenological extractions of $h_1$ from data. The leftmost (blue) point is from our global fit at 90\% confidence level, the next three points are obtained by analyzing the Collins effect from deep-inelastic scattering data only. From left to right: results from Ref.~\citenum{Kang:2015msa_51} (red), Ref.~\citenum{Anselmino:2013vqa_51} (magenta), and Ref.~\citenum{Lin:2017stx_137} (green), respectively. The other (black) points labeled $1, \ldots, 8$ are some recent lattice computations. Moving from the most recent to the oldest: label 1 from Ref.~\citenum{Alexandrou:2019brg_99_93}, label 2 from Ref.~\citenum{Harris:2019bih},  label 3 from Ref.~\citenum{Hasan:2019noy_51}, label 4 from Ref.~\citenum{Yamanaka:2018uud}, label 5 from Ref.~\citenum{Gupta:2018lvp}, label 6 from Ref.~\citenum{Alexandrou:2017qyt}, label 7 from Ref.~\citenum{Bali:2014nma}, label 8 from Ref.~\citenum{Green:2012ej}. All the lattice points are computed at $Q^2 = 4$ GeV$^2$ as well as our result. The point from Ref.~\citenum{Lin:2017stx_137} (green) is obtained by constraining the fit to reproduce the average lattice $g_T$. However, this comes at the price of breaking any  agreement with lattice for separate $\delta u$ and $\delta d$, as shown in right panel. Here, the result for $\delta d$ vs. $\delta u$ from Ref.~\citenum{Lin:2017stx_137} is indicated by the bottom (green) cross, while the lattice results from Refs.~\citenum{Gupta:2018lvp,Alexandrou:2017qyt} are indicated by the left and right small (black) crosses on the right, respectively. Our calculation is represented by the top (blue) cross, which is in agreement with lattice for $\delta d$ although within a very large error bar. The middle (brown) cross is the outcome of our global fit when constrained to reproduce the average lattice value for $g_T$. We also show the scatter plot of all 600 replicas' points with the shaded area corresponding to the 90\% confidence level. Indeed, forcing the compatibility with lattice $g_T$ comes at the price of destroying the agreement with lattice $\delta d$. 

We explored also the possibility of further constraining the fit to simultaneously reproduce all the average lattice values for $g_T, \, \delta u$, and $\delta d$. We observe a marked worsening of the fit quality: from the $\chi^2$/d.o.f. $= 1.76\pm 0.11$ of the original global fit to $\chi^2$/d.o.f. $= 2.29\pm 0.25$. This indicates that there is indeed a tension between lattice and phenomenology in the simultaneous comparison of results for the various components of the tensor charge. 

\section{Acknowledgments}
This work is supported by the European Research Council (ERC) under the European Union's Horizon 2020 research and innovation program (Grant n. 647981, 3DSPIN).



\newpage 
%

\renewcommand*{\FigPath}{./WeekII/16_Signori/Figs}
\def\figsubcap#1{\par\noindent\centering\footnotesize(#1)}

\title{Unpolarized TMDs: extractions and predictive power}
\wstoc{Unpolarized TMDs: extractions and predictive power}{Andrea Signori}

\author{Andrea Signori}
\index{author}{Signori, A.}

\address{Dipartimento di Fisica, Universit\`a di Pavia, via Bassi 6, I-27100 Pavia, Italy \\
INFN, Sezione di Pavia, via Bassi 6, I-27100 Pavia, Italy \\
Theory Center, Thomas Jefferson National Accelerator Facility \\ 12000 Jefferson Avenue, Newport News, VA 23606 USA \\
E-mail: asignori@jlab.org}

\begin{abstract}
We discuss the predictive power of transverse-momentum-dependent distributions as a function of the kinematics and comment on recent extractions from experimental data. 
\end{abstract}


\bodymatter

\section{Unpolarized TMDs}
\label{s:unp_TMDs}

The unpolarized transverse-momentum-dependent parton distribution function (TMD PDF) for a parton with specific flavor $a$, $F_a(x, k_T^2; \mu, \zeta)$, represents the probability density to pick a parton with collinear momentum fraction $x$ and transverse momentum $k_T$ inside a hadron.  
The variables $\mu$ and $\zeta$ are the UV-renormalization and rapidity renormalization scales, respectively. 
No spin dependence is present in this case. 
The TMD PDF in the $k_T$- and $b_T$-space are connected via a Fourier transform in the transverse variables: 
\begin{align}
F_a(x, k_T^2; \mu, \zeta) = 
\int_0^{\infty} \frac{db_T}{2\pi} \, b_T J_0(k_Tb_T) F_a(x, b_T^2; \mu, \zeta)\, .
\label{e:FT_kT_bT}
\end{align}
In principle, eq.~\eqref{e:FT_kT_bT} implies that in order to determine the TMD PDF at a specific $k_T$ value, the knowledge of $F_a(x, b_T^2; \mu, \zeta)$ is needed both at small $b_T$ ($b_T^{-1} \gg \Lambda_{\rm QCD}$) and at large $b_T$ ($b_T^{-1} \ll \Lambda_{\rm QCD}$). 
In the small-$b_T$ region, the $b_T$-dependence of the TMD PDF is well determined by perturbative QCD. 
On the other hand, in the large-$b_T$ region $F_a(x, b_T^2; \mu, \zeta)$ is sensitive to the non-perturbative hadron structure and needs to be extracted from experimental data (or accessed by other techniques). 
Moreover, it comes as a natural question to understand in which kinematic regions the TMD formalism has {\em predictive power}, namely when the TMD distributions are dominated by the perturbative physics at small $b_T$ and can accurately predict physical observables.

\section{Extractions from data}
\label{s:extractions}


In a recent global analysis of experimental data from Semi-Inclusive Deep-Inelastic Scattering (SIDIS), Drell-Yan and $Z$-boson production in TMD factorization~\cite{Bacchetta:2017gcc_46}, the non-perturbative, large $b_T$, part of the unpolarized quark TMD PDFs and TMD fragmentation functions (FFs) has been fit to the data as the weighted sum of two Gaussian functions, with the same (different) width in the case of the TMD PDFs (FFs). TMD evolution effects have been taken into account up to next-to-leading logarithmic (NLL) accuracy, with a matching of the TMD distributions onto the collinear ones at leading order (LO) in the strong coupling $\alpha_s$. 
Within that scheme and at the input scale of the fit ($Q=1$ GeV), the average squared transverse momentum of the quark in the hadronic target as a function of the light-cone momentum fraction $x$, $\langle k_T^2 \rangle(x)$, is a well-defined quantity and can be analytically calculated in terms of the fit parameters. The same applies to the average squared transverse momentum acquired by the detected hadron during the hadronization process, $\langle P_\perp^2 \rangle(z)$, where  $z$ is the light-cone momentum fraction of the produced hadron in SIDIS (see Fig.~\ref{f:kin_dep_av2tm}).   
The kinematic dependence of the non-perturbative functions at large $b_T$, together with the matching of the TMD distributions onto the collinear ones at small $b_T$, reflects a non-trivial interplay between the collinear and the transverse structure of hadrons. In principle, also a flavor dependence can be taken into account~\cite{Signori:2013mda}, but precise flavor-sensitive data are needed to constrain that, for example from SIDIS at an Electron-Ion Collider. 

More recent extractions of the unpolarized quark TMDs~\cite{Bertone:2019nxa,Bacchetta:2019sam,Scimemi:2019cmh} consider also higher-order corrections neglected in previous analyses. 
This has the advantage to separate the truly non-perturbative hadron structure from the perturbative effects, but renders the definition of the average transverse momenta more problematic.

\begin{figure}[h]%
\begin{center}
\parbox{2.4in}{\includegraphics[width=2.3in]{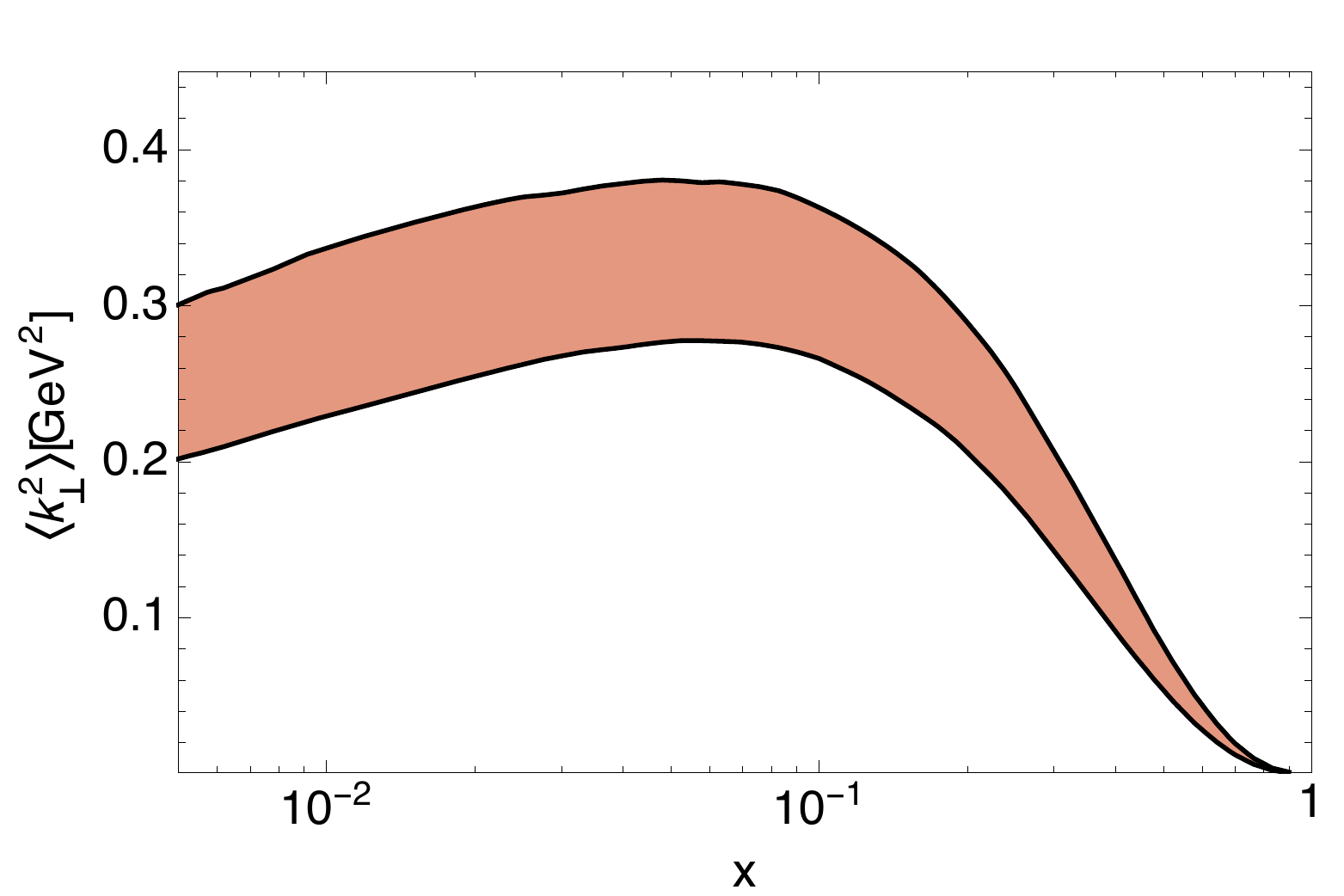}
\figsubcap{a}}
\hspace*{2pt}
\parbox{2.4in}{\includegraphics[width=2.3in]{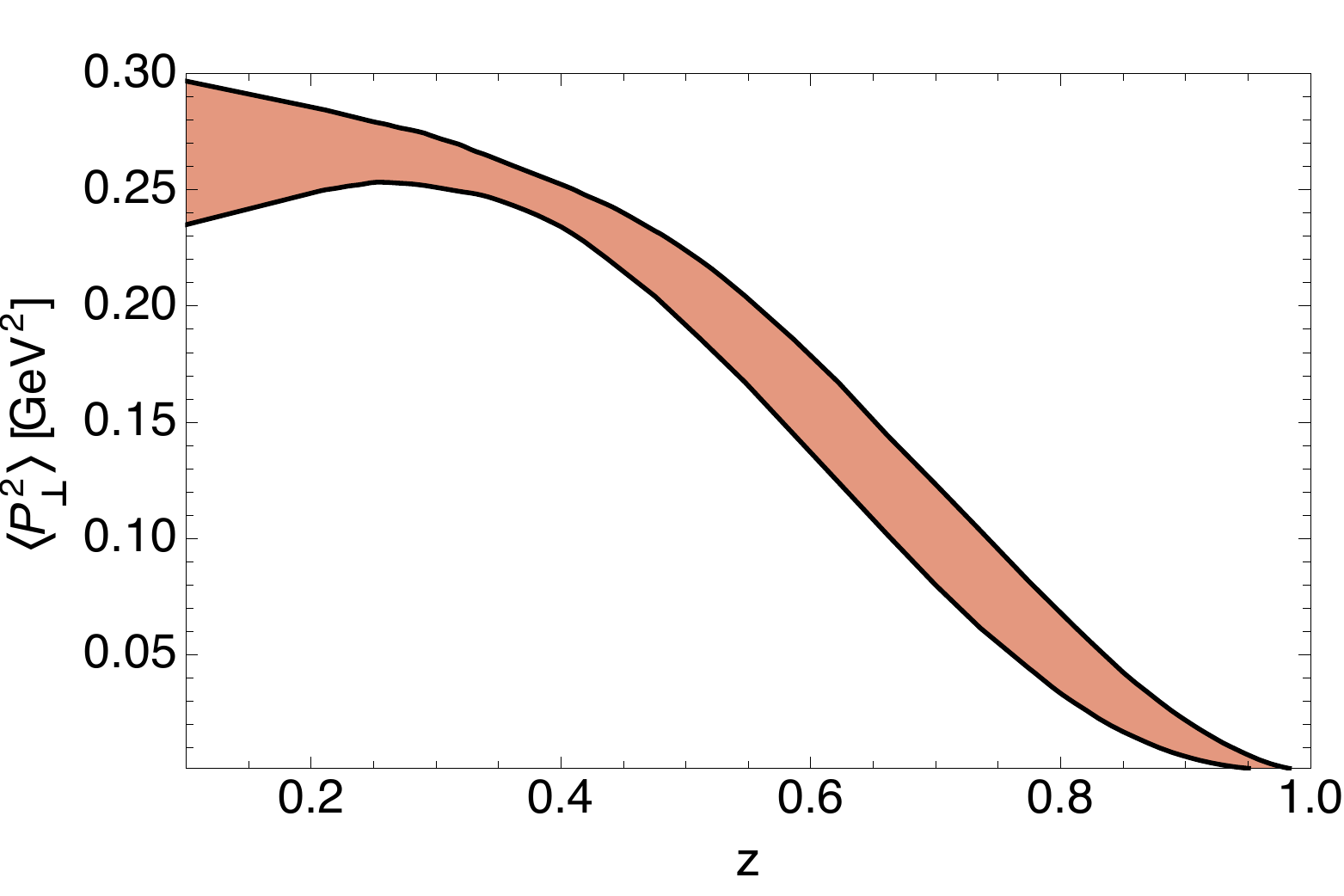}
\figsubcap{b}}
\caption{(a) Average squared transverse momentum from the TMD PDF and (b) from the TMD FF at $Q=1$ GeV, and their dependence on the light-cone momentum fractions. The red band represents the $68\%$ confidence level~\cite{Bacchetta:2017gcc_46}.}
\label{f:kin_dep_av2tm}
\end{center}
\end{figure}

\section{Predictive power}
\label{s:predictive_power}

The predictive power of the TMD formalism can be addressed by studying the behavior of TMD PDFs and cross sections through a saddle-point approximation~\cite{Grewal:2020inprep,Qiu:2000hf_46}.   
This approximation allows one to calculate a specific class of integrals by evaluating the integrand at the so-called ``saddle point''.   
Considering eq.~\eqref{e:FT_kT_bT}, we determine which $b_T$ region the saddle point belongs to as a function of the kinematics. 
%

By choosing the final and initial scales respectively as $\mu_f^2 = \zeta_f = Q^2$ and $\mu_i^2 = \zeta_i = \mu_b^2$, where $\mu_b \equiv 2e^{-\gamma_E}/b_T$, and working at leading logarithmic accuracy, the saddle point for the TMD PDF at $k_T=0$ reads:
\begin{equation}
\label{e:sp_sol_LL}
b_T^{sp}(x,Q) = \frac{2e^{-\gamma_E}}{\Lambda_{QCD}} \bigg(  \frac{Q}{\Lambda_{QCD}}  \bigg)^{-  \Gamma_1^{\text{c}} / \big[\Gamma_1^{\text{c}} + 8\pi b_0 \big(1 - {\cal X}(x,\mu_b^\star) \big)  \big]  } \, , 
 \ \ \ \ 
 {\cal X}(x,\mu) = \frac{d \ln f_a(x,\mu)}{d \ln \mu^2} \, ,
\end{equation}
where 
$\Gamma_1^{\text{c}}$ is the first term in the expansion of the cusp anomalous dimension, $b_0$ is the one-loop coefficient of the QCD beta function,  
and $\mu_b^\star \equiv 2e^{-\gamma_E}/b_T^{sp}$.  


The function ${\cal X}(x,\mu)$ quantifies the impact of the DGLAP evolution on the position of the saddle point. 
Its sign changes according to the value of the collinear momentum fraction $x$, and determines the $x$-dependence of the saddle point.
In general, the function is positive for $x \lesssim 0.1$ and negative for $x \gtrsim 0.1$. 
Thus it suppresses the value of the saddle point $b_T^{sp}(x,Q)$ 
for $x \lesssim 0.1$ and it enhances it for $x \gtrsim 0.1$. 
Thus the saddle point of $F_a(x,k_T=0,Q,Q^2)$ drifts towards the small $b_T$ perturbative region when $Q$ is large and $x$ is small. 

Nonetheless, one can check~\cite{Grewal:2020inprep} that the saddle point of the TMD PDF for an up quark at $k_T=0$, $Q=M_Z = 91.18$ GeV always lies outside of the small $b_T$ region (defined as $b_T< b_{\rm max} = 0.5$ GeV$^{-1}$, where $b_{\rm max}$ is the largest $b_T$ value for which a perturbative calculation is performed) for any $x > 10^{-4}$. 
This means that at large $Q$ and for any $x$, the TMD PDF of an up quark is contributed to a certain extent by the non-perturbative physics at large $b_T$. 
However, cross sections are calculated as the convolution of two TMD distributions, and, for the $Q=M_Z$ case, 
the saddle-point 
of the cross section is further shifted to the perturbative region~\cite{Qiu:2000hf_46} with respect to a single TMD distribution.
In support of this, in Fig.~\ref{f:sigma_Z} (a) we describe CMS data~\cite{Chatrchyan:2011wt} for $Z$-boson production at low transverse momentum $q_T \ll M_Z$ and central rapidity $|y|<2.1$ without fitting any non-perturbative parameter~\cite{Grewal:2020inprep}. The only two parameters in the functional form at large $b_T$ are determined by imposing the continuity of the first and second derivatives of the TMD PDF at $b_T = b_{\rm max} = 0.5$ GeV$^{-1}$. 
The calculation of the cross section includes terms up to NNLO in $\alpha_s$ in the hard part and in the collinear expansion of the TMD PDFs, and TMD evolution is implemented at NNLL accuracy. 
In Fig.~\ref{f:sigma_Z} (b), instead, the normalized integrand  in $b_T$-space of the differential cross section is displayed for $q_T=0$, $y=\pm 2.1,\, 0$. 
In both cases the peak of the integrand lies well in the perturbative region ($b_T < 0.5$ GeV$^{-1}$), and the same holds true for the saddle point of the cross section.  

This confirms that at large $Q$ and small $x$ the TMD formalism is fully predictive. 
One must anyway take into account that the predictive power is not an absolute concept, but is related to the precision of the experimental observable under consideration~\cite{Bacchetta:2018lna,Bozzi:2019vnl}. 
In the kinematic regions where the predictive power is lower the physical observables have a significant dependence on the non-perturbative part of the TMD distributions. These regions, accessible in principle at collider and fixed-target experiments, present an excellent opportunity to study hadron structure and hadronization in a multidimensional momentum space. 

\begin{figure}[h]%
\begin{center}
\parbox{2.4in}{\includegraphics[width=2.3in]{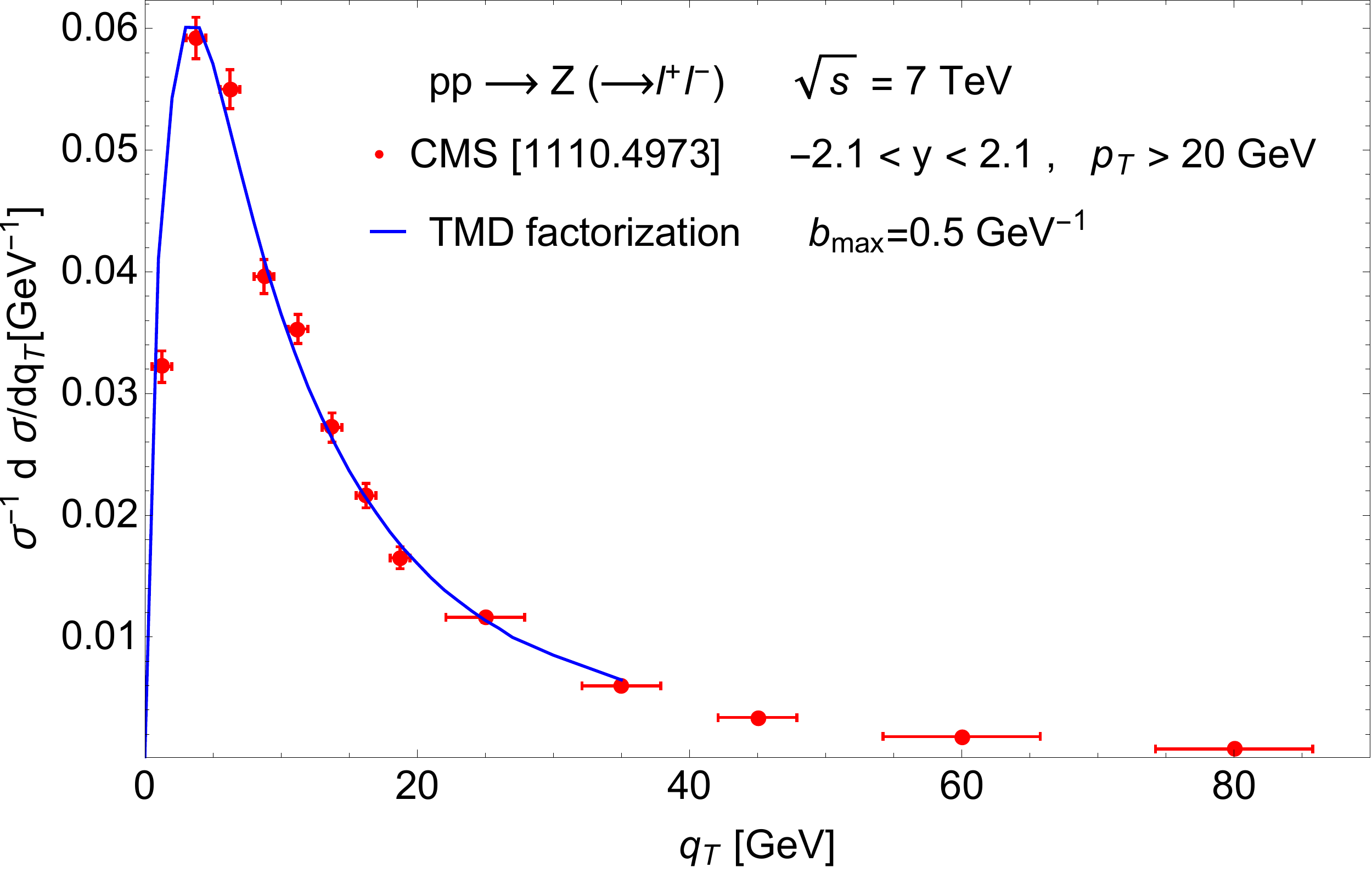}
\figsubcap{a}}
\hspace*{2pt}
\parbox{2.4in}{\includegraphics[width=2.3in]{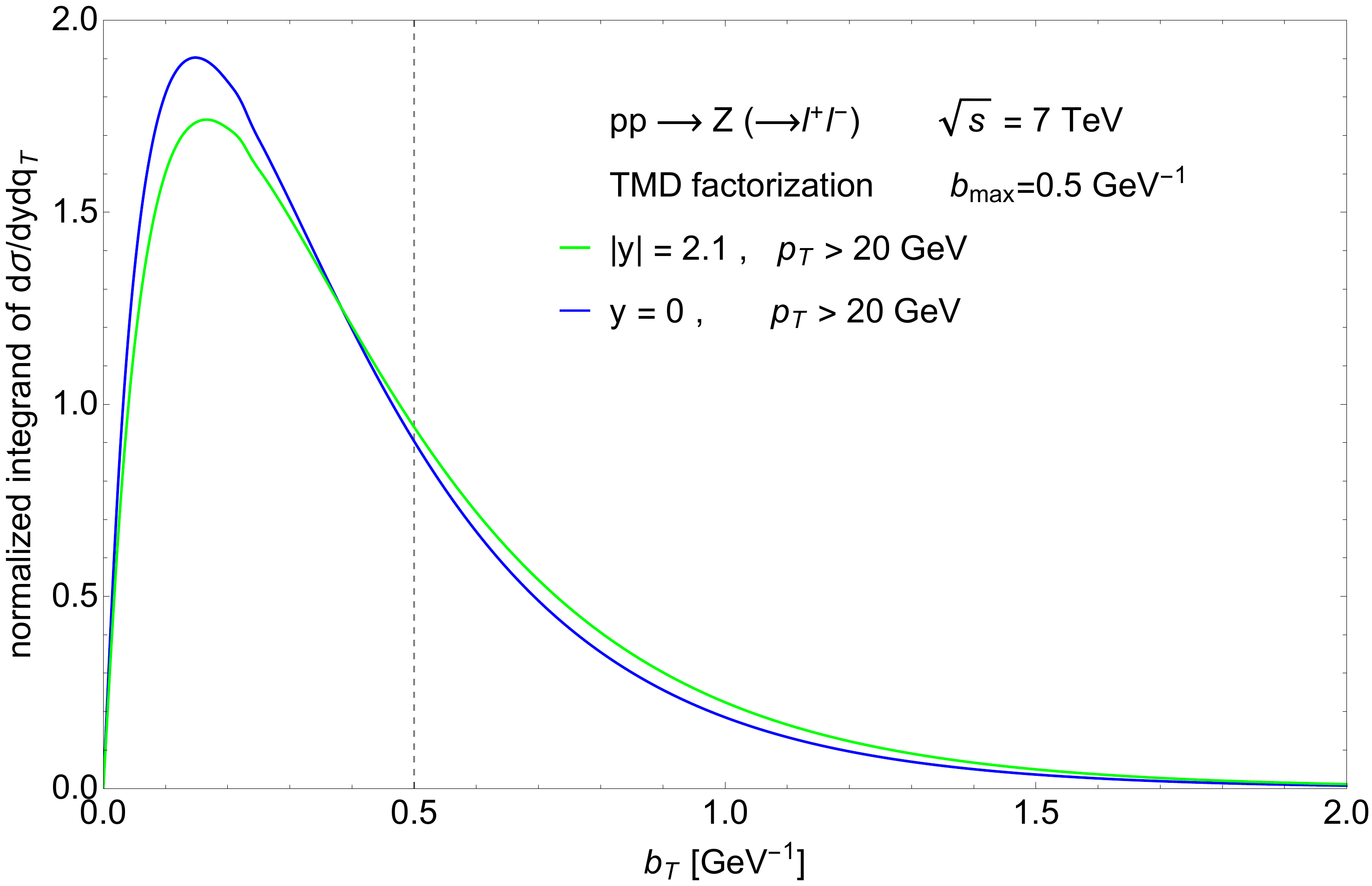}
\figsubcap{b}}
\caption{(a) Differential cross section for $Z$ boson production in TMD factorization ($q_T \ll Q$) vs CMS data. (b) Normalized integrand for the cross section in $b_T$ space at $q_T=0$, $y=\pm 2.1, 0$.}
\label{f:sigma_Z}
\end{center}
\end{figure}

{\em Acknowledgments:} the author thanks A. Bacchetta, F. Delcarro, M. Grewal, Z. Kang, C. Pisano, J.W. Qiu, M. Radici for the collaboration on these topics. 
This work was supported by the  U.S. DoE contract DE-AC05-06OR23177, under which JSA LLC manages and operates Jefferson Lab and by the E.U. Commission through the MSCA SQuHadron (grant ID: 795475). 




\newpage
\renewcommand*{\FigPath}{./Logo/} 

\begin{tcolorbox}[colframe=white]
\begin{minipage}{0.2\textwidth}
\includegraphics[width=1.\textwidth]{\FigPath/INT_Workshop_Logo_Final_Black.png}
\end{minipage}
\begin{minipage}{0.7\textwidth}
\wstoc{\bf Week III}{}
\title{Week III}
\end{minipage}
\end{tcolorbox}

\bodymatter

\wstoc{Introduction for Week III}{Elke Aschenauer,
Yoshitaka Hatta, 
Yuri Kovchegov, 
Keh-Fei Liu,
C\'edric Lorc\'e, 
Cyrille Marquet, 
Alexei Prokudin}

\index{author}{Aschenauer, E.}
\index{author}{Hatta, Y.}
\index{author}{Kovchegov, Y.}
\index{author}{Liu, K.}
\index{author}{Lorc\'e, C.}
\index{author}{Marquet, C.}
\index{author}{Prokudin, A.}

\title{Introduction for Week III}

~\vspace{3cm}

The composition and origin of the proton spin has been an important question in QCD since the dawn of the proton spin puzzle. In the late 1980s, the European Muon Collaboration (EMC) \cite{Ashman:1987hv,Ashman:1989ig} measured the proton spin $S_q$ carried by the quark spins with $0.01<x<0.7$ to be significantly smaller than $1/2$, contrary to what had been expected based on the constituent quark model. This observation started the proton spin puzzle. Further developments in theory led to the helicity sum rules in the Jaffe-Manohar \cite{Jaffe:1989jz} and Ji \cite{Ji:1996ek} forms, see also the recent reviews \cite{Leader:2013jra,Wakamatsu:2014zza}. These helicity sum rules show that the proton spin consists of contributions of the quark and gluon spins and orbital angular momenta. The EMC measurement \cite{Ashman:1987hv,Ashman:1989ig} can be interpreted as implying that a significant fraction of the proton spin is carried by the gluon spin, or by the quark and gluon orbital motion, or by the quark spins with the values of Bjorken $x$ outside of the measured range. This conclusion was further reinforced by the discovery of the non-zero contribution of the gluon spins $S_g=\Delta G$ at RHIC \cite{Adamczyk:2014ozi}, which moved us closer to resolving the spin puzzle. Nevertheless, all the possible sources of the proton spin need to be explored and clarified before one can confidently claim understanding of the proton spin. The EIC would help us by providing experimental data which may prove to be decisively important in establishing the origin of the proton spin. 

Week III of the program was chiefly dedicated to the proton spin puzzle. In the last decade, developments in this field went along several important directions. Experimental tools and helicity PDF extraction methods using perturbative QCD approaches have significantly improved over the last decade. Jet measurements may appear to be particularly useful for the extraction of helicity PDFs from the upcoming EIC data.

Progress in lattice QCD was driven in part by the improvements in the calculations of the moments of PDFs, and, in part, by the efforts to calculate the PDFs on the lattice directly using quasi- and pseudo-PDFs. These
 new techniques have been recently applied to the helicity PDF sector. In addition, the components of the Ji spin decomposition have been calculated on the lattice \cite{Deka:2013zha,Alexandrou:2017oeh}, while those of the Jaffe-Manohar decomposition remain
 a challenge. 

Research continues on trying to measure the quark and gluon orbital angular momentum (OAM). It has been realized that the proper definition of the Jaffe-Manohar OAM involves certain twist-three GPDs. While it is challenging to experimentally access such twist-three effects, approaches based on measuring Wigner distributions of partons and employing it for inferring the parton OAM appear to be particularly promising.  

The contributions to helicity sum rules coming from small-$x$ quarks and gluons may also be potentially important. Some theoretical progress leading to estimates of the amount of proton spin coming from low $x$ happened in the last decade as well. 

Spin decomposition is not the only fundamental question about the proton structure that was addressed in week III. An important new topic gaining momentum in the recent years is the proton mass decomposition, see \cite{Lorce:2017xzd} for a review of the various sum rules. Questions involve understanding how much of the proton mass is due to the gluon condensate and how much of it is due to the chiral condensate. These are deep questions, progress on which would allow us to shed light on fundamental QCD topics such as quark confinement and chiral symmetry breaking. It may be possible to individually measure the quark and gluon contributions to the proton mass at the EIC.

~\\

\begin{flushright}
Elke Aschenauer \\
Yoshitaka Hatta \\
Yuri Kovchegov \\
Keh-Fei Liu \\
C\'edric Lorc\'e \\
Cyrille Marquet \\
Alexei Prokudin

\end{flushright}



%



\newpage  
%

\renewcommand*{\FigPath}{./WeekIII/01_Hatta/}

\renewcommand{\nn}{\nonumber \\}
\renewcommand{\beq}{\begin{eqnarray}}
\renewcommand{\eeq}{\end{eqnarray}}

\wstoc{$J/\psi$ photo-production near threshold and the proton mass problem}{Yoshitaka Hatta}
\title{$J/\psi$ photo-production near threshold and the proton mass problem}


\author{Yoshitaka Hatta}
\index{author}{Hatta, Y.}

\address{Brookhaven National Laboratory, Upton, NY 11973, USA \\
E-mail: yhatta@bnl.gov\\ }

\begin{abstract}
The photo-production of $J/\psi$ near threshold is a promising process to access the contribution from the gluon condensate to the proton mass. We study this process in a holographic model and compare our result with the latest experimental data from the GlueX collaboration. 
\end{abstract}


\bodymatter

\section{Introduction}

Recently, there has been renewed interest in the mass structure of the proton, mainly triggered by a report\cite{nas} from  the National Academy of Science (NAS) in which the origin of the proton mass has been identified as one of the key scientific questions to be addressed at the future Electron-Ion Collider (EIC). This is a multi-faceted question which ultimately touches fundamental issues such as confinement and chiral symmetry breaking. The question can also be addressed at different conceptual levels. On one hand, first-principle lattice QCD calculations can reproduce the masses of the proton and other baryon resonances.  On the other hand, the detailed mechanisms by which massless quarks and gluons bind themselves and convert their interaction energy into the observed hadron masses is poorly understood. It is known that collider experiments can measure the kinetic energy contributions of quarks and gluons, as the second moment of the ordinary parton distribution functions. But the full understanding of the problem will require more information than just the kinetic energy, and it is a challenge to what extent the EIC, or more generally, lepton-proton scattering experiments can explore the nonperturbative aspect of the proton mass. In this contribution to the proceedings, I report on our recent, `holographic' approach to this problem.\cite{Hatta:2018ina_137,Hatta:2019lxo_134}

\section{QCD trace anomaly and near-threshold $J/\psi$ production}

One way to understand the origin of the proton mass is to look at the trace of the energy momentum tensor
\beq
T^\alpha_\alpha= \frac{\beta(g)}{2g}F^{\mu\nu}F_{\mu\nu} + m(1+\gamma_m(g)) \bar{\psi}\psi,
\eeq
where $\beta$ is the QCD beta function and $\gamma_m$ is the mass anomalous dimension. This is the QCD trace anomaly which breaks the approximate conformal symmetry of the QCD Lagrangian and gives hadrons a nonperturbatively generated mass. By taking the  forward proton matrix element in the proton single-particle state, one finds the relation 
\beq
\langle P|T^\alpha_\alpha|P\rangle = 2M^2.
\eeq
In particular, in the chiral limit $m=0$, the proton mass is entirely explained by the gluon condensate $\langle P|F^{\mu\nu}F_{\mu\nu}|P\rangle$. Then an interesting question arises as to whether one can experimentally  measure the matrix element $\langle P|F^{\mu\nu}F_{\mu\nu}|P\rangle$. Such a possibility was first studied by Kharzeev et al. \cite{Kharzeev:1998bz_131}
in the context of the near-threshold, exclusive photo-production of $J/\psi$ in lepton-proton scattering $eP \to e'\gamma^{(*)} P \to e'J/\psi P'$. By assuming vector dominance, one can relate the physical amplitude $\gamma^{(*)} P \to J/\psi P'$ to the forward amplitude $J/\psi P \to J/\psi P$.  This is sensitive to the matrix element of the gluon condensate $\langle P| F^2|P\rangle$ since  heavy quarkonia such as $J/\psi$ primarily interact with the proton via two-gluon operators including $F^2$. However,  near the threshold, the momentum transfer $\sqrt{-t}=\sqrt{-(P-P')^2}\approx 1.5$ GeV is large, on the order of the charm quark mass, and the extrapolation $t\to 0$ has to be done more carefully. As observed by Frankfurt and Strikman,\cite{Frankfurt:2002ka} the $t$-dependence of this process must come from that of `two-gluon' form factors. While they did not specify what exactly these form factors are, there is in fact only one class of `two-gluon' form factors in QCD, namely, the gravitational form factors
\beq
\langle P'|(T^R_{q,g})^{\alpha\beta}|P\rangle = \bar{u}(P')\Bigl[ A^R_{q,g}\gamma^{(\alpha}\bar{P}^{\beta)} + B^R_{q,g}\frac{\bar{P}^{(\alpha}i\sigma^{\beta)\lambda}\Delta_\lambda}{2M}  \nn 
+ C^R_{q,g}\frac{\Delta^\alpha\Delta^\beta-g^{\alpha\beta}\Delta^2}{M} + \bar{C}^R_{q,g}M\eta^{\alpha\beta} \Bigr] u(P) , \label{jid}
\eeq
where $T_{q,g}$ are the quark and gluon parts of the energy momentum tensor, and $A,B,C,\bar{C}$ depend on $t=(P-P')^2=\Delta^2$ and the renormalization scale. It can be shown\cite{Hatta:2019lxo_134} that the nonforward matrix element $\langle P'|F^2_R|P\rangle$ (subscript $R$ stands for `renormalized') can be entirely expressed in terms of the gluon gravitational form factors 
\beq
\langle P'|F^2_R|P\rangle &=& \bar{u}(P') \Bigl[ (K_g A_g^R + K_q A_q^R)M+\frac{K_g B_g^R+K_q B_q^R}{4M}\Delta^2  \nonumber \\ 
&& \qquad \qquad  -3\frac{\Delta^2}{M}(K_g C_g^R + K_q C_q^R) +4(K_g\bar{C}_g^R + K_q \bar{C}_q^R)M\Bigr] u(P), \label{im}
\eeq
where the coefficients $K_{q,g}$ has been calculated in perturbation theory up to three loops in the $\overline{\rm MS}$ scheme.\cite{Hatta:2018sqd,Tanaka:2018nae} The problem of extrapolation $t\to 0$ reduces to knowing the $t$-dependence of these gravitational form factors. The latter can be studied, for example, in lattice QCD. 

\section{Holographic model}

Our next task is to find the relation between the scattering amplitude $\langle J/\psi P'|\gamma P\rangle$ and the matrix element $\langle P'|F^2|P\rangle$. Unfortunately, at the moment this cannot be done in the usual framework of QCD factorization, and one has to rely on nonperturbative approaches/models. In Refs.~[\refcite{Hatta:2018ina_137,Hatta:2019lxo_134}], we have proposed a holographic model based on the AdS/CFT correspondence. Previously, the AdS/CFT correspondence has been mainly applied to high energy scattering with limited success. The main problem is that, at high energy and at strong coupling, the scattering amplitude is mostly real and dominated by the gravition exchange which predicts too strong rise of the cross section with increasing energy. However, these features become attractive for the present, low-energy process. Near the threshold, the scattering amplitude is purely real, and the graviton is dual to  the energy-momentum tensor. One can thus relate the graviton exchange amplitude to the gravitational form factors. An interesting new feature at low energy is that the scattering amplitude becomes sensitive to the dilaton exchange which is dual to the operator $F^2$. At high energy, such a contribution is suppressed (higher twist effect) compared to the graviton exchange. 

The details of our model and calculation can be found in the original references.\cite{Hatta:2018ina_137,Hatta:2019lxo_134} Here we show our fit of the latest experimental result from the GlueX collaboration  at Jefferson Lab.\cite{Ali:2019lzf_128} In this plot, the overall normalization of the total cross section is adjusted to the data and the shape of the curve is our prediction. The solid and dashed curves are for $b=0$ and $b=1$, respectively, where $b$ is defined by 
\beq
b \equiv \frac{\langle P|m(1+\gamma_m)(\bar{\psi}\psi)_R|P\rangle}{2M^2}, \qquad 1-b=  \frac{\langle P|\frac{\beta}{2g}(F^2)_R|P\rangle}{2M^2}. \label{defb}
\eeq
$b=1$ is an extreme scenario where the proton mass entirely comes from the quark condensate, and $b=0$ the other extreme where it entirely comes from the gluon condensate. One can see already by inspection that a better fit is obtained by $b=0$ (solid curve), suggesting that the gluons play a more dominant role  for generating  the proton mass. (Note that the empirical value is $b\sim 0.1$ from the nucleon sigma term.) We however note that, if we allow for negative $b$-values, we actually obtained better $\chi^2$ fits. In this regard, it is thus interesting to consider more realistic AdS/QCD models. 

\begin{figure}[!h]
\centering
\includegraphics[width=0.97\textwidth]{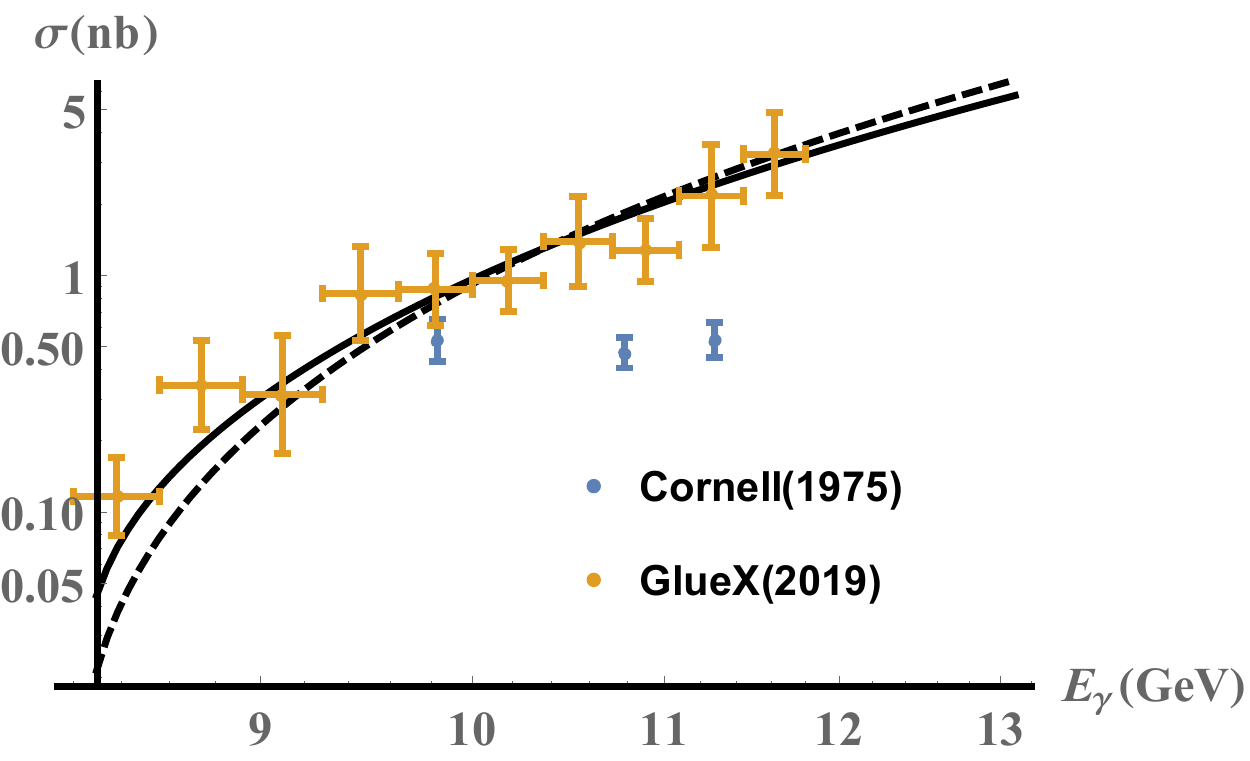}
\caption{\label{pgt} The total cross section as a function of the photon energy $E_\gamma$ in the proton rest frame. The threshold is at $E_\gamma=8.2$ GeV.}
\end{figure}

\section{Conclusions}

In view of the recent NAS report, the proton mass problem should be seriously addressed at the EIC. As far as the measurement of the trace anomaly (gluon condensate) is concerned, the high energy of EIC may be thought of as a disadvantage. However, this is not necessarily the case. It has been shown\cite{Hatta:2019lxo_134} that the near-threshold production of $J/\psi$ and $\Upsilon$ can be measured at RHIC in ultraperipheral collisions (UPCs). It is also a very interesting subject to be pursued at the EIC in China (EIcC) where  the collision energy is lower than the US EIC. More theoretical efforts are also needed, especially towards the first-principle calculation of the process within QCD factorization. A combined effort of theory and various experiments (JLab, RHIC, EIC, EIcC) will be necessary to answer this difficult problem.

\section*{Acknowledgments}
I thank Dilun Yang, Kazuhiro Tanaka and Abha Rajan for collaboration. 
This work is supported in part by  the U.S. Department of Energy, 
Office of Science, Office of Nuclear Physics, under contract number  DE-SC0012704  and the LDRD program of Brookhaven 
National Laboratory.




\newpage
%

\renewcommand*{\FigPath}{./WeekIII/03_Kovchegov/}

%
%

%
%

\def\eq#1{{Eq.~(\ref{#1})}}
\def\fig#1{{Fig.~\ref{#1}}}
\def\eqref#1{{(\ref{#1})}}
\newcommand{\ben}{\begin{eqnarray*}}
\newcommand{\een}{\end{eqnarray*}}
\newcommand{\un}[1]{\underline{#1}}
\newcommand{\amu}{\alpha_\mu}
\newcommand{\pd}{\partial}

\renewcommand{\tr}{\mbox{tr}}
\newcommand{\thalf}{\tfrac{1}{2}}
\newcommand{\llangle}{\Big\langle \!\! \Big\langle}
\newcommand{\rrangle}{\Big\rangle \!\! \Big\rangle}

\newcommand{\stackeven}[2]{{{}_{\displaystyle{#1}}\atop\displaystyle{#2}}}
\renewcommand{\lsim}{\stackeven{<}{\sim}}
\renewcommand{\gsim}{\stackeven{>}{\sim}}
\renewcommand{\as}{\alpha_s}

\newcommand{\half}{\frac{1}{2}}

\renewcommand{\bra}[1]{\left\langle #1 \right|}
\renewcommand{\ket}[1]{\left| #1 \right\rangle}

\newcommand{\ul}[1]{\underline{#1}}

\renewcommand{\cc}{\mbox{c.c.}}

\wstoc{Quark and Gluon Helicity at Small $x$ }{Yuri V. Kovchegov}
\title{Quark and Gluon Helicity at Small $x$ }
%
%

\author{Yuri V. Kovchegov}
\index{author}{Kovchegov, Y.}

\address{Department of Physics, The Ohio State
           University, Columbus, OH 43210, USA \\
Email: kovchegov.1@osu.edu
          }

\begin{abstract}
 We determine the small-$x$ asymptotics of the quark and gluon
        helicity distributions in a proton at leading order in
        perturbative QCD at large $N_c$. To achieve this, we first
        evaluate the quark and gluon helicity TMDs at small $x$, simplifying them and relating them to the so-called polarized dipole amplitudes. We then construct and solve novel small-$x$ large-$N_c$ evolution
        equations for the polarized dipole amplitudes. Our main results are the small-$x$ asymptotics
        for the quark helicity distribution
\begin{align}\label{dq_final}
  \Delta q \sim \left(
    \frac{1}{x} \right)^{\alpha_h^q} \ \ \ \mbox{with} \ \ \ \alpha_h^q =
  \frac{4}{\sqrt{3}} \, \sqrt{\frac{\as \, N_c}{2 \pi}} \approx 2.31 \,
  \sqrt{\frac{\as \, N_c}{2 \pi}}
\end{align}
and the small-$x$ asymptotics of the gluon helicity distribution
\begin{align}\label{dG_final}
  \Delta G \sim \left(
    \frac{1}{x} \right)^{\alpha_h^G} \ \ \ \mbox{with} \ \ \ \alpha_h^G =
  \frac{13}{4 \sqrt{3}} \, \sqrt{\frac{\as \, N_c}{2 \pi}}
  \approx 1.88 \, \sqrt{\frac{\as \, N_c}{2 \pi}},
\end{align}
both in the large-$N_c$ limit.
\end{abstract}


\bodymatter


\section{Introduction}

The main long-term goal of this work is to reliably determine the small-$x$ asymptotics
of the quark and gluon helicity PDFs and TMDs, along with the small-$x$ asymptotics of the quark and gluon orbital angular momentum (OAM) distributions. These asymptotics, once established theoretically, can be compared with the data on helicity distributions and OAM to be collected at the EIC: if the comparison is successful, one can use the theoretical predictions to assess the net amount of spin coming from the entire small-$x$ region, thus obtaining the contribution from an essential part of the spin puzzle. 

Below we will review our determination of the small-$x$ asymptotics of the quark and gluon helicity,
and then present the determination of the small-$x$ asymptotics for
the quark and gluon helicity distributions. This talk and these proceedings are
based on \cite{Kovchegov:2015pbl,Kovchegov:2016weo,Kovchegov:2016zex,Kovchegov:2017jxc,Kovchegov:2017lsr,Kovchegov:2018znm}.


\section{Quark Helicity Distribution}

In \cite{Kovchegov:2018znm}, it was shown that the SIDIS quark helicity TMD can be simplified at small-$x$ and written as 
\begin{align} 
  \label{e:qTMD2}
  g_{1L}^{q , S} (x, k_T^2) = \frac{8 N_c}{(2\pi)^6} \sum_f
  \int\limits^1_{\Lambda^2/s} \! \frac{dz}{z} \int \! d^2 x_{01} \, d^2
  x_{0'1} \, e^{- i \ul{k} \cdot (\ul{x}_{01} - \ul{x}_{0'1})} \:
  \frac{\ul{x}_{01} \cdot \ul{x}_{0'1}}{x_{01}^2 \, x_{0'1}^2} \:
  G \left(x_{10}^2 , \frac{z}{x} Q^2 \right)
\end{align}
in the flavor-singlet case.  In the above and
throughout this proceedings contribution, we use light-front coordinates $x^\pm \equiv
\frac{1}{\sqrt{2}} (x^0 \pm x^3)$, denote transverse vectors
$(x_\bot^1 , x_\bot^2)$ by $\ul{x}$ and their magnitudes by $x_T
\equiv | \ul{x} |$, and indicate differences in transverse coordinates
by the abbreviated notation $\ul{x}_{10} \equiv \ul{x}_1 -
\ul{x}_0$ with $x_{10} = |\ul{x}_{10}|$. The center-of-mass energy squared for the scattering process is
$s$, the infrared (IR) transverse momentum cutoff is $\Lambda$, and
$z$ is the smallest of the fractions of the light-cone momentum of the dipole carried
by the quark and anti-quark.  

The impact-parameter integrated ``polarized dipole amplitude" is
\begin{align}
  \label{eq:Gint}
  G(x_{10}^2 , z s) = \int d^2 b_{10} \ G_{10} (z s)
\end{align}
with $\un{b}_{10}  = (\un{x}_1 + \un{x}_0)/2$, where the polarized dipole
scattering amplitude $G_{10} (z s)$ is
\begin{align} 
  \label{eq:Gdef} 
  G_{10} (z s) &\equiv \frac{1}{2 N_c} \,
  \llangle \mbox{T} \, \tr \left[ V_{\ul 0} V_{\ul 1}^{pol \, \dagger} \right] +
  \mbox{T} \, \tr \left[V_{\ul 1}^{pol} V_{\ul 0}^\dagger \right] \rrangle (z s)
\nonumber \\ &\equiv
\frac{z s}{2 N_c} \, \left\langle \mbox{T} \, \tr \left[ V_{\ul 0} V_{\ul 1}^{pol
      \, \dagger} \right] + \mbox{T} \, \tr \left[V_{\ul 1}^{pol} V_{\ul
      0}^\dagger \right] \right\rangle (z s) ,
\end{align}
where the double-angle brackets are defined to scale out the
center-of-mass energy $z s$ between the polarized (anti)quark and the
target.  The ingredients in \eq{eq:Gdef} are the light-cone Wilson lines (in the fundamental representation),
\begin{align} 
  \label{e:Wlineunp} V_{\ul 0} \equiv V_{\ul{x}_0} [+\infty, -\infty]
  \equiv \mathcal{P} \exp\left[ i g \int\limits_{-\infty}^{+\infty}
    dx^- A^+ (x^+ =0, x^- , \ul{x}_0) \right] ,
\end{align}
and the ``polarized Wilson lines" $V_{\ul 1}^{pol} = V_{{\ul x}_1}^{pol}$ with \cite{Kovchegov:2017lsr,Kovchegov:2018znm}
\begin{align} 
  \label{eq:Wpol_all} 
  & V^{pol}_{\un x} = \frac{i g p_1^+}{s} \,
  \int\limits_{-\infty}^\infty d x^- \, V_{\ul x} [+\infty, x^-] \: 
  F^{12} (x^-, \un{x}) \: V_{\ul x} [x^- , -\infty]  - \frac{g^2 \, p_1^+}{s}
  \int\limits_{-\infty}^\infty d x_1^- \, \int\limits_{x_1^-}^\infty d x_2^- \notag  \\ & \times V_{\ul x} [+\infty, x_2^-] \:  t^b \, {\psi}_\beta (x_2^-, {\un x})  \, U_{\ul x}^{ba} [ x_2^-,  x_1^-] \left[ \frac{1}{2} \, \gamma^+ \, \gamma^5 \right]_{\alpha\beta} \, {\bar \psi}_\alpha (x_1^-, {\un x}) \, t^a  \: V_{\ul x} [x_1^- , -\infty]. 
\end{align}

The polarized dipole amplitude obeys the evolution equations which close in the large-$N_c$ and large-$N_c \& N_f$ limits. The large-$N_c$ evolution equations are illustrated diagrammatically in \fig{fig:Gevol_v2} and read \cite{Kovchegov:2015pbl,Kovchegov:2016weo,Kovchegov:2016zex,Kovchegov:2017jxc,Kovchegov:2017lsr,Kovchegov:2018znm}
\begin{subequations}\label{GNc}
\begin{align}\label{GNc1}
& G_{10} (z) = G_{10}^{(0)} (z) + \frac{\alpha_s \, N_c}{2 \pi} \int\limits_{\frac{1}{s \, x_{10}^2}}^{z}
\frac{dz'}{z'} \int\limits^{x_{10}^2}_\frac{1}{z' s} \frac{d x_{21}^2}{x_{21}^2} \: \left[ \Gamma_{10,21} (z') + 3 \, G_{21} (z')  \right], \\
\label{GNc2}
& \Gamma_{10,21} (z') = \Gamma_{10,21}^{(0)} (z') + \frac{\alpha_s \, N_c}{2 \pi} \int\limits_{\min \{ \Lambda^2, \frac{1}{x_{10}^2} \} / s }^{z'}
\frac{dz''}{z''} \int\limits^{\min \{ x_{10}^2, x_{21}^2 z'/z'' \} }_\frac{1}{z'' s} \frac{d x_{32}^2}{x_{32}^2} \\ & \hspace*{5cm} \times \left[ \Gamma_{10,32} (z'') + 3 \, G_{32} (z'')  \right]. \notag
\end{align}
\end{subequations}
Here we had to introduce an auxiliary function $\Gamma$, termed the ``neighbor
dipole amplitude'', in which further evolution is constrained by the
lifetime of an adjacent dipole \cite{Kovchegov:2015pbl}.
The equations \eqref{GNc} resum powers of $\as \, \ln^2 (1/x)$: this is the double logarithmic approximation (DLA).

\begin{figure*}
\centering
\includegraphics[width= 0.6 \textwidth]{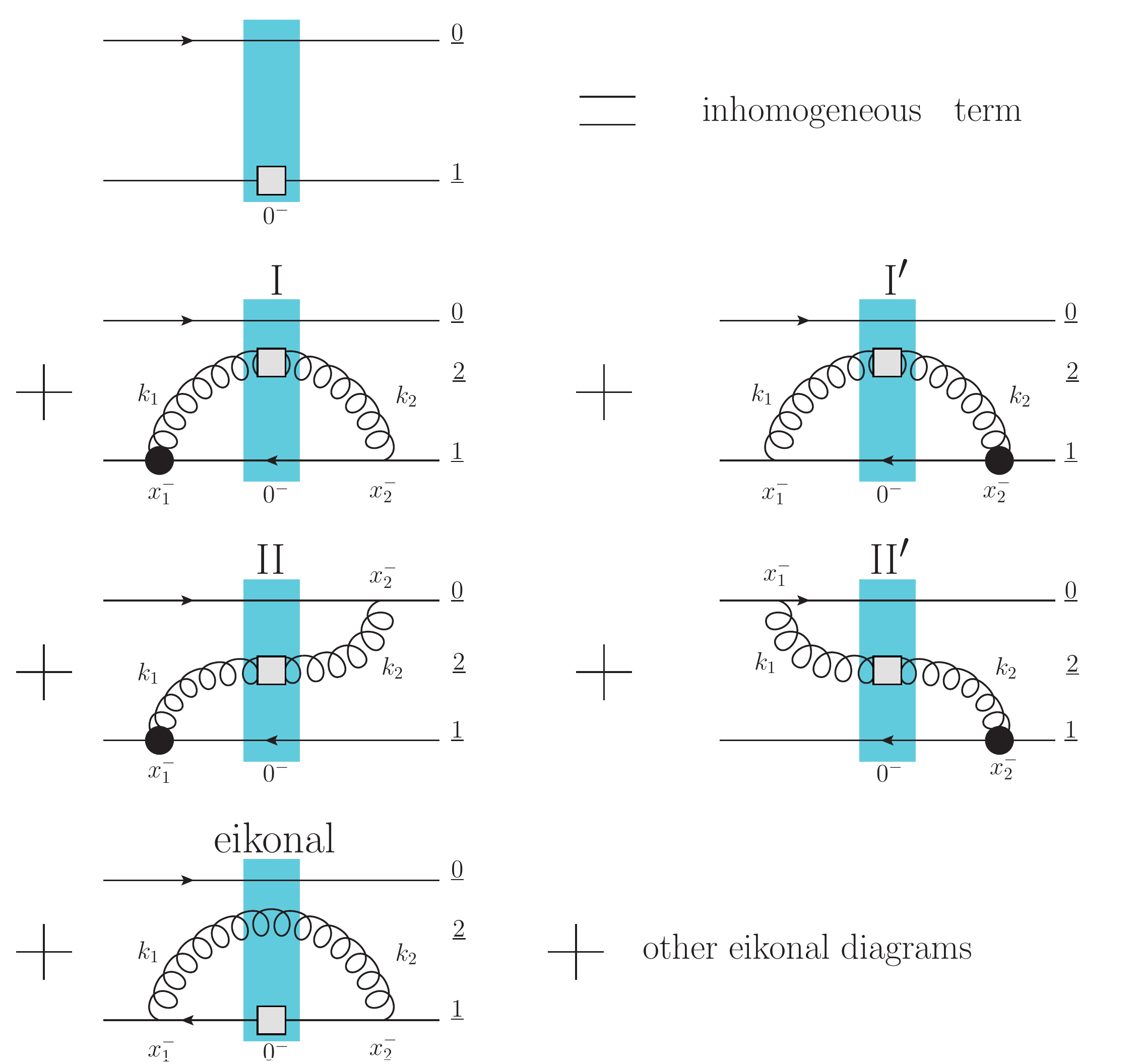} 
\caption{Diagrams illustrating the evolution of the polarized dipole
  amplitude $G_{10} (z s)$ at large $N_c$  The blue band represents the shock wave (the proton), the black vertex represents the sub-eikonal
  operator insertion(s) from \protect\eqref{eq:Wpol_all}, and
  the gray box represents the polarized Wilson line.}
\label{fig:Gevol_v2}
\end{figure*}

Equations~\eqref{GNc} were solved numerically in
\cite{Kovchegov:2016weo} and analytically in \cite{Kovchegov:2017jxc} yielding
\begin{align}
  G(x_{10}^2 , z s) \propto  (z s \, x_{10}^2)^{\alpha_h^q} 
\end{align}
with
\begin{align} 
  \label{M:ahel}
  \alpha_h^q = \frac{4}{\sqrt{3}} \sqrt{\frac{\alpha_s N_c}{2\pi}}
  \approx 2.31 \sqrt{\frac{\alpha_s N_c}{2\pi}} .
\end{align}
This leads to \eq{dq_final} above. 


\section{Gluon Helicity Distribution}

The small-$x$ asymptotics of the gluon helicity can be determined in a
similar way \cite{Kovchegov:2017lsr}. Starting with the operator
definition of the gluon dipole helicity TMD and simplifying it at small $x$ we arrive at \cite{Kovchegov:2017lsr}
\begin{align}   
  g_{1L}^{G \, dip} (x, k_T^2) &= \frac{- 8 i \, N_c}{g^2 (2\pi)^3} \,
  \int d^2 x_{10} \, e^{i \un{k} \cdot \un{x}_{10}} \: k_\bot^i
  \epsilon_T^{i j} \: \left[ \int d^2 b_{10} \, G_{10}^j (z s =
    \tfrac{Q^2}{x}) \right] 
\end{align}
where we have defined another dipole-like polarized operator
\begin{align}   \label{eq:Gidef}
G^i_{10} (z s) \equiv \frac{1}{2 N_c} \,  
\left\langle \tr \left[ V_{\ul 0} (V_{\ul 1}^{pol \, \dagger})_\bot^i
\right] + \cc \right\rangle (z s)
\end{align}
with a different ``polarized Wilson line"  
\begin{align} 
\label{M:Vpol2} 
(V_{\ul x}^{pol})_\bot^i &\equiv
  \int\limits_{-\infty}^{+\infty} dx^- \, V_{\ul x} [+\infty, x^-] \:
  \left( i g \, p_1^+ \, A_\bot^i (x) \right) \: V_{\ul x} [x^- ,
  -\infty].
\end{align}
Again one can write evolution equations for $G^i_{10} (z s)$ which close in the large-$N_c$ and large-$N_c \& N_f$ limits. The solution of the large-$N_c$ equations gives \cite{Kovchegov:2017lsr}
\begin{align} 
    \label{G3}
    G_2 ( x_{10}^2 , z s) \propto (z s \, x_{10}^2)^{\alpha^G_h}
\end{align}
with    
\begin{align} \label{e:Gint} \alpha^G_h = \frac{13}{4 \sqrt{3}} \,
  \sqrt{\frac{\as \, N_c}{2 \pi}} \approx 1.88 \, \sqrt{\frac{\as \,
      N_c}{2 \pi}} ,
\end{align}
leading to \eq{dG_final} above. 


\section{Conclusions}

In conclusion let us point out that we expect the asymptotics \eqref{dq_final} and \eqref{dG_final} to be modified qualitatively in the large-$N_c \& N_f$ limit, i.e., when quarks are included. For OAM distributions at small $x$ the analysis \cite{Kovchegov:2019rrz} similar to the above gives $L_{q + \bar{q}} (x, Q^2)  = - \Delta \Sigma (x, Q^2)$ and $L_G (x, Q^2)  = \left( \frac{\alpha_h^q}{4} \, \ln \frac{Q^2}{\Lambda^2} \right) \, \Delta G (x, Q^2)$ in the large-$N_c$ limit. 

%
\section*{Acknowledgments}
%

The author would like to thank Daniel Pitonyak and Matthew Sievert for their collaboration on this project. 
This material is based upon work supported by the U.S. Department of
Energy, Office of Science, Office of Nuclear Physics under Award
Number DE-SC0004286.





\newpage

\renewcommand*{\FigPath}{./WeekIII/02_Lin/Figs}

\wstoc{Progress and Prospects of Lattice-QCD Parton Distribution Functions}{Huey-Wen Lin}
\title{
Progress and Prospects of Lattice-QCD Parton Distribution Functions 
}

\author{Huey-Wen Lin$^*$}
\index{author}{Lin, H.}

\address{Department of Physics and Astronomy, and Computational Mathematics, Science \& Engineering, Michigan State University, \\
East Lansing, MI 48824, USA\\
$^*$E-mail: hwlin@pa.msu.edu\\
https://web.pa.msu.edu/people/hwlin/}

\begin{abstract}
There have been rapid developments in parton distribution functions (PDFs) using lattice QCD for both precision moments and direct calculation of the Bjorken-$x$ dependence. 
In this talk, I show some progress along these directions and show some examples of how lattice-QCD calculations can play a significant role in improving our understanding of PDFs in the future.  
\end{abstract}


\bodymatter

Parton distribution functions (PDFs) are not only fundamental properties of quantum chromodynamics (QCD) but also are key inputs to predict cross sections in high-energy scattering experiments and to aid new-physics searches at the Large Hadron Collider. A PDF, say $q(x)$, describes
the probability of finding the parton (such as 
quarks or gluons) carrying a fraction $x$ of the longitudinal momentum
within a hadron.
Calculating the $x$-dependence of PDFs from first principles has long been a holy grail for nuclear and high-energy physics. In modern parton physics, the PDFs are defined from the lightcone correlations of quarks and gluons in the hadron, so they involve strong infrared dynamics and can only be solved by nonperturbative methods such as lattice QCD (LQCD). However, the direct calculation of PDFs on a Euclidean lattice has been extremely difficult, because the real-time dependence of the lightcone makes it infeasible to extract them from lattice simulations with imaginary time. Early lattice-QCD studies based on the operator product expansion (OPE) could only access the lower moments of the PDF.
 A similar situation also occurs in lattice calculations of other parton observables, such as the distribution amplitudes (DAs) and generalized parton distributions (GPDs). For many years, progress in LQCD hadron structure was done by pushing moment calculations to the physical pion mass and expanding the study of LQCD systematics.

Recently, a new idea has
been proposed that circumvents the limitations of the moment approach~\cite{Ji:2013dva_171,Ji:2014gla_24}: ``large-momentum effective theory'' (LaMET). 
In this approach, one computes a time-independent spatially displaced matrix element that can be connected to the PDF. A convenient choice for leading-twist PDFs is to take the
hadron momentum and quark-antiquark separation to be along the $z$ direction 
%
$h_\Gamma(z,p_z) 
= 
\frac{1}{4 p_z}\sum_{s=1}^2\left\langle p,s\right| \bar{\psi}(z)\Gamma e^{ig\int_0^z
A_z(z^\prime) dz^\prime} \psi(0) \left| p,s\right\rangle,
$
%
where $p_z$ is the hadron momentum boosted in the $z$ direction, $s$ its spin, and $z$ is the separation of the quark and antiquark fields $\bar\psi$ and $\psi$. 
There are multiple choices of operator in this framework that will recover the same lightcone PDFs when the large-momentum limit is taken. 
%

\begin{figure}[htbp]
\includegraphics[width=.95\textwidth]{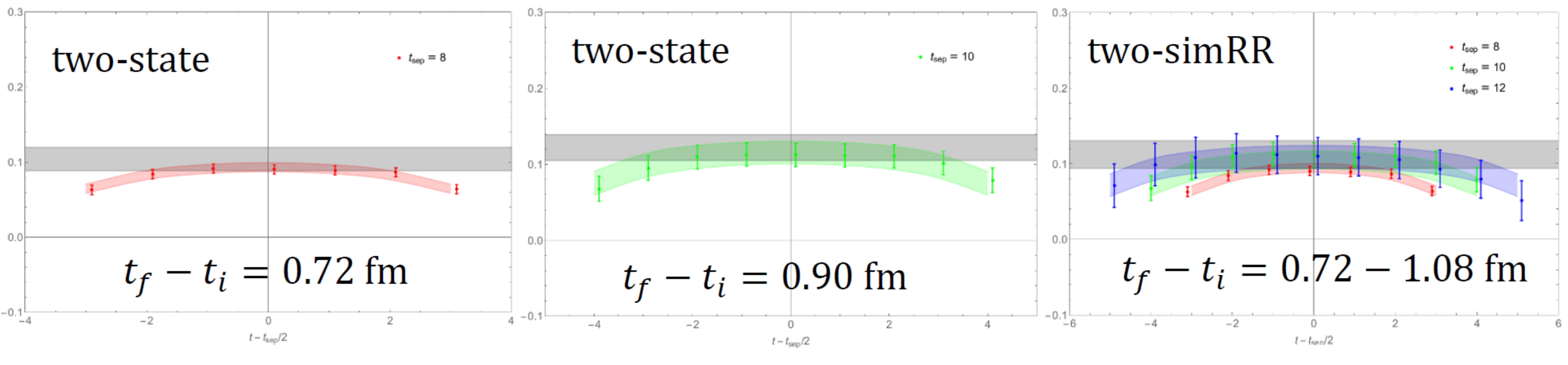}
\caption{
This figure shows the consistency between two-state fits truncated at different terms and different $t_\text{sep}$ data for the case of transversity matrix elements at $z=3$, $P_z=2.6$~GeV on physical pion mass ensembles.}
\label{fig:fit-checks}
\end{figure}

We perform lattice calculations of the bare isovector quark unpolarized, helicity, and transversity quasi-PDFs using clover valence fermions~\cite{Rajan:2017lxk,Bhattacharya:2015wna,Bhattacharya:2015esa,Bhattacharya:2013ehc} on an ensemble of gauge configurations with lattice spacing $a=0.09$~fm, box
size $L\approx 5.8$~fm, and with pion mass $M_\pi \approx 135$~MeV and $N_f=2+1+1$ (degenerate up/down, strange and charm) flavors of highly improved staggered dynamical quarks (HISQ)~\cite{Follana:2006rc} generated by MILC Collaboration~\cite{Bazavov:2012xda}. 
We use Gaussian momentum smearing~\cite{Bali:2016lva} for the quark field to increase the overlap of the lattice sources with the ground state of the large-boost nucleon. 
%
For the nucleon matrix elements of $\hat{O}(z,a)$ at a given boost momentum, 
$\widetilde{h}(z,P_z,a)$, 
we extract the ground-state matrix elements from each three-point correlator, $C_\Gamma^{(3\text{pt})} (P_z, t, t_\text{sep})$ by fitting the following form:
\begin{multline}
C^\text{3pt}_{\Gamma}(P_z,t,t_\text{sep}) =
   |{\cal A}_0|^2 \langle 0 | \mathcal{O}_\Gamma | 0 \rangle  e^{-E_0t_\text{sep}}+|{\cal A}_1|^2 \langle 1 | \mathcal{O}_\Gamma | 1 \rangle  e^{-E_1t_\text{sep}} \nonumber\\
   +|{\cal A}_1||{\cal A}_0| \langle 1 | \mathcal{O}_\Gamma | 0 \rangle  e^{-E_1 (t_\text{sep}-t)} e^{-E_0 t} +|{\cal A}_0||{\cal A}_1| \langle 0 | \mathcal{O}_\Gamma | 1 \rangle  e^{-E_0 (t_\text{sep}-t)} e^{-E_1 t} + \ldots \,,
  \label{eq:3ptfit}
\end{multline}
where 
the operator is inserted at time $t$, and the nucleon state is annihilated at the sink time
$t_\text{sep}$, which is also the source-sink separation (after shifting source time to zero).  
The state $|0\rangle$ represents the ground state and $|n\rangle$ with $n
> 0$ the excited states.
In our two-state fits,
the amplitudes ${\cal A}_i$ and the energies $E_i$ are functions of $P_z$ and can be obtained from the corresponding two-point correlators.  
Fig.~\ref{fig:fit-checks} shows one of the many studies of excited-state contamination, performing fits with and without the 
$\langle 1 | \mathcal{O}_\Gamma | 1 \rangle$  contribution (labeled as ``two-simRR'' and ``two-sim'', respectively) and 
using data from different source-sink separations $t_\text{sep}$~\cite{Chen:2018xof,Lin:2018qky_24,Liu:2018hxv_24}.

\begin{figure}[htbp]
\includegraphics[width=.32\textwidth]{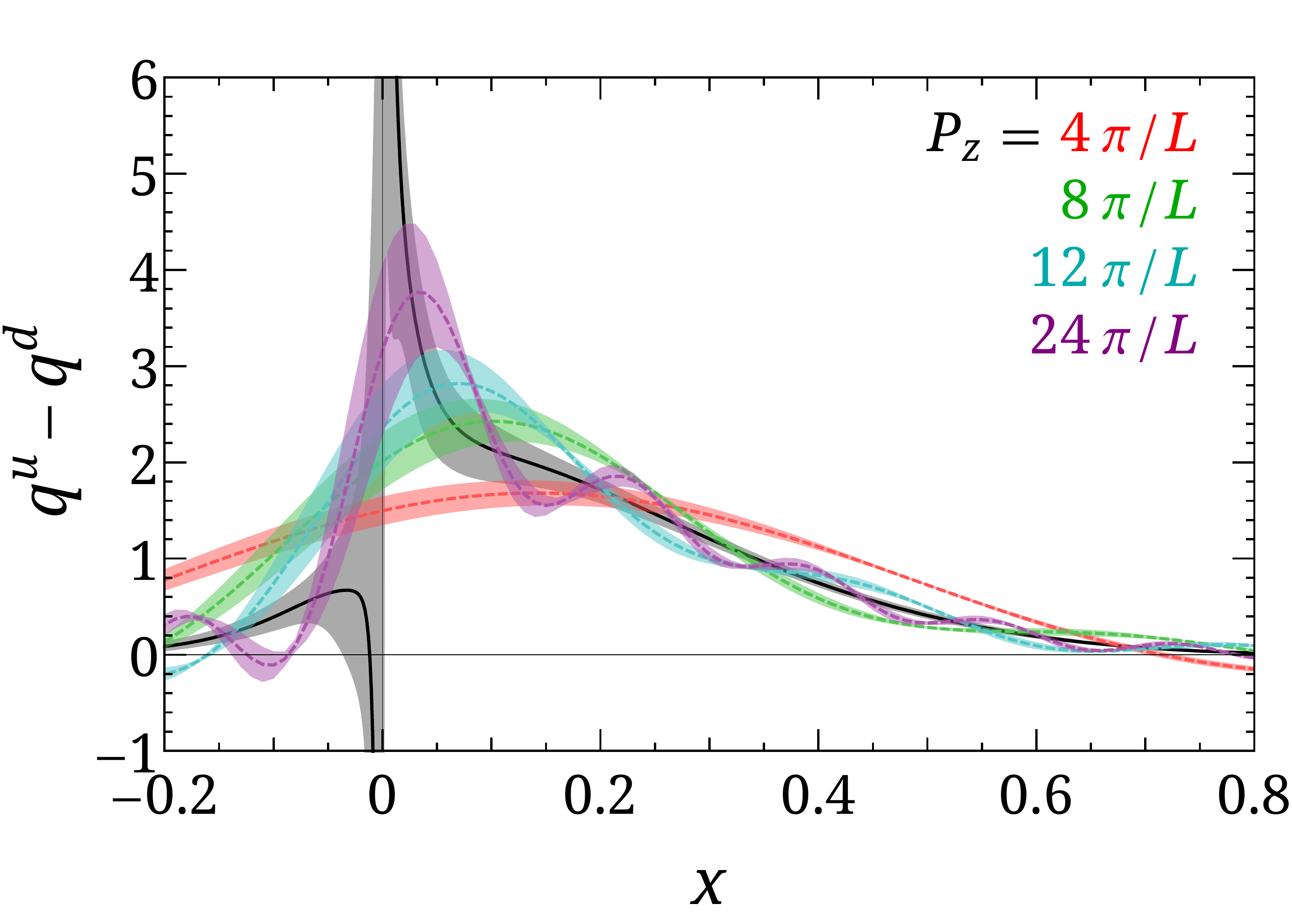}
\includegraphics[width=.32\textwidth]{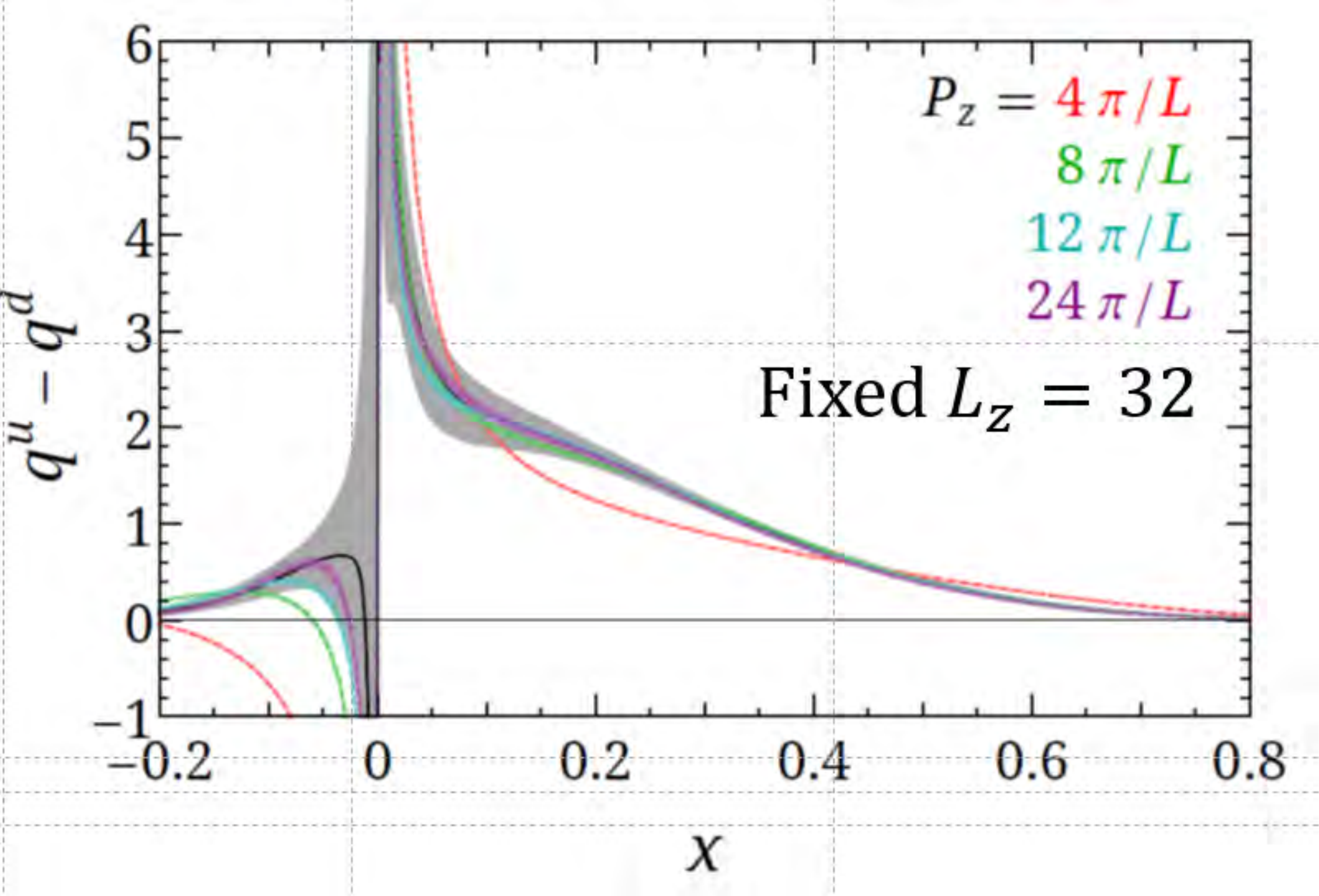}
\includegraphics[width=.32\textwidth]{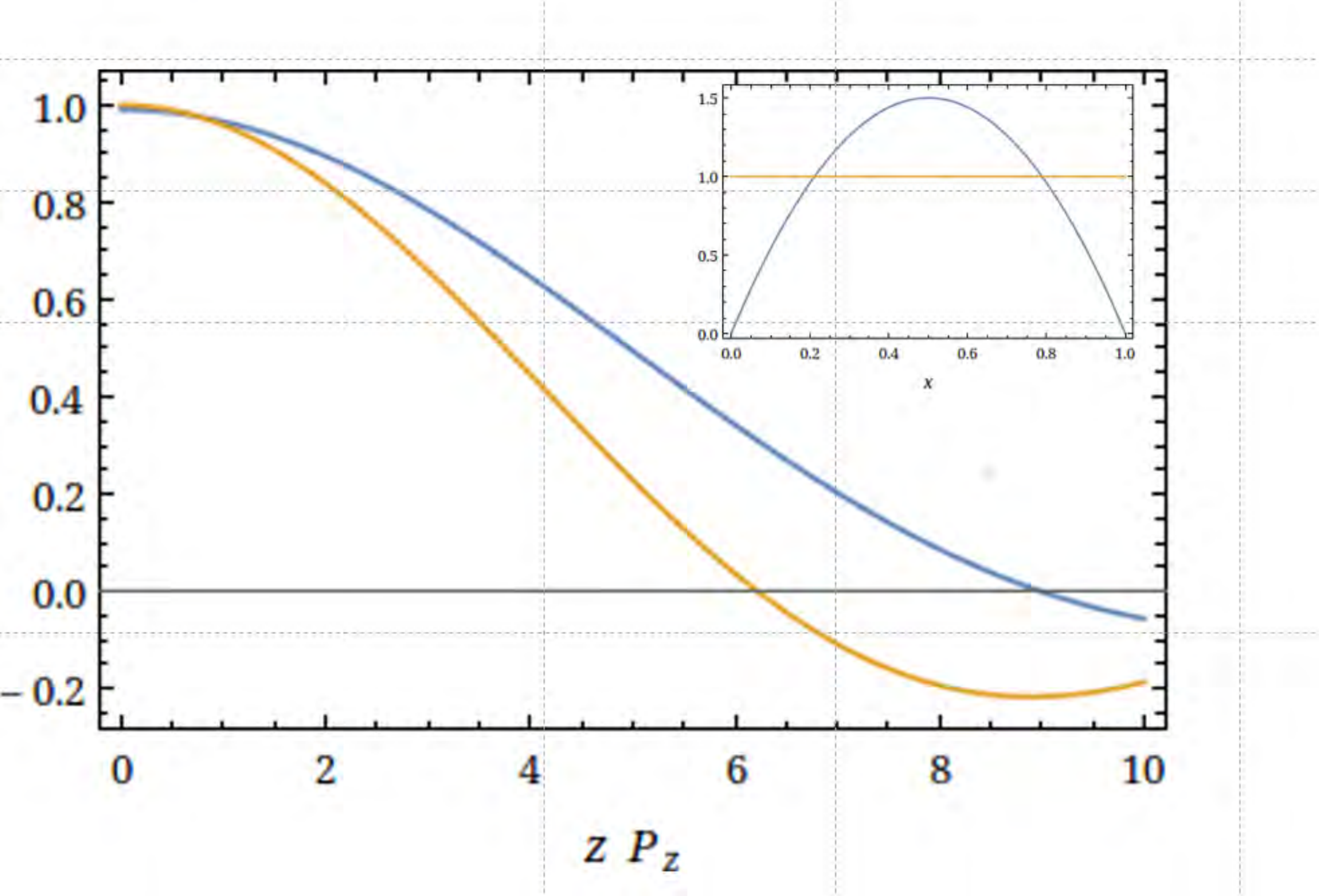}
\caption{
(Left) Fourier truncation at finite $z P_z$ can create oscillation artifacts in the PDF reconstruction. 
(Middle) Boost-momentum dependence in reconstruction of the PDF. Any line deviating from the original input PDF indicates the systematic error in the PDF calculation. 
(Right) Real matrix elements in coordinate ($zP_z$) space with 2 very distinct PDF distributions in $x$ space. 
}
\label{fig:PDFerrors}
\end{figure}

We then calculate the nonperturbative renormalization (NPR) factor $\widetilde{Z}(z,p^R_z,\mu_R,a)$ from the amputated Green function of $\hat O(z,a)$ with a similar procedure~\cite{Liu:2018uuj_24,Liu:2018hxv_24}, where $p^R_z$ and $\mu_R$ are the Euclidean quark momentum in the $z$-direction and the off-shell quark momentum, respectively.
The bare matrix element of $\hat{O}(z,a)$, $\widetilde{h}(z,P_z,a)$, 
has ultraviolet (UV) power and logarithmic divergences as $a\to 0$ and must be nonperturbatively renormalized to have a well defined continuum limit. 
%
Next, we need to Fourier transform the $h_R(z,P_z, p_z^R,\mu_R)$ into $x$-space to obtain the quasi-distribution $\widetilde{q}(x, P_z, p_z^R, \mu_R)$:
%
$\widetilde{f}(x,1/a,p_z)  
= \int \frac{dz}{2\pi} e^{-i x z p_z} p_z h_\Gamma(z,p_z).$
%
As originally pointed out in 2017 and demonstrated using CT14 NNLO~\cite{Dulat:2015mca} at 2~GeV, a naive Fourier transform from momentum-space $x$ to coordinate space $z$ and then back suffers an inverse problem~\cite{Lin:2017ani} (see the left-hand side of Fig.~\ref{fig:PDFerrors}). 
The oscillation is less noticeable if calculation stays in the small-$P_z$ region, as shown in the pink band in the figure. 
This means that since the lattice calculation has finite displacement $z$ in the nonlocal operator and cannot actually use infinitely boosted momentum, a straightforward Fourier transform produces truncation effects, resulting in unphysical oscillatory behavior, as observed in earlier works~\cite{Chen:2017mzz_24,Green:2017xeu_24}. 
The antiquark and small-$x$ regions suffer the maximum deformation.  
Two ideas (``filter'' and ``derivative'' methods)~\cite{Lin:2017ani} were originally proposed to remove this biggest systematic uncertainty in the LaMET approach to studying $x$-dependent hadron structure: Fourier-transformation truncation. When not assuming a parametrization form, this determines the shape of the PDF. The first lattice PDF at physical pion mass was used to demonstrate how the proposed methods improve real-world lattice calculations. 
A third method was proposed in late 2017, modifying the Fourier transformation in LaMET using a single-parameter Gaussian weight~\cite{Chen:2017lnm}. In 2019, another three methods were proposed~\cite{Karpie:2019eiq}.
Following the recent work~\cite{Chen:2018xof,Lin:2018qky_24,Liu:2018hxv_24}, we adopt the simple but effective ``derivative'' method:
%
$\tilde{Q}(x,P_z,p_z^R,\mu_R) = i\int_{-z_\text{max}}^{+z_\text{max}} \!\!\! dz \ e^{i x P_z z}  \tilde{h}'_R(z,P_z,p_z^R,\mu_R)/x,
$
%
where $\tilde{Q}$ is quasi-PDF ($q(x)$, $\Delta q(x)$ and $\delta q(x)$ respectively), and 
$\tilde{h}'_R$ is the derivative of the renormalized matrix elements for the corresponding operator. 
One immediately notices that when $P_z$ is small, the sea quark asymmetry would come out of lattice calculation with the wrong sign, which is exactly what was seen in the low-$P_z$ PDF calculations~\cite{Lin:2017ani,Alexandrou:2018pbm_24}, in addition to missing the small-$x$ region. 
There are a number of proposals to avoid the Fourier transformation by working in position space; this would work in an ideal world when there is sufficiently precise data throughout the large-$zP_z$ region. However, in reality, the lattice data taken in the small-$zP_z$ region is not precise enough to even discern whether the parton distribution is flat across all $x$; one loses sensitivity to two very distinct distributions in $x$-space, which now become very similar in $zP_z$ space. 
Furthermore, one still need large $zP_z$ to reliably obtain the distribution in the small-$x$ region. It would be great to have a systematic way to demonstrate the lattice data inputs in $zP_z$ space.


\begin{figure}[htbp]
\includegraphics[width=.32\textwidth]{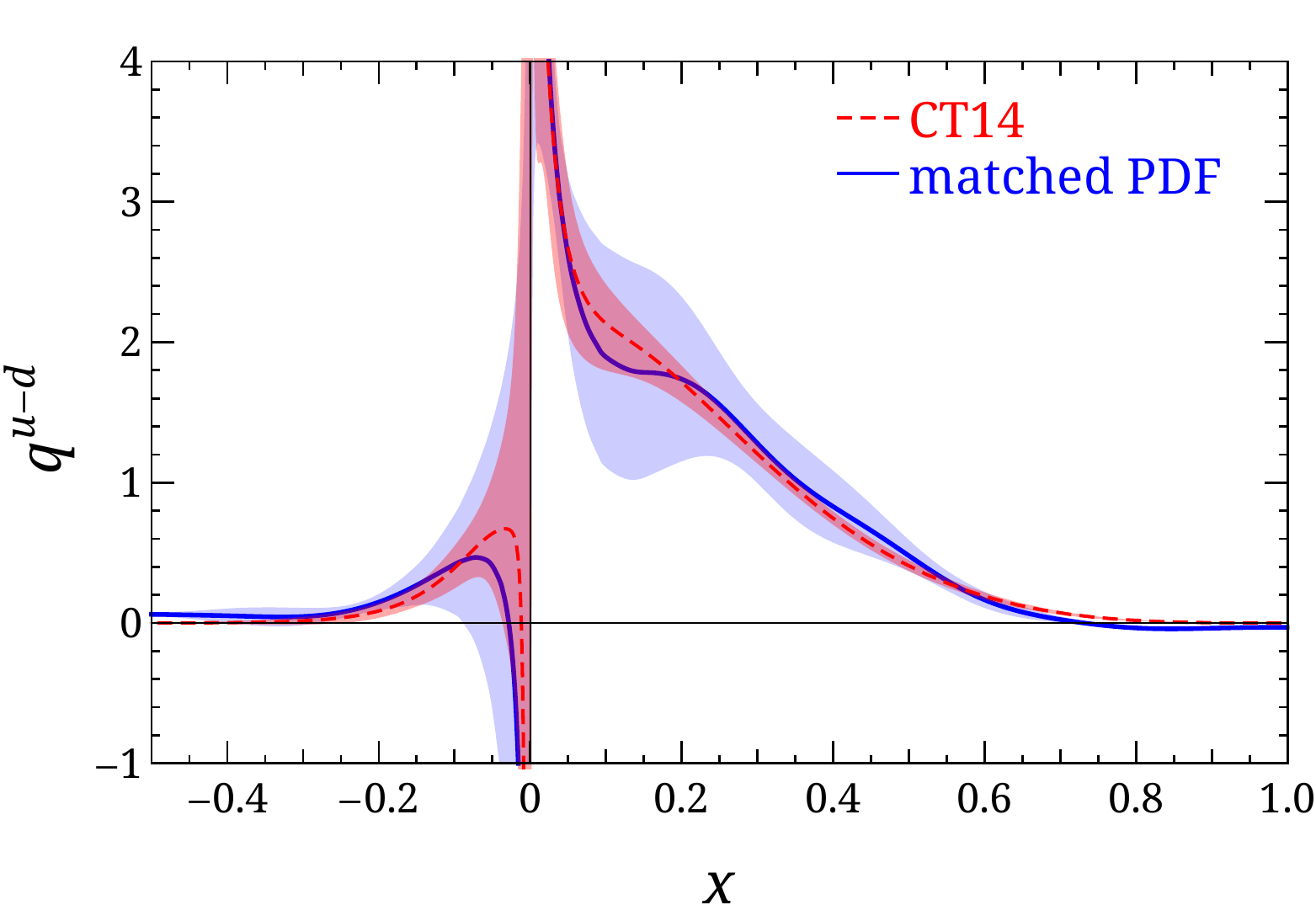}
\includegraphics[width=.32\textwidth]{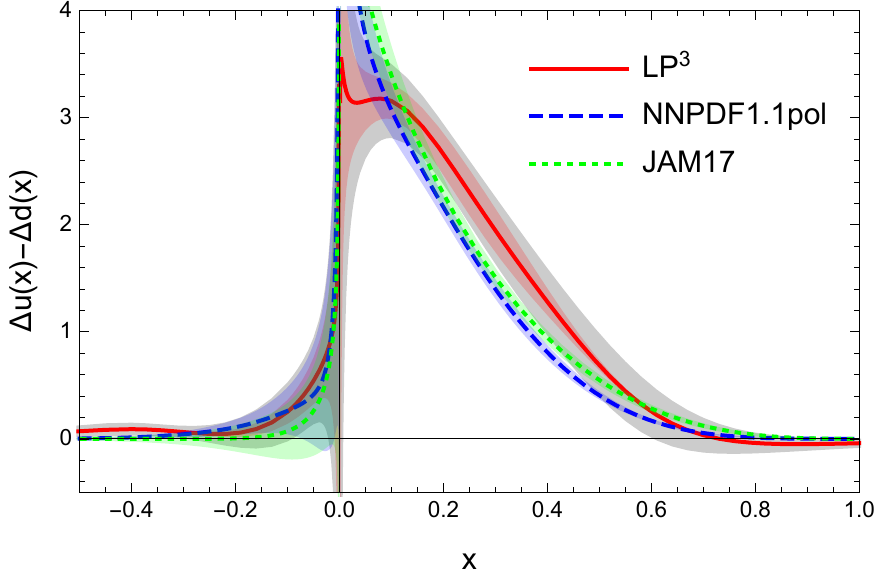}
\includegraphics[width=.32\textwidth]{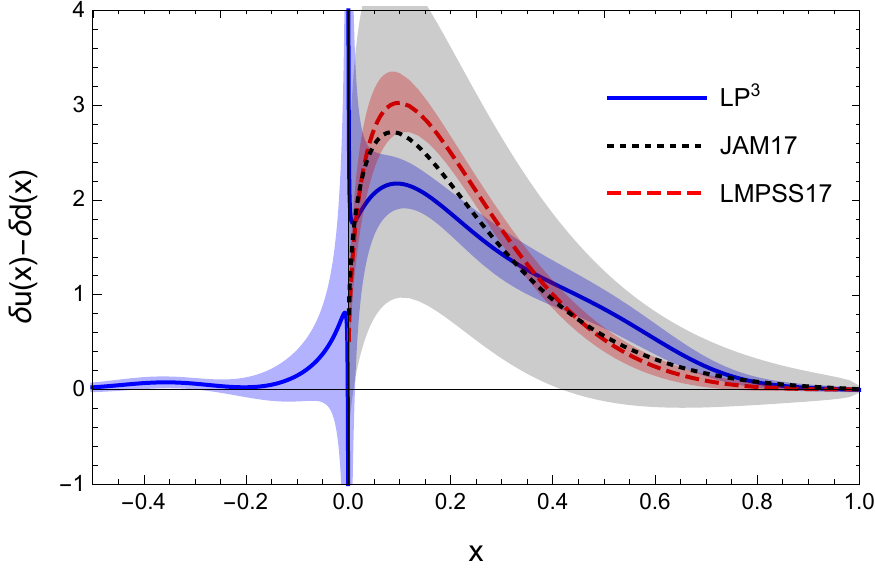}
\caption{
LP$^3$'s isovector unpolarized (left), helicity (middle) and transversity (right) PDFs renormalized at 3~GeV 
in comparison with global-fit PDFs}
\label{fig:finalPDF}
\end{figure}

Once we have the quasi-PDFs, we can relate them to the true lightcone PDFs using the matching condition
%
$\widetilde{q}(x,\Lambda ,p_z) = 
  \int_{-1}^1 \frac{dy}{\left\vert y\right\vert} 
    Z\left( \frac{x}{y}, \frac{\mu}{p_z}, \frac{\Lambda}{p_z}\right)_{\mu^2 = Q^2} q(y,Q^2) +
  \mathcal{O}\left( \frac{\Lambda_\text{QCD}^2}{p_z^2},\frac{M^2}{p_z^2}\right), 
$
%
where $\mu$ is the renormalization scale,
$Z$ is a matching kernel and $M$ is the hadron mass.
Here the $\mathcal{O}\left(M^2/p_z^2\right)$ terms are target-mass corrections 
and the $\mathcal{O}\left(\Lambda_\text{QCD}^2/p_z^2\right)$ terms are 
higher-twist effects, both of which are suppressed at large hadron momentum. 
Early exploratory works have shown great promise in obtaining quantitative results for the unpolarized, helicity and transversity quark and antiquark distributions~\cite{Lin:2014gaa,Lin:2014yra,Lin:2014zya,Chen:2016utp}.
The first LaMET PDFs at physical pion mass~\cite{Lin:2017ani} were done using small momentum ($P_z^\text{max}\approx 1.3$~GeV); as described earlier, one expects the wrong sign of sea-flavor asymmetry to be seen in the Fourier transformation.  
Figure~\ref{fig:finalPDF} shows our newer PDF results at larger momenta, calculated above 2~GeV, and then renormalized at 3~GeV. 
The errorband displayed here includes the systematic error coming from variations in the renormalization scale, $zP_z$ in the Fourier transformation, estimation of lattice-spacing and finite-volume effects from other nucleon matrix studies and the matching formula approximation. 
As expected from the Fourier transformation study, the sea-flavor asymmetry is recovered with increased momentum. 
In the positive isovector quark region, our results agree nicely with CT14~\cite{Dulat:2015mca}, which is consistent with NNPDF3.1~\cite{Ball:2017nwa} and CJ15~\cite{Accardi:2016qay}, up to the small-$x$ region where even larger $zP_z$ data is required for lattice calculation to have control over these regions. 
The middle plot of Fig.~\ref{fig:finalPDF} shows LP$^3$'s isovector quark helicity PDF~\cite{Lin:2018qky_24} matched in $\overline{\text{MS}}$-scheme at
scale $\mu=3$~GeV, extracted from LaMET at the largest proton momentum (3~GeV),
compared with fits by NNPDFpol1.1~\cite{Nocera:2014gqa} and JAM~\cite{Ethier:2017zbq}. The red band contains
statistical error, while the gray band also includes systematics.
The right-hand side of Fig.~\ref{fig:finalPDF} shows LP$^3$'s proton isovector transversity PDF~\cite{Liu:2018hxv_24} at renormalization scale $\mu=\sqrt{2}$~GeV ($\overline{\text{MS}}$ scheme), extracted from lattice QCD and LaMET at $P_z=3$~GeV, compared with global fits by JAM17 and LMPSS17~\cite{Lin:2017stx_69}.
The blue error band includes statistical errors (which fold in the excited-state uncertainty) and systematics mentioned in the unpolarized PDF cases.



\newpage
%


\wstoc{Transverse Force Tomography}{Fatma Aslan, Matthias Burkardt, Marc Schlegel}

\title{Transverse Force Tomography}

\author{Fatma Aslan}
\index{author}{Aslan, F.}
\address{Physics Department, New Mexico State University,\\
Las Cruces, NM 88003, U.S.A.\\
$^*$E-mail: fpaslan@nmsu.edu\
}

\author{Matthias Burkardt}
\index{author}{Burkardt, M.}

\address{Physics Department, New Mexico State University,\\
Las Cruces, NM 88003, U.S.A.\\
$^*$E-mail: burkardt@nmsu.edu}

\author{Marc Schlegel}
\index{author}{Schlegel, M.}

\address{Physics Department, New Mexico State University,\\
Las Cruces, NM 88003, U.S.A.\\
$^*$E-mail: schlegel@nmsu.edu}

\begin{abstract}
While twist-2 GPDs allow for a determination of the distribution of partons on the transverse plane, twist-3 GPDs contain quark-gluon correlations that provide information about the average transverse color Lorentz force acting on quarks. As an example, we use the nonforward generalization of $g_T(x)$, to illustrate how twist-3 GPDs can  provide transverse position information about that force.

\end{abstract}


\bodymatter

\section{Transverse Imaging}\label{aba:sec1}
For a transversely localized nucleon state, such as (${\cal N}$ is a normalization factor)
\begin{equation}
|{\bf R}_\perp =0,p^+,\Lambda\rangle \equiv
{\cal N} \int d^2{\bf p}_\perp |{\bf p}_\perp ,p^+,\Lambda\rangle,
\end{equation}
which has its transverse center of longitudinal momentum at the (transverse) origin, one can define transverse charge distributions as
\begin{eqnarray}\rho_{\Lambda^\prime\Lambda} ({\bf b}_\perp) 
&\equiv& \langle {\bf R}_\perp =0,p^+\!,\Lambda^\prime| \bar{q}({\bf b}_\perp)
\gamma^+ q({\bf b}_\perp)
|{\bf R}_\perp =0,p^+\!,\Lambda\rangle
\label{eq:F}\\
&=&\left|{\cal N}\right|^2 
\int\! \!\!d^2{\bf p}_\perp\!\!\!\int \!\!\!d^2{\bf p}_\perp^\prime
\langle {\bf p}_\perp^\prime,p^+\!,\Lambda^\prime| \bar{q}(0)
\gamma^+q(0)
|{\bf p}_\perp,p^+\!,\Lambda\rangle e^{i{\bf b}_\perp\cdot({\bf p}_\perp
-{\bf p}_\perp^\prime)}
\nonumber\\
&=& \int d^2{\bf \Delta_\perp} F_{\Lambda^\prime \Lambda}(-{\bf \Delta_\perp^2}) e^{-i{\bf b}_\perp \cdot {\bf \Delta_\perp}}.\nonumber
\end{eqnarray}
Here $\Lambda$, $\Lambda^\prime$ are the polarization of the target states, and $F_{\Lambda^\prime,\Lambda}$ is a superposition of the Dirac and Pauli form factors - details are depending on the polarizations. Note that in the $2^{nd}$ step in Eq. (\ref{eq:F}) it was crucial that the matrix element only depends in the $\Delta_\perp$, but not on the overall ${\bf P}_\perp=\frac{1}{2}({\bf p}_\perp+{\bf p}^\prime_\perp)$ - otherwise it would not be possible to factor out the ${\bf P}_\perp$ integration and cancel it against $|{\cal N}|^2$.

As a side remark, when one tries to localize a state in 3 dimensions, a factorization of the ${\vec P}$-integration is not possible due to various relativistic factors. As a result, a similar procedure in 3 dimension fails and the physical interpretation of 3-dimensional Fourier transforms of form factors as charge distribution in position space is flawed due to relativistic corrections when one looks at details smaller than the Compton wavelength of the target.

Very similar steps can be repeated for $x$-dependent distributions resulting in the position space interpretation for Generalized Parton Distributions (GPDs) \cite{mb:GPD}.

\section{Transverse Force}
The $x^2$ moments of the genuine twist-3 part of twist-3 PDFs are related to forward matrix elements of quark-gluon-quark correlations. 
For example, the polarized twist 3 PDF $g_2(x)$ can be cleanly separated from the leading twist contribution $g_1$ by measuring
the longitudinal (beam) - transverse (target) double-spin asymmetry in DIS. After subtracting the Wandzura-Wilczek contribution
$g_2^{WW}(x) \equiv -g_1(x) +\int_x^1 \frac{dy}{y}g_1(y)$, one is left with the genuine twist 3 part $\bar{g}_2(x)=g_2(x)-g_2^{WW}(x)$ (here we neglect quark mass contributions), whose $x^2$ moment reads \cite{qGq}
\begin{equation}
d_2\equiv 3\int dx\,x^2{ \bar{g}_2(x)} =
\frac{1}{2M{P^+}^2S^x} \left\langle P,S \left|
\bar{q}(0) \gamma^+gG^{+y}(0)q(0)\right|P,S\right\rangle
\label{eq:qFq}
\end{equation}
To understand the physical meaning of this correlator, we decompose the light-cone component of the gluon field strength tensor appearing in (\ref{eq:qFq}) in terms of color electric and magnetic fields 
\begin{equation}
\sqrt{2}G^{+y} = G^{0y}+G^{zy} = -{ E^y}+{ B^x}
=-\!\left({{\vec E}} + {\vec v}
\times {{\vec B}}\right)^y\end{equation}
for a quark that moves with the velocity of light in the $-\hat{z}$ direction - which is exactly what the struck quark does in a DIS experiment after having absorbed the virtual photon${\vec v}=(0,0,\!\!-1)$. Since the Gluon field is correlated with the quark density, this means that $d_2$ has the physical interpretation as the average color-Lorentz force acting on a quark in a DIS experiment right after (since the matrix element is local) having absorbed the virtual photon\cite{mb:force}. This is the same final state interaction (FSI) force that also
produces single-spin asymmetries.

\section{Transverse Force Tomography}
Since $d_2$ arises as the expectation value of a $\bar{q}Gq$ corellator in a plane wave state it can only provide volume-averaged information. In order to obtain position information, a momentum transfer must be involved. This is one of the motivations for
studying twist 3 GPDs. After subtracting Wandzura-Wilzek type terms and surface terms (terms that involve a twist 2 contribution multiplied by a momentum transfer),  $x^2$ moments of twist 3 GPDs allow determining non-forward matrix elements of $\bar{q}Fq$ correlators. This motivates parameterizing these matrix elements in terms of Lorentz invariant generalized form factors.

Lorents invariance implies that the matrix elements of $\bar{q}(0)\gamma^\rho G^{\mu \nu}(0)q(0)$ can be parameterized in terms of 8 generalized form factors \cite{abs2}

In the relevant case of $\bar{q}(0)\gamma^+ G^{+i}(0)q(0)$ this reduces to five form factors \cite{abs1}

\begin{eqnarray}\label{FormFactors}
\langle p',\lambda'|\bar{q}(0)\gamma^{+}igG^{+i}(0) q(0)|p,\lambda\rangle
&=&\overline{u}(p',\lambda')\Big\{\dfrac{1}{M^2}[P^{+}\Delta_{\perp}^i-P^{\perp}\Delta^{+})]\gamma^+\Phi_1(t)\\ \nonumber +\dfrac{P^+}{M}i\sigma^{+i}\Phi_2(t)
&+&\dfrac{1}{M^3} i \sigma ^{+\Delta}\big[P^+\Delta_{\perp}^i\Phi_3(t)-P^\perp \Delta^+\Phi_4(t)\big]\nonumber\\ 
&+&\dfrac{P^+\Delta^+}{M^3}i \sigma ^{i\Delta}\Phi_5(t) \Big\}u(p,\lambda)\nonumber
\end{eqnarray}
As was the case for twist 2 GPDs and charge form factors, a position space interpretation requires a vanishing longitudinal momentum transfer $\Delta^+=0$. 
The transverse Fourier transform of these generalized form factors have the following interpretation
\begin{itemize}
\item $\Phi_1$ describes an axially symmetric transverse force in an unpolarized target
\item $\Phi_2$ describes a force field perpendicular to the transverse polarization of the target, to a $\perp$ position resolved
Sivers force \cite{Sivers}
\item $\Phi_3$ describes a tensor type force similar to what one would expect from a color magnetic dipole field correlated with the target tranverse spin. 
\item $\Phi_4$ \& $\Phi_5$ involve a factor $\Delta^+=0$ and thus do not contribute to $\perp$ force tomography
\end{itemize}

\section{summary}

The Fourier transform of twist 2 GPDs w.r.t. the transverse momentum transfer provides transverse images of quark distributions. Taking $x^2$ moments of twist 3 PDFs allows determining the average transverse force that acts on a quark in a DIS experiments. Combining these two ideas, the transverse Fourier transform of the $x^2$ moment of twist 3 GPDs allows determining the spatially resolved transverse force that acts on a quark in a DIS experiment.

{\bf Acknowledgements:}
This work was supported by the DOE under grant number 
DE-FG03-95ER40965


\newpage
%

\renewcommand*{\FigPath}{./WeekIII/05_Vogelsang/Figs}
\def\beq{\begin{equation}}
\def\eeq{\end{equation}}
\def\beeq{\begin{eqnarray}}
\def\eeeq{\end{eqnarray}}
\def\nn{\nonumber}
\renewcommand\as{\alpha_{\mathrm{S}}}
\def\ket#1{|{#1}\ra}
\def\bra#1{\la{#1}|}
\renewcommand{\la}{\langle}
\newcommand{\ra}{\rangle}

\wstoc{Spin content of the proton at higher orders}{Daniel de Florian, Werner Vogelsang}
\title{Spin content of the proton at higher orders}

\author{Daniel de Florian}
\index{author}{de Florian, D.}

\address{International Center for Advanced Studies (ICAS), ECyT-UNSAM,
Campus Miguelete, \\ 25 de Mayo y Francia, (1650) Buenos Aires, Argentina\\
E-mail: deflo@unsam.edu.ar}

\author{Werner Vogelsang}
\index{author}{Vogelsang, W.}

\address{Institute for Theoretical Physics, T\"ubingen University, \\
Auf der Morgenstelle 14, 72076 T\"ubingen, Germany\\
E-mail: werner.vogelsang@uni-tuebingen.de}

\begin{abstract}
We study the scale evolution of the quark and gluon spin contributions to the proton spin,
using the recently derived three-loop results for the helicity evolution kernels. We find that the evolution of the 
quark spin contribution may actually be extended to four-loop order. 
We investigate the scale dependence of $\Delta \Sigma$ and $\Delta G$ numerically, 
both to large and down to lower ``hadronic'' scales.
\end{abstract}


\bodymatter

\section{Introduction}

To determine the decomposition of the proton spin in terms of the contributions by quarks and anti-quarks,
gluons, and orbital motion is a key goal of the EIC. The two physically most relevant spin sum rules for the 
proton are the Ji decomposition~\cite{Ji:1996ek_21_141},
which ascribes the proton spin to gauge-invariant contributions by quark spins and orbital 
angular momenta, and total gluon angular momentum, and the Jaffe-Manohar decomposition~\cite{Jaffe:1989jz_21_63_54_249},
in which there are four separate pieces corresponding to quark and gluon spin and orbital contributions,
respectively. The spin parts in the Jaffe-Manohar sum rule are related to parton distributions 
measurable in high-energy processes. The sum rule reads
\beq\label{JM}
\frac{1}{2}\,=\,\frac{1}{2}\Delta \Sigma(Q^2)+\Delta G(Q^2)+L_q(Q^2)+L_g(Q^2)\,,
\eeq
where 
\beq\label{singsec}
\hspace*{-2mm}
\Delta \Sigma(Q^2)\,=\, \sum_q^{N_f} \int_0^1 dx \,
\Big( \Delta q(x,Q^2)+ \Delta \bar{q}(x,Q^2)\Big)\,,\;\Delta G(Q^2) \,=\, \int_0^1 dx \,\Delta g(x,Q^2).
\eeq

\section{Evolution equations and first moments of the splitting functions}

As indicated in Eq.~(\ref{JM}), the contributions to the proton spin are all scale dependent,
although the dependence cancels in their sum. The dependence on $Q^2$ is given by 
spin-dependent QCD evolution equations. The kernels relevant for the evolution of the first 
moments $\Delta q(Q^2)$, $\Delta \bar{q}(Q^2)$, $\Delta G(Q^2)$ have been derived 
to lowest order (LO) in Refs.~\cite{Ahmed:1976ee,Altarelli:1977zs_210}, to next-to-leading order (NLO) 
in Refs.~\cite{Kodaira:1979pa,Mertig:1995ny,Vogelsang:1995vh,Vogelsang:1996im}, 
and recently to next-to-next-to-leading order (NNLO) in 
Refs.~\cite{Vogt:2008yw,Moch:2014sna,Moch:2015usa}. The kernels for the separate evolution 
of $L_q$ and $L_g$ in the Jaffe-Manohar decomposition 
are known only to LO~\cite{Ji:1995cu,Hagler:1998kg,Harindranath:1998ve},
although the evolution of their sum is known from Eq.~(\ref{JM}) to the same order as that of 
$\frac{1}{2}\Delta\Sigma+\Delta G$, that is, to NNLO. As has been shown in 
Refs.~\cite{Hatta:2012cs_243,Boussarie:2019icw,Hatta:2019csj},
the separate evolution of $L_q$ and $L_g$ involves higher-twists.  

The generic evolution equation for the first moment of a spin-dependent parton distribution 
$a,b\equiv u,\bar{u},d,\bar{d},s,\bar{s},\ldots,G$ reads:
\beq \label{evN}
\frac{d\Delta a(Q^2)}{d\ln Q^2} = \sum_b \,
\Delta P_{ab}\big(a_s(Q^2)\big) \;\Delta b(Q^2) \;,
\eeq
where $\Delta P_{ab}$ describes the splitting $b\to a$. The $\Delta P_{ab}$ are perturbative 
in the strong coupling $\as$; their perturbative series starts at ${\cal O}(\as)$:
\beq
\label{eq:split}
\Delta P_{ab}=a_s\Delta P_{ab}^{(0)}+
a_s^2 \Delta P_{ab}^{(1)}+
a_s^3 \Delta P_{ab}^{(2)}+ {\cal O}\big(a_s^4\big)\, .
\eeq
with $a_s\equiv \as/(4\pi)$. 
The evolution equations may be simplified by introducing non-singlet and singlet combinations
of the quark and antiquark distributions; see e.g. Ref.~\cite{deflo}. 
In the singlet sector we have coupled evolution equations
for $\Delta \Sigma$ and $\Delta G$: 
\begin{eqnarray}
\label{sieq}
\frac{d}{d\ln Q^2} \left( \begin{array}{c}
\Delta\Sigma (Q^2)\\[0mm]
\Delta G(Q^2) \end{array}\right) \,=\, 
\left( \begin{array}{cc}
\Delta P_{\Sigma\Sigma}\big(a_s(Q^2)\big) & 2 N_f \,\Delta P_{qG}\big(a_s(Q^2)\big) \\[0mm]
\Delta P_{Gq}\big(a_s(Q^2)\big) & \Delta P_{GG}\big(a_s(Q^2)\big) \end{array}\right)\,
\left( \begin{array}{c}
\Delta\Sigma (Q^2)\\[0mm]
\Delta G(Q^2) \end{array}\right) ,
\end{eqnarray}
with the singlet anomalous dimension $\Delta P_{\Sigma\Sigma}$ and the 
first moments of the splitting functions involving gluons, $\Delta P_{qG}$, $\Delta P_{Gq}$, 
$\Delta P_{GG}$. At lowest order,
\cite{Ahmed:1976ee,Altarelli:1977zs_210}
\beq\label{LOP}
\Delta P_{\Sigma\Sigma}^{(0)}\,=\, 0 \,,\quad \Delta P^{(0)}_{qG}\,=\,0 \,,\quad
\Delta P^{(0)}_{Gq}\,=\,3 C_F \,,\quad\Delta P^{(0)}_{GG}\,=\,\beta_0 \,.
\eeq
The second-order results in the $\overline{\mathrm{MS}}$ scheme 
may be found in Refs. \cite{Mertig:1995ny,Vogelsang:1995vh,Vogelsang:1996im}:
\beeq\label{NLOP}
\Delta P_{\Sigma\Sigma}^{(1)}&=& - 2 N_f \,  \Delta P^{(0)}_{Gq} \,,\quad \Delta P^{(1)}_{qG}\,=\,0\,,\nn\\[0mm]
\Delta P^{(1)}_{Gq}&=&\frac{71}{3} C_F C_A -9 C_F^2-\frac{2}{3} C_F N_f \,,\quad
\Delta P^{(1)}_{GG}\,=\, \beta_1\,.
\eeeq
Finally, at NNLO we have from Refs. \cite{Kodaira:1979pa,Moch:2014sna, Moch:2015usa}:
\beeq\label{NNLOP}               
\Delta P_{\Sigma\Sigma}^{(2)} &=& - 2 N_f \,  \Delta P^{(1)}_{Gq}\,,\quad\Delta P^{(2)}_{qG}\,=\,0 \,, \quad
\Delta P^{(2)}_{GG}\,=\, \beta_2\,, \nn \\[0mm]
\Delta P^{(2)}_{Gq}&=& \frac{1607}{12} C_F C_A^2 - \frac{461}{4} C_F^2 C_A + \frac{63}{2} C_F^3 \nn \\[0mm]
               &+& \left(\frac{41}{3}-72 \zeta_3\right) C_F C_A N_f
       -\left( \frac{107}{2}-72 \zeta_3\right) C_F^2 N_f-\frac{13}{3} C_F N_f^2\,.
\eeeq
There are systematic patterns among these results which may be understood from 
general arguments. The explicit results shown in the above equations suggest that
\beq\label{gen1}
\Delta P_{\Sigma\Sigma} \,=\, - 2 N_f \, a_s\,\Delta P_{Gq}\,.
\eeq
Furthermore, we deduce from Eqs.~(\ref{NLOP}),(\ref{NNLOP})
\beq\label{gen2}
\Delta P_{qG}\,=\,0\,,\quad
\Delta P_{GG}\,=\,-\beta(a_s)/a_s\,.
\eeq
As discussed in Refs.~\cite{Altarelli:1990jp,deflo}, these all-order results may be readily understood from 
the fact that the quark singlet combination $\Delta \Sigma$ 
corresponds to the proton matrix element of the flavor-singlet axial current, 
\beq
S^\mu\,\Delta \Sigma\,=\,\bra{P,S\,}\,\bar{\psi} \,\gamma^\mu \gamma_5 \,\psi\,\ket{\,P,S}
\,\equiv\,\bra{P,S\,}\,j_5^\mu \,\ket{\,P,S}\,,
\eeq
where $S$ is the proton's polarization vector. Because of the axial anomaly, the singlet axial current is not 
conserved, $\partial_\mu \,j_5^\mu=2N_f\,a_s\,\partial_\mu K^\mu$, with the ``anomalous current'' $K$. 
This implies that in fact $\partial_\mu(j_5^\mu-2N_f\,a_s K^\mu)=0$.
In perturbation theory we may relate matrix elements of $K^\mu$ to the gluon spin contribution:
$S^\mu\,\Delta G=-\bra{P,S\,}\,K^\mu \,\ket{\,P,S}$. Although $K$ depends on the choice of gauge,
its forward proton matrix element is gauge invariant, except for topologically nontrivial gauge 
transformations that change the winding number. The latter feature makes the identification
of $\Delta G$ with the matrix element of $K$ impossible beyond perturbation theory~\cite{Jaffe:1989jz_21_63_54_249}.
We may thus conclude in perturbation theory that
\beq
\frac{d}{d\ln Q^2} \left( \Delta\Sigma (Q^2)+2N_f\,a_s(Q^2)\,\Delta G(Q^2)\right)\,=\,0\,.
\eeq
Inserting the general evolution equations for $\Delta \Sigma$ and $\Delta G$ in~(\ref{sieq}),
as well as the renormalization group equation for $a_s(Q^2)$, we recover the results in Eqs.~(\ref{gen1}),(\ref{gen2}).

It is now clear that in the $\overline{\rm MS}$ scheme a {\it single} anomalous dimension, 
$\Delta P_{\Sigma\Sigma}$, resulting from the axial anomaly, governs the evolution of the 
quark and gluon spin contributions. Defining~\cite{Altarelli:1990jp} $\Delta \Gamma(Q^2)\equiv a_s(Q^2)\Delta G(Q^2)$,
we obtain
\beq
\label{sieq2}
\frac{d}{d\ln Q^2} \left( \begin{array}{c}
\Delta\Sigma\\[0mm]
\Delta \Gamma \end{array}\right) \,=\, 
\left( \begin{array}{cc}
\Delta P_{\Sigma\Sigma}(a_s)& 0 \\[0mm]
-\frac{1}{2N_f} \, \Delta P_{\Sigma\Sigma}(a_s)& 0 \end{array}\right)\,
\left( \begin{array}{c}
\Delta\Sigma \\[0mm]
\Delta \Gamma \end{array}\right)  \;.
\eeq
Thanks to Eq.~(\ref{gen1}) we may now determine~\cite{deflo} the {\it four-loop} (N$^3$LO) contribution to 
$\Delta P_{\Sigma\Sigma}$ from the three-loop value 
$\Delta P_{Gq}^{(2)}$ computed in Ref. \cite{Moch:2015usa}:
\beeq
\Delta P_{\Sigma\Sigma}^{(3)}&=&- 2 N_f \,  \Delta P^{(2)}_{Gq}\;=\;
 - 2 N_f C_F \left[\,\frac{1607}{12} C_A^2 - \frac{461}{4} C_F C_A + \frac{63}{2} C_F^2\right. \nn \\[0mm]
               &+&\left.  \left(\frac{41}{3}-72 \zeta_3\right) C_A N_f
       -\left( \frac{107}{2}-72 \zeta_3\right) C_F N_f-\frac{13}{3} N_f^2\,\right].
\eeeq

\section{Numerical solutions in the singlet sector}

We may now solve~\cite{deflo} the singlet evolution equation~(\ref{sieq2}). We note that 
such solutions were presented to LO in Refs.~\cite{Stratmann:2007hp,Thomas:2008ga,Wakamatsu:2009gx}.
The paper~\cite{Altenbuchinger:2010sz} considered the evolution of $\Delta\Sigma$ up to NNLO.
The solution for the singlet at scale $Q$ in terms of its boundary value at the ``input'' scale $Q_0$ reads
 \beeq
\label{eq:singsol}
\frac{ \Delta\Sigma (Q^2)}{ \Delta \Sigma (Q_0^2)} &=& 
\exp
 \Bigg[  - \frac{ a_Q -a_{0}}{\beta_0} \,\Delta P_{\Sigma\Sigma}^{(1)} \Bigg] \times
 \exp
\Bigg[   \frac{ a_Q^2 -a_{0}^2}{2\beta_0^2} \,  \left( \beta_1 \,\Delta P_{\Sigma\Sigma}^{(1)}  -\beta_0\,
\Delta P_{\Sigma\Sigma}^{(2)}\right) \Bigg] \nn\\[0mm]
&&\hspace*{-1.6cm}\times\exp
\Bigg[   \frac{ a_Q^3 -a_{0}^3}{3\beta_0^3} \left(
- \beta_1^2 \, \Delta P_{\Sigma\Sigma}^{(1)}+\beta_0 \beta_2 \, \Delta P_{\Sigma\Sigma}^{(1)}+\beta_0\beta_1 \,
\Delta P_{\Sigma\Sigma}^{(2)}- \beta_0^2 \,\Delta P_{\Sigma\Sigma}^{(3)}\right)\Bigg]
\,,
\eeeq
where $a_Q\equiv a_s(Q^2)$ and $a_0\equiv a_s(Q_0^2)$.
\begin{figure}[t!]
\begin{center}
 \parbox{2.1in}{\includegraphics[width=2in]{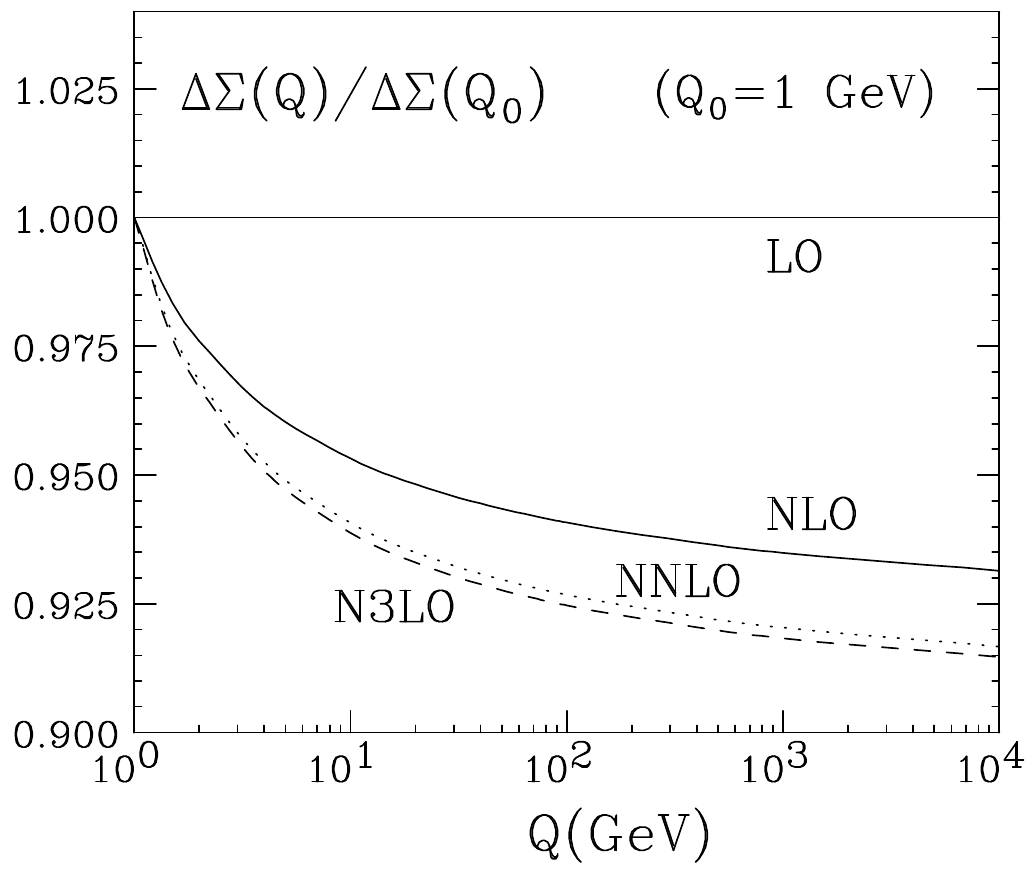}}
 \hspace*{4pt}
 \parbox{2.1in}{\includegraphics[width=1.9in]{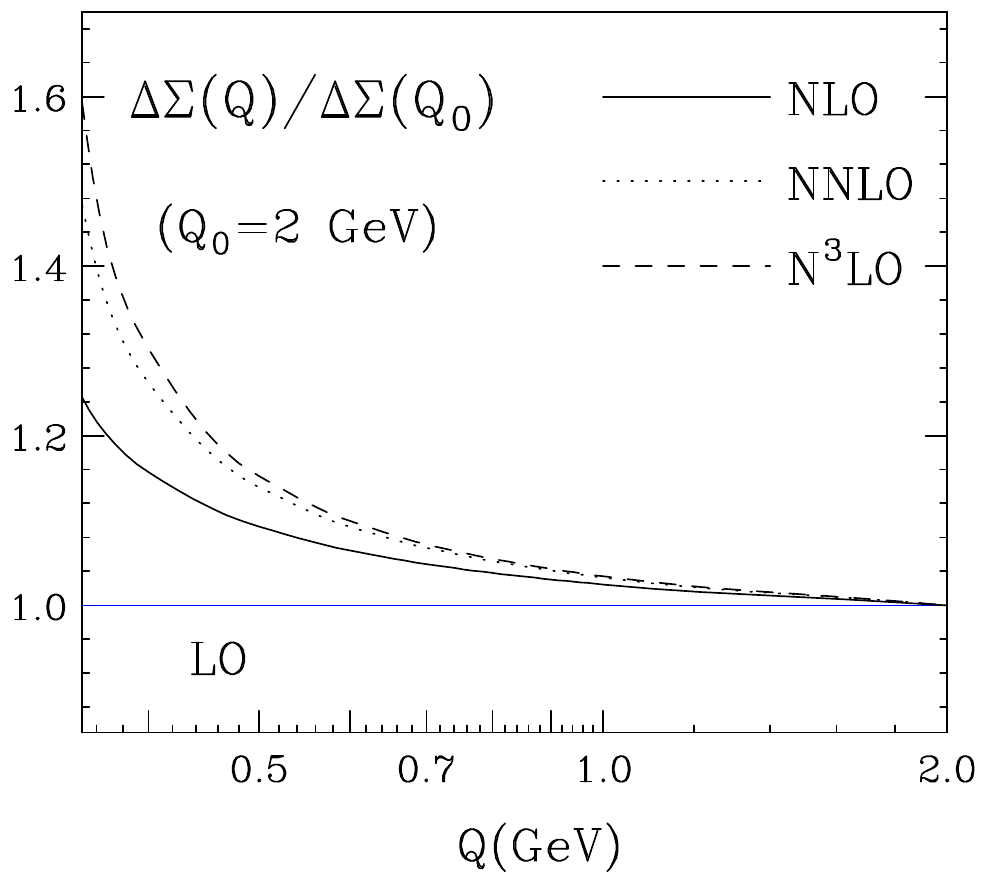}}
 \caption{Left: Evolution of the first moment of the polarized singlet distributions at LO, NLO, 
NNLO and N$^3$LO, starting from the initial scale $Q_0=1$ GeV. Right:
Backward evolution of the first moment of the polarized singlet distributions at LO, NLO, NNLO and N$^3$LO, 
starting from the initial scale $Q_0=2$ GeV. From Ref.~\cite{deflo}.}
\label{singlet}
\end{center}
\end{figure}

The left part of Fig.~\ref{singlet} shows the quark singlet evolution factor on the right-hand-side of Eq.~(\ref{eq:singsol}), 
assuming a fixed number $N_f=3$ in the anomalous dimensions and the beta function,
and using the full NNLO evolution of the coupling constant. We have chosen a relatively low input scale  $Q_0=1$,
with a value $\alpha_s(Q_0)=0.404$. One can see that the NLO evolution affects the quark spin content of the proton by up to $7\%$
while NNLO evolution adds an extra $\sim 1-2\%$ effect. The numerical impact of the four-loop
term $\Delta P_{\Sigma\Sigma}^{(3)}$ reaches only ${\cal{O}}(0.2\%)$ at the highest scale.

Ultimately, as discussed in Ref.~\cite{Jaffe:1987sx,Altenbuchinger:2010sz}, one may want to compare helicity
parton distribution functions extracted from experiment or computed on the lattice~\cite{Lin:2017snn_117} with 
calculations performed in QCD-inspired models of nucleon structure. The latter are typically 
formulated at rather low momentum scales of order of a few hundred MeV. Given the high
order of perturbation theory now available for evolution, it is therefore interesting to evolve
the singlet spin contributions not only to large perturbative scales, but also ``backward'' 
towards the limit of validity of perturbation theory~\cite{Altenbuchinger:2010sz}. 
In the right part of Fig.~\ref{singlet} we show the evolution of $\Delta\Sigma$ at LO, NLO, NNLO and N$^3$LO
down to $Q\sim 0.35$ GeV, starting from the initial scale $Q_0=2$ GeV. 
As can be observed, and as is expected, the higher order 
terms affect the evolution of the singlet in a significant way, much more strongly than what we 
found for the evolution to larger scales. On the other hand, a striking feature is that the 
evolution remains relatively stable even down to scales as low as $Q=0.35$~GeV where 
the coupling constant becomes $\alpha_s\sim 1.3$. In addition, all higher orders
(NLO, NNLO, N$^3$LO) go in the same direction. We note that the upturn of $\Delta\Sigma$
toward small scales -- in the direction of large quark and anti-quark spin contributions to the proton spin --
was already observed to NLO and NNLO in Refs.~\cite{Jaffe:1987sx} and~\cite{Altenbuchinger:2010sz},
respectively. We also remark that results on high-loop evolution may be useful for lattice-QCD
studies of nucleon structure, possibly allowing cross-checks of the nonperturbative renormalization
carried out on the lattice.  

The solution of the evolution equation for the gluon spin contribution now follows directly~\cite{deflo}. 
From the lower row in Eq.~(\ref{sieq2}) we have by simple integration
\beq\label{DGamsol}
a_Q\Delta G(Q^2)\,=\,a_0\Delta G(Q_0^2)-\int_{a_0}^{a_Q} da_s\,
\frac{\Delta P_{\Sigma\Sigma}(a_s)}{2N_f \beta(a_s)}\,\Delta\Sigma(Q^2)\,.
\eeq
An immediate observation is that the integral on the right-hand-side of~(\ref{DGamsol}) starts at order
$a_s(Q^2)$ and $a_s(Q_0^2)$. Therefore, we arrive at the well-known result~\cite{Altarelli:1988mu} 
that the leading term in $\Delta\Gamma$ is a constant in $Q^2$, so that the 
first moment of the gluon spin contribution evolves as the inverse of the strong coupling.
An explicit solution for $\Delta G(Q^2)$ to NNLO is obtained by inserting the solution for $\Delta\Sigma(Q^2)$ 
from Eq.~(\ref{eq:singsol}) into~(\ref{DGamsol}) and carrying out the integration. The result is 
given in Ref.~\cite{deflo}. 

\begin{figure}[h]
\begin{center}
 \parbox{2.1in}{\includegraphics[width=2in]{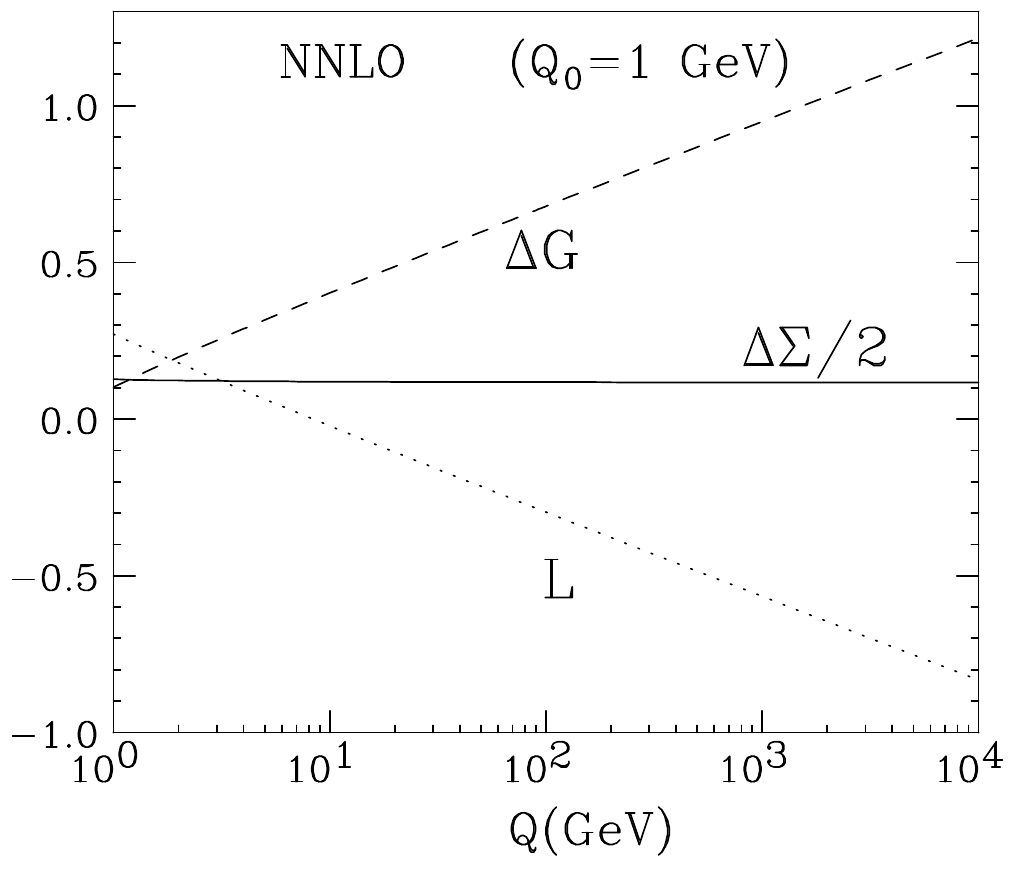}}
 \hspace*{4pt}
 \parbox{2.1in}{\includegraphics[width=2in]{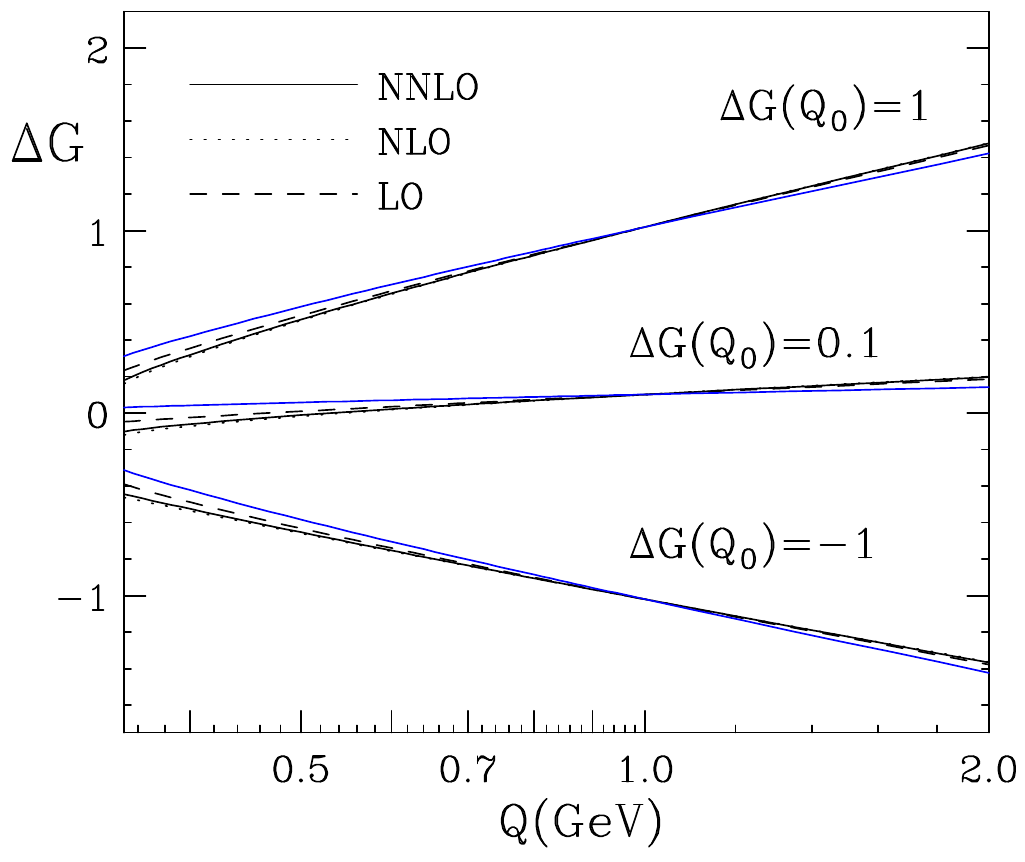}}
 \caption{Left: Evolution of the quark and gluon spin contributions $\frac{1}{2}\Delta\Sigma$
and $\Delta G$ at NNLO, starting from the inital scale $Q_0=1$ GeV. We also show the evolution of
$L_q+L_g$. Right: Backward evolution of the gluon spin contribution $\Delta G$ 
at LO (dashes), NLO (dots) and NNLO (solid line), starting from three different scenarios at the inital scale 
$Q_0=1$ GeV: $\Delta G(Q_0)=+1,0.1,-1$. The blue solid line corresponds to the leading $1/a_s$ term in the solution.}
\label{fig:all}
\end{center}
\end{figure}

Figure \ref{fig:all} shows the NNLO evolution of the gluon spin contribution to the proton spin, starting from 
the values $\Delta G=0.102$ and $\Delta\Sigma=0.254$ at $Q_0=1$ as realized in the 
global analysis~\cite{deFlorian:2014yva,deFlorian:2008mr}. We also show the evolution of $\frac{1}{2}\Delta\Sigma$
and the evolution of the total orbital angular momentum $L_q+L_g$. 
Notice that both $\Delta G$ and $L_q+L_g$ have a divergent behaviour at large scales, 
resulting in a rather unphysical cancellation of two very large contributions to fulfill the spin sum rule.
Again it is also interesting to analyze the behavior of the gluonic spin contribution
at lower scales. In the right part of Fig. \ref{fig:all} we show the backward evolution of $\Delta G$ at 
LO (dashes), NLO (dots) and NNLO (solid line) for three different scenarios, corresponding to  setting $\Delta G(Q_0)=+1,0.1,-1$ 
at the initial scale  $Q_0=1$ GeV. For each scenario, we observe a striking convergence of the fixed order results down to very low scales, always towards small gluonic contributions. Our findings set a strong constraint on the proton spin content carried by gluons at {\it hadronic} scales.
Within the rather extreme scenarios analyzed here (for which the gluon contribution accounts for 
as much as twice the spin of the proton at  $Q_0=1$ GeV!), we obtain the requirement $ |\Delta G(Q\sim0.35 \,{\rm GeV})| \lesssim 0.3$.  

We finally note that, depending on the input values of $\Delta G(Q_0^2)$ 
and $\Delta \Sigma(Q_0^2)$ the evolution can be towards large positive or negative values. 
This implies that there is a specific input, a ``critical point'', for which $\Delta G(Q^2)$ actually remains 
almost constant and tends to a finite asymptotic value as $Q^2\to\infty$.
This ``static'' value of $\Delta G$ is expected to change from order to order in perturbation theory.
The explicit calculation shows~\cite{deflo} that $\Delta G_{\mathrm{stat}}^{{\mathrm{LO}}}(Q_0^2)  \,\simeq \,-0.113$, 
$\Delta G_{\mathrm{stat}}^{{\mathrm{NLO}}}(Q_0^2) \simeq\, -0.13$, $\Delta G_{\mathrm{stat}}^{{\mathrm{NNLO}}}(Q_0^2)  \simeq \,-0.125$,
at $Q_0=1$~GeV. Beyond LO, the ``static'' solutions are no longer completely constant in $Q^2$; however, by construction
they converge asymptotically to a finite value. We believe that these solutions, especially because of the fact that they 
have a well behaved asymptotic limit at large scales, deserve further attention since they arise as  strong boundaries 
on non-perturbative physics from almost purely perturbative considerations.


\newpage

\wstoc{Quark-gluon-quark contributions to twist three GPDs}{Abha Rajan, Simonetta Liuti}
\title{Quark-gluon-quark contributions to twist three GPDs}

\author{Abha Rajan$^*$ }
\index{author}{Rajan, A.}
\address{Physics Department, Brookhaven National Laboratory,\\ Bldg. 510A, Upton, NY 11973, USA \\
$^{*}$ Email: arajan@bnl.gov}


\author{Simonetta Liuti$^{**}$}
\index{author}{Liuti, S.}

\address{Physics Department, University of Virginia,\\  382 McCormick Road, VA 22903, USA \\and\\Laboratori Nazionali di Frascati, INFN, Frascati, Italy\\
$^{**}$ Email: sl4y@virginia.edu}


\begin{abstract}
Twist three Generalized Parton Distributions and Generalized Transverse Momentum Distributions are central to the description of partonic orbital angular momentum. We discuss the origin of the genuine twist three contribution, described by the quark gluon quark correlator, to the twist three Generalized Parton Distribution $\widetilde{E}_{2T}$ and how it connects to calculations of the $k_T^2$ moments of the Generalized Transverse Momentum Distribution $F_{14}$.  \end{abstract}

\bodymatter


\section{Introduction}
Higher twist effects are of interest due to multiple reasons, while on the one hand they provide a direct window into non-perturbative effects explicitly involving quark gluon interactions,\cite{Kiptily,Hatta,phd1,phd2} on the other, they also hold the promise of connecting to observables accessible in current experiments.\cite{Kriesten} We focus on twist three Generalized Parton Distributions (GPDs) in the chiral even sector that describe partonic orbital angular momentum and spin orbit correlations. However, this work can also be extended to the chiral odd sector highlighting completely different aspects of the partonic structure of the proton. In essence, this work is an extension of similar studies on the twist three PDF $g_T$ \cite{WW,Tanger} to the off forward sector which facilitates the existence of polarization states of the initial and final hadrons that are not permissible in the forward case.   \\

Although twist three GPDs are defined by a quark-quark correlator, they do receive a contribution from a genuine twist term involving an explicit quark gluon quark correlator. The other piece comes from leading twist GPDs. The way to derive this decomposition essentially involves two steps, first, to study the dependence of $k_T$ dependent distributions called Generalized Transverse Momentum Dependent Distributions (GTMDs)\cite{MMS} and collinear distributions on functions that parameterize the completely unintegrated correlator to derive the so called Lorentz Invariance Relations (LIRs) and second, to use the Equations of Motion (EoM) to derive a relation describing the Wandzura-Wilczek contribution to the twist three GPD and the genuine twist three contribution. \cite{phd1,phd2}\\

In this contribution, we discuss in section 2 the derivation of LIRs and EoM relations and finally the steps involved in singling out the genuine twist three piece. In section 3, we focus on the role of the gauge link. As the derivation involves going through the $k_T$ dependent distributions, one is free to choose between a staple gauge link (which describes the final state interactions in an actual experiment) and a straight gauge link simply connecting the quark field operators entering the quark-quark operator by a straight line. If one starts from the premise that the shape of the gauge link is immaterial to the GPDs (which are collinear) entering the LIRs and EoM relations, one finds the genuine twist three terms contributing to the twist three GPD in either case to be the same. However, a more interesting exploration would be if we had a way of independently measuring quantities entering the relations for the two cases. We are interested in the twist three GPD $\tilde{E}_{2T}$ and the GTMD $F_{14}$ which enter the description of partonic orbital angular momentum.\cite{Lorce} While the lattice provides actual calculations of the $k_T^2$ moment of the GTMD $F_{14}$ both in the straight and staple cases, \cite{Engel} the extraction of twist three GPDs from experiment would provide a clean comparison of the two cases.\cite{Kriesten} Recently, there have also been suggestions to directly measure the multiparton distribution or the form factor that would connect to the genuine twist three piece. \cite{Prokudin,Aslan} 

\section{Quark-gluon-quark contribution to twist three GPDs}
To single out the contribution from the quark-gluon-quark correlator to twist three GPDs we need to first work out the LIRs and EoM relations. We focus on the twist three GPD $\widetilde{E}_{2T}$, described by the matrix element of quark fields with the  projection operator $\gamma_T^i$ in a longitudinally polarized proton, and the GTMD $F_{14}$ describing unpolarized quarks in a longitudinally polarized proton. LIRs are derived by expressing both $\widetilde{E}_{2T}$ and $F_{14}$ in terms of Generalized Parton Correlation Functions (GPCFs) \cite{MMS} that parameterize the completely unintegrated quark-quark correlator. As both $\widetilde{E}_{2T}$ and $F_{14}$ describe vector distribution functions (some component of $\gamma^\mu$ between the quark field operators), the same set of GPCFs describe both functions. As a result, one arrives at the following LIR~\cite{phd1},
\begin{equation}
\frac{d F_{14}^{(1)}}{ d x}  = \widetilde{E}_{2T} +H +E + {\cal A}_{F_{14}}.  
\end{equation}
Here, $H$ and $E$ are the vector twist two GPDs. ${\cal A}_{F_{14}}$ essentially signifies the effect of the shape of the gauge link used, going to zero for a straight gauge link connecting the quark field operators in the quark-quark correlator describing these functions. The source of this term is the introduction of new GPCFs in the case of a staple gauge link because of the inclusion of another vector $v^-$ describing the gauge link either going to $+ \infty$ or $- \infty$ on the light cone. $F_{14}^{(1)}$ is defined as follows~\cite{phd1},

\begin{equation}
\label{ktmoment}
F_{14}^{(1)} = 2\int d^2 k_T  \frac{k_T^2}{M^2}
\frac{k_T^2 \Delta_{T}^{2} - (k_T \cdot \Delta_{T} )^2 }{k_T^2 \Delta_{T}^{2} } F_{14} (x, 0, k_T^2, k_T\cdot\Delta_T, \Delta_T^2).
\end{equation}
Note that, in the forward limit, this reduces to the standard $k_T^2 $-moment,
\begin{equation}
\left. F_{14}^{(1)} \right|_{\Delta_{T} =0} =
\int d^2 k_T \frac{k_T^2 }{M^2 } F_{14} (x,0,k_T^2,0,0) \, .
\end{equation}

A completely independent set of relations is obtained by using the equations of motion. Central to the construction of these relations is the observation that, taken between physical particle states, matrix elements of operators that vanish according to the classical field equations of motion vanish in the quantum theory. One obtains a relation involving the twist three GPD $\widetilde{E}_{2T}$, the twist two GPDs $H$, $E$ and $\widetilde{H}$, $F_{14}^{(1)}$ and a genuine twist three term ${\cal M}_{F_{14}}$ explicitly involving the gluon field which originates from the gluon field in the covariant derivative and the derivatives acting on the Wilson line, 

\begin{equation}
0 = x \widetilde{E}_{2T} + \widetilde{H} - F_{14}^{(1)} + {\cal M}_{F_{14}} \, .
\end{equation}

Using the LIR to eliminate $F_{14}^{(1)}$ from the EoM relation one obtains~\cite{phd1},

\begin{equation}
\label{F14_WW1}
\widetilde{E}_{2T} =
- \int_x^1 \frac{dy}{y}(H + E) - \left[ \frac{\widetilde{H}}{x} -\int_x^1 \frac{dy}{y^2} \widetilde{H}\right]  - \left[ \frac{1}{x}\mathcal{M}_{F_{14}} - \int_x^1 \frac{dy}{y^2} \mathcal{M}_{F_{14}}  \right]
- \int_x^1 \frac{dy}{y} {\cal A}_{F_{14} } \, .
\end{equation}

In the above, there are two distinct contributions - one that comes from the leading twist GPDs also known as the Wandzura-Wilczek contribution and the genuine twist three contribution described by ${\cal M}_{F_{14} }$ and  ${\cal A}_{F_{14} }$. In the next section we explore how the choice of the gauge link affects these terms.

\section{Role of the Gauge link}
As GPDs are collinear objects one expects them to be unaffected by the choice of the shape of the gauge link. Hence, evaluating the LIRs and EoM relations in equations (1) and (4) for the two cases allows one to obtain a relation between ${\cal A}_{F_{14} }$ and ${\cal M}_{F_{14} }$. Here, $|_{v=0}$ denotes the straight gauge link scenario.\\

\noindent From the LIR,
\begin{equation}
\label{eq:LIRviolf14}
\left. \frac{d F_{14}^{(1)}}{dx} - \frac{d F_{14}^{(1)}}{dx} \right|_{v=0} =  {\cal A}_{F_{14} }. 
\end{equation}

\noindent From EoM,
\begin{equation}
F_{14}^{(1)} -
\left. F_{14}^{(1)} \right|_{v=0} =
{\cal M}_{F_{14} } - \left. {\cal M}_{F_{14} } \right|_{v=0}.
\end{equation}

\noindent Thus ${\cal M}_{F_{14}}$ and ${\cal A}_{F_{14} }$ are related as follows,

\begin{equation}
{\cal A}_{F_{14} } (x) = \frac{d}{dx} \left(
{\cal M}_{F_{14} } - \left. {\cal M}_{F_{14} } \right|_{v=0} \right).
\label{amrel}
\end{equation}

To check if this framework does indeed describe the physics of twist three objects, one should calculate the pieces separately and compare. $F_{14}$ depends on the intrinsic transverse momentum $k_T$ and any experimental measurement of it in future will involve final state interactions described by a staple shaped gauge link. Hence, from this viewpoint, $\left.F_{14}^{(1)} \right|_{v=0}$ is more abstract but, even so, it can be accessed on the lattice as can $F_{14}^{(1)}$ with a staple gauge link.\cite{Engel} Hence, the experimental extraction of $\widetilde{E}_{2T}$ will provide an independent check of these ideas and a deeper understanding of the role played by the gauge link.

\section{Conclusions}
We have discussed the decomposition of the twist three GPD $\widetilde{E}_{2T}$ into two components, one involving leading twist quantities, the Wandzura-Wilczek piece, and the other explicitly involving the gluon field which is referred to as the genuine twist three piece. As $k_T$ dependent distribution functions are involved in the derivation, the gauge link plays a key role. Although the final result involves well defined collinear quantities, independent calculations, on the lattice and in models, and experimental measurements of the separate pieces involved will provide deeper understanding of higher twist effects and verification of the framework used to describe them.    

\section*{Acknowledgements}The authors are indebted to Michael Engelhardt for discussions on this topic and contributions to \cite{phd1, phd2}. This work was supported by DOE grants
DE-SC0016286, by the Jefferson Science Associates grant (A.R.), and
by the DOE Topical Collaboration on TMDs. A.R. was supported for this meeting by DOE grant DE-SC0012704 and acknowledges the
LDRD grant from Brookhaven Science Associates.


\newpage
\renewcommand*{\FigPath}{./Logo/} 

\begin{tcolorbox}[colframe=white]
\begin{minipage}{0.2\textwidth}
\includegraphics[width=1.\textwidth]{\FigPath/INT_Workshop_Logo_Final_Black.png}
\end{minipage}
\begin{minipage}{0.7\textwidth}
\wstoc{\bf Week IV}{}
\title{Week IV}
\end{minipage}
\end{tcolorbox}

\bodymatter

\wstoc{Introduction for Week IV}{Yoshitaka Hatta, 
Yuri Kovchegov, 
Cyrille Marquet, 
Alexei Prokudin}
\vspace{1cm}
\begin{center}
{\bf Introduction for Week IV}
\end{center}
~\vspace{3cm}

In this Chapter we present the proceedings of a five-day symposium held in the middle of our INT program, which covered all the major EIC physics topics. The symposium involved researchers from BNL and JLAB, and from the national and international EIC communities, representing all the major topics of the program. The contributions in this Chapter represent a sample of the symposium's program. 

~\\

\begin{flushright}
Yoshitaka Hatta \\
Yuri Kovchegov \\
Cyrille Marquet \\
Alexei Prokudin \\~\\

\newpage


\end{flushright}

%



 \newpage
  
\renewcommand*{\FigPath}{./WeekIV/01_Pasquale_DiNezza/}

\wstoc{The LHCspin project}{Pasquale~Di~Nezza}
\title{The LHCspin project}


\author{Pasquale~Di~Nezza}
\index{author}{Di~Nezza, P.}

\address{INFN Laboratori Nazionali di Frascati, Frascati (Rome), Italy \\
E-mail: Pasquale.DiNezza@lnf.infn.it\\ }

\author{V. Carassiti$^a$, G. Ciullo$^{a,b}$, P. Lenisa$^{a,b}$, L.~L.~Pappalardo$^{a,b}$}

\address{$^a$INFN Ferrara, Italy  \\
$^b$Dipartimento di Fisica e Scienze della Terra, Universit\'a di Ferrara, Italy \\
E-mail: vito@fe.infn.it, ciullo@fe.infn.it, lenisa@fe.infn.it, pappalardo@fe.infn.it \\ }

\author{E. Steffens}

\address{Physics Dept., FAU Erlangen-N\"{u}rnberg, Erlangen, Germany  \\
E-mail: Erhard.Steffens@fau.de\\ }

\begin{abstract}
The LHCspin project aims to bring polarized physics at the LHC through the installation of a gaseous fixed target at the upstream end of the LHCb detector. The forward geometry of the LHCb spectrometer ($2<\eta<5$) is perfectly suited for the reconstruction of particles produced in fixed-target collisions. The fixed-target configuration, with center-of-mass energy at $\sqrt{s}=115$ GeV, allows to cover a wide backward center-of-mass rapidity region, corresponding to the poorly explored high x-Bjorken and high x-Feynman regimes. The project has several ambitious goals, in particular  regarding nucleon's internal dynamics in terms of both quarks and gluons degrees of freedom. The use of transversely polarized H and D targets will allow to study the quarks TMDs in pp collisions at unique kinematic conditions.
\end{abstract}


\bodymatter

\section{Introduction}
Fixed-target collisions with a proton beam at the TeV scale provide unique laboratories for the study of the nucleon's internal dynamics and, more in general, for the investigation of the complex phenomena arising in the non-perturbative regime of QCD. In particular, due to the substantial boost of the reaction products in the laboratory frame, fixed-target collisions measured with a forward spectrometer such as LHCb, allow one to access the backward center-of-mass rapidity region ($-3<y^*<0$), corresponding to the poorly explored high $x$-Bjorken regime. These measurements will thus allow to open the way to innovative and fundamental measurements in regions of the kinematic plane which are still essentially unexplored \cite{after}, Fig.\ref{Kineplane} left. Furthermore, the use of a gas target has the advantage of allowing for a broad variety of nuclear targets, thus providing novel probes for the study of the nucleon and nuclear structure, and for measurements of great interest ranging from heavy-ion physics to cosmic rays physics and dark matter search. A first step in this direction has been already done at LHCb with the installation, during the 
LHC Long Shutdown II 2019-2020, of an unpolarised fixed gas system called SMOG2 \cite{tp}.

\begin{figure}[h!]
   \begin{minipage}{12pc}
      \includegraphics[width=13pc]{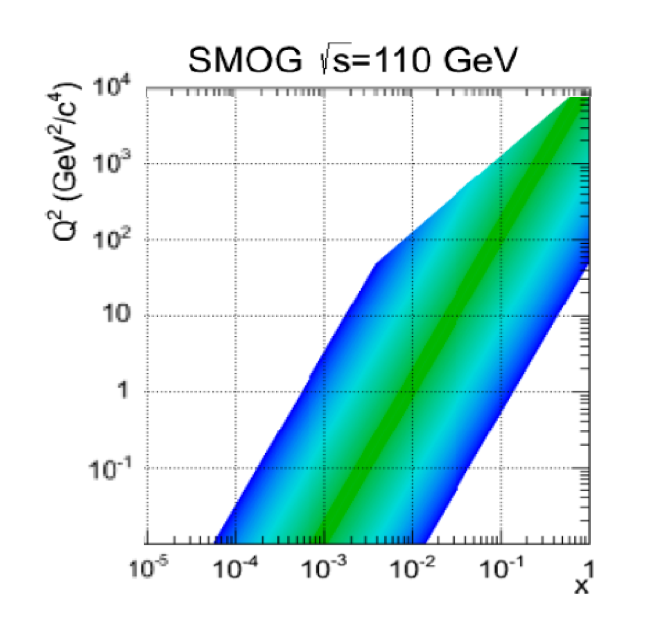}
    \end{minipage}\hspace{1pc}%
    \begin{minipage}{12pc}
      \includegraphics[width=16.5pc]{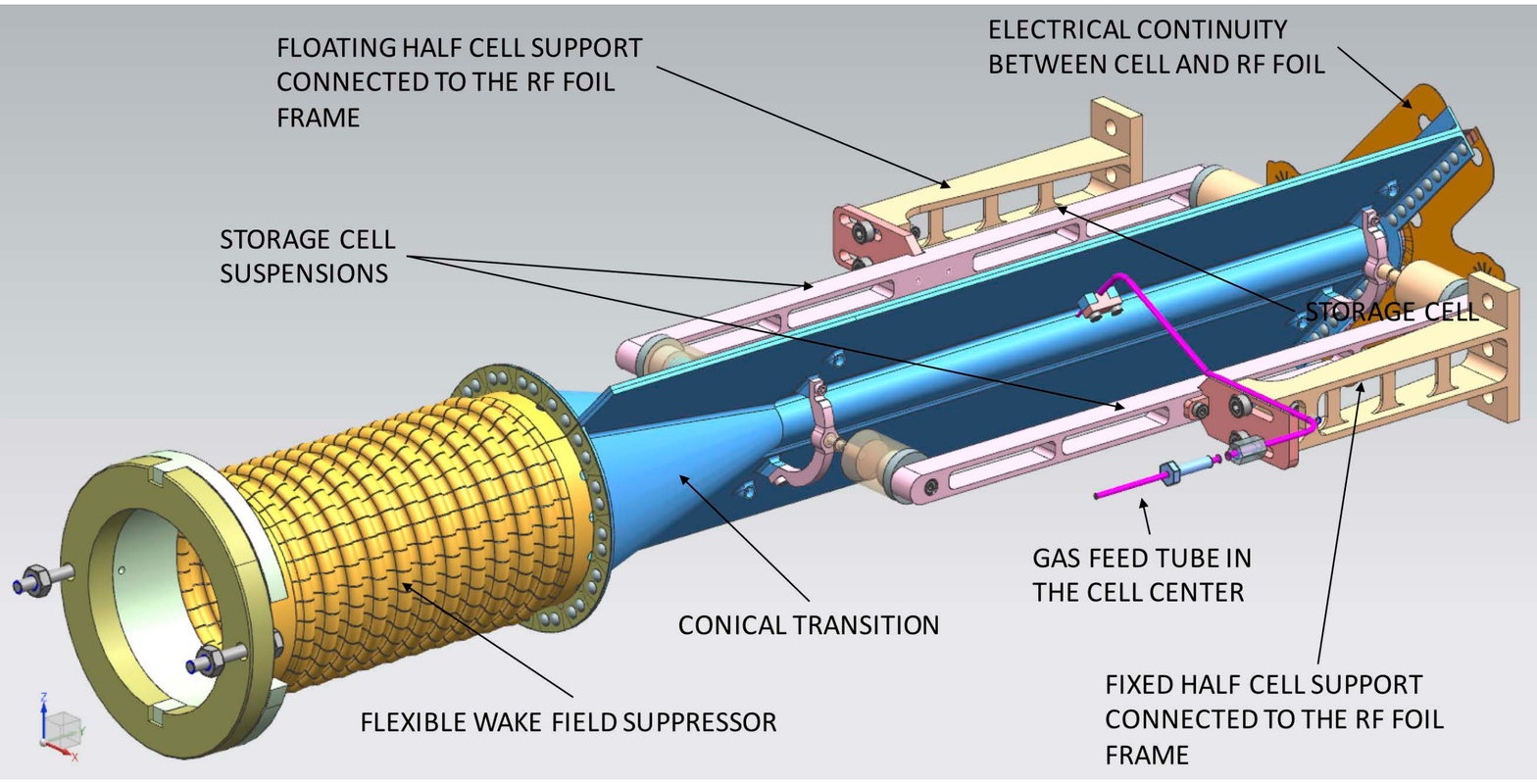}
    \end{minipage}\hspace{1pc}%
    \caption{\label{Kineplane} Left: Accessible range in the ($x$, Q$^2$) plane with fixed-target collisions at LHCb with SMOG2 \cite{smog2}. Right: details of the SMOG2 storage cell.}
\end{figure}

\section{The physics case}

Although tremendous advances have been made over the past decades in defining the quark and gluon dynamical substructure of the nucleon, the present knowledge of the PDFs still suffers from large uncertainties, especially at very-high and very-low $x$, leaving open fundamental questions about QCD and confinement. In many cases, the PDF uncertainties have become the limiting factor in the accuracy of the predictions for LHC measurements, especially concerning  measurements of SM and BSM observables. Considering also the explicit dependence of PDFs on the parton transverse momenta, radically new perspectives in the exploration of the structure of the nucleon can be opened. Transverse Momentum Dependent PDFs
(TMDs) decribe to spin-orbit correlations inside the nucleon, and are indirectly sensitive to the still unknown parton orbital angular momentum, the main missing piece in the proton spin puzzle. In addition, they provide the possibility to map the parton densities in the 3-dimensional momentum space, allowing for a nucleon tomography \cite{3dd}. Quark TMDs can now be studied at LHC in fixed-target hadron-hadron collision (Drell-Yan) at unique kinematic conditions, providing a complementary approach and an ideal test-bench for testing universality, factorization and evolution of QCD. In contrast to the quark TMDs, the experimental access to the gluon TMDs is still extremely limited. In high-energy hadronic collisions, heavy quarks are dominantly generated by gluon-gluon interactions. As a consequence, the most efficient way to access the gluon distributions is through the study of inclusive heavy-flavour production in fixed-target measurements at $\sqrt{s}=$115 GeV. 

Relativistic heavy-ion collisions allow to investigate the high-density and high-temperature regime of QCD. The production of heavy quarks is particularly suited for the study of the phase transition between ordinary hadronic matter and the Quark-Gluon Plasma (QGP). Furthermore, studying the production of quarkonia and heavy mesons with different nuclear targets will allow to set stringent constraints on the nuclear PDFs and to investigate their peculiar features. Furthermore, by exploiting the use of the LHC Pb beams, one can also study QGP formation in fixed-target PbA collisions at $\sqrt{s_{NN}}=72 {\rm ~GeV}$ through the measurement of flow observables and correlations for a variety of collision systems. In particular, by measuring the production of different charmonia excited states ($J/\psi, \psi',\chi_c$, etc.) a study of the intriguing phenomenon of sequential suppression is also possible.

\section{Unpolarized Target}

Among the main LHC experiments, LHCb is the only detector that can already run both in collider and fixed-target mode. By using the SMOG system\cite{smog2}, a low flow rate of noble gas was injected into the vacuum vessel of the LHCb VErtex LOcator (VELO) detector. The resulting beam-gas collision allowed to study proton-nucleus and nucleus-nucleus collisions on various target types and at different center-of-mass energies. Several dedicated runs have already been performed since 2015 using He, Ar, or Ne targets with proton and Pb beams \cite{cr}. New perspectives will be open with the more complex SMOG2 target system. Among several advantages, this new setup increases by up to two orders of magnitude the target areal density (and thus the luminosity), by injecting the same amount of gas of SMOG and to run simultaneously with the collider mode, with a negligible impact on the beam life-time and on the LHCb physics program at $\sqrt{s}=14 {\rm ~TeV}$. The final design of the storage cell is shown in Fig.~\ref{Kineplane} right. 


\section{Polarized Target}

The implementation of a polarized gaseous target at the upstream end of the LHCb spectrometer \cite{spin18}, together with the SMOG2 system, constitutes the core of the LHCspin project \cite{espplhcspin}. The concept of the apparatus is based on the polarized target system used at the HERMES experiment \cite{hermes}. The setup will consist of four main components: an Atomic Beam Source (ABS), a Target Chamber (TC) hosting a T-shaped storage cell with a length of the order of 30 cm, a diagnostic system, and an additional tracking detector. The ABS generates a beam of polarized atomic gas (H or D) that is injected into a storage cell, in order to maximize the target areal density. The cell is placed inside the TC, into the LHC primary vacuum. The diagnostic system, including a Breit-Rabi polarimeter and a Gas Target Analyzer, allows to monitor both the fraction of atomic gas into the cell and the degree of polarization. The target chamber also hosts a transverse magnet ($\sim 300~{\rm mT}$), needed to define and keep the transverse polarization of the target gas. The HERMES polarized target has been successfully operated over a decade, with very high performances \cite{hermes}. The arrangement of the Polarized Gas Target (PGT) in the beam line upstream of the VELO is shown schematically in Fig.\ref{pgt}. Due to the distance between the target cell and the VELO detector, an additional tracker has to be installed inside the TC, in order to supplement the tracking capabilities of the VELO in this upstream region. To maximize the acceptance, this new tracker must be located as close as possible to the beam. Considering the geometry of the cell, assuming an ABS intensity of $6.5 \cdot 10^{16}$ atoms/s into the cell feed tube and a conservative LHC proton beam intensity of $3.8 \cdot 10^{18}$ p/s for the LHC Run4, one obtains an instantaneous luminosity for pH collisions of the order of {\bf $L_{pH}= 2.7\cdot 10^{32}~{\rm cm}^{-2}s^{-1}$} with the cell at room temperature, or {\bf $L_{pH}= 4.6\cdot 10^{32}~{\rm cm}^{-2}s^{-1}$} with the cell cooled at 100 K.

\begin{figure}[!h]
\centering
\includegraphics[width=0.97\textwidth]{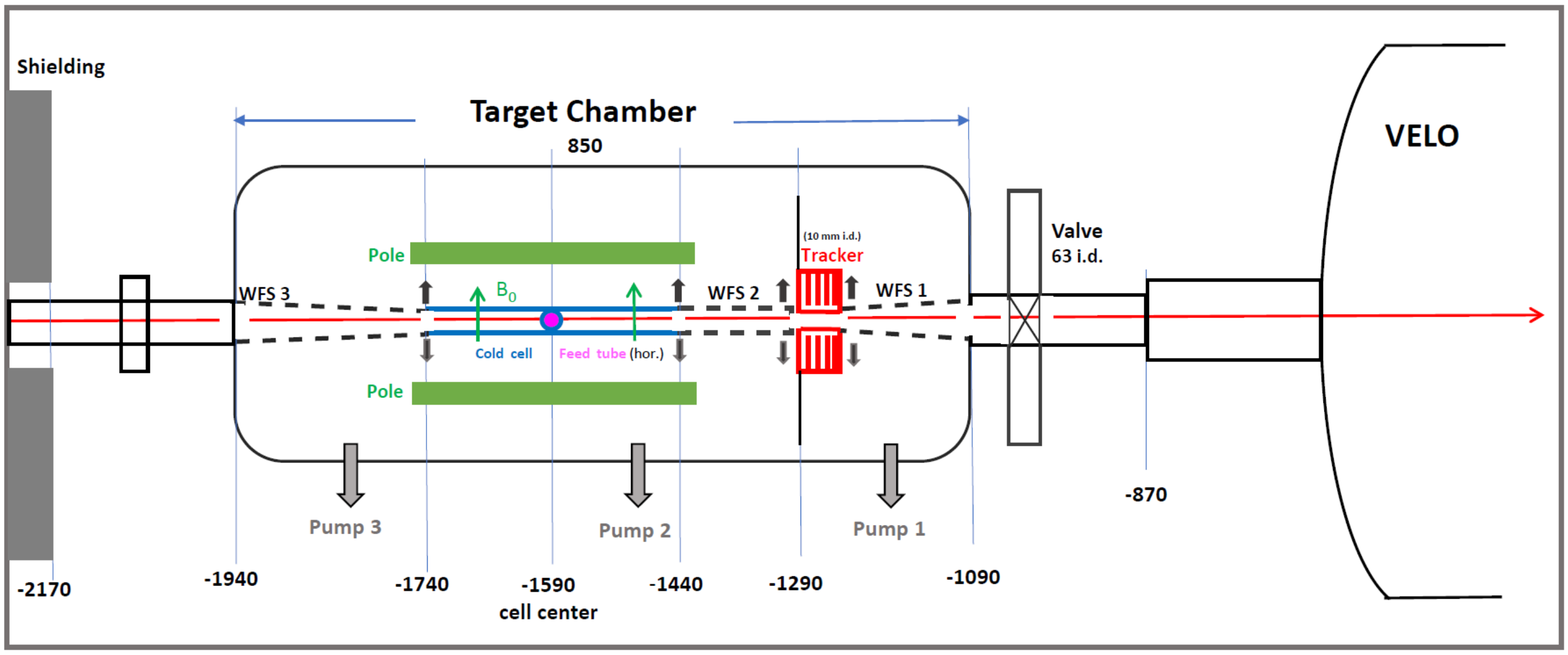}
\caption{\label{pgt} Schematic drawing of the proposed PGT arrangement upstream of VELO detector.}
\end{figure}

\section{Conclusions}

Fixed target collisions at the LHC offers unique opportunities. The LHCb unpolarized gas target SMOG2 will be installed by the end of 2019, while, if approved, the polarized gas target in front of the LHCb spectrometer will bring for the first time spin physics to the LHC, and LHCb will become the first experiment simultaneously running in collider and fixed-target mode with polarized targets, opening a whole new range of explorations.   



\newpage


\renewcommand*{\FigPath}{./WeekIV/02_Royon/}
 
\wstoc{Probing BFKL dynamics, saturation and diffraction at hadronic colliders}{Christophe Royon} 
\title{Probing BFKL dynamics, saturation and diffraction at hadronic colliders}

\author{Christophe Royon}
\index{author}{Royon, C.}

\address{The University of Kansas, Lawrence, USA}


\begin{abstract}
We discuss how we can probe BFKL dynamics and saturation effects at the LHC and the EIC. We also present prospects concerning 
photon induced processes at the LHC and the EIC.  
\end{abstract}

\bodymatter

\section{Proton and heavy ion structures}
Following the proton structure determined using fixed target data from the SLAC, BCDMS, NMC, E665, etc, experiments,
the H1 and ZEUS experiments at HERA completed the knowledge by measuring the proton structure function $F_2$ down to 
$x \sim$ $3\times 10^{-5}$ that led to the discovery of the rise of the gluon density at small $x$~\cite{h1zeus}. The Dokshitzer Gribov Lipatov Altarelli Parisi (DGLAP)~\cite{dglap} evolution equation can
describe data over 6 orders of magnitude in $Q^2$. The D0 and CDF experiments at the Tevatron and then the ATLAS and 
CMS experiments at the LHC completed our knowledge of the gluon density at high $x$ by measuring the jet inclusive cross
section~\cite{inclusivejet}. No clear sign of Balitsky Fadin 
Kuraev Lipatov (BFKL) resummation effects~\cite{bfkl} have been shown so far. 

\begin{figure}
\begin{center}
\epsfig{figure=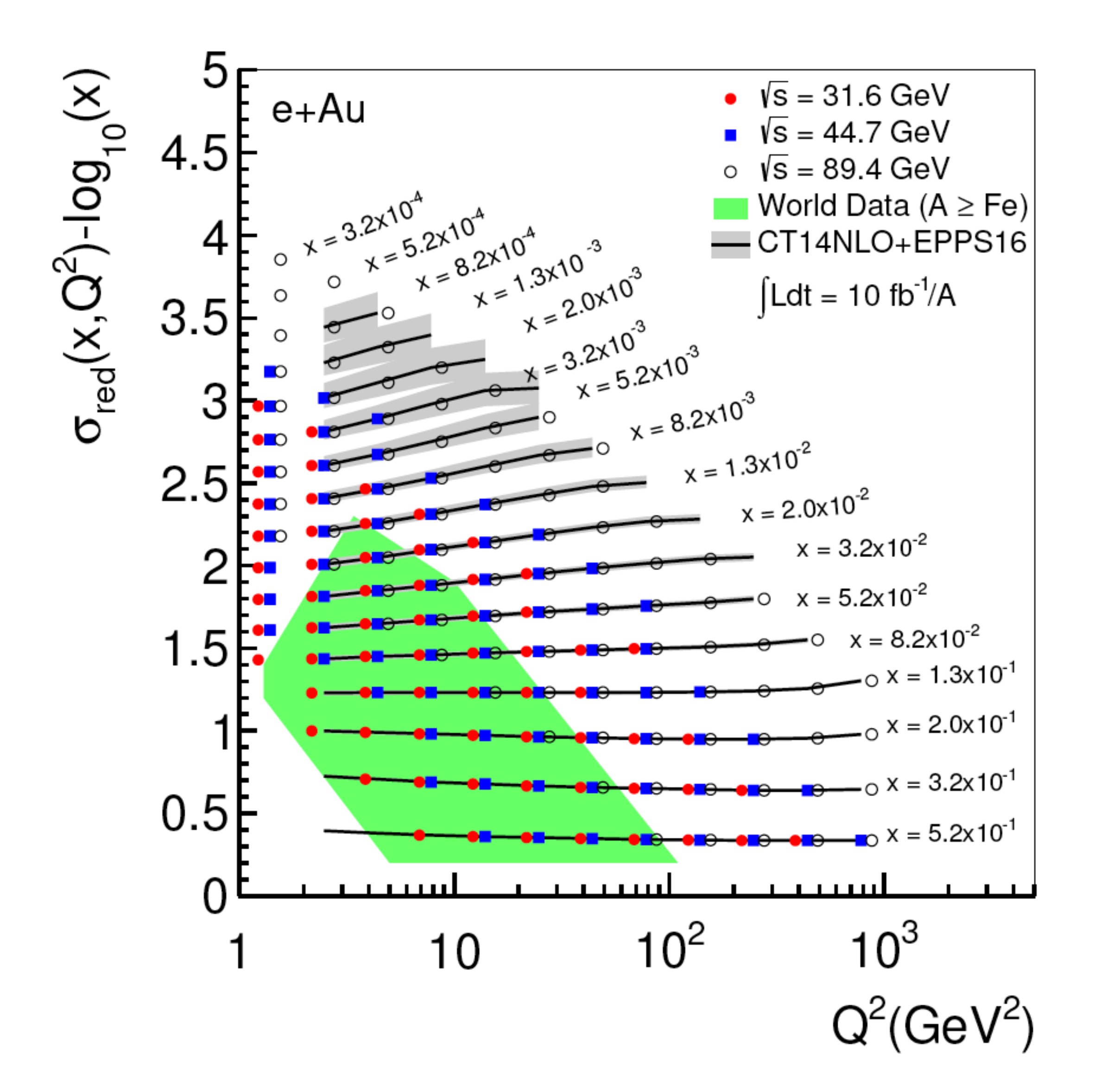,height=3.4in} 
\end{center}
\caption{
\label{ion} 
$e Au$ possible cross section measurements at the EIC for three different center-of-mass energies than can reach high precision at low $x$.}
\end{figure}

The knowledge of the heavy ion structure is much worse and the situation is similar to the one before HERA. The measurements
cover mainly the region above $x> 10^{-3}$ and no data is available at small $x$. The Electron-Ion Collider (EIC) will 
allow to reach $x$ values down to $\sim 10^{-5}$ where saturation effects are expected to appear in heavy ions. As an
example, we give the precision that can be reached on the measurement of the $e Au \rightarrow eX$ cross section~\cite{elke}
down to $x \sim 3 \times 10^{-4}$  in Fig.~\ref{ion} for three center-of-mass energies at the EIC.

In order to look for BFKL resummation effects, some dedicated experimental observables are needed. The measurement
of very forward jet at HERA is directly sensitive to these effects. The idea is to measure the scattered electron and 
jets in the very forward region with a large rapidity interval between the electron and a jet. If the transverse momentum
of the jets and the $Q^2$ of the exchanged virtual photon are similar, the DGLAP cross section is suppressed because
of the $k_T$ ordering of the gluon emission whereas the BFKL cross section is enhanced~\cite{usbfklhera}. The H1
collaboration measured the triple differential jet cross section as a function of $x$ in different bins of $Q^2$ and
jet $p_T^2$. A good agreement was found between the BFKL NLL calculation and the measurement where the DGLAP
NLO prediction undershoots data at low $x$~\cite{usbfklhera}. 

A similar observable called Mueller-Navelt jets~\cite{mnjets} relies on the observation of dijet events with a large 
rapidity interval between them. The DGLAP cross section is small and the BFKL cross section enhanced. The 
azimuthal decorrelation between the two jets~\cite{mnus} is larger for BFKL than for DGLAP because of the additional
gluons that can be emitted along the ladder but unfortunately suffers from higher order corrections. Many additional observables were 
proposed to study Mueller-Navelet jets but they did not lead to a clear observation of BFKL resummation effects since
higher order QCD calculation lead to similar effects.  New less inclusive variables such as the average rapidity of soft gluon
emissions between the Mueller Navelet jets are needed in order to see some evidence of BFKL resummation effects.

\begin{figure}
\begin{center}
\epsfig{figure=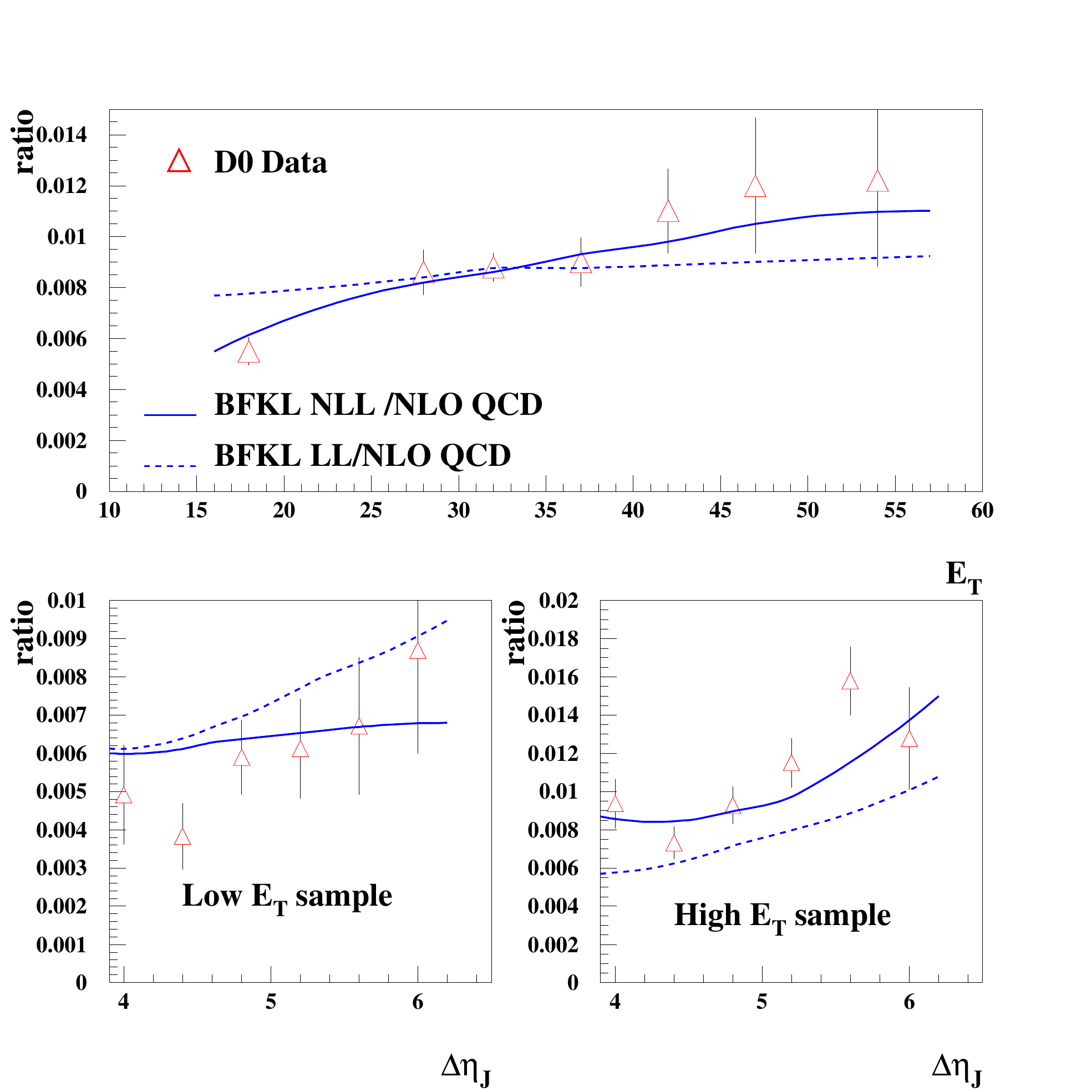,height=3.in} 
\end{center}
\caption{
\label{jetgapjet} 
Gap between jets measured by the D0 collaboration as a function of jet $E_T$ and $\Delta \eta_J$ compared to BFKL LL and NLL
calculations. }
\end{figure}

Another observable sensitive to BFKL resummation effects is the so-called jet-gap-jet events where two jets are 
separated by a gap, a region devoid of any activity.  The BFKL NLL cross section was implemented in the HERWIG 
Monte Carlo~\cite{ourjgj} in order to take into account the difference between the gap size and the difference in rapidity
between the two jets. This is due to the fact that the gap is between the edge of the jets whereas the rapidity difference
is between the jet centers. A good agreement is found between the BFKL NLL calculation and the jet gap jet event 
ratio measured by the D0 collaboration~\cite{d0jgj} as a function of jet $E_T$, or $\Delta \eta$ difference between the 
two jets for the low and high $E_T$ samples as shown in Fig.~\ref{jetgapjet}. It is worth noticing that this calculation
did not include the NLO impact factors but a full NLO calculation (including kernel and impact factors) is in progress.

Looking for BFKL resummation effects and saturation will also be an important topic at the EIC where forward jet measurements 
will also be performed for different heavy ions, A precise measurement of the longitudinal structure function will also be
of great interest since the impact of saturation is different on $F_L$ and $F_2$. A nice test of saturation models
is also the clear dependence of the saturation scale as a function of $A$ for different heavy ions~\cite{raju} that can be
probed at the EIC.

\section{Inclusive and exclusive diffraction}
Diffractive events and the measurement of the parton distributions in the pomeron have been performed at HERA with
high precision~\cite{heraf2d}. Two different methods are mainly used to detect and measure diffractive events, namely
the presence of a rapidity gap in the direction of the outgoing proton or the direct detection of the intact proton in the
final state. This led to the first measurement of the parton distributions in the pomeron and to the evidence that the Pomeron is gluon-dominated~\cite{heraf2d}. Diffractive measurements have also been performed at the
Tevatron and the LHC and the percentage of diffractive events at the Tevatron was measured to be around 1\% compared
to about 10\% at HERA. This difference is due to the additional soft gluon exchanges than can suppress the gap at 
hadronic colliders. The measurement of the gluon and quark densities in the pomeron (and the comparison with the results at HERA in order to understand if the same mechanism is responsible for diffraction) can be achieved at the LHC
using as examples dijets, $\gamma+$jet or $W+X$ events at the LHC~\cite{usdiff}. Jet gap jet events in diffraction 
also lead to a clean test of BFKL resummation~\cite{jgjdif}. Looking for diffractive events at the EIC for different
heavy ions, especially in vector meson production, will be an important part of the program following the experience at previous colliders in order to understand better
the mechanism of diffraction and look for saturation effects.

The LHC can also be seen at a $\gamma \gamma$ collider and photon exchanges can be probed by tagging the intact
protons at the LHC. Measuring for instance two photon production in photon induced processes allows to probe the
quartic anomalous four photon coupling with unprecedented precision~\cite{usgammagamma}. The dominating background at high diphoton 
masses (typically about 400 GeV) is typically due to pile up events where the protons originate from 
secondary interactions with respect to the two high $p_T$ photons. In order to get rid of this background, a good
matching in mass and rapidity between the diphoton and the intact protons is requested that leads to a negligible~\cite{usgammagamma}
background for 300 fb$^{-1}$. The reach on the 4 $\gamma$ anomalous coupling using this
method of detecting intact protons in the final state is typically two orders of magnitude better than usual methods 
at the LHC. The same method can be used to look for axion like particle (ALP) at the LHC at high mass that might
appear as a photon-induced produced resonance decaying into two photons. The reach in the coupling versus
axion mass is shown in Fig~\ref{axion} and leads to unprecedented sensitivities at high masses~\cite{alps}. The same method
can also be used to look for $\gamma \gamma WW$, $\gamma \gamma ZZ$, $\gamma \gamma \gamma Z$ couplings~\cite{usgammaw}.
We gain two or three orders of magnitude on the anomalous couplings with respect to more
usual methods at the LHC (looking for instance into the $Z$ boson decay into three photons). 

To conclude, we presented different measurements at HERA, Tevatron or the LHC that lead to a better understanding
of the proton structure or for new phenomena such as quartic anomalous couplings that might appear because of
composite Higgs bosons, extra-dimensions or the existence of ALPs. The EIC will allow extending our knowledge on 
heavy ion physics in the same way as we now understand better the proton and probably discovering for instance
saturation phenomena in the high gluon density regime.

\begin{figure}[th]
\centering
\includegraphics[width=0.65\textwidth]{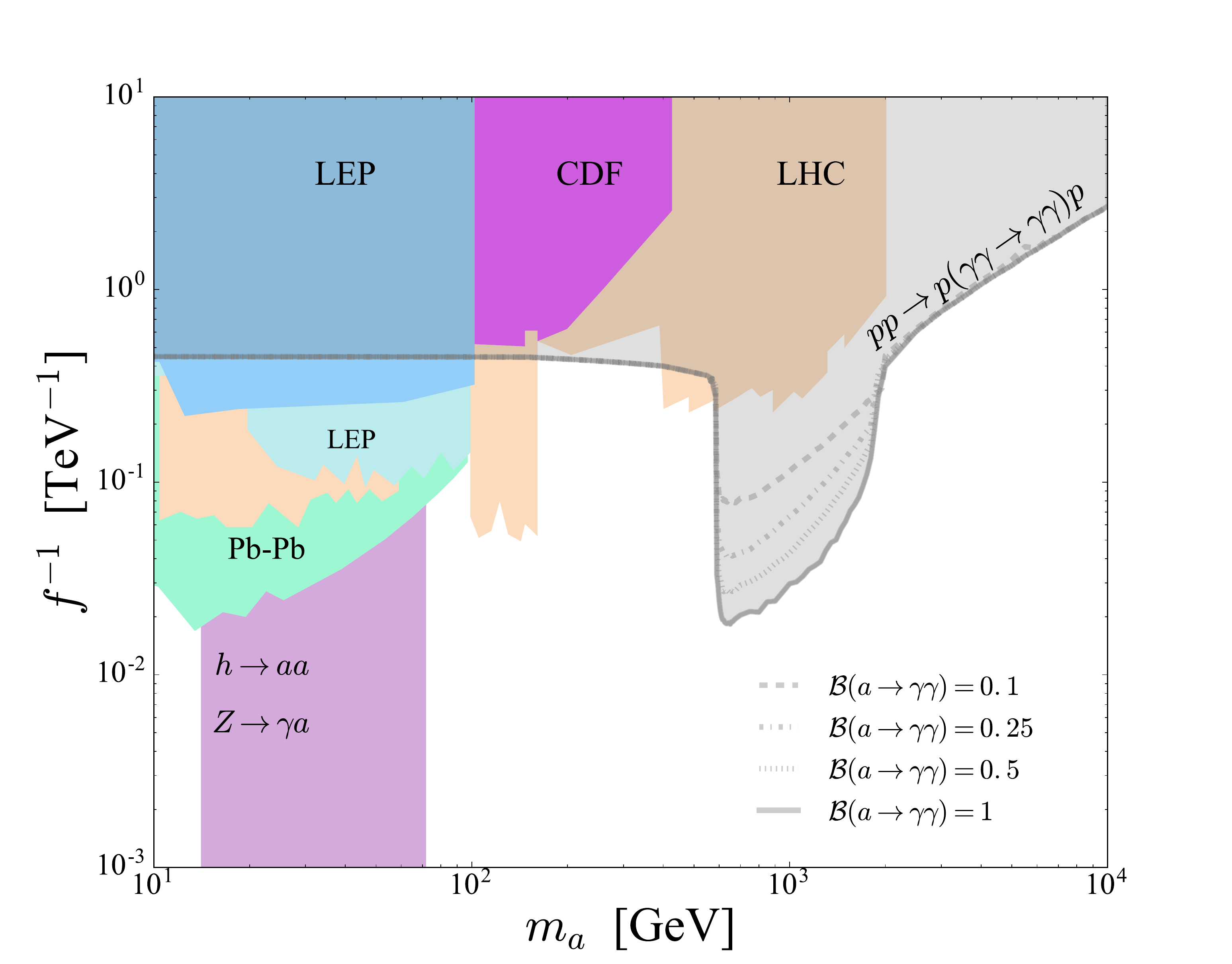}
\caption{
\label{axion} 
Sensitivity to axion like particle production at the LHC in photon-induced processes by tagging the intact protons in the final state in
the coupling versus axion mass plane.}
\end{figure}


\newpage
%

\renewcommand*{\FigPath}{./WeekIV/03_Lorce/}

\wstoc{Mass structure and pressure forces inside the nucleon}{C\'edric Lorc\'e}

\title{Mass structure and pressure forces inside the nucleon}

\author{C\'edric Lorc\'e}
\index{author}{Lorc\'e, C.}

\address{CPHT, CNRS, Ecole Polytechnique, Institut Polytechnique de Paris,\\ Route de Saclay, 91128 Palaiseau, France\\
E-mail: cedric.lorce@polytechnique.edu}

\begin{abstract}
We summarize recent works on the question of the nucleon mass decomposition and the 2D relativistic distribution of pressure forces on the light front. All these mechanical properties are encoded in the energy-momentum tensor of the system which can be constrained using various types of high-energy lepton-nucleon scatterings. Some further developments for targets with spin $>1/2$ are also reported.
\end{abstract}


\bodymatter

\section{Introduction}
Elastic scattering on the nucleon provided during the last 60 years key information about how electric charge and magnetization are distributed in position space within the nucleon~\cite{Perdrisat:2006hj,Miller:2010nz_79}. 20 years ago, other exclusive reactions like e.g. deeply virtual Compton scattering and meson production were shown to give access to generalized parton distributions (GPDs), which are the mother distributions of both ordinary parton distributions and electromagnetic form factors~\cite{Diehl:2003ny,Belitsky:2005qn,Kumericki:2016ehc_75}. On top of providing tomographic pictures of the internal structure of the nucleon, GPDs also give access to the gravitational form factors (GFFs) which characterize the energy-momentum tensor (EMT)~\cite{Ji:1996ek_28_187}. Just like the Fourier transform of electromagnetic form factors can be interpreted in terms of spatial distribution of electric charge and magnetization, the Fourier transform of GFFs can be interpreted in terms of spatial distribution of energy, momentum and pressure forces~\cite{Polyakov:2002yz_31_3,Polyakov:2018zvc_39}. 

We present in this contribution a short summary of some recent developments providing a new look on the nucleon internal structure.

\section{Mass decomposition and balance equations}
Because of Poincar\'e symmetry, one can write in general for the matrix element of the EMT $T^{\mu\nu}(0)$ as follows
\begin{equation}\label{Poincare}
\langle P|T^{\mu\nu}(0)|P\rangle=2P^\mu P^\nu
\end{equation}
for a spin-$1/2$ state with relativistic normalization $\langle p'|p\rangle=2p^0(2\pi)^3\delta^{(3)}(\vec p'-\vec p)$. It is then clear that the mass $M$ of the nucleon can be expressed in an explicitly covariant way in terms of the trace of the EMT~\cite{Shifman:1978zn_183,Jaffe:1989jz_28_83_71_331}
\begin{equation}
\langle P|T^{\mu}_{\phantom{\mu}\mu}(0)|P\rangle=2M^2.
\end{equation}
The trace of the QCD EMT tensor being given by
\begin{equation}
T^{\mu}_{\phantom{\mu}\mu}=\frac{\beta(g)}{2g}\,G^2+(1+\gamma_m)\,\overline\psi m\psi,    
\end{equation}
it is tempting to interpret $\langle P|\frac{\beta(g)}{2g}\,G^2|P\rangle$ and $\langle P|(1+\gamma_m)\,\overline\psi m\psi|P\rangle$ as the gluon and quark contributions to the nucleon mass, respectively. This is however incorrect since the partial (quark or gluon) EMT is not conserved and reads in general~\cite{Ji:1994av_199,Lorce:2017xzd_195}
\begin{equation}
\langle P|T^{\mu\nu}_{q,G}(0)|P\rangle=2P^\mu P^\nu A_{q,G}(0)+2M^2\eta^{\mu\nu}\bar C_{q,G}(0),
\end{equation}
where $A_{q,G}(t)$ and $\bar C_{q,G}(t)$ are GFFs depending on the squared four-momentum transfer $t=\Delta^2=(p'-p)^2$. Unlike electromagnetic form factors, GFFs also depend on the renormalization scale and scheme. The extra term accounts for the non-conservation of the partial EMT. From a more physical point of view, it is related to the pressure-volume work exerted by the quark and gluon subsystems. The nucleon being a stable object, the total pressure-volume work has to vanish and therefore disappears once the EMT is summed over quark and gluon contributions like in Eq.~\eqref{Poincare}. Paying attention to properly distinguish contributions to energy and pressure-volume work, one arrives at a proper mass decomposition and a balance equation~\cite{Lorce:2017xzd_195}
\begin{equation}
M=U_q+U_G,\qquad W_q+W_G=0
\end{equation}
with $U_{q,G}=\left[A_{q,G}(0)+\bar C_{q,G}\right]M$ and $W_{q,G}=-\bar C_{q,G}M$. Using current phenomenological estimates~\cite{Gao:2015aax}, one finds that $U_q\approx 0.44 M$ and $W_q\approx0.11 M$. Contrary to what is sometimes claimed in the literature, the large value of $\langle P|\frac{\beta(g)}{2g}\,G^2|P\rangle/2M=U_G-3W_G\approx 0.89 M$ does not indicate that gluons are responsible for most of the nucleon mass, but comes from the fact that the gluon pressure-volume work is large (reflecting the relativistic nature of the nucleon) and negative (i.e. attractive).

\section{Relativistic 2D distributions of pressure forces}
By analogy with electromagnetic form factors, one can interpret the Fourier transform of GFFs in the Breit frame $\vec P=(\vec p^{\,\prime}+\vec p)/2=\vec 0$ in terms of 3D spatial distribution of mass, momentum and pressure forces~\cite{Polyakov:2002yz_31_3}. Distributions defined in this way are however known to be plagued by relativistic corrections~\cite{Burkardt:2000za_67}. The latter can be avoided by considering instead transverse 2D spatial distributions within the light-front formalism. The ones associated with the EMT are defined in the symmetric Drell-Yan frame $\vec P_\perp=\vec 0_\perp$, $\Delta^+=0$ as~\cite{Lorce:2017wkb_63,Lorce:2018egm_191}
\begin{equation}\label{EMTdist}
\langle T^{\mu\nu}_{q,G}\rangle(\vec b_\perp)=\int\frac{\textrm{d}^2\Delta_\perp}{(2\pi)^2}\,e^{-i\vec\Delta_\perp\cdot\vec b_\perp}\,\frac{1}{2P^+}\langle P^+,\tfrac{\vec\Delta_\perp}{2}|T^{\mu\nu}_{q,G}(0) |P^+, -\tfrac{\vec\Delta_\perp}{2}\rangle
\end{equation}
with $a^+=(a^0+a^3)/\sqrt{2}$. The $T^{++}$ component plays the role of gallilean mass in the transverse plane and has been studied in Refs.~\citenum{Abidin:2008sb,Lorce:2018zpf}. The longitudinal orbital angular momentum of quarks and gluons can be obtained from the $T^{+i}$ components~\cite{Lorce:2017wkb_63} and the transverse $T^{ij}$ components can be interpreted in terms of 2D pressure forces~\cite{Lorce:2018egm_191}. Remarkably, in the last two cases the relativistic 2D distributions appear to coincide with the projection of the 3D distributions defined in the Breit frame onto the transverse plane.

One can write in the transverse plane~\cite{Lorce:2018egm_191}
\begin{equation}
\langle T^{ij}_{q,G}\rangle(\vec b_\perp)=\sigma(b)\,\delta^{ij}_\perp+\Pi(b)\left(\frac{b^i_\perp b^j_\perp}{b^2}-\delta^{ij}_\perp\right),
\end{equation}
where $b=|\vec b_\perp|$, $\sigma(b)$ represents 2D isotropic pressure and $\Pi(b)$ represents 2D pressure anisotropy. In non-relativistic systems, pressure anisotropy is confined to a very thin region at the boundary and described by a surface tension. In relativistic systems like the nucleon and compact stars, pressure anisotropy extends over a larger region in the bulk. Using a multipole parametrization for the nucleon GFFs~\cite{Lorce:2018egm_191}, it appeared that the quark contribution to the EMT is mostly repulsive and short range, while the gluon contribution is mostly attractive and long range. This configuration ensures naturally the mechanical stability of the nucleon as a whole.

\section{Higher-spin targets}
Studies of the EMT mostly focused on the nucleon, or more generally spin-$1/2$ targets. Higher-spin targets are however of deep interest in both hadronic and nuclear physics. Recently, a detailed study of the spin-$1$ case has been conducted in Ref.\citenum{Cosyn:2019aio_59}. It indicated in particular that higher-spin targets simply involve additional contributions associated with higher spin-multipoles, suggesting an alternative approach based on a covariant multipole expansion (still under development). Poincar\'e symmetry has been used to constrain some of the GFFs and to derive the Ji relation for arbitrary spin targets~\cite{Cotogno:2019xcl,Lorce:2019sbq}.

\section{Conclusion}
We presented a short summary of recent developments about the energy-momentum tensor of hadrons. A tomography of the origin of mass and spin along with pressure forces is now possible. Stability conditions may provide new constraints on the observables and hints about the mechanism of confinement. In the coming years, further constraints on gravitational form factors using both exclusive high-energy experiments and Lattice QCD are expected, bringing our understanding of the hadron internal structure to a whole new and exciting level.

\section*{Acknowledgements}
Some of the works presented here have been supported by the Agence Nationale de la Recherche (ANR-16-CE31-0019 and ANR-18-ERC1-0002), the P2IO LabEx (ANR-10-LABX-0038) in the framework ``Investissements d'Avenir'' (ANR-11-IDEX-0003-01) managed by the Agence Nationale de la Recherche, the CEA-Enhanced Eurotalents Program co-funded by FP7 Marie Sklodowska-Curie COFUND Program (No. 600382), the U.S. Department of Energy, Office of Science, Office of Nuclear Physics (No. DE-AC02-06CH11357), and an LDRD initiative at Argonne National Laboratory (No. 2017-058-N0).



 \newpage
%

\renewcommand*{\FigPath}{./WeekIV/04_Metz/}

\wstoc{Generalized TMDs and Wigner Functions}{Andreas~Metz}

\title{Generalized TMDs and Wigner Functions}

\author{Andreas~Metz}
\index{author}{Metz, A.}

\address{Department of Physics, Temple University, Philadelphia, PA 19122, USA \\
E-mail: metza@temple.edu}

\begin{abstract}
Generalized TMDs (GTMDs) of hadrons are the most general two-parton correlation functions. 
The Fourier transforms of GTMDs are partonic Wigner functions.
Both GTMDs and Wigner functions were only briefly mentioned in the EIC White Paper~\cite{Accardi:2012qut_3}, but in the meantime, several interesting developments have happened in this field.
In this write-up, we give a very brief overview of these objects and address the question whether they can play an important role for the EIC science case.
To answer this question we discuss the presently known physics content of GTMDs and Wigner functions, as well as observables that are sensitive to these quantities. 
\end{abstract}


\bodymatter

\section{Definition of GTMDs and Wigner Functions}
GTMDs can be considered generalizations of either GPDs or TMDs.
In the case of quarks they are defined through the off-forward correlator~\cite{Ji:2003ak_26, Belitsky:2003nz_38, Meissner:2008ay_26, Meissner:2009ww_38}
\begin{equation} 
W^{q \, [\Gamma]} (P,\Delta,x,\vec{k}_\perp) = 
\int \frac{dz^- \, d^2\vec{z}_\perp}{2 (2\pi)^3} \, e^{i k \cdot z} \, 
\langle p' | \, \bar{q}(- \tfrac{z}{2}) \, \Gamma \, {\cal W}(- \tfrac{z}{2}, \tfrac{z}{2}) \, q(\tfrac{z}{2}) \, | p \rangle \Big|_{z^+ = 0} \,,
\label{e:gtmd_corr}
\end{equation}
where `$q$' indicates a quark field operator, $\Gamma$ a generic gamma matrix, and $\cal W$ a Wilson line.
(We have suppressed spin labels for the hadrons, and subtleties related to the Wilson line that runs along the light-cone~\cite{Echevarria:2016mrc_26}.)
The average hadron momentum is denoted by $P = (p + p')/2$, and the momentum transfer by $\Delta = p' - p$.
For leading twist, the correlator in Eq.~(\ref{e:gtmd_corr}) can be parameterized in terms of 16 (complex-valued) quark GTMDs~\cite{Meissner:2009ww_38}.
The same number of GTMDs exists for gluons~\cite{Lorce:2013pza_26}.
Each GTMD depends on the longitudinal $(x)$ and transverse $(\vec{k}_\perp)$ parton momentum, as well as the longitudinal $(\xi)$ and transverse $(\vec{\Delta}_\perp)$ momentum transfer to the target.  

Partonic Wigner functions are typically considered for $\xi = 0$ only, where the defining correlator is the Fourier transform of the correlator in Eq.~(\ref{e:gtmd_corr})~\cite{Ji:2003ak_26, Belitsky:2003nz_38, Lorce:2011kd_38},
\begin{equation}
{\cal W}^{q \, [\Gamma]}(x, \vec{k}_\perp, \vec{b}_\perp) =
 \int\frac{d^2\vec{\Delta}_\perp}{(2\pi)^2} \,
 e^{-i \, \vec{\Delta}_\perp \! \cdot \vec{b}_\perp} \, W^{q \, [\Gamma]} (x, \vec{k}_\perp, \vec{\Delta}_\perp) \Big|_{\xi = 0} \,.
\end{equation}
They share important features of Wigner functions in non-relativistic quantum mechanics, which are the counterpart of classical phase space distributions and contain as much information as the wave function~\cite{Wigner:1932eb}.
In particular, also in field theory one can compute an observable like in statistical mechanics according to 
\begin{equation}
\langle O(x, \vec{k}_\perp, \vec{b}_\perp) \rangle =  \int dx \, d^2\vec{k}_\perp \, d^2\vec{b}_\perp \, O(x, \vec{k}_\perp, \vec{b}_\perp) \, {\cal W}^{q [\Gamma]}(x, \vec{k}_\perp, \vec{b}_\perp) \,,
\label{e:observable}
\end{equation}
which represents a very appealing feature of Wigner functions.

\section{Physics Content of GTMDs and Wigner Functions}
We now discuss four aspects related to the very rich physics content of GTMDs and Wigner functions.
First, they can be considered ``mother functions'' of both GPDs and TMDs.
This is illustrated by the equations
\begin{eqnarray}
{\cal F}^{q \, [\Gamma]}(x,\vec{b}_\perp) & = & \int d^2 \vec{k}_\perp \, {\cal W}^{q \, [\Gamma]}(x, \vec{k}_\perp, \vec{b}_\perp) \,,
\label{e:W_proj_1} \\
\Phi^{q \, [\Gamma]}(x,\vec{k}_\perp) & = & \int d^2 \vec{b}_\perp \, {\cal W}^{q \, [\Gamma]}(x, \vec{k}_\perp, \vec{b}_\perp) \,,
\label{e:W_proj_2}
\end{eqnarray}
where ${\cal F}^{q \, [\Gamma]}$ is the correlator defining parton distributions that depend on the impact parameter $(\vec{b}_\perp)$ (which in turn are Fourier transforms of GPDs), while $\Phi^{q \, [\Gamma]}$ defines TMDs.
Equations similar to~(\ref{e:W_proj_1}), (\ref{e:W_proj_2}) hold for the GTMD correlator.
GTMDs and Wigner functions therefore contain all the information encoded in GPDs and TMDs, and additional physics which drops out upon the  projections in Eqs.~(\ref{e:W_proj_1}), (\ref{e:W_proj_2}). 
This feature alone makes their exploration worthwhile.

Second, Wigner functions may allow for 5D imaging of hadrons~\cite{Lorce:2011kd_38}.
In fact, for different polarizations of the nucleon and/or the quark, interesting results in models have been obtained which are consistent with expectations from confinement~\cite{Lorce:2011kd_38}.
But, like in non-relativistic quantum mechanics, partonic Wigner functions are quasi-distributions only and as such can be negative.
Indeed, negative results for Wigner functions were found in a quark target model~\cite{Hagiwara:2014iya}.
Husimi distributions were therefore proposed as alternative tool for quantifying the 5D structure of hadrons~\cite{Hagiwara:2014iya}.
While most likely partonic Husimi distributions are positive definite, their connection to GPDs and TMDs is less clean than for Wigner functions.
Clearly, further studies seem necessary in this field.

Third, GTMDs and Wigner functions have attracted considerable attention in relation to the spin sum rule of the nucleon.
Specifically, based on Eq.~(\ref{e:observable}), the orbital angular momentum (OAM) of quarks in a longitudinally polarized nucleon can be computed according to~\cite{Lorce:2011kd_38, Hatta:2011ku_38, Ji:2012sj_26}
\begin{eqnarray}
L_z^q & = & \int dx \, d^2\vec{k}_\perp \, d^2\vec{b}_\perp \, (\vec{b}_\perp \times \vec{k}_\perp)_z \, {\cal W}_L^{q \, [\gamma^+]}(x, \vec{k}_\perp, \vec{b}_\perp)
\nonumber \\
& = & - \, \int dx \, d^2\vec{k}_{\perp} \, \frac{\vec{k}_{\perp}^{\,2}}{M^2} \, F_{1,4}^q(x,\vec{k}_\perp^{\,2}) \Big|_{\Delta = 0} \,,
\label{e:OAM}
\end{eqnarray}
where the subscript `$L$' indicates longitudinal nucleon polarization, $\gamma^+$ implies unpolarized quarks, and $F_{1,4}$ is a specific twist-2 GTMD~\cite{Meissner:2009ww_38}.
It is remarkable that the definition in Eq.~(\ref{e:OAM}) holds for both the Jaffe-Manohar OAM $L_{\textrm{JM}}$~\cite{Jaffe:1989jz_26} and the Ji OAM $L_{\textrm{Ji}}$~\cite{Ji:1996ek_26}.
This representation of OAM also leads to an intuitive interpretation of the difference $L_{\textrm{JM}} - L_{\textrm{Ji}}$~\cite{Burkardt:2012sd}, which arises due to a different gauge link structure for the Wigner functions in the two cases.
Moreover, Eq.~(\ref{e:OAM}) allowed for the first calculation of $L_{\textrm{JM}}$ in lattice QCD~\cite{Engelhardt:2017miy_38, Engelhardt:2018zma}.
We note in passing that Wigner functions could also be used to define densities of OAM~\cite{Lorce:2011kd_38, Hatta:2011ku_38, Ji:2012ba, Hatta:2012cs, Lorce:2012ce, Rajan:2016tlg}.
Generally, the developments described in this paragraph can be considered a milestone in high-energy spin physics.

Forth, GTMDs and Wigner functions give access to spin-orbit correlations~\cite{Lorce:2011kd_38, Lorce:2014mxa_26, Lorce:2015sqe_26}.  
One example is the GTMD $G_{1,1}$ which contains information about the correlation between the longitudinal spin and OAM of a parton.
Such spin-orbit correlations are very similar to the ones in atomic systems like the hydrogen atom. 
Also in this area more work seems needed in order to fully reveal the underlying physics.
\begin{figure}[t]
\begin{center}
\includegraphics[width = 6.0cm]{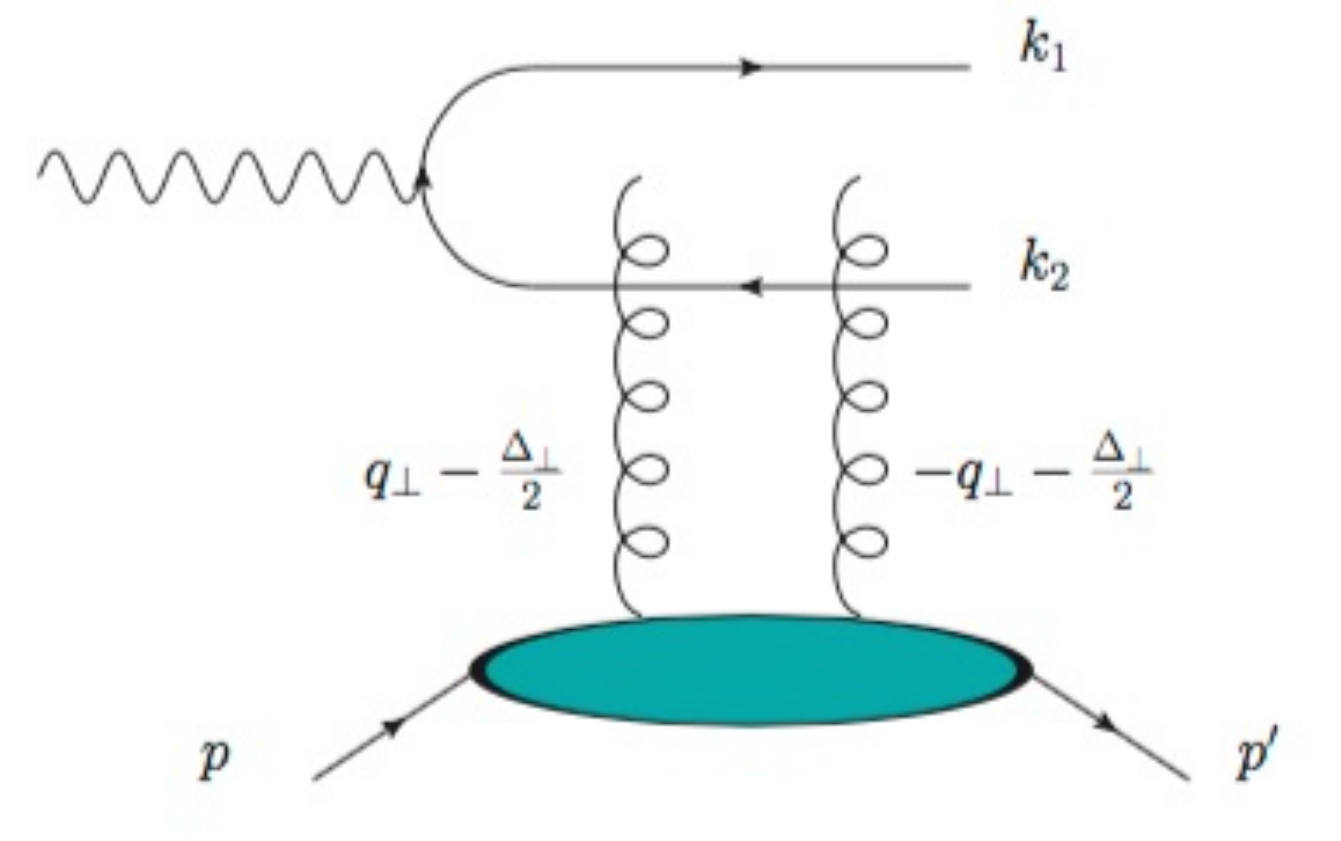}
\hspace{0.8cm}
\includegraphics[width = 5.5cm]{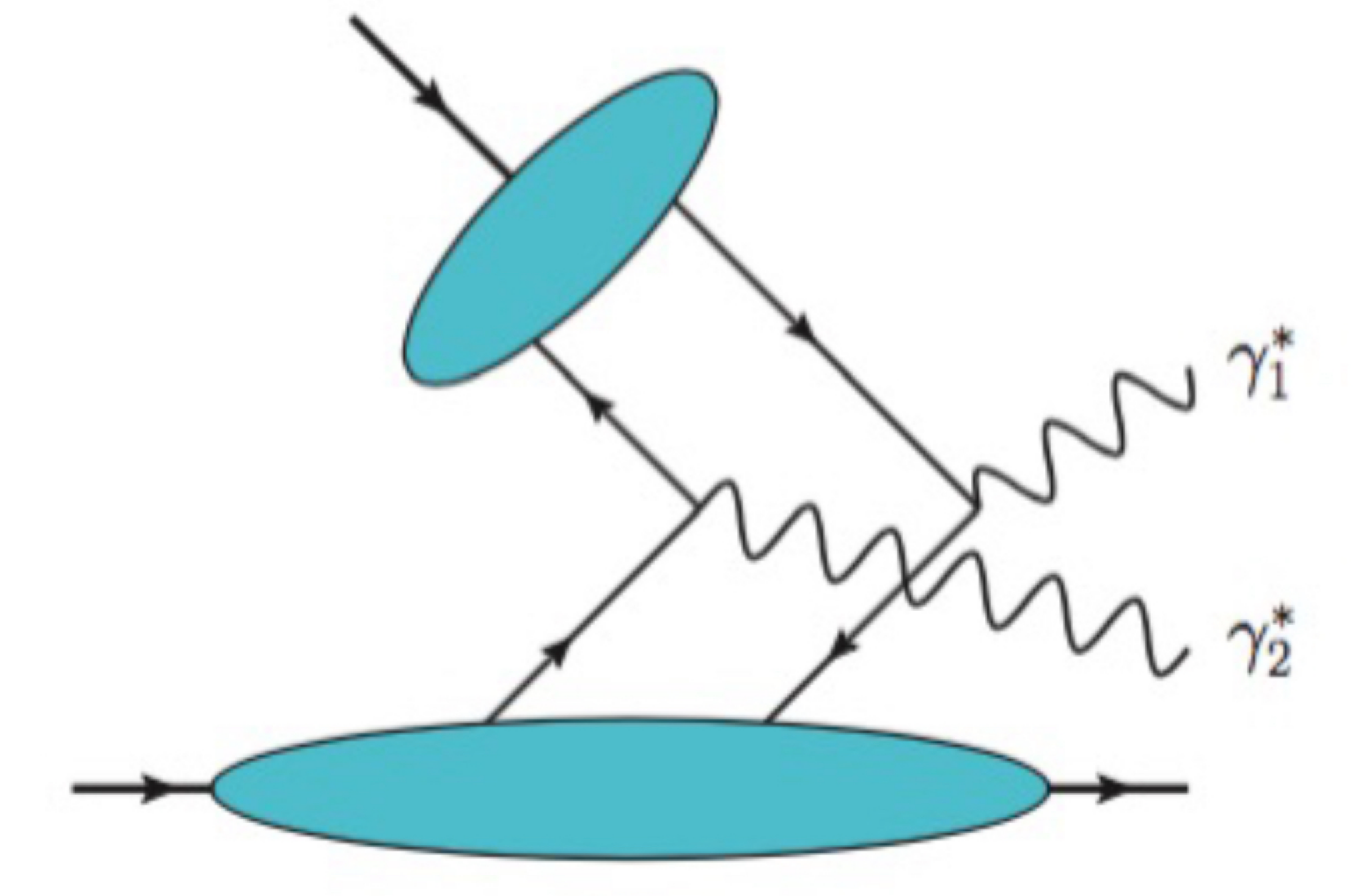}
\end{center}
\caption{Left panel: Schematic representation of diffractive exclusive dijet production in lepton-nucleon/nucleus scattering~\cite{Hatta:2016dxp_38}.
Right panel: One lowest-order diagram for the exclusive pion-nucleon double Drell-Yan process~\cite{Bhattacharya:2017bvs_38}.
The di-lepton pairs are not shown.}
\label{f:processes}
\end{figure}

\section{Observables for GTMDs and Wigner Functions}
After the first publications on partonic Wigner functions~\cite{Ji:2003ak_26, Belitsky:2003nz_38} it took about one decade until a related observable was identified.
Specifically, it was shown that diffractive exclusive dijet production in deep-inelastic lepton-nucleon/nucleus scattering (see the l.h.s.~of Fig.$\,$\ref{f:processes}), which could be measured at a future electron-ion collider, is sensitive to GTMDs of gluons~\cite{Hatta:2016dxp_38}.
The first work on the unpolarized cross section was followed by studies of a particular polarization observable that is related to the OAM of gluons~\cite{Hatta:2016aoc_26,Ji:2016jgn_26}.
One could also measure the same process in ultra-peripheral proton-nucleus and nucleus-nucleus collisions~\cite{Hagiwara:2017fye_26}.
The available studies on exclusive dijet production deal with the gluon sector only and are largely confined to the small-$x$ region.
At present, it is unclear whether in lepton-nucleon/nucleus scattering one can also measure GTMDs for moderate $x$ and/or for quarks.

Further opportunities for measuring Wigner functions exist for hadronic collisions.
In particular, the exclusive (pion-nucleon) double Drell-Yan process --- see the r.h.s.~of Fig.$\,$\ref{f:processes} --- could give access to quark GTMDs~\cite{Bhattacharya:2017bvs_38}.
In fact, double Drell-Yan is so far the only known process with a direct sensitivity to quark GTMDs.
Closely related studies showed that gluon GTMDs (at any $x$) could be addressed via double quarkonium production in, e.g., proton-proton collisions, either for the fully exclusive process~\cite{Bhattacharya:2018lgm_26} or for the case that one proton breaks up~\cite{Boussarie:2018zwg_26}.

\section{Summary and Outlook}
GTMDs and partonic Wigner functions contain a wealth of interesting physics.
Though not all questions about their physics content have been answered, it is worth trying to measure them --- and calculate them in lattice QCD and in suitable models.
From a theoretical point of view it is known by now that observables for GTMDs exist, and a future electron-ion collider could well play an important role in this field.
With regard to lepton-nucleon/nucleus scattering one would hope that (theoretically) clean access to gluon GTMDs is possible also for the region of intermediate $x$ (as opposed to the small-$x$ saturation regime) and, in particular, that a process can be identified which is sensitive to GTMDs of quarks.

\section*{Acknowledgments}
This work has been supported by the National Science Foundation under grant number  PHY-1812359, and by the U.S.~Department of Energy, Office  of  Science, Office of Nuclear Physics, within the framework of the TMD Topical Collaboration.



 \newpage




%
%






\renewcommand*{\FigPath}{./WeekIV/05_Roy_Venugopalan/}

\wstoc{NLO impact factor for inclusive photon+dijet production in $e+A$ DIS at small $x$}{Kaushik Roy, Raju Venugopalan}
\title{NLO impact factor for inclusive photon+dijet production in $e+A$ DIS at small $x$}

\author{Kaushik Roy$^*$ }
\index{author}{Roy, K.}

\address{Department of Physics and Astronomy, Stony Brook University,\\
Stony Brook, NY 11794-3800, USA\\
and \\ Physics Department, Brookhaven National Laboratory,\\ Bldg. 510A, Upton, NY 11973, USA\\
$^*$E-mail: kaushik.roy.1@stonybrook.edu}

\author{Raju Venugopalan$^{**}$}
\index{author}{Venugopalan, R.}

\address{Physics Department, Brookhaven National Laboratory,\\ Bldg. 510A, Upton, NY 11973, USA \\
$^{**}$ Email: raju@bnl.gov}


\begin{abstract}
We summarize the results of a recent first computation~\cite{Roy:2019hwr} in the Color Glass Condensate effective field theory (CGC EFT) of the next-to-leading order (NLO) impact factor for inclusive photon+dijet production in electron-nucleus (e+A) deeply inelastic scattering (DIS) at small $x$. Our computation simultaneously provides the ingredients to compute fully inclusive DIS, inclusive photon, inclusive dijet and inclusive photon+jet channels to the same accuracy. We outline the Wilsonian renormalization group (RG) procedure by virtue of which extant results for the next-to-leading log (in $x$)  JIMWLK evolution, can be integrated with our results to improve the precision to $O(\alpha_s^3 \ln(1/x))$ accuracy. This paves the way towards quantitative studies of saturation at a future Electron-Ion Collider.
\end{abstract}


\bodymatter

\section{Introduction}

The collective dynamics of saturated gluons is described by a weak coupling, classical effective field theory (EFT) framework  known as the Color Glass Condensate (CGC)~\cite{McLerran:1993ni_80,McLerran:1993ka_85,McLerran:1994vd_350,Iancu:2003xm,Gelis:2010nm_180,Kovchegov:2012mbw_150,Blaizot:2016qgz}. In this contribution, we will briefly discuss the first  computation~\cite{Roy:2019hwr,Roy:2019cux} in the CGC EFT of the next-to-leading order (NLO) (in the strong coupling $\alpha_{S}$) ``impact factor" for inclusive production of photon in association with a quark-antiquark dijet in deeply inelastic electron-nucleus ($e+A$) scattering at high energies. This is a timely computation considering the prospect of such a measurement at the luminosities of a future high energy Electron Ion Collider (EIC)~\cite{Accardi:2012qut_4,Aschenauer:2017jsk}. As we will outline here, our computation allows predictions to $O(\alpha_{S}^{3} \, \ln(1/x))$ accuracy which will be instrumental for precision tests of saturation at an EIC. Embedded in our computation are also the first CGC results to the same accuracy for inclusive dijet, inclusive photon+jet and inclusive photon measurements. The momentum space methods developed in this work are very efficient and allow us to explore even higher order contributions, starting with next-to-next-to-leading order (NNLO) and beyond.


\section{CGC essentials and the LO computation}

At leading order (LO) in the CGC power counting, we obtain the following result~\cite{Roy:2018jxq} for the inclusive photon+dijet ($\gamma+q{\bar q}$) differential cross-section:
\begin{equation}
\frac{\mathrm{d}^{3} \sigma^{{\rm LO};\gamma+q\bar{q}+X}}{\mathrm{d}x \,  \mathrm{d}Q^{2} \mathrm{d}^{6} K_{\perp} \mathrm{d}^{3} \eta_{K} }= \frac{\alpha_{em}^{2}q_{f}^{4}y^{2}N_{c}}{512 \pi^{5} Q^{2}} \, \frac{1}{(2\pi)^{4}} \,  \frac{1}{2} \,  L^{\mu \nu} \tilde{X}_{\mu \nu}^{\text{LO}} \, .
\label{eq:triple-differential-CS-LO}
\end{equation}
Here $\alpha_{em}=e^{2}/4\pi$ is the electromagnetic fine structure constant,  $y=\frac{q\cdot P_{N}}{\tilde{l}\cdot P_{N}}$ is the inelasticity, $Q^{2}= -q^{2} >0$ is the virtuality of the exchanged photon, and $\mathrm{d}^{6} K_{\perp} \mathrm{d}^{3} \eta_{K}$ collectively denotes the phase space density of final state quark, antiquark and photon. Likewise, $L^{\mu \nu}$ is the familiar lepton tensor in DIS~\footnote{For notations, conventions and detailed expressions, see~\cite{Roy:2018jxq,Roy:2019hwr}.}. 

We are interested in the hadronic subprocess which can be visualized as a fluctuation of the virtual photon into a $q\bar{q}$ dipole which interacts with the gluon dominated nuclear matter via electric/color charges and emits a real photon either before or after this interaction. The details of this interaction is contained in the hadron tensor which, at LO, is given by,~\footnote{Here $\int \mathrm{d} \Pi^{\text{LO}}_{\perp} = \int_{\bm{l}_{\perp}} \int_{\bm{x}_{\perp},\bm{y}_{\perp}} e^{i \bm{l}_{\perp}.(\bm{x}_{\perp}-\bm{y}_{\perp})-i(\bm{k}_{\perp}+\bm{k}_{\gamma \perp}).\bm{x}_{\perp}-i\bm{p}_{\perp}.\bm{y}_{\perp}} $.} 
\begin{align}
\tilde{X}^{\text{LO}}_{\mu \nu}&=  \int [ \mathcal{D} \rho_{A} ] \, W_{\Lambda_0^-} [\rho_{A}] \, \hat{X}^{\text{LO}}_{\mu \nu} [\rho_{A}]  \nonumber \\
&=  2\pi  \, \delta(1-z_{q}-z_{\bar{q}}-z_{\gamma}) \!\! \int \mathrm{d} \Pi_{\perp}^{\text{LO}} \!\! \int  \! \mathrm{d} {{\Pi_{\perp}^{\prime \text{LO}}}}^{\star} \!  \tau^{q\bar{q},q\bar{q}}_{\mu \nu}(\bm{l}_{\perp},\bm{l'}_{\perp}) \times \nonumber \\
& \Xi(\bm{x}_{\perp},\bm{y}_{\perp};\bm{y'}_{\perp},\bm{x'}_{\perp}) \, ,
\label{eq:H-tensor-LO}
\end{align}
where $\tau^{q\bar{q},q\bar{q}}_{\mu \nu}(\bm{l}_{\perp},\bm{l'}_{\perp})$ denotes the spinor trace in the LO cross-section~\cite{Roy:2018jxq}. The first line in the above equation represents the CGC averaging procedure: a classical statistical (weighted) average over the quantum expectation value, $\hat{X}[\rho_{A}]$. Here $\rho_{A}$ represents the color charge density of large momentum (large $x$) partons that play the role of static color sources for small $x$ dynamical classical gluon fields; this Born-Oppenheimer separation of parton modes is the essence of the CGC EFT. A knowledge of the initial distribution of such sources at the separation scale $\Lambda_{0}^{-}$, contained in the nonperturbative gauge invariant weight functional $W_{\Lambda_{0}^{-}} [\rho_{A}]$  completely specifies the EFT.

In our momentum space computation, the information about $\rho_{A}$ enters through the quark/antiquark  propagators computed in the background classical field of the nucleus, given in the light cone gauge $A^{-}=0$~\footnote{For a right moving nucleus with $P_{N}^{+} \rightarrow \infty$, this is what we call the ``wrong" light cone gauge. The conventional choice $A^{+}=0$ follows from the number density interpretation of parton distribution functions. See~\cite{Roy:2018jxq} for details.} by
\begin{equation}
S_{ij}(p,p') =  S_0(p)\,\mathcal{T}_{q;ij}(p,p')\,S_0 (p')\,,
\label{eq:dressed-quark-mom-prop}
\end{equation}
where $S_0 (p) = \frac{i \slashed{p}}{p^{2}+i \varepsilon}$ is the free massless fermion propagator, and 
\begin{equation}
\mathcal{T}_{ij}(p,p') =  (2 \pi)\, \delta(p^{-}-p'^{-}) \gamma^{-} \text{sign}(p^{-}) \, 
 \int \mathrm{d}^{2} \bm{z}_{\perp} \, e^{-i(\bm{p}_{\perp} - \bm{p'}_{\perp}).\bm{z}_{\perp}}\,\, \tilde{U}^{\text{sign}(p^{-})}_{ij} (\bm{z}_{\perp}) \, ,
\label{eq:fermion-vertex}
\end{equation}
represents the effective vertex denoted by a cross-hatch circle in  Fig.~\ref{fig:dressed-fermion-propagator}.
\begin{figure}[!htbp]
\begin{center}
\includegraphics[scale=0.2]{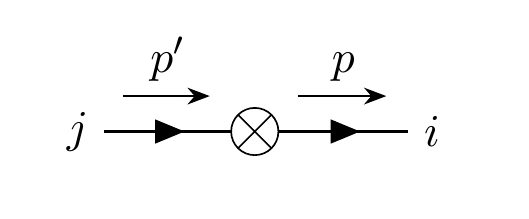}
\caption{``Dressed'' fermion propagator in the background classical color field of the nucleus. The $\otimes$ symbol denotes all multiple scattering insertions {\it including} as well,  the possibility of no scattering. $i$ and $j$ denote color indices in the fundamental representation of $SU(N_{c})$. \label{fig:dressed-fermion-propagator}}
\end{center}
\end{figure}
The lightlike Wilson lines given in the fundamental representation of $SU(N_{c})$ by $
\tilde{U}(\bm{x_{ \perp}})=\mathcal{P}_{-} \text{exp}\left[-ig^{2} \int_{-\infty}^{+ \infty} \mathrm{d}z^{-} \frac{1}{\nabla^{2}_{\perp}}  \rho_{A}^{a} (z^{-},\bm{x_{ \perp}})t^{a} \right] $, efficiently resum all higher twist contributions $\frac{\rho_{A}}{\nabla_\perp^2}\rightarrow \frac{Q_S}{Q^2}$ from the multiple scattering of the $q\bar{q}$ pair off the color field of the nucleus. The CGC averaging introduces non-trivial correlations between the small $x$ classical gluon fields and this is contained in the object $\Xi$ appearing in the second line of Eq.~\ref{eq:H-tensor-LO}, defined as 
\begin{equation}
\Xi(\bm{x}_{\perp},\bm{y}_{\perp};\bm{y'}_{\perp},\bm{x'}_{\perp})=1-D_{xy}-D_{y'x'}+Q_{y'x';xy} \, .
\label{eq:LO-cross-section-color-structure}
\end{equation}
Here 
\begin{align}
D_{xy} &=\frac{1}{N_{c}} \left \langle \text{Tr}\Big( \tilde{U}(\bm{x}_{\perp}) \tilde{U}^{\dagger}(\bm{y}_{\perp}) \Big) \right \rangle  \,  , \enskip \nonumber \\
Q_{xy;zw}  &=\frac{1}{N_{c}} \left \langle \text{Tr} \Big( \tilde{U}(\bm{x}_{\perp}) \tilde{U}^{\dagger}(\bm{y}_{\perp})  \tilde{U}(\bm{z}_{\perp}) \tilde{U}^{\dagger}(\bm{w}_{\perp}) \Big)\right \rangle\, , 
\label{eq:dipole-quadrupole-Wilson-line-correlators}
\end{align} 
represent respectively the dipole and quadrupole Wilson line correlators which are the ubiquitous building blocks of high energy QCD. By taking appropriate limits, one can also recover~\cite{Roy:2018jxq} the nuclear gluon distribution and unintegrated gluon distributions~\cite{Dominguez:2011wm,Dominguez:2011br} from these correlators.

\section{Structure of higher order computations: the NLO impact factor}

To match the accuracy of theoretical predictions to the anticipated precision of experimental data, we need to extend our computation to higher orders which include contributions from quantum fluctuations in the $q\bar{q}$ projectile and the target. In the CGC EFT, leading (large)  logarithms in $x$ ($\alpha_{S}^{k} \ln^{k} (1/x)\sim 1$ for small $x$) which are generated by ``slow" or semi-fast gluon modes are absorbed into the renormalization (RG) evolution of the weight functional $W[\rho_{A}]$; the corresponding RG equation it satisfies is known as the JIMWLK equation~\cite{JalilianMarian:1997gr_110,JalilianMarian:1997dw,Iancu:2000hn_105,Ferreiro:2001qy_151}. The JIMWLK equation resums all powers of $\alpha_{S} \ln (1/x)$ and $Q_{s}^2/Q^2$ arising in loop corrections. This procedure allows us to factorize the cross-section into an ``impact factor'' contribution (free of large logarithms in $x$) that is convoluted with generalized unintegrated nuclear distributions that have the RG evolution of the weight functional embedded in them. In our NLO computation~\cite{Roy:2019hwr} of the $\gamma+q\bar{q}$ process, we have demonstrated an independent first principles derivation of the JIMWLK RGE.

 In addition we have also computed the $x$-independent genuinely $\alpha_{S}$ suppressed pieces that constitute the NLO impact factor. This allows us to go  one step further and consider relevant NNLO processes that contain next-to-leading logarithms (in $x$) (NLL$x$), $\alpha_{S}^{2}\ln (1/x)$ contributions that are effectively of ``size'' $\alpha_{S}$ for $\alpha_{S} \ln(1/x) \sim 1$. By an appropriate matching (see~\cite{Roy:2019hwr,Roy:2019cux} for details) of such contributions between the projectile and target modes, characterized respectively by their longitudinal momenta, $l^{-}>\Lambda_{0}^{-}$ and $l^{-} < \Lambda_{0}^{-}$, we can write the hadron tensor for inclusive photon+dijet production to NLO+NLL$x$ accuracy as
\begin{align}
& {\tilde X}_{\mu\nu}^{\text{NLO}+\text{NLL$x$}} = \!\!  \int [\mathcal{D} \rho_{A}] \, \Big\{ W^{NLLx}[\rho_{A}]\, \hat{X}_{\mu \nu}^{\text{LO}} [\rho_{A}]  + W^{LLx}[\rho_{A}] \, \hat{X}_{\mu \nu}^ {\text{NLO;finite}} [\rho_{A}] \Big\}  \nonumber \\
& \simeq  \int [\mathcal{D} \rho_{A}] \,  \Big( W^{NLLx} [\rho_{A}] \, \Big\{ \hat{X}_{\mu \nu}^{\text{LO}} [\rho_{A}] + \hat{X}_{\mu \nu}^ {\text{NLO;finite}} [\rho_{A}]  \Big\}  + O(\alpha_{S}^{3} \ln(\Lambda^{-}_{1}/\Lambda^{-}_{0}) ) \Big) \, ,
\label{eq:dsigma-NLO-NLLx}
\end{align}
where $W^{LLx}$ and $W^{NLLx}$ denote respectively the weight functionals that contain resummation of leading and next-to-leading logarithms (in $x$). The latter, defined as $
W_{\Lambda_1^-}^{NLLx}  [\rho_{A}] =\Big\{  1+ \ln(\Lambda_{1}^{-}/ \Lambda_{0}^{-}) (\mathcal{H}_{\text{LO}}+\mathcal{H}_{\text{NLO}})  \Big\}\,W_{\Lambda_{0}^{-}}[\rho_{A}] $, requires input about the 
 NLL$x$ JIMWLK evolution which is contained in the $O(\alpha_{S}^{2})$ NLO Hamiltonian, $\mathcal{H}_{\text{NLO}}$ computed in \cite{Balitsky:2013fea,Kovner:2013ona,Balitsky:2014mca,Lublinsky:2016meo,Caron-Huot:2013fea} (see also  \cite{Kovchegov:2006vj_90,Braun:2007vi}).

A knowledge of the NLO impact factor and NLL$x$ JIMWLK evolution can therefore be combined, as shown above, to extend the scope of the computation to $O(\alpha_S^3 \ln(1/x))$ accuracy. However as the $\simeq$ symbol indicates, this knowledge is insufficient to capture all the diagrams (for example, processes at N$^{3}$LO) that contribute to this accuracy.    
\begin{figure}[!htbp]
\begin{center}
\includegraphics[scale=0.2]{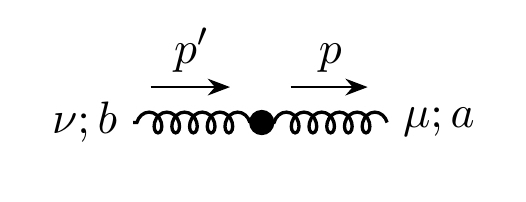}
\caption{Dressed gluon propagator with the dot representing the eikonal interaction vertex with the background classical field. This includes all possible scatterings including the case of ``no scattering''.\label{fig:dressed-gluon-propagator}}
\end{center}
\end{figure} 

 We shall now highlight key features of the computation of the inclusive NLO photon+dijet impact factor in \cite{Roy:2019hwr}. An important ingredient in our computation is the gluon ``small fluctuations" propagator in the $A^-=0$ gauge classical shock wave background field~\cite{McLerran:1994vd_350,Ayala:1995hx,Ayala:1995kg,Balitsky:2001mr,Roy:2018jxq}: 
\begin{equation}
G_{\mu \nu;ab}(p,p') =  G^{0}_{\mu \rho;ac}(p)\,\mathcal{T}_{g}^{\rho \sigma;cd}(p,p')\,G^{0}_{\sigma \nu;db} (p') \, ,
\label{eq:dressed-gluon-mom-prop}
\end{equation}
where $G^{0}_{\mu \rho;ac}(p)=\frac{i}{p^{2}+i\varepsilon}\Big( -g_{\mu \rho}+\frac{p_{\mu}n_{\rho}+n_{\mu}p_{\rho}}{n.p} \Big)\delta_{ac}$ is the free propagator with  Lorentz indices 
$\mu,\rho$, color indices $a,c$ and $n^{\mu}=\delta^{\mu +}$ .
The effective gluon vertex 
\begin{align}
\mathcal{T}^{\mu \nu;ab}_{g}(p,p')  &=-2\pi \delta(p^{-}-p'^{-})  (2p^{-}) g^{\mu \nu} \, \text{sign}(p^{-}) \times
\nonumber \\
& \, \int \mathrm{d}^{2} \bm{z}_{\perp} \, e^{-i(\bm{p}_{\perp}-\bm{p'}_{\perp}).\bm{z}_{\perp} }\,\, \Big( U^{ab} \Big)^{\text{sign}(p^{-})} \!\!\!\!\!\!\!\!\!\!\!\!\! (\bm{z}_{\perp}) \, ,
\label{eq:vertex-gluon}
\end{align}
corresponding to multiple scattering of the gluon off the shock wave background field, is represented by the filled blobs in Fig.~\ref{fig:dressed-gluon-propagator}. 

 The quantum fluctuations (with $l^{-}>\Lambda_{0}^{-}$)  that contribute towards the NLO impact factor can be broadly classified into the modulus squared of real gluon emission amplitudes and the interference of virtual gluon exchange processes with LO diagrams; in each case, the real or virtual gluon can scatter off the shock wave or propagate freely without scattering (see~\cite{Roy:2019hwr} for the complete set of real and virtual graphs contributing at NLO). These can be categorized systematically by their color structures, allowing one to clearly observe the cancellation of the soft, collinear and ultraviolet (UV) divergences that arise in the intermediate steps of our computation. Soft singularities arising from the $l^{-}=0$ gluon pole manifest as logarithms in $\Lambda_{0}^{-}$ (or equivalently $x$) which are absorbed into the RG evolution of $W_{\Lambda_{0}^{-}}[\rho_{A}]$; as already mentioned, this  provides an explicit proof of high energy JIMWLK factorization for a non-trivial process other than fully inclusive DIS. 

In the limit of massless quarks, the NLO computation at fixed coupling $\alpha_{S}$ requires no renormalization. As such the UV divergences cancel between graphs but because we are not integrating over the momenta of the $\gamma + q\bar{q}$ final state, there are collinear divergences that survive real-virtual cancellations. These are absorbed into a jet algorithm. We work in the approximation of narrow jets~\cite{Ivanov:2012ms} of jet cone radius $R \ll 1$. The dominant contribution is of the form $\alpha_{S} (A \, \ln (R) +B)$, where $A, B$,  spelled out in \cite{Roy:2019hwr}, are of $O(1)$; all non-collinearly divergent contributions are phase space suppressed by powers of $R^{2}$.

After careful consideration of all divergences, our final result for the triple differential cross-section for the $\gamma+q\bar{q}$ jet production in e+A DIS is~\cite{Roy:2019hwr,Roy:2019cux}
\begin{align}
\frac{\mathrm{d}^{3} \sigma^{\text{LO+NLO+NLL$x$};\rm jet}}{\mathrm{d}x \mathrm{d}Q^{2} \mathrm{d}^{6} K_{\perp} \mathrm{d}^{3} \eta_{K} }& = \frac{\alpha_{em}^{2}q_{f}^{4}y^{2}N_{c}}{512 \pi^{5} Q^{2}} \, \frac{1}{(2\pi)^{4}} \,  \frac{1}{2} \, L^{\mu \nu}  {\tilde X}_{\mu \nu}^{\text{LO+NLO+NLL$x$}; \rm jet}  \,,
\end{align}
where the hadron tensor at $O(\alpha_{S}^{3} \ln (1/x))$ accuracy can be written as~\cite{Roy:2019hwr,Roy:2019cux}
\begin{align}
{\tilde X}_{\mu \nu}^{\text{LO+NLO+NLL$x$}; \rm jet} &=   \int [\mathcal{D} \rho_A] \, W_{x_{\rm Bj}}^{ NLLx}[\rho_A] \Bigg[\Bigg( 1+  \frac{2\alpha_{S}C_{F}}{\pi} \, \Bigg\{ -\frac{3}{4} \ln \Big(  \frac{R^{2} \vert \bm{p}_{J\perp} \vert \, \vert \bm{p}_{K\perp} \vert  }{4 z_{J} z_{K} Q^{2} e^{\gamma_{E}}} \Big)  \nonumber \\
& +\frac{7}{4} -\frac{\pi^{2}}{6} \Bigg\} \, \Bigg) {\tilde X}_{\mu\nu}^{\rm LO; jet} [\rho_{A}]  + {\tilde X}_{\mu \nu; \rm finite}^{\rm NLO; \rm jet} [\rho_{A}] \, \Bigg] \,.
\label{eq:Xmunu-final}
\end{align}
In Eq.~\ref{eq:Xmunu-final}~\footnote{$\bm{p}_{J \perp,K\perp}$ are the transverse momenta carried by the two jets and $z_{J},z_{K}$ are their respective momentum fractions relative to the projectile momentum. $\gamma_{E}$ is the Euler-Mascheroni constant.}, the finite terms ${\tilde X}_{\mu \nu; \rm finite}^{\rm NLO; jet} $ are of order $\alpha_S$ relative to the leading term and constitute the NLO impact factor; their explicit evaluation is the principal goal of~\cite{Roy:2019hwr}. While some of these expressions can be obtained analytically, others have to be numerically evaluated.

\section{Summary and outlook}

We provided here a brief summary of the computation of the differential cross-section for inclusive photon production in small $x$ $e+A$  DIS in the CGC EFT. We discussed the ingredients, in particular the simple momentum space structures of the dressed quark and gluon propagators in the LC gauge $A^{-}=0$, that allow for efficient higher order computations. The numerical computation of the NLO impact factor, along with NLO BK/JIMWLK evolution is a challenging task, but it is clearly feasible and will be crucial for precision phenomenology of the aforementioned production channels that can be accessed at a future EIC. Note that photon+jet measurements were previously performed for $e+p$ collisions at HERA at fairly large values of $Q^{2}$ and jet, photon momenta~\cite{Abramowicz:2012qt}. Since the integrated luminosity at HERA is likely several orders of magnitude lower than the EIC, photon+dijet measurements are clearly feasible at EIC despite the somewhat lower energies.

One may also consider employing this framework in p+A collisions, beyond the current state-of-the art for inclusive hadron~\cite{Chirilli:2012jd_50,Chirilli:2011km_65,Stasto:2011ru,Altinoluk:2014eka,Ducloue:2019ezk_75}, quarkonium~\cite{Kang:2013hta,Ma:2014mri} and photon production~\cite{Benic:2016uku_60,Benic:2016yqt_146,Benic:2018hvb_55}, to NNLO in the CGC power counting and beyond. These efforts will therefore pave the way towards global analyses of  multiparton correlators and the study of their evolution with energy, through quantitative global analyses of data.

\section*{Acknowledgments}

This material is based on work supported by the U.S. Department of Energy, Office of Science, Office of Nuclear Physics, under Contracts No. de-sc0012704 and within the framework of the TMD Theory Topical Collaboration. K. R is supported by an LDRD grant from Brookhaven Science Associates and by the Joint BNL-Stony Brook Center for Frontiers in Nuclear Science (CFNS).


 \newpage
%

\renewcommand*{\FigPath}{./WeekIV/06_Hautmann/}

\def\ltap{\raisebox{-.6ex}{\rlap{$\,\sim\,$}} \raisebox{.4ex}{$\,<\,$}} 
\def\gtap{\raisebox{-.6ex}{\rlap{$\,\sim\,$}} \raisebox{.4ex}{$\,>\,$}} 

\renewcommand{\as}{\alpha_\mathrm{s}}

\wstoc{Hadron structure and parton branching beyond\\ collinear approximations}{Francesco Hautmann}
\title{Hadron structure and parton branching beyond\\ collinear approximations
}

\author{Francesco Hautmann}
\index{author}{Hautmann, F.}

\address{University of Oxford / University of Antwerp\\
E-mail: francesco.hautmann@physics.ox.ac.uk}

\begin{abstract}
We briefly illustrate recent developments in the parton branching formulation of 
TMD evolution and their impact on precision measurements in high-energy hadronic collisions.  
\end{abstract}


\bodymatter

\vskip 0.6 cm 

The impact of hadron structure on precision studies of fundamental interactions and 
searches for new physics 
plays an essential role at high-energy colliders of the present and next generation. The 
QCD theoretical framework based on 
collinear parton distribution functions (PDFs) and parton showers, in particular,  is 
extremely successful in describing a wealth of collider data.~\cite{Kovarik:2019xvh}  

However, PDFs are the result of a 
strong reduction of information and tell us only about the longitudinal momentum of partons in a fast moving hadron. 
This restriction is lifted in  the transverse momentum dependent (TMD) parton distribution functions ---  more general distributions which provide 
``3-dimensional imaging" of hadron structure.~\cite{Angeles-Martinez:2015sea}  Such distributions are needed to obtain  QCD factorization 
formulas for  collider  observables in  ``extreme"  kinematic regions characterized by multiple momentum scales. These  will be  relevant  
both for experiments at the high-energy frontier and for exploring the region of   the  highest masses  accessible  at the high-luminosity frontier. 

A large body of knowledge has been built  about 
collinear PDFs  over the last three decades  from the analysis of  high-energy experimental data in hadronic collisions, 
greatly aided by the development of realistic  Monte Carlo  (MC) event 
simulations~\cite{Bengtsson:1987kr,Marchesini:1987cf}  for the parton cascades associated with  PDF evolution. 
TMDs, on the other hand,  are much less known.  Hadronic 3D imaging, with  its 
implications for high-energy physics, will constitute the subject of 
intensive studies in the forthcoming decade.    
 The construction of MC   event 
 generators incorporating 
TMDs and 3D hadron structure effects~\cite{Hautmann:2019rvr}  
is 
thus 
a central objective of 
physics programs for future hadron colliders (HL-LHC, LHeC, EIC, FCC).

Steps toward TMD MCs have recently been  taken in the works~\cite{Martinez:2019mwt,Martinez:2018jxt,Hautmann:2017fcj,Hautmann:2017xtx}, in which a parton branching 
formalism is proposed for TMD evolution, and applications to deep inelastic scattering (DIS) and Drell-Yan (DY) processes are presented. 
In the following we give a brief account of these studies.

The parton branching (PB) approach gives TMD evolution equations of the schematic form~\cite{Hautmann:2017fcj,Hautmann:2017xtx}  
\begin{eqnarray}
\label{eq:tmdevol}
 {A}_a\left( x, {\bm k}, \mu^2\right) &=& 
 \Delta_a\left(\mu^2, \mu_0^{2}\right)
 {A}_a\left( x, {\bm k}, \mu_0^2\right)
  \\ 
 &+& 
 \sum_b\int \frac{\textrm{d}^2{\boldsymbol \mu}^{\prime}}{\pi {\mu}^{\prime 2}} 
   \int  \textrm{d}z 
\   {\cal K}_{a b}    \left(  x, {\bm k}, \mu^2 ;   z , z_M ,  {\mu}^{\prime 2}     \right) \times \nonumber \\
& & {A}_b\left( x / z ,  {\bm k} + a(z){\boldsymbol \mu}^\prime, \mu^{\prime 2}\right) \; , 
\nonumber   
\end{eqnarray} 
where $
A_a\left( x, {\bm k}, \mu^2\right)$ is the 
TMD distribution of flavor $a$, carrying the longitudinal momentum  fraction $x$ of the hadron's momentum and  transverse momentum ${\bm k}$ 
at the evolution scale $\mu$;  $z$ and ${\boldsymbol \mu}^\prime$ are the branching variables, with $z$ being the longitudinal momentum transfer  
at the  branching, and  $ \mu^\prime = \sqrt{ {\boldsymbol \mu}^{\prime 2}}$ the momentum scale at which the branching occurs; 
$z_M$ is the soft-gluon resolution scale;  the function $a(z)$ specifies the ordering condition  in the branching; 
 $ {\cal K}_{a b} $    are evolution kernels, computable in terms of  
 Sudakov form factors,   real-emission splitting functions and phase-space constraints.    The initial  
evolution scale is denoted by $\mu_0$;      the   distribution  $ {A}_a\left( x, {\bm k}, \mu_0^2\right)$ at scale $\mu_0$ in the first term on the right hand side of 
Eq.~(\ref{eq:tmdevol})   is the intrinsic   $k_T$  distribution. 

The PB evolution equations  are designed to be applicable over a wide kinematic range from low to high energies and implementable in 
MC generators.  
By taking  the ordering function  $a(z)$, soft-gluon resolution scale $z_M$ and strong coupling $\as$  of  the form prescribed 
by  angular ordering~\cite{Catani:1990rr,antw1908},  
  Eq.~(\ref{eq:tmdevol})   gives, once it is integrated over transverse momenta,  
  the CMW coherent-branching equation.~\cite{Marchesini:1987cf,Catani:1990rr}
On the other hand, for  soft-gluon resolution  
$z_M \to 1$ and strong coupling $\as \to \as (\mu^{\prime 2})$,   integrating  Eq.~(\ref{eq:tmdevol}) over transverse momenta 
gives    collinear PDFs satisfying  
DGLAP evolution equations.~\cite{Gribov:1972ri,Altarelli:1977zs,Dokshitzer:1977sg}
The convergence to DGLAP 
 at leading order (LO) and  next-to-leading order (NLO)  has been 
verified numerically in~\cite{Hautmann:2017xtx} against the evolution program~\cite{Botje:2010ay} 
  at the 
   level of better   than 1\% over a range of five orders of magnitude both in $x$ and in $\mu$.
Besides the collinear limits, Eq.~(\ref{eq:tmdevol}) can be used at unintegrated level for event simulation 
of TMD physics effects.~\cite{jcc-book}

In Figs.~\ref{TMD_pdfs3} and \ref{Zpt-TMD_uncertainty} we give examples of PB-TMD applications to DIS and DY processes. 
Fig.~\ref{TMD_pdfs3}~\cite{Martinez:2018jxt}  shows results for TMDs from PB fits at NLO 
 to the HERA high-precision inclusive 
DIS data~\cite{Abramowicz:2015mha},  performed 
using the   fitting platform   \verb+xFitter+~\cite{Alekhin:2014irh} and the numerical techniques  
developed in~\cite{Hautmann:2014uua} to treat the transverse momentum dependence in the fitting procedure. 
 In~\cite{Martinez:2018jxt}    two fitted TMD sets are presented,  differing  by the 
treatment of the momentum scale in the  coupling $\as$, so that one can 
compare the effects of $\as$ evaluated at the transverse momentum scale  
prescribed by the angular-ordered branching~\cite{Catani:1990rr,Martinez:2018jxt} with $\as$ evaluated at the evolution scale.  
The TMDs are extracted including a determination of experimental and theoretical uncertainties. The lower panels in 
Fig.~\ref{TMD_pdfs3}    show these uncertainties for $\bar u$ and gluon 
 distributions, at fixed values of $x$ and $\mu$, as a function of transverse momentum. 

The  $k_T$ dependence in Fig.~\ref{TMD_pdfs3} results from  intrinsic transverse momentum and  evolution.   
The intrinsic  $k_T$ in  Fig.~\ref{TMD_pdfs3} is taken for simplicity to be 
described by a gaussian at $\mu_0 \sim {\cal O}$ (1 GeV)  
with (flavor-independent and $x$-independent) width $\sigma = k_0 / \sqrt{2}$,  $k_0 = 0.5$ GeV. 
This is to be compared with higher values of intrinsic $k_T \sim$ 2  GeV obtained  from tuning in shower MC event  
generators (see e.g.~\cite{Khachatryan:2015pea}).

\begin{figure}[htb]
\begin{center} 
\includegraphics[width=0.405\textwidth]{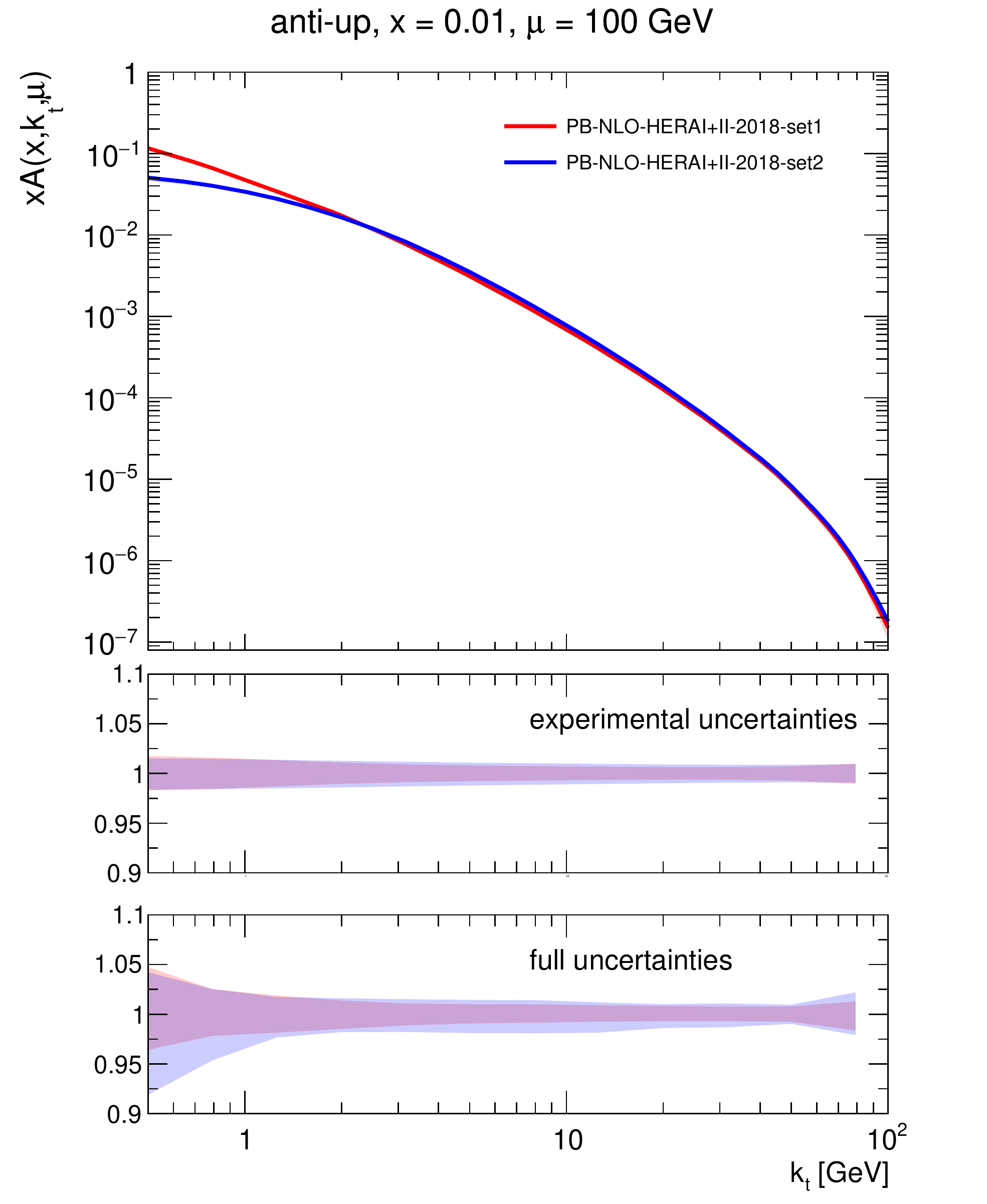}
\includegraphics[width=0.405\textwidth]{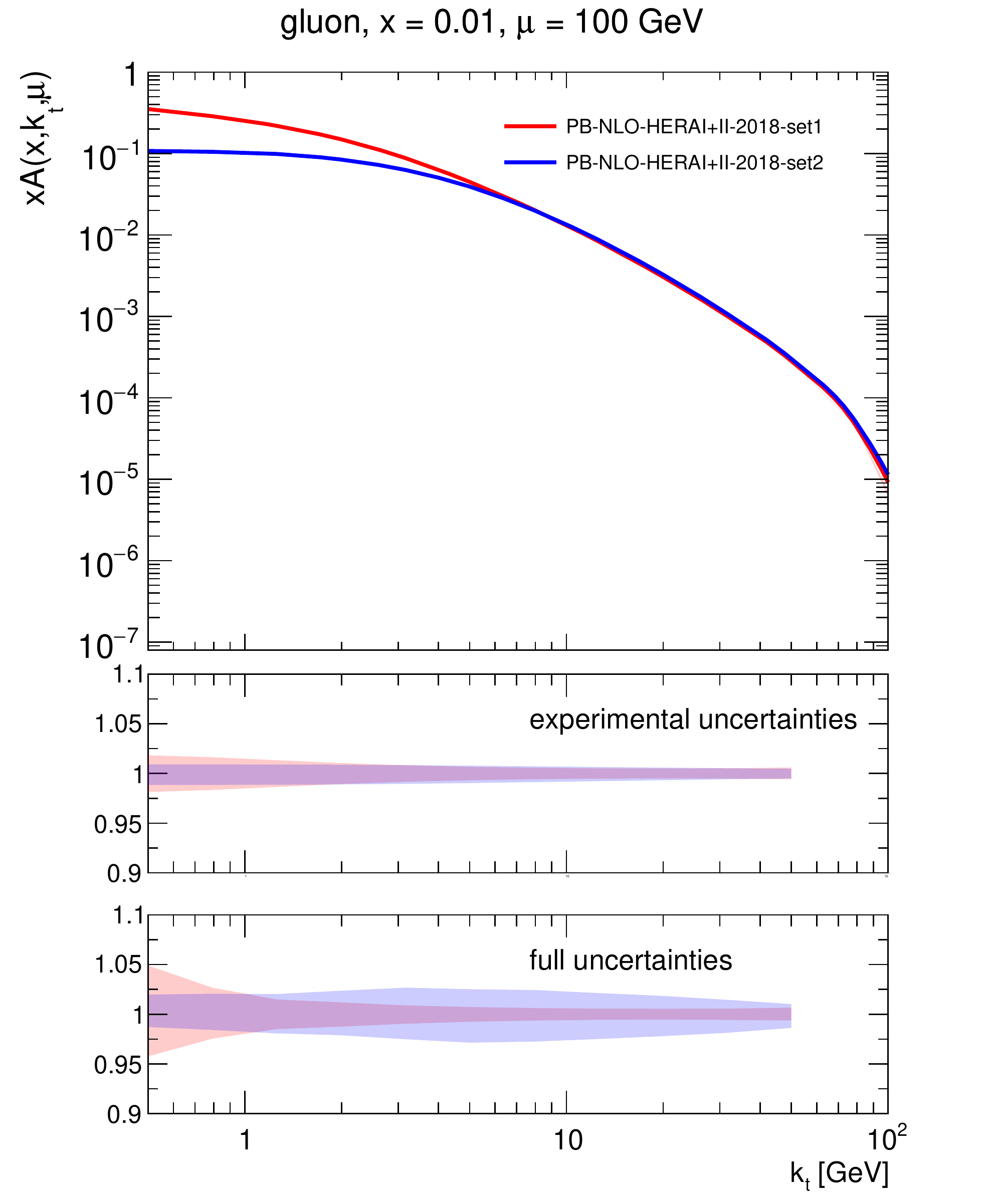}
  \caption{\small TMD   $\bar u$ and gluon  distributions as a function of $k_T$ for  $\mu=100$~GeV at $x=0.01$~\protect\cite{Martinez:2018jxt}. In the lower panels 
 the relative uncertainties are shown,  coming  from experimental uncertainties and the total of experimental and model uncertainties.
  }
\label{TMD_pdfs3}
\end{center}
\end{figure}

In Fig.~\ref{Zpt-TMD_uncertainty}~\cite{Martinez:2019mwt}   the PB  TMDs are combined with the 
NLO calculation of 
 DY  $Z$-boson production to determine predictions  for  the lepton-pair transverse momentum $p_T$  spectrum. 
 These  are  compared   with 
  LHC measurements.~\cite{Aad:2015auj}  
The computation in Fig.~\ref{Zpt-TMD_uncertainty} 
 requires addressing  issues of matching~\cite{jcc-fh-jhep}  analogous to those that arise in the case of parton showers. The 
matching is accomplished in the aMC@NLO framework.~\cite{alwall14}       
The calculations are performed via  
 {\sc Cascade}~\cite{Jung:2010si}   to read LHE~\cite{Alwall:2006yp}  files, perform  TMD evolution,~\cite{Hautmann:2017fcj} produce output files,  
 and {\sc Rivet}~\cite{Buckley:2010ar} to analyze the outputs.

The behaviors in the DY spectrum in Fig.~\ref{Zpt-TMD_uncertainty}  
 can be understood in terms of the  $k_T$  distributions in  Fig.~\ref{TMD_pdfs3}. 
The uncertainties on the DY predictions 
come from TMD uncertainties and scale variations, with the latter dominating the overall uncertainty. 
We see from the left panel in Fig.~\ref{Zpt-TMD_uncertainty}  that the spectrum at low $p_T$ is sensitive to the 
angular ordering effects embodied in the  different treatment of $\as$ in the  PB Set 1 and Set 2. 
The bump in the $p_T$ distribution for intermediate $p_T$ is an effect of  the matching and choice of 
the matching scale~\cite{Martinez:2019mwt,alwall14}   --- a similar effect is seen when using parton showers instead of PB  TMD. 
The deviation in  the spectrum at higher $p_T$ is due to including only  ${\cal O} (\as) $ corrections but missing higher orders. We see from 
 the right panel of Fig.~\ref{Zpt-TMD_uncertainty}  that the contribution from DY + 1 jet at NLO plays an important role at 
 larger $p_T$.  The  merging of higher jet multiplicities~\cite{bermudeztalk} in the PB TMD framework is one  of the 
  ongoing  developments  needed for MC event generators including 3D hadron  structure effects.

\begin{figure}[htb]
\begin{center} 
\includegraphics[width=0.405\textwidth]{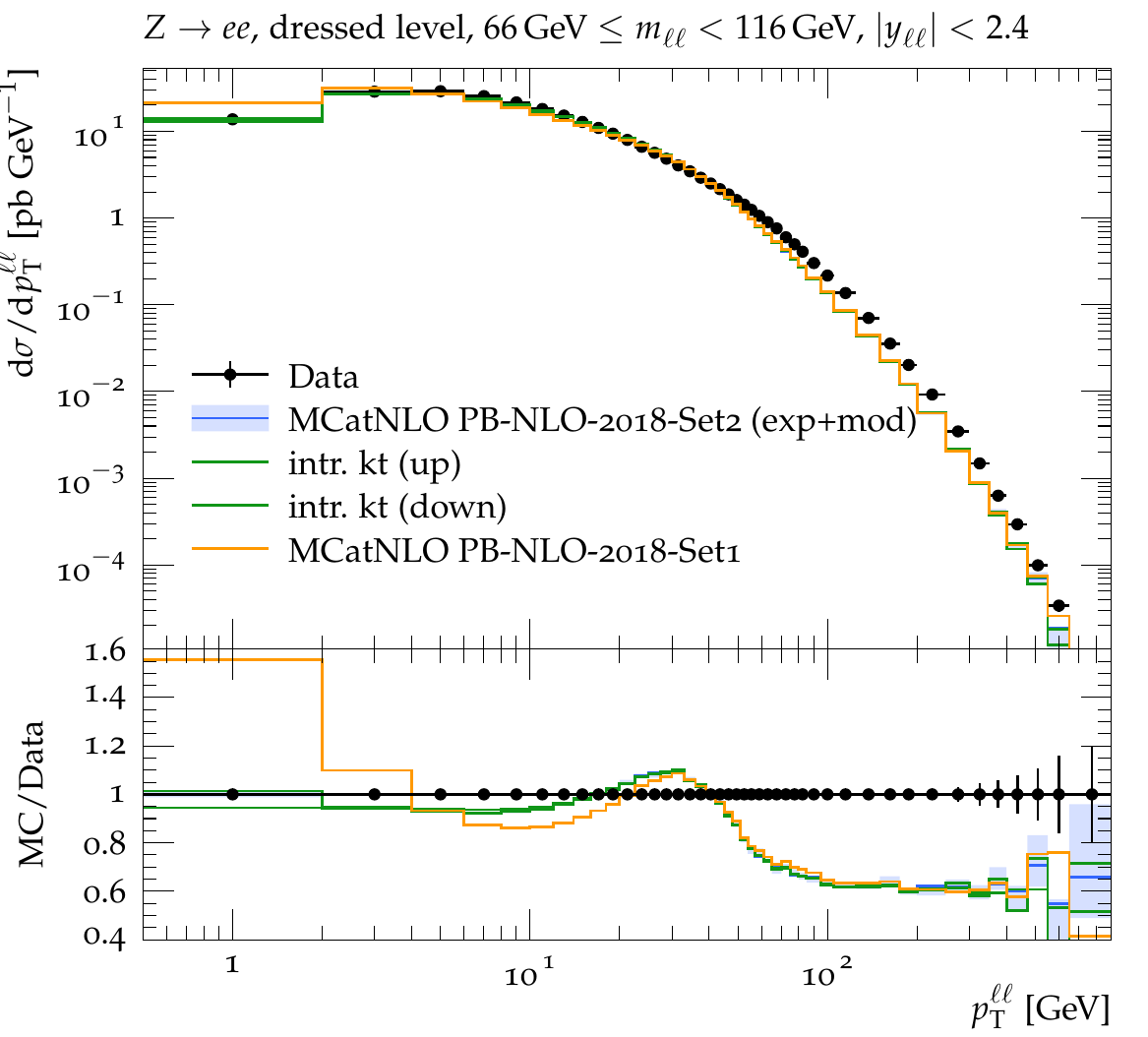} 
\includegraphics[width=0.405\textwidth]{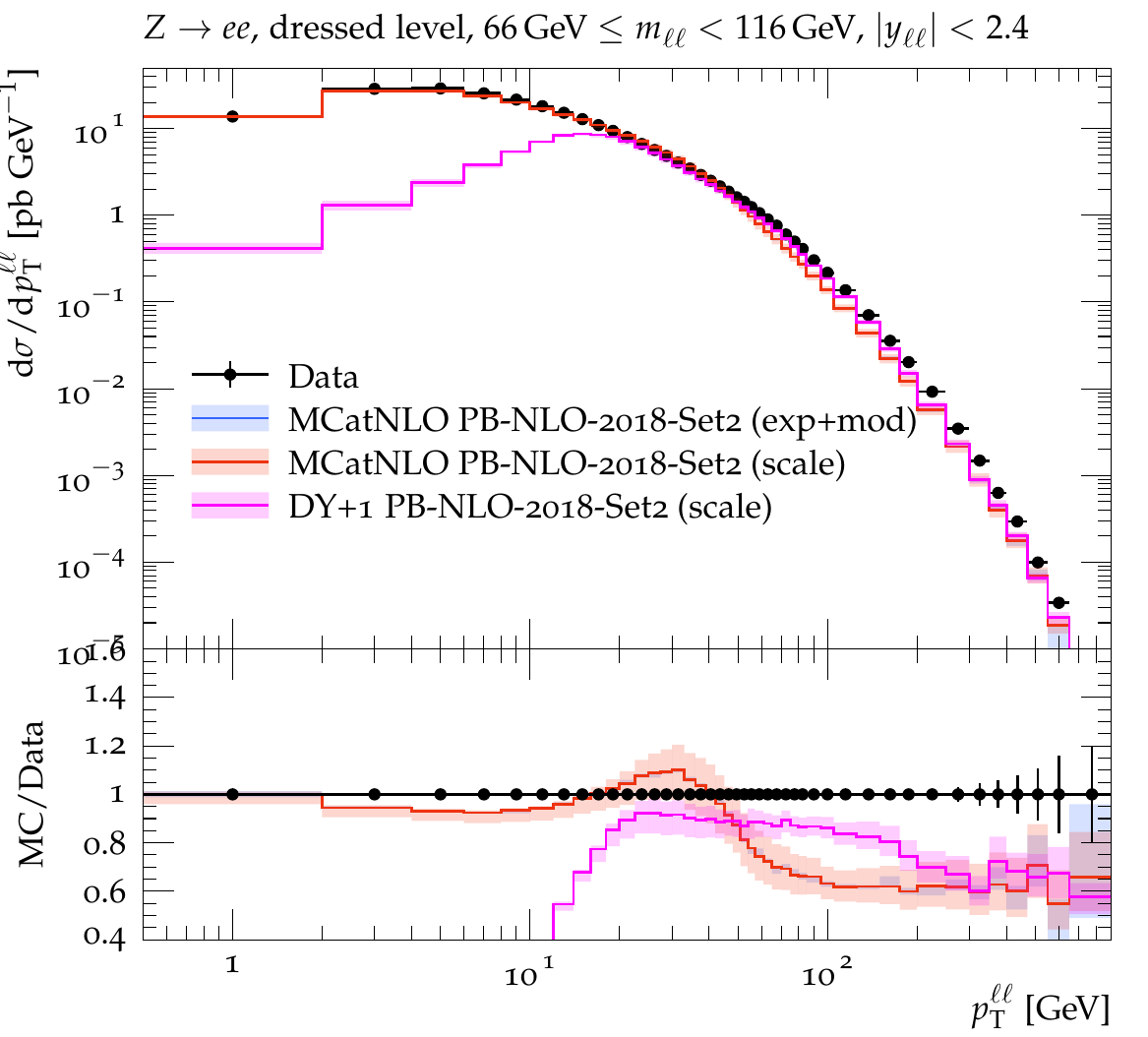} 
\caption{\small Transverse momentum $p_T$ spectrum of Z -bosons as measured by \protect\cite{Aad:2015auj} at $\sqrt{s}=8 $ TeV compared to the 
prediction~\protect\cite{Martinez:2019mwt}  using aMC@NLO and NLO PB -TMD. 
Left: uncertainties from the PB -TMD and 
 from changing the width of the intrinsic gaussian distribution by a factor of two. Right: with uncertainties from the TMDs  and scale variation combined. 
  }
\label{Zpt-TMD_uncertainty} 
\end{center}
\end{figure}

\vspace*{0.3cm}  
\noindent {\bf Acknowledgments}. 
I thank  the INT staff and  workshop organizers   for the invitation and hospitality.


 \newpage
%
\renewcommand*{\FigPath}{./WeekIV/07_Meziani/}

\wstoc{Origin of the Proton Mass? Heavy Quarkonium Production at Threshold from Jefferson Lab to an Electron Ion Collider}{Zein-Eddine Meziani, Sylvester Joosten}
\title{Origin of the Proton Mass? Heavy Quarkonium Production at Threshold from Jefferson Lab to an Electron Ion Collider}

\author{Zein-Eddine Meziani$^*$ and Sylvester Joosten$^\dagger$}
\index{author}{Meziani, Z.}
\index{author}{Joosten, S.}

\address{Argonne National Laboratory,\\
Argonne, IL 60439, USA\\
$^*$E-mail: zmeziani@anl.gov\\
$^\dagger$E-mail:sjoosten@anl.gov}

\begin{abstract}
In this article we discuss the science enabled by heavy quarkonium production in electro- and photo-production on a proton, its possible relation to the determination of the trace anomaly contribution to the proton mass, and the search of LHCb charm pentaquark. We describe the elastic threshold $J/\psi$ production experiments carried out or planned at Jefferson Lab. We also discuss a possible  elastic threshold $\Upsilon$ electroproduction measurement on a proton at a future electron ion collider (EIC).
\end{abstract}

\bodymatter

\section{Motivation}\label{aba:sec1_270}
\subsection{Origin of the proton mass}
In quest for a deeper understanding of the structure of the proton, there has been a renewed interest in the proton mass decomposition in terms of its constituents\cite{Ji:1994av_198,Ji:1995sv,Lorce:2017xzd_194,Lorce:2018egm_190}, similar to that of its spin\cite{Ji:1996ek_27_186}, using the energy-momentum tensor. While the proton mass decomposition is frame dependent and not unique, a judicious choice of frame might offer a unique insight into the structure of the proton with the added advantage that all the terms of the decomposition can be measurable. For the sake of simplicity, we will focus here on the``trace decomposition", that is of the matrix element of the trace of the energy-momentum tensor (EMT) at zero momentum transfer \cite{Shifman:1978zn_182,Kharzeev:1995ij} a direct consequence the anomalous breaking of scale invariance in a quantum field theory, Quantum Chromodynamics (QCD) in this case\cite{Peskin:1995ev}.
The trace decomposition of the EMT reads\cite{Kharzeev:1995ij}:
\begin{equation}
\Theta^{\mu}_{\mu} = \frac{\beta(g)}{2g}G^{\alpha \beta a} G^{a}_{\alpha \beta} + \sum_{l=u,d,s}m_l (1+\gamma_{m_l}) \bar q_lq_l + \sum_{h=c,b,t}m_h (1+\gamma_{m_h}) \bar q_hq_h 
\label{tr:01}
\end{equation}
where  $\beta(g) = -b g^3/16\pi^2 +...$ and $b=9-2/3n_h$
At small momentum transfer the heavy quarks decouple\cite{Shifman:1978zn_182}  
\begin{equation}
\sum_{h=c,b,t} \bar q_hq_h \rightarrow -\frac{2}{3}n_h\frac{g^2}{32\pi^2} G^{\alpha \beta a} G^{a}_{\alpha \beta} + ...
\label{tr:02}
\end{equation}
inserting (\ref{tr:02}) in  (\ref{tr:01}) by replacing $\beta (g)$ by $\tilde \beta (g)$ we obtain an expression of the trace with two terms only, the light quarks contribution to the trace of the EMT and the so-called "trace anomaly" piece which is purely gluonic in nature.
\begin{equation}
\Theta^{\mu}_{\mu} = \frac{\tilde\beta(g)}{2g}G^{\alpha \beta a} G^{a}_{\alpha \beta} + \sum_{l=u,d,s}m_l (1+\gamma_{m_l} )\bar q_lq_l + 
\label{tr:03}
\end{equation}
The matrix element of the trace of the EMT in the proton state is proportional to  the mass of the proton.
An important question  one might ask is whether each of the two contributions to the proton mass can be measured and evaluated using ab initio calculations. The light quarks mass term is known as the $\pi N$ $\sigma$ term and has been extensively studied~\cite{Alarcon:2011px,MartinCamalich:2011py}. The trace anomaly piece has received little attention,  since to this day no direct calculation using lattice QCD have been carried out  nor measurements,  sensitive directly to this matrix element,  performed. Elastic electro- or photo- production of $J/\psi$ and $\Upsilon$ at threshold measurements with high precision could pave a path to access the trace anomaly matrix element. Some work in understanding this non perturbative region of the strong interaction  has already begun~\cite{Mamo:2019mka,Hatta:2018ina_178,Hatta:2019lxo_174,Gryniuk:2016mpk,Kharzeev:1998bz_170}

\subsection{LHCb pentaquark search}

To confirm or refute the resonance nature of the  recent LHCb charm pentaquark discovery~\cite{Aaij:2015tga,Aaij:2019vzc}several authors~\cite{Wang:2015jsa,Karliner:2015voa,Kubarovsky:2015aaa,Blin:2016dlf} suggested to produce such resonances  in the $s$-channel in direct photo-production. Such experiments have been recently carried out at Jefferson Lab in three different halls, hall D, B and C. The GlueX collaboration at Jefferson Lab measured photo-production of $J/\psi$ near threshold using the GlueX detector in Hall D and published its results~\cite{Ali:2019lzf_166}. Experiment E12-16-007 ~\cite{Meziani:2016lhg} in Hall C was designed to perform a direct search of the higher mass narrow width $P_c^+(4450)$. It ran during the Spring of 2019 and the analysis will soon be completed. CLAS12 in Hall B has taken also the relevant data on a hydrogen target and the analysis is underway.

\section{A threshold $J/\psi$ production  with SoLID at Jefferson Lab}
A more dedicated program of high precision $J/\psi$ electro- and photo-production  measurements on the proton  using the Solenoidal Large Intensity Device (SoLID) is  planned in Hall A. In Jefferson Lab experiment E12-12-006~\cite{SoLIDjpsi:proposal} we will be able to measure differential cross sections in the four momentum transfer $t$  as well as determine the total cross section  very close to the threshold region on a nucleon. This experiment will be able to address some of the interesting questions where the gluonic contribution of the strong interaction is manifest. Whether we are interested in exploring the trace anomaly contribution to the nucleon mass, the existence of charm pentaquarks or the binding of the $J/\psi$-nucleon system through  Van der Waals color forces, the threshold region  is  the right fertile region for these investigations. In Fig.~\ref{fig1.2}(a) we show projections of what can be achieved in experiment E12-12-006 in both elastic electro- and photo-production for the total cross section as a function of the virtual/real photon energy.

\section{A Threshold $\Upsilon$ production at a future Electron Ion Collider}
Given the lower center of mass energy of the threshold region for the $J/\psi$ production on a nucleon it is best to consider a higher center of mass energy provided by the production of $\Upsilon$, hence reachable by the current EIC designs and with sufficient luminosity for a precision measurement. In this case, a more potent way to address the origin of the proton mass question or the existence of bottom pentaquarks is through elastic production of $\Upsilon$ in the threshold region again. Here the mass of the bottom quark as well as  the probe resolution in electroproduction gives two independent knobs in the investigation of the gluonic interaction between the $\Upsilon$ and the nucleon.  
In Fig.~\ref{fig1.2}(b) we illustrate what can be achieved in 100 days at a luminosity of 100fb$^{-1}$ at an EIC in quasi-real $\Upsilon$ production kinematics.

\def\figsubcap#1{\par\noindent\centering\footnotesize(#1)}
\begin{figure}[h]%
\begin{center}
  \parbox{2.4in}{\includegraphics[width=2.3in]{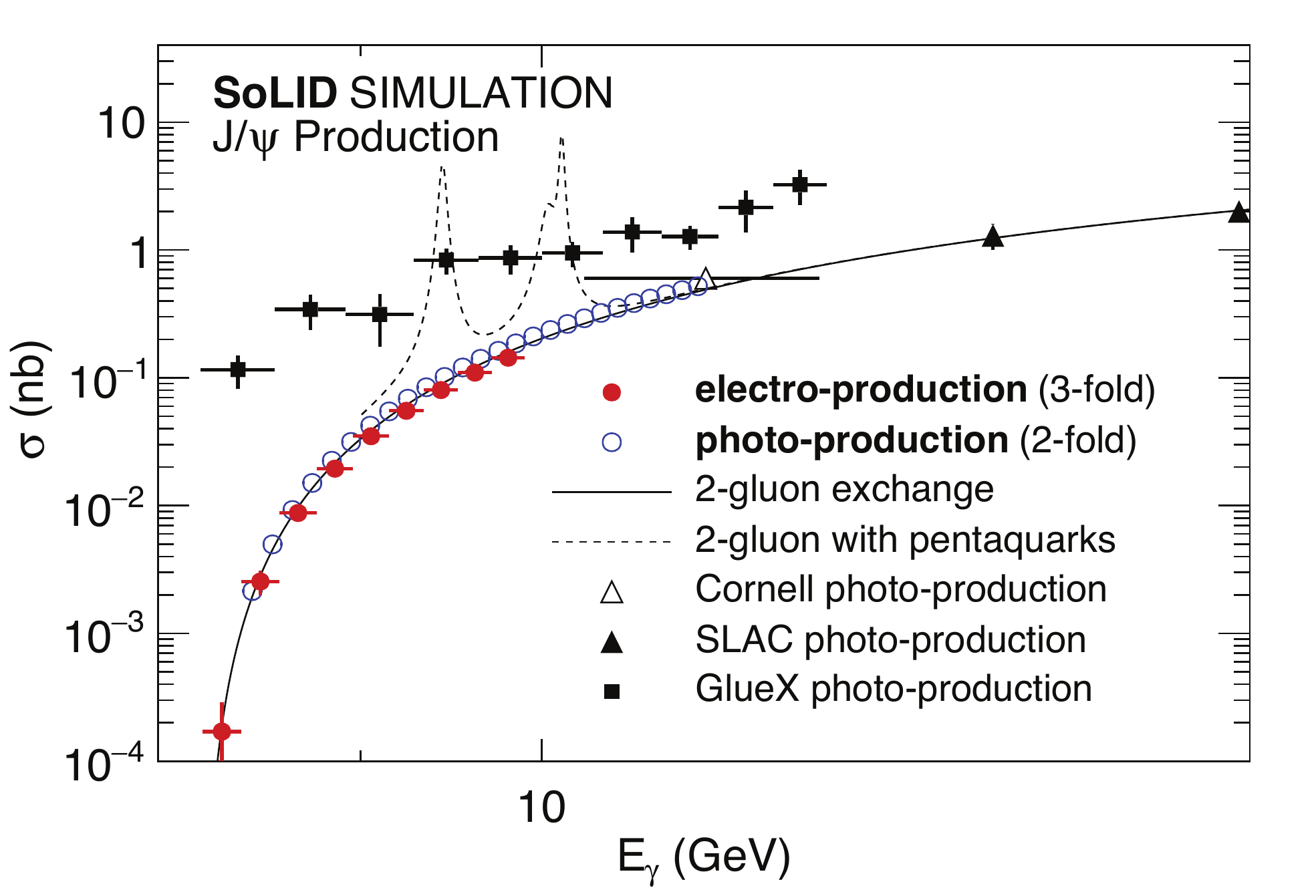}\figsubcap{a}}
  \hspace*{4pt}
  \parbox{2.4in}{\includegraphics[width=2.3in]{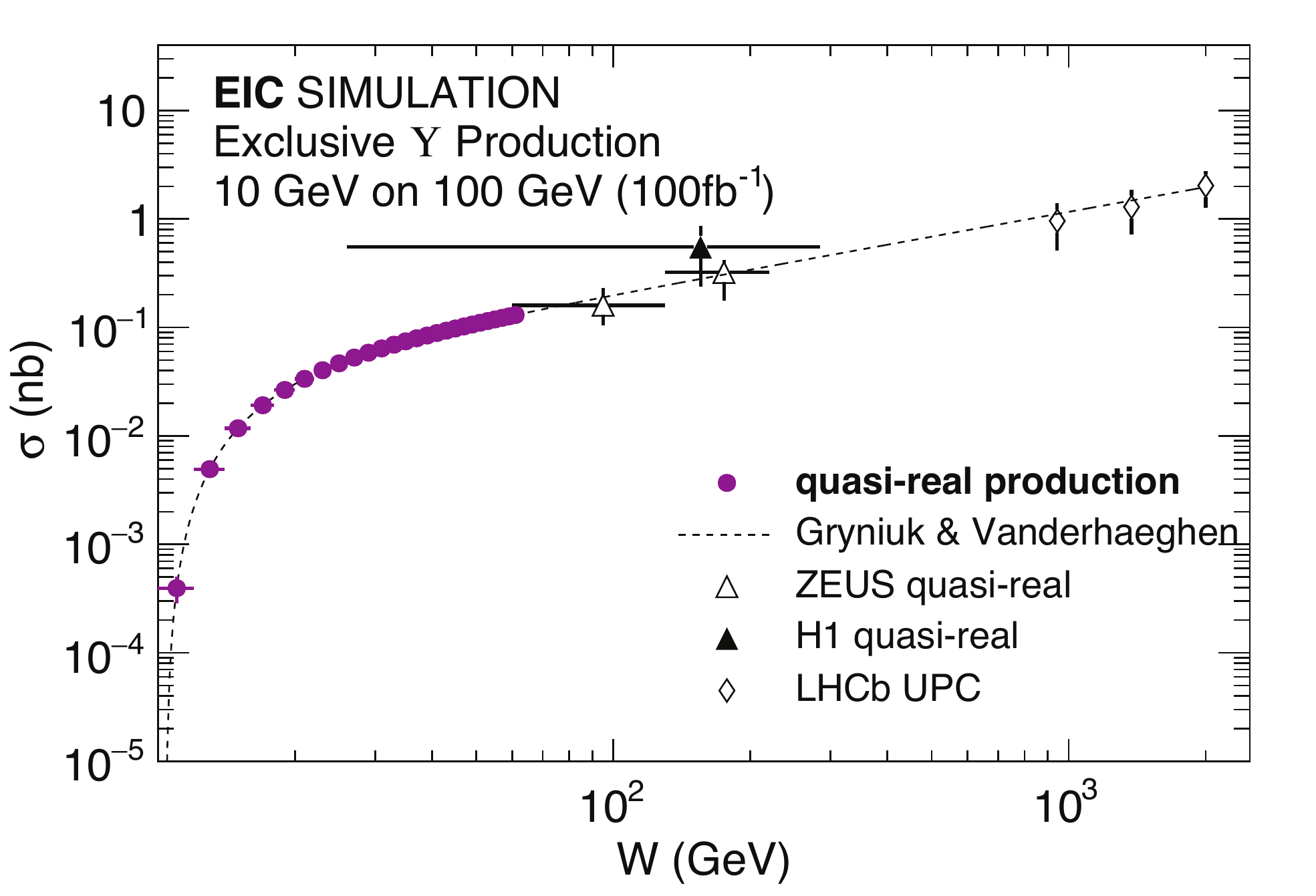}\figsubcap{b}}
  \caption{ (a) Projection of the total cross section using a 2-gluon exchange model~\cite{Brodsky:2000zc} along with the world data. Note that model  significantly underestimate the GlueX data which would lead to larger rates, thus increased sensitivity close to threshold.  (b) 
 Projection of the total quasi real production of $\Upsilon$ using model~\cite{Gryniuk:2016mpk} along with the sparse world data.}%
  \label{fig1.2}
\end{center}
\end{figure}
 \section*{Acknowledgments}
The authors thank the organizers for the opportunity to present this work. This work is supported by  the Department of Energy contracts DC-AC02-06CH11357 and DE-FG02-94ER40844


\



 \newpage
\wstoc{Probing Nuclear Structure with Future Colliders}{
Timothy~J. Hobbs, 
Pavel~M. Nadolsky, 
Fredrick~I. Olness, 
Bo-Ting Wang 
}
\title{Probing Nuclear Structure with Future Colliders}

\author{
Timothy~J. Hobbs\rlap,${}^{a,b}$ 
Pavel~M. Nadolsky\rlap,${}^{a}$ 
Fredrick I. Olness\rlap,${}^{a}$\footnote{Presenter.}
Bo-Ting Wang${}^{a}$ 
}
\index{author}{Hobbs, T. 
}
\index{author}{Nadolsky, P.
}
\index{author}{Olness, F.
}
\index{author}{Wang, B.
}

\address{${}^{a}$Department of Physics, Southern Methodist University,\\
 Dallas, TX 75275-0175, U.S.A. }

\address{${}^{b}$Jefferson Lab, EIC Center, Newport News, VA 23606, U.S.A.}


\begin{abstract}

Improved knowledge of the nucleon structure is a crucial pathway toward a deeper understanding 
of the fundamental nature of the QCD interaction, and will enable important future discoveries. 
The  experimental facilities proposed for the next decade offer a tremendous opportunity
to advance the precision of our theoretical predictions to unprecedented levels.  
In this report we briefly highlight some of the recently developed tools and techniques 
which, together with data from these new colliders, have the potential to revolutionize our understanding of the QCD theory in the next decade.
\end{abstract}
%

\bodymatter


\section{Introduction}
\label{sec:intro}
%
%
%
The recent discovery of the Higgs boson at the Large Hadron Collider (LHC) was a remarkable endeavour which culminated in the  2013 Nobel Prize.
As we look ahead to the next decade,   a number of new facilities on 
the horizon 
will provide new insights about fundamental microscopic phenomena.
This includes the high-luminosity upgrade of the LHC, (HL-LHC), 
a proposed electron ring for the LHC (LHeC),\footnote{A Large Hadron electron Collider at CERN
({http://lhec.web.cern.ch/}).
} and a proposed Electron-Ion Collider (EIC).\footnote{Electron-Ion Collider (EIC) User Group
({http://www.eicug.org/}).}

For all these facilities, the route to new discoveries will be  high-precision comparisons between theory and data that can validate the features of the Standard Model and search for discrepancies which may signal undiscovered phenomena. 

These comparisons between data and theory rely crucially on the Parton Distribution Functions (PDFs) which connect the theoretical quarks and gluons with experimental observations.  Unfortunately, the PDFs are often the limiting factor in these comparisons.
Thus, our ability to fully characterize the Higgs boson and constrain physics Beyond the Standard Model (BSM) ultimately comes down to how accurately we determine the underlying PDFs. If you cannot distinguish signal from background, you cannot make discoveries. 

In this brief report, we will examine how new data and new tools might facilitate these discoveries.\footnote{The articles cited here are limited due to space; please also see references therein.}
\begin{figure*}[t]
\centering
\includegraphics[clip,width=0.95\textwidth]{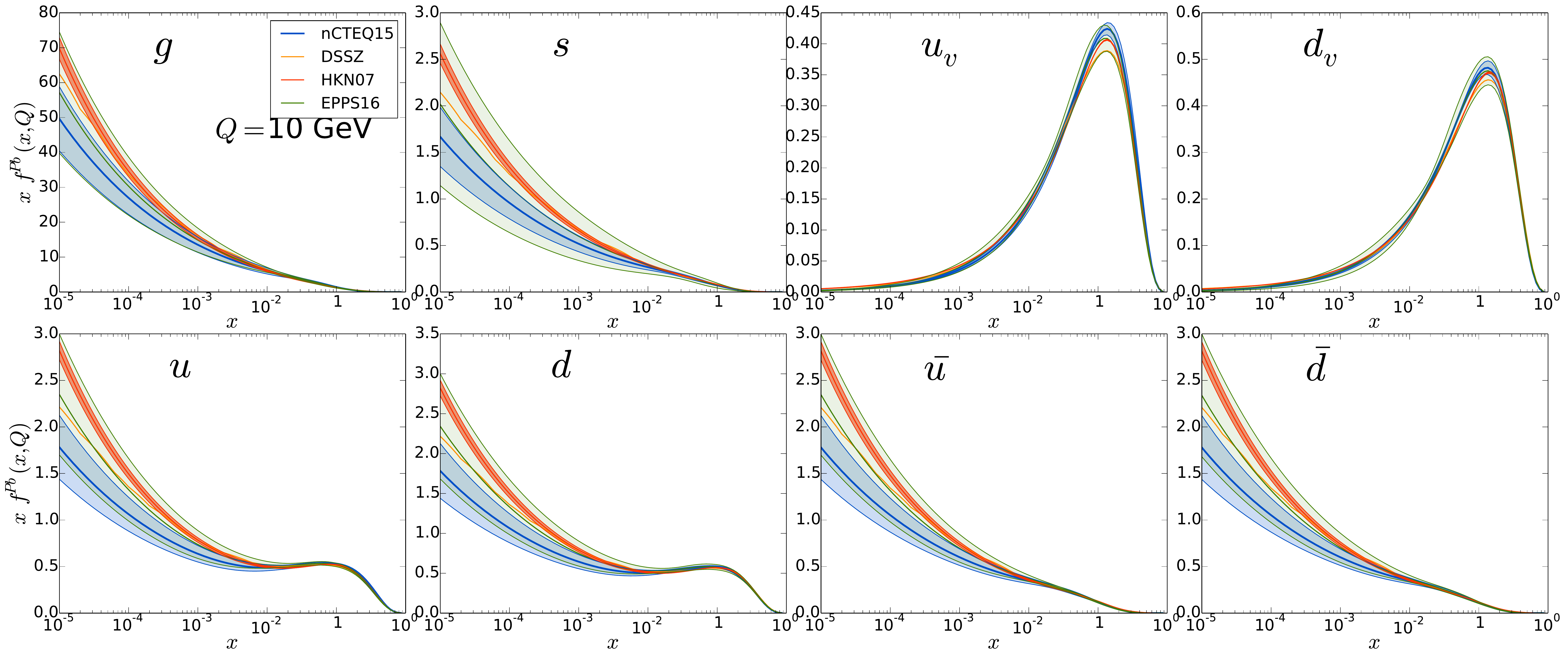}
	\caption{
	Comparison of selected nuclear PDF flavors and uncertainties including 
	HKN07, DSSZ,
	nCTEQ15.\cite{Kovarik:2015cma_201}
	and
	EPPS16\cite{Eskola:2016oht_196},
	}
\label{fig:npdf}
\end{figure*}

\section{Nuclear PDFs}
Although the fits to the proton PDFs have been quite successful, 
it is crucial to extend the framework to consider the nuclear degrees of freedom. For example, the neutrino-induced Deeply Inelastic Scattering (DIS) process provides essential information for PDF flavor differentiation. 
In Fig.~\ref{fig:npdf} we display a comparison of recent nuclear PDFs with uncertainties. While there has been impressive progress in recent years 
constraining the nPDFs, there is certainly room for improvement, especially compared to the proton PDF efforts; the new experimental facilities 
(including LBNF)
will provide important new PDF constraints which in turn will constrain 
potential BSM signals. 

\section{PDF Sensitivity \& PDFSense\cite{Wang:2018heo,Hobbs:2019gob}}
In order to quickly and efficiently determine the potential impact of 
new data sets on the PDFs and other observables,  we introduce a generalization of the PDF-mediated correlations called the sensitivity $S_f$.
This is a combination of the correlation coefficient and the residuals (data minus theory scaled by the uncertainty) of the PDF uncertainty,
and it  identifies those experimental data points that tightly 
constrain the PDFs. 
We find the sensitivity is useful for identifying regions of the $\{x,Q^2\}$ kinematic plane in which the
PDFs are particularly constrained by physical observables.
The details of the sensitivity can be found in Ref.~\cite{Wang:2018heo} and the implementation is available in the public package
PDFSense ({https://metapdf.hepforge.org/PDFSense/}).

\begin{figure*}[th]
\includegraphics[clip,width=0.66\textwidth]{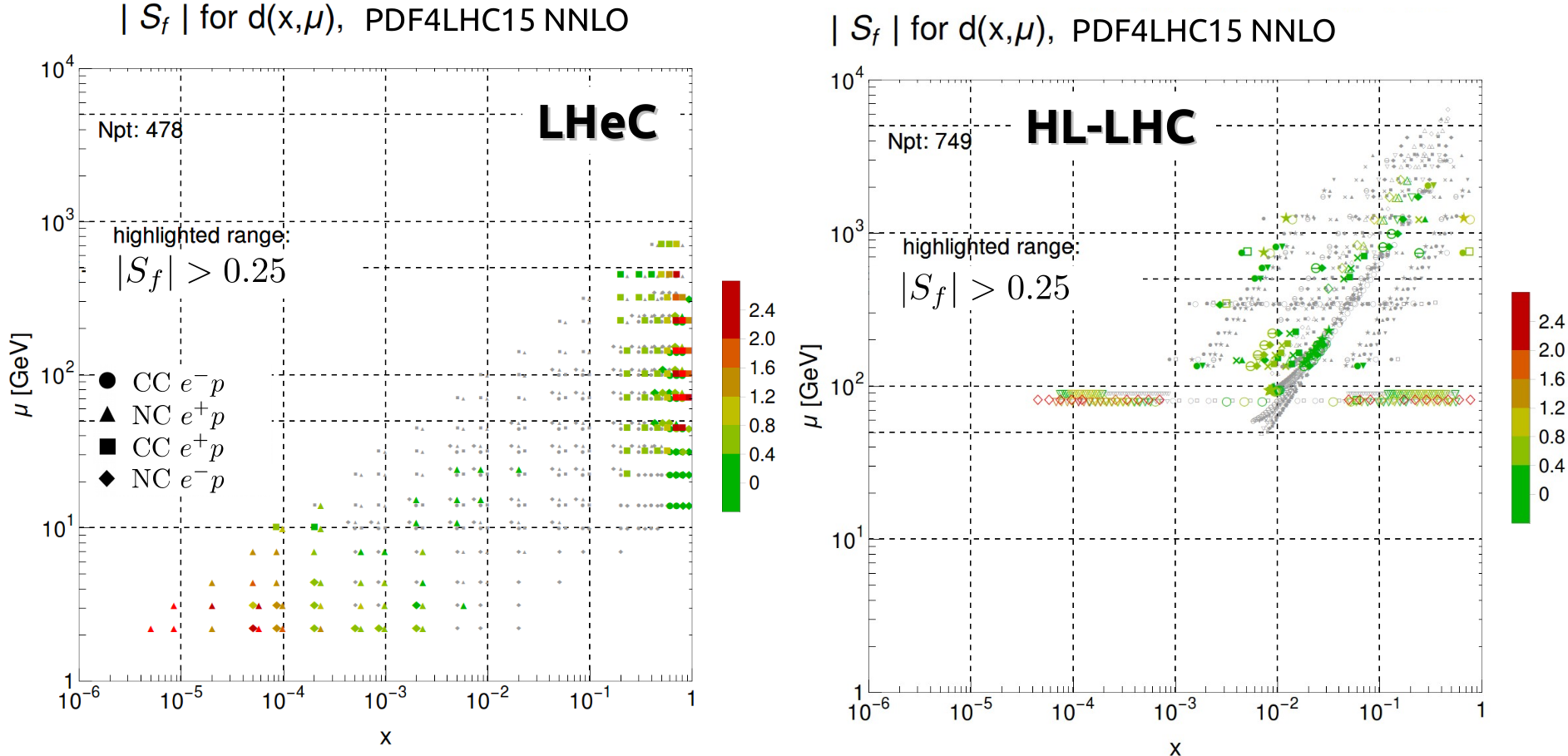}
\hfil 
\includegraphics[clip,width=0.33\textwidth]{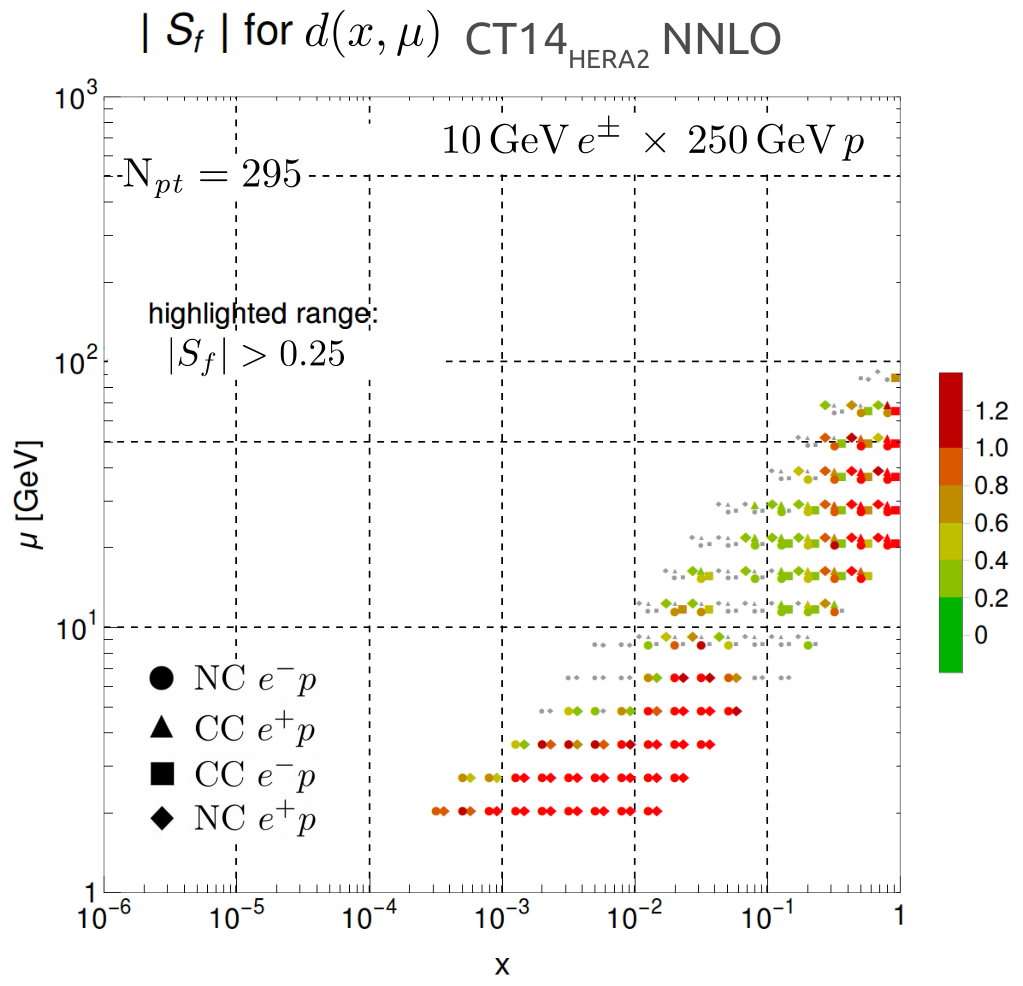}
	\caption{
	Future HEP experiments such as the  LHeC, HL-LHC, and EIC  can have substantial PDF
	sensitivity $S_f$ as shown here for the   $d(x,\mu)$ distribution computed according to the
	conventions in Refs.~\citenum{Wang:2018heo},~\citenum{Hobbs:2019sut}. 
	}
\label{fig:sens}
\end{figure*}

In Fig.~\ref{fig:sens} we display the sensitivities $S_f$ for
the LHeC, HL-LHC, and EIC using pseudodata sets for a sample PDF flavor, $d(x,Q)$, {\it c.f.,} Ref.~\cite{Hobbs:2019sut}.
We observe the LHeC shows strong sensitivity in both the very high- and low-$x$ regions, while the HL-LHC covers the intermediate-$x$ region out to very large $Q^2$, and the  EIC complements these in the high-to-medium-$x$ region at 
lower $Q^2$ values. 
The combination of these measurements can provide very strong constraints on
the various PDF flavors across the broad  $\{x,Q^2\}$ kinematic plane.


In Fig.~\ref{fig:higgs}, we plot the sensitivity of EIC pseudodata to the
high-energy LHC  Higgs production cross section, $\sigma_H$.
The constraints that a medium-energy machine like an EIC would place
on Higgs phenomenology stem from the predominance of the $gg \to  H$
fusion channel at the LHC, and the 
sensitivity of the DIS data to the gluon via QCD evolution, which  
then connects the high-$x$ and low-$Q$ gluon PDF probed by the EIC to the lower-$x$ and high-$Q$ PDF of the LHC.
The sensitivity of the EIC pseudodata generally surpasses that of
the fixed-target experiments that currently dominate the constraints
on high-$x$ PDFs.
Therefore, the EIC will  strongly constrain the PDF dependence of HEP
observables at moderate to  large-$x$, including several in the Higgs
and electroweak sectors, like $M_W$ and $\sin (2 \theta_W)$.


\section{PDFs from Lattice QCD}
Additional information on PDFs may also come from Lattice QCD
calculations\cite{Lin:2017snn_191} which generally compute the Mellin moments of PDFs, as illustrated in
Fig.~\ref{fig:moment}, where we display the sensitivity, $S_f$, of the first moment for the
strange quark.\cite{Hobbs:2019gob}
Despite
additional data from HERA and LHC, 
it is interesting to note that some of the strongest
sensitivities come from the fixed-target neutrino DIS experiments
(CCFR, NuTeV) in the high-$x$ and low-$Q$ region.  
At present, the strange PDF still has relatively
large PDF uncertainties, and this can limit the precision of ``standard candle" benchmark
process such as $W/Z$ production; hence, additional constraints from
Lattice QCD would be welcome.

\section{Borrowing from Machine Learning}
Finally, in  Fig.~\ref{fig:tsne} we display a  distribution of 
residuals using a
t-distributed Stochastic Neighbor Embedding (t-SNE) which 
is a machine-learning algorithm for visualization of high-dimensional data.
In this case, we have taken the 56-dimensions of the PDF uncertainty eigenvectors and reduced this to a 3D projection of the  4000+ data from the CT14 fit. The algorithm identifies points with similar characteristics and 
groups these together. We have color coded the points for the DIS, 
Drell-Yan, and Jets+$t\bar{t}$ sets.
In this case, the fact that the algorithm has grouped similar experiments together suggests that such machine-learning techniques can help us identify subtle relations that may not be apparent via other methods.\cite{Wang:2018heo}

\begin{figure}[t]
\begin{minipage}{0.32\textwidth}
\centering
\includegraphics[clip,width=0.99\textwidth]{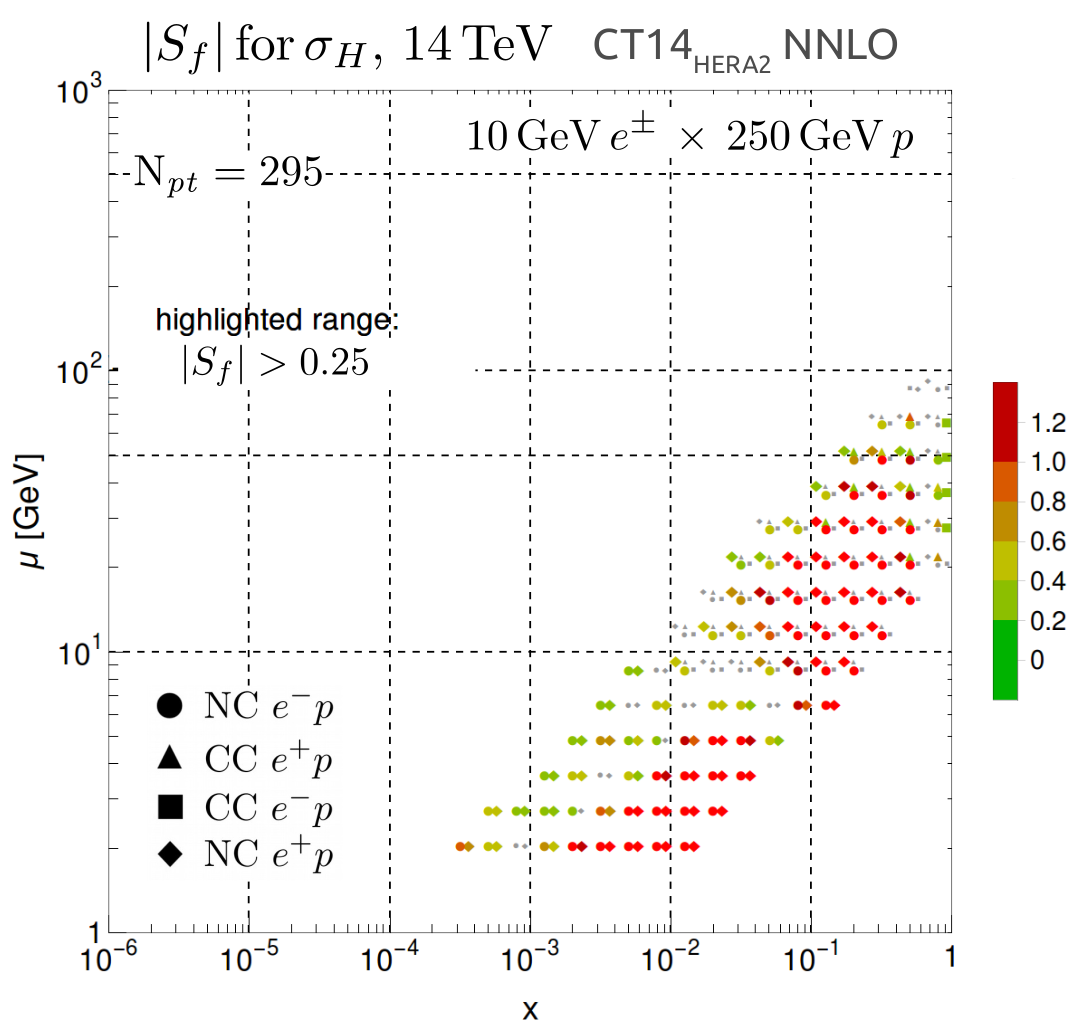}
	\caption{The EIC  pseudodata
sensitivity  for Higgs
production\cite{Hobbs:2019sut} with  $\int L = 100~fb^{-1}$.
}
\label{fig:higgs}
\end{minipage}
\hfil
\begin{minipage}{0.32\textwidth} 
\centering
\includegraphics[clip,width=0.99\textwidth]{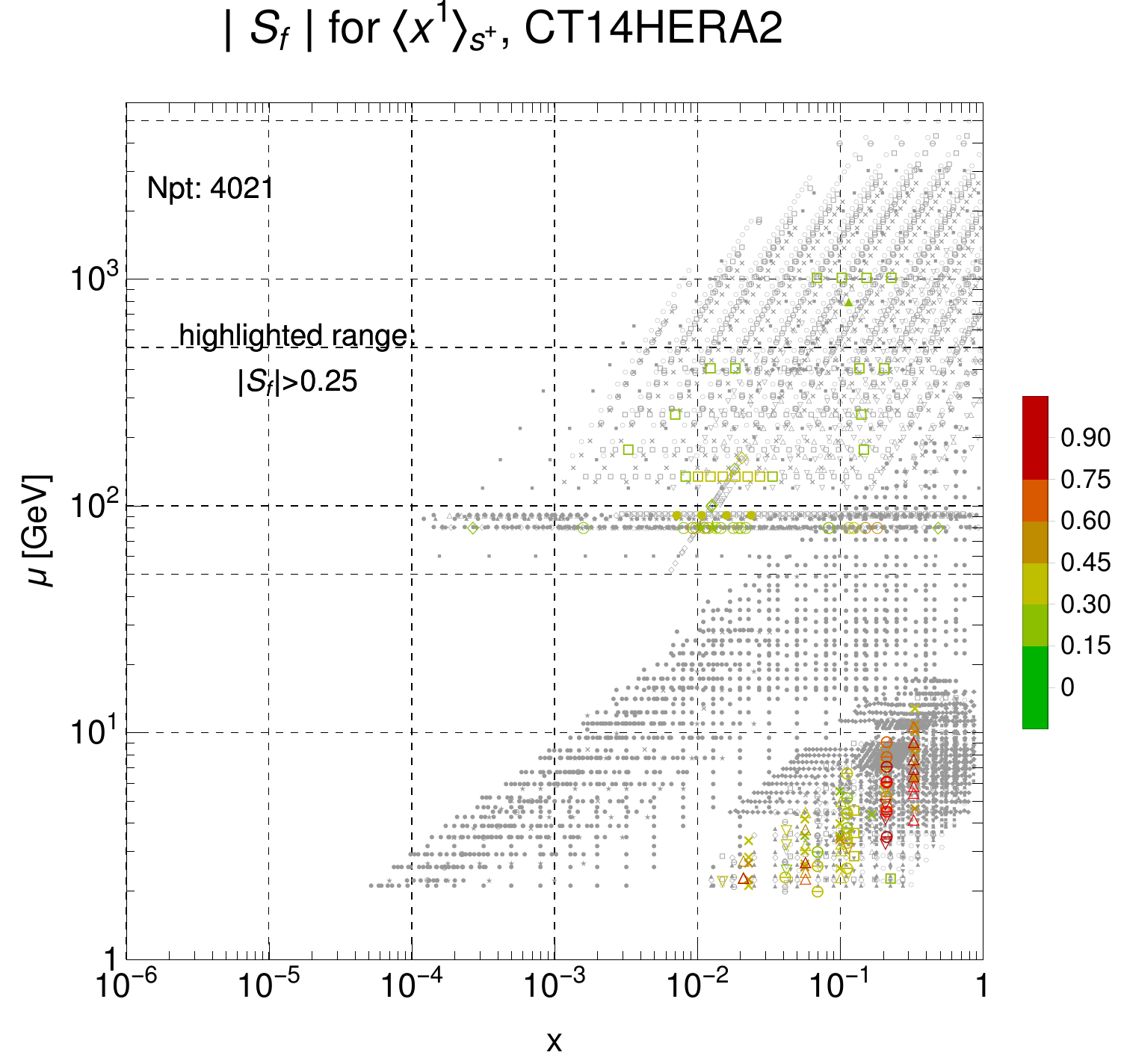}
	\caption{Sensitivity to the 1st-order Mellin moments of the $s^+$ distribution;\cite{Hobbs:2019gob} $\mu_F=2$~GeV.
	}
\label{fig:moment}
\end{minipage}
\hfil
\begin{minipage}{0.32\textwidth}
\centering
\includegraphics[clip,width=0.99\textwidth]{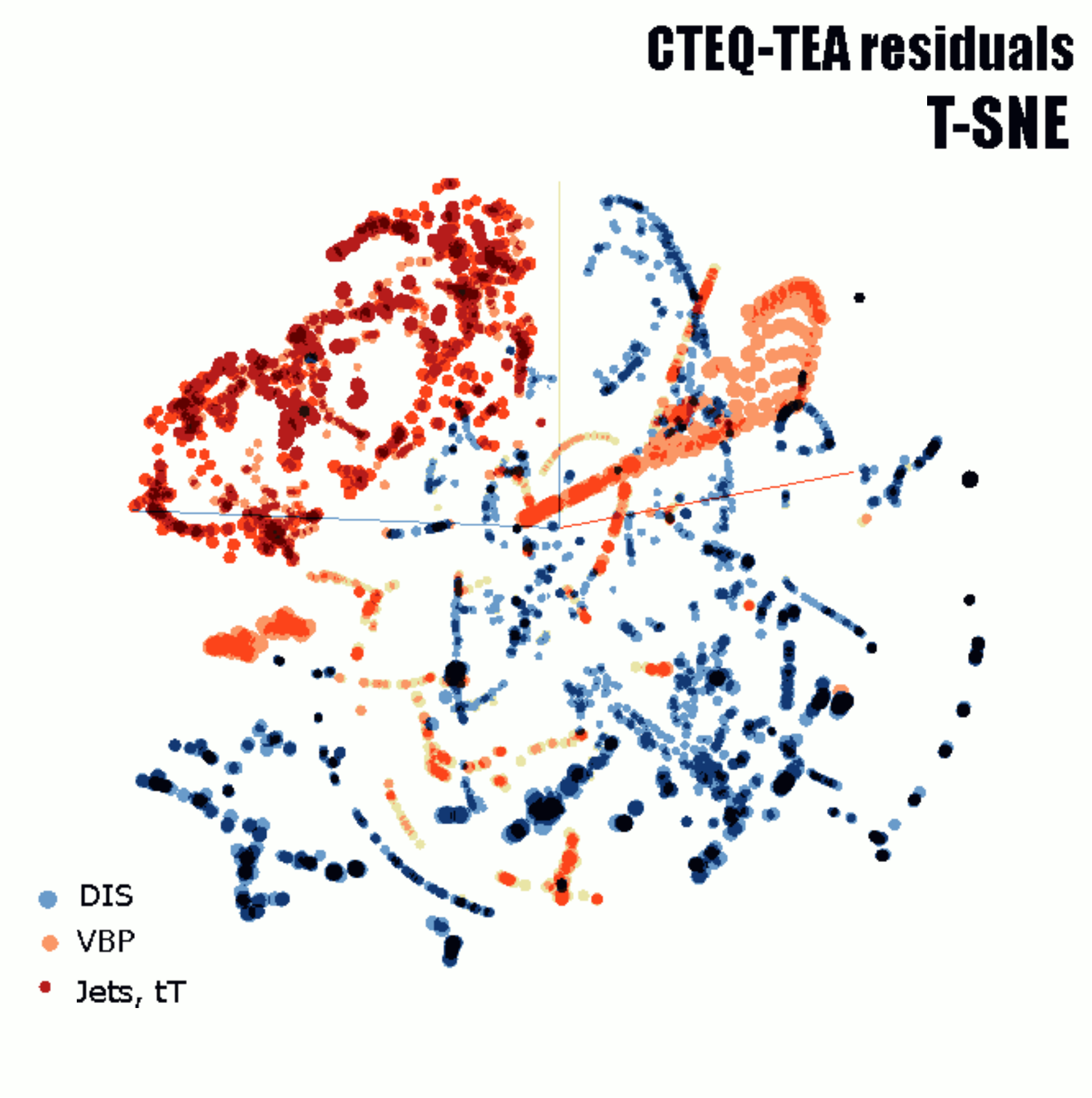}
	\caption{
 Distribution from the t-SNE
clustering method.\cite{Wang:2018heo}  
\vspace{20pt}
	}
\label{fig:tsne}
\end{minipage}
\label{fig:HIGGS}
\end{figure}

\section{Conclusions}
We have highlighted a few of the tools and techniques that will allow us to most effectively 
capitalize on the wealth of new data from these proposed colliders to advance our knowledge
of the structure of the nucleon, the associated nuclear PDFs, and the underlying QCD theory. 
Additionally, these tools demonstrate how 
each of these experimental facilities occupies a unique place in the kinematical parameter space. 
As such, these complementary data sets will provide us with unprecedented understanding of  strong interaction physics, 
and provide the key for many new discoveries.



\subsection*{Acknowledgments}
We are grateful to our colleagues of the CTEQ-TEA, nCTEQ, and xFitter collaborations, and also the INT for sponsorship of this workshop.
This work was supported by the U.S. Department of Energy under Grant No. DE-SC0010129.
The research of TJH is supported by an EIC Center@JLab Fellowship.



\newpage

\renewcommand*{\FigPath}{./WeekIV/09_Vitev/}

\def\eq#1{{Eq.~(\ref{#1})}}
\def\fig#1{{Fig.~\ref{#1}}}

\renewcommand{\tr}{\mbox{tr}}
\renewcommand{\thalf}{\tfrac{1}{2}}
\renewcommand{\llangle}{\Big\langle \!\! \Big\langle}
\renewcommand{\rrangle}{\Big\rangle \!\! \Big\rangle}

\renewcommand{\stackeven}[2]{{{}_{\displaystyle{#1}}\atop\displaystyle{#2}}}
\renewcommand{\lsim}{\stackeven{<}{\sim}}
\renewcommand{\gsim}{\stackeven{>}{\sim}}

\renewcommand{\sh}[1]{#1\hskip-7pt \diagup}
\newcommand{\Sh}[1]{#1\hskip-11pt \diagup}
 
\newcommand{\revtex}{REV\TeX\ }
\newcommand{\classoption}[1]{\texttt{#1}}

\renewcommand{\bra}[1]{\left\langle #1 \right|}
\renewcommand{\ket}[1]{\left| #1 \right\rangle}
\newcommand{\tvec}[1]{\vec{#1}_\bot}
\renewcommand{\ul}[1]{\underline{#1}}
\newcommand{\psibar}{\overline{\psi}}
\newcommand{\ubar}[1]{\overline{U}_{#1}}
\newcommand{\vbar}[1]{\overline{V}_{#1}}
\newcommand{\Tr}{\mathrm{Tr}}
\newcommand{\braket}[2]{\left\langle #1 \right| \left. #2 \right\rangle}
\newcommand{\footalert}[1]{\footnote{\textcolor{red}{ #1 }}}
\newcommand{\ord}[1]{\mathcal{O}\left( #1 \right)}
\newcommand{\gPM}{g^{+-}}
\newcommand{\gpm}{g_{+-}}
\renewcommand{\half}{\frac{1}{2}}

\wstoc{Radiative processes and jet modification at the EIC}{Ivan Vitev}

\title{Radiative processes and jet modification at the EIC}

\author{Ivan Vitev}
\index{author}{Vitev, I}

\address{Los Alamos National Laboratory, Theoretical Division, Mail Stop B283\\
Los Alamos, NM 87545, USA\\
$^*$E-mail: ivitev@lanl.gov\\
}

\begin{abstract}
A U.S.-based Electron-Ion Collider will provide the ultimate capability to determine both the structure and properties of nucleons and nuclei, as well as 
how matter and energy can be transported through a strongly interacting quantum mechanical environment.  The production 
and propagation of long-lived heavy subatomic particles is a unique and critical part of this planned decade-long research program.
In these proceedings we report the derivation of all branching processes  in nuclei that  lead to a modification of semi-inclusive hadron production, jet cross sections, and  jet substructure  when compared to the vacuum. This work allows for their evaluation to any desired order in opacity. As an example, we show an application to the modification of light hadron and open heavy flavor fragmentation functions at the EIC. We discuss how this observable can shed light on the physics of hadronization and parton energy loss in large nuclei.
\end{abstract}


\bodymatter

\section{Introduction}
The EIC is the top priority for new construction for DOE Office of Science, Office of Nuclear Physics and is expected to come online around 2028.
There is a short window of opportunity to sharpen its physics program and ensure that the experimental capabilities to carry out the required measurements  are available. To understand how partons propagate in cold nuclear matter and how hadrons form in the QCD environment is  an important pillar of the EIC science portfolio. Existing measurements from the HERMES Collaboration~\cite{Airapetian:2007vu}  have not been able to differentiate between competing models of light hadron attenuation~\cite{Kopeliovich:2003py,Arleo:2003jz} in semi-inclusive deep inelastic scattering (SIDIS) on nuclear targets. As we will show below, $D$-mesons and $B$-mesons production in such reactions is a sensitive probes of the nuclear matter transport properties and the suppression patterns these heavy particles  show are distinctly different for different theoretical models. This, in turn,
requires better understanding of medium-induced branching processes for light partons and heavy quarks~\cite{Kang:2016ofv} and the process of
hadronization.  
 
 \section{Radiative processes in DIS on large nuclei}
Recently, we developed a formalism~\cite{Sievert:2018imd} that allows us to compute the gluon spectrum of a quark jet  to an arbitrary order in opacity, the average number of scatterings in the nuclear medium.  This calculation goes beyond the simplifying limit in which the gluon radiation is soft and  significantly extends previous work, which computes the full gluon spectrum only to first order in opacity, see Fig.~\ref{f:Jet_Kinematics}.  As all branching processes in matter have the same topology, the calculation can be generalized to obtain the full set of in-medium splittings~\cite{Sievert:2019cwq}.
The parton flavor and mass dependence are elegantly captured by the lightcone wavefunction formalism.

We can express the  medium-induced radiative processes as a sum of Initial/Initial,  Initial/Final,  Final/Initial and Final/Final
contributions, where the nomenclature refers to an interaction with a scattering center in matter before or after the splitting in the
amplitude and the conjugate amplitude, respectively. By evaluating Feynman diagrams corresponding to an interaction between the propagating parton system and the medium we derive both the initial conditions and kernels that relate $N$ and $N-1$  orders in opacity. 
The recursion relations between these four functions can be cast in the form of a matrix equation, with a particularly simple 
triangular form due to their causal structure:
\begin{align} \label{e:reactmtx}
\begin{bmatrix*}[l]
f_{F / F}^{(N)}  \\
f_{I / F}^{(N)}  \\
f_{F / I}^{(N)}  \\
f_{I / I}^{(N)} 
\end{bmatrix*}
= \int\limits_{x_0^+}^{\min[ x^+ , y^+ , R^+ ]} \frac{dz^+}{\lambda^+} 
\int\frac{d^2 q}{\sigma_{el}} \frac{d\sigma^{el}}{d^2 q}
\begingroup
\renewcommand*{\arraystretch}{1.4}
\begin{bmatrix}
\mathcal{K}_1^{} & \mathcal{K}_2 & \mathcal{K}_3 & \mathcal{K}_4 \\
0 & \mathcal{K}_5 & 0& \mathcal{K}_6 \\
0 & 0 & \mathcal{K}_7 & \mathcal{K}_8 \\
0 & 0 & 0 & \mathcal{K}_9
\end{bmatrix}
\endgroup
\begin{bmatrix*}[l]
f_{F / F}^{(N-1)}  \\
f_{I / F}^{(N-1)}  \\
f_{F / I}^{(N-1)}  \\
f_{I / I}^{(N-1)} 
\end{bmatrix*} ,
\end{align}
where the matrix of integral kernels $\mathcal{K}_{1-9}$ is an explicit representation of the reaction operator, $\lambda$ is the 
parton scattering length, and  ${d\sigma^{el}}/{d^2 q}$ describes the distribution of transverse momentum transfers.

We have written a Mathematica code to solve the recursion relation Eq.~(\ref{e:reactmtx}) and obtained explicit results to second 
order in opacity~\cite{Sievert:2018imd}. The fully differential spectrum of in-medium branchings has been evaluated 
numerically~\cite{Sievert:2019cwq}.

\begin{figure}[t]
\begin{center}
\includegraphics[width= 3.5in]{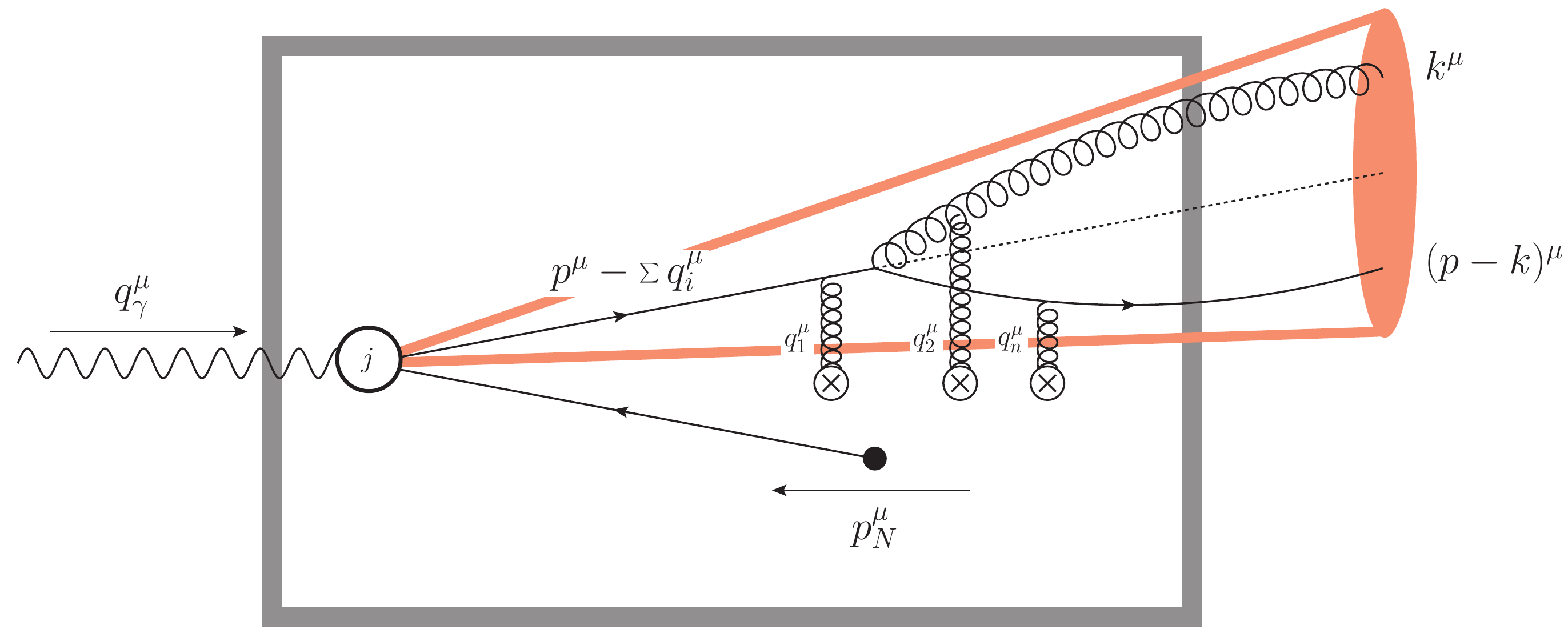} 
\caption{Illustration of the indicated jet kinematics for SIDIS in the Breit frame.  The dark box represents the medium (nucleus) and the red cone represents the jet.}
\label{f:Jet_Kinematics}
\end{center}
\end{figure}

\section{Modification of light and heavy flavor mesons in SIDIS on nuclei}
\begin{figure}[h]
\begin{center}
\includegraphics[width=3.in]{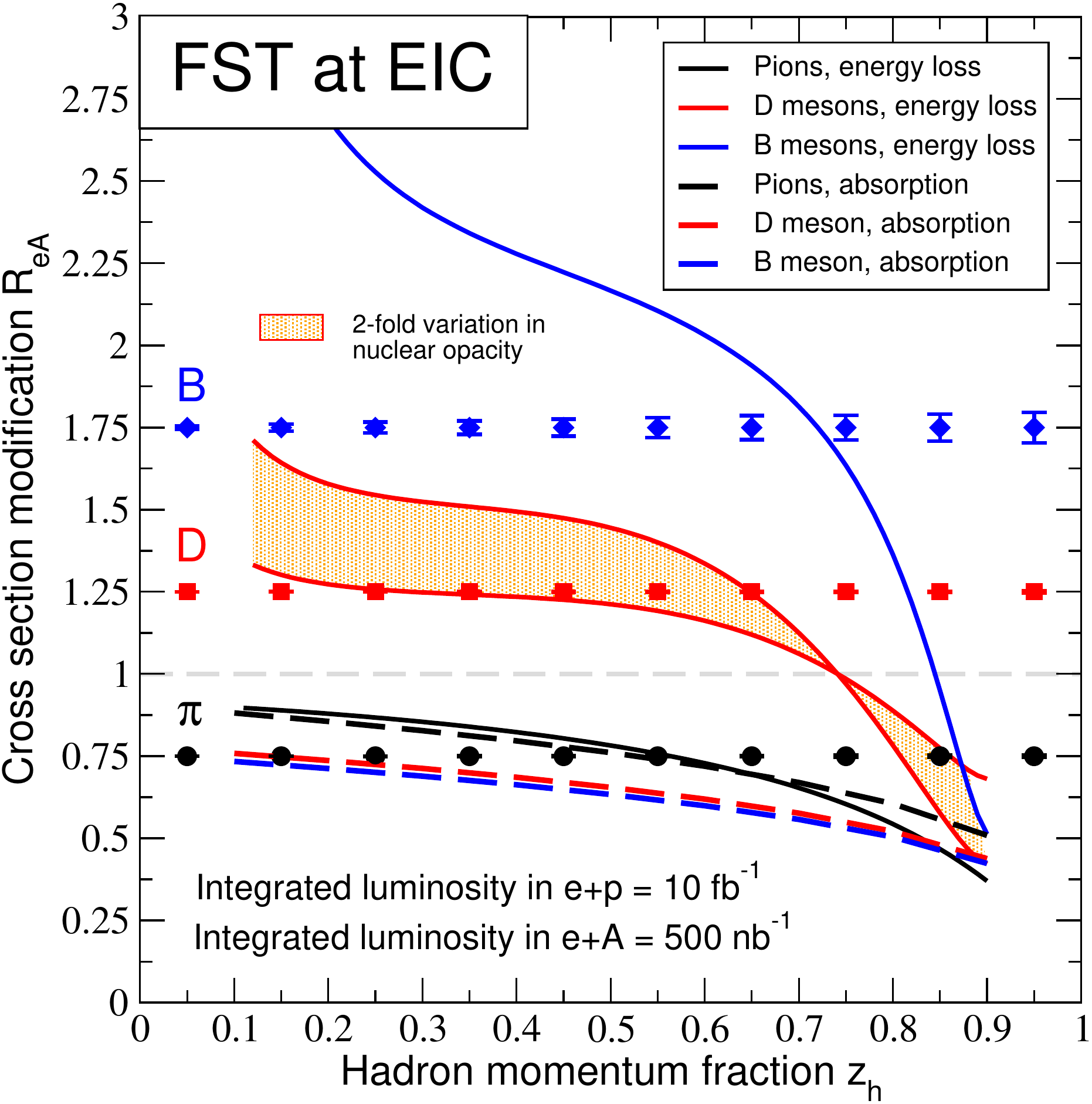}
\end{center}
\caption{Heavy quarks ($B$- and $D$-mesons) provide unparalleled discrimination power for the nature of relativistic transport through nuclear matter. Expected statistical precision for $\pi^\pm$ (black), $D^\pm$ (red), $B^\pm$ (blue) with conservative FST and other EIC detector efficiencies compared to models of parton energy  loss (solid lines) and in-medium hadronization (dashed lines). Projected measured shape is set to be flat for simplicity.}
\label{fig:money}
\end{figure}

Let us now turn to one physics example at the EIC. The tremendous discriminating power of heavy flavor  with respect to models of energy loss (solid lines) and hadronization in nuclear matter (dashed lines) is shown in Fig.~\ref{fig:money}. We use the soft gluon emission energy loss limit of the full splitting kernels described above. The observable 
\begin{equation}
R_{eA} = \frac{d\sigma^{eA}}{d\nu dz_h} \bigg/ A \cdot \frac{d\sigma^{ep}}{d\nu dz_h} 
\end{equation}
 is the hadron cross sections ration in $e+A$ to $e+p$ reactions, appropriately normalized, versus the hadron momentum fraction $z_h$. 
 The calculation is performed for $\nu =20$~GeV and $Q^2$ integrated in the relevant kinematic range.
  $R_{eA}$ shows strong enhancement for $D$-mesons and $B$-mesons if hadronization takes place outside of the large nucleus, where
  the heavy quark-to-meson fragmentation functions are taken from~\cite{Sharma:2009hn}.
  $R_{eA}$ is, however,   suppressed if hadrons are formed early and absorbed in nuclear matter. Such differentiation is  not possible with light hadrons (shown by solid and dashed black lines) since both models predict similar cross section attenuation.  We note that the absorption model shown here relies on the fact that  formation time of heavy hadrons decreases with mass.
  
To perform this measurement, a forward silicon tracker (FST) in the proton or nucleus going direction at the EIC is necessary. Conceptual design for such
tracker has been developed by Los Alamos National Laboratory (LANL) scientists.  Our simulations show that  with  one year of low luminosity running at center-of-mass energy  
$\sqrt{s}=69.3$~GeV, the statistical precision of the measurement is  high, as illustrated by the small statistical error bars in Fig.~\ref{fig:money}. Future heavy flavor results from the EIC will allow us to  pin down the opacities of cold nuclear matter and transport properties of large nuclei.  For reference, the yellow band in  Fig.~\ref{fig:money} shows the effect of a factor of two uncertainty in nuclear opacity  on the  distribution of $D$-mesons in $e+A$ collisions. It is clear that data taken with the FST can constrain such quantities at the $ \sim20\%$ level.  Decorrelation of hadrons due to multiple scattering in nuclear matter~\cite{Qiu:2003pm} can complement cross section attenuation measurements.

\section{Conclusions}
In summary, inclusive hadron and jet production  will play an important role at the EIC in determining the transport properties of nuclei. With this in mind,  we presented solutions for the $q \rightarrow qg$, $g\rightarrow gg$, $q \rightarrow g q$, $g \rightarrow q\bar{q}$  in-medium splitting kernels  beyond the soft gluon approximation  to  ${\cal O}(\alpha_s)$ and to  any desired order in opacity.
We also demonstrated that the full process-dependent in-medium splitting kernels can be evaluated numerically to higher orders in opacity in a  realistic nuclear  medium. The splitting functions are ready for broad phenomenological applications, such as jet and heavy flavor production at the EIC.  
 
 The way in which the cross sections of heavy mesons are modified in reactions with  nuclei of known size (such as  Cu or  Au) relative to the ones measured in reactions with protons provides an experimental handle on the space-time dynamics of hadronization -- the process that describes how elementary particles are formed from the fundamental quark and gluon constituents.  The way in which heavy quarks lose energy as they propagate in the nucleus is a key diagnostic of the dynamic response and not yet known transport properties of nuclear matter. Knowledge of such properties is essential to understand particle transport in extremely dense environments, for example neutron stars. For these measurements to become reality, 
 silicon tracking with high spatial and timing resolution at the EIC is necessary.



\newpage
%

\renewcommand*{\FigPath}{./WeekIV/10_Burkardt}

\renewcommand*{\bi}{\begin{itemize}}
\newcommand*{\ei}{\end{itemize}}
\renewcommand*{\be}{\begin{eqnarray}}
\renewcommand*{\ee}{\end{eqnarray}}
\renewcommand*{\bea}{\begin{eqnarray}}
\renewcommand*{\eea}{\end{eqnarray}}
\newcommand*{\itmh}{\item[{$\hookrightarrow$}]}
\newcommand*{\itmhr}{\item[{$\hookrightarrow$}]}

\wstoc{Aspects of Quark Orbital Angular Momentum}{Matthias Burkardt}

\title{Aspects of Quark Orbital Angular Momentum}

\author{Matthias Burkardt}
\index{author}{Burkardt, M.}

\address{Physics Department, New Mexico State University,\\
Las Cruces, NM 88003, U.S.A.\\
$^*$E-mail: burkardt@nmsu.edu}

\begin{abstract}
The difference between the quark orbital angular momentum (OAM) defined in light-cone gauge (Jaffe-Manohar) compared to
defined using a local manifestly gauge invariant operator (Ji) is interpreted in terms of the change in quark OAM as the quark leaves the 
target in a DIS experiment. We also discuss the possibility to measure quark OAM directly using twist 3 GPDs, and to calculate quark OAM in lattice QCD.

\end{abstract}


\bodymatter
\section{Introduction}
GPDs provide information about the longitudinal momentum and the transverse
position of partons\cite{mb1}. This is reminiscent or orbital angular momentum (OAM), which also requires momentum and position in orthogonal directions. It is thus not completely surprising that GPDs can be related to angular momentum as\cite{JiPRL}
\be
J_q=\frac{1}{2}\int_0^1 dx\, x\left[H^q(x,0,0)+E^q(x,0,0)\right].
\ee
Here $J_q=L_q+\frac{1}{2}\Delta q$ where $\frac{1}{2}\Delta q$ is the quark spin contribution and
\begin{equation}
{ L}_q^z= \int d^3r \langle PS| q^\dagger \left({\vec r} \times \frac{1}{i}{\vec D}
\right)^zq |PS\rangle /\langle PS|PS\rangle
\label{M012}
\end{equation}
in a nucleon state polarized in the $+\hat{z}$ direction. Here
${\vec D}={\vec \partial}-ig{\vec A}$ is the gauge-covariant
derivative.

\section{Angular Momentum Decompositions}
Unfortunately, there are complications due to issues related to the definition of OAM.
To illustrate this point, let us start from a definition of photon total angular momentum based on the Poynting vector. Performing some integration by parts and making use of $\vec \nabla \cdot \vec E =e\psi^\dagger\psi$ one can
rewrite $\vec J_\gamma$ as a term that can be interpreted as photon OAM, a term that cancels the contribution involving
$\vec A$ in $\vec L_q$, and a term that can be interpreted as photon spin:
\begin{eqnarray}\label{eq:Jg}
{\vec J}_\gamma &=& \int d^3r \,{\vec r} \times \left({\vec E}\times{\vec B}
\right) =  \int d^3r \,\left[E^j \left({\vec r}\times {\vec \nabla}
\right) A^j
+ \left({\vec r}\times {\vec A}\right) e\psi^\dagger \psi
+ {\vec E}\times {\vec A}\right]
\end{eqnarray}
Eq. (\ref{eq:Jg}) illustrates that decomposing ${\vec J}_\gamma$ into spin and orbital also
shuffles angular momentum from photons to electrons!

Jaffe and Manohar have proposed an alternative decomposition of the
nucleon spin, which does have a partonic interpretation, and in which also two terms, 
$\frac{1}{2}\Delta q$ and $\Delta G$,
are experimentally accessible \cite{JM}
\begin{equation}
\frac{1}{2}=\frac{1}{2}\sum_q\Delta q + \sum_q {\cal L}^q+
\Delta G + {\cal L}^g.
\label{eq:JJM}
\end{equation}
In this decomposition the quark OAM is defined as 
\begin{equation}
{\cal L}^q \equiv \int d^3r \langle PS|q^\dagger_+\!\left({\vec r}\times \frac{1}{i}{\vec \partial}
\right)^z \!\!q_+  |PS\rangle / \langle PS|PS\rangle,
\label{M+12}
\end{equation}
where light-cone gauge $A^+=0$ is implied. Although Eq. (\ref{M+12}) is not manifestly gauge invariant as written, gauge invariant extensions can be defined
\cite{lorce,hatta}. Indeed, manifestly gauge invariant definitions for each of the terms
in Eq. (\ref{eq:JJM}) exist,  which, with the exception of $\Delta q$, involve matrix elements of
nonlocal operators. In light-cone gauge those nonlocal operators reduce to a local 
operator, such as Eq. (\ref{M+12}).

\section{OAM from Wigner Distributions}

Given the fact that different spin decompositions and corresponding definitions for quark OAM are possible raises the question about the physical interpretation of those differences. In this effort significant progress has been made based on 5-dimensional Wigner distributions.

Wigner distributions can be defined as 
off forward matrix elements of non-local
correlation functions\cite{jifeng,Metz,lorce} with $P^+=P^{+\prime}$, $P_\perp = -P_\perp^\prime = \frac{q_\perp}{2}$
\begin{eqnarray}\label{eq:wigner}
\!\!\!\!\!\!\!W^{\cal U}\!(x,\!{\vec b}_\perp,\! {\vec k}_\perp)\!
\equiv \!\!\!
\int \!\!\frac{d^2{\vec q}_\perp}{(2\pi)^2}\!\!\int \!\!\frac{d^2\xi_\perp d\xi^-\!\!\!\!}{(2\pi)^3}
e^{-i{\vec q}_\perp \!\!\cdot {\vec b}_\perp}\!
e^{i(xP^+\xi^-\!\!-{\vec k}_\perp\!\!\cdot{\vec \xi}_\perp)}
\langle P^\prime S^\prime |
\bar{q}(0)\Gamma {\cal U}_{0\xi}q(\xi)|PS\rangle .
\end{eqnarray}
Throughout this paper, we will chose ${\vec S}={\vec S}^\prime = \hat{\vec z}$. Furthermore, we will focus on the 'good' component by selecting $\Gamma=\gamma^+$.
To ensure manifest gauge invariance, a Wilson line gauge link 
${\cal U}_{0\xi}$ connecting the quark field operators at position $0$ and $\xi$ is included. The issue of choice of path
for the Wilson line will be addressed below. 

In terms  of  Wigner distributions,  TMDs and OAM can be defined 
as \cite{lorce}
\begin{eqnarray}
f(x,{\vec k}_\perp) &=& \int d^2{\vec b}_\perp 
W^{\cal U}(x,{\vec b}_\perp,{\vec k}_\perp)\\
L_{\cal U}&=& \int dx d^2{\vec b}_\perp d^2{\vec k}_\perp \left({\vec b}_\perp \times {\vec k}_\perp \right)^z
W^{\cal U}(x,{\vec b}_\perp,{\vec k}_\perp).
\nonumber
\end{eqnarray}
No issues with Heisenberg's uncertainty principle arise here: only perpendicular combinations of position ${\vec b}_\perp$ and momentum ${\vec k}_\perp$ are
needed simultaneously in order to evaluate the integral for
$L_{\cal U}$.

A straight line connecting $0$ and $\xi$ for the Wilson line in ${\cal U}_{0\xi}$ results in
\cite{jifeng}
\begin{eqnarray}
L^q_{straight}
&=&
L^q_{Ji}.
\label{eq:LJi}
\end{eqnarray}
However, depending on the context, other choices for the path in the Wilson link ${\cal U}$ should be made. Indeed for TMDs probed in SIDIS the path should be taken to be a straight line to $x^-=\infty$
along (or, for regularization purposes, very close to) the light-cone. This particular choice ensures proper inclusion of the FSI experienced by the struck quark as it leaves the nucleon
along a nearly light-like trajectory in the Bjorken limit. However, a Wilson line to
$\xi^-=\infty$, for fixed ${\vec \xi}_\perp$ is not yet sufficient to render Wigner distributions
manifestly gauge invariant, but a link at $\xi^-=\infty$ must be included to ensure manifest
gauge invariance. While the latter may be unimportant in some gauges, it is crucial in
light-cone gauge for the description of TMDs relevant for SIDIS. 

Let ${\cal U}^{+LC}_{0\xi}$ be the Wilson path ordered exponential obtained by first taking
a Wilson line from $(0^-,{\vec 0}_\perp)$ to $(\infty,{\vec 0}_\perp)$, 
then to $(\infty,{\vec \xi}_\perp)$, and then to $(\xi^-,{\vec \xi}_\perp)$, with each segment being a straight line (Fig. \ref{fig:staple}) \cite{hatta}. 
\begin{figure}
\includegraphics[scale=0.5]{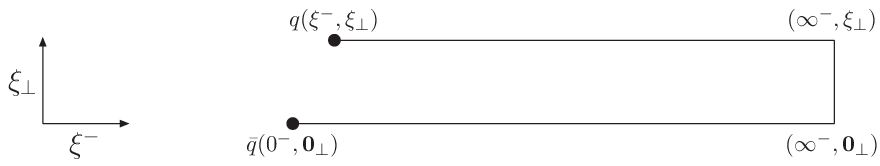}
\caption{Illustration of the path for the Wilson line gauge link ${\cal U}^{+LC}_{0\xi}$ entering  $W^{+LC}$ }
\label{fig:staple}
\end{figure}
The shape of the segment at $\infty$ is irrelevant as the gauge field is pure gauge there, but it is still necessary to include a connection at $\infty$ and for
simplicity we pick a straight line. 
In light-cone gauge $A^+=0$, only the segment at $\xi^-=\pm \infty$ contributes and 
the OAM looks similar to the local manifestly gauge invariant expression, except
\be
{\vec r}\times {\vec A}({\vec r}) \longrightarrow {\vec r}\times {\vec A}(r^-=\pm \infty, {\bf r}_\perp).
\ee

This observation is crucial for understanding the difference 
between the Ji vs. Jaffe-Manohar OAM, which in light-cone gauge\footnote{As $L^q$ involves a manifestly gauge invariant local operator, it can be evaluated in any gauge.}
involves only the replacement ${ A}_\perp^i({\vec r}) \longrightarrow {A}_\perp^i(r^-=\pm \infty, {\bf r}_\perp)$.
Using
\begin{eqnarray}
{A}^i_\perp (r^-\!\!=\infty,{\bf r}_\perp)-{ A}^i_\perp (r^-,{\bf r}_\perp)
=\!\!\int_{r^-}^\infty\!\!\!\!\! dz^-
\partial_- {A}^i_\perp (z^-,{\vec r}_\perp)
= \!\!\int_{r^-}^\infty\!\!\!\!\! dz^- G^{+i}(z^-,{\vec r}_\perp)
\label{eq:kp}
\end{eqnarray}
where $G^{+\perp}=\partial_-A^\perp$ is the gluon field strength tensor in $A^+=0$ gauge. 
\be
{\cal L}^q-L^q = -g \!\!\int \!\!d^3x\!\left\langle P\!,\!S\right|\!
\bar{q}({\vec x})\!\gamma^+\!\!
\left[{ {\vec x}\! \times\! \!
\int_{x^-}^\infty \!\!\!\!\!dr^- F^{+\perp}(r^-,{\bf x}_\perp)
}\right]^z\!\!\!\!
q({\vec x}) \!\left| P\!,\!S\right\rangle/ \langle PS|PS\rangle
\label{eq:torque}
\ee
Note that 
\begin{equation}
-\sqrt{2}gG^{+y}\equiv -gG^{0y}-gG^{zy} = g\left(E^y-B^x
\right)
=g\left({\vec E}+{\vec v}\times {\vec B}\right)^y
\end{equation}
yields the $\hat{y}$ component of the color Lorentz force acting on a particle that moves with the velocity of light in the $-\hat{z}$ direction (${\vec v}=(0,0,-1)$) --- which is the direction of the 
momentum transfer in DIS. Thus the difference between the Jaffe-Manohar and Ji OAMs
has the semiclassical interpretation of the change in OAM due to the torque from the FSI as the quark leaves the target:\cite{mb:torque}
while $L^q$ represents the local and manifestly gauge invariant OAM of the
quark {\it before} it has been struck by the $\gamma^*$, ${\cal L}^q$ represents 
the gauge invariant OAM {\it after} it has left the nucleon and moved to $r^-=\infty$.

To convince oneself that the net effect from such a torque can be nonzero, one can consider an ensemble of quarks ejected from 
a nucleon with a color-magnetic field aligned with its spin.

\section{Torque in Spectator Models}
Given the fact that the Jaffe-Manohar and Ji definitions of quark orbital angular momentum are differ by the potential angular momentum raises the question as how significant that difference actually is.
In Ref. \cite{JiElectron} it was shown that to ${\cal O}(\alpha)$ in QED
$L_{Ji}={\cal L}_{JM}$. This corrects an earlier result \cite{BC} where the contribution from states with longitudinally polarized Pauli-Villars photons had been omitted.

In order to assess the significance of FSI effects for quark OAM we thus considered the effects of the vector potential in the scalar diquark model. While we do not consider this model a good approximation for QCD, it has been very useful in several respects:
Most importantly, the model allows for a fully Lorentz invariant calculation of 'nucleon' matrix elements --- which is not the case for almost all other models for nucleon structure.
Furthermore, this was the first model that clearly illustrated the role of FSI and the Sivers effect\cite{Sivers1} in SIDIS and DY \cite{BHS}. 

We found that a nonzero potential angular momentum arises at the same order
as transverse single-spin asymmetries \cite{BHS}, i.e. one photon/gluon exchange \cite{mb:torque}.

{\bf Acknowledgements:}
This work was supported by the DOE under grant number 
DE-FG03-95ER40965


\newpage
\renewcommand*{\FigPath}{./Logo/} 

\begin{tcolorbox}[colframe=white]
\begin{minipage}{0.2\textwidth}
\includegraphics[width=1.\textwidth]{\FigPath/INT_Workshop_Logo_Final_Black.png}
\end{minipage}
\begin{minipage}{0.7\textwidth}
\wstoc{\bf Weeks V \& VI}{}
\title{Weeks V \& VI}
\end{minipage}
\end{tcolorbox}

%
\wstoc{Summary of Weeks 5 \& 6}{Giovanni A. Chirilli, Anna M. Stasto, Thomas Ullrich, Bo-Wen Xiao}
\title{Summary of Weeks 5 \& 6}

\author{Giovanni A. Chirilli$^*$}
\address{Institut f\"ur Theoretische Physik, Universit\"at Regensburg,\\
	D-93040 Regensburg, Germany,\\
$^*$E-mail: giovanni.chirilli@ur.de}

\author{Anna M. Stasto}
\address{Department of Physics, The Pennsylvania State University, University Park, PA 16802, USA}

\author{Thomas Ullrich}
\address{Department of Physics, Brookhaven National Laboratory,\\ Upton, NY 11973, USA}

\author{Bo-Wen Xiao}
\address{Key Laboratory of Quark and Lepton Physics (MOE) and Institute
of Particle Physics,\\ Central China Normal University, Wuhan 430079, China}

\index{author}{Chirilli, G.}
\index{author}{Stasto, A.}
\index{author}{Ullrich, T.}
\index{author}{Xiao, B.}

\begin{abstract}
Weeks 5 and 6 of the INT program 2018 were dedicated to physics
opportunities at a future Electron-Ion Collider.
Discussions were wide ranging and included topics such as short-range
correlation, jets physics in deep inelastic scattering, nuclear PDFs,
TMDs, and GPDs as well as small-$x$ helicity distributions, non-linear
small-$x$ evolution equations, diffraction and particle production and
correlations.
\end{abstract}


\bodymatter

\section{Jet physics}

Since the beginning of high-energy collider physics, jets have been an important tool in the exploration 
of the Standard Model and especially of QCD and have provided discoveries and insights in $e^+e^-$, $ep$, $pA$, 
and $AA$ collisions. Jet computations have become highly advanced in the high energy physics community, 
and are an important ingredient in searches for the physics beyond the standard model. In heavy-ion physics, 
jet quenching provides a quantitative measure of the transport properties of the strongly interacting 
quark-gluon plasma (QGP). Because jets have proven to be such a powerful tool, physicists have invested 
greatly in improving the understanding and simulation of jets over the last decade. 
These sophisticated tools are now available for use as probes of the partonic structure of protons and nuclei 
in DIS collisions at an EIC. In the extensive report of the 2010 INT program on ``Gluons and the quark 
sea at high energies'' in Fall of 2010, jets were listed as a golden measurement in $eA$ collisions at an 
EIC although the subsequent EIC White Paper did not address this topic largely due to a  lack of follow-up studies. 
While still in the early stages, studies have now begun on a number electron-nucleus and electron-hadron 
topics where jets could play an important role. Examples relevant for the $eA$ program are the study of 
hadronization to shed light on the nature of color neutralization and confinement, parton shower evolution 
in strong color fields to measure cold nuclear matter transport coefficients, and the study of 
diffractive dijet production, which can provide direct access to gluon distribution 
functions in nuclei and possibly even the nuclear Wigner function. 

During this INT program the current status of jet finding at EIC energies was prominently discussed, 
including overviews of general jet properties, underlying event characteristics, and preliminary substructure 
investigations. It was emphasized that jets at an EIC will have substantially lower $p_T$ and fewer particles 
per jet than at hadron colliders but will also experience relatively little underlying event contamination, 
allowing for jet finding even with large cone radii not possible in $pp$, or $AA$ at the LHC or RHIC. 
The importance of high center-of-mass energies to jet yields and kinematic reach was also underscored. 
Two goals of the EIC program will be the exploration of cold nuclear matter and study of the 
hadronization process. Substructure observables, which quantify how energy is distributed within 
the jet may be useful in the study of these topics. The comparison of the substructure in $eA$ with 
 that in $ep$ is expected to be sensitive to details of hadronization in nuclear matter and possibly to 
certain properties of the matter such as transport coefficients. Theoretical and simulation studies of angularity, 
a specific substructure variable, at EIC energies are currently ongoing and the latest progress was presented. 

The current theoretical understanding and especially the modeling of energy loss in matter has made 
substantial progress in the past few years, due in part to concerted efforts such as the former JET 
and more recently the JETSCAPE collaboration.  While these groups focused on the modification of 
jet substructure in relativistic heavy-ion collisions, the available expertise and new insight can now 
be deployed for DIS in $eA$. Multistage jet evolution in JETSCAPE, a publicly available software 
package containing a framework for Monte Carlo event generators, provides an integrated description 
of jet quenching by combining multiple models, with each becoming active at a different stage of the 
parton shower evolution. Jet substructure modification due to different aspects of jet quenching can 
be studied using jet shape and jet fragmentation observables and various combinations of jet energy 
loss models are exploited. Going from $AA$ (hot) to $eA$ (cold QCD) physics implies various changes. 
There is little to no hadronic energy loss in $eA$, the hadronic response of the medium is substantially 
smaller, and may require the simulation of TMDs and GPDs. Transport coefficients that quantify how 
the medium changes the jet and those that quantify the space-time structure of the
deposited energy are smaller in $eA$ than in $AA$ and studies will have to quantify what key 
measurements are needed to determine them with the highest precision. First attempts to include $eA$ into 
JETSCAPE are underway but will take some time and effort until  it will be completed.

Besides the study of energy loss and fragmentation in matter, jet physics in $eA$ has the potential 
to greatly expand the experimental toolbox and sensitivity to nuclear effects, thereby allowing a 
close connection to first principles calculations in QCD. One example is the Weiz\"acker-Williams 
TMD gluon distribution, and in particular the distribution of linearly polarized gluons that appears 
in a variety of processes but most promisingly in dijets. New NLO calculations of exclusive diffractive 
dijet production show that these processes are perfectly suited for precision saturation physics and 
gluon tomography with $b_T$ dependence at an EIC. Furthermore, there is an increasing body of 
theoretical work on probing the small $x$ gluon Wigner function in both $ep$ and $eA$ collisions 
by quantifying the angular correlation between the nucleon recoil momentum and the dijet transverse 
momentum in diffractive dijet production. So far many of these exciting measurements are based on 
theoretical work alone. Feasibility studies of these measurements at an EIC need to be conducted in 
the near future before they can be regarded as part of the EIC physics program. 

\section{Nuclear TMDs and GPDs}

Many contributions and discussion at this INT program focused on imaging of the proton, i.e., 
establishing its complete momentum and spatial structure in many dimensions and the dependence 
of this structure on the spin of the proton. On the one hand, the three-dimensional structure of the 
proton in momentum space can be described via Transverse Momentum Dependent (TMD) parton 
distribution functions that encode a dependence on the intrinsic transverse momentum of the 
parton, $k_T$, in addition to the longitudinal momentum fraction, $x$, 
of the parent proton carried by the parton. On the other hand, the structure 
of partons in a proton and their correlations in the impact parameter space is encoded 
in the Generalized Parton Distributions (GPDs), which lead to tomographic imaging of 
a proton in spatial coordinates. TMDs and GPDs are three-dimensional functions that 
have been introduced as a suitable theoretical tool to study the partonic structure of the nucleon 
in $k_T$ or $b_T$ space. However, accessing their mother functions, the five-dimensional 
Wigner distributions, would provide the most complete information. The Wigner functions are 
not calculable from first principles in perturbative QCD,  thus possibility of experimental constraints are of 
crucial importance. It was only recently that experimental means to access the elliptic gluon 
Wigner function were proposed focusing on exclusive dijet production measurements. 

While many studies have been carried out in the past decade to demonstrate how 
an EIC could constrain TMDs and GPDs of a free proton, very few such studies exist for nuclei. 
Our knowledge of the modifications of TMDs and GPDs induced by the nuclear environment 
is rudimentary, providing an enormous opportunity for the EIC to improve our knowledge 
of the multidimensional partonic structure of nuclei. 

Current knowledge of quark TMDs typically comes from polarized semi-inclusive DIS (SIDIS) 
experiments, where a final state hadron is observed. Today, very little is known about gluon 
TMDs, because they typically require higher-energy scattering processes and are harder to isolate 
as compared to quark TMDs. Nuclear gluon TMDs can be accessed through measuring heavy-quark meson pairs, 
quarkonia, or through the analysis of azimuthal asymmetries for heavy quark pair and dijet production in $eA$ 
collisions.  These studies are quite crucial since important connections have been established between the framework 
of TMDs and the theory of Color Glass Condensate (CGC) at small $x$ that need to be explored.

GPDs of the proton were extensively discussed in the first week of the program. 
As with TMDs, little is known about nuclear GPDs (nGPDs) and few data exists. 
To-date no feasibility study  has been conducted to establish how well an EIC could constrain 
nGPDs. GPDs can be experimentally constrained in exclusive DIS processes, such as Deeply 
Virtual Compton Scattering (DVCS), which is the production of a real photon, and Vector Meson 
Production (VMP). In $ep$ the scattered proton can be measured with Silicon tracker detectors 
installed along the beam line on moving vessels, known as Roman pots (RPs). Such spectrometer 
enables a precise measurement of the momentum transfer by the scattered proton, $t$. 
For light nuclei (D to He4) this might be still possible, while heavy nuclei stay within 
the beam envelope and are not detectable. An alternative way is to derive $t$ from the 
event kinematics since the angle and momentum of the scattered electron and the produced 
real photon (DVCS) or vector meson (VMP) are known. This method was also used at HERA but 
has the downside that it is less precise and incoherent diffractive events can contaminate the sample. 
At an EIC, incoherent processes might be detected and suppressed by tagging the fragments of the 
nuclear breakup using Zero-Degree Calorimeters (ZDC) and Roman Pots combined. 

Studying DVCS and VMP in light nuclei is also interesting because existing EIC designs make 
possible to polarize beams from D up to Li. Polarized He3 beams for example will allow us for 
the simultaneous measurement of both, the tagged neutron structure and coherent diffraction, 
which is especially interesting since the spin of He3 is dominated by the neutron. 
Measuring DCVS on the neutron compared to the proton is key to understand the underlying 
$u/d$ quark-flavor separation in GPDs. While experimentally relatively easy, He3 
is theoretically more challenging. In that aspect it is probably convenient to also measure DVCS on 
deuteron via applying a double tagging method by tagging the neutron with the ZDC and 
measuring the recoil proton momentum in the RPs. This way, providing a measurement of the 
neutron structure not affected by final state interactions. Again, more simulations are currently 
needed to establish the degree of feasibility and accuracy if these measurements at an EIC.

\section{Diffraction at an EIC}

An important discovery at HERA electron-proton collider was the observation that about a significant fraction 
of the events, about 10\%, were diffractive. 
These were the events where the proton stays intact or dissociates into a state with proton 
quantum numbers despite highly energetic collision. The proton, or the proton remnant, 
was separated from the rest of the event by the rapidity gap, the large region of the 
detector which was void of any activity.  Both HERA experiments, H1 and ZEUS 
performed detailed measurements of the diffractive events, both inclusive and exclusive. 
The measurements of the inclusive diffraction were performed using two complementary 
experimental methods, using the rapidity gap method and using the forward (or leading) 
proton spectrometer to detect the elastically scattered proton.  The measurement 
of diffraction is crucial as it tests our understanding of the transition between the soft and hard QCD dynamics.  
The fact that the event contains the rapidity gap, and the proton stays intact in the violent 
interaction means that the interaction is likely to proceed through the exchange of the color neutral object, 
and detailed measurements of such reaction can provide vital information about the confinement phenomena. 
Theoretical studies of the QCD in the high energy limit also point to the link between the diffraction 
and the low $x$ as well as saturation phenomena. In the case of the scattering off nuclei, there exists 
known relation between the nuclear shadowing and diffraction in ep scattering. The study of diffraction 
in DIS allows to investigate the range and limits of the factorization in the perturbative QCD. 
Exclusive diffraction of vector mesons in  electrons scattering off protons and nuclei,  
is especially sensitive to the details of the 3D structure of hadrons and nuclei.

\section{Inclusive diffraction on protons and nuclei}

 EIC would be able to measure the inclusive diffraction off protons and nuclei. 
In the proton case the kinematics covered will be heavily overlap with HERA kinematics, 
and will allow for the in-dependent measurement of the $F_2^D(3)$. A novelty at an EIC, 
would be a possibility of the measurement of the longitudinal diffractive structure function $F_L^D(3)$. 
This observable was extracted at HERA, but the possibility of the variation of the electron energy at an EIC, would likely 
allow to measure this important quantity to a precision competitive with HERA. 
This quantity could provide a vital information about the diffractive gluon distribution in the proton, 
as well as provide the information about the importance of the higher twists in diffraction. 

An EIC would be able to extract the diffractive structure functions and diffractive parton distribution functions 
in nuclei for the first time. The large luminosity and wide range of nuclei injected at the EIC, 
would allow to perform the extraction of the nuclear diffractive parton densities with the precision 
comparable to the precision of the diffractive parton densities that were 
measured at HERA in electron-proton collisions. One important aspect, that needs to be studied 
carefully is that of distinguishing between a coherent and incoherent diffraction in electron-nucleus collisions. 
The coherent diffraction is the case when the nucleus stays intact, and it is expected that it will dominate 
for very low values of the momentum transfer. In the case of the incoherent diffraction, the nucleus can 
easily break-up into nucleons, while still the rapidity gap can be present between the nuclear remnants 
and the rest of the event. Still, the incoherent diffraction off of the nucleus can be a coherent event 
on the nucleon, as there could be a simple break up of the nucleus without the pion production. 
This has to be distinguished from the incoherent-inelastic diffraction where the event is inelastic 
scattering off the nucleons. A dedicated instrumentation in the forward region of the EIC detector, 
with the Zero Degree Calorimeter, will need to be implemented to clearly distinguish between these different scenarios. 
Another challenging problem is the experimental separation of events in which the scattered nucleus is in excited state as opposed to the non-excited one for low momentum transfers. 
An EIC would also provide a possibility of unique tests of the Gribov’s relation between the shadowing 
and diffraction. This can be achieved by measuring precisely diffractive structure functions in 
electron -- proton and comparing them with the inclusive structure functions in the electron -- deuteron scattering. 
Extension of this relation to higher number of scatterings can be effectively tested by the 
measurements of the inclusive structure functions and the shadowing for heavier nuclei.

Recent study shows that there exists an analogy between diffractive electron-nucleus scattering events and realizations of
one-dimensional branching random walks selected according to the height of the genealogical tree of
the particles near their boundaries in statistical physics. This allows the prediction of rapidity gap distribution in diffractive DIS which can be potentially measured and tested in the future EIC. It may lead to more fundamental and quantitative understanding of diffraction in DIS. 

\section{Exclusive diffraction on protons and nuclei}

In addition to the inclusive diffraction, an EIC would be an excellent machine to perform detailed 
measurements of the exclusive diffraction, where the final states involve elastically scattered protons 
(or nuclei) as well as vector mesons. Such measurements were performed at HERA, and sparked 
a lot of interest, due to their unique sensitivity to the spatial distribution of gluons in-side the proton. 
The process of exclusive diffractive vector meson production can be theoretically described using a dipole model, 
which assumes a factorization of the process into two steps: the fluctuation of the virtual 
photon in to a dipole and formation of the vector meson wave function and the dipole interaction with the target. 
The latter one encodes all the information about the interaction with the hadronic target. By measuring the 
momentum transfer dependence of this process, one can learn about the spatial distribution of gluons 
inside the target, through the Fourier transform into the coordinate space. Simulations were performed 
using various models, which include the small $x$ effects and demonstrate that such a process offers 
a sensitivity to the small $x$ dynamics and can potential be a smoking gun to determine the onset of the 
parton saturation in hadrons and in nuclei. The presence and the position of the minima in the t distribution 
of this process, and in particular the dependence of the minima on the virtuality, type of vector mesons 
and the energy can provide the valuable information about the change of the density distribution inside 
the hadron with respect to these variables. The same process can be studied in the case of the nuclear targets. 
In the nuclear case, the coherent and incoherent diffractive production of vector mesons, have to be clearly 
distinguished, as they would provide different constraints on the internal structure of the nuclei. 
The incoherent diffraction, in particular, can provide the information about the fluctuations 
of the density in-side the hadrons and nuclei.

Exclusive diffractive productions of dijets and $\rho$ mesons are known to be sensitive to the gluon generalized-TMD 
and the gluon GPDs at small-$x$, respectively. They provide important channels for the multiple-dimensional 
tomography of low-$x$ region of partons inside hadrons.  Recently, these two processes have been computed 
at the NLO level in the small-$x$ formalism, which paves the way for NLO quantitative studies of small-$x$ 
phenomenologies in diffractive processes, and for precision tests of saturation effects at an EIC.

\section{Small-$x$ behavior of quark and gluon helicities}

Since many years ago, DIS experiments have revealed that proton is a composite particle made up 
by quarks and gluons, and proton spin is expected to be the helicity sum of the spin and orbital 
angular momentum of quarks and gluons. Small-$x$ behavior of quark and gluon helicities also 
play an important role in understanding the outstanding proton spin puzzle, since the helicity 
contributions from the small-$x$ regime may be significant. Recently, a series of work on the 
small-$x$ asymptotics and evolution of helicity distributions can provide better constraints on 
the quark and gluon spin contributions. Also, one should be able to measure related spin contributions 
and spin dependent TMDs at an EIC with high precision.

\section{TMD factorization and rapidity evolution}

Despite the fact that the small-$x$ gluon distribution and the gluon TMD are usually used in different 
kinematical region, there have been a lot of theoretical interests in the intimate connection between 
the gluon distribution in small-$x$ formalism and the gluon TMD in the TMD formalism. In particular, 
they have the same operator definition before any quantum evolution. Although the gluon distributions 
in these formalisms follow two different evolution equations, recent work demonstrates how 
the rapidity evolution of gluon TMD distribution changes from nonlinear evolution at small-$x$ 
to the linear TMD evolution at moderate $x$. Moreover, it has been suggested that the small-$x$ inspired rapidity scheme for TMD
is an alternative choice to the known ultraviolet-rapidity Collins-Soper-Sterman formalism.

\section{Small-$x$ physics and gluon saturation}

\subsection{Non-linear small-$x$ evolution}
The non-linear small-$x$ evolution equation, which is known as the Balitsky-Kovchegov-JIMWLK 
evolution equation, provides a vital description of the QCD dynamics at small-$x$ region. 
It predicts the gluon saturation phenomenon which could be tested and measured at the EIC. 
The leading order version of this evolution equation has been studied extensively, however, 
the impact of the next-to-leading order correction turn out to be quite subtle, due to instability 
of the large and negative double collinear logarithms appeared at NLO. The resummation 
of these double logarithms can cure the instability and match to the full NLO small-$x$ evolution.

\subsection{Particle production in dilute-dense collisions}

In recent years, the outstanding negativity problem in single forward rapidity hadron productions 
in proton-nucleus collisions has attracted a lot of theoretical attention. It was found that the 
NLO corrections in the small-$x$ formalism for single hadron production overwhelm the 
LO contribution and make the total contribution negative in the high $p_T$ region of produced hadron. 
The essential step of the NLO calculation involves subtractions of collinear and rapidity divergences. 
A new factorization scheme without the rapidity subtraction for this process can avoid the inherent 
fine-tuning issue in the rapidity subtraction, which might be the origin of the negativity problem.

Similar to the above-mentioned single inclusive forward hadron productions, the DIS structure function 
at NLO also receives large and negative contribution. The total DIS cross section in the small-$x$ 
dipole formalism can be computed at the NLO accuracy, which in principle can allow us to go beyond 
leading order in the era of the EIC. Again, due to the instability caused by large and negative NLO contributions, 
the NLO cross section depends on the factorization scheme used to resum large logarithms of energy 
into renormalization group evolution with the BK equation. It has been shown to be plausible to 
use a factorization scheme consistent with the one proposed in single inclusive production 
to obtain meaningful results for the DIS structure function.

\subsection{Gluon TMDs at an EIC and in pA collisions}

There has been significant process in the understanding of the relation between the process 
dependence of gluon TMDs and the corresponding gauge links in recent years. This allows to clarify 
the two gluon distributions puzzle in small-$x$ physics and make predictions for various gluon TMDs 
which can be measured at the future EIC. In addition to the usual unpolarized gluon TMD, the linearly 
polarized gluon distribution and the gluon generalized-TMD as well as odderon effects can be also 
probed in various processes in electron-ion collisions. The interaction region design of the EIC with the 
forward detector system can cover these interesting physics needs for wide ranges of nucleon energies. 
Together with complementary studies of other observables such as dijet productions in proton-nucleus collisions, 
one can test the universality of gluon distributions and its gauge link dependence predicted by theoretical calculations. 

\subsection{Gluon saturation dynamics and long-range rapidity correlations}

Long-range two-particle rapidity correlations have been observed at RHIC and the LHC in both 
heavy ion collisions and small systems created in high-multiplicity proton-proton and proton-nucleus collisions. 
The origin of such correlation, in particular the odd azimuthal harmonics, is one of the most interesting topics 
in heavy ion physics. Usually, the direct production of two particles involving gluons gives zero contribution 
to the odd harmonics due to the vanishing imaginary amplitude. Usually, odd azimuthal harmonics appear 
only in the terms contributing at least three interactions with the projectile to the two-gluon production cross section. 
This may resolve the ambiguity and allow for further successes of the saturation approach to the long-range correlation phenomenology.

\section{Summary}

In summary, as outlined above, there is quite a lot of interesting physics that one can explore in a future EIC especially via $eA$ collisions. The synergy between 
the EIC experiment and theory can enable us to obtain the multi-dimensional tomography of nucleon and heavy nucleus and to understand intriguing phenomena such as short range correlations. 

\newpage
%

 \renewcommand*{\FigPath}{./WeekVI/01_mantysaari}

 \newcommand{\rt}{{\mathbf{r}}}
\newcommand{\xt}{{\mathbf{x}}}
\newcommand{\yt}{{\mathbf{y}}}
\newcommand{\nc}{N_\mathrm{c}}
\newcommand{\bt}{{\mathbf{b}}}
\newcommand{\bti}{{\mathbf{b}_{i}}}
\renewcommand{\gev}{\ \textrm{GeV}}
\newcommand{\der}{\mathrm{d}}
\newcommand{\xpom}{{x_\mathbb{P}}}
\newcommand{\A}{\mathcal{A}}

\wstoc{Exclusive vector meson production at the EIC}{Heikki M\"antysaari}
\title{
Exclusive vector meson production at the EIC
}

\author{Heikki M\"antysaari$^*$}
\index{author}{M\"antysaari, H.}

\address{Department of Physics, University of Jyv\"askyl\"a, %
 P.O. Box 35, 40014 University of Jyv\"askyl\"a, Finland, and \\
Helsinki Institute of Physics, P.O. Box 64, 00014 University of Helsinki, Finland \\
$^*$E-mail: heikki.mantysaari@jyu.fi}

\begin{abstract}
We discuss how vector meson production at the future Electron Ion Collider can be used to probe  non-linear dynamics in heavy nuclei. Additionally, the potential to study the evolution of proton and nuclear geometries with event-by-event fluctuations is illustrated.
\end{abstract}


\bodymatter
\vspace{-1em}
\section{Introduction}
Exclusive vector meson production is a very powerful tool to probe the internal structure of protons and nuclei. In these processes no color string between the target and the produced particle can exist (as it would break and produce many particles). Thus, there can not be net color transfer to the target, which in perturbative QCD requires at least two gluons to be exchanged. Consequently, the cross section is approximatively proportional to the \emph{squared} gluon distribution~\cite{Ryskin:1992ui_223}.

Good sensitivity on the small-$x$ gluonic structure makes exclusive processes potentially very powerful in studies of non-linear saturation effects. As the gluon densities rise towards small $x$ due to the enhanced radiation of soft gluons in QCD, at some point it is expected that one reaches the limit where non linear effects tame this growth. As a result, anew state of matter where color fields are as strong as allowed in the nature is formed. An effective field theory describing QCD in this regime is known as the Color Glass Condensate~\cite{Gelis:2010nm_217}.

Additionally, in exclusive process the total momentum transfer can be measured by reconstructing the momentum of the produced particle and that of the outgoing lepton. As the transverse momentum transfer is the Fourier conjugate to the impact parameter, these processes at high energies provide access to the spatial distribution of small Bjorken-$x$ gluons in the target wave function, as well as to  the event-by-event geometry fluctuations as discussed below.

\section{Proton structure}
  \begin{figure}[tb]
    \centering
    \begin{minipage}{.48\textwidth}
        \centering
        \includegraphics[width=\textwidth]{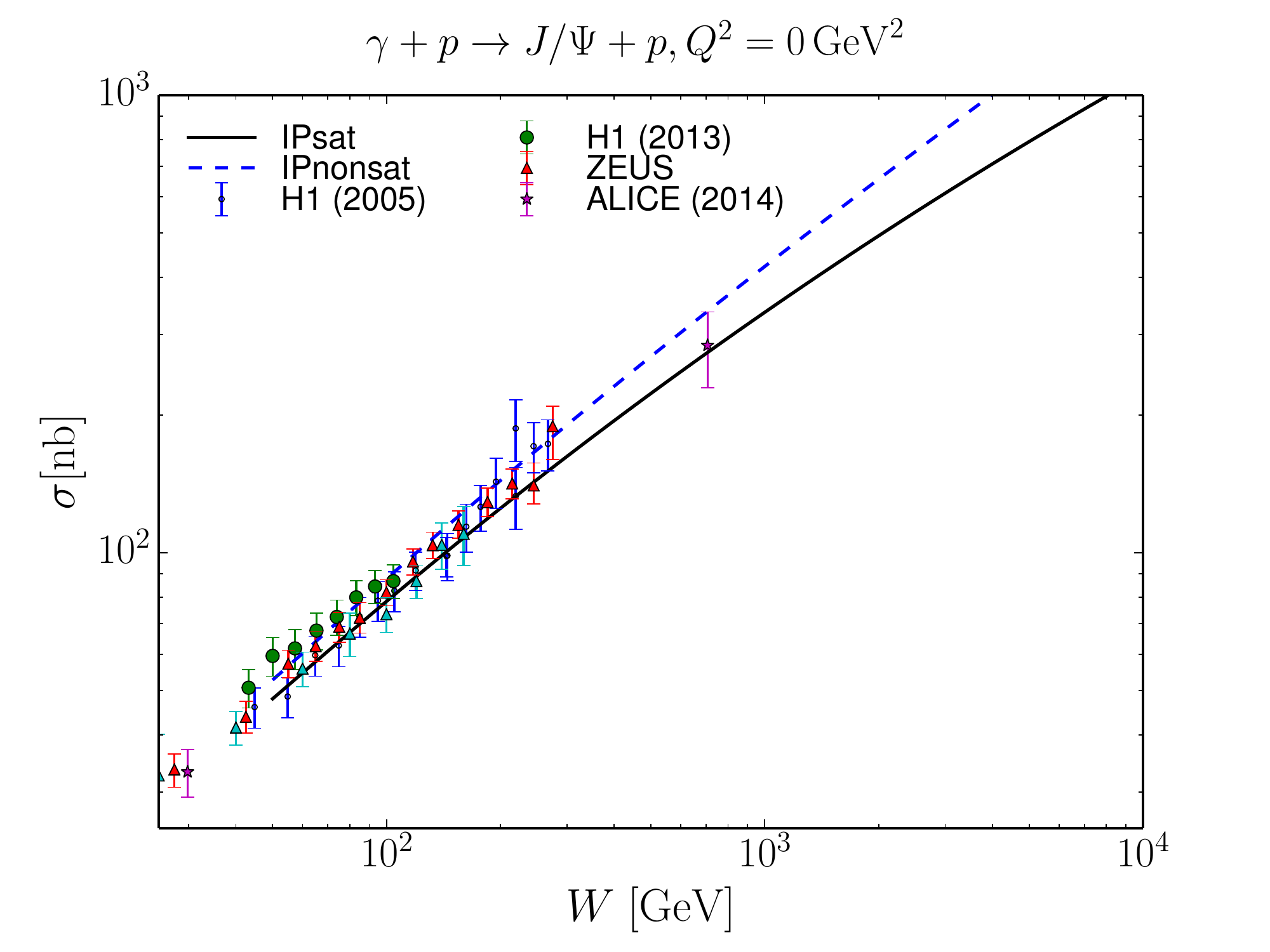} 
				\caption{Total cross section for $J/\Psi$ photoproduction as a function of collision energy from~\cite{Mantysaari:2018nng}.}
		\label{fig:jpsi_wdep}
    \end{minipage}
    \quad
     \begin{minipage}{0.48\textwidth}
        \centering
      \includegraphics[width=\textwidth]{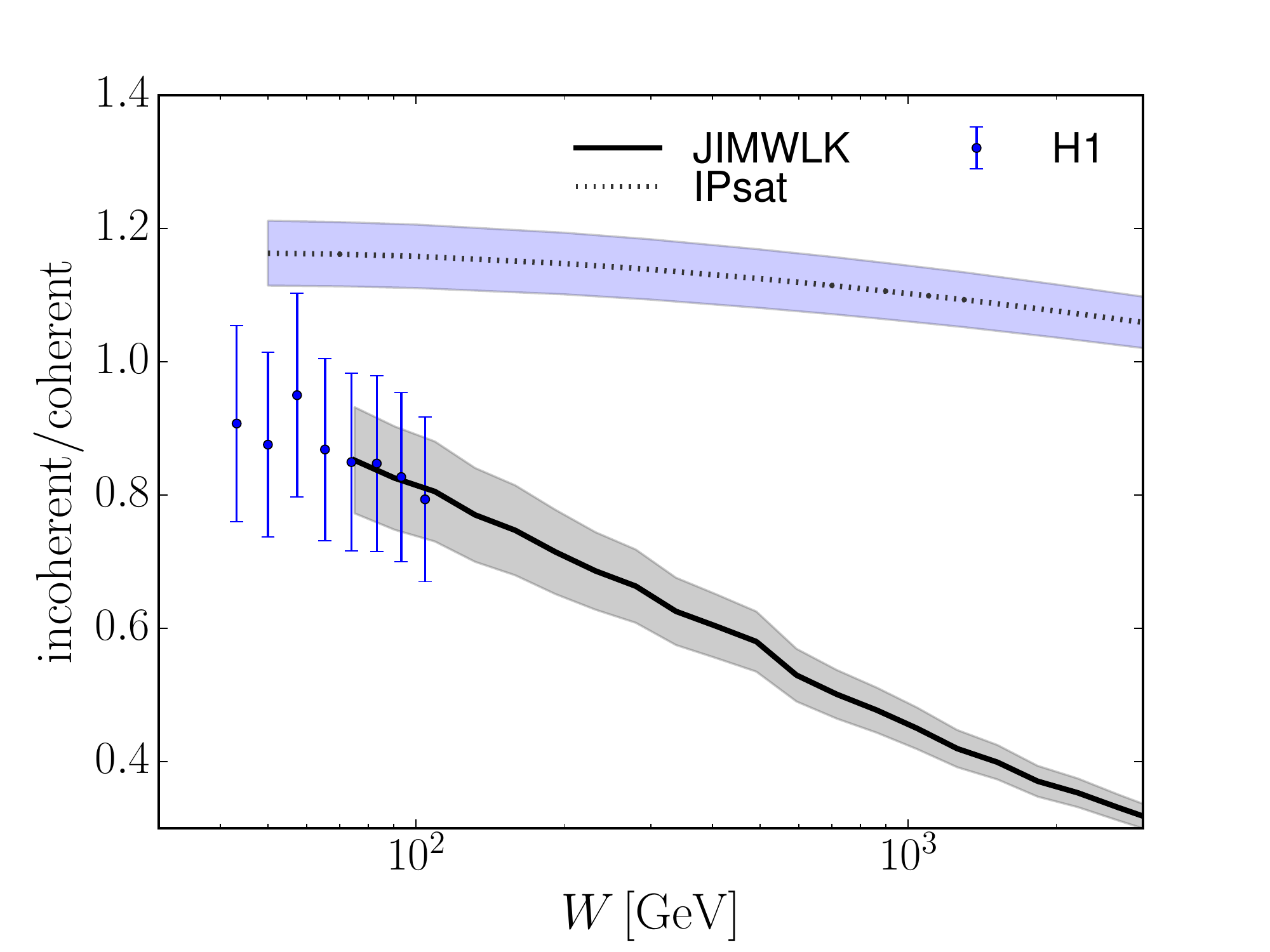} 
				\caption{Incoherent-to-coherent cross section ratio as a function of  $W$ from~\cite{Mantysaari:2018zdd_85}.}
		\label{fig:incohcohratio}
    \end{minipage}
\end{figure}
At high energies, the cross section for the exclusive production of the vector meson $V$ can be written as~\cite{Kowalski:2006hc}
\begin{equation}
\label{eq:diff_amp}
 \A^{\gamma^* p \to V p}_{T,L} = i\int \der^2 \rt \int \der^2 \bt \int \frac{\der z}{4\pi}  
  (\Psi^*\Psi_V)_{T,L}(Q^2, \rt,z) 
 e^{-i[\bt - (1-z)\rt]\cdot \boldsymbol{\Delta}}  2N(\rt,\bt,\xpom),
\end{equation}
where $\Psi^*\Psi_V$ is the overlap between the virtual photon wave function (computed from QED) and the vector meson wave function (modelled), and the dipole amplitude $N$ describes the scattering of a dipole with size $\rt$ off the target, with the impact parameter $\bt$. Here $T$ and $L$ refer to the photon polarization.

When the target remains intact, we talk about coherent diffraction and the cross section can be written as
$\frac{\der \sigma^{\gamma^* p \to V p}}{\der t} = \frac{1}{16\pi} \left| \langle  \A^{\gamma^* p \to V p}_{T,L} \rangle \right|^2.$
Here the avarage $\langle \rangle$ is taken over the possible configurations of the target, and consequently the cross section is sensitive on the average $\bt$ dependence of the dipole-target interaction. On the other hand, if one subtracts the coherent contribution from the total diffractive cross section, the incoherent contribution where the target is required to break up  remains:
$\frac{\der \sigma^{\gamma^*p \to Vp^*}}{\der t} = \frac{1}{16\pi} \left( \left \langle \left|   \A^{\gamma^* p \to V p}  \right|^2 \right \rangle -  \left | \left \langle   \A^{\gamma^* p \to V p}  \right \rangle \right|^2 \right).
$
As this cross section is a variance, it measures the amount of event-by-event fluctuations in the impact parameter dependence of $N$, and thus the amount of density fluctuations.

The Bjorken-$x$ evolution of the dipole amplitude $N$ can be computed perturbatively (see e.g.~ \cite{Balitsky:1995ub_121}) 
However, the initial condition for the evolution is non-perturbative and is usually obtained by fitting the HERA structure function data.
When considering exclusive processes additional complications arise as one has to describe also the geometry evolution, which is sensitive to infrared dynamics. For example in Ref.~\cite{Mantysaari:2018zdd_85} the geometry evolution from JIMWLK evolution equation was included in the analysis of the HERA data. An alternative approach is to parametrize the $x$ dependence and geometry as e.g. in the IPsat model.

Different experiments at HERA and at the LHC have measured $J/\Psi$ production in photon-proton interaction at different center-of-mass energies $W$. As this cross section is generally expected to be sensitive to saturation effects, in Ref.~\cite{Mantysaari:2018nng} the IPsat model parametrization for the dipole-proton interaction, and its linearized version (IPnonsat), were fitted to the HERA structure function data. In both cases, equally good description of the total cross section data was obtained. Then, the vector meson production as a function of $W$ was calculated and compared with the available data. The results are shown in Fig.~\ref{fig:jpsi_wdep}, where it is seen that the non-linear effects are small in the currently available energy range with proton targets.

The geometry evolution in Ref.~\cite{Mantysaari:2018zdd_85} makes it possible to study how the event-by-event fluctuations evolve towards small $x$. As discussed in Refs.~\cite{Mantysaari:2016ykx,Mantysaari:2016jaz}, the event-by-event fluctuating proton shape can be constrained by requiring a simultaneous description of the coherent and incoherent $J/\Psi$ production $t$ spectra from HERA. This analysis shows that at $x\approx 10^{-3}$ the fluctuations are significant. Then, one can perform a JIMWLK evolution down to small $x$. What is found in Ref.~\cite{Mantysaari:2018zdd_85} is that the evolution makes  protons smoother at long distance scales. This causes the incoherent cross section to grow more slowly than the coherent cross section (note that in the black disk limit the incoherent cross section is suppressed as it gets contributions only from the edges of the target). This observation is compatible with the HERA measurements as shown in Fig.~\ref{fig:incohcohratio}. In the IPsat model, on the other hand, there is no geometry evolution and the cross section ratio is constant. 

\section{From protons to heavy nuclei}
  \begin{figure}[tb]
    \centering
    \begin{minipage}{.48\textwidth}
        \centering
        \includegraphics[width=\textwidth]{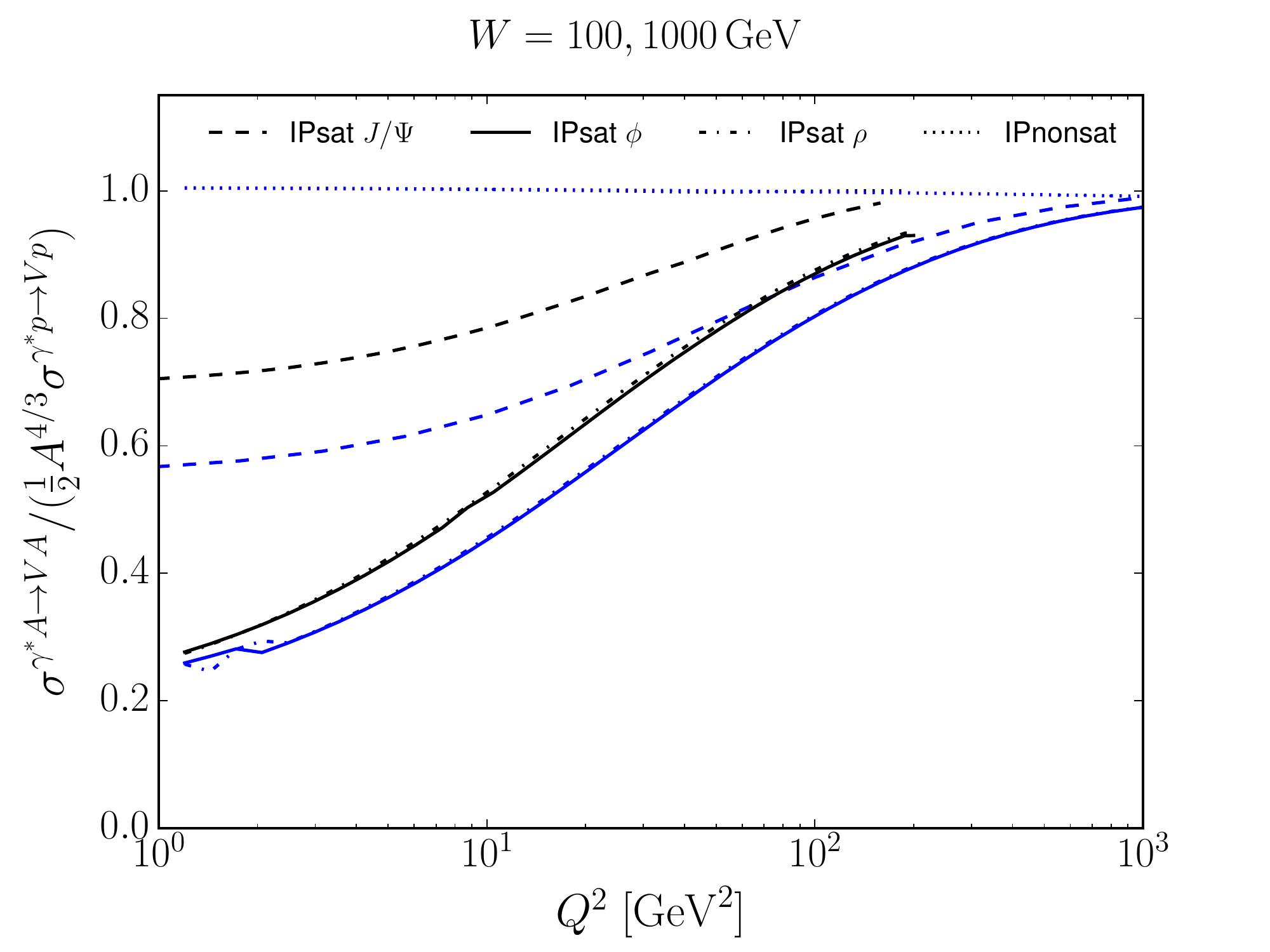} 
				\caption{Nuclear suppression factor for coherent vector meson production from~\cite{Mantysaari:2018nng}.}
		\label{fig:vm_suppression}
    \end{minipage}
    \quad
     \begin{minipage}{0.48\textwidth}
        \centering
      \includegraphics[width=\textwidth]{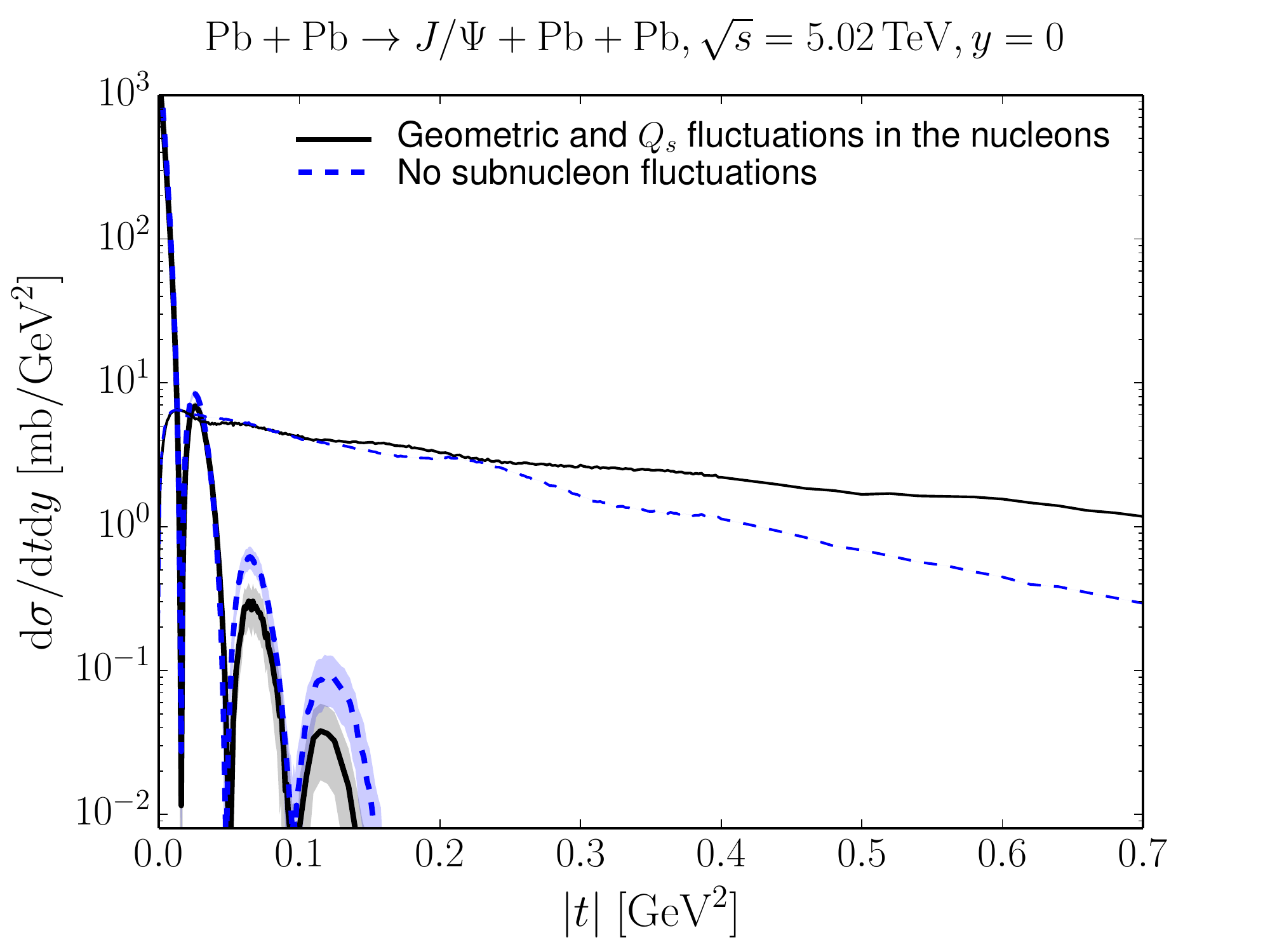} 
				\caption{$J/\Psi$ production cross section as a function of squared momentum transfer from~\cite{Mantysaari:2017dwh}.}
		\label{fig:nuke_t}
    \end{minipage}
   
\end{figure}

As gluon densities at given transverse point are enhanced by roughly $A^{1/3}$,  vector meson production off nuclei is a promising  process to look for saturation effects~\cite{Mantysaari:2017slo}.

The size $\rt$ of the interacting dipole is set by the vector meson mass $M_V$, such that $|\rt| \sim 1/M_V$. Very small dipoles have vanishing interaction probability due to the color transparency. Similarly, larger dipoles interact more strongly and as such are more sensitive to saturation effects. In Fig.~\ref{fig:vm_suppression} the cross section for exclusive vector meson production off gold nucleus is shown, divided by the same cross section off protons normalized such that in the absence of non-linear effects the ratio would be unity. This indeed is the case when the linear IPnonsat model is used to calculate dipole-nucleus interaction. 

In the IPsat parametrization which includes saturation, the suppression factor is found to depend strongly on the meson mass, the light $\rho$ and $\phi$ mesons being more heavily suppressed than $J/\Psi$. In Fig.~\ref{fig:vm_suppression} the suppression is shown as a function of photon virtuality $Q^2$, which also controls the dipole size. At large virtualities the dipoles are in general smaller and suppression vanishes. However, this transition is relatively slow, due to the significant contribution from the large dipoles to the transverse cross section even at large $Q^2$, see Ref.~\cite{Mantysaari:2018nng} for more details.

Finally in Fig.~\ref{fig:nuke_t} the coherent and incoherent $J/\Psi$ photoproduction cross section off lead are shown in the LHC kinematics. The results are shown in two cases: with proton shape and normalization ($Q_s$) fluctuations, and with nucleons having no substructure (dashed lines). As the substructure has little effect on average geometry, the coherent cross sections are compatible. The incoherent cross sections, on the other hand, are very different at $|t|\gtrsim 0.2\gev^2$, which corresponds to the distance scale of the hot spots in nucleons. 

\section{Outlook}
New nuclear DIS data at high energies, both from the ultra peripheral heavy ion collisions at the LHC and from the future EIC, will provide us a new window for the studies of non-linear dynamics in the dense QCD matter. 
In addition to saturation effects, it will also be possible to study the evolution of the fluctuating transverse geometry, which is fundamentally interesting and provides crucial input for the modelling of Quark Gluon Plasma production.
\vspace{-1em}
\subsubsection*{Acknowledgments}
This work was supported by the Academy of Finland, project 314764, and by the European Research Council, Grant ERC-2015-CoG-681707.

%


\newpage 
%
\renewcommand*{\FigPath}{./WeekVI/01_Balitsky/}

\renewcommand{\be}{\begin{eqnarray}}
\renewcommand{\ee}{\end{eqnarray}}

\newcommand{\dhd}{{\textstyle d}
\lower.03ex\hbox{\kern-0.38em$^{\scriptstyle-}$}\kern-0.05em{}}
\newcommand{\dbar}{{\textstyle \delta}
\lower.03ex\hbox{\kern-0.38em$^{\scriptstyle-}$}\kern-0.05em{}}
\renewcommand{\half}{{1\over 2}}
\newcommand{\bu}{{\bullet}}
\newcommand{\bara}{{\bar a}}
\newcommand{\barb}{{\bar b}}
\newcommand{\barc}{{\bar c}}
\newcommand{\bard}{{\bar d}}
\newcommand{\bare}{{\bar e}}
\newcommand{\barf}{{\bar f}}
\newcommand{\barg}{{\bar g}}
\newcommand{\barh}{{\bar h}}
\newcommand{\barj}{{\bar j}}
\newcommand{\bark}{{\bar k}}
\newcommand{\barn}{{\bar n}}
\newcommand{\barp}{{\bar p}}
\newcommand{\barr}{{\bar r}}
\newcommand{\bars}{{\bar s}}
\newcommand{\bart}{{\bar t}}
\newcommand{\baru}{{\bar u}}
\newcommand{\barv}{{\bar v}}
\newcommand{\barw}{{\bar w}}
\newcommand{\barx}{{\bar x}}
\newcommand{\bary}{{\bar y}} 
\newcommand{\barz}{{\bar z}}

\newcommand{\balfa}{{\bar \alpha}}
\newcommand{\barDe}{{\bar \Delta}}
\newcommand{\bartial}{{\bar \partial}}
\newcommand{\blambda}{{\bar \lambda}}
\newcommand{\bfi}{{\bar \phi}}
\newcommand{\bsi}{{\bar \psi}}
\newcommand{\Bsi}{{\bar \Psi}}
\newcommand{\bsigma}{{\bar \sigma}}
\newcommand{\bhi}{{\bar \chi}}
\newcommand{\bxi}{{\bar \xi}}
\newcommand{\bzeta}{{\bar \zeta}}
\newcommand{\barA}{{\bar A}}
\newcommand{\barB}{{\bar B}}
\newcommand{\barC}{{\bar C}}
\newcommand{\barD}{{\bar D}}
\newcommand{\barG}{{\bar G}} 
\newcommand{\barF}{{\bar F}}   
\newcommand{\barK}{{\bar K}} 
\newcommand{\barP}{{\bar P}} 
\newcommand{\barS}{{\bar S}}
\newcommand{\barU}{{\bar U}}
\newcommand{\barX}{{\bar X}} 
\newcommand{\barY}{{\bar Y}}  

\newcommand{\Bxi}{{\bar \Xi}}
\newcommand{\bLa}{{\bar \Lambda}}
\newcommand{\Bomega}{{\bar \aleph^\dagger}}
\newcommand{\barUps}{{\bar \Upsilon}} 

\newcommand{\cala}{{\cal A}}
\newcommand{\calc}{{\cal C}}
\newcommand{\cald}{{\cal D}}  
\newcommand{\calf}{{\cal F}}
\newcommand{\calg}{{\cal G}} 
\newcommand{\calj}{{\cal J}} 
\newcommand{\kal}{{\cal L}} 
\newcommand{\calh}{{\cal H}} 
\newcommand{\caln}{{\cal N}}  
\newcommand{\calo}{{\cal O}}    
\newcommand{\calr}{{\cal R}}  
\newcommand{\calp}{{\cal P}}  
\newcommand{\calq}{{\cal Q}}  
\newcommand{\cals}{{\cal S}} 
\newcommand{\calt}{{\cal T}} 
\newcommand{\calu}{{\cal U}} 
\newcommand{\calv}{{\cal V}} 
\newcommand{\calw}{{\cal W}} 
\newcommand{\calz}{{\cal Z}} 
\newcommand{\calx}{{\cal X}} 
\newcommand{\dotx}{{\dot x}} 
\newcommand{\dodin}{{\dot 1}} 
\newcommand{\ddva}{{\dot 2}} 
\newcommand{\dalpha}{{\dot \alpha}} 
\newcommand{\dbeta}{{\dot \beta}} 
\newcommand{\dgamma}{{\dot \gamma}}
 
\newcommand{\halo}{{\hat{\cal O}}}
\newcommand{\hate}{{\hat e}} 
\newcommand{\hatk}{{\hat k}} 
\newcommand{\hatp}{{\hat p}} 
\newcommand{\hatx}{{\hat x}} 
\newcommand{\haty}{{\hat y}} 
\newcommand{\hatz}{{\hat z}} 
\newcommand{\hatA}{{\hat A}}
\newcommand{\hatB}{{\hat B}}
\newcommand{\hatC}{{\hat C}}
\newcommand{\hatD}{{\hat D}}
\newcommand{\hatF}{{\hat F}}
\newcommand{\hatK}{{\hat K}}
\newcommand{\hatM}{{\hat M}}
\newcommand{\hatP}{{\hat P}} 
\newcommand{\hatX}{{\hat X}} 
\newcommand{\hatY}{{\hat Y}}
\newcommand{\hatZ}{{\hat Z}} 
\newcommand{\hatH}{{\hat H}} 
\newcommand{\hatO}{{\hat O}} 
\newcommand{\hartial}{{\hat \partial}} 
\newcommand{\hsi}{{\hat \psi}} 
\newcommand{\hbsi}{\hat {\bar\psi}} 
\newcommand{\halef}{{\hat \aleph}} 
\newcommand{\haleph}{{\hat \aleph}}

\newcommand{\hamma}{{\breve \gamma}}

\newcommand{\veck}{{\vec k}}
\newcommand{\vecp}{{\vec p}}
\newcommand{\veq}{{\vec q}}
\newcommand{\vecx}{{\vec x}}
 \newcommand{\vecy}{{\vec y}} 
 \newcommand{\vecz}{{\vec z}} 
\newcommand{\vehp}{{\hat {\vec p}}} 

\newcommand{\notA}{{\not\! A}}
\newcommand{\notp}{{\not\! p}}

\newcommand{\tile}{{\tilde e}}
\newcommand{\tilf}{{\tilde f}}
\newcommand{\tilh}{{\tilde h}}
\newcommand{\tilj}{{\tilde j}} 
\newcommand{\tilk}{{\tilde k}} 
\newcommand{\tiln}{{\tilde n}} 
\newcommand{\tilp}{{\tilde p}}
\newcommand{\tils}{{\tilde s}}
\newcommand{\tildet}{{\tilde T}}
\newcommand{\tildelta}{{\tilde \Delta}}
\newcommand{\tilu}{{\tilde u}} 
\newcommand{\tilx}{{\tilde x}} 
\newcommand{\tilz}{{\tilde z}} 
\newcommand{\tilA}{{\tilde A}}
\newcommand{\tilB}{{\tilde B}}
\newcommand{\tilC}{{\tilde C}}
\newcommand{\tilD}{{\tilde D}}
\newcommand{\tilF}{{\tilde F}}
\newcommand{\tilG}{{\tilde G}}
\newcommand{\tilJ}{{\tilde J}}
\newcommand{\tiL}{{\tilde L}}
\newcommand{\tilP}{{\tilde P}}
\newcommand{\tilS}{{\tilde S}}
\newcommand{\tilT}{{\tilde T}}
\newcommand{\tilU}{{\tilde U}}
\newcommand{\tilV}{{\tilde V}}
\newcommand{\tilW}{{\tilde W}}
\newcommand{\tilX}{{\tilde X}}
\newcommand{\tikal}{\tilde{\cal L}} 
\newcommand{\tilcaf}{\tilde{\cal F}} 
\newcommand{\ticalf}{\tilde{\cal F}} 
\newcommand{\ticala}{\tilde {\cal A}} 
\newcommand{\ticalc}{\tilde {\cal C}} 
\newcommand{\ticalo}{\tilde {\cal O}} 
\newcommand{\ticalg}{\tilde {\cal G}} 
\newcommand{\ticalp}{\tilde {\cal P}} 
\newcommand{\ticalq}{\tilde {\cal Q}} 
\newcommand{\tibeta}{\tilde {\beta}} 
\newcommand{\tamma}{\tilde {\gamma}} 
\newcommand{\tilga}{\tilde {\Gamma}} 
\newcommand{\tilO}{\tilde {\Omega}} 
\newcommand{\tiLa}{\tilde {\Lambda}} 
\newcommand{\tFi}{{\tilde \Phi}}

\newcommand{\talfa}{\tilde {\alpha}} 
\newcommand{\tfi}{\tilde {\phi}} 
\newcommand{\tigma}{\tilde {\sigma}} 
\newcommand{\tsi}{\tilde {\psi}} 
\newcommand{\tipsi}{\tilde {\psi}} 
\newcommand{\tups}{\tilde {\Upsilon}} 
\newcommand{\talef}{\tilde {\aleph}} 
\newcommand{\taleph}{\tilde {\aleph}} 
\newcommand{\btups}{\bar {\tilde\Upsilon}} 

\newcommand{\breal}{\breve {\alpha}}
\newcommand{\brebe}{\breve {\beta}}
\newcommand{\bre}{\breve {e}} 
\newcommand{\bref}{\breve {f}} 
\newcommand{\breh}{\breve {h}} 
\newcommand{\brek}{\breve {k}} 
\newcommand{\bres}{\breve {s}} 
\newcommand{\brex}{\breve {x}} 
\newcommand{\bry}{\breve {y}} 
\newcommand{\brez}{\breve {z}} 

\newcommand{\che}{{\check e}}
\newcommand{\chek}{{\check k}}
\newcommand{\ches}{{\check s}}
\newcommand{\chex}{\check {x}} 
\newcommand{\chez}{{\check z}}

\newcommand{\matA}{\mathbb{A}} 
\newcommand{\matB}{\mathbb{B}} 
\newcommand{\matC}{\mathbb{C}} 
\newcommand{\matD}{\mathbb{D}} 
\newcommand{\matF}{\mathbb{F}} 
\newcommand{\matG}{\mathbb{G}} 
\newcommand{\matP}{\mathbb{P}} 
\newcommand{\mato}{\mathbb{O}} 
\newcommand{\matY}{\mathbb{Y}} 
\newcommand{\barmatY}{\bar{\mathbb{Y}}} 

\newcommand{\scra}{\mathscr{A}}
\newcommand{\scrd}{\mathscr{D}}
\newcommand{\scrf}{\mathscr{F}}
\newcommand{\scrg}{\mathscr{G}}
\newcommand{\scrp}{\mathscr{P}}

\newcommand{\fra}{\mathfrak{A}}
\newcommand{\frad}{\mathfrak{D}}
\newcommand{\fraf}{\mathfrak{F}}
\newcommand{\frag}{\mathfrak{G}}
\newcommand{\frap}{\mathfrak{P}}

\newcommand{\bfA}{{\bf A}}
\newcommand{\bfD}{{\bf D}}
\newcommand{\bfG}{{\bf G}}
\renewcommand{\bfP}{{\bf P}}
\newcommand{\bfPsi}{{\bf \Psi}}


\newcommand{\IM}{{\rm Im}\,}
\newcommand{\card}{\#}
\renewcommand{\la}[1]{\label{#1}}
\renewcommand{\eq}[1]{(\ref{#1})}
\newcommand{\figref}[1]{Fig. \ref{#1}}
\newcommand{\abs}[1]{\left|#1\right|}
\newcommand{\comD}[1]{{\color{red}#1\color{black}}}

\makeatletter
     \@ifundefined{usebibtex}{\newcommand{\ifbibtexelse}[2]{#2}} {\newcommand{\ifbibtexelse}[2]{#1}}
\makeatother


\newcommand{\footnoteab}[2]{\ifbibtexelse{%
\footnotetext{#1}%
\footnotetext{#2}%
\cite{Note1,Note2}%
}{%
\newcommand{\textfootnotea}{#1}%
\newcommand{\textfootnoteab}{#2}%
\cite{thefootnotea,thefootnoteab}}}

\def\e{\epsilon}
     \def\bT{{\bf T}}
    \def\bQ{{\bf Q}}
    \def\wT{{\mathbb{T}}}
    \def\wQ{{\mathbb{Q}}}
    \def\ttQ{{\bar Q}}
    \def\tQ{{\tilde \bP}}
        \def\bP{{\bf P}}
    \def\CF{{\cal F}}
    \def\cC{\CF}
     \def\Tr{\text{Tr}}
     \def\l{\lambda}
\def\hbZ{{\widehat{ Z}}}
\def\bZ{{\resizebox{0.28cm}{0.33cm}{$\hspace{0.03cm}\check {\hspace{-0.03cm}\resizebox{0.14cm}{0.18cm}{$Z$}}$}}}
\newcommand{\rb}{\right)}
\newcommand{\lb}{\left(}

\wstoc{Conformal properties of rapidity evolution of TMDs}{Ian Balitsky, Giovanni A. Chirilli}
\title{Conformal properties of rapidity evolution of TMDs}

\author{Ian Balitsky$^*$}

\address{Department of Physics, Old Dominion Univ.,
Newport News, VA23529\\
and
Theory Group, JLab, 12000 Jefferson Ave, Newport News, VA 23606\\
$^*$E-mail: balitsky@jlab.org}

\author{Giovanni A. Chirilli}

\address{Institut f\"ur Theoretische Physik, Universit\"at Regensburg, D-93040 Regensburg, Germany\\}

\begin{abstract}
We discuss  conformal properties of TMD operators and present the 
result of the conformal rapidity evolution of TMD operators in the Sudakov region.
\end{abstract}

\index{author}{Balitsky, I.}
\index{author}{Chirilli, G.}


In recent years, the  transverse-momentum dependent parton distributions (TMDs) 
\cite{Collins:1981uw_208,Collins:1984kg_202,Ji:2004wu_196,GarciaEchevarria:2011rb_190}  have been widely used in the analysis of 
 processes like semi-inclusive deep inelastic scattering 
or particle production in hadron-hadron collisions (for a review, see Ref. \cite{Collins:2011zzd_184}). 

The TMDs are defined as matrix elements of quark or gluon operators with attached light-like gauge links
(Wilson lines) going to either $+\infty$ or $-\infty$ depending on the process under consideration. 
It is well known that these TMD operators exhibit rapidity divergencies due to infinite light-like gauge links 
and the corresponding rapidity/UV divergences  should be regularized. There are two schemes on the market: 
the most popular is based on CSS  \cite{Collins:1984kg_202} or SCET \cite{Rothstein:2016bsq} formalism and the second one is 
adopted from the small-$x$ physics \cite{Lipatov:1996ts,Kovchegov:2012mbw_178}.
The obtained evolution equations differ even at the leading-order level and
 need to be reconciled, especially in view of the future EIC accelerator which will probe the 
TMDs at values of Bjorken $x$ between small-$x$ and $x\sim 1$ regions. 

In our opinion, a good starting point is to obtain conformal leading-order evolution equations. 
It is well known that  at the leading order pQCD is conformally invariant so there is a hope to get 
any evolution equation without explicit running coupling from conformal considerations. 
In our case, since TMD operators are defined with attached light-like Wilson lines,
formally they will transform covariantly under the subgroup of full conformal group 
which preserves this light-like direction. 
However, as we mentioned, the TMD operators contain rapidity divergencies which need to be regularized. 
At present, there is no rapidity cutoff which preserves conformal invariance so the best one can do is to find
the cutoff which is conformal at the leading order in perturbation theory. In higher orders, one should not 
expect conformal invariance since it is broken by running of QCD coupling. However, if one considers corresponding 
correlation functions in  ${\cal N}=4$ SYM, one should expect conformal invariance. 
After that, the results obtained in ${\cal N}=4$ SYM theory can be used as a starting point of QCD calculation.
Thus, the idea is to find TMD operator conformal in ${\cal N}=4$ SYM and use it in QCD.

\section{Conformal invariance of TMD operators}

For definiteness, we will talk first about gluon operators with light-like Wilson lines stretching to $-\infty$ in ``+'' direction.
The gluon TMD (unintegrated gluon distribution) is defined as \cite{Mulders:2000sh_172}
\begin{eqnarray}
&&\hspace{-0mm}
\cald(x_B,k_\perp,\eta)~=~\!\int\!d^2z_\perp~e^{i(k,z)_\perp}\cald(x_B,z_\perp,\eta),
\label{TMDg}\\
&&\hspace{-0mm}
g^2\cald(x_B,z_\perp,\eta)~\stackrel{z_-=0}{=}~{-x_B^{-1}\over 2\pi p^-}\!\int\! dz^+ ~e^{-ix_Bp^-z^+}
\langle P|\calf^a_\xi(z)
[z-\infty n,-\infty n]^{ab}\calf^{b\xi}(0)|P\rangle
\nonumber
\end{eqnarray}
where $|P\rangle$ is an unpolarized target with momentum $p\simeq p^-$ (typically proton) and $n=({1\over\sqrt{2}},0,0,{1\over\sqrt{2}})$ is a light-like vector 
in ``+'' direction.
Hereafter we use the notation
\begin{equation}
\hspace{-0mm}
\calf^{\xi,a}(z_\perp,z^+)~\equiv~gF^{- \xi,m}(z)[z,z-\infty n]^{ma}\Big|_{z^-=0}
\label{kalf}
\end{equation}
where $[x,y]$ denotes straight-line gauge link connecting points $x$ and $y$:
\begin{equation}
~[x,y]~\equiv~{\rm P}e^{ig\int\! du~(x-y)^\mu A_\mu(ux+(1-u)y)}
\label{defu}
\end{equation}
To simplify  one-loop evolution
we multiplied $F_{\mu\nu}$ by coupling constant. Since the $gA_\mu$ is renorm-invariant we do not need to consider self-energy diagrams (in 
the background-Feynman gauge). Note that $z^-=0$ is fixed  by
the original factorization formula for particle production  \cite{Collins:2011zzd_184} (see also the discussion in 
Ref. \cite{Balitsky:2017flc_166,Balitsky:2017gis_160}).

The algebra of full conformal group $SO(2,4)$  consists of four operators $P^\mu$, six $M^{\mu\nu}$, four 
special conformal generators $K^\mu$, and dilatation operator $D$.
It is easy to check that in the leading order the following 11 operators act on gluon TMDs covariantly
\begin{equation}
P^i,P^-,M^{12},M^{-i},D,K^i,K^-,M^{-+}
\label{generators}
\end{equation}
while the action of operators $P^+,M^{+i}$, and $K^+$ do not preserve the  form of the operator (\ref{kalf}).
The corresponding group consists of transformations which leave the hyperplane $z^-=0$
and vector $n$ 
invariant. Those include 
shifts in transverse and $``+''$ directions, rotations in the 
transverse plane, Lorentz rotations/boosts created by $M^{-i}$, dilatations, and special conformal
transformations
\begin{equation}
z'_\mu~=~{z_\mu -a_\mu z^2\over 1-2a\cdot z+a^2z^2}
\label{speconf}
\end{equation}
with $a=(a^+,0,a_\perp)$.

As we noted, infinite Wilson lines in the definition (\ref{kalf}) of TMD operators make 
them divergent. As we discussed above, it is very advantageous to have a cutoff 
of these divergencies compatible with approximate conformal invariance of tree-level QCD.
The evolution equation with such cutoff should be invariant with respect to transformations 
described above.

In the next Section we demonstrate that the ``small-x'' rapidity cutoff enables us to
get a conformally invariant evolution of TMD in the so-called Sudakov region. 

\section{TMD factorization in the Sudakov region}

The rapidity evolution of TMD operator (\ref{TMDg})
is very different in the region of large and small longitudinal separations $z^+$. 
The evolution at small $z^+$ is linear and double-logarithmic
while  at large $z^+$ the evolution become non-linear
due to the production of color dipoles typical for small-$x$ evolution. 
It is convenient to consider as a starting point the simple case of
TMD evolution in the so-called Sudakov region corresponding to
small longitudinal distances. 

First,  let us specify what we  call a Sudakov region. 
A typical  
factorization formula for the differential cross section  of particle production 
in hadron-hadron collision is \cite{Collins:2011zzd_184, Collins:2014jpa_154}
\begin{eqnarray}
&&\hspace{-2mm}
{d\sigma\over  d\eta d^2q_\perp}~=~
\sum_f\!\int\! d^2b_\perp e^{i(q,b)_\perp}
\cald_{f/A}(x_A,b_\perp,\eta)
\nonumber\\
&&\hspace{-2mm}
\times~
\cald_{f/B}(x_B,b_\perp,\eta)\sigma(ff\rightarrow H)~+~...
\label{TMDf}
\end{eqnarray}
where $\eta=\half\ln{q^+\over q^-}$ is the rapidity, $\cald_{f/h}(x,z_\perp,\eta)$ is the 
TMD density of  a parton $f$  in hadron $h$, and $\sigma(ff\rightarrow H)$ is the cross section of production of particle $H$ 
of invariant mass $m_H^2=q^2\equiv Q^2$ in the scattering of two partons.  
(One can keep in mind Higgs production in the approximation of point-like gluon-gluon-Higgs vertex).  
The Sudakov region is defined by  $Q\gg q_\perp\gg 1$GeV since at such kinematics 
there is a double-log evolution for transverse momenta between $Q$ and $q_\perp$. In the coordinate space,
TMD factorization  (\ref{TMDf})  looks like
\begin{eqnarray}
&&\hspace{-0mm}
\langle p_A,p_B| g^2F^a_{\mu\nu}F^{a\mu\nu}(z_1)  g^2F^b_{\lambda\rho}F^{b\lambda\rho}(z_2)|p_A,p_B\rangle
\label{facoord}\\
&&\hspace{-1mm}
=~{1\over N_c^2-1}\langle p_A|\ticalo_{ij}(z_1^-,z_{1_\perp};z_2^-,z_{2_\perp}) |p_A\rangle^{\sigma_A}
 \langle p_B|\calo^{ij}(z_1^+,z_{1_\perp};z_2^+,z_{2_\perp}) |p_B\rangle^{\sigma_B}~+...
\nonumber
\end{eqnarray}
where
\begin{eqnarray}
&&\hspace{-2mm}
\calo_{ij}(z_1^+,z_{1_\perp};z_2^+,z_{2_\perp})
=~\calf^a_i(z_1)
[z_1-\infty n,z_2-\infty n]^{ab}\calf^b_j(z_2)\Big|_{z_1^-=z_2^-=0}\,,
\label{kalo}
\\
&&\hspace{-2mm}
\ticalo_{ij}(z_1^-,z_{1_\perp};z_2^-,z_{2_\perp})
=~
\calf^a_i(z_1)
[z_1-\infty n',z_2-\infty n']^{ab}\calf^b_j(z_2)\Big|_{z_1^+=z_2^+=0}\,,
\nonumber\\
&&\hspace{-2mm}
\calf^{i,a}(z_\perp,z^-)~\equiv~gF^{+i,m}(z)[z,z-\infty n']^{ma}\Big|_{z^+=0}\,.
\label{tikalo}
\end{eqnarray}
Here 
$p_A=\sqrt{s\over 2}n+{p_A^2\over\sqrt{2s}}n',~p_B=\sqrt{2\over s}n'+{p_B^2\over\sqrt{2s}}n$
and $n'=\big({1\over\sqrt{2}},0,0,-{1\over\sqrt{2}}\big)$. Our metric is $x^2=2x^+x^- - x_\perp^2$.

As we mentioned, TMD operators exhibit rapidity divergencies due to infinite light-like gauge links.
The ``small-$x$ style'' rapidity cutoff for longitudinal divergencies is imposed
as the upper limit of $k^+$ components of gluons emitted from the 
Wilson lines. As we will see below, to get the conformal invariance of the leading-order evolution
we need to impose the cutoff of $k^+$ components of gluons correlated with transverse size of TMD in the following way:
\begin{eqnarray}
&&\hspace{-1mm} 
\calf^{i,a}(z_\perp,z^+)^\sigma
\equiv gF^{- i,m}(z)\big[{\rm P}e^{ig\!\int_{-\infty}^{z^+}\!\! dz^+ A^{-,\sigma}(up_1+x_\perp)}\big]^{ma},
\nonumber\\
&&\hspace{-0mm} 
A^\sigma_\mu(x)~=~\int\!{d^4 k\over 16\pi^4} ~\theta\Big({\sigma\sqrt{2}\over z_{12_\perp}}-|k^+|\Big)e^{-ik\cdot x} A_\mu(k)
\label{cutoff}
\end{eqnarray}
Similarly, the operator $\ticalo$ in Eq. (\ref{tikalo}) is defined with 
with the rapidity cutoff for $\beta$ integration imposed as $\theta\big({\tigma\sqrt{2}\over z_{12_\perp}}-|k^-|\big)$.

The Sudakov region $Q^2\gg q_\perp^2$ in the coordinate space corresponds to
\begin{equation}
z_{12_\parallel}^2~\equiv~2z_{12}^-z_{12}^+~\ll~z_{12_\perp}^2
\label{sudkoord}
\end{equation}
where $z_{12}\equiv z_1-z_2$.
In the leading log approximation, 
the upper cutoff for $k^+$ integration in the target matrix element in Eq. (\ref{facoord}) is 
$\sigma_B={1\over \sqrt{2}}{z_{12_\perp}\over z^-_{12}}$ and  similarly the $\beta$-integration 
cutoff in projectile matrix element is $\sigma_A={1\over \sqrt{2}}{z_{12_\perp}\over z^+_{12}}$
\footnote{Hereafter we use the simplified notation $z_{12_\perp}\equiv |z_{12_\perp}|$. }.

In the next Section we demonstrate that rapidity cutoff (\ref{cutoff}) enables us to
get a conformally invariant evolution of TMD in the Sudakov region (\ref{sudkoord}). 

\section{One-loop evolution of TMDs}
\subsection{Evolution of gluon TMD operators in the Sudakov region}
In this Section we derive the evolution of gluon TMD operator (\ref{kalo}) with respect to cutoff $\sigma$
in the leading log approximation.
%
\begin{figure}[htb]
\begin{center}
\includegraphics[width=55mm]{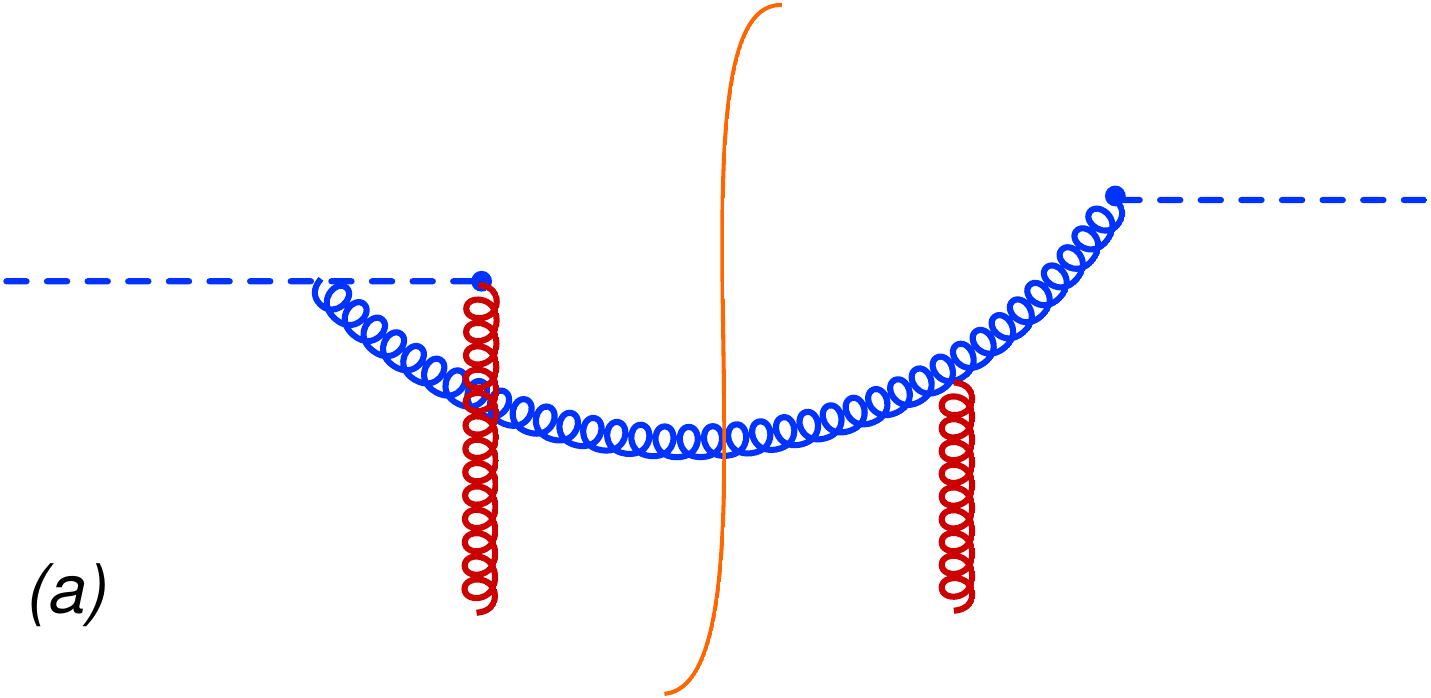}
\hspace{12mm}
\includegraphics[width=55mm]{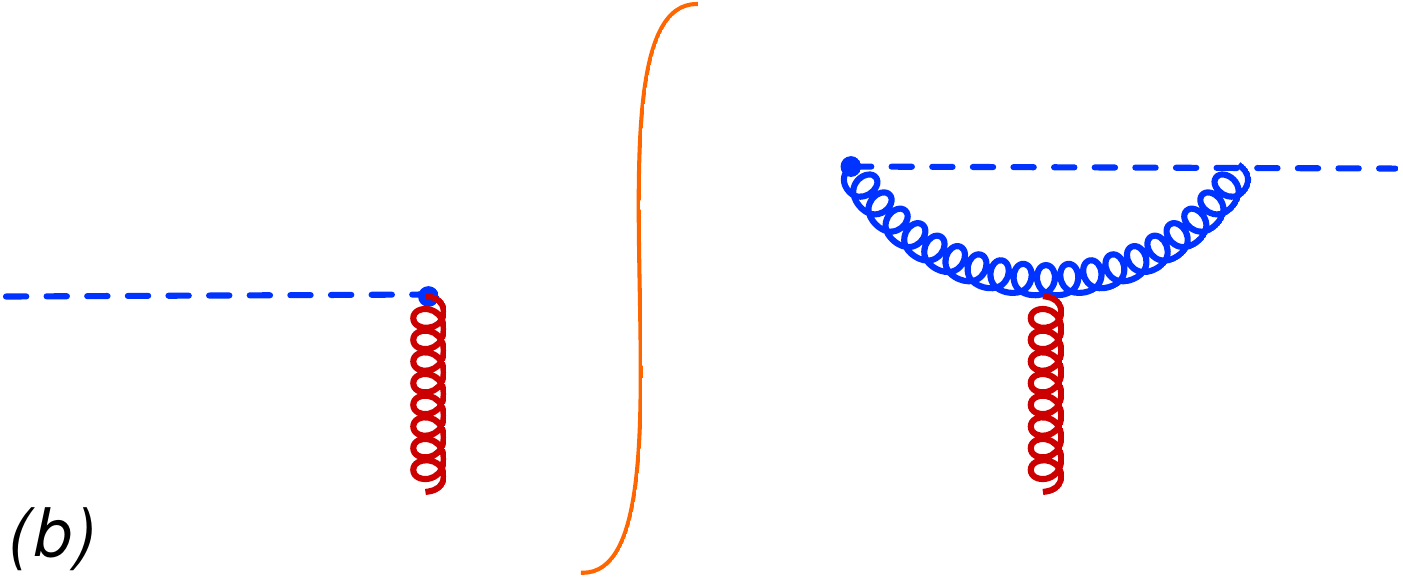}
\end{center}
\caption{Typical diagrams for production (a) and virtual (b) contributions to the evolution kernel. The dashed lines denote gauge links.\label{fig:2}}
\end{figure}
%
As usual, to get an evolution equation we integrate over momenta ${\sigma_2\sqrt{2}\over z_{12_\perp}}>k^+>{\sigma_1\sqrt{2}\over z_{12_\perp}}$. 
To this end, we calculate diagrams shown in Fig. 1 in the background field of gluons with $k^+<{\sigma_1\sqrt{2}\over z_{12_\perp}}$.
The calculation is easily done by method developed in Refs. \cite{Balitsky:2015qba,Balitsky:2016dgz} 
and the result is
\begin{equation}
\calo^{\sigma_2}(z_1^+,z_2^+)
~=~{\alpha_sN_c\over 2\pi}\!\int\limits_{\sigma_1\sqrt{2}\over z_{12_\perp}}^{\sigma_2\sqrt{2}\over z_{12_\perp}}\!
{dk^+\over k^+}~K\calo^{\sigma_1}(z_1^+,z_2^+)
\label{sudeqn1}
\end{equation}
where the kernel $K$ is given by
\begin{eqnarray}
&&\hspace{-1mm}
K\calo(z_1^+,z_2^+)~
\label{kernel}\\
&&\hspace{-1mm}
=~\calo(z_1^+,z_2^+)
\!\int_{-\infty}^{z_1^+}\! {dz^+\over z_2^+-z^+}e^{-i{z_{12_\perp}\sigma\over \sqrt{2}(z_2-z)^+}}
+~\calo(z_1^+,z_2^+)
\!\int_{-\infty}^{z_2^+}\! {dz^+\over z_1^+-z^+}e^{i{z_{12_\perp}\sigma\over \sqrt{2}(z_1-z)^+}}
\nonumber\\
&&\hspace{-1mm}
~-\!\int^{z_1^+}_{-\infty} \! dz^+ 
 {\calo(z_1^+,z_2^+)-\calo(z^+,z_2^+)\over z_1^+ -z^+}
~-\!\int^{z_2^+}_{-\infty} \! dz^+ 
 {\calo(z_1^+,z_2^+)-\calo(z_1^+,z^+)\over z_2^+-z^+ }
 \nonumber
\end{eqnarray}
where we suppress arguments $z_{1_\perp}$ and $z_{2_\perp}$ since they do not change 
during the evolution in the Sudakov regime.
The first two terms in the kernel $K$ come from the ``production'' diagram in Fig. 1a while the last two terms from 
``virtual'' diagram in Fig. 1b. 
The result (\ref{kernel}) can be also obtained from Ref. \cite{Balitsky:2016dgz}  by Fourier transformation of Eq. (5.9) 
with the help of Eqs. (3.12) and (3.30) therein. The approximations  for diagrams in Fig. 1 leading to Eq. (\ref{kernel}) are valid as long as 
\begin{equation}
k^+\gg {z^+_{12}\over z_{12_\perp}^2}
\label{sudapprox}
\end{equation}
which gives the region of applicability of Sudakov-type evolution.

Evolution equation (\ref{sudeqn1}) can be easily integrated using Fourier transformation. 
Since 
\begin{eqnarray}
&&\hspace{-0mm}
Ke^{-ik^- z_1^+ + ik '^-z_2^+}~
=~\Big[
-2\ln \sigma z_{12_\perp}-\ln(ik^-)
\label{Kexp}\\
&&\hspace{-0mm}
-~\ln(-ik'^-)+\ln 2-4\gamma_E~+~O\big({z^+_{12}\over z_{12_\perp}\sigma}\big)\Big]e^{-ik^- z_1^+ +ik '^-z_2^+}
\nonumber
\end{eqnarray}
one easily obtains
\begin{eqnarray}
&&\hspace{-1mm}
\calo^{\sigma_2}(z_1^+,z_2^+)~=~
e^{-2\balfa_s\ln{\sigma_2\over\sigma_1}[\ln\sigma_1\sigma_2+4\gamma_E-\ln 2\big]}
\!\int\! dz'^+_1 dz'^+_2 ~\calo^{\sigma_1}(z'^+_1,z'^+_2)~z_{12_\perp}^{-2\balfa_s\ln{\sigma_2\over\sigma_1}}
\nonumber\\
&&\hspace{-1mm}
\times~{1\over 4\pi^2}\bigg[{i\Gamma\big(1-2\balfa_s\ln{\sigma_2\over\sigma_1}\big)
\over (z_1^+ - z'^+_1 + i\epsilon)^{1-2\balfa_s\ln{\sigma_2\over\sigma_1}}}
+c.c.\bigg]
\bigg[{i\Gamma\big(1-2\balfa_s\ln{\sigma_2\over\sigma_1}\big)
\over (z_2^+ - z'^+_2 + i\epsilon)^{1-2\balfa_s\ln{\sigma_2\over\sigma_1}}}
+c.c.\bigg]
\label{result}
\end{eqnarray}
where we introduced notation $\balfa_s\equiv{\alpha_sN_c\over 4\pi}$.
It should be mentioned that the factor $4\gamma_E$ is ``scheme-dependent'': if one introduces to $\alpha$-integrals 
smooth cutoff $e^{-\alpha/a}$ instead of rigid cutoff $\theta(a>\alpha)$, the value $4\gamma_E$ changes
to $2\gamma_E$.

It is easy to see that the r.h.s. of Eq. (\ref{result}) transforms covariantly under all transformations 
(\ref{generators}) except Lorentz boost generated by $M^{+-}$. The reason is that the Lorentz boost
in $z$ direction changes cutoffs for the evolution. To understand 
that, note that  Eq. (\ref{Kexp}) is valid until $\sigma>{z^+_{12}\over z_{12_\perp}^2}$ 
so the  linear evolution (\ref{result}) is applicable in the region between
 \begin{equation}
\sigma_2=\sigma_B= {z_{12_\perp}\over z^-_{12}\sqrt{2}}~~~~~{\rm and}~~~~~\sigma_1={z^+_{12}\sqrt{2}\over z_{12_\perp}} 
 \label{fla1a}
 \end{equation}
 From Eq. (\ref{result}) it is easy to see that
 Lorentz boost  $z^+\!\rightarrow \!\lambda z^+, ~z^-\!\rightarrow\! {1\over\lambda} z^-$ changes the value of target
matrix element $\langle p_A|\calo|p_B\rangle$ by $\exp\{4\lambda\balfa_s\ln{z_{12\parallel}^2\over z_{12_\perp}^2}\}$ but simultaneously it will change the 
result of similar evolution for projectile matrix element $\langle p_A|\ticalo|p_A\rangle$ by $\exp\{-4\lambda\balfa_s\ln{z_{12\parallel}^2\over z_{12_\perp}^2}\}$ 
so the overall result for the amplitude (\ref{facoord}) remains intact.

To compare with conventional TMD analysis let us write down the evolution of ``generalized TMD''\cite{Meissner:2009ww_96_142_58_364_148,Lorce:2013pza_38_46_142}
\begin{eqnarray}
&&\hspace{-0mm}
D^\sigma(x,\xi)~=~
\int\! dz^+e^{-ix\sqrt{s\over 2}z^+}\langle p'_B|\calo^\sigma\big(-{z^+\over 2},{z^+\over 2}\big)|p_B\rangle
\nonumber
\end{eqnarray}
where $\xi=-{p'_B-p_B\over \sqrt{2s}}$.
From Eq. (\ref{result}) one easily obtains
\begin{equation}
\hspace{-0mm}
{D^{\sigma_2}(x,\xi)\over D^{\sigma_1}(x,\xi)}~=~e^{-2\balfa_s\ln{\sigma_2\over\sigma_1}
[\ln\sigma_2\sigma_1(x^2-\xi^2)sz_{12_\perp}^2+4\gamma_E-\ln 2]}
\label{momresult}
\end{equation}
For usual TMD at  $\xi=0$ with the limits of Sudakov evolution set by Eq. (\ref{fla1a}) one obtains
\begin{equation}
\hspace{-0mm}
{D^{\sigma_2}(x,q_\perp)\over D^{\sigma_1}(x,q_\perp)}
~=~e^{-2\balfa_s\ln{Q^2\over q_\perp^2}\big[\ln{Q^2\over q_\perp^2}+4\gamma_E-\ln 2\big]}
\label{vperesult}
\end{equation}
which coincides with usual one-loop evolution of TMDs \cite{Aybat:2011zv} up to replacement $4\gamma_E-2\ln 2\rightarrow4\gamma_E-4\ln 2$.
As we discussed, such constant depends on the way of cutting $k^-$-integration which should be
coordinated with the cutoffs in the ``coefficient function'' $\sigma(ff\rightarrow H)$ in Eq. (\ref{TMDf}).
Thus,  the discrepancy is just like using two different schemes for usual renormalization. It should be mentioned, however,
that at $\xi\neq 0$ the result (\ref{momresult}) differs from conventional one-loop result which does not depend on $\xi$ , see e.g. \cite{Echevarria:2016mrc_38_52_136}.

\section{Conclusions and Outlook}
The first conclusion is that the 11-parameter subgroup of  $SO(2,4)$ formed by generators (\ref{generators}) formally leaves TMD operators invariant. 

The second result is related to the fact that conformal invariance is violated by the rapidity cutoff (even in  ${\cal N}=4$ SYM). 
We have studied the TMD evolution in the Sudakov region of intermediate $x$ and demonstrated that the rapidity cutoff used in 
small-$x$ literature preserves all generators of our subgroup except 
the Lorentz boost which is related to the change of that cutoff.  

Our main outlook is to try to connect to small-$x$ region, first in ${\cal N}=4$ SYM and then in QCD.
As we mentioned above, although the TMD evolution in a small-$x$ 
region is conformal with respect to $SL(2,C)$ group, and our evolution (\ref{result}) is also conformal (albeit with respect to different
group of which $SL(2,C)$ is a subgroup),
the transition between Sudakov region and small-$x$ region is described by rather 
complicated interpolation formula \cite{Balitsky:2015qba} which is not conformally invariant. 
Our hope is that in a conformal theory one can simplify that transition using the conformal 
invariance requirement. 
 The study is in progress.

\section*{Acknowledgments}
\label{sec:acknowledgments}
We thank  V.M. Braun, A. Vladimirov and A. Tarasov for  discussions. 
The work of I.B.  was supported by DOE contract
 DE-AC05-06OR23177  and by the grant DE-FG02-97ER41028.





\newpage 
%



\renewcommand*{\FigPath}{./WeekVI/02_IancuLappiTriantafyllopoulos}

\long\def\comment#1{ }
\newcommand{\eqn}[1]{Eq.~\eqref{#1}}
\renewcommand{\beq}{\begin{equation}}
\renewcommand{\eeq}{\end{equation}}
\renewcommand{\nn}{\nonumber\\}
\newcommand{\dif}{{\rm d}}
\newcommand{\rmd}{{\rm d}}
\newcommand{\rme}{{\rm e}}
\newcommand{\rmi}{{\rm i}}
\newcommand{\del}{\partial}

\newcommand{\bk}{\bm{k}}
\newcommand{\bq}{\bm{q}}
\newcommand{\bp}{\bm{p}}
\newcommand{\bx}{\bm{x}}
\newcommand{\by}{\bm{y}}
\renewcommand{\bu}{\bm{u}}
\newcommand{\bv}{\bm{v}}
\newcommand{\bz}{\bm{z}}
\newcommand{\bw}{\bm{w}}
\newcommand{\br}{\bm{r}}
\newcommand{\bb}{\bm{b}}
\renewcommand{\pt}{p_\perp} 
\renewcommand{\kt}{k_\perp} 
\renewcommand{\qt}{q_\perp} 
\newcommand{\lt}{\ell_\perp} 
\newcommand{\atpi}{\frac{\bar{\alpha}}{2 \pi}}
\newcommand{\abar}{\bar{\alpha}_s}
\newcommand{\order}[1]{\mcal{O}{(#1)}}
\newcommand{\mcal}{\mathcal}
\newcommand{\sdla}{{\rm \scriptscriptstyle DLA}}
\newcommand{\ssl}{{\rm \scriptscriptstyle SL}}
\newcommand{\bfkl}{{\rm \scriptscriptstyle BFKL}}
\newcommand{\nnlo}{{\rm \scriptscriptstyle NNLO}}
\newcommand{\nlo}{{\rm \scriptscriptstyle NLO}}
\newcommand{\lo}{{\rm \scriptscriptstyle LO}}
\newcommand{\slog}{{\rm \scriptscriptstyle STL}}
\newcommand{\gbw}{{\rm \scriptscriptstyle GBW}}
\newcommand{\uv}{{\rm \scriptscriptstyle UV}}
\newcommand{\mv}{{\rm \scriptscriptstyle MV}}
\newcommand{\CXY}{{\rm \scriptscriptstyle CXY}}
\newcommand{\rcBK}{{\rm \scriptscriptstyle rcBK}}
\newcommand{\fcBK}{{\rm \scriptscriptstyle fcBK}}
\newcommand{\calS}{\mathcal{S}}
\newcommand{\calN}{\mathcal{N}}
\newcommand{\calA}{\mathcal{A}}
\newcommand{\calT}{\mathcal{T}}
\newcommand{\calK}{\mathcal{K}}
\newcommand{\minus}{\!-\!}

\newcommand{\qvec}{\vec{q}}
\newcommand{\pvec}{\vec{p}}
\newcommand{\kvec}{\vec{k}}
\newcommand{\qpvec}{{\vec{q}{\, '}}}
\newcommand{\ppvec}{{\vec{p}{\, '}}}
\newcommand{\kpvec}{{\vec{k}{\, '}}}
\newcommand{\cf}{C_\mathrm{F}}
\renewcommand{\nc}{N_\mathrm{c}}
\newcommand{\epst}{\boldsymbol{\varepsilon}}

\wstoc{Small-$x$ physics in the dipole picture at NLO accuracy}{Edmond Iancu, Tuomas Lappi, Dionysios Triantafyllopoulos}
\title{Small-$x$ physics in the dipole picture at NLO accuracy}

\author{Edmond Iancu}

\address{Institut de Physique Th\`eorique, Universit\'e Paris-Saclay, CNRS, CEA, F-91191 Gif-sur-Yvette, France}

\author{Tuomas Lappi}

\address{
Department of Physics, %
 P.O. Box 35, 40014 University of Jyv\"askyl\"a, Finland
and 
Helsinki Institute of Physics, P.O. Box 64, 00014 University of Helsinki,
Finland
}

\author{Dionysios Triantafyllopoulos}

\index{author}{Iancu, E.}
\index{author}{Lappi, T.}
\index{author}{Triantafyllopoulos, D.}

\address{European Centre for Theoretical Studies in Nuclear Physics and Related Areas (ECT*)
and Fondazione Bruno Kessler, Strada delle Tabarelle 286, I-38123 Villazzano (TN), Italy}

\begin{abstract}
We  review recent progress in NLO calculations for dilute-dense processes in the CGC picture. In particular, we focus here on recent steps  in understanding high energy renormalization group evolution (BK/JIMWLK), the total DIS cross section at small $x$ and forward particle production in proton-nucleus collisions at next-to-leading order. 
\end{abstract}


\bodymatter

\section{Introduction}\label{aba:sec1_345}

The increasing accuracy of experimental data for small $x$ processes, both from the EIC and from ongoing LHC experiments, calls for a corresponding increase in the accuracy of theoretical calculations. This requires going to higher orders in perturbation theory also for processes where nonlinear gluon saturation phenomena are dominant. In order for the theory to have predictive power, it is crucial to have a consistent treatment of both inclusive and exclusive processes in DIS, and in forward rapidity proton-nucleus collisions and ultraperipheral collisions of heavy ions.  In the high energy limit such a consistent, systematically perturbatively improveable, framework for describing the physics of gluon saturation  is provided by the Color Glass Condensate (CGC) effective theory formulation. Here, the gluon fields of the target are described in terms of Wilson lines, path ordered exponentials in the strong color fields, which are measurable as the eikonal scattering amplitudes of a light projectile passing through the field. 
 
Here we will review recent progress on  NLO calculations in the CGC picture, concentrating on three different parts of the program towards a consistent framework for small-$x$ physics. We will first, in Sec. \ref{sect:hee} discuss high energy (BK/JIMWLK) renormalization group evolution of the Wilson lines and their correlators, i.e. eikonal scattering amplitudes, with the energy scale of the process. We will then, in Sec.~\ref{sec:dis} move to calculations of the total DIS cross section in the dipole picture that naturally emerges in the CGC framework.  Finally, in Sec.~\ref{sec:pasinc}, we discuss single inclusive particle production in forward rapidity proton-nucleus collisions. Exclusive processes will be discussed in other contributions to this volume.

\section{High energy evolution}
\label{sect:hee}

Starting at next-to-leading order in perturbation theory, calculations of scattering processes at very high energy contain large logarithms of the center of mass energy $\sqrt{s}$ which can be resummed into a RG evolution. In terms of Wilson lines, which are appropriate for describing the scattering of small projectiles off a generic target,    this evolution is given by the JIMWLK equation\cite{JalilianMarian:1997gr_131,Iancu:2000hn_125,Ferreiro:2001qy_195}. In most applications one only needs the scattering of a color dipole, whose evolution can be derived from the JIMWLK equation in a mean field approximation. The corresponding closed equation is known as the Balitsky-Kovchegov (BK) equation\cite{Balitsky:1995ub_119,Kovchegov:1999yj}. 

\subsection{Double logarithms and instabilities}
\label{sect:Yevol}

 The BK equation has now been derived up to  NLO accuracy\cite{Balitsky:2006wa,Kovchegov:2006vj_107,Balitsky:2008zza}. For the moment we shall take into account only the most dominant of the NLO terms, which also require a special treatment in order to lead to physically meaningful results. With $(\bx,\by)$ the transverse coordinates of the dipole, the equation of interest for the $S$-matrix $S_{\bx\by}(Y)$ reads 
\begin{align}
\hspace*{-0.1cm}
	\label{dsdydl}
	\frac{\del S_{\bx\by}}{\del Y}  = 
	\frac{\abar}{2\pi} 
	\int\frac{\dif^2\bz \, (\bx \minus\by)^2}{(\bx\minus\bz)^2(\bz\minus\by)^2}
	\left[1 \minus \abar\ln\frac{(\bx\minus\bz)^2}{(\bx\minus\by)^2} 
	\ln\frac{(\bz\minus\by)^2}{(\bx\minus\by)^2}\right]
	\left(S_{\bx\bz}S_{\bz\by} \minus S_{\bx\by} \right), 
\end{align}
where $\abar = \alpha_s\nc/\pi$ and with $\nc$ the number of colors. We stress that the variable, for which the above equation has been derived, is the \emph{projectile rapidity} $Y$. The dipole is a right moving object with large plus longitudinal momentum $q^+$ and the target hadron a left moving object with minus longitudinal momentum, so that $s=2 q^+ P^-$ is the COM energy squared. If a typical parton in the target carries a fraction $x_0$ of $P^-$ and if $Q_0$ is its typical transverse momentum, we define $p^+ =Q_0^2/2 x_0 P^-$. Then the rapidity variable is defined as the boost invariant ratio 
\begin{equation}
	\label{Ydef}
	Y = \ln\frac{q^+}{p^+} = \ln \frac{x_0 s}{Q_0^2} = 
	\ln\frac{x_0}{x} \frac{Q^2}{Q_0^2},  
\end{equation} 
where $Q^2 \sim 1/r^2$ with $r=|\bx-\by|$ the dipole size and $x=Q^2/s$ as traditionally defined. In the kinematical limit of interest, $Y$ is assumed to be parametrically large. In general, the scattering amplitude $T_{\bx\by}= 1- S_{\bx\by}$ is small when the target is dilute, while it approaches the unitarity limit $T_{\bx\by}(Y)= 1$ when the target is dense or ``saturated''. The two regimes are separated by a dynamically generated scale, the saturation momentum $Q_s(Y)$, which should be an increasing function of the rapidity $Y$. In other words, the scattering amplitude should increase with increasing $Y$ and fixed dipole size $r$, or with increasing $r$ and fixed rapidity $Y$. This is the physical expectation and this is indeed what happens when we keep only the LO term in \eqn{dsdydl} as one can verify both numerically and analytically.

The next natural step is to solve \eqn{dsdydl} including the ``non-conformal'' NLO term. In the regime where the daughter dipoles are large, i.e.~for $|\bx-\bz| \simeq |\bz - \by| \gg r$,  the double logarithm gets significantly large. For any reasonable value of the coupling $\abar$, the total contribution of the NLO term is larger in magnitude than the one of the LO term, and because of its negative sign it leads to a particularly awkward solution which exhibits an unphysical oscillating behavior. These features have been confirmed by analytical studies and numerical solutions \cite{Avsar:2011ds,Lappi:2015fma,Iancu:2015vea}.

Such big \emph{collinear} logarithms spoil the convergence of the perturbative expansion in $\abar$ and an all-orders resummation in the regime where large daughter dipoles are emitted, can be the only resolution to the problem. As is always the case in a Quantum Field Theory, large logarithmic contributions are associated with a certain physical mechanism. Here, they are generated by diagrams in light-cone perturbation theory in which the successive gluon (or dipoles at large-$\nc$) emissions in the projectile are not only strongly ordered towards smaller (plus) longitudinal momenta and larger dipoles sizes, as clear from the above discussion, but are also ordered towards smaller lifetimes or equivalently larger light-cone energies \cite{Beuf:2014uia,Iancu:2015vea}.  

Simply by enforcing this time-ordering in the projectile wavefunction, we are led to a well-defined evolution which goes beyond a fixed order expansion in $\abar$. In fact there are two possibilities on how to proceed, and they are equivalent to the order of accuracy. In the one approach, the resummation of the double collinear logarithms is explicit in the evolution kernel and the ensuing equation reads\cite{Iancu:2015vea} 
\begin{align}
	\label{resbk}
	\frac{\del S_{\bx\by}}{\del Y}  = 
	\frac{\abar}{2\pi} 
	\int\frac{\dif^2\bz \, (\bx-\by)^2}{(\bx-\bz)^2(\bz-\by)^2}\,
	\mathcal{K}_{\sdla}(\rho_{\bx\by\bz})
	\left(S_{\bx\bz} S_{\bz\by} - S_{\bx\by} \right), 
\end{align}
where the kernel is given by
\begin{equation}
	\label{kdla}
	\mathcal{K}_{\sdla}(\rho) \equiv 
	\frac{\rm{J}_1(2 \sqrt{\abar \rho^2})}{\sqrt{\abar \rho^2}}
	 \quad \textrm{with} \quad
	\rho_{\bx\by\bz}^2 = \ln\frac{(\bx\minus\bz)^2}{(\bx\minus\by)^2} 
	\ln\frac{(\bz\minus\by)^2}{(\bx\minus\by)^2}.
\end{equation}
Eq.~\eqref{resbk} remains a local equation in $Y$, like Eq.~\eqref{dsdydl}. When expanding to first non-trivial order in $\abar$ we recover the NLO double logarithm in Eq.~\eqref{dsdydl}. But the all-orders kernel exhibits a completely different behavior at large distance: the Bessel function oscillates strongly when its argument gets large, and thus eliminates the contribution of large daughter dipoles which could violate time-ordering.

 In the other approach the respective evolution equation reads\cite{Beuf:2014uia}
 \begin{align}
	\label{shiftbk}
	\frac{\del S_{\bx\by}(Y)}{\del Y}  = 
	\frac{\abar}{2\pi} 
	\int &\frac{\dif^2\bz \, (\bx-\by)^2}{(\bx-\bz)^2(\bz-\by)^2}\,
	\times \Theta(Y \minus \rho_{\rm min})
	\nn
	& \left[S_{\bx\bz}(Y \minus \Delta_{\bx\bz;r})
	S_{\bz\by}(Y \minus \Delta_{\bz\by;r}) - S_{\bx\by}(Y) \right], 
\end{align}
where the two quantities $\rho_{\rm min}$ and $ \Delta_{\bx\bz;r}$ are given by
\begin{equation}
	\label{rhomin}
	\rho_{\rm min} = \ln \frac{1}{r^2_{\min} Q_0^2}
\quad \textrm{and} \quad \Delta_{\bx\bz;r} = 
\max \left\{0,\ln\frac{|\bx\minus\bz|^2}{r^2} \right\},
\end{equation}
and $r_{\min}$ is the size of the smallest of the three dipoles. Although we shall not show the details here, which require to write Eq.~\eqref{shiftbk} in an integral form, the shifts in the arguments of the $S$-matrices are a direct consequence of the time-ordering constraint. Again there is a change w.r.t.~the LO BK equation only regarding the contribution of the sufficiently large daughter dipoles. Contrary to Eq.~\eqref{resbk}, the above keeps the LO dipole kernel but is a non-local equation in $Y$. Equations~\eqref{resbk} and \eqref{shiftbk} both reduce to the LO BK equation when expanded to LO in $\abar$, while they agree to all orders in $\abar$ so long as one is interested in the double logarithmic contributions arising from the emission of very large dipoles. The pure $\abar^2$ terms (i.e.~terms of order $\mcal{O}(\abar^2)$ not enhanced by large collinear logarithms) in the two equations are not the same, but they can be made to match each other and also match the respective terms of the full NLO BK equation. 

Eqs.~\eqref{resbk} and \eqref{shiftbk} have been solved as initial value problems\cite{Iancu:2015joa,Lappi:2016fmu}, using typical initial conditions like the MV model\cite{McLerran:1993ka_101,McLerran:1993ni_95}. The solutions seem to be stable, but there are many issues in the results which are puzzling. The solution to the BK equation is always characterized by the speed $\lambda = \dif \ln Q_s^2/\dif Y$ of the evolution and the slope $\gamma = \dif \ln T/\dif \ln r^2$ in the perturbative side of the amplitude. The aforementioned equations lead to a somewhat low speed (such that after taking into account all the NLO corrections, it is difficult to cope with the phenomenology) and a significantly larger slope when compared to the LO dynamics. For fixed coupling evolution one finds that $\gamma$ is close to 1 or even larger than 1, except for values of $\abar$ which are unnaturally small. Given that at LO the asymptotic value $\gamma\simeq 0.63$ is a hallmark of BFKL dynamics and saturation, one may wonder why this feature is lost when moving to NLO and beyond. 

The first thing to realize, is that $Y$ is not the correct variable to use. It was merely introduced since in the presence of saturation it is much easier to calculate using the kinematics of the projectile which is a simple object. Instead, we should use the \emph{target rapidity} $\eta$ defined as the logarithm of the ratio of the minus longitudinal momenta, namely
\begin{equation}
	\label{eta}
	\eta= \ln\frac{x_0 P^-}{q^-}=  
	\ln  \frac{x_0 s}{Q^2} = Y - \rho.
\end{equation}
We have defined the minus longitudinal momentum of the dipole according to $2 q^-= Q^2/q^+$ and the logarithmic variable $\rho = \ln 1/r^2 Q_0^2$. When doing DIS of a photon off the hadronic target, typically $Q^2$ is also the virtuality of the photon and thus $\eta$ is closely related to the logarithm of the Bjorken variable. Moreover, since we are mostly interested in a situation where $Q^2 \gg Q_0^2$, the rapidities $Y$ and $\eta$ can be very different. We express the results in terms of the \emph{physical} variable $\eta$ defining   
\begin{equation}
\label{sbar}
	\bar{S}_{\bx\by}(\eta) \equiv S_{\bx\by}(Y=\eta+\rho), 
\end{equation}  
from which we extract the saturation momentum $Q_s^2(\eta)$. We find that indeed the evolution in terms of $\eta$ is faster, since the new speed reads $\bar{\lambda} = \lambda/(1-\lambda)$, while the shape of the amplitude is less steep since the new slope is given by $\bar{\gamma} = \gamma (1-\lambda)$.  
   
Still, there are two serious issues, which seem too difficult to deal with using the current procedure. First, we have not been very careful about the initial condition at $Y=0$ when solving Eqs.~\eqref{resbk} or \eqref{shiftbk}. In fact, these two equations are \emph{boundary} value problems, more precisely a condition must be given at $Y=\rho$, but such a problem looks very hard to solve. A workaround, would be to write an initial condition at $Y=0$ which reproduces via evolution the desired amplitude, e.g.~the MV model, at $Y=\rho$. This has been exactly done only at the level of linear evolution in the DLA\cite{Ducloue:2019ezk_89}, but it looks almost impossible to do it for the non-linear equation. Second, the shift $\Delta$ in Eq.~\eqref{shiftbk}, is not really uniquely defined as we have written in Eq.~\eqref{rhomin}, rather it can be only specified within double-log accuracy. This arbitrariness in the choice of $\Delta$ leads to a small reasonable uncertainty in the results terms of $Y$. However, such a scheme dependence becomes very strong when expressing the results in terms of $\eta$ and leaves the whole approach without any predictive power\cite{Ducloue:2019ezk_89}. For example, for certain schemes which at a first sight look fair, the speed $\bar{\lambda}$ turns out to be physically non-acceptable even for typical values of $\abar$.

\subsection{Evolution in the target rapidity}
\label{sect:etaevol}
  
The way to avoid the difficulties encountered at the end of Sect.~\ref{sect:Yevol} is rather simple. Instead of making the change of variables given in Eq.~\eqref{sbar} on the solution, we shall do it directly on the evolution equation. Focusing on the non-local version in Eq.~\eqref{shiftbk}, it is an easy exercise to show that going from $(Y,\rho)$ to $(\eta,\rho)$ leads to\cite{Ducloue:2019ezk_89}
\begin{align}
	\label{shiftbketa}
	\frac{\del \bar{S}_{\bx\by}(\eta)}{\del \eta}  = 
	\frac{\abar}{2\pi} 
	\int&\frac{\dif^2\bz \, (\bx-\by)^2}{(\bx-\bz)^2(\bz-\by)^2}\,
	\nn
	&\times  \Theta(\eta \minus \delta_{\bx\by\bz})
	\left[\bar{S}_{\bx\bz}(\eta \minus \delta_{\bx\bz;r})
	\bar{S}_{\bz\by}(\eta \minus \delta_{\bz\by;r}) - \bar{S}_{\bx\by}(\eta) \right], 
\end{align}
where the shifts are given by
\begin{equation}
	\label{delta}
	\hspace*{-0.22cm}
	\delta_{\bx\bz;r} =\ln\frac{|\bx\minus\bz|^2}{r^2}  - \Delta_{\bx\bz;r} =\max \left\{0,\ln\frac{r^2}{|\bx\minus\bz|^2} \right\}, \quad
	\delta_{\bx\by\bz} = \max\{\delta_{\bx\bz;r},\delta_{\bz\by;r}\}.
\end{equation}
We are still looking at the evolution from the projectile point of view, since the transverse sector involves the splitting of a dipole into two. The equation remains non-local where now the shift is effective only when one of the daughter dipoles is very small. When evolving in $\eta$ time ordering is trivially satisfied, while it is only due to the shift $\delta$ in Eq.~\eqref{delta} that $k^+$ ordering is also guaranteed to hold. As before, the prescription for the shift is not unique beyond double-logarithmic accuracy. Finally, let us mention that we can truncate Eq.~\eqref{shiftbketa} to second order in $\abar$ by expanding  to first order in the shift. Not surprisingly, the ensuing equation contains a large double logarithm when a small daughter dipole is formed. Although such an emission is \emph{atypical} in the course of evolution, it can give rise to instabilities when $\eta$ gets large due to BFKL diffusion, and thus it is mandatory that one use Eq.~\eqref{shiftbketa} which effectively resums such logarithmic contributions to all orders in $\abar$.

The most essential feature of Eq.~\eqref{shiftbketa} is that, contrary to Eq.~\eqref{shiftbk}, it is an \emph{initial} value problem. We can specify a physical condition $\bar{S}_{\bx\by}(\eta=0) = \bar{S}^{(0)}_{\bx\by}$ and proceed to solve the equation. The solution shows a weak dependence on the detailed form of the shift $\delta$, e.g.~the uncertainty in the value of the speed $\bar{\lambda}$ is of order $\mcal{O}(\abar^2)$, which is precisely what we should expect to the order of accuracy. Also the speed does not show any weird behavior, and the ratio $\bar{\lambda}/\abar$ is a monotonic function of $\abar$, namely it decreases when the coupling increases. Thus, we have a procedure which contains the proper physics in a controlled approximation.

\subsection{Beyond double logarithms}
\label{sect:other}

So far we have focused on the higher order terms enhanced by double collinear logarithms, but there are other large corrections to be taken into account. The NLO BK equation also contains single transverse logarithms\cite{Iancu:2015joa}, for which it is intuitively clear that they stand for DGLAP corrections on top of the small-$x$ evolution. That is, we can have a sequence of two emissions where only the first one is enhanced in the longitudinal sector, while both are strongly ordered in dipole sizes, either towards large daughter dipoles or towards dipoles of which one is much smaller. To a first approximation, such effects can be naturally resummed by inserting a power law suppression in the evolution kernel. More precisely, we introduce the factor
\begin{equation}
	\label{ka1}
	\mathcal{K}_{A_1} = 
	\left[\ 
	\frac{(\bx-\by)^2}{\min\{(\bx-\bz)^2,(\bz-\by)^2\}}
	\right]^{\pm A_1},
\end{equation}
with $A_1 = 11/12$ the familiar coefficient in the relevant DGLAP anomalous dimension. The sign in the exponent is such that $\mathcal{K}_{A_1}$ is never larger than one.

The last corrections that are mandatory to be included, are those related to the running coupling. We recall that on the r.h.s.~of the NLO BK equation there is a term proportional to the first coefficient of the QCD $\beta$-function and which reads\cite{Balitsky:2006wa,Kovchegov:2006vj_107}
\begin{equation}
	\label{nlobeta}
	\frac{\bar{b}\abar^2}{2\pi}
	\int\frac{\dif^2\bz \, (\bx\minus\by)^2}{(\bx\minus\bz)^2(\bz\minus\by)^2}
	\left[
	\ln(\bx\minus\by)^2\mu^2
	\minus \frac{(\bx\minus\bz)^2 \minus (\by\minus\bz)^2}{(\bx\minus\by)^2}
	\ln \frac{(\bx\minus\bz)^2}{(\by\minus\bz)^2} 
	\right].
\end{equation}
Here $\bar{b} = (11 \nc - 2 N_{\rm f})/12 \nc$, $N_f$ is the number of flavors and $\mu$ is a renormalization scale at which the coupling is evaluated. The logarithms can give large contributions and spoil the convergence of the perturbative expansion when either $r$ is very small or very large compared to $1/\mu$ or when the gluon at $\bz$ is close to any of the legs of the parent dipole. Thus $\mu$ must be chosen in such a way to guarantee that there are no large logarithms for any size of the three dipoles involved in the splitting process. Usually in QCD it is the hardest scale that determines the scale for the running coupling and one readily sees that this is true also here. The choice of $\mu$ is not unique and we shall list two different prescriptions which are equivalent to the order of accuracy. The first one is just the \emph{smallest dipole scheme} defined by
\begin{equation}
	\label{amin}
	\bar{\alpha}_{\min} = \abar(r_{\min})
	\quad \textrm{with} \quad
	r_{\min} = \min\{|\bx-\by|,|\bx-\bz|,|\bz-\by|\}.
\end{equation}
The second one is the BLM scheme, which in the current context means to choose $\mu$ so that the whole integrand in Eq.~\eqref{nlobeta} vanishes. This leads to 
\begin{equation}
	\label{ablm}
	\bar{\alpha}_{\rm \scriptscriptstyle BLM} = 
	\left[ 
	\frac{1}{\abar(|\bx \minus\by|)}
	+ \frac{(\bx\minus\bz)^2 - (\bz\minus\by)^2}{(\bx\minus\by)^2}
	\frac{\abar(|\bx-\bz|)-\abar(|\bz-\by|)}{\abar(|\bx-\bz|)\abar(|\bz-\by|)}
	\right]^{-1},
\end{equation}
and it is an elementary exercise to show that $\bar{\alpha}_{\rm \scriptscriptstyle BLM}$ reduces to $\bar{\alpha}_{\min}$ for all the configurations in which one of the three dipoles is much smaller than the other two.

Finally, we point out that there are other NLO corrections, which however are not enhanced by large logarithms in any kinematic regime. It suffices to say that the resummations which we reviewed here, can be matched to the full NLO BK equation. Thus, we have a stable formalism which includes the necessary resummations and is accurate to order $\mcal{O}(\abar^2)$. In particular one must revise here the discussion given in Sect.~\ref{sect:etaevol}, where the matching was done only at LO (cf.~Eq.~\eqref{shiftbketa}) and this is why the uncertainty due to the choice of $\delta$ was of order $\mcal{O}(\abar^2)$. When the matching is lifted to NLO accuracy, we can show that such an uncertainty due to the choice of a scheme is reduced to order $\mcal{O}(\abar^3)$, as it should.

\section{Total DIS cross section}
\label{sec:dis}

The simplest scattering observable where the dipole $S$-matrix element $ \calS_{\bx\by}$ appears is the total deep inelastic scattering cross section. Here the leading order picture is that the quark-antiquark dipole discussed in the context of the evolution equation is a quantum fluctuation of the virtual photon emitted by the lepton. Depending on the polarization state of the $\gamma^*$ one must consider separately transversally and longitudinally polarized photons. Experimentally the polarization states are separated using the kinematics of the scattered lepton and varying $\sqrt{s}$. Measuring both cross sections will be a central part of the physics program at the EIC.

In this ``dipole factorization'' one separates the process into an impact factor describing the fluctuation of the virtual photon into a partonic state, at leading order a dipole and at NLO also a $q\bar{q}g$ state, and the scattering amplitude of this state with the target, which at leading order is  $1 - S_{\bx\by}$. The term ``impact factor'' is most often used for calculations in momentum space: for deep inelastic scattering at small-$x$ there is a  momentum space calculation in Refs.~\citenum{Balitsky:2012bs,Balitsky:2010ze}. However, for the eikonal scattering picture in general, and for nonlinear high energy evolution in the CGC picture as discussed in Sec.~\ref{sect:hee}, it is more convenient to work in mixed space, with transverse coordinates and longitudinal momentum. The calculation of the total DIS cross section in this formulation has been completed more recently in Refs.~\citenum{Beuf:2016wdz,Beuf:2017bpd,Hanninen:2017ddy}. This is the formulation we will discuss in more detail here.

The theoretical tool of choice for these calculations is light cone perturbation theory~\cite{Kogut:1969xa,Bjorken:1970ah,Lepage:1980fj,Brodsky:1997de}. Here one calculates diagrammatically the perturbative expansion in terms of bare Fock states of the incoming particle (in this case $\gamma^*$) state. The coefficients of this expansion are known as the \emph{light cone wave functions}. For the NLO DIS cross section the calculation is thus based on obtaining the virtual photon wavefunction to NLO accuracy.

\subsection{Leading order}

In general, the total DIS cross section is obtained, via the optical theorem, from the forward limit of the elastic photon-target scattering amplitude
\begin{equation}\label{eq:dissigma}
\sigma^{\gamma^{\ast}}[A] = \left.
 \frac{2}{2q^+(2\pi)\delta(q'^+-q^+)} \mathrm{Re} \biggl [{}_{\text{i}}\langle \gamma^{\ast}(\qpvec,Q^2,\lambda')\vert 1- \hat{S}_E\vert \gamma^{\ast}(\qvec,Q^2,\lambda)\rangle_{\text{i}}\biggr ] \right|_{\qpvec \to \qvec}
\end{equation}
Here the photon states $\vert \gamma^{\ast}(\qvec,Q^2,\lambda)\rangle_{\text{i}}$ are interacting theory states, that must be developed in a perturbation theory expansion in terms of the bare Fock states of the theory. The eikonal scattering operator $\hat{S}_E$ is diagonal in transverse position space, and in the CGC expressed in terms of Wilson lines in the target gluon field. At leading order, the only contributing photon Fock state is the quark-antiquark dipole, and the corresponding tree-level virtual photon wave functions are easily calculated. The resulting leading order expression 
\begin{equation}\label{eq:sigmagammastar}
	\sigma_{T,L}^{\gamma^*p}(x,Q^2) = 2\sum_f \int \rmd z  \rmd^2 \bb \rmd^2 \br \left|\Psi_{T,L}^{\gamma^* \to q\bar q}\right|^2 \left(1-S(\bb, \br, x) \right),
\end{equation}
has been widely used to describe HERA DIS data, with the dipole amplitude $S$ obtained from BK or JIMWLK evolution or from different more phenomenological parametrizations.

\subsection{Next to leading order}

\begin{figure}[tbh!] 
\centerline{
\includegraphics[width=0.25\textwidth]{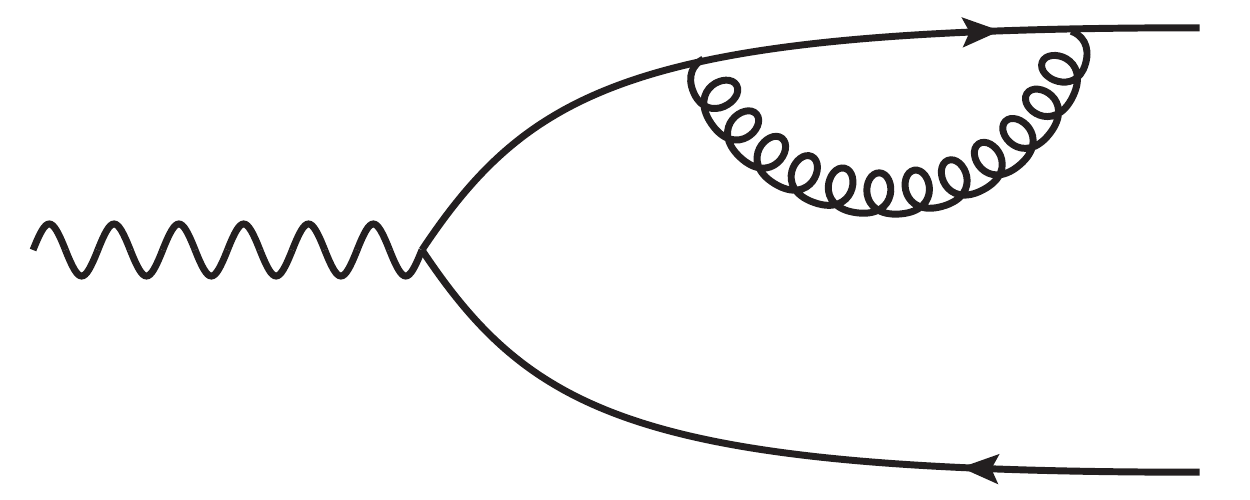} 
\includegraphics[width=0.25\textwidth]{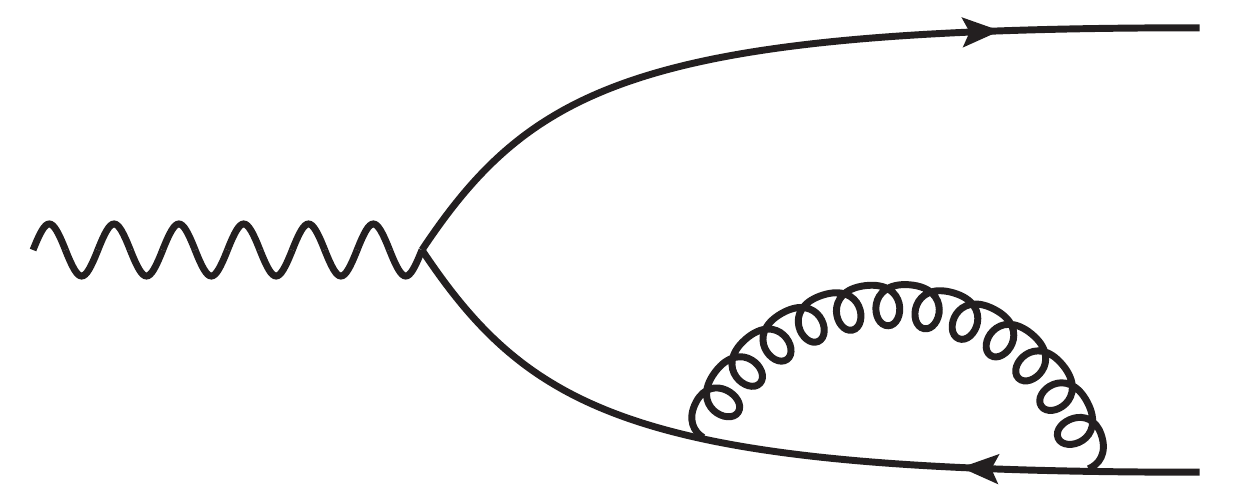}
\includegraphics[width=0.25\textwidth]{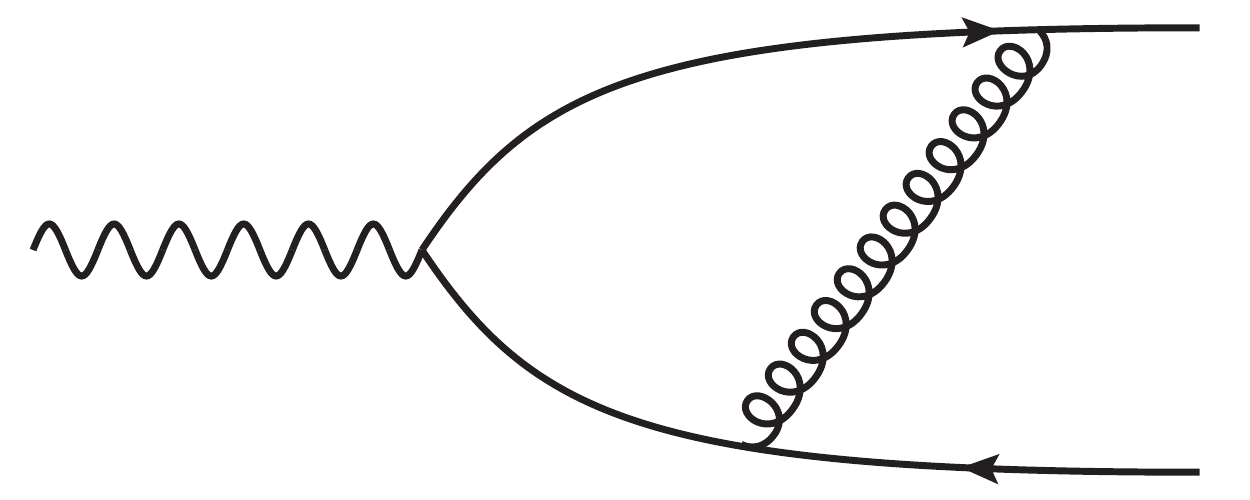}
\includegraphics[width=0.25\textwidth]{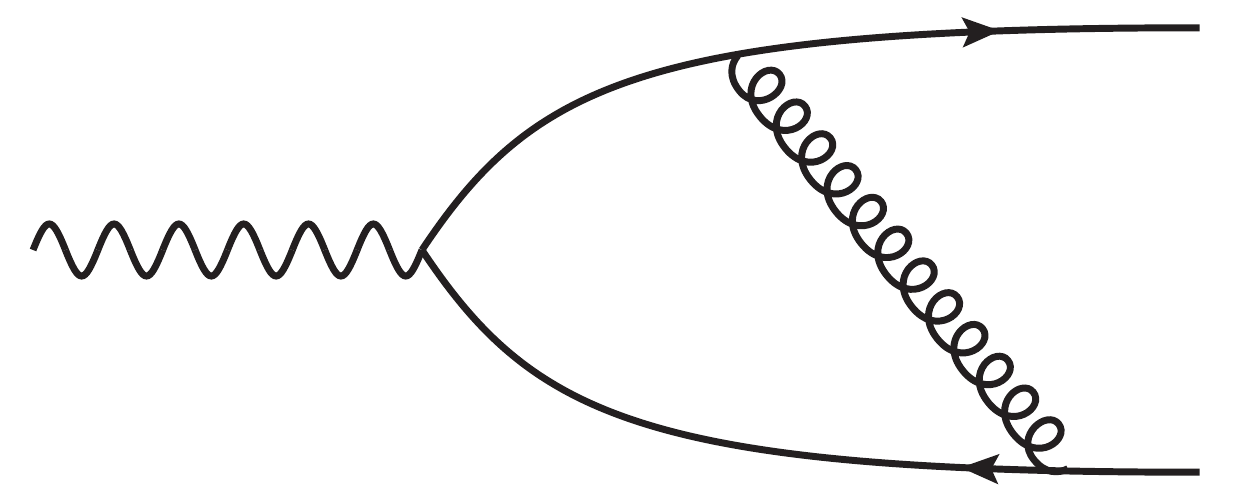}
}
 \caption{Loop diagrams needed for the NLO $\gamma^* \to q\bar{q}$ wavefunction. In addition to these diagrams one must also include the corresponding ``instantaneous interaction'' contributions.}
 \label{fig:disloopdiags}
\end{figure}

\begin{figure}[tbh!] 
\centerline{
\includegraphics[width=0.33\textwidth]{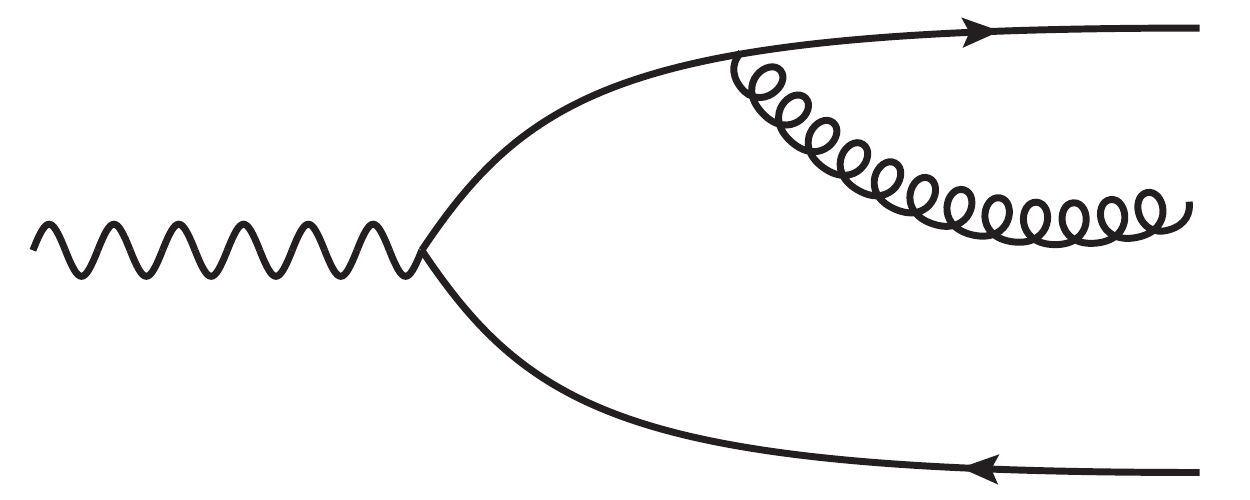}
\includegraphics[width=0.33\textwidth]{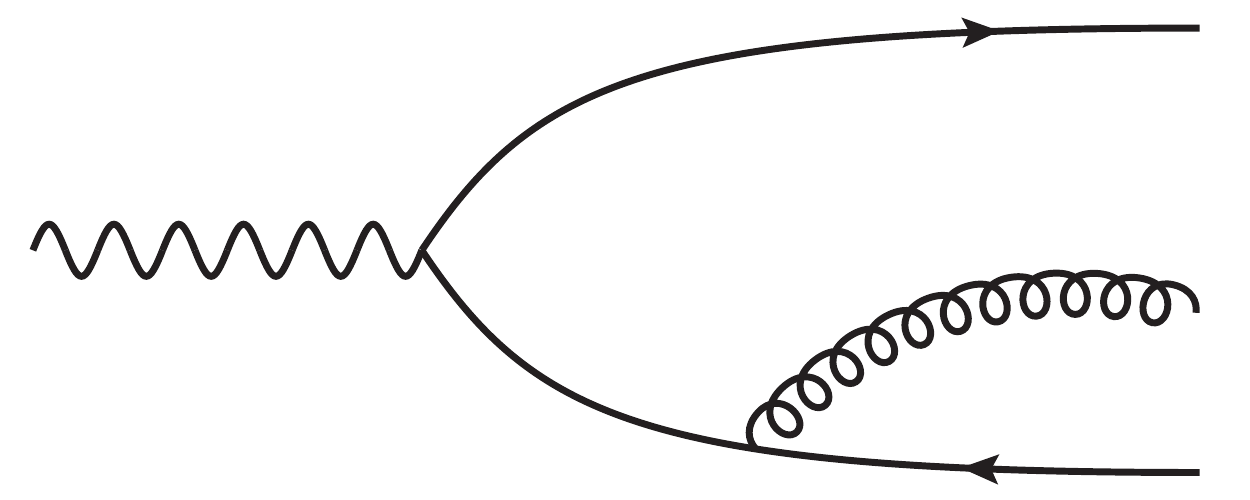}
}
 \caption{Gluon emission diagrams needed for the $\gamma^* \to q\bar{q}g$ wavefunction. In addition to these diagrams one must also include the corresponding ``instantaneous interaction'' contributions.}
 \label{fig:disraddiags}
\end{figure}

At NLO accuracy one must develop the interacting photon state in \eqn{eq:dissigma} to include one loop corrections to the $\gamma^* \to q\bar{q}$ wavefunction, shown in Fig.~\ref{fig:disloopdiags}, and the contribution of the $q\bar{q}g$ state, shown in Fig.~\ref{fig:disraddiags}. This calculation has been done in Refs.~\citenum{Beuf:2016wdz,Beuf:2017bpd} in the conventional dimensional regularization (CDR) scheme and independently  in the four-dimensional helicity (FDH) scheme in Ref.~\citenum{Hanninen:2017ddy}.
The result can be written, using the notations of Ref.~\citenum{Hanninen:2017ddy}, as 
\begin{equation}
\label{eq:finalresult}
\sigma^{\gamma^{\ast}_{\rm T,L}}[A] = \sigma^{\gamma^{\ast}_{\rm T,L}}\bigg\vert_{q\bar{q}} +  \sigma^{\gamma^{\ast}_{\rm T}}\bigg\vert_{q\bar{q}g},
\end{equation}
where the $q\bar{q}$ term is   
\begin{multline}
\sigma^{\gamma^{\ast}_{\rm T,L}}\bigg\vert_{q\bar{q}}  = 
2\sum_f \int \rmd z  \rmd^2 \bb \rmd^2 \br \left|\Psi_{T,L}^{\gamma^* \to q\bar q}\right|^2 
\\
\times 
\biggl \{1 + \left ( \frac{\alpha_s\cf}{\pi}\right )\biggl [\frac{1}{2}\log^2\left ( \frac{z}{1-z}\right )-\frac{\pi^2}{6} + \frac{5}{2} \biggr ]\biggr \}
\left(1-S(\bb, \br, x) \right)
\end{multline}
The $q\bar{q}g$ terms have slightly more complicated expressions\cite{Beuf:2016wdz,Beuf:2017bpd,Hanninen:2017ddy} that differ for the trasverse and longitudinal polarizations, and involve the $S$-matrix elements for the $q\bar{q}g$ Fock state scattering on the target
\begin{equation}
S_{\bx \by \bz} = \frac{\nc}{2\cf}\biggl [S_{\bx\bz} S_{\bz\by} - \frac{1}{\nc^2} S_{\bx\by}\biggr ].
\end{equation}
 where the nonlinear term is the same as in the BK equation as written in \eqn{resbk}, and the second, $\nc$-suppressed one, a part of the linear term.

Further details about the calculations leading to the NLO result can be found in the original references, so let us here make a few remarks on the slight differences between the two independent calculations. It is important to note that the split into the two terms \eqref{eq:finalresult} is not unique. Both the virtual (Fig.~\ref{fig:disloopdiags}) and real (Fig.~\ref{fig:disraddiags}) separately contain divergences that, at least in light cone quantization, appear as logarithmic (transverse) UV ones. In coordinate space they correspond to the configurations where the gluon (at $\bz$) is very close to its parent quark (at $\bx$) or antiquark (at $ \by$). In the calculation  these divergences are regularized using transverse dimensional regularization, and cancel between the real and virtual terms in the end. This cancellation relies on the coincidence limit of the $q\bar{q}$ and $q\bar{q}g$ scattering amplitudes
$S_{\bx \by \bz} \underset{\bz \to \bx}{\longrightarrow } S_{\bx \by}$, which is always satisfied by their definitions in terms of Wilson line correlators. In practice the cancellation  is effectuated by subtracting from the real part a divergent term proportional to  $S_{\bx \by}$.  This subtraction term must have the correct form in the limits $\bz \to \bx$ and $\bz \to \by$, but there is freedom in choosing its functional form for other values of $\bz$.  In addition to the different variant of the dimensional regularization scheme, the calculation in  Refs.~\citenum{Beuf:2016wdz,Beuf:2017bpd} on one hand and in Ref.~\citenum{Hanninen:2017ddy} on the other hand, differ by the choice of this subtraction term. Thus the form in which the result is quoted is different, but the actual expressions equivalent.

The NLO impact factor contains a large logarithm of the energy (or of Bjorken $x$ in the context of DIS). In the DIS case this large logarithm resides in the  $\left.\sigma^{\gamma^{\ast}_{\rm T}}\right|_{q\bar{q}g}$ term in \eqn{eq:finalresult}, which contains an explicit integral over the longitudinal momentum fraction of the gluon in the diagrams of Figs.~\ref{fig:disloopdiags} and~\ref{fig:disraddiags}. When the gluon becomes very soft (i.e. the momentum fraction $z_g \to 0$), this integral yields a large logarithm of the lower cutoff, which kinematically is $\sim 1/x_\textrm{Bj}$. This contribution must  be subtracted from the calculation of the cross section and absorbed into the BK/JIMWLK evolution of the target
 in order to extract the finite genuinely NLO contribution.  This factorization procedure is much less straightforward in the case of small $x$ than for collinear factorization. It  is tied in with the issues of collinear logarithms or kinematical constraints in the NLO evolution discussed in \ref{sect:hee}. For DIS, there is a preliminary numerical implementation of the cross section formulae\cite{Ducloue:2017ftk}. This work shows that indeed the NLO corrections are sizeable (several tens of \%), but controllable. In particular, there is a large cancellation between the 
two different terms in \eqn{eq:finalresult}, making the overall NLO correction smaller than the two separate terms individually. However, a more systematical comparison with experimental data has not yet been achieved, and it is not obvious if the factorization procedure here is the optimal one. The issues here are largely the same as in forward particle production, which has been discussed more in the recent literature. Thus we will not discuss this further here, but return to the issue in that context in Sec.~\ref{sec:pasinc}.

\subsection{Massive quarks}

The calculation of the NLO DIS cross section discussed above was only performed with massless quarks. The total charm quark cross section is likely to be an important observable at the EIC. In particular, it is a more safely perturbative probe of weak coupling physics than the total cross section, being less sensitive to very large dipoles in the ``aligned jet'' (large $r$, small $z(1-z)$) configurations allowed even at large $Q^2$ for massless quarks~\cite{Mantysaari:2018zdd_83}. Currently the work to extend the calculation described here to massive quarks is ongoing, and we shall briefly discuss the new issues involved. 

In principle what one should calculate are precisely the same diagrams as for massless quarks. The presence of quark masses of course complicates somewhat the kinematics and the algebra in the loop integrals. More important, however, is that one must confront the known\cite{Mustaki:1990im,Burkardt:1991tj,Zhang:1993dd} issues with quark mass renormalization in light cone perturbation theory. 

To start, let us examine in more detail the structure of the elementary fermion-gauge boson vertex  of QED or QCD, for exaxmple for the emission of a gauge boson with momentum $k$ from a quark with momentum $p$, with the quark after the emission having 3-momentum $\ppvec = \pvec - \kvec$:
\begin{equation}
\biggl [ \bar{u}_{h'}(p') {\varepsilon \!\!/} ^*_\lambda(k) u_h(p)\biggr ] 
\end{equation}
Using the properties of the free spinors $u,\bar{u},v, \bar{v}$ and polarization vectors $\varepsilon(k)$, and 3-momentum conservation, this vertex can be expressed in terms of three independent Lorentz structures: 
$\bar{u}_{h'} \gamma^+ u_h \delta^{ij} q^i\epst_{\lambda}^{\ast j}$, 
$\bar{u}_{h'} \gamma^+  [\gamma^i,\gamma^j] u_h q^i\epst_{\lambda}^{\ast j}$ and $\bar{u}_{h'}  \gamma^+ \gamma^j  u_h
\, m_q \epst_{\lambda}^{\ast j}$. The first two are light cone  helicity conserving ones  ($\sim \delta_{h,h'}$) that are present independently of the quark mass. The third one is a light cone helicity flip term $\sim \delta_{h,-h'}$, and is explicitly proportional to the quark mass. Note also that the helicity flip vertex has one less power of the transverse momentum than the nonflip one, thus resulting in less UV divergent contributions. 

The inclusion of the additional helicity flip structure introduces two new kinds mass-dependent UV divergent contributions. Firstly, in the  first two diagrams of Fig.~\ref{fig:disloopdiags} one can have a flip vertex at both ends of the quark line. Since for massless quarks the transverse momentum integral in the loop is quadratically divergent, this yields a logarithmic UV divergence proportional to $m_q^2$. This divergence is absorbed into a renormalization of the quark mass squared appearing in the energy denominator of the leading order wave function. The corresponding counterterm is the ``kinetic mass'' counterterm, since it is associated with the kinetic property of the mass as a parameter in the dispersion relation relating the energy $p^-$ to the momentum $p^+,\bp$. Secondly, in the second two (vertex correction) diagrams of  Fig.~\ref{fig:disloopdiags} one can take one out of the three vertices to have a flip, and get a logarithmically divergent contribution proportional to the quark mass $m_q$. This divergence, on the other hand, separately from the other one, is absorbed into a renormalization of the quark mass appearing in the helicity flip part of the leading order vertex. The associated counterterm is referred to as the ``vertex mass'' one, since it is related to the role of the quark mass as the coefficent of the helicity flip amplitude in an interaction with gauge bosons. 

In a covariant formulation of the theory rotational invariance guarantees that both the kinetic and vertex masses remain the same at all orders in perturbation theory. However, in light cone quantization one chooses a specific coordinate axis as the longitudinal one. If the regularization method used in loop calculations is not rotationally invariant, the two counterterms can become different. This is indeed the case for the scheme of transverse dimensional regularization and a cutoff in the longitudinal ($p^+$) momentum that has been used in NLO calculations of DIS for mass. There are two possible ways to remedy this problem. One way is to introduce an additional renormalization condition  to separately determine the two mass counterterms by enforcing rotational invariance order by order in perturbation theory. A convenient possibility or the case of DIS is to require that the decay amplitudes of transversally and longitudinally polarized timelike virtual photons are equal. The other option is to examine more closely the regularization procedure in the case of the problematic diagrams, which in this case are in fact the propagator correction ones. By carefully combining them with the corresponding instantaneous interaction diagrams before integration, and evaluating a specific subset of one-body phase space integrals in a rotationally invariant manner, one can restore rotational invariance without an additional renormalization condition. The full details of this procedure will be explained in more detail in a forthcoming publication.

\section{Forward particle production in $pA$ collisions}
\label{sec:pasinc}

Particle production at forward rapidities and semi-hard transverse momenta in proton (or deuteron)-nucleus
collisions at RHIC and the LHC is an important source of information about the small-$x$ part of the nuclear
wavefunction, where gluon occupation numbers are high and non-linear effects like gluon saturation and
multiple scattering are expected to be important. On the theory side, the cross-section 
for single-inclusive particle production has been computed \cite{Chirilli:2012jd_59} 
 up to next-to-leading order (NLO) in the framework of the so-called ``hybrid factorization'' \cite{Dumitru:2005gt}, 
but the result is problematic: the cross-section suddenly turns negative when increasing the transverse
momentum of the produced hadron, while still in the semi-hard regime \cite{Stasto:2013cha}.
Various proposals to fix this difficulty, by modifying the scale for the rapidity subtraction, have
only managed to push the problem to somewhat larger values of the 
transverse momentum\cite{Stasto:2016wrf}.

In a recent paper\cite{Iancu:2016vyg}, it has have argued that this negativity problem is an
artefact of the approximations used within hybrid factorization in order to obtain
a result which looks local in rapidity. On that occasion, it has also been proposed a more general
factorization scheme, which is non-local in rapidity but yields a manifestly positive 
cross-section to NLO accuracy. Subsequently, this whole strategy has been extended to the 
calculation of the DIS structure functions at NLO~\cite{Ducloue:2017ftk}.  

Another subtle issue refers to the use of a running coupling within the non-local factorization. Indeed, 
the cross-section is written in momentum space, but it involves the solution to the BK equation, which is
most naturally solved in coordinate space; the mismatch between the respective prescriptions 
for the running of the coupling can lead into trouble\cite{Ducloue:2017mpb}. A solution 
to this difficulty that was recently proposed\cite{Ducloue:2017dit} will be briefly reviewed here.


\subsection{Leading order formalism}
\label{sec:1}

For simplicity, we focus here on the $q \to q$ channel and do not consider the fragmentation functions. 
To leading order (LO) in the ``hybrid factorization'', quark production at forward rapidities in $pA$ collisions 
proceeds as follows: a quark which is initially collinear with the incoming proton
scatters off the dense gluon distribution in the nuclear target and thus acquires a transverse 
momentum $\bk$. The  LO quark multiplicity is computed as follows:
\beq\label{LO}
 \frac{\rmd N^{{\rm LO}}}{\rmd^2\bk\, \rmd \eta}=\frac{x_p q(x_p)}{(2\pi)^2}\,
{\calS}(\bk,X_g)\,,\qquad {\calS}(\bk,X_g)=\int \rmd^2\br\, \rme^{-\rmi \bk \cdot \br} {S}(\br,X_g),
\eeq
where $\eta$ is the rapidity of the produced quark in the center-of-mass frame, $x_p q(x_p)$ is the quark distribution of the proton, and $x_p=({\kt}\rme^\eta/\sqrt{s})$ 
and $X_g=({\kt}\rme^{-\eta}/\sqrt{s})$ are the longitudinal momentum fractions carried by the partons
participating in the collision --- a quark from the proton and a gluon from the nucleus.
The {\em forward kinematics} corresponds to $\eta$ positive and large, 
which implies $X_g\!\ll\! x_p \!< \! 1$. 

Furthermore,  $\calS(\bk,X_g)$ is the relevant  unintegrated gluon distribution 
(the ``dipole TMD''), obtained as the Fourier transform of the $S$-matrix 
${S}(\br,X_g)$ for the  elastic 
scattering between a quark-antiquark color dipole with transverse size $\br$ and the nucleus. 
This quantity depends upon $X_g$ via the high energy evolution responsible for the rise in the
gluon distributions with decreasing $X_g$. To the LO accuracy at hand, this evolution is
described by (the LO version of the) the BK equation  \cite{Balitsky:1995ub_119,Kovchegov:1999yj},
which resums to all orders the radiative corrections $\propto (\abar Y_g)^n$, 
with $Y\equiv \ln(1/X_g)$.
These corrections are associated with successive emissions of soft gluons, which are strongly ordered in longitudinal momenta and hence can be computed in the eikonal approximation. 

The LO BK equation is boost invariant --- it equivalently describes the high-energy evolution of the
dipole projectile, or of the nuclear target. For what follows it is suggestive to vizualize
this evolution in a Lorentz frame in which 
the ``primary'' gluon (the one which is closest in rapidity to the dipole) is emitted by the dipole,
whereas all the other ``soft'' gluons belong to the nuclear wavefunction (see Fig.~\ref{QtoQG} for an
illustration). Then the LO BK equation can be written in  integral form,
\begin{align}\label{BKint}
 S\big(\bx,\by; X_g\big)=S(\bx,\by;X_0) &+
 \frac{\abar}{2\pi} \int_{X_g/X_0}^{1}\frac{\rmd x}{x}\int 
\frac{\rmd^2\bz \,(\bx-\by)^2}{(\bx-\bz)^2(\bz-\by)^2}
\nn& \times
\big[
 S\big(\bx,\bz; X(x)\big) S\big(\bz, \by; X(x)\big)
 -S\big(\bx,\by; X(x)\big)\big]\,,
 \end{align}
where $\bx$ and $\by$ are the transverse coordinates of the quark and antiquark legs of the
dipole (so, $\br=\bx\minus \by$) and $\bz$ is the transverse position of 
the primary gluon, which carries a fraction $x\ll 1$ of the longitudinal momentum of the incoming quark.
Furthermore, $X_0$ is the value of $X$ at which one starts the high-energy evolution 
of the target, $S(\bx,\by;X_0)$ is the corresponding initial condition (say, as given by
the McLerran-Venugopalan (MV) model \cite{McLerran:1993ni_95,McLerran:1993ka_101}),
and $X(x)\equiv X_g/x$ is the longitudinal momentum fraction of the gluons in the target which 
are probed by the scattering. Notice that $\ln(1/X(x))=\ln(x/X_g)=Y_g- \ln(1/x)$ is the rapidity 
separation between the primary gluon and the valence partons in the nucleus.

\begin{figure}[t] \centerline{
\includegraphics[width=.5\textwidth]{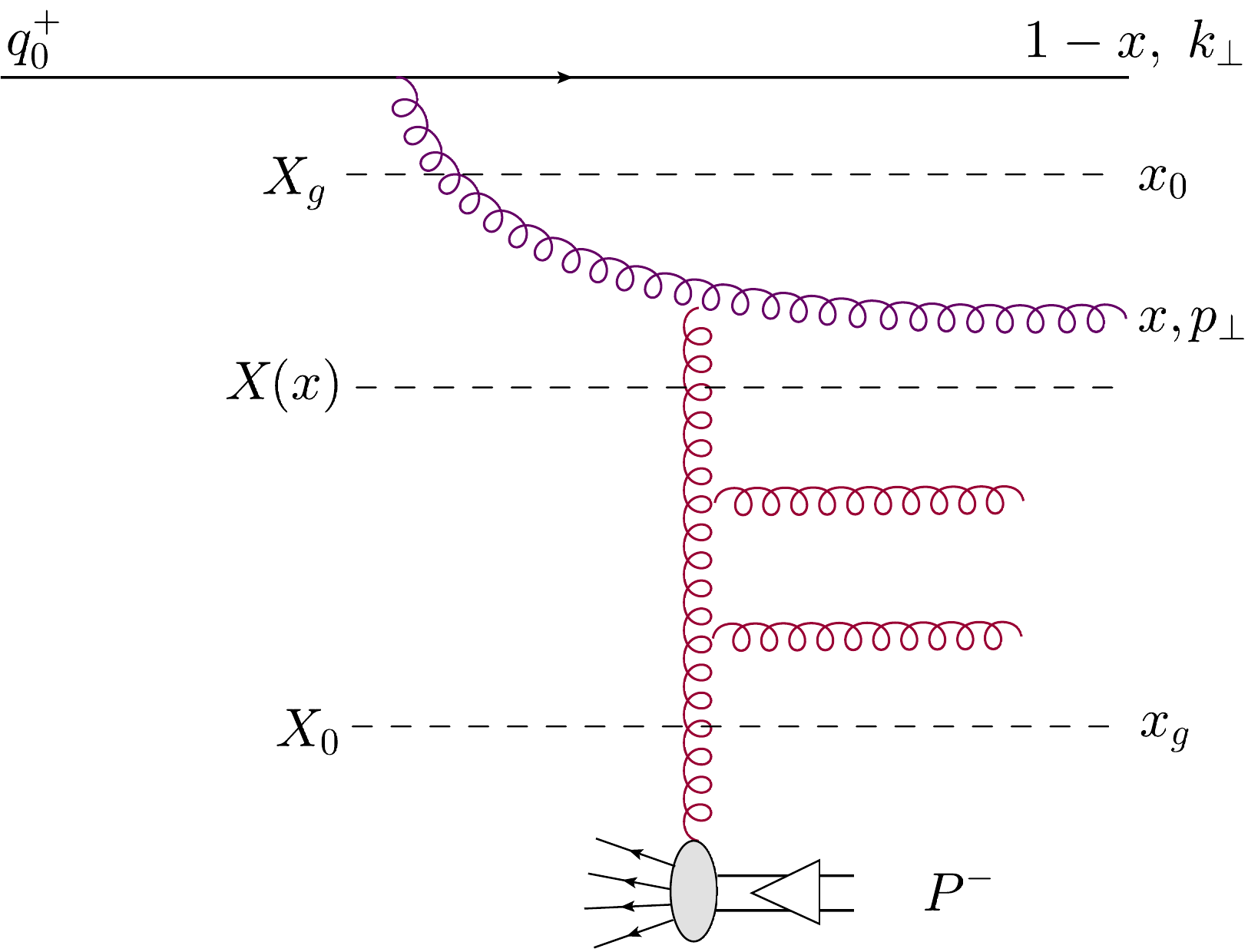}}
 \caption{\small A graph contributing to the amplitude for forward quark. 
 When the longitudinal fraction $x$ of the ``primary gluon''
 (the gluon directly emitted by the quark) is small, $x\ll 1$, this graph  is a part of the high-energy
 evolution of the LO multiplicity. But for generic values $x\sim 1$ (non-eikonal emission), 
 it contributes to the NLO impact factor. }
 \label{QtoQG}
\end{figure}

\subsection{The NLO impact factor}
\label{sec:2}

At NLO, one needs to also include the ``pure-$\alpha_s$'' corrections, i.e. the radiative corrections of $\order{\alpha_s}$ which are not enhanced by $Y_g$. These can be divided into two classes: \texttt{(i)} NLO corrections to the high-energy evolution, i.e. to the kernel (more generally, to the structure) of the BK equation, and  \texttt{(ii)} NLO corrections to the ``impact factor'', i.e., to the ``hard'' matrix element which describes the quark-nucleus scattering in the absence of any evolution, that is for $X_g\sim X_0$. 

The LO impact factor describes the scattering between a bare quark collinear with the proton and the nucleus.
At NLO, the wavefunction of the incoming quark may also contain a ``primary'' gluon with longitudinal fraction $x$ (cf.  Fig.~\ref{QtoQG}). For $x\ll 1$, this primary emission was already 
 included in the LO evolution, as manifest  in \eqn{BKint}. The NLO correction 
to the impact factor is rather associated with a relatively hard primary emission, with $x\sim \order{1}$, which must be computed {\it exactly} (i.e. beyond the eikonal approximation). In practice though, it turns out that separating the LO evolution from the NLO  correction to the impact factor is quite subtle. For this reason, we shall  first present an ``unsubtracted'' expression for the NLO quark multiplicity  \cite{Iancu:2016vyg} in which these two effects are mixed with each other. 


 The LO multiplicity~(\ref{LO}) receives NLO corrections proportional to the $\nc$ and $\cf$ color factors which have been computed in Refs.~\citenum{Chirilli:2011km_77,Chirilli:2012jd_59}. To keep the discussion simple, we shall only consider the $\nc$ terms which are\cite{Ducloue:2016shw} the origin of the negativity problem observed in Ref.~\citenum{Stasto:2013cha}. Besides, we shall treat the dipole evolution to LO  (the running coupling corrections will be later added).  Finally, we shall focus on relatively hard momenta $k_\perp \gtrsim Q_s(X_g)$ for the produced quark, since this is the most interesting case for the phenomenology and also the regime where the negativity problem has been observed in the literature~\cite{Stasto:2016wrf}.
 The sum of the LO and  NLO contributions proportional to $\nc$ can be written as \cite{Iancu:2016vyg}
  \beq
 \label{nlounsub}
 \frac{\dif N^{{\rm LO} + \nc}}{\dif^2\bk\, \dif \eta}  = 
 \frac{x_p q(x_p)}{(2\pi)^2}\,
 \mathcal{S}(\bk,X_0)
 +  \frac{\abar}{2\pi} \int_{X_g/X_0}^{1}\frac{\rmd x}{x}\, \mathcal{K}(\bk,x,X(x)),
 \eeq
where the first term in the r.h.s. is the tree-level contribution (or equivalently the initial condition at $X_g=X_0$), whereas the second term encodes, in compact but rather formal notations, all the quantum corrections which are
relevant to the accuracy of interest:  the $\nc$ piece of the NLO corrections to the impact factor and the LO BK evolution of the dipole $S$-matrix. The  kernel $\mathcal{K}(\bk,x,X(x))$ is built with vertices for the (generally, non-eikonal) emission of the primary gluon and with dipole $S$-matrices  --- evolved from $X_0$ down to $X(x)=X_g/x$ --- describing the scattering between the nuclear target and the 2-parton projectile (the original quark plus the primary gluon). Explicit expressions can be 
found in Refs.~\citenum{Chirilli:2011km_77,Chirilli:2012jd_59,Iancu:2016vyg,Ducloue:2017dit}.

\begin{figure}[t]
\centerline{
\includegraphics[width=.48\textwidth]{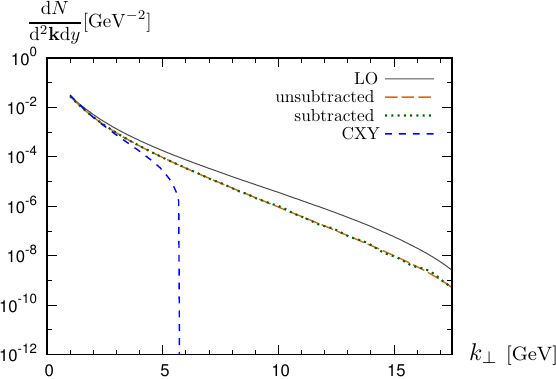}\qquad
\includegraphics[width=.48\textwidth]{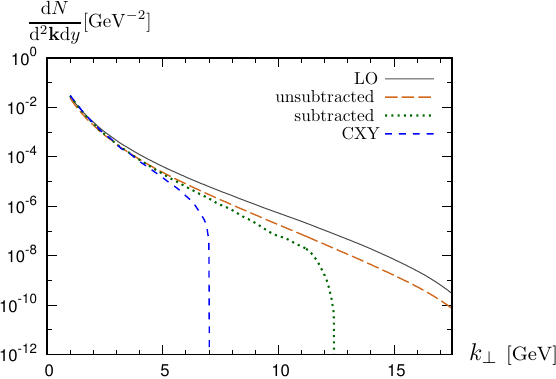}}
\caption{\sl Numerical results comparing the LO result, \eqn{LO},
 with different formulations of the NLO factorization:
``unsubtracted'', cf. \eqn{nlounsub}, ``subtracted'', cf.  \eqn{nlosub},  and
``CXY'', cf.  \eqn{nlocxy}. Left: fixed coupling. Right: Running coupling with mixed 
RC prescriptions: transverse-space prescription $\abar(\kt)$ for the
primary gluon emission, but coordinate-space (Balitsky) prescription for the BK equation.
These results are taken from Ref.~\cite{Ducloue:2017mpb}, to which we refer for more
details.}
\label{fig:num}
\end{figure}

We shall refer to the formula  in \eqn{nlounsub} as  ``unsubtracted'', since the NLO corrections are not explicitly
separated from the LO result.  This formula has been numerically evaluated in Ref.~\citenum{Ducloue:2017mpb}, 
with the result shown  in the left panel of Fig.~\ref{fig:num} (the curve denoted as ``unsubtracted''). 
This result is seen to be positive, as expected on physical grounds, and also smaller than the respective LO result --- meaning that the NLO corrections are negative. In order to disentangle these corrections from the LO contribution and also make contact with the original calculations\cite{Chirilli:2011km_77,Chirilli:2012jd_59,Stasto:2013cha}, it is useful to observe that the LO result \eqref{LO}  
can be recovered from \eqn{nlounsub} by taking the eikonal limit $x\to 0$ in the emission vertices
 (while keeping the $x$-dependence in the rapidity arguments $X(x)$ of the dipole $S$-matrices); that is,
\begin{align}
	\label{nlolo}
	\frac{\dif N^{{\rm LO}}}{\dif^2\bk\, \dif \eta} 
	=  &
	\frac{x_p q(x_p)}{(2\pi)^2}
	\mathcal{S}(\bk,X_0)	+
	\frac{\abar}{2\pi} \int_{X_g/X_0}^{1}\frac{\rmd x}{x}\,  \mathcal{K}(\bk,x=0,X(x)).
\end{align}
This is indeed consistent with equations \eqref{LO} and \eqref{BKint}, since the integral term above is the same as the Fourier transform of the respective term in \eqn{BKint} times ${x_p q(x_p)}/{(2\pi)^2}$. 
By subtracting \eqn{nlolo} from \eqn{nlounsub}, one finds
 \beq
 \label{nlosub}
 \frac{\dif N^{{\rm LO} + \nc}}{\dif^2\bk\, \dif \eta}  = \frac{\dif N^{\rm LO}}{\dif^2\bk\, \dif \eta}
 +\frac{\abar}{2\pi} \int_{X_g/X_0}^{1}\frac{\rmd x}{x} \, \big[\mathcal{K}(\bk,x,X(x))
 -\mathcal{K}(\bk,x=0,X(x))\big].
 \eeq
The equivalence between Eqs.~\eqref{nlounsub} and \eqref{nlosub} is confirmed by the numerical study in Ref.~\citenum{Ducloue:2017mpb}:  the ``subtracted'' curve in Fig.~\ref{fig:num} (left), 
as obtained by numerically computing the r.h.s. of  \eqn{nlosub}, perfectly matches the ``unsubtracted'' curve obtained from  \eqn{nlounsub}. But from the previous discussion,  it should be clear that this equivalence
 relies in an essential way on the fact that the dipole $S$-matrix is an exact solution to the LO BK equation. 
 Any approximation in solving this equation or in evaluating the NLO correction
(the integral term) in the r.h.s. of \eqn{nlosub} would lead to differences with potentially dramatic consequences.
This observation is important in view of the fact that the ``subtracted'' formula  \eqref{nlosub}
 is not quite the same as the NLO prediction~\cite{Chirilli:2011km_77,Chirilli:2012jd_59} 
 of the hybrid factorization  (or the closely related $k_T$-factorization).
 The latter is local in rapidity, that is, the NLO correction to the impact factor is fully factorized from the high-energy evolution, which is evaluated at the scale $X_g$ (corresponding to the maximal rapidity separation between the projectile and the target). By contrast, the NLO correction in \eqn{nlosub} is {\em non-local in rapidity}~:  it involves the dipole evolution at all the intermediate scales $X_0 > X(x) > X_g$.
 
In order to arrive at the ``$k_T$-factorized'' formula presented in \cite{Chirilli:2011km_77,Chirilli:2012jd_59}, which is local in $X$, certain approximations need to be made. First, one observes that due to the subtraction in \eqn{nlosub}, the integral is dominated by large values $x\sim 1$. Hence, to the NLO accuracy of interest, it is justified to \texttt{(i)}  replace the rapidity argument of the dipole $S$-matrices  by its value at $x=1$, i.e.~$X(x) \to 
X(1)=X_g$, and \texttt{(ii)}  ignore the lower limit $X_g/X_0\ll 1$ in the integral over $x$.  One thus obtains
 \beq
 \label{nlocxy}
 \frac{\dif N^{{\rm LO} + 
c}}{\dif^2\bk\, \dif \eta}\bigg|_{\rm CXY}  = 
 \frac{\dif N^{\rm LO}}{\dif^2\bk\, \dif \eta} + \int_{0}^1
 \frac{\dif x}{x} \, \big[\mathcal{K}(\bk,x,X_g)
 -\mathcal{K}(\bk,x=0,X_g)\big],
 \eeq
where all the $S$-matrices implicit in the r.h.s. are evaluated at the scale $X_g$.

\eqn{nlocxy} is not anymore equivalent to Eqs.~\eqref{nlounsub} and \eqref{nlosub} and, despite the seemingly reasonable approximations, it is rather pathological, as it rapidly becomes negative when increasing the transverse momentum of the produced quark. This is demonstrated by the curve ``CXY'' in the left panel of 
Fig.~\ref{fig:num},  obtained~\cite{Ducloue:2017mpb} by numerically evaluating the r.h.s. of \eqn{nlocxy}. The reason is that the replacement $X(x)\to X_g$ in the argument of the dipole $S$-matrix leads to an over-subtraction: the negative contribution proportional to $\mathcal{K}(\bk,x=0,X_g)$ becomes too large in magnitude and overcompensates for the LO piece. Moreover, the replacement $X_g/X_0\to 0$ in the lower limit is not physically motivated, since it violates constraints imposed from the correct kinematics, and thus it introduces spurious contributions.

\subsection{Adding a running coupling}

So far, we have considered the academic case of a fixed coupling, 
but it is quite clear that, both for the consistency
of the NLO calculation and for realistic applications to phenomenology, 
it is necessary to include the effects of the running of the coupling. Since the
transverse momentum $k_\perp$ of the produced quark is the largest scale in
the problem, it is intuitively clear that this is also the scale which controls
the running of the coupling. It therefore looks reasonable to generalize the previous
results by simply replacing $\abar\to\abar(k_\perp)$ in the primary emission vertices
in equations \eqref{nlounsub} and \eqref{nlosub}, while simultaneously using rcBK
(the LO BK equation with a running coupling) for the evolution of the dipole
$S$-matrix.  Yet, a moment of thinking reveals that such a procedure is not exempt of
ambiguities, which in some cases may lead to serious problems. 
We now present several examples in that 
sense~\cite{Iancu:2016vyg,Ducloue:2017mpb,Ducloue:2017dit}.

In practice, it is preferable (for many good reasons~\cite{Ducloue:2017mpb})
to solve rcBK in the transverse coordinate space, meaning that the respective running coupling 
(RC) prescription must be formulated in coordinate space as well. Clearly, the scale dependence of
the RC does not ``commute'' with the Fourier transform (FT), e.g.
\beq\label{noncom}
\abar(k_\perp)\mathcal{K}(\bk,x,X(x)) \,\ne\,
\int \rmd^2\br\, \rme^{-\rmi \bk \cdot \br} \abar(r_\perp) {K}(\br,x,X(x))\,,
\eeq
where $ {K}(\br,x,X(x))$ is the FT of $\mathcal{K}(\bk,x,X(x))$. One may interpret
this mismatch as merely a part of our scheme dependence, but in some cases it may have 
dramatic consequences.

First, it spoils the equivalence between the ``unsubtracted'' and ``subtracted''  expressions for the NLO 
multiplicity. Recall indeed that in going from \eqn{nlounsub}  to \eqn{nlosub}, we have used the fact
that  the integral term in \eqn{nlolo} coincides with the Fourier transform of the r.h.s. of the LO BK equation.
Clearly, this property is spoilt after replacing $\abar\to\abar(k_\perp)$ in \eqn{nlolo}, while at the same
time using the coordinate-space version of the rcBK equation.  Due to the fine-tuning inherent in the derivation
of the ``subtracted'' expression, any such a mismatch could lead to a resurgence of the negativity problem.
This is indeed observed by the numerical study in~\cite{Ducloue:2017mpb} (see
the right plot of  Fig.~\ref{fig:num}): whereas the  ``unsubtracted'' result remains positive and shows
a similar trend as at fixed coupling, the ``subtracted'' one eventually turns negative, albeit at some 
larger value for $\kt$ than for the ``$k_T$-factorized'' expression \eqref{nlocxy} (now extended to a RC).

In order to cope with such issues, an alternative
numerical implementation in which the calculation is fully performed in coordinate space was suggested\cite{Ducloue:2017mpb}. In particular,
the integral term in Eqs.~\eqref{nlounsub} or \eqref{nlosub} is constructed as in the r.h.s. of \eqn{noncom},
that is, as the FT of a quantity originally computed in coordinate space. To get more insight on the role
of the RC in this context, it is instructive to consider the eikonal limit $x\to 0$, in which  \eqn{nlounsub} 
reduces to the LO result in \eqn{nlolo}. In coordinate space and with a RC, the integrand there should be
understood as
\beq\label{locoord}
``\abar(r_\perp) {K}(\br,x=0)'' \,=\,\frac{x_p q(x_p)}{(2\pi)^2}
 	 	 \int \frac{\dif^2 \bx}{2\pi} 
	\frac{\abar(\br)\br^2}{\bx^2(\bx \minus \br)^2}
	\left[S(\bx) S(\br - \bx) - S(\br) \right],
\eeq
where $\abar(\br)$ within the integrand generically refers to any coordinate-space RC prescription which 
also depends upon the size $\br$ of the parent dipole; e.g., the smallest dipole prescription $
\abar(r_{\rm min})$, with $r_{\rm min} \equiv {\rm min}\{|\br|, |\bx|,|\br-\bx|\}$. The generalization of
\eqn{locoord} to generic  (non-eikonal) values of $x$ can be found in Refs.~\citenum{Ducloue:2017mpb,Ducloue:2017dit}.

The final results for the NLO quark multiplicity obtained via this procedure\cite{Ducloue:2017mpb}
turned out to be extremely peculiar and physically unacceptable: not only they are dramatically different from
the results obtained with the momentum-space prescription $\abar(k_\perp)$, but they also show an unphysical
trend: the NLO corrections are very large and positive, and rapidly increase
 with $k_\perp$ (see the comparison between
the curves ``rcBK$(r_{\rm min})$'' and ``$\abar(k_\perp)$'' in the left plot in Fig.~\ref{fig:NLOratio}).

\begin{figure*}[t]
\begin{center}
\begin{minipage}[b]{0.49\textwidth}
\begin{center}
\includegraphics[width=0.95\textwidth,angle=0]{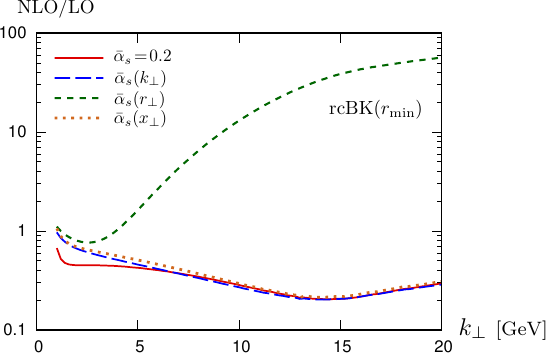}
\end{center}
\end{minipage}
\begin{minipage}[b]{0.49\textwidth}
\begin{center}
\includegraphics[width=0.95\textwidth,angle=0]{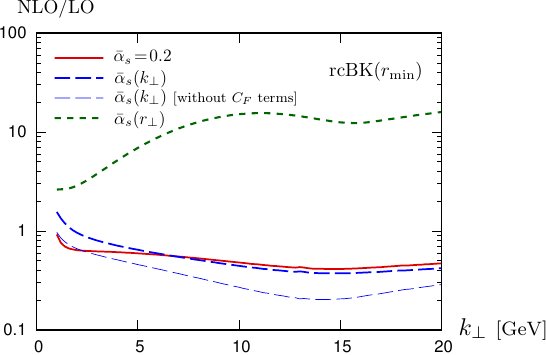}
\end{center}
\end{minipage}
\end{center}
\caption{\label{fig:NLOratio} Left: Ratio of the NLO multiplicity (including only the $\nc$ terms) and the LO one for different prescriptions of the running coupling. Right: Ratio of the total NLO quark multiplicity (including both the $\nc$ and $\cf$ terms) and the LO one for three running coupling prescriptions. For comparison, we also show the results for $\abar(k_{\perp})$ when including only the $\nc$ terms (same as the curve ``$\abar(k_\perp)$'' in the left panel). For both figures $\sqrt{s}=500~$GeV, $\eta=3.2$ and the evolution of the color dipoles is obtained by solving the Balitsky-Kovchegov equation with the smallest dipole prescription using an MV~\cite{McLerran:1993ni_95} initial condition at $X_0=0.01$.}
\end{figure*}

The origin of this problem and also a solution to it have been identified\cite{Ducloue:2017dit}.
Once again, this is related to the non-commutativity of the RC with the FT. We have no place here to
explain the precise technical problem in the context of the NLO multiplicity, but we can illustrate it
with a simpler example\cite{Ducloue:2017dit}: this is \eqn{noncom} with the kernel
$K$ replaced by the dipole $S$-matrix, evaluated at tree-level and in the single scattering approximation:
$S=1-T$ with $T={r_{\perp}^2 Q_s^2}\ln \frac{1}{r_{\perp}^2 \Lambda^2}$. The FT of $S$ (i.e. the
analog of the l.h.s. of  \eqn{noncom}) is controlled by the logarithmic singularity of $T$ as $r_\perp\to 0$,
which yields
\beq
 \label{nkres}
 \abar(k_{\perp}) \int \dif^2 \br\, \rme^{-\rmi \bk \cdot \br}
  \left( 
 -{r_{\perp}^2 Q_s^2}\,\ln \frac{1}{r_{\perp}^2 \Lambda^2}
 \right)= \frac{16\pi \abar(k_{\perp})Q_s^2}{k_{\perp}^4}.
 \eeq
This is the expected result for the high-$k_{\perp}$ tail of the quark multiplicity within the MV model.
On the other hand, when inserting the RC $\abar(\br)$ inside the integrand, the mathematics goes very differently:
the RC itself  has a logarithmic singularity as $r_\perp\to 0$, due to the asymptotic freedom, 
which now dominates the FT:
\beq\label{nrres} 
\int \dif^2 \br\,\rme^{-\rmi \bk \cdot \br}  \abar(r_{\perp}) S(\br)\simeq
\int \dif^2 \br\,\rme^{-\rmi \bk \cdot \br}  \abar(r_{\perp}) \simeq
- \frac{4 \pi}{\bar{b} [\ln (k_{\perp}^2/\Lambda^2)]^2}\,\frac{1}{k_{\perp}^2}\,.\eeq
(We have used $\abar(r_{\perp}) = \big[\bar b \ln \frac{1}{r_{\perp}^2 \Lambda^2}\big]^{-1}$.) Both the sign
and the power law tail in \eqn{nrres} are different from the correct ones in \eqn{nkres}. This is very similar
with the results for the coordinate-space calculation ``rcBK$(r_{\rm min})$'' in  Fig.~\ref{fig:NLOratio}.

This example suggests that the difficulty encountered with the coordinate-space calculation in Ref.~\citenum{Ducloue:2017mpb} is due to the fact that the argument of the RC depends upon
the parent dipole size $\br$ (the coordinate involved in the FT). Accordingly, the solution to this
problem as suggested in Ref.~\citenum{Ducloue:2017dit} consists in using a different RC prescription,
which is independent of $\br$ and hence commutes with the FT: in the notations of \eqn{locoord},
this is the {\em daughter dipole} prescription $\abar(x_\perp)$. With this prescription, the 
coordinate-space calculation becomes remarkably close to that using the momentum-space 
prescription $\abar(\kt)$ (see the left plot in Fig.~\ref{fig:NLOratio}).
 
This being said, the daughter dipole prescription is not ideal either: First, it is so finely-tuned that one
cannot study the scheme-dependence of the calculation. Second, it cannot be extended to the NLO
corrections proportional to the quark Casimir $\cf$. We thus consider that the most physical choice
in the general case
is the momentum space prescription $\abar(k_{\perp})$. In Fig.~\ref{fig:NLOratio}~(right) we show the results we obtain when including both the $\nc$ and $\cf$ NLO corrections with fixed, momentum-space and 
coordinate-space  RC. For comparison we also show the results obtained with the momentum space prescription including only the $\nc$ NLO terms. This allows us to see that the inclusion of the $\cf$ terms has a sizable effect and, being opposite in sign compared to the $\nc$ terms, they reduce the size of the NLO corrections to the cross-section.  This cancellation is similar to the one for the DIS cross section  discussed in Sec.~\ref{sec:dis} between the $q\bar{q}$-term (with color factor $\cf$) and the $q\bar{q}g$-term that has the color factor $\nc$ of the BK equation.

{\bf Acknowledgments}
The work of E.I. was supported in part by the Agence Nationale de la Recherche project 
 ANR-16-CE31-0019-01.  T.~L. has been supported by the Academy of Finland, project 321840, and by the European Research Council, grant ERC-2015-CoG-681707. The content of this article does not reflect the official opinion of the European Union and responsibility for the information and views expressed therein lies entirely with the authors.

\newpage

%
%
%
\renewcommand*{\FigPath}{./WeekVI/03-Yulia_Charm/Figs}
%
%
%
\wstoc{Probing nuclear gluons with heavy flavor production at EIC}{Yulia~Furletova, Nobuo~Sato, Christian~Weiss}

\title{Probing nuclear gluons with heavy flavor production at EIC}
\author{Yulia~Furletova$^*$, Nobuo~Sato, Christian~Weiss}
\index{author}{Furletova, Yu.}
\index{author}{Weiss, C.}
\index{author}{Sato, N.}

\address{Jefferson Lab, Newport News, VA 23606, USA \\
$^*$E-mail: yulia@jlab.org}
\begin{abstract}
The nuclear modifications of the parton densities in different regions of $x$
(EMC effect, antishadowing, shadowing) reveal aspects of the fundamental QCD 
substructure of nucleon interactions in the nucleus. 
We study the feasibility of measuring nuclear gluon densities at large $x$
using open heavy flavor production (charm, beauty) in DIS at EIC.
This includes (a) charm production rates and kinematic dependences; 
(b) charm reconstruction at large $x_B$ using exclusive and inclusive modes, 
enabled by particle identification
and vertex detection; (c) impact of inclusive charm data on nuclear gluon density.
\end{abstract}


\bodymatter

\section{Introduction}\label{aba:sec1_346}
Measurements of nuclear parton densities are an essential part of the EIC physics program.\cite{Accardi:2011mz} 
The nuclear modifications in different regions of $x$ (EMC effect, antishadowing, shadowing) reveal aspects 
of the fundamental QCD substructure of nucleon interactions in the nucleus. Of particular interest are 
the modifications of the nuclear gluon densities at large $x$, i.e., their possible suppression at $x > 0.3$ 
(gluonic EMC effect) or enhancement at $x\sim  0.1$ (gluon antishadowing). Nuclear gluon densities 
at $x > 0.1$ have so far been determined only indirectly, through the $Q^2$ dependence of inclusive 
nuclear DIS cross sections (DGLAP evolution). 

Open heavy flavor production (charm, beauty) in DIS provides a direct probe of the gluon density 
in the target. At leading order (LO) in the pQCD expansion the heavy quark pair is 
produced through the photon–gluon fusion process (see Fig.~\ref{fig:charm_production}a);
higher-order QCD corrections have been and are under good theoretical control
(uncertainties, stability).\cite{Baines:2006uw} Extensive measurements of open charm and beauty 
production have been performed at HERA in $ep$ DIS at $x_B < 10^{-2}$ and found good agreement 
with the QCD predictions.\cite{hera_inclusive,Abramowicz:2013eja} 

The EIC would make it possible to use heavy flavor production as a probe of nuclear gluon densities. 
The EIC luminosity of $\sim 10^{34}$ cm$^{-2}$ s$^{-1}$ 
per nucleon ($\sim 10^2$ times higher than HERA) would substantially increase the 
heavy flavor production rates and allow one to extend such measurements to $x_B$ $\gtrsim$ 0.1.
Next-generation detection capabilities (particle identification or PID, vertex detection) would enable
new methods of charm reconstruction at large $x_B$ using exclusive and inclusive modes.
Altogether such measurements could significantly advance knowledge of nuclear gluon densities at large $x_B$.
Their feasibility and impact should therefore be studied with high priority and
inform the detector design.\cite{LD1601,Aschenauer:2017oxs_567} Here we report some results 
of the study of Ref.~\cite{LD1601} and on-going efforts.

\section {Charm production at large $x_B$.}
Charm production rates in $eN$ DIS have been estimated using QCD cross
sections and phase space integration (LO formulas,
HVQDIS LO/NLO code\cite{Harris:1997zq}).
Figure~\ref{fig:charm_production}b shows the charm rates differentially
in $x_B$ (5 bins per decade) and integrated over $Q^2$ (two different
lower limits), for an integrated luminosity of 10 fb$^{-1}$;
it also shows the total DIS rates in the same bins.
One observes: (a) The charm production rates decrease
rapidly above $x_{\rm B} \sim 0.1$, due to the drop of the gluon
density. Nevertheless charm rates of few $\times 10^{5}$ are
achieved at $x_{\rm B} \sim 0.1$ with 10 fb$^{-1}$ integrated
luminosity. (b) The fraction of DIS events with charm production at 
$x_{\rm B} \sim 0.1$ is $\sim 1\%$ for $Q^2 > 5$ GeV$^2$
($\sim 2\%$ for $Q^2 > 20$ GeV$^2$). These observations define the baseline 
requirements for charm reconstruction at large $x_B$ with EIC: 
the charm reconstruction efficiency should be $\gtrsim$10\%, 
and the reconstruction methods have to work in an environment 
where charm events constitute only $\sim$1\% of DIS events.

Fig.~\ref{fig:charm_production}c shows the rapidity distributions of the
produced $c\bar c$ pairs in collider experiments. As an example it shows
the distributions for two different electron/nucleon beam energies, 5$\times$100 GeV 
and 10$\times$50 GeV, corresponding to the same CM energy $s =$ 2000 GeV$^2$. 
One sees that charm pairs at $x_B \sim 0.1$ are produced at central rapidities, 
where good PID and vertex detection is provided by the central detector.
%
%
\begin{figure*}[t]
\begin{tabular}{ll}
\parbox[c]{0.45\textwidth}{\includegraphics[width=0.45\textwidth]{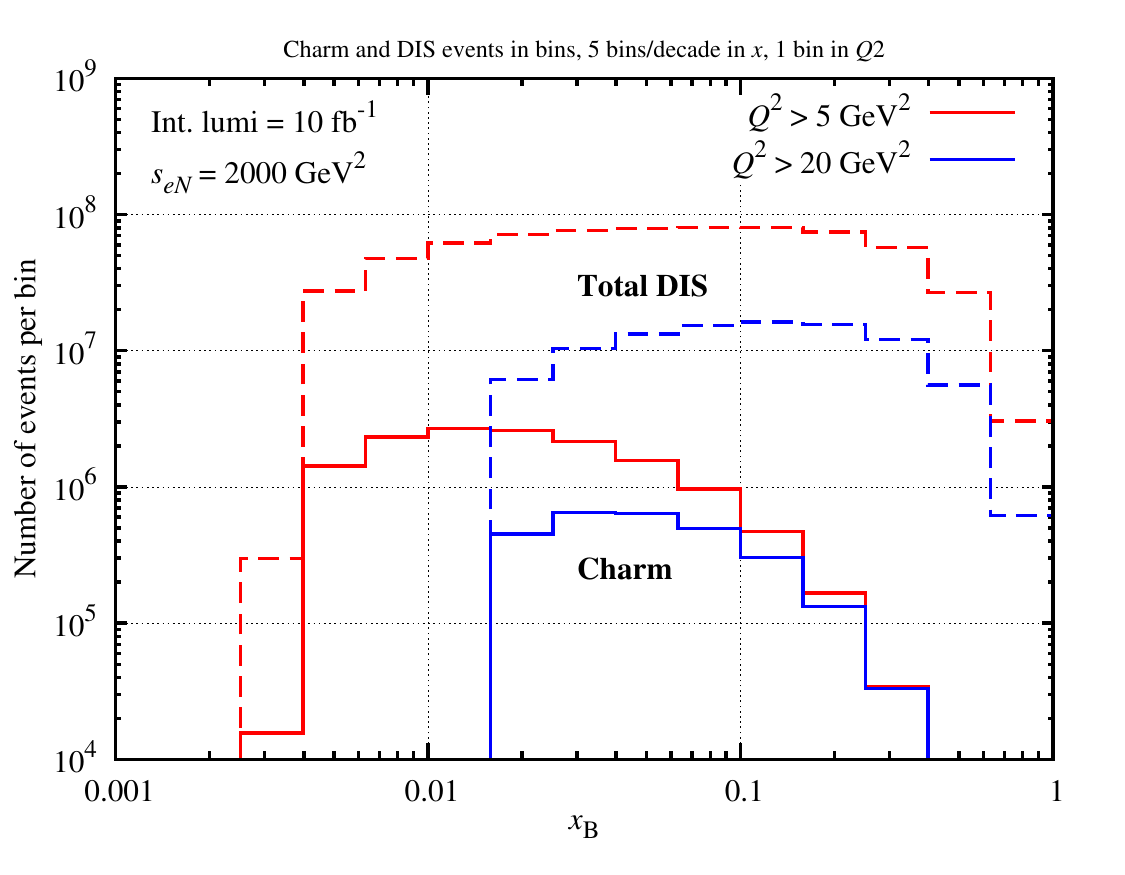}}
&
\parbox[c]{0.52\textwidth}{\includegraphics[width=0.52\textwidth]{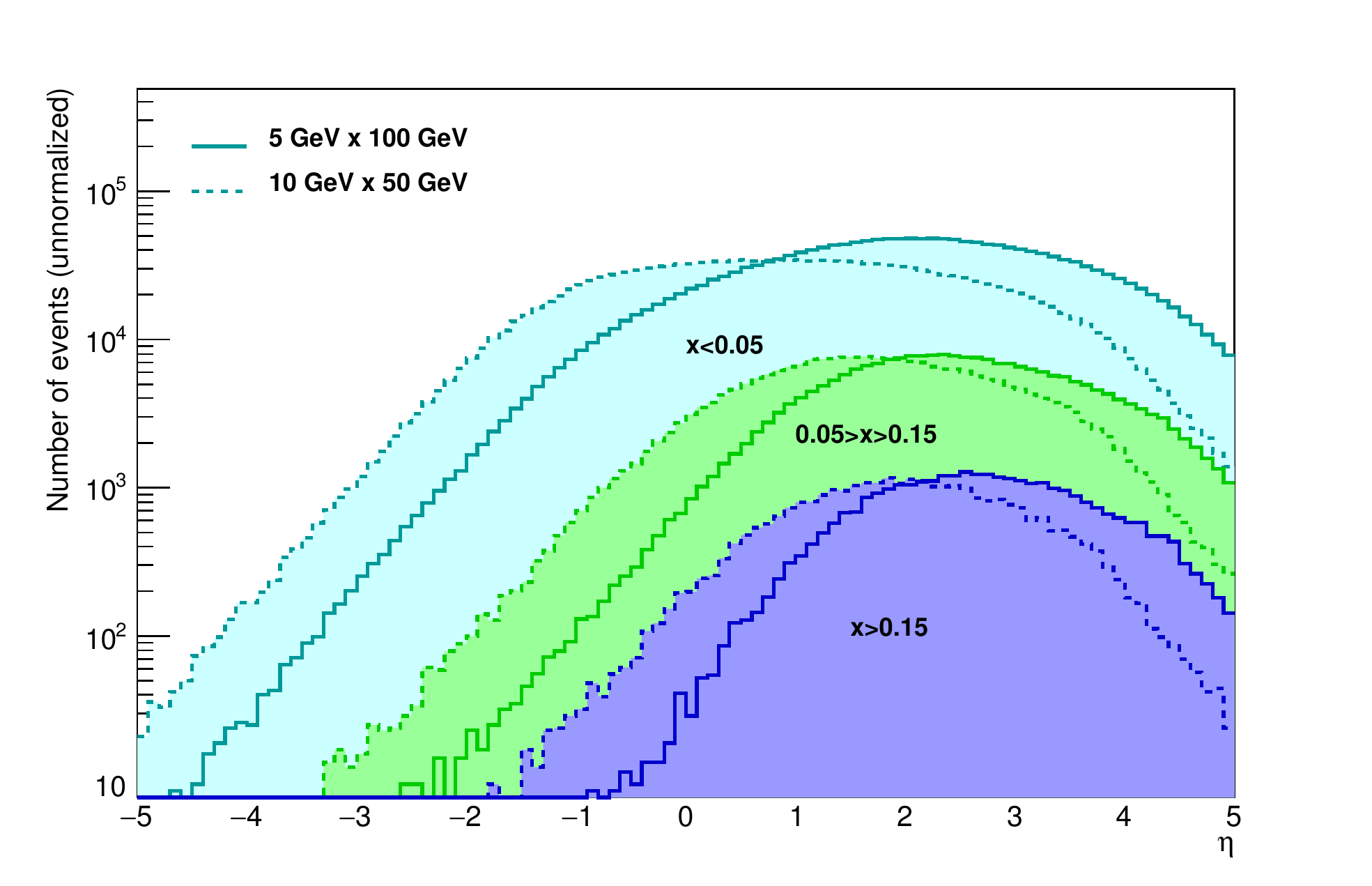}}
\\[-1ex]
{\small (b)} & {\small (c)}
\end{tabular}
\\[0ex]
\begin{tabular}{ll}
\parbox[c]{0.25\textwidth}{\includegraphics[width=0.25\textwidth]{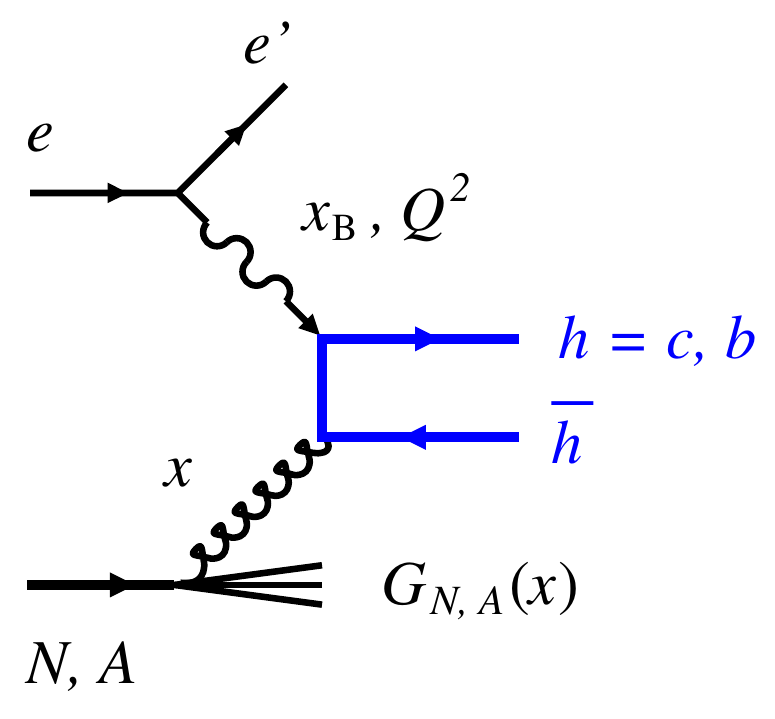}}
\hspace{0.05\textwidth}
\parbox[c]{0.65\textwidth}{
\caption[]{(a) Charm production in DIS in LO QCD. (b)
Estimated number of DIS events (dashed lines) and charm events (solid lines) 
in DIS at EIC (CM energy $s_{eN} = 2000$ GeV$^2$, integrated nucleon luminosity 
10 fb$^{-1}$). The bins in $x_{\rm B}$ are 5 per decade as indicated on the plot. 
$Q^2$ is integrated from the lower value indicated (5 or 20 GeV$^2$) to the 
kinematic limit at the given $x_B$.
(c) Rapidity distribution of $c\bar c$ pairs produced in in DIS events with
$x_B = 0.1$, for electron/nucleon beam energies
5$\times$100 GeV and 10$\times$50 GeV. \label{fig:charm_production}
}}
\\[-1ex]
{\small (a)} &
\\[-2ex]
\end{tabular}
\end{figure*}
\section {Charm reconstruction at large $x_B$ with EIC}
Charm events are identified by reconstructing the $D$ mesons that are produced 
by charm quark fragmentation and subsequently decay into $\pi$ and $K$.
Charm reconstruction at large $x_B$ relies essentially on the PID and vertex 
detection capabilities of the EIC detectors and has been simulated using 
a schematic detector model \cite{LD1601}. Two different methods 
are being considered: \\
(a) {\it Exclusive method}, in which
individual $D$ mesons are reconstructed from exclusive decays into
charged hadrons (see Fig.~\ref{fig:charm_exclusive}a). Experiments at 
HERA-I made extensive use of 
the $D^\ast$ channel, which exhibits a distinctive two-step decay 
$D^\ast \rightarrow D^0 \pi^+({\rm slow}), D^0 \rightarrow K^- \pi^{+}$, 
and can be reconstructed without PID or vertex detection, but provides a
reconstruction efficiency of only $\sim$1\%.\cite{Abramowicz:2013eja} 
 At EIC the PID and vertex capabilities
allow one to use also other $D$-meson decays ($D^0, D^+, D^+_s$);
summing these channels increases the overall efficiency to $\sim$6\%.\cite{Abramowicz:2013eja} 
\\
(b) {\it Inclusive method}, in which $D$
mesons are identified through inclusive decays with a displaced vertex
using the decay length significance distributions (hadronic and
semileptonic decays, see Fig.~\ref{fig:charm_exclusive}.\cite{hera_inclusive} 
This technique was used at HERA-II for charm reconstruction at $x_B < 10^{-2}$. 
At EIC the combination with PID and improved vertex detection would
make it possible to use this method at larger $x_B$. The overall efficiency 
achievable with this method is estimated at $\sim$25\%. 

Both charm reconstruction methods are expected to be applicable
in the high-background environment of DIS at $x_B \gtrsim 0.1$. 
The overall efficiency would be sufficient for $F_{2c}$ measurements 
and large-$x$ gluon density extraction. Systematic uncertainties need to be 
studied with a detailed detector design.
%
%
\begin{figure*}[t]
\begin{tabular}{ll}
\includegraphics[width=0.63\textwidth]{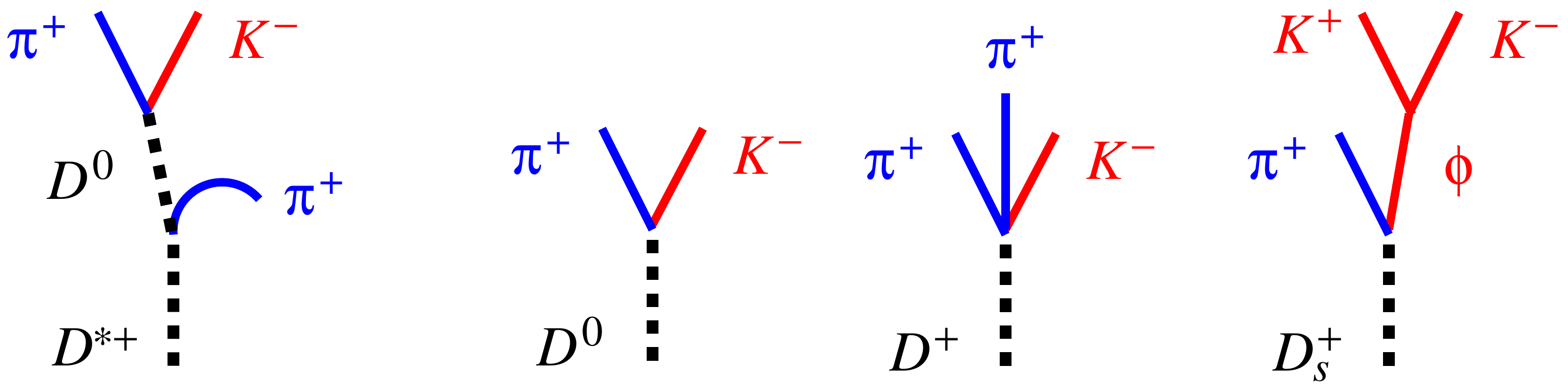}
&
\includegraphics[width=0.24\textwidth]{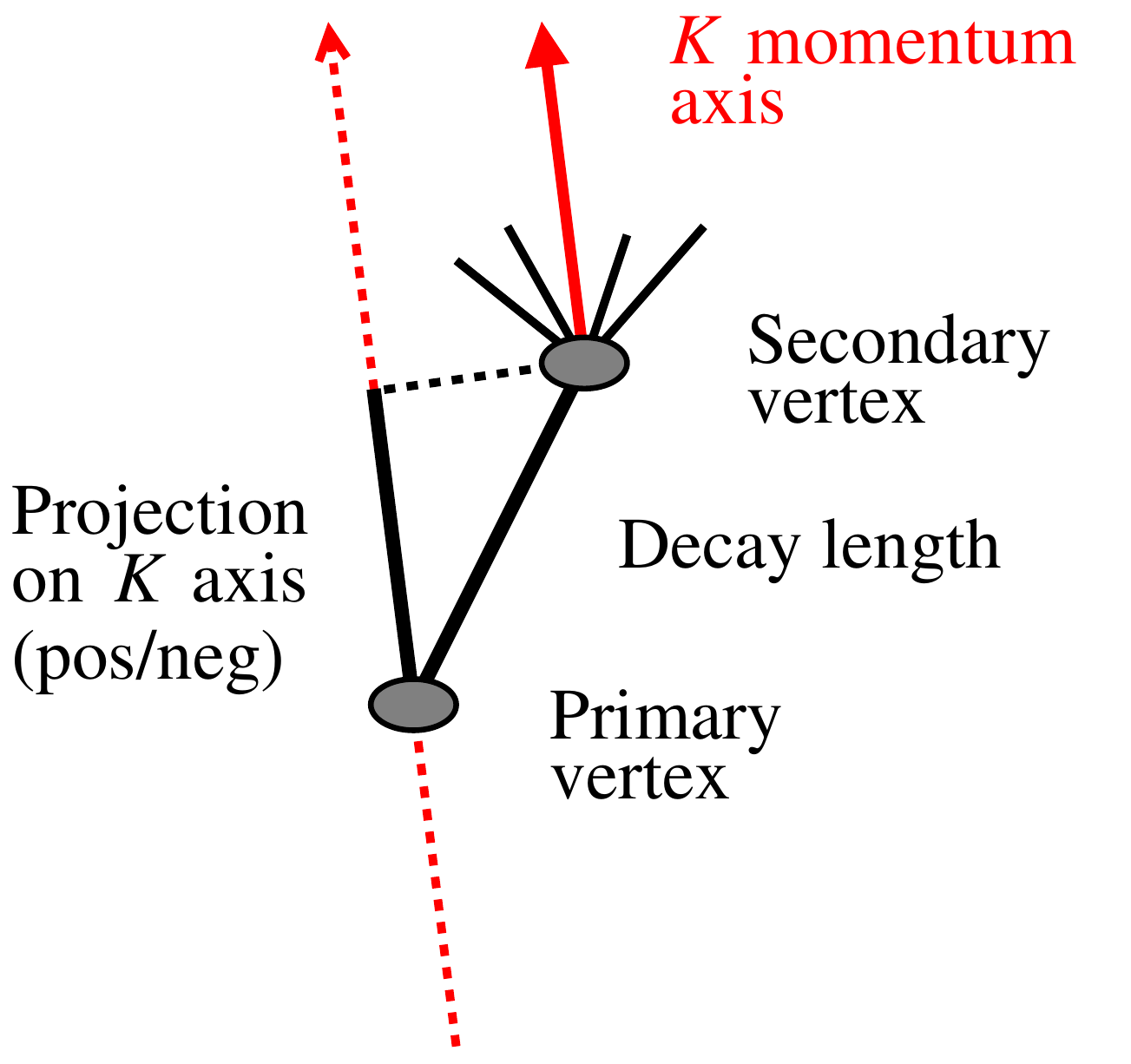}
\\[-.5ex]
{\small (a)} & {\small (b)}
\end{tabular}
\caption[]{(a) Exclusive $D$ meson decays into charged $\pi/K$ final states.
(b) Charm reconstruction using the decay length significance distribution in inclusive $D$-meson decays.
\label{fig:charm_exclusive}}
\end{figure*}
\section{Impact on large-$x$ nuclear gluons} 
The impact of open charm measurements at EIC on the nuclear gluon densities has been 
studied using a Monte-Carlo reweighting method~\cite{Sato:2016tuz}. Figure~\ref{fig:charm_impact}a
shows a sample set of pseudodata for the nuclear charm structure function $F_{2A}^c(x, Q^2)$; 
the errors in the measured region are dominated by systematics and estimated at $\sim$10\%. 
(The $F_{2A}^c$ measurements could be extended to larger $x_B$, where statistical errors dominate.) 
Figure~\ref{fig:charm_impact}b shows the impact of the pseudodata on the nuclear gluon density 
(here EPS09 LO;\cite{Eskola:2009uj_567} NLO simulations are in progress). 
One sees that the charm data substantially reduce the gluon uncertainties at $x > 0.1$ 
and would allow one to establish the presence of a ``gluonic EMC effect.'' 
%
%
\begin{figure*}[t]
\begin{tabular}{ll}
\includegraphics[width=0.45\textwidth]{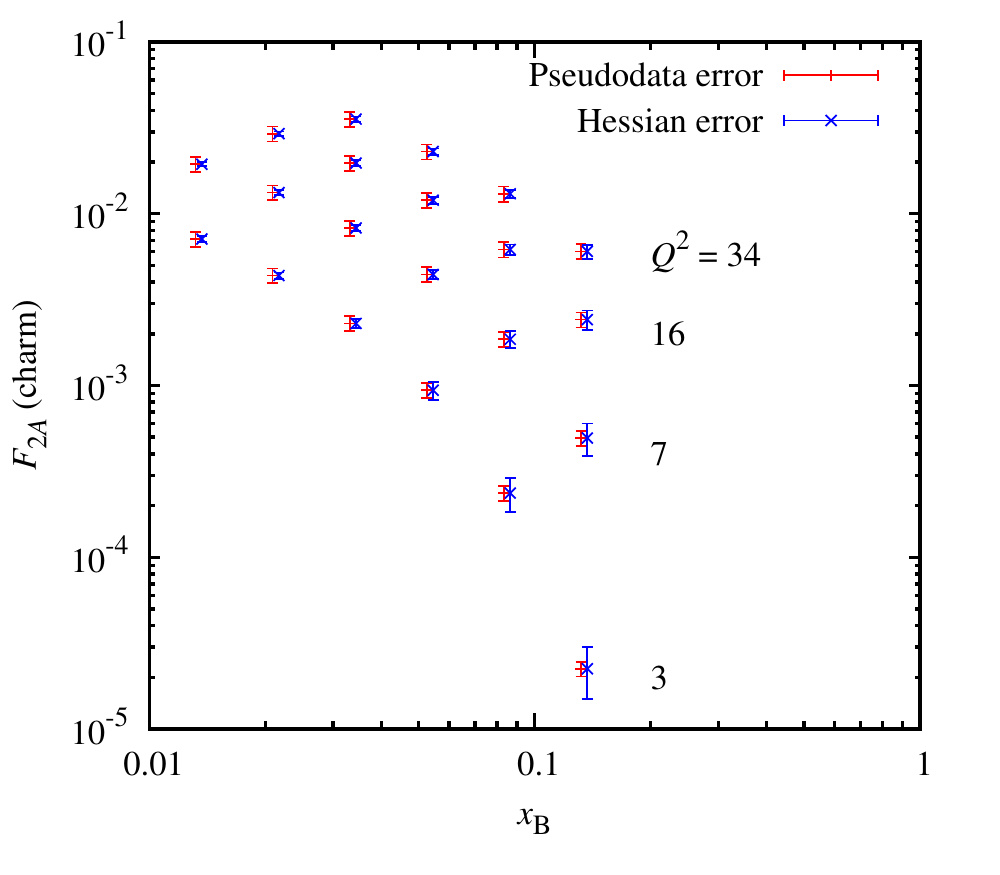}
&
\includegraphics[width=0.44\textwidth]{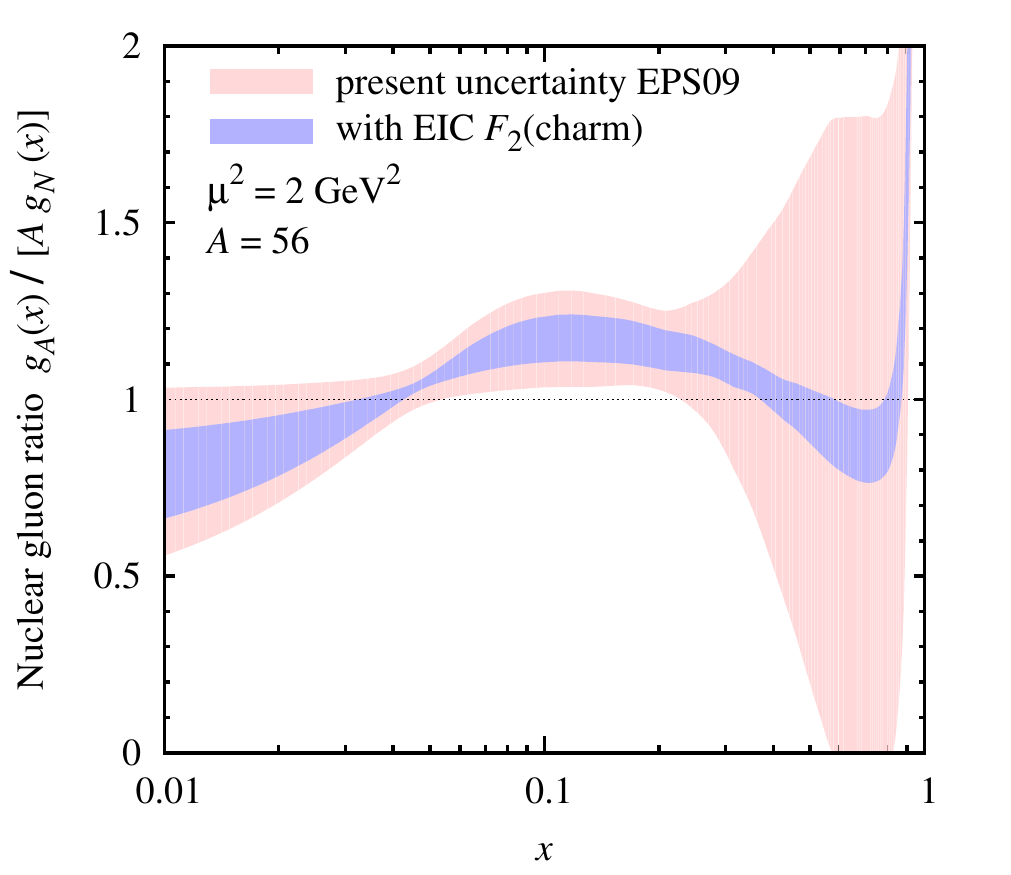}
\\[-2ex]
{\small (a)} & {\small (b)}
\end{tabular}
\caption[]{(a) Sample set of pseudodata in the nuclear charm structure function $F_{2A}^{c}$
used in the impact study. The two error bars show the assumed pseudodata errors 
and the theory error estimated with the Hessian uncertainty of the nuclear gluon
PDF parametrization (EPS09 LO\cite{Eskola:2009uj_567}). (b) Impact of charm pseudodata 
on the nuclear gluon PDF.
\label{fig:charm_impact}}
\end{figure*}
\section{Summary}
A high-luminosity EIC with next-generation detectors would offer excellent opportunities 
for measurements of open heavy flavor production in $ep/eA$ scattering. The charm production
rates appear sufficient to constrain nuclear gluons at $x > 0.1$, if charm reconstruction 
could be performed with an overall efficiency $> 10\%$. Heavy quark production at EIC 
could also be used for other physics studies, such as heavy quark fragmentation functions, 
jets physics, and heavy quark propagation and hadronization in nuclei. \\
{\small 
This material is based upon work supported by the U.S. Department of Energy, Office of Science, 
Office of Nuclear Physics under contract DE-AC05-06OR23177.}
\bibliographystyle{ws-procs961x669}


%

\newpage
\renewcommand*{\FigPath}{./Logo/} 

\begin{tcolorbox}[colframe=white]
\begin{minipage}{0.2\textwidth}
\includegraphics[width=1.\textwidth]{\FigPath/INT_Workshop_Logo_Final_Black.png}
\end{minipage}
\begin{minipage}{0.7\textwidth}
\wstoc{\bf Week VII}{}
\title{Week VII}
\end{minipage}
\end{tcolorbox} 
 
 
%
\wstoc{Summary of Week 7}{Adrian Dumitru, Fran\c{c}ois Gelis, Tuomas Lappi, Yacine Mehtar-Tani}
\title{Summary of Week 7}

\author{Adrian Dumitru}
\address{Department of Natural Sciences, Baruch College, CUNY, \\
17 Lexington Avenue, New York, NY 10010, USA}
\author{Fran\c{c}ois Gelis}
\address{ Institut de physique th\'eorique, Universit\'e Paris Saclay, \\
CNRS, CEA, F-91191 Gif-sur-Yvette, France}

\author{Tuomas Lappi}
\address{Department of Physics, P.O. Box 35, FI-40014 \\
University of Jyv\"askyl\"a, Finland and \\ Helsinki Institute of Physics, P.O. Box 64, FI-00014 \\
University of Helsinki, Finland }

\author{Yacine Mehtar-Tani}
\address{Physics Department, Brookhaven National Laboratory, Upton, NY 11973, USA}

\index{author}{Dumitru, A.}
\index{author}{Gelis, F.}
\index{author}{Lappi, T.}
\index{author}{Mehtar-Tani, Y.}

\begin{abstract}
Week 7 of the INT program 2018 ``Probing Nucleons and Nuclei in High
Energy Collisions'' was dedicated to topics at the interface of the
electron-ion collider (EIC), heavy ion and proton-nucleus
collisions. The EIC will provide complementary tools to investigate
and constrain the initial state in HIC collisions, as well as
transport properties of QCD matter which can be extracted from
observables that are sensitive to final states interactions such as
pt-broadening and energy loss.  The contributed talks and discussions
covered a variety of physics topics from saturation physics and the
origin of multi-particle correlations in HIC to jet quenching and the
strong coupling regime of high energy scattering.

\end{abstract}


\bodymatter

\section{Saturation physics and the Color Glass Condensate}
At high energy the nuclear wave function is characterized by a large
number of soft gluons that is bound to saturate as a consequence of
unitary. The phenomenon of gluon saturation is expected to take place
at values of $x$ of order $10^{-2}$ or smaller and is accessible at
RHIC and the LHC, and will be probed at the future EIC.

In this regime, standard perturbation theory breaks down and the
relevant degrees of freedom are strong classical fields that scale
parametrically like $A\sim 1/g$, where $g \ll 1$ is the strong
coupling constant.  The theory of the Color Glass Condensate (CGC) consists
in separating in rapidity the forward moving color charges which are
treated as stochastic classical sources and the strong gluonic fields
generated by them. It provides an initial condition for heavy ion
collisions characterized by high occupation number of gluons. It was
shown in several simulations based on solving Yang-Mills equations and
kinetic theory that such highly occupied systems tend to exhibit
hydrodynamic behavior before they reach thermal equilibrium.

Saturation physics in the context of the EIC was extensively discussed
in previous weeks of the program. In week 7, K.~Watanabe presented
work on quarkonium production in the CGC and argues that factorization
breaking effects due to soft gluon exchanges between the $c\bar c$
pair factorization may be enhanced in pA collisions as compared to
AA. Hence, it will be interesting to investigate these effects in eA
collisions at an EIC. S.~Benic presented a NLO calculation of photon
production in pA in the CGC framework. This is part of the recent
effort to compute higher order corrections to less inclusive
observables at small-x which will play an important role for precision
predictions at the future EIC to pin down saturation physics. Finally,
L.~McLerran discussed the problem of a fast moving particle
interacting with a Color Glass sheet in order to understand the
fragmentation region of nucleus-nucleus collisions and the
``transport'' of baryon number.

\section{Multi-particle correlations}

One of the most intensely debated questions in recent years in the HI
communities is to what extent the particle correlations measured via a
Fourier decomposition of multi-particle distributions, the so-called
$v_n$'s, emerge due to collective behavior of the produced particles
in the final state (``flow''), or from the initial state. Although it
is commonly accepted that in heavy ion collisions collective flow is
dominant, in smaller systems such as proton-proton and proton-nucleus
collisions at relatively small multiplicities the applicability of
hydrodynamics is not evident. The EIC will shed light on the origin of
correlations in small systems. Is a droplet of QGP created in such
systems? How important are final state interactions? These are among
the questions that an EIC will help answer. In week 7 M.~Mace described the 
physics of multiparticle correlations originating in the wavefunctions of the
incoming nuclei. In small collision systems it is important to
model simultaneously the effect of both these initial state correlations
and ones originating from hydrodynamical flow. The talk of B.~Schenke
discussed the interplay of these two effects in proton-proton and
proton-nucleus collisions in a model that combines a CGC description
of the initial state with a hydrodynamical evolution of the plasma.
In a session in week~7
preliminary data from ZEUS at HERA was also discussed. It indicates that
at HERA long-range in rapidity two-particle azimuthal correlations are not visible
within the statistical uncertainties in $\gamma^* -p$ collisions at high $Q^2$, even
in high multiplicity
events. However, after the INT 2018 program the ATLAS collaboration
presented data on two-particle azimuthal correlations in photo-nuclear
$\gamma-$Pb events at the LHC (ATLAS-CONF-2019-022), which appear
somewhat similar to those in $p-p$ and $p-A$ collisions. Thus, it will be
very interesting to study such correlations with much higher luminosities
at an EIC as a function of $Q^2$, nuclear mass number $A$, event multiplicity etc.

\section{Jets and jet quenching}

In the early 80's Bjorken suggested that the suppression of high
energy jets in high multiplicity proton-proton collisions would
signal the creation of the quark-gluon plasma. Such a suppression is a
result of collisional and radiative energy loss suffered by high-$p_T$
partons as they pass trough the hot QCD matter.

This phenomenon dubbed ``jet quenching'' was successfully observed in
the suppression of high pt hadrons in nucleus-nucleus collisions at
RHIC as compared to proton-proton collisions scaled by the number of
binary collisions. At the LHC, fully reconstructed jets were
measured. Remarkably their suppression persists up to 1 TeV as a
result of substantial final state interactions.

How energy is transported from the TeV jet scale to the scale given by
the temperature of the plasma, of order few hundred MeV, and how it is
dissipated in the plasma, is one of the central questions in jet
quenching studies.

The theory of jet quenching treats jet fragmentation as a perturbation
on top of the strong classical field that describes the QGP. Owing to
the fact that interactions in quantum field theory are not point like
the fragmentation of jets in a plasma is characterized by two regimes:
a coherence regime where the splitting of a parton is not resolved by
the medium which rather sees the jet a single color charge, a
manifestation of color transparency; and a decoherence regime where
the jet fragments decohere from the parent due to rapid color
randomization.

The all order picture of jet quenching to leading logarithmic accuracy
was discussed by E.~Iancu. In addition to analytic calculation of jet
substructure observables such as the jet Fragmentation Function a
Monte Carlo phenomenological study was presented.  It was pointed out
that color decoherence would lead to an excess of soft particles
within the jet cone in qualitative agreement with data. However, it
was recently suggested that medium response to the jet propagation may
yield a similar effect.

The EIC would provide a much cleaner environment to address the
question of color decoherence. In effect, soft particles in high
energy collisions are entangled throughout the entire event, in eA
collisions, final state interactions of jets with cold nuclear matter
may alter the color flow. Furthermore, a recent feasibility study has
shown that jets up to 20~GeV can be reconstructed at the EIC. This
will allow to study IR-collinear safe jet observables to investigate
final state interactions as well as nuclear and proton structure.

Moreover, jet substructure studies, emphasized by K.~Lee, E.~Iancu
and B.~Jacak in their respective contributions, will open a new
channel for addressing the aforementioned questions by taking
advantage of the versatility and wealth of substructure observables
such as angularities, and groomed observables, developed by the high
energy community in the context of new particle searches. By looking
inside jets one may learn about the role of non-pertubative physics in
ep and eA collisions, in particular the effects of color flow and
hadronization. One could also probe the mechanisms that underly the
interaction of a coherent multi partonic system, that is, the QCD jet,
with the extended nuclear matter.

\section{Transport properties of QCD matter}

In general, probing transport properties of hot nuclear matter is a
challenging task. The transport coefficient $\hat q$ was extracted
recently with a value of order a couple GeV$^2$/fm. The uncertainties
associated with this extraction are substantial. One may hope,
by investigating momentum broadening and jet energy loss in eA
collisions, to mitigate the theoretical uncertainties in the
determination of $\hat q$.

B.~Jacak discussed these questions from an experimentalist
perspective. In this contribution, the connexion and complementarity
of observables such as forward di-hadron and di-jet production, that
probe saturation physics, and momentum broadening in single inclusive
hadron/jet production in probing transport properties of cold nuclear
matter has been emphasized.

Medium-induced radiation is a building block of in-medium parton
cascades in many jet-quenching models. It is a function of the transport
coefficient $\hat q$ and thus constitutes another measurement of the
transport coefficient in addition to elastic processes.  This
elementary process is characterized by the suppression of large
frequency gluons as a result of the Landau-Pomerantchuk-Migdal effect. A
direct measurement of this coherence phenomena is difficult in a HI
environment and suffers from various theoretical uncertainties related
to the initial state of the collision, medium back reaction, multiple
gluon emissions. At an EIC, these uncertainties will be reduced and it
would be interesting to investigate this process in more details.

\section{Strong coupling aspects of high energy scattering}
In the framework of AdS/CFT correspondence a minimal bound for the
shear viscosity to entropy ratio for a plasma at infinite coupling was
predicted. The phenomenologically extracted values from RHIC and LHC
flow harmonics appear to be close to the holographic prediction
leading to the picture of the perfect liquid QGP. Holography provides
a powerful tool to investigate the strong coupling regime of field
theories. This approach may be of phenomenological relevance at the
EIC as well. Two talks addressed this subject.  K.~Mamo presented
a calculation for DIS based on the AdS/CFT correspondence where the
nucleus is described by a black hole in the gravitational dual. The
resulting R ratio is found to exhibit shadowing and anti-shadowing.
In the contribution by H.-U. Yee, non-perturbative QED
effects in pA collisions and their interplay with QCD within a QCD
string model were discussed.



 \newpage 
  \renewcommand*{\FigPath}{./WeekVII/01_Yee/}

\wstoc{Interplay between Reggeon and Photon in Proton-Nucleus Collisions}{Ho-Ung Yee}
 \title{Interplay between Reggeon and Photon in Proton-Nucleus Collisions}

\author{Ho-Ung Yee$^*$}

\address{Department of Physics, University of Illinois,
Chicago, Illinois 60607, USA\\
$^*$E-mail: hyee@uic.edu}
\index{author}{Yee, H.}

\begin{abstract}
In high energy collisions of proton and nucleus of charge $Z\gtrsim 100$, the Reggeon process that involves an exchange of a pair of quark and anti-quark, is subject to non-pertubative QED interaction with the electromagnetic field of the nucleus projectile, with the effective coupling constant $Z\alpha_{\rm EM}\sim{\cal O}(1)$. We study the interplay between non-perturbative QCD and QED in the Reggeon process of proton-nucleus collisions, using an effective theory of confined QCD strings. We find a qualitative change of large $s$ behavior of the amplitude.

\end{abstract}


\bodymatter

\section{Introduction}

Collision of elementary particles with high center-of-mass energy ($\sqrt{s}\gg \Lambda_{QCD}$), but low momentum transfer ($\sqrt{-t}\ll \Lambda_{QCD}$) is a non-perturbative process in QCD, involving a large spatial impact parameter $b\sim 1/\sqrt{-t}\gg 1/\Lambda_{QCD}$.
In this regime, an effective theory of confined QCD strings becomes useful, which can give important insights to the problem, and even quantitative predictions for some observables that rely on only a few universal features of the effective theory.
The useful concepts such as Pomeron and Reggeon naturally emerge in the effective theory, as the exchanges of virtual QCD strings between the projectiles across the transverse distance which is far beyond the non-perturbative scale of QCD. 
Some important observables such as the slope and intercept of Regge trajectory can be explained 
by this effective string theory, but there are also several limitations of the theory: for example, the inability of including the Pomeron interactions and all higher Pomeron loops as string branchings, which is one reason why the Froissart bound is not yet explained in the theory.

The effective string theory as a non-perturbative QCD finds its more rigorous justification in the 
AdS/CFT correspondence, a holographic duality between a strongly coupled gauge theory and a string theory in five dimensional AdS space that also includes gravity as a new degree of freedom.
The fifth extra dimension in the string theory side of the duality maps to the length/energy scale of the gauge theory, which gives a powerful treatment of scale variations of physical observables in the gauge theory. See Ref.\citenum{kiminad} for an interesting application of this idea to the picture of dipole density evolution in high energy collisions.
When $\sqrt{-t}\ll \Lambda_{QCD}$ in high energy scattering problems, it is shown that most of the exchanged strings in the string theory side resides in the four dimensional bottom of the five dimensional space \cite{Brower:2006ea,Janik:2001sc,Basar:2012jb}, which reduces the description of these strings to the original four dimensional effective theory of QCD strings.

A neutral Pomeron that carries no quark charges is represented by a closed string exchange in the effective theory. A process that involves an exchange of a pair of quark and anti-quark is described by an open string exchange, which we call the Reggeon.
In a minimal model, the action for these objects is given by the area of the string world-sheet multiplied by the string tension $1/2\pi\alpha'$.
In quantum theory, we do the usual path integral over all possible shapes of the exchanged strings with the boundary condition that is specified by the projectiles.
In the semi-classical approximation that is leading order in $\hbar$, one looks for the saddle point that extremizes the classical action. As we will review, this is enough to find the slope parameter of the Regge behavior of the scattering amplitude in large $s$ limit.

The exchanged pair of quark and anti-quark in Reggeon process carries electric charges and is subject to
QED interactions with the electromagnetic field of the projectiles, which
 is normally suppressed by the fine-structure constant $\alpha_{\rm EM}\sim 0.01$. In a proton-nucleus collision with a nucleus charge $Z\gtrsim 100$, the electromagnetic field of the nucleus is enhanced by the factor of $Z$, and the QED interaction between the exchanged Reggeon and the electromagnetic field of the nucleus is strong, $Z\alpha_{\rm EM}\sim {\cal O}(1)$.
This warrants non-perturbative study of the QED interaction in the Reggeon exchange process, that possibly interplays with the QCD dynamics captured by the effective string theory. This is the subject of this presentation, based on Ref.\citenum{HUY}. We will see that there is a qualitative change in the large $s$ behavior of the Reggeon total cross section in proton-nucleus collisions.

\section{String Instantons}

A fruitful way of thinking of high energy collision with a large impact parameter $b$ is to consider it
as a time-dependent quantum tunneling process. The QCD confinement puts a linear potential barrier for a string from one projectile to stretch to the other projectile, with the barrier height $E_m={b\over 2\pi\alpha'}$.
When the end point of a quantum string tunnels through the barrier and meets the other projectile right on the time of collision, there will be an enormous acceleration that initiates the process of string breaking and particle production. 

Using something called the world-sheet T-duality, we can map the high energy scattering problem to the problem of Schwinger pair production of string in an effective electric field $E\sim \chi/\alpha'$ acting on the end points of the string stretched over the distance $b$, where $\chi=\log s$ is the rapidity gap of the two projectiles.\cite{Basar:2012jb} This physics underlies a striking resemblance between the Regge amplitude
$\exp[-b^2/4\alpha'\chi]$ and the Schwinger pair production amplitude $\exp[-m^2/eE]$, under a map $m\to b/2\pi\alpha'$ and $E\to \chi/\alpha'$. The Schwinger pair production is understood as a quantum tunneling phenomenon.

\def\figsubcap#1{\par\noindent\centering\footnotesize(#1)}
\begin{figure}[t]
\begin{center}
 \parbox{2.1in}{\includegraphics[width=1.5in]{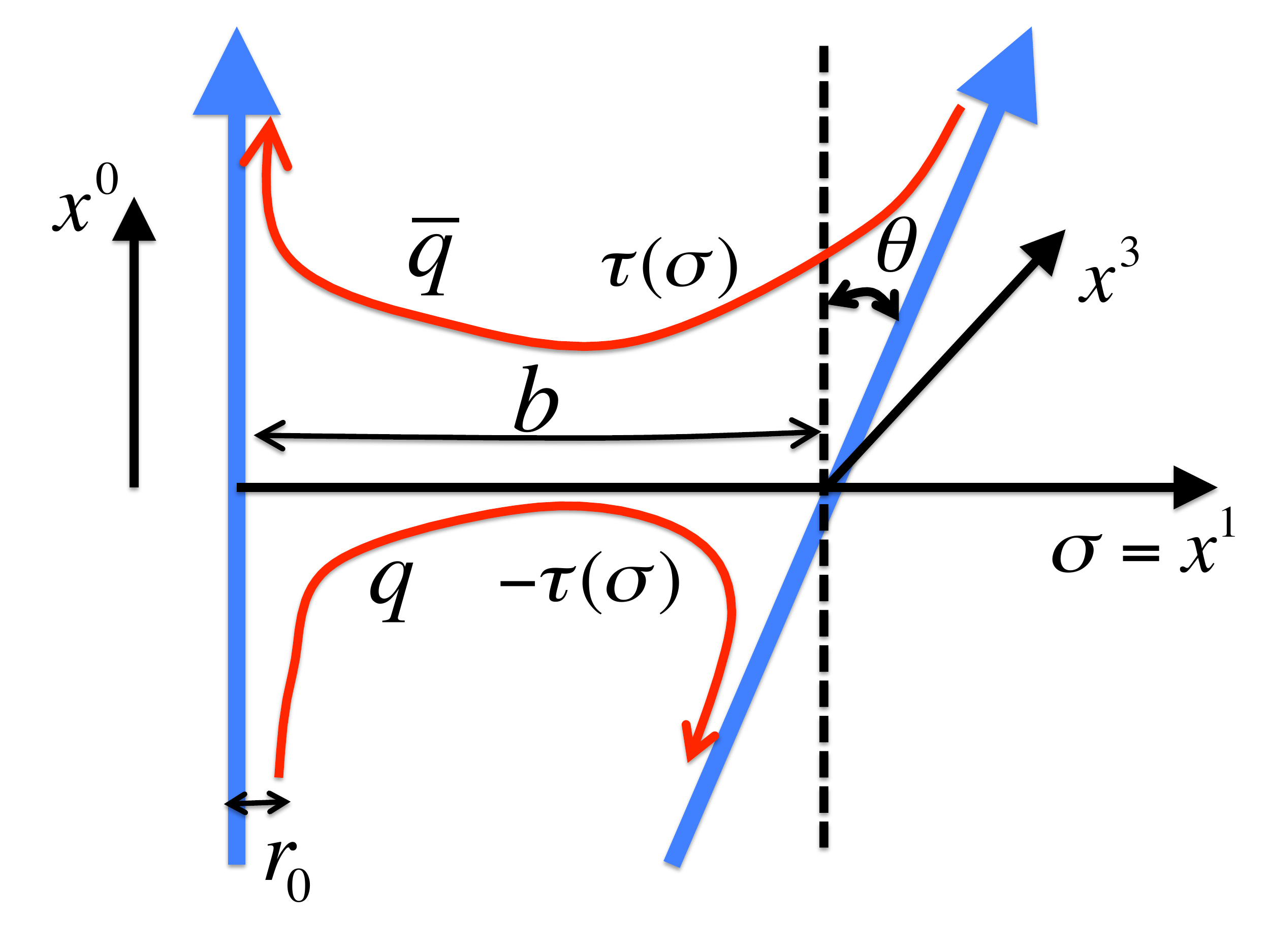}
 \figsubcap{a}}
 \hspace*{4pt}
 \parbox{2.1in}{\includegraphics[width=1.5in]{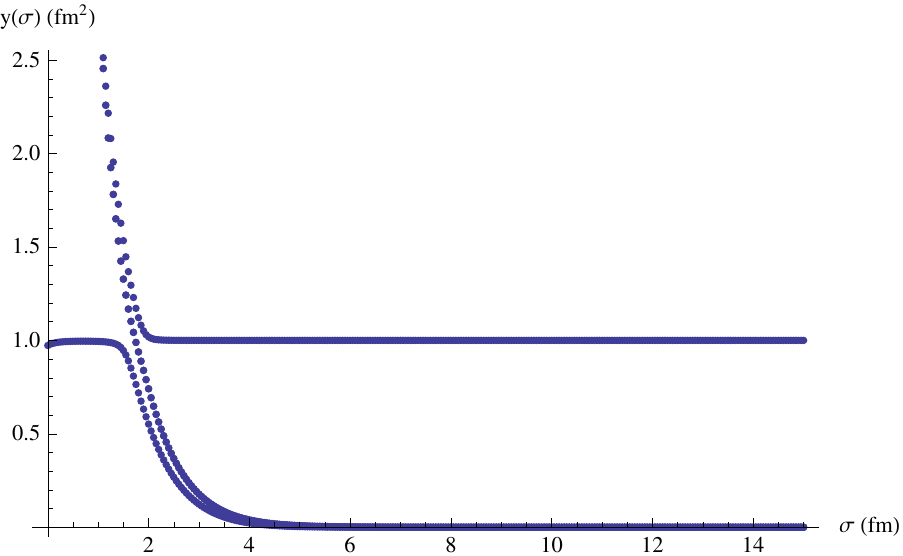}
 \figsubcap{b}}
 \caption{
 (a) The geometry of exchanged open string in the Reggeon process. (b) A typical string instanton in the Reggeon process of proton-nucleus collision.}
\label{fig}
\end{center}
\end{figure}
Guided by quantum mechanics of tunneling, one looks for a classical saddle point
of the string action in Euclidean space time, an instanton of string, with the boundary condition set by the projectiles \cite{Janik:2001sc,Gross:1987kza}. Comparing the Rindler space $ds^2=-d\tau^2+\tau^2 d\chi^2$ and the Euclidean space $ds^2=dr^2+r^2d\theta^2$, the rapidity gap of the projectiles maps to an angle between the two world-lines of the projectiles in Euclidean space by an analytic continuation $\theta=-i\chi$.\cite{MEGGIOLARO}
The geometry of an open string instanton for the Reggeon process is shown in Fig.\ref{fig}(a).
The string world-sheet lies on a helicoid surface bounded by the two world-lines of the projectiles making the angle $\theta$ and separated by the distance $b$.\cite{Janik:2001sc} The shape is parametrized by a single function $\tau(\sigma)$ ($0<\sigma<b$) that is the width of the string world-sheet at a coordinate distance $\sigma$. The string action becomes
\begin{equation}
S={1\over 2\pi\alpha'}\int_0^b d\sigma\int_{-\tau(\sigma)}^{\tau(\sigma)}d\tau \sqrt{1-{\chi^2\over b^2}\tau^2}\,,\quad \chi=i\theta\,,
\end{equation}
with the saddle point $\tau(b)=b/\chi$, and $S_{\rm instanton}(\chi,b)=b^2/4\alpha'\chi$. The instanton in the T-dual picture of Schwinger pair production can also be constructed with the same value of the action at the saddle point \cite{Basar:2012jb,SCHUBERT}.

The scattering amplitude of the Reggeon as a quantum tunneling process in leading order of $\hbar$ will be 
$
{\cal T}(s,b)\sim {i\over s}\exp[-S_{\rm instanton}(\chi,b)]
$,
where the $i/s$ factor in front is due to the quark wave function overlap in the Reggeon process.
The amplitude in momentum space is
$
{\cal T}(s,t)=\int d^2 b\,\, e^{-iq\cdot b} \,{\cal T}(s,b)\,\,(t\equiv -q^2)
$,
and the total cross section from the optical theorem is 
\begin{equation}
\sigma_{\rm tot}={1\over s}\int d^2 b\,\,{\rm Re}\,[e^{-S_{\rm instanton}(\chi,b)}]\,.
\end{equation}
With the above result, we obtain ${\cal T}(s,t)\sim i(\log s) s^{\alpha' t-1}$, which is the Regge trajectory with a slope $\alpha'$. It has been shown that the 1-loop fluctuation contribution around the instanton
makes the intercept value $\sigma_{\rm tot}\sim s^{\alpha_0}$ with $\alpha_0\approx 0.25$ \cite{Basar:2012jb,Janik:2001sc}.

\section{String Instantons in Proton-Nucleus Collisions}

With $Z\alpha_{\rm EM}\sim {\cal O}(1)$ in proton-nucleus collisions, the QED interaction between the exchanged pair of quark and anti-quark in the Reggeon process and the electromagnetic field of the nucleus projectile is no longer negligible.
The QED interaction is given by the QED Wilson line of the electromagnetic field of the nucleus, computed along the world-lines of the exchanged quark and anti-quark pair, that is, the boundaries of the exchanged open string for a Reggeon as shown in Fig.\ref{fig}(a).
As the QED interaction is equally non-perturbative compared to QCD string dynamics,
we search for a new saddle point of the total action of the exchanged string, given by the sum of the usual string action of world-sheet area and the QED Wilson line described as above \cite{HUY}. 
The electromagnetic field of the nucleus is given by the Coulomb gauge field in Euclidean space time, $A_0=i Ze/4\pi r$, where $r=|\vec x|$ is the spatial distance from the nucleus, and the QED contribution to the action is 
\begin{equation}
\exp\left[ie\sum_i \,q_i\int_{C_i}\,\,A_\mu dx^\mu\right]\equiv \exp[-S_{\rm QED}]\,,
\end{equation}
where $i=u,d$ labels the exchanged quark or anti-quark with a charge $q_i$, whose world-line is denoted as $C_i$. The saddle point equation is modified as ($y(\sigma)\equiv \tau(\sigma)^2$)
\begin{equation}
{1\over 2\pi\alpha'}\sqrt{1-{\chi^2\over b^2}y(\sigma)}-{q_iZe^2\over 4\pi}{\left({\cosh}({\chi\sigma\over b})(\sigma+r_0)-{\chi\over b}{\sinh}({\chi\sigma\over b})y(\sigma)\right)\over \left((\sigma+r_0)^2-{\sinh}^2({\chi\sigma\over b})y(\sigma)\right)^{3/2}}=0\,,
\end{equation}
which we solve numerically for the instanton shape $\tau(\sigma)$ ($r_0$ is the size of the nucleus). In Fig.\ref{fig}(b) we show a typical situation of two branches of solutions: the upper curve is the ``perturbative QED branch" which smoothly connects to the solution in $Z\alpha_{\rm EM}\to 0$ limit, but ceases to exist in the range $\sigma<\sigma_c\approx 0.23 \,b/\chi$. One finds that $S_{\rm QED}$ for this branch is purely imaginary. The lower curve is the new type of instanton as a result of non-perturbative interplay between QED and QCD string. It gives rise to a real part of
$S_{\rm QED}$ with the total action ${\rm Re}[S_{\rm tot}]\sim1/4\pi\alpha'(Z)\,b^2/\chi^2$, with a $Z$ dependent numerical constant $\alpha'(Z)$.

From this, the Reggeon amplitude in proton-nucleus collisions behaves as 
${\cal T}(s,t)\sim i(\log s)^2 e^{\alpha'(Z)(\log s)^2 t}$, with the total cross section 
$\sigma_{\rm tot}\sim (\log s)^2/s$ in large $s$ limit. This is consistent with
the Froissart bound $\sigma_{\rm tot}\le(\log s)^2$.

\section*{Acknowledgments}

I thank Gokce Basar, Dima Kharzeev, and Ismail Zahed for fruitful collaborations.
This work is supported by the U.S. Department of Energy, Office of Science, Office of Nuclear Physics, with grant Nos. DE-SC0018209 and DE-FG0201ER41195.

 \newpage


\renewcommand*{\FigPath}{./WeekVII/02_Mamo/}

\wstoc{Holographic Approaches to DIS on a Nucleus}{Kiminad A. Mamo}

\title{Holographic Approaches to DIS on a Nucleus}

\author{Kiminad A. Mamo$^*$}
\index{author}{Mamo, K.}

\address{Department of Physics and Astronomy, Stony Brook University,\\
Stony Brook, New York 11794-3800, USA\\
$^*$E-mail: kiminad.mamo@stonybrook.edu}

\begin{abstract}

We consider deep inelastic scattering (DIS) on a dense nucleus described as an extremal RN-AdS black hole with holographic quantum fermions in the bulk. We find that the R-ratio (the ratio of the structure function of the black hole to proton) exhibit  shadowing for $x < 0.1$, anti-shadowing for $0.1 < x < 0.3$, EMC-like effect for $0.3 < x < 0.8$ and Fermi motion for $x > 0.8$ in a qualitative agreement with the experimental observation of the ratio for DIS on nucleus for all range of $x$. We also take the dilute limit of the black hole and show that its R-ratio exhibits EMC-like effect for $0.2 < x < 0.8$ and the Fermi motion for $x > 0.8$, and no shadowing is observed in the dilute limit for both bottom-up (using Thomas-Fermi approximation for the nucleon distribution inside the dilute nucleus), and top-down (considering the dilute nucleus to be a Fermi gas in AdS) approaches.

\end{abstract}


\bodymatter

\section{Introduction}
We use holography or the AdS/CFT correspondence to model deep inelastic scattering (DIS) on both dilute and dense nuclei \cite{Mamo:2018eoy,Mamo:2018ync,Mamo:2019jia}. See Ref. 4 for application of holography in proton-nucleus collisions. We model a dilute nucleus in a bottom-up and top-down approaches. In the bottom-up approach, we consider a nucleus to be made of nucleons distributed according to the Thomas-Fermi approximation, and DIS on the dilute nucleus is dominated by the incoherent scattering on individual nucleons. Therefore, in this bottom-up approach, we use the structure function of each nucleons calculated in holography as input to determine the structure function of the nucleus. We find that the R-ratio exhibits an EMC effect in the large-x regime.

In the top-down approach, we consider the nucleus to be a Fermi gas in AdS spacetime. And, by carrying out the DIS directly on the Fermi gas, we determine the structure function of the nucleus from the current-current correlation function computed by using a one-loop Witten diagram in pure AdS where the fermions are running in the loop forming a Fermi gas. We find that the R-ratio exhibits EMC effect, similar to the bottom-up approach.

We model the dense nucleus by an extremal RN-AdS black hole, and identify the structure function of the extremal RN-AdS black hole to the structure function of a nucleus. At leading order in $1/N_c$ expansion, the structure function of the nucleus, is extracted from the current-current correlation function of the classical extremal RN-AdS black hole background without matter contribution. We find that the structure function of the classical extremal RN-AdS black hole is mostly at small-x, and exhibits shadowing.

At sub-leading order in $1/N_c$ expansion, the highly damped bulk Dirac fermions (protons) due to the background RN-AdS black hole, also contribute to the extremal RN-AdS black hole's or the dense nucleus's structure function. We find that the quantum correction contributes mostly at large-x regime.

\section{DIS on a Dilute Nucleus}\label{dilute}

In holography, Compton scattering on a single nucleon at the boundary maps onto
the scattering of a U(1) current onto a bulk Dirac fermion which, at large-x, is
dominated by s-channel exchange of bulk Dirac fermion resonances, while at small-x
the same scattering is dominated by the t-channel exchange of spin-j glueball
resonances, with the interpolating result for the structure function of the proton
(modeled as a bulk Dirac fermion) as\cite{Polchinski:2002jw}
\begin{equation}
\label{F2p}
F_2^p(x, q^2)=\tilde{\mathbb C}\left(\frac{m_N^2}{-q^2}\right)^{\tau-1}
\left(x^{\tau+1}(1-x)^{\tau-2}+{\mathbb C}\left(\frac{m_N^2}{-q^2}\right)^{\frac 12}\frac 1{x^{\Delta_{\mathbb P}}}\right)\,,
\end{equation}
with $mR=3/2$ or $\tau=\Delta-1/2=3$.
\subsection{Bottom-up Approach to Dilute Nucleus}
In our bottom-up approach to DIS on a dilute nucleus, we consider DIS on a nucleus
described using a density expansion where the leading density contribution to the current-current correlation of a nucleus is \cite{Mamo:2018eoy}
\begin{eqnarray}
\label{W11}
\frac{{\cal G}_A^{\mu\nu}}{\left<P_A|P_A\right>}
&\approx & \rho_0\frac {4\pi} 3 R_A^3\int\, \frac{d^3p}{2V_3E_p}\,\frac{\theta(p_F-|\vec p|)}{\frac 43 \pi p_F^3}\,
{\cal G}^{\mu\nu}_p\nonumber\\
&+&16\pi\int_{R_A}^{R_A+\Delta}r^2dr\int\, \frac{d^3p}{(2\pi)^3}\frac 1{2V_3E_p}{\theta(p_F(r)-|\vec p|)}\, {\cal G}^{\mu\nu}_p\,,
\end{eqnarray}
where ${\cal G}_p^{\mu\nu}$ is the current-current correlation function for DIS  scattering on a single nucleon.
The R-ratio (the structure function of the dilute nucleus normalized by the structure
function of each constituent nucleon) is plotted in Fig.~\ref{dilute-ratio}a.

\begin{figure}[t]
\begin{center}
\begin{tabular}{cc}
\includegraphics[width=5cm]{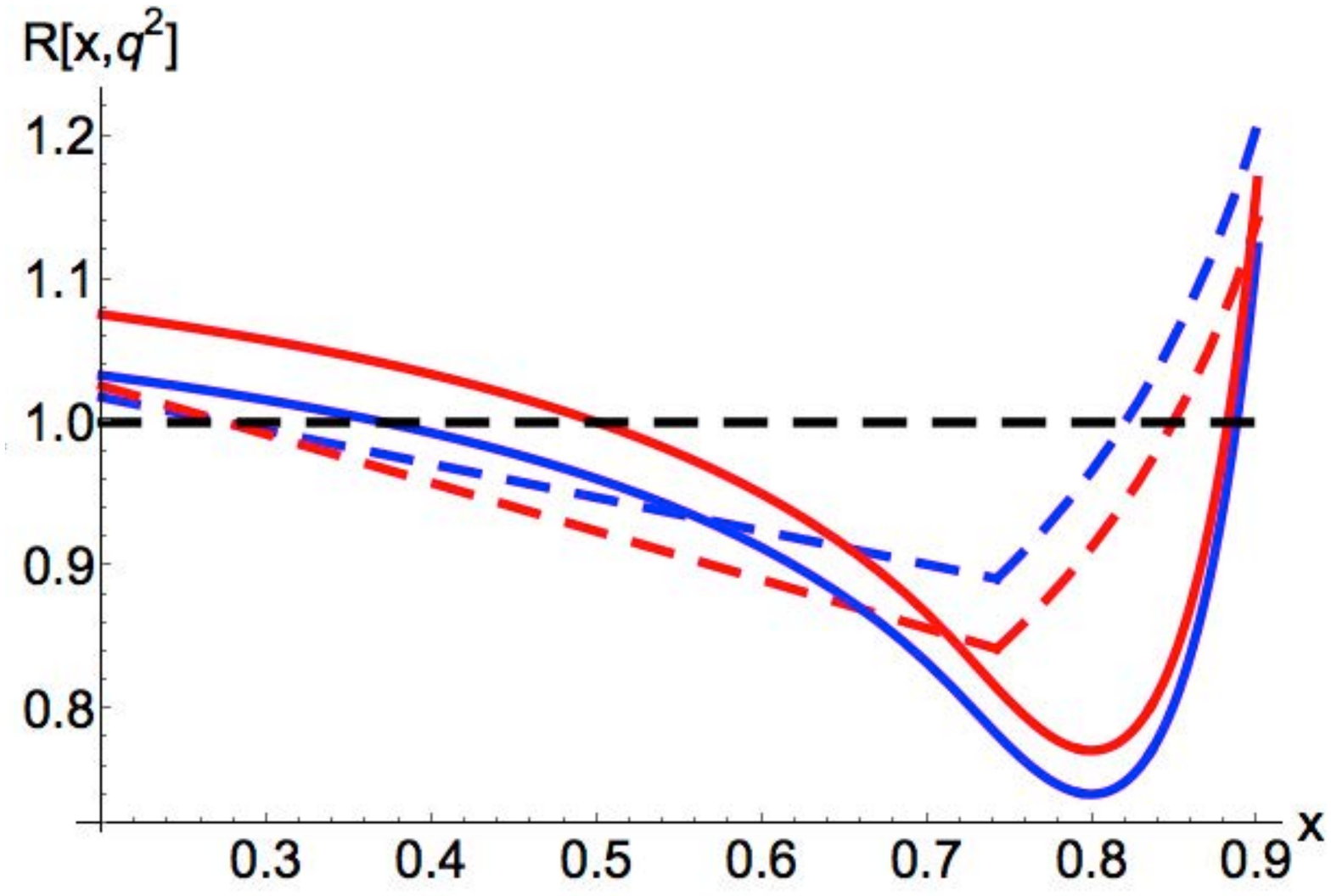}

&
\includegraphics[width=5cm]{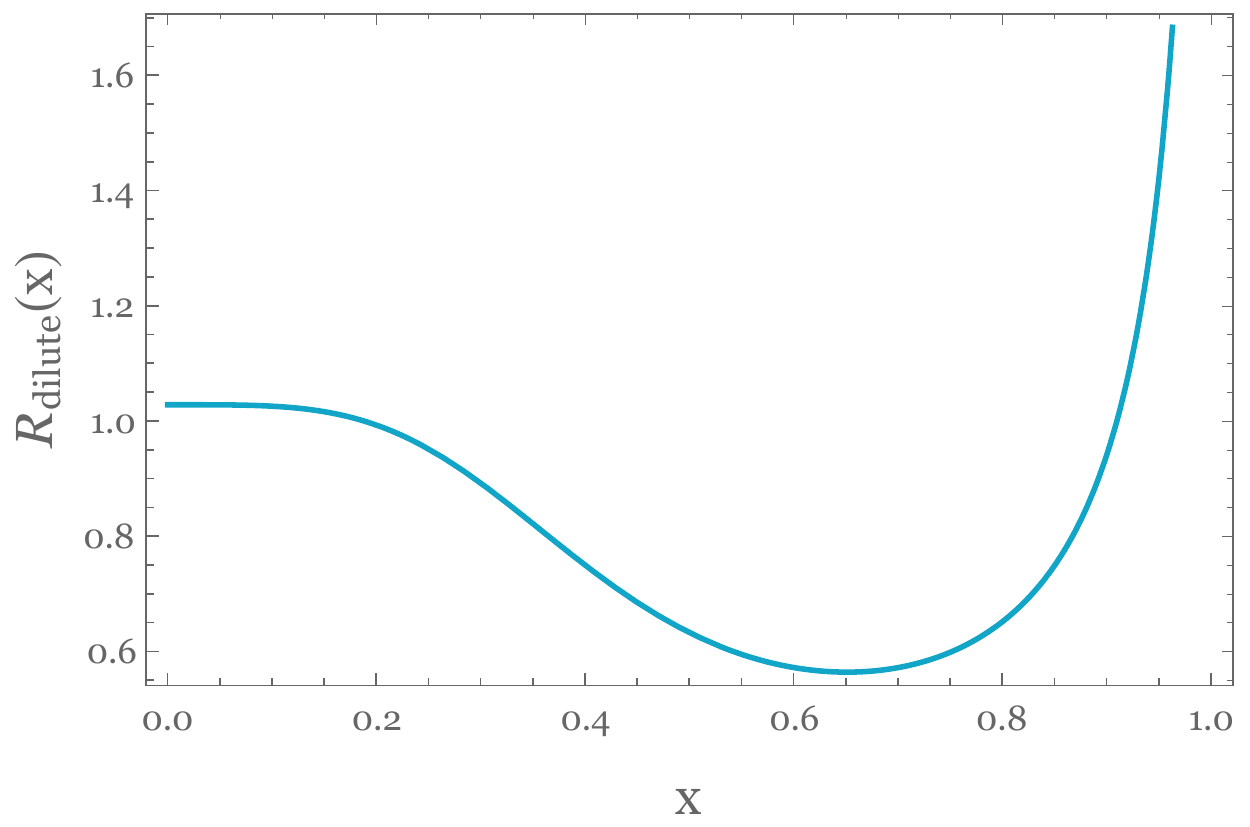}

\\
(a) & (b)
\end{tabular}
\caption{(a) R-ratio at large-x using the leading density contribution and the holographic nucleon structure function
  (solid curves), versus the parametrized empirical ratio (dashed curves), for atomic number $A=12$ (blue curves) and $A=42$ (red curves). (b) Dilute R-ratio for $k_F/m_N=1$, $e_R=0.3$ (U(1)-charge of the bulk fermion), $2\pi^2c_5/\sqrt{4\pi\lambda}=0.01$ (strong coupling), $\tau=3$ (hard scaling exponent),
$j=0.08$ (Pomeron intercept).}\label{dilute-ratio}
\end{center}
\end{figure}

\subsection{Top-down Approach to Dilute Nucleus}

In our top-down construction for DIS on a dilute nucleus, we directly do DIS on a
gas of bulk Dirac fermions (protons) undergoing Fermi motion with momentum $k$. In
other words, we compute the one-loop correction to the current-current correlation
function in AdS where the bulk Dirac fermions are running in the loop forming a
Fermi gas. In the large-x regime, we find \cite{Mamo:2019jia}
\begin{equation}
\frac{F_{2}(x,q^2)}{A}\approx 8\pi^2(\tau-1)^2e_R^2\Big(\frac{\beta m_N^2}{q^2}\Big)^{\tau-1}
\,x_{{F}}^{\tau+1}(1-x_{{F}})^{\tau-2}\,,
\end{equation}
where $x_F = \frac{xm_N}{E_F}$ is defined in terms of the Fermi energy $E_F=(k_F^2+m_N^2)^{\frac 12}$ of a single
nucleon with Fermi momentum $k_F$ . Similarly for small-x we have
\begin{equation}
\frac{F_{2}(x,q^2)}{A}\approx {\pi C_{\lambda}}
\Big(\frac{\beta m_N^2}{q^2}\Big)^{\tau-1}\,\frac{1}{x_{{F}}}[I_{0,2\tau+3} + I_{1,2\tau+3}]\,.
\end{equation}
We have plotted the R-ratio in Fig.~\ref{dilute-ratio}b.

\section{DIS on a Dense Nucleus}
We model DIS on a dense nucleus by the corresponding DIS on an extremally
charged (RN)-AdS black hole.

\subsection{DIS on a Classical RN-AdS Black Hole}
The structure function extracted from DIS on a classical extremal RN-AdS black
hole is given by $F^A_2(x,Q^2) = F_T(x,Q^2) + F_L(x,Q^2)$ where (computed in \cite{Mamo:2018ync}, by closely following \cite{Hatta:2007cs} for uncharged thermal AdS black holes)
\begin{equation}
  \label{NEW}
F^A_T(x,Q^2)=\tilde C_{T}\,\frac{A}{x}\,
\left(\frac{3x^2Q^2}{4m^2_N}\right)^{\frac 23}\,,\,\,\,\,\,
F^A_L(x,Q^2)=\tilde C_{L}\,\frac{3A}{4x}
\left(\frac{3x^2Q^2}{4m^2_N} \right)\,,
\end{equation}
with $\tilde C_{T,L}/C_{T,L}=\pi^5{(48\alpha)^2}/{2N_{c}^2}$. We have plotted the R-ratio in Fig.~\ref{dense-ratio}a.

\begin{figure}[t]
\begin{center}
\begin{tabular}{cc}
\includegraphics[width=5cm]{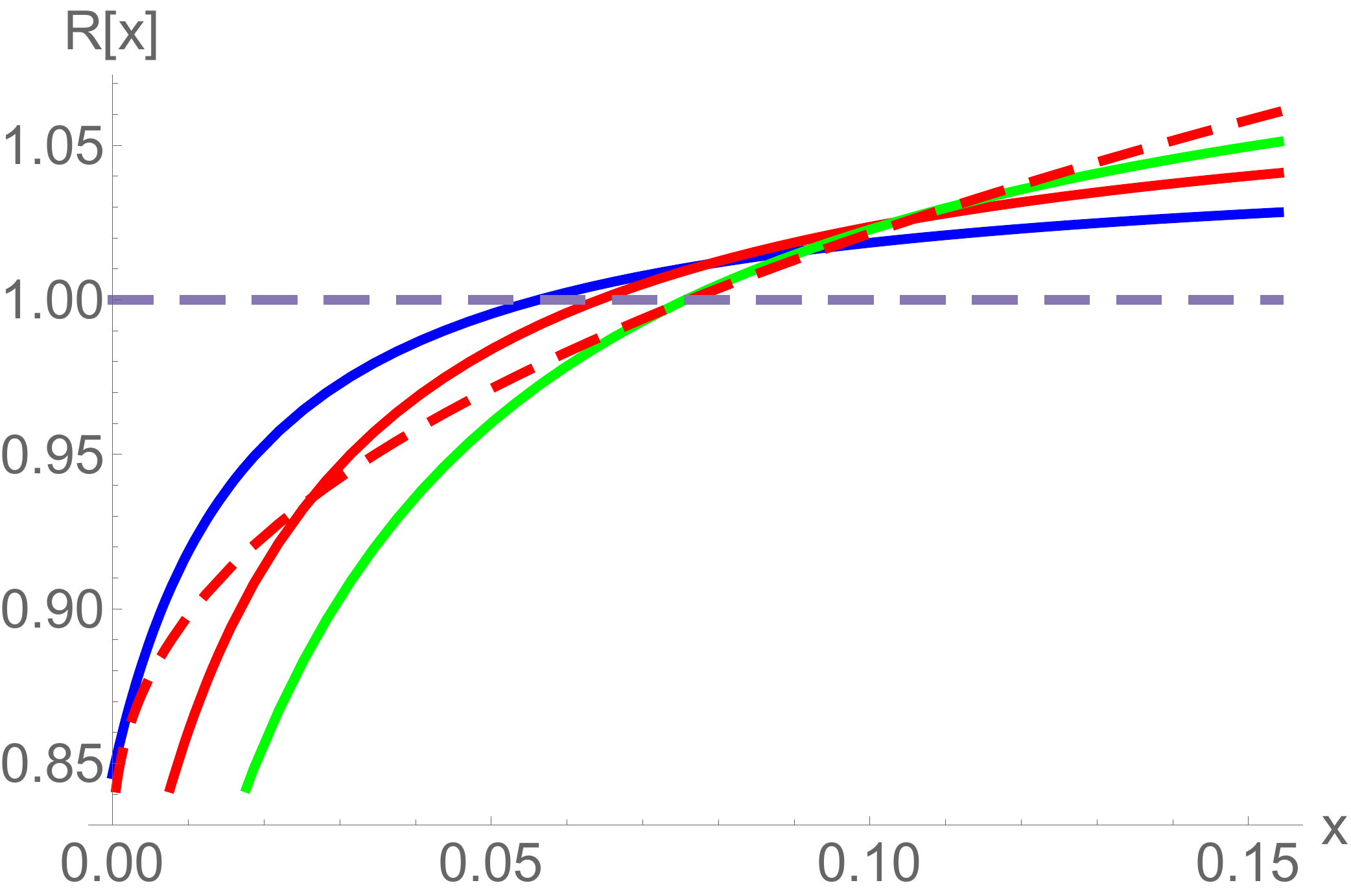}

&
\includegraphics[width=5cm]{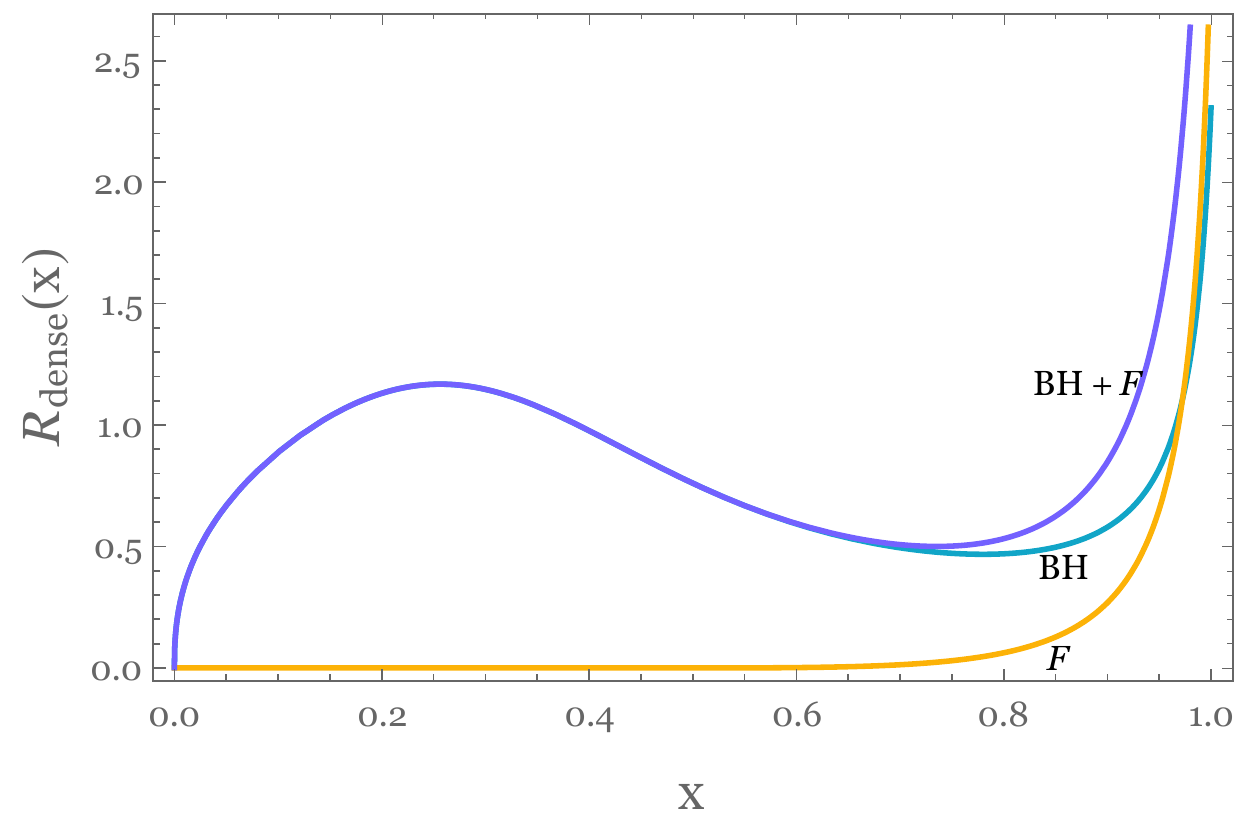}

\\
(a) & (b)
\end{tabular}
\caption{(a) Parametrized DIS data on nuclei (solid curves)  vs holography (dashed curve) in the shadowing region.  (b) Dense R-ratio for $k_F/m_N=3.5$, BH $\equiv$ black hole, F $\equiv$ quantum fermions.}\label{dense-ratio}
\end{center}
\end{figure}

\subsection{DIS on a Quantum RN-AdS Black Hole}
The leading order structure function (in $1/N_c$ expansion of the current-current correlation
function) due to the classical RN-AdS black hole receives one-loop correction from the quantum fermions hovering around the extremal RN-AdS black hole. Therefore, by including the quantum correction due to the bulk Dirac fermions
(protons), the total structure function of the dense nucleus, modeled as a quantum RN-AdS black hole is given by\cite{Mamo:2019jia}
\begin{eqnarray}
\frac{F_2^{\rm dense}(x,q^2)}{A}&\approx & C_T\left(\frac{3q^2}{4m_N^2}\right)^{\frac 23}\,x^{\frac 13}+C_{AdS2}\,e_R^2\,\left(\frac{\mu^2}{q^2}\right)^{\nu_{k_F}+2}\nonumber\\
&\times & x_{k_F}^{\nu_{k_F}+5}(1-x_{k_F})^{\tau-\frac 32}\,{}_2F_1^2\left(\tau_+, \tau_-, \tau-1, 1-x_{k_F}\right)\,,
\end{eqnarray}
where the first contribution (which is the leading and dominant contribution) stems
from DIS on the classical black hole, and the second and subleading contribution
stems from DIS scattering of the emerging holographic Fermi liquid near the black
hole horizon which is a quantum correction that is vanishingly small at small-x.
We have plotted the R-ratio in Fig.~\ref{dense-ratio}b.

\section{Summary and Conclusion}
We have shown that the quantum extremal RN-AdS black hole exhibits shadowing in small-x regime, see Fig.~\ref{dense-ratio}, and EMC effect in the large-x regime where it is dominated by a Fermi gas in AdS spacetime, see Fig.~\ref{dilute-ratio}.

\section{Acknowledgments}
I thank Ismail Zahed for collaboration on this work. This work was supported by the U.S. Department of Energy under Contract No. DE-FG-88ER40388.


\newpage
%

\renewcommand*{\FigPath}{./WeekVII/03_Benic/Figs}
\renewcommand{\rmd}{\mathrm{d}}
\newcommand{\kgp}{\boldsymbol{k}_{\gamma\perp}}
\newcommand{\etag}{\eta_\gamma}
\newcommand{\qp}{\boldsymbol{q}_\perp}
\renewcommand{\pp}{\boldsymbol{p}_\perp}
\newcommand{\Pp}{\boldsymbol{P}_\perp}
\renewcommand{\calN}{\mathcal{N}}
\newcommand{\kp}{\boldsymbol{k}_{\perp}}
\newcommand{\khp}{\boldsymbol{k}_{1\perp}}
\newcommand{\kAp}{\boldsymbol{k}_{2\perp}}



\wstoc{CGC photon production at NLO in pA collisions}{Sanjin~Beni\' c, Kenji Fukushima, Oscar Garcia-Montero, Raju Venugopalan}
\title{CGC photon production at NLO in pA collisions}

\author{Sanjin~Beni\' c}

\address{Yukawa Institute for Theoretical Physics, Kyoto University,\\
Kyoto 606-8502, Japan}

\author{Kenji Fukushima}

\address{Department of Physics, The University of Tokyo,
7-3-1 Hongo, Bunkyo-ku,\\
Tokyo 113-0033, Japan\\
Institute for Physics of Intelligence (IPI), The University of Tokyo,
7-3-1 Hongo, Bunkyo-ku,\\ 
Tokyo 113-0033, Japan}

\author{Oscar Garcia-Montero}

\address{Institut f\" {u}r Theoretische Physik, Universit\" {a}t Heidelberg, Philosophenweg 16,\\ 
69120 Heidelberg, Germany}

\author{Raju Venugopalan}

\address{Physics Department, Brookhaven National Laboratory, Bldg. 510A, Upton,\\
New York 11973, USA}

\index{author}{Beni\' c, S.}
\index{author}{Fukushima, K.}
\index{author}{Garcia-Montero, O.}
\index{author}{Venugopalan, R.}

\begin{abstract}
We summarize our analytical and numerical results obtained up to now on CGC photon production at NLO in high energy hadronic collisions. This includes the full analytic NLO CGC formula based on which inclusive photon cross section is considered.
\end{abstract}


\bodymatter

\section{On using photons as a probe of the Color Glass Condensate}

Photons are a precious tool in high energy hadronic collisions. In AA collisions, one refers to thermal photons as a probe of the temperature of the formed quark-gluon plasma, while in pp collisions isolated photons are an important benchmark of perturbative QCD calculations. Here isolated means that within a cone $R = \sqrt{\Delta\eta^2 + \Delta \phi^2}$ around the photon, hadronic activity should be below some experimentally set energy threshold. This in particular serves to reduce the contribution from the poorly constrained quark-to-photon or gluon-to-photon fragmentation. Finally, note that direct photons, also often measured in hadronic collisions, take into account the fragmentation component as well. 

Final states including photons are under better theoretical control than purely hadronic final states due to the absence of final state interactions. This is an important, and often stated, advantage in considering photon over hadron, but there are a couple of noteworthy remarks, nevertheless. Photon cross sections are smaller by a factor of $\alpha_e$ with respect to hadronic cross sections. Direct photons have a large background from decay photons. Low energy photons typically fragment off hadrons and are more difficult to measure and isolate.

The potential of photons to elucidate the CGC gluon distributions in pA is recognized by the Born level partonic cross section $q g \to q\gamma$, see \cite{Gelis:2002ki} and references within. In \cite{Benic:2016yqt_3,Benic:2016uku_3,Benic:2017znu,Benic:2018hvb_3} we emphasized a purely gluonic channel $gg \to q\bar{q}\gamma$ that induces significant corrections to the Born process already at RHIC energies while it is dominating at the LHC. At the same order the $qg \to qg\gamma$ channel contributes in the forward region \cite{Altinoluk:2018uax}. 

At LHC, the following data is currently available. ATLAS and CMS results in pp at $k_{\gamma\perp} \gtrsim 20$ GeV, but this is too hard $k_{\gamma\perp}$ for CGC. Preliminary ALICE data for isolated photon in pp at $\sqrt{s} = 7$ TeV ($k_{\gamma\perp} > 10$ GeV) and for direct photon in pPb at $\sqrt{s} = 5.02$ TeV ($k_{\gamma\perp} \gtrsim 0.5$ GeV) \cite{Acharya:2019jkx} and \cite{Schmidt:2018ivl}, respectively, but without nuclear modification factor. Isolated photon-jet and photon-hadron angular correlations from ALICE in pp and pPb \cite{Arratia:2018nra}: the results show no nuclear effects.

\section{Theoretical framework}

We are using a dilute-dense framework, appropriate for particle productions in pA collisions, but also in pp, where higher twist contributions from the dense target (t) become small, quantifiable, corrections. Apart from the Born process, our results are based on the NLO $gg \to q\bar{q} \gamma$ channel. The projectile gluon is described by an UGD $\varphi_p(Y_p, \khp)$, with $Y_p$ the rapidity and $\khp$ the finite transverse momentum of the projectile, $\khp^2 \ll (Q_S^p)^2$ ($Q_S^p$ is the saturation scale in the projectile proton). The target description in general involves not only 2-point, but also 3-point and 4-point Wilson line correlators. In the large $N_c$ limit adopted here, the inclusive photon cross section simplifies to \cite{Benic:2016uku_3,Benic:2018hvb_3}
\be
\begin{split}
  \frac{d \sigma^{p t\to \gamma X}}{d^2 \kgp d \eta_\gamma} &=
  (\pi R_t^2) \sum_f \frac{\alpha_e \alpha_S N_c^2 q_f^2}{64\pi^4 (N_c^2 - 1)} \int_{\eta_q \eta_p}\int_{\qp \pp \khp \kp}\frac{\varphi_p(Y_p,\khp)}{\khp^2}\\
  &\times \tilde{\calN}_{t,Y_t}(\kp)\tilde{\calN}_{t,Y_t}(\kgp + \qp + \pp - \khp-\kp)\Theta^{gg\to q\bar{q} \gamma}(\kp,\khp)
\end{split}
\label{eq:cs}
\ee
where the $\tilde{\calN}_{t,Y_t}(\kp)$ is the target gluon fundamental dipole. The cross section in the $k_\perp$-factorization approximation is obtained through a replacement
\be
\tilde{\calN}_{t,Y_t}(\kp)\tilde{\calN}_{t,Y_t}(\kAp-\kp) \to \frac{1}{2}(2\pi)^2\left[\delta(\kp) + \delta^{(2)}(\kAp - \kp)\right] \calN_{t,Y_t}(\kAp)\,.
\ee
In the actual computation we will be using
\be
\varphi_p(Y_p,\khp) = (\pi R_p^2) N_c \khp^2 \mathcal{N}_{p,Y_p}(\khp)/4\alpha_S\,,
\ee
a convenient replacement for a number of reasons ($\mathcal{N}_{p,Y_p}(\khp)$ is the adjoint dipole). By resumming soft gluon radiation from the projectile our result becomes of broader relevance as it picks up part of the contribution that would be present in a dense-dense collision. Furthermore, as now $\khp^2 \sim (Q_S^p)^2$, in addition to $\kAp^2 \sim (Q_S^t)^2$, the saturation scale tempers the infrared divergences that this process would inherit from its collinear counterpart, see e.~g. \cite{Chirilli:2012jd_3}.

\section{Results}

We have performed numerical computation of the inclusive photon cross section in pp by summing up the Born channel with valence quarks only and the NLO channel \eqref{eq:cs}, see \cite{Benic:2018hvb_3} for details. Our results show that the NLO channel gives a significant up to $40\%$ correction to Born cross section at RHIC, dominating by more than a $90\%$ contribution at LHC, see Fig.~\ref{fig:inc0} (left). The ratio of $k_\perp$-factorized cross section to CGC cross section is shown on Fig.~\ref{fig:inc0} (right). 
Shown on Fig.~\ref{fig:inc} is the prediction for the inclusive cross section in pp collisions at $\sqrt{s} = 7$ TeV and $13$ TeV  considering isolated photons with an isolation cut $R = 0.4$.

\begin{figure}[h]
\begin{center}
\includegraphics[width=2.0in]{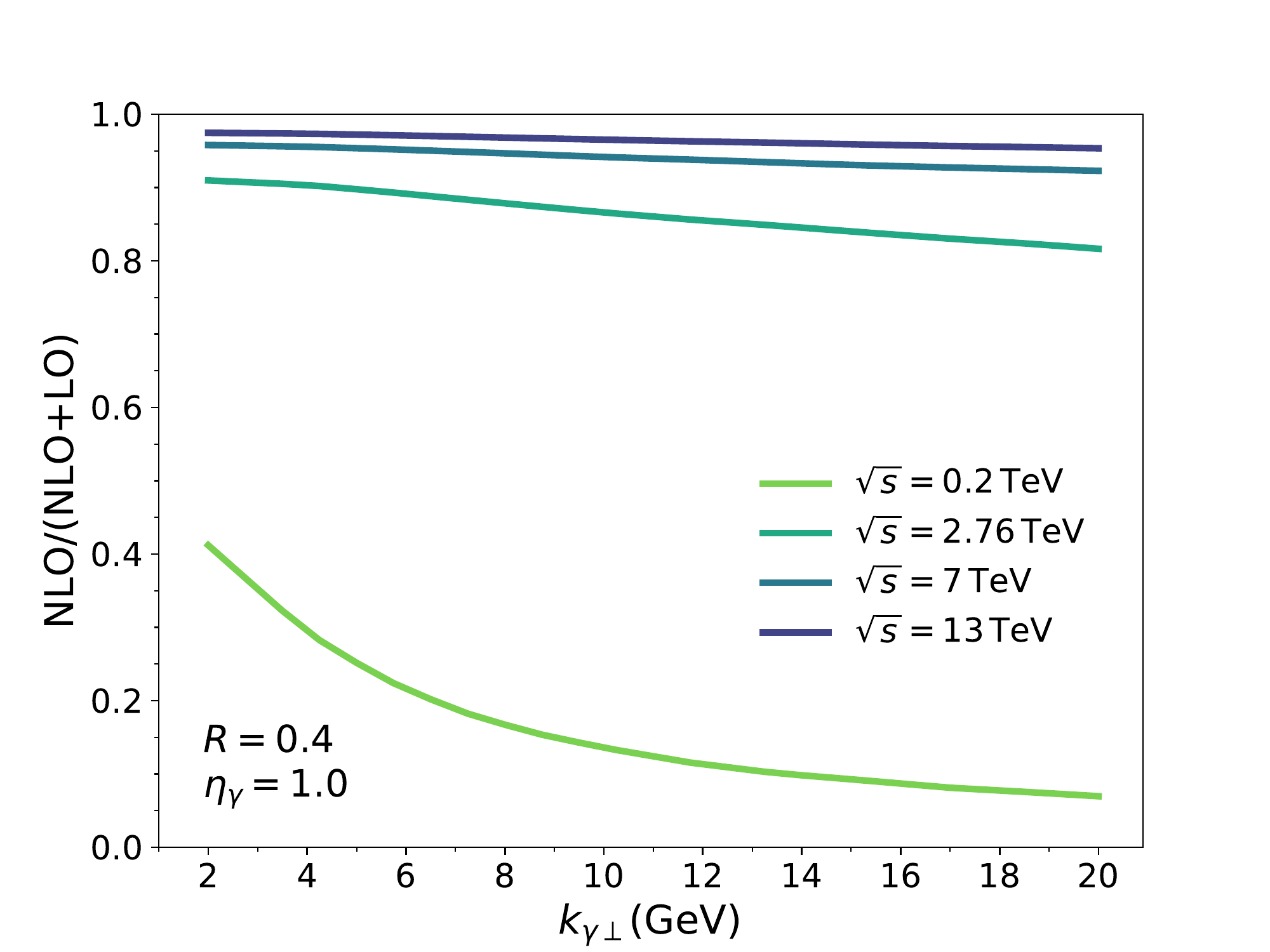}
\includegraphics[width=2.0in]{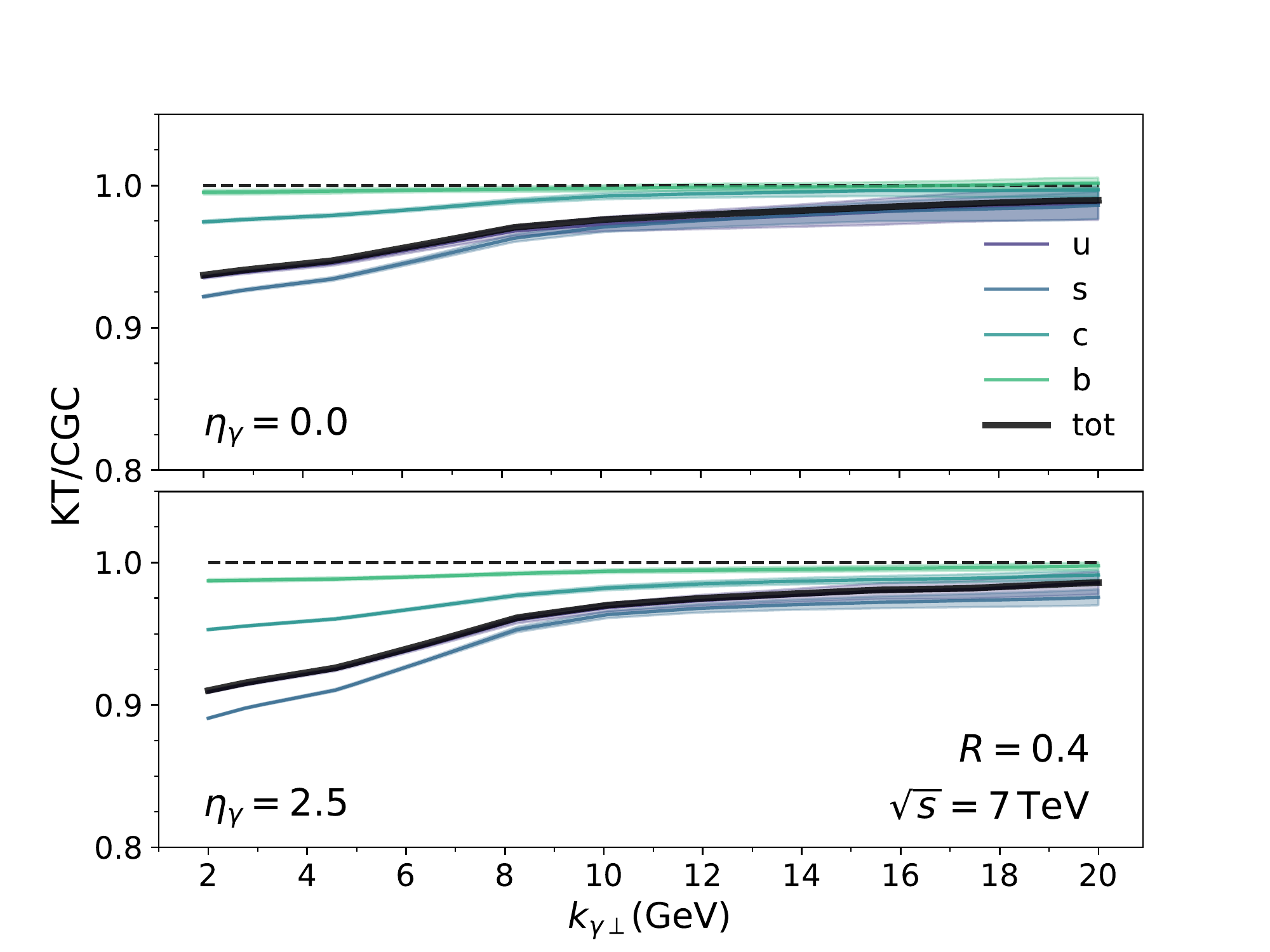}
\end{center}
\caption{Left: Fraction of the NLO $gg \to q\bar{q}\gamma$ channel contribution to $pp \to \gamma X$ for different collision energies. Right: ratio of $k_\perp$-factorized vs CGC cross section per flavor. From \cite{Benic:2018hvb_3}.}
\label{fig:inc0}
\end{figure}

\begin{figure}[h]
\begin{center}
\includegraphics[width=2.0in]{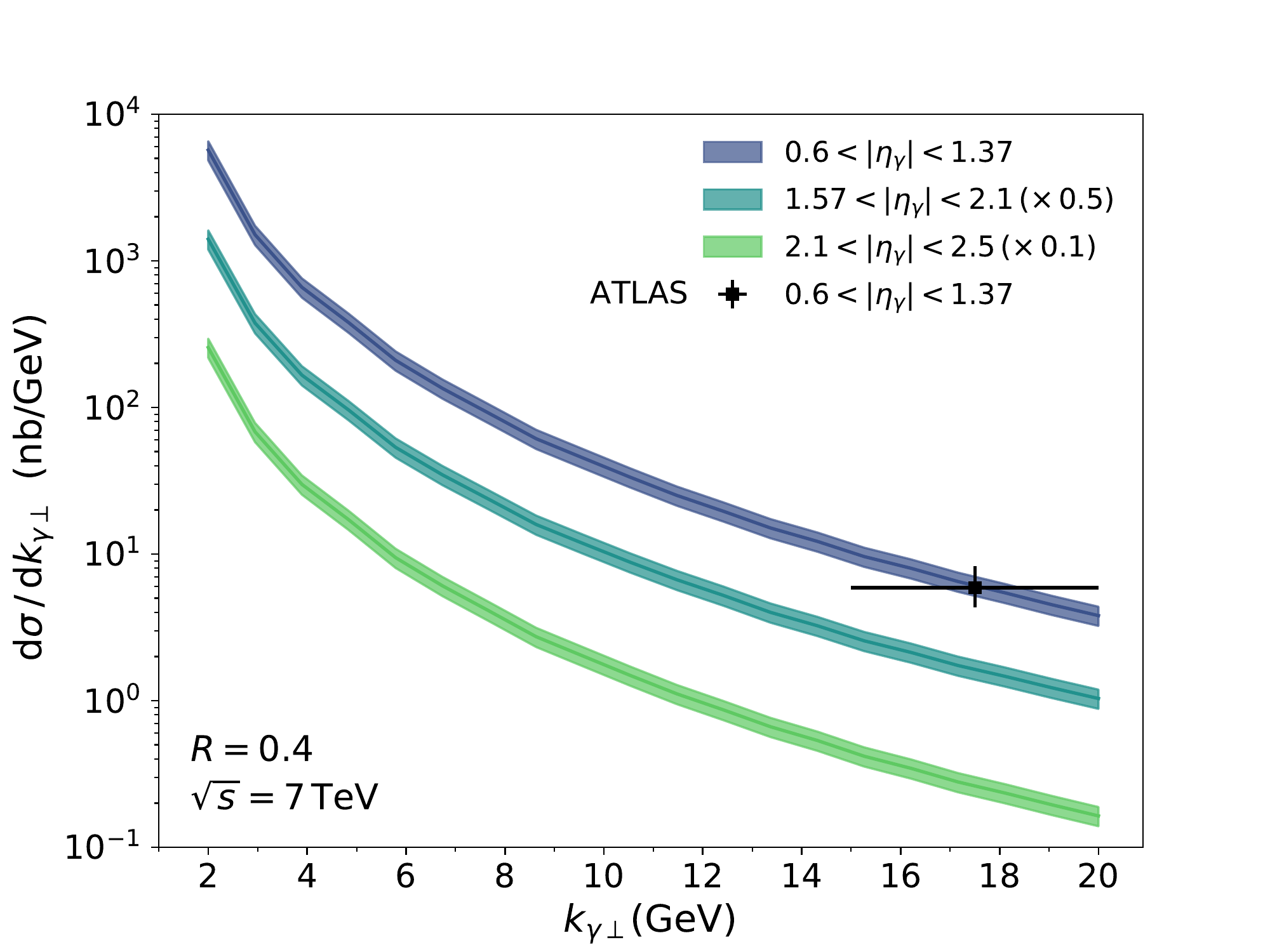}
\includegraphics[width=2.0in]{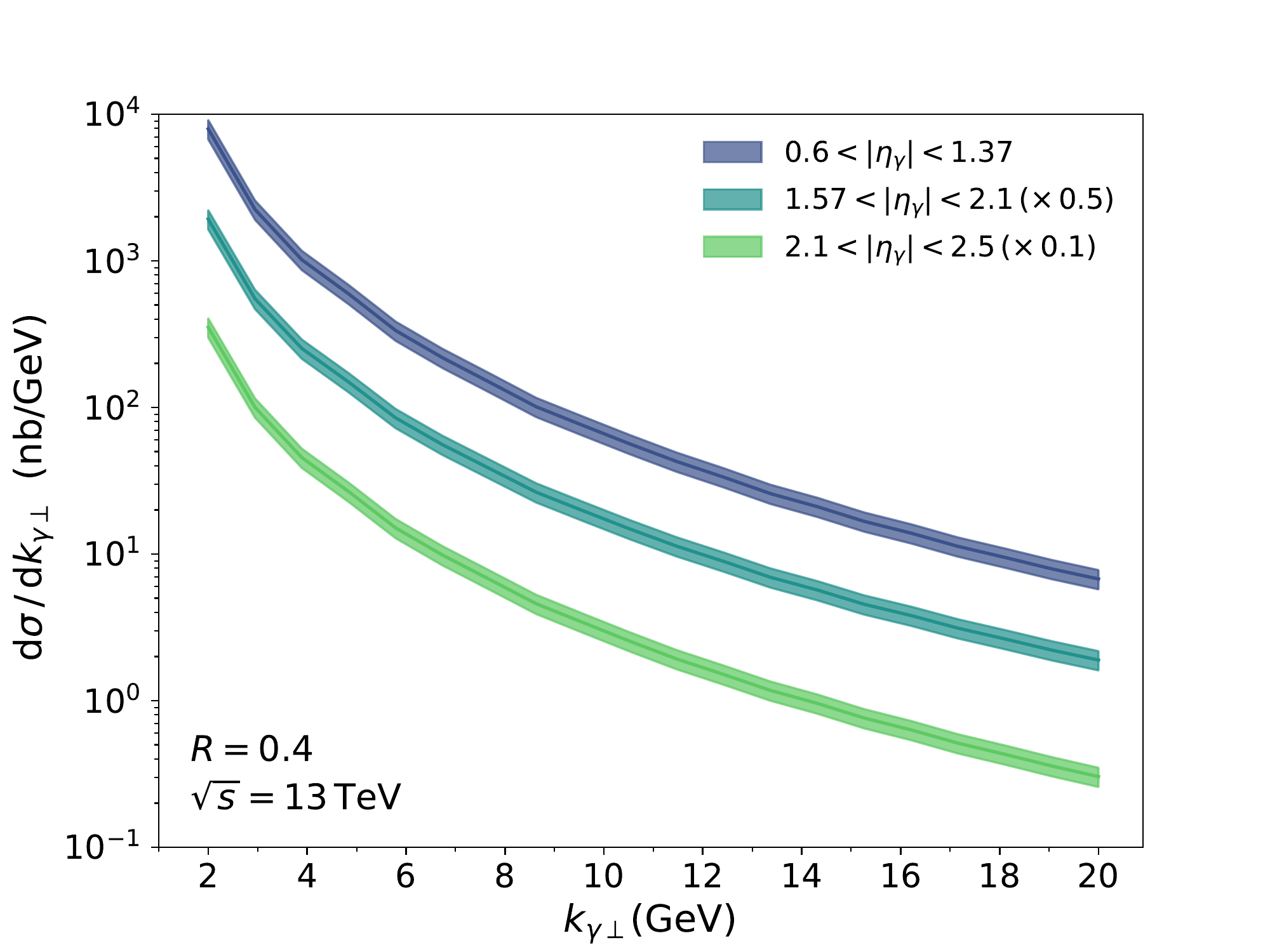}
\end{center}
\caption{Numerical results for inclusive isolated photon production in $pp$ at $\sqrt{s} = 7$ TeV (left) and $\sqrt{s} = 13$ TeV (right) with $R = 0.4$ in several rapidity bins. From \cite{Benic:2018hvb_3}.}
\label{fig:inc}
\end{figure}

\section{Conclusions}

Violations of $k_\perp$ factorization, found to be $10\%$ already in pp, Fig.~\ref{fig:inc0} (right), are expected to be enhanced in pA. This will be important for the nuclear modification factor as a part of our future work. There are promising future upgrades at the LHC \cite{N.Cartiglia:2015gve} aimed at measuring this quantity.

As a final remark, note that photons are almost always produced in association with quarks (a rare exception is \cite{Benic:2016yqt_3}) and so, especially in an inclusive process $h_1 h_2 \to \gamma X$, part of the signature of CGC correlations is always carried away by quarks.
It is therefore interesting to consider photon-jet/hadron correlations. In an almost back-to-back configuration individual particle momenta can be hard and so we can place isolation cuts on hard photons. Furthermore, back-to-back configurations mark the TMD limit in which various gluon TMDs can be explored at small-x.
\section{Acknowledgments}

S.~B.~ is supported by the JSPS
fellowship and JSPS Grant-in-Aid for JSPS fellows 17F17323. S.~B.~ acknowledges HRZZ
Grant No. 8799 at the University of Zagreb for computational resources. K.~F. was supported by JSPS KAKENHI Grant No. 18H01211. R.~V+s work is supported by the U.S. Department of
Energy, Office of Science, Office of Nuclear Physics, under Contracts No. DE-SC0012704 and within the framework
of the TMD Theory Topical Collaboration. This work is part of and supported by the DFG Collaborative Research
Centre ``SFB 1225 (ISOQUANT)''.



 \newpage
%

\renewcommand*{\FigPath}{./WeekVII/04_Watanabe/}

\wstoc{Factorization for quarkonium production in proton-proton and proton-nucleus collisions}{Kazuhiro Watanabe}

\title{Factorization for quarkonium production in proton-proton and proton-nucleus collisions
}

\author{Kazuhiro Watanabe}
\index{auhor}{Watanabe, K.}

\address{Theory Center, Jefferson Laboratory,\\ 
Newport News, Virginia 23606, USA.\\
E-mail: watanabe@jlab.org}

\begin{abstract}
This proceeding article is aimed at drawing attention to selected theoretical issues on quarkonium production models and its factorization in proton-proton and proton-ion nucleus collisions within the color-glass-condensate framework.
\end{abstract}


\bodymatter

\section{Background}\label{sec1}

Quarkonium has been playing a central role in various research grounds to expose rich QCD dynamics, that yields fundamental properties of visible matters and even early universe. Specifically, quarkonium could be an invaluable probe into gluon's 3-dimensional structure inside hadrons\,\cite{Accardi:2012qut_1} and provide us important aspects of an extreme state of matter created in heavy-ion collisions\,\cite{Andronic:2015wma}. More profound insights into quarkonium production mechanisms in elementary processes would make such expectations more compelling.

Nowadays, numerous precise data of quarkonium production in proton-proton ($p+p$) and proton-nucleus ($p+A$) collisions make it possible to investigate quarkonium production mechanism in a vacuum and nuclear medium. Of particular interest is that quarkonium production of low transverse momentum $P_\perp$ can test the color-glass-condensate (CGC) framework, incorporating the so-called gluon saturation effect. As an extensive search for the gluon saturation is a likely prospect in future Electron-Ion-Colliders(EICs), at this time, it should be valuable to recapitulate issues on quarkonium production in the CGC framework. In this article, we will focus on factorization for quarkonium production in $p+p$ and $p+A$ collisions in the CGC framework.

\section{Models of quarkonium production at near threshold}\label{sec2}

\begin{figure}[t]
\centering
\includegraphics[width=0.45\textwidth]{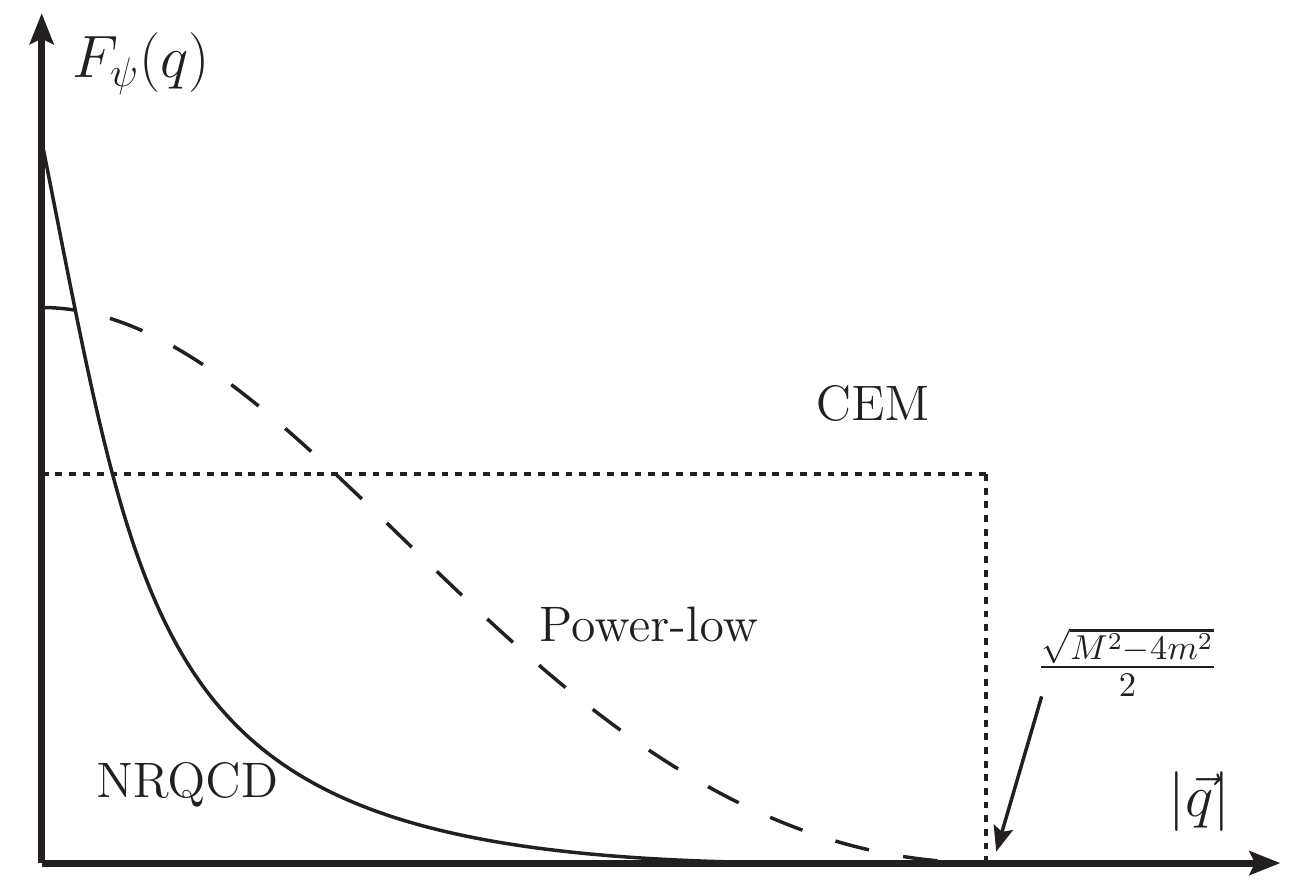}
\includegraphics[width=0.495\textwidth]{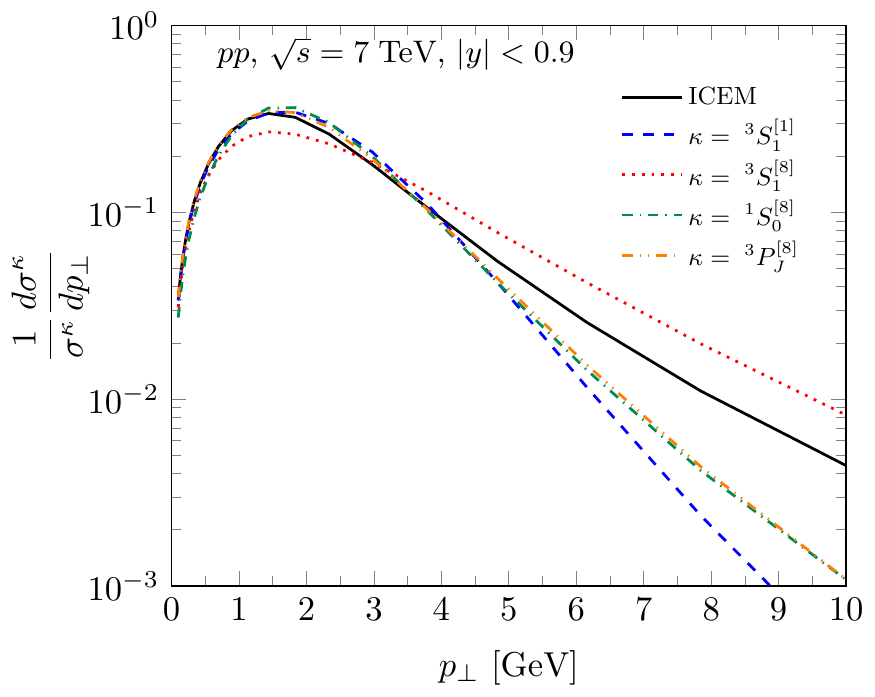}
\caption{(Left): A schematic view of $F_\psi$ for different models. (Right): Normalized differential cross section of $c\bar{c}$ production for each intermediate channel $\kappa$ in the CGC+NRQCD and the CGC+CEM (solid line)\,\cite{Ma:2018bax}.  }
\label{fig1}
\end{figure}

Quarkonium transmutation from a heavy quark pair ($Q\bar{Q}$) takes place at long distance, far from a hard scattering point. Once if we can factorize such a nonperturbative part from a hard scattering part in the amplitude level, we may describe the quarkonium formation with suitable models whatever we want. There are two useful models in this regard; Color Evaporation Model (CEM) is simple, but the spin and color of the $Q\bar{Q}$ are blinded there. 
Nonrelativistic QCD (NRQCD) factorization approach is more sophisticated and allows us to take into account the quark pair's spin and color dependence explicitly. However, as a big issue, there is no proof of the NRQCD factorization at all order in quark velocity $v$ yet.

The above models can be expressed in a unified form of $d\sigma_\psi \propto \sum_\kappa \int dq^2 F_\psi(\kappa,q) \frac{d\sigma_{Q\bar{Q}[\kappa]}}{dq^2}$ with $\kappa$ and $q$ intermediate states and a relative momentum between $Q\bar{Q}$. The transition factor $F_\psi$ includes nonperturbative information on the dynamics of the $Q\bar{Q}$ fragmentation into a bound quarkonium $\psi$. Figure\,\ref{fig1}(left) shows a schematic view of $F_\psi$. It depends on the production model. However, the CGC calculations tell us qualitatively that the model dependence is hardly noticeable for quarkonium production of low $p_\perp$ as displayed in Fig.\,\ref{fig1}(right). This may mean that both of the production models capture an important point about quarkonium production at near-threshold; that is, the color octet $Q\bar{Q}$ are important intermediate states at low $P_\perp$.

\section{Low-$P_\perp$ quarkonium production in $p+p$ collisions}\label{sec3}

Now we restrict our attention into low $P_\perp$ quarkonium production at forward rapidity. The CGC framework is robust in forward $p+p$ and $p+A$ collisions because the gluon saturation scale is enhanced. In the far-forward rapidity region, the quantum interference between the $Q\bar{Q}$ pair and spectators in the direction of the beam proton can be suppressed by powers in $1/P_+$ expansions, even if $P_\perp=\mathcal{O}(\Lambda_\mathrm{QCD})$. So the quarkonium formation part can be effectively factorized from the $Q\bar{Q}$ production part. Nevertheless, in the CGC framework, the gluon distribution function inside the proton is no longer a leading twist one and cannot be factorized clearly from a hard part because the saturation effect should inherit infinite twist effects.

The problem we must ask ourselves is whether the saturation effect can be read from the $P_\perp$ spectrum of quarkonium. This is a delicate issue. If $P_\perp$ is of the order $\mathcal{O}(M)$ with $M$ quarkonium's mass, a large logarithmic correction like $\ln^2(M^2/P_\perp^2)$ (Sudakov factor) becomes significant and comparable with the saturation effect. Figure\,\ref{fig2} (left) presents that the information of the gluon saturation may be covered with the Sudakov effect for $\Upsilon$ production ($M\sim10\,\mathrm{GeV}$). As a valuable benchmark, we show in Fig.\,\ref{fig2} (right) forward $\Upsilon$ production in the collinear factorization + Sudakov factor. These phenomenological results tell us that the search for the saturation would be obscure unless lighter quarkonium, e.g., $J/\psi$, is used as a probe.

\begin{figure}[t]
\centering
\includegraphics[width=0.475\textwidth]{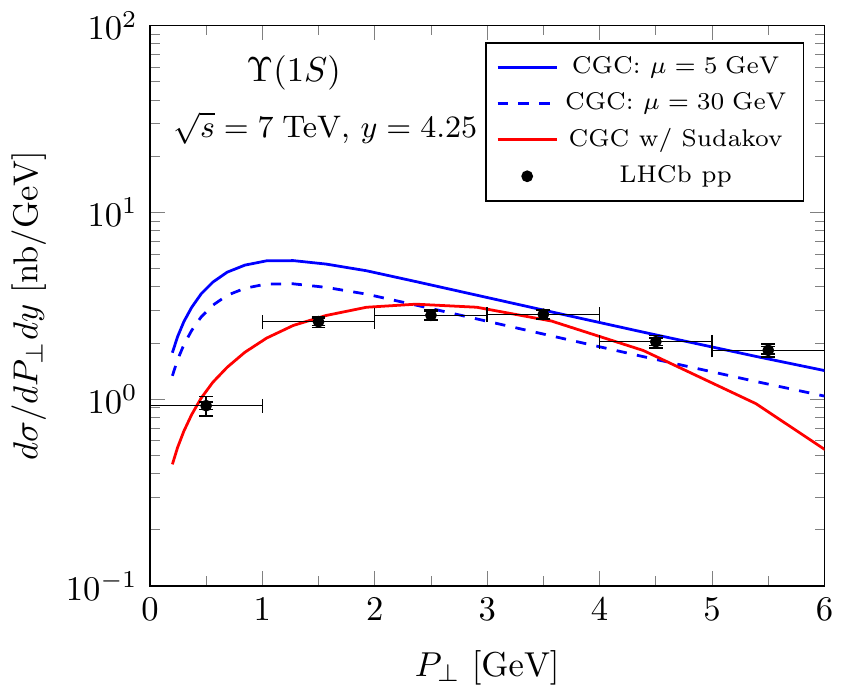}
\includegraphics[width=0.495\textwidth]{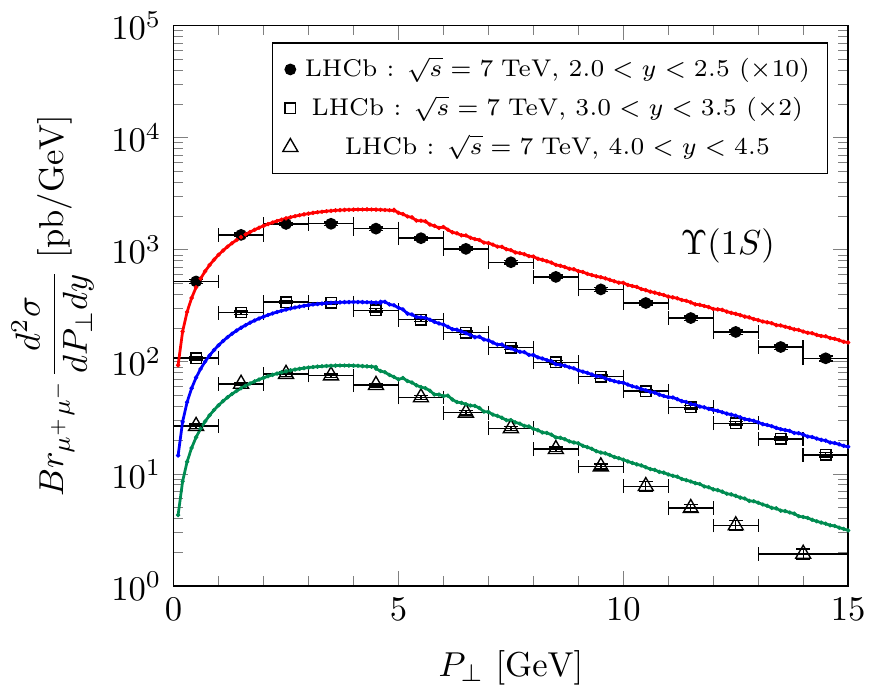}
\caption{Demonstrations of the $\ln^2(M^2/p_\perp^2)$-type resummation for forward $\Upsilon(1S)$ production in $p+p$ collisions in the CGC framework (left)\,\cite{Watanabe:2015yca} and collinear factorization framework (right)\,\cite{Qiu:2017xbx}.
}
\label{fig2}
\end{figure}

\section{Quarkonium production and multi-parton dynamics}\label{sec4}


Some qualitative discussions given above remain unchanged in $p+A$ collisions. However, the interference between the $Q\bar{Q}$ pair and spectator partons in the final state may not be so suppressed in $p+A$ collisions because particle multiplicity is high. Figure\,\ref{fig3} (left) demonstrates nuclear modification factor $R_{pA}$ for $J/\psi$ and $\psi(2S)$ production in the CGC framework with a final state effect (parton comover effect)\,\cite{Ma:2017rsu}. The parton comover effect in $p+A$ collisions can be implemented by implementing $M \to M+\Lambda$. $\Lambda$ denotes the average momentum kick by comover partons moving along with the $Q\bar{Q}$. The comover partons at forward rapidity partially include the spectators. The substantial suppression of $\psi(2S)$ can be attributed to the comover effect on top of the saturation effect. Consequently, the final state factorization for $\psi(2S)$ production breaks due to the final state effect.

High multiplicity events in $p+p$/$p+A$ collisions, which are extreme rare phenomena compared to minimum bias events, provide unique opportunities to test quarkonium production mechanism. It expects that such rare events should be related to rare parton configurations inside the hadron/nucleus. So, the CGC framework can model that picture naturally by taking into account the initial fluctuation effect of the saturation scales\,\cite{Ma:2018bax}. Figure\,\ref{fig3} (right) shows the CGC prediction about normalized $N_{J/\psi}$ vs normalized $N_{ch}$ in $p+p$ collisions. Remarkably, one can see the CGC+CEM framework can describe the LHC data, albeit we must investigate this problem further and clarify whether the final state factorization breaks. Recently, there are very active debates on the existence of the QCD medium at the last stage in small colliding systems. Quarkonium production potentially provides us with a hint to pin down what phenomena happen there.

\section{Concluding remark}\label{sec5}

\begin{figure}[t]
\centering
\includegraphics[width=0.495\textwidth]{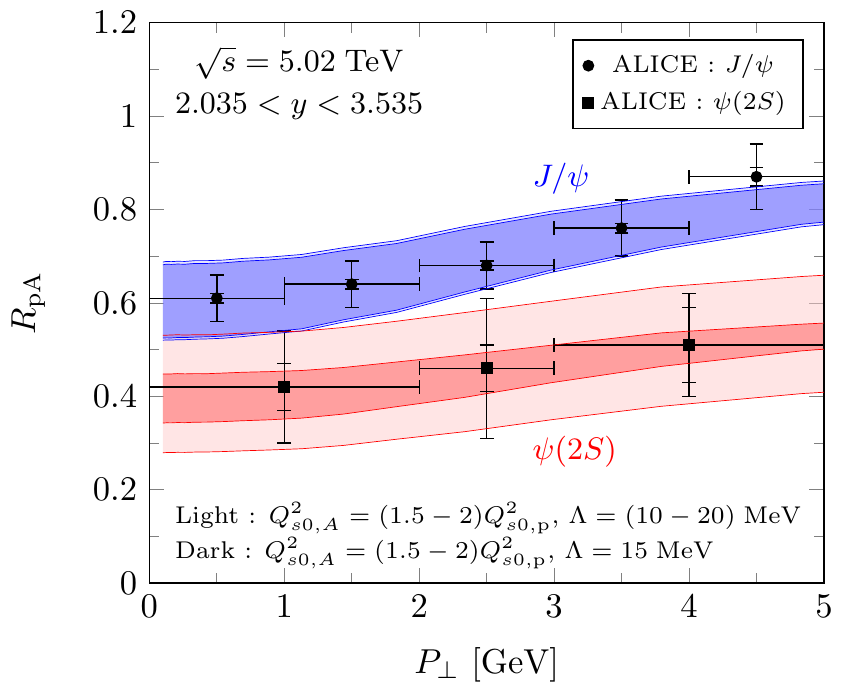}
\includegraphics[width=0.495\textwidth]{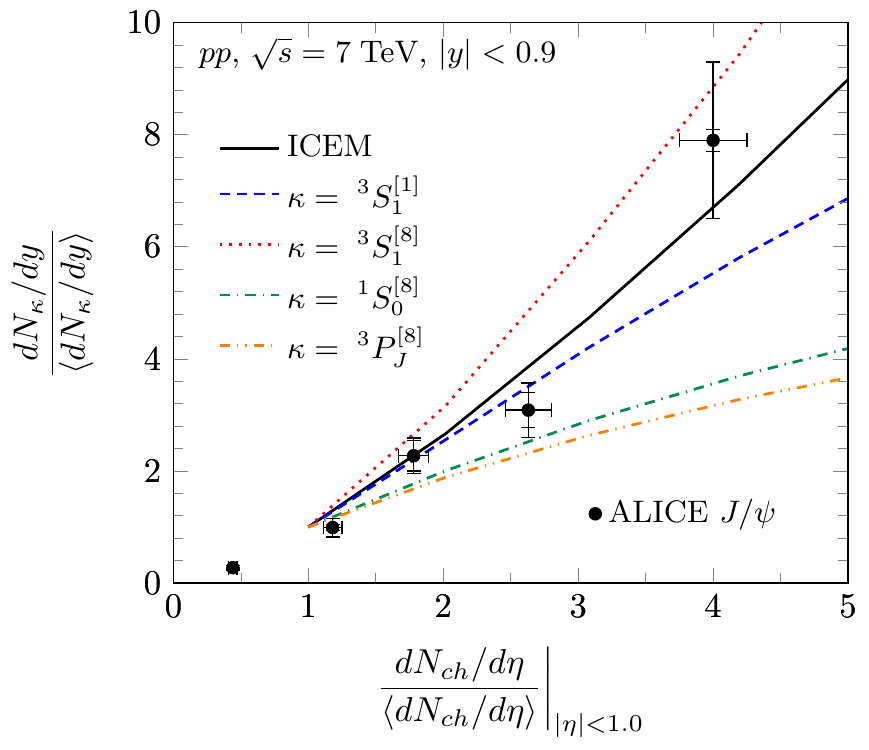}
\caption{(Left): $p_\perp$ dependence of nuclear modification factor for $J/\psi$ and $\psi(2S)$ at the LHC\,\cite{Ma:2017rsu}. (Right): $N_{J/\psi}$ vs $N_{ch}$ in $p+p$ collisions at the LHC\,\cite{Ma:2018bax}.}
\label{fig3}
\end{figure}

It is interesting to extend our analyses to photoproduction of quarkonia in $e+p$ collisions as well as ultraperipheral $p+A$/$A+A$ collisions. Clarifying how much the factorization for quarkonium production in extensive processes is controllable would pave the way for resolving the issues we have seen. Such research would help us to explore the gluon saturation in future EICs.

The author is supported by Jefferson Science Associates, LLC under U.S. DOE Contract \#DE-AC05-06OR23177.





\end{document}